\pdfoutput=1
\documentclass{ucalgthes1}   

\usepackage{graphics}
\usepackage{url}
\usepackage{psfrag}
\usepackage{comment}
\usepackage{longtable}
\usepackage{hyperref}
\usepackage[letterpaper,top=1in, bottom=1.22in, left=1.40in, right=0.850in]{geometry}
\usepackage{fancyhdr}
\usepackage{amsfonts}
\title{Evolution and Regularisation of Vacuum Brill Gravitational Waves
 \\ \bigskip in Spherical Polar Coordinates }
\author{Andrew Masterson}
\thesisyear{2014}
\thesis{thesis}    % the word dissertation can be inserted between {}
\newcommand{\thesistitle}{Evolution and Regularisation of Vacuum Brill Gravitational Waves in Spherical Polar Coordinates}
\monthname{September}
\dept{PHYSICS AND ASTRONOMY}
\degree{DOCTOR OF PHILOSOPHY IN PHYSICS}
\begin{document}
\makethesistitle
\pagenumbering{roman}     % resets page counter to one
\setcounter{page}{1}
\chapter*{UNIVERSITY OF CALGARY \\ FACULTY OF GRADUATE STUDIES}
\thispagestyle{empty}
The undersigned certify that they have read, and recommend
to the Faculty of Graduate Studies for acceptance, a \Thesis\ entitled
``\thesistitle'' submitted by \Author\
in partial fulfillment of the requirements for the degree of
\Degree.\\

%
%                 Substitute  List of Examiners
%
\begin{signing}{Department of Physics and Astronomy}
\signline
Supervisor, Dr.~David Hobill \\
Department of Physics and Astronomy \\
\signline
Dr.~Saurya Das \\
Department of Physics and Astronomy, University of Lethbridge  \\
\signline
Dr.~David Feder \\
Department of Physics and Astronomy \\
\signline
Dr.~Denis Leahy \\
Department of Physics and Astronomy \\
%\newsigncolumn         use this command to start a new column if necessary
\newsigncolumn
\signline
Dr.~Gilad Gour \\
Department of Mathematics and Statistics  \\
\signline
Dr.~W.E. Couch \\
Department of Mathematics and Statistics \\

%\signline
%Dr.~A.~B.~Brown \\
%Department of Academic Computing  \\
\end{signing}
\newpage
\phantomsection
\altchapter{Abstract} 
This thesis presents the mathematical and numerical methods necessary to regularise and evolve Brill Gravitational Waves in spherical polar coordinates.  A Cauchy ADM formulation is used for the time evolution.

We find strong evidence that all IVP formulations of pure vacuum Brill gravitational waves collapse to form singularities/black holes, and we do not observe critical black hole mass scaling phenomena in the IVP parameter phase space that has been characterised in non-vacuum systems.  A theoretical framework to prove this result analytically is presented.  We discuss the meaning of Brill metric variables, the topology of trapped surfaces for various scenarios, and verify other results in the field related to critical values of initial value parameters and black hole formation approaching spatial infinity.

The instability of Minkowski (flat) space under Brill wave and more general perturbations is demonstrated.

The main numerical tools employed to achieve a stable evolution code are (1) derivation of appropriate regularity conditions on the lapse function $\alpha$ and metric function $q$, (2) the move to a 4th order correct discretisation scheme with appropriate boundary conditions, (3) the use of exponential metric terms, (4) an understanding of the right mix of free versus constrained evolution and (5) the development of appropriate numerical techniques for discretisation and differencing to reduce numerical error, along with a characterisation of condition numbers.

\newpage
\phantomsection
\altchapter{Acknowledgements}
Any large project requires the input, consideration and understanding of many different individuals.

Firstly I would like to thank my wife Melanie for her years of patience, support and understanding while I was in the process of excising numerical daemons from this code.  She has always stood by and supported this endeavour, even when it seemed its foolhardiest.

Thanks also goes to David Hobill for his mentorship, insights, and patience during this process.  Neither of us imagined the road being so long when we set out on this journey, but here we are.  This has been a tough nut to crack.

To my daughters Moira, Fionnuala and Oonagh; may you come in time to understand what drives minds to do the things they do.  Deep thought has its own rewards as well.

And lastly, to those friends who did not abandon hope and instead let the journey guide itself, I say thanks for being there when you were needed.  It is appreciated.

\newpage
\phantomsection
\altchapter{List of Variables Used in this Thesis}
This table is to aid readers by providing a summary of the notation used in this thesis.
\begin{table}[h]\begin{center}
\begin{tabular}{|cl|}\hline
Symbol & Usage in this thesis \\ \hline
\multicolumn{2}{|c|}{Dynamic Variables} \\ \hline
$\alpha$ & lapse scalar; 4-metric variable \\
$\beta^i,v_1,v_2$ & shift vector; 4-metric variable \\
$\chi,\Phi$ & shift vector potentials \\
$\gamma_{ij}$ & 3-spatial metric \\
$\phi=\log\psi$ & conformal factor (metric variable) \\
$q$ & determines IVP and gravitational ``wave'' shape/strength (metric variable)\\
$K^i_{\;j}$ & 3-spatial extrinsic curvature \\
$H_{\{a,b,c,d\}}$ & 3-spatial extrinsic curvature \\
$\Psi_n,I,J$ & Weyl scalar (4D curvature invariant) \\
$R_{ijkl}$ & 3-spatial Riemann curvature \\
$R_{ij}$ & 3-spatial Ricci curvature \\
$R$ & Ricci scalar \\
$C_{\alpha\beta\gamma\delta}$ & Weyl curvature \\
\hline \multicolumn{2}{|c|}{Coordinates} \\ \hline
$\eta,\theta,\varphi$ & radial, polar angle and azimuthal angle coordinates \\
$f=f(\eta)=r$ & radial function \\
\hline \multicolumn{2}{|c|}{Miscellaneous} \\ \hline
$A$ & ``Amplitude'' of IVP wave for $q$ \\
$s_0$ & ``Width'' of IVP wave for $q$ \\
$\bar{\theta},\sigma$ & expansion and shear of geodesic congruences \\
$k$ & iteration counter for time evolution ($t_k=k\Delta t$ in general)\\
$D_a$ & 3-spatial covariant derivative operator \\
$K_n$ & $n$-condition number of matrix \\
$\tilde{M}$ & Mass Aspect \\
$i,j$ & radial and angular grid indices respectively \\
${\{\}}_{i},{\{\}}_{j}$ & 3-spatial covariant tensor indices \\
${\{\}}_{\alpha},{\{\}}_{\beta}$ & 4D covariant tensor indices \\
\hline
\end{tabular}\caption{List of variables}
\end{center}\end{table}

\begin{singlespace}
\newpage
\phantomsection
\tableofcontents
\pagestyle{plain}
\newpage
\phantomsection
\listoftables
\pagestyle{plain}
\newpage
\phantomsection
\listoffigures
\pagestyle{plain}
\clearpage
\end{singlespace}
\clearpage          % otherwise tables will be numbered wrong

\pagenumbering{arabic}
\fancyhead[RO,LE]{\thepage}
\fancyfoot{} 
\chapter{Introduction}\label{chap:intro}
\bigskip
Numerical Relativity presents the opportunity to mesh physics, mathematics and numerical analysis for the purpose of investigating one of the more famous results from 20th century physics: Einstein's field equations for gravitation.

Einstein's equations are a mathematical formulation for some physical phenomena that cannot be described or explained with Newtonian gravity.  In the weak field limit Einstein's equations reduce to familiar Newtonian gravity, and in the stronger field limit we encounter many interesting new phenomena.  Some of the phenomena that are encountered as a result of Einstein's equations include black holes, gravitational lensing\footnote{Observed in many astrophysical systems.}, orbital precession of gravitationally bound astrophysical objects\footnote{First observed in Mercury's perihelion shift, later in binary pulsars.}, gravitational redshift and gravitational waves.

One major distinction between the results of Newtonian gravity and Einstein's gravity (often called General Relativity or GR) is that Newtonian gravity is completely dependent on the matter distribution being studied.  Mass is present in the form of matter (dust, fluids, neutrons, plasma, etc.) and it \emph{creates} a gravitational field.  In General Relativity, it is possible to have gravity with no matter present as gravity is linked to the geometry of spacetime (which is always present) and any matter distribution that may or may not be present.

While a mass distribution in a General Relativity spacetime will still create gravitational effects, the new ``relativistic'' phenomena mentioned above do not require any mass present in the spacetime to manifest themselves.  A black hole can exist with no matter or non-gravitational fields present and gravitational waves only require a curved Riemannian geometry to propagate.  This presence of \emph{vacuum dynamics} in GR is a crucial distinction between Newtonian gravity and GR as Newtonian gravity has nothing to say when there is no matter present.  Pure vacuum dynamics is not a well studied area within the whole realm of GR\footnote{Excepting perhaps vacuum Bianchi cosmologies.}; in the numerical world, this can probably be attributed to the difficulty of ensuring you have zero energy-momentum present as any finite error is infinitely far from zero. This makes discussions around error measures very difficult.

Gravitational wave spacetimes therefore present an excellent opportunity to explore a key area where GR provides dynamics and Newtonian theory does not: the vacuum.

To do so, in a general sense, we wish to formulate an initial value problem and evolution in the same way that hydrodynamics, electrodynamics, reservoir engineering, weather simulations, flight simulators, etc. examine physical systems.  We must specify an initial configuration for the system in question, usually parametrised by a few key values that completely describe the initial configuration.  We then use a time evolution algorithm to propagate our solution forward or backwards in time\footnote{GR adds a huge layer of complexity here - our time coordinate and slicing methods are freely specifiable, as is our choice of coordinates really - we don't have to evolve through time - we can evolve through null cones or many other methods.  The principle of general covariance ensures that they are all physically equivalent.} and examine the resulting physical system.

In 1959, Brill published his seminal paper \cite{Brill} on gravitational waves detailing one set of conditions under which the \emph{vacuum} initial value problem (IVP) formulation of the Einstein equations is well posed; i.e. the ``mass/energy'' of the system is positive definite.  This is a remarkable result in that it proves that pure gravitational curvature, without any matter, electromagnetic radiation, etc. can have positive energy.

Since then, there has been much effort put into determining what the gravitational ``waves'' as detailed by Brill and others mean, how they are produced, if it is possible that they carry ``energy'' and can collapse to form black holes.  Further work has been done to determine the physical properties of gravitational waves produced by astrophysical systems so that we can detect them with some of the large detection arrays\footnote{These are large interferometric arrays looking for deviations on the order of $<10^{-18}$m - so thermal noise, radiation pressure, trucks on the highway, etc. all pose significant challenges.} like LIGO, VIRGO, GEO 600, TAMA 300 and LISA, with most of the recent effort being done on binary black hole inspiral/coalescence.

In addition, many astrophysical matter collapse simulations have generated gravitational wave signatures, and the fundamental study of gravitational waves decoupled from matter distributions should provide insight into what they are and how they behave\footnote{Analogous to studying electromagnetic waves in a vacuum, versus those coupled to charged sources.}.

As the underlying Einstein field equations that govern the evolution (and IVP) of gravitational waves are impossible to solve analytically except in the most simple cases, we are forced to employ a numerical analysis framework to see what the solutions to the equations look like, and what they do.  To this end, we will be investigating vacuum Brill gravitational waves and associated time dependent properties using numerical computational methods.

The numerical results presented in this thesis allow for an exploration of the physical properties of Brill wave spacetimes, including (i) trapped surface/apparent horizon structure, (ii) location of singularities, (iii) curvature behaviour close to singularity formation, (iv) asymptotic structure of the spacetime and (v) curvature propagation/behaviour in the highly non-linear interior region.

The analytic results presented in this thesis based on the Raychaudhuri equations show that all Brill waves spacetimes will encounter singularities\footnote{But that is all those equations tell us - numerical simulation is required to explore the physics more completely.}. This analytic result supports the universal numerical discovery of singularities in this thesis, which is in contrast to the vacuum Brill wave literature.  It also indicates that previous numerical results in the field\footnote{Specific references for vacuum scenarios are
% \cite{Alcubierre,alcubierre:3dbrill,eppley,garfinkle,miyama,RinneCQG,sorkin,sorkin:code,Thornburg:cartesian} and are also 
discussed in section \ref{sec:brillcollapse_historic}.} are either incomplete or incorrect.  As such, the numerical and mathematical techniques presented in this thesis can be valuable for numerical relativity groups as these techniques have been successfully employed to discover a previously unknown phenomena.  It seems that numerical dissipation is frequently used to mask or smooth out regularity problems and the mathematical regularity conditions derived herein (especially for the lapse function $\alpha$) are therefore key conditions that can be employed where appropriate instead of \emph{ad hoc} methods.

This also indicates that we can explore previously unseen physics with the numerical code and simulations presented in this thesis.

The analytic results from the Raychaudhuri equations further generalise to a larger class of spacetimes, and have important implications for the (in)stability of flat space (and more general spacetimes), which is another new surprising result.

For the numerical simulation we choose a spherical polar coordinate system as these coordinates provide a natural method of matching radially radiative outer boundaries in the asymptotic limit and as the steady-state solution for non-rotating / charged black holes is the spherically symmetric Schwarzschild solution.  Far out from the strongly non-linear regime of gravitational interaction we would expect radially propagating information, so having spherical polar coordinates allows the use of more intuitive outer boundary conditions.  This is especially important for elliptic equation solvers, as their very nature means that information can propagate across the grid instantaneously due to the linked nature of all grid points.  We also know that the boundaries are a key determining factor in any solution to an elliptic partial differential equation as the interior region is determined very strongly by even small fluctuations in the boundaries for many systems.  Therefore it is very important to have rigorous boundary conditions in place for all elliptic equations, and spherical polar coordinates offer the best candidate for this\footnote{\cite{garfinkle}, for example, discusses the difficulties associated with outer boundary conditions in cylindrical coordinates.}.

We will also split our spacetime into $3$ spatial coordinates and $1$ time coordinate, which is commonly called an ``$3+1$'' splitting.  This involves creating a spatial hypersurface $\Sigma$ and a time-like normal vector to the hypersurface.

\section{Outline}
The remainder of this thesis is organised as follows to discuss the various mathematical, physical and numerical choices that must be explored to arrive at a final working code.

Chapter \ref{chap:mathrel} provides a mathematical framework for GR, discusses the details of how one formulates a Cauchy problem approach to solve Einstein's field equations, and gives an overview of the Brill Gravitational Wave problem.

Chapter \ref{chap:nummethod} gives an overview of some methods for converting continuous equations and variables to discrete ones for use in computer modeling.

Chapter \ref{chap:coordgauge} discusses coordinate, metric and gauge choices used to simplify the 2+1 equations into a more tractable format.

Chapter \ref{chap:changes}'s focus is on the variety of numerical methods used to discretise the coupled non-linear PDEs for use over a grid, as well as boundary conditions, fitting methods and regularisation techniques.  We also discuss the lessons learned in the course of trying several different numerical methods to solve this system of coupled, non-linear PDEs.  Given some historical constraints on computing power it has been tremendously beneficial to revisit some of the basic tenets with modern computing power now at our disposal.

Chapter \ref{chap:numcode} discusses the structure of the numerical code that was used to investigate the axisymmetric Brill Wave problem, as well as various techniques employed.

Chapter \ref{chap:results} presents the results of the running the 2+1 code in various situations and presents a discussion of those results, as well as theoretical backing for observations.

Chapter \ref{chap:errs} presents error analysis, convergence tests and an examination of alternate code parameters and evolution schemes to justify the numerical results.

Appendix \ref{appendix:maxima} discusses the Maxima code used to generate the symbolic equations for discretisation.

Appendix \ref{appendix:eqns} presents some additional equations used for various checks in the code.

Appendix \ref{appendix:misc_algorithms} contains some miscellaneous algorithms and numerical methods that were employed throughout the thesis.

Appendix \ref{chap:testgridcoord} presents the mathematical framework for the 1+1 formalism and will cover some work done on alternate gridding and coordinate systems, as well as results associated with those investigations.

\chapter{Mathematical Basis of General Relativity}\label{chap:mathrel}
\bigskip

\section{Differential Geometry and Fundamental Forms}
Fundamental to an understanding of GR is a foundation in differential geometry, as this is the mathematical language that Einstein's field equations are cast in.  Einstein himself spent several years studying differential geometry to allow him to formulate a theory that incorporated general covariance - the ability to transform to alternate coordinate systems and have the same physically invariant laws.

We first introduce the concept of a metric, which is used to measure distances and angles on a differentiable Riemannian manifold.  The metric is also known as the First Fundamental Form in differential geometry.

In Riemannian geometry we introduce a metric tensor, $g_{\alpha\beta}$, that measures invariant distances on a manifold and gives the Riemannian line element $ds$ by:
$${ds}^2=\sum_{\alpha,\beta=1}^N g_{\alpha\beta}{dx}^{\alpha}{dx}^{\beta}$$
where the ${dx}^{\xi}$ are the coordinate differentials associated with the coordinate system we are using, $N$ is the number of dimensions and our metric $g_{\alpha\beta}=g_{\alpha\beta}(x^\xi)$.  One important property of $ds$ is that it is a scalar and therefore \emph{invariant}, so it produces the same result independent of the coordinate system we are using.

For example, on a simple 2D Euclidean manifold in familiar $(x,y)$ Cartesian coordinates, this reduces to the familiar Pythagorean formula
$${ds}^2={dx}^2+{dy}^2$$
where our metric is given by
\begin{equation}
g_{\alpha\beta} = \left( \begin{array}{cc}
1 & 0 \\
0 & 1\\
\end{array} \right),\end{equation}

The second fundamental form in differential geometry combines the notion of intrinsic and extrinsic curvature, where the extrinsic curvature is a measure of a manifold's curvature in an embedding manifold, and the intrinsic curvature measures the deviation from Euclidean geometry as measured on the manifold.

To envision the meaning of these two types of curvature, imagine a flat piece of paper with a triangle drawn on it.  This piece of paper has no intrinsic or extrinsic curvature.  If we fold the paper into a cylinder, it has now acquired extrinsic curvature, but no intrinsic curvature.  The paper is curved in a higher-dimensional embedding manifold (3D space), but the triangle's internal angles still sum to $180^\circ$, and parallel lines stay parallel.  As there is no effect that is measurable on the paper's surface to indicate that it is curved, its intrinsic curvature is zero.

The surface of a sphere, however, has intrinsic curvature as ``parallel'' lines converge, which is a condition that can be measured on the manifold itself.  The sphere also has extrinsic curvature in a 3D embedding Euclidean manifold.

%  A simple example of extrinsic curvature would be a circle of radius $R$ on a flat plane, which would have the curvature
%$$\kappa=\frac{1}{R}$$
%and also produces the familiar result that for a flat line (radius of curvature $R \rightarrow \infty$)
%$$\kappa=0$$

The ADM\footnote{Short for \emph{Arnowitt, Deser and Misner} \cite{ADM}.} formulation of the Cauchy IVP for General Relativity makes heavy use of the extrinsic curvature, so we mention it for future consideration.

\subsection{Tensors Fields and the Algebra of Tensor Components}
Tensors are central to the formulation of General Relativity, so we provide a definition:  a tensor represents a geometric or physical object and obeys the rules of multi-linear algebra.  Often a tensor is represented by a multi-dimensional array whose components transform in a particular manner under a change of basis.
A corollary of this is that a tensor is independent of which coordinate system it is represented in.  The tensor ``type'' indicates how many covariant ($p$) and contravariant ($s$) indices it has in the form $(s,p)$, and the rank of a tensor is defined as $p+s$.

Scalars are the simplest form of tensors (type $(0,0)$, rank $0$) as they are single-valued at a point $P$ on the manifold, and vectors are the simplest non-trivial example of tensors (type $(1,0)$ or $(0,1)$, both rank $1$).
Given $x^i$ and ${x'}^{i}$ which are coordinates that are defined on the same manifold\footnote{For example Cartesian $x^i=(x,y,z)$ and spherical polar ${x'}^i=(r,\theta,\phi)$ coordinates on a 3D manifold.} that have a defined transformation $x^i \mapsto {x'}^{i}$ and the inverse transformation exists we can discuss tensor transformation rules.

A \emph{covariant} vector (or type $(0,1)$ tensor) \emph{component} is one that transforms from unprimed to primed coordinates using the transformation (here we represent the covariant components with a subscripted index):
$$X'_a=\frac{\partial x^b}{\partial {x'}^{a}} X_b$$
Where we have used the \emph{Einstein summation notation} to simplify the display of the equations.  Einstein summation notation implies that all repeated indices are summed over the entire range of the indices (i.e. the total number of coordinates). For example, if the dimension of the manifold is $N$:
$$g^{ij}T^{a}_{\ jid} \equiv \sum_{i,j=1}^N g^{ij}T^{a}_{\ jid} $$
and from this point forward we will use Einstein summation notation unless otherwise noted.

A \emph{contravariant} vector (or type $(1,0)$ tensor) \emph{component} is one that transforms from unprimed to primed coordinates using the transformation (here we represent the contravariant components with a superscripted index):
$${X'}^a=\frac{\partial x^{'a}}{\partial x^b} X^b$$
Higher rank tensors can have a mix of covariant and contravariant components and transform like
$${X'}^{a\ldots}_{b\ldots}=\frac{\partial x^{'a}}{\partial x^c}\frac{\partial x^d}{\partial {x'}^{b}} X^{c\ldots}_{d\ldots}$$

The metric has the special property that
$$g^{ab}g_{bc}=\delta^a_{\ c}$$
where $\delta^a_{\ c}$ is the Kronecker delta
\begin{eqnarray}\delta^a_{\ c} & = & \left\{ \begin{array}{cc} 1 & a=c \\ 0 & a \ne c \end{array} \right. \nonumber \end{eqnarray}
From which it follows that one can convert a contravariant tensorial component into a covariant one by using the metric in the following manner (called ``lowering'' a component)
$$X_a=g_{ab}X^b$$
and similarly one can convert a covariant tensorial component into a contravariant one by using the following transformation (called ``raising'' a component)
$$X^a=g^{ab}X_b$$

We define a \emph{contraction} on a tensor by performing a summation over a pair of contravariant and covariant indices.  For example: letting $T^{a\ \ }_{\ bc}$ be a tensor, we can define the contraction
$$g^{ij}T^{a\ \ }_{\ ji}  = T^{ai\ }_{\ \ i} = T^{a} $$
and it is important to keep track of which indices are which.
 
\subsection{Differentiation on manifolds}\label{sec:covdiff}
In general Riemannian and pseudo-Riemannian geometries, partial derivatives of tensors do not transform as higher rank tensors, so we now investigate the nature of differentiation of tensors on these manifolds.

Consider the vector $X_a$ at two points on the manifold $P$ and $Q$ that we wish to perform a differentiation operation on with respect to a set of coordinates.  On a Euclidean manifold it is sufficient to ``transport'' the vector $X_a(P)$ to the point $Q$ and find the difference between components in question\footnote{And dividing by the difference in coordinate values and taking the limit as that difference goes to zero of course.}.

When a manifold is non-Euclidean, this transporting of $X_a(P)$ to the point $Q$ introduces a shift in the vector that causes differencing between them to be non-tensorial.  (i.e. it is not invariant under the coordinate transformations listed above)

In General Relativity physical laws must remain invariant under general coordinate transformations (the Principle of General Covariance).  Thus we are dealing with tensorial quantities and we need a differential operator for tensors that transforms like a tensor under coordinate transformations.

The way to remedy this problem is to define the \emph{covariant derivative} in such a way that it preserves the tensorial nature of the equations.  As part of this we need a connection coefficient $\Gamma^a_{bc}$ which describes how one ``parallel transports'' a tensor from point $Q$ to point $P$ on the manifold.  This leads to the definition of the covariant derivative of a contravariant vector as:
\begin{equation}\label{eqn:covderivconvec}\nabla_cX^a = \partial_cX^a + \Gamma^a_{bc}X^b\end{equation}
Where we use the notation
$$\partial_c \equiv \frac{\partial}{\partial x^c}$$
The covariant derivative $\nabla_cX^a$ in (\ref{eqn:covderivconvec}) itself transforms like a second-rank \emph{tensor}, as desired.  This generalises to a covariant tensor $X_a$ as
$$\nabla_cX_a = \partial_cX_a - \Gamma^b_{ac}X_b$$
and to a mixed second-rank tensor $X^i_j$ as
$$\nabla_cX^i_j = \partial_cX^i_j + \Gamma^i_{ac}X^a_j - \Gamma^a_{jc}X^i_a $$
with the obvious extension to a general mixed, rank-$n$ tensor
\begin{equation}\nabla_dX^{abc\ldots}_{\ \ \ \ ijk\ldots} =
 \partial_dX^{abc\ldots}_{\ \ \ \ ijk\ldots}
+ \Gamma^a_{ed}X^{ebc\ldots}_{\ \ \ \ ijk\ldots}
+ \Gamma^b_{ed}X^{aec\ldots}_{\ \ \ \ ijk\ldots} + \ldots
- \Gamma^e_{id}X^{abc\ldots}_{\ \ \ \ ejk\ldots}
- \Gamma^e_{jd}X^{abc\ldots}_{\ \ \ \ iek\ldots} - \ldots
\nonumber \end{equation}
The covariant derivative of the rank $n$ tensor $X^{abc\ldots}_{\ \ \ \ ijk\ldots}$ itself transforms like a rank $n+1$ tensor.

The covariant derivative of a scalar is just the partial derivative, i.e.
$$\nabla_c\alpha = \partial_c\alpha$$
We restrict ourselves to the consideration of torsion-free manifolds, which means mathematically that our connection coefficients are symmetric in their covariant components:
$$\Gamma^a_{bc}=\Gamma^a_{cb}$$
and means geometrically that the order in which we apply coordinate ``transports'' for differentiation is unimportant.

For Riemannian and pseudo-Riemannian geometries the connection coefficients (also called Christoffel symbols of the second kind) are related to the metric by:
\begin{equation}\Gamma^a_{bc}=\frac{1}{2} g^{ad}(\partial_cg_{bd}+\partial_bg_{cd}-\partial_dg_{bc})\label{eqn:christoffel}\end{equation}
from which it follows that
$$\nabla_c g^{ab} = \nabla_c g_{ab} = 0$$

\subsection{Lie Derivatives}
We can define the derivative of a tensor field along a vector field $X^a$ (also called the Lie Derivative) in the following manner:
\begin{eqnarray}
\pounds_Xf & \equiv & X^a\nabla_a f \nonumber \\ \mbox{}
\pounds_X S^a & \equiv & X^b \nabla_b S^a - S^b \nabla_b X^a \nonumber \\ \mbox{}
\pounds_X S_a & \equiv & X^b \nabla_b S_a + S_b \nabla_a X^b \nonumber \\ \mbox{}
\pounds_X T^a_b & \equiv & X^c \nabla_c T^a_b + T^a_c \nabla_b X^c - T^c_b \nabla_c X^a \nonumber
\end{eqnarray}
Which is used in the derivation of the ADM equations.

\subsection{Curvature}
One property of interest of covariant derivatives is that they are generally non-commutative, unlike partial derivatives.  If we calculate the commutator of covariant differentiation on a vector $X^a$, we find that
$$\nabla_{\left[c\right.} \nabla_{\left.d\right]} X^a = \nabla_c \nabla_d X^a - \nabla_d \nabla_c X^a= \frac{1}{2}R^a_{\ bcd}X^b $$
where $R^a_{\ bcd}$ is the Riemann tensor, which is defined by
\begin{equation}R^a_{\ bcd}=\partial_c \Gamma^a_{bd} - \partial_d \Gamma^a_{bc} + \Gamma^e_{bd} \Gamma^a_{ec} - \Gamma^e_{bc} \Gamma^a_{ed}\label{eqn:riemann}\end{equation}
Using this equation and (\ref{eqn:christoffel}) we see that the Riemann tensor is non-linear in the metric and its first and second derivatives.  This is the source of non-linearity in general relativity, and arises from mathematical/geometrical considerations only (i.e. is independent of any physical source terms).

The Ricci Tensor (also known as the intrinsic curvature) is defined by a contraction on the Riemann tensor
\begin{equation}R_{ab}=R^c_{\ acb}\label{eqn:riccitensor}\end{equation}
and the Ricci (or curvature) Scalar is defined by one further contraction
\begin{equation}R=g^{ab}R_{ab} = R^a_{\ a}\label{eqn:ricciscalar}\end{equation}
The Einstein Tensor $G_{ab}$ can now be defined from these quantities
\begin{equation}G_{ab}=R_{ab} - \frac{1}{2} g_{ab} R \label{eqn:einsteintensor}\end{equation}
One property of the Einstein tensor is that it satisfies
\begin{equation} \nabla_a G_b^a \equiv 0 \end{equation}
which are referred to as the contracted Bianchi identities, which will be useful to us later.  Another feature of the Einstein tensor is that it is symmetric, i.e.
\begin{equation}G_{ab}=G_{ba}\label{eqn:einsteintsymm}\end{equation}

\subsection{Geodesics and the Raychaudhuri equations}
Geodesics represent the generalisation of Euclidean ``straight lines'' in Riemannian geometry as they minimize the distance between two points on a curved manifold.  As test particles in General Relativity follow geodesics, a study of the properties of geodesics is key to understanding the physics of curved spacetimes.  Notably, a curved manifold causes the deviation $\eta^\alpha$ between neighbouring geodesics to be altered due to the Riemannian curvature, and this relationship can be expressed mathematically via the equation of geodesic deviation:
\begin{equation}\label{eqn:raychearly}\frac{D^2\eta^\alpha}{D\tau^2} - R^\alpha_{\beta\gamma\delta}\xi^\beta \xi^\gamma \eta^\delta=0\end{equation}
where $\xi^\alpha$ is the tangent (velocity) vector to the geodesic and $\tau$ is the proper time along the geodesic.

By similarly defining
$$B_{\alpha\beta}=\nabla_\beta \xi_\alpha$$
we find that the geodesic equation (\ref{eqn:raychearly}) becomes\footnote{See \cite{Blau}, chapter 11 for a detailed derivation and discussion of this result.}:
$$\frac{D B^\alpha_\beta}{d\tau} + B^{\alpha}_{\gamma}B^{\gamma}_{\beta} = R^\alpha_{\gamma\delta\beta} \xi^\gamma \xi^\delta$$
Taking the trace of this equation we arrive at the Raychaudhuri equation for the expansion $\bar{\theta}$ of a timelike geodesic congruence in 4 dimensions:
\begin{equation}\label{eqn:raychaudhuri_early} \frac{d\bar{\theta}}{d\tau} = -\frac{1}{3}\bar{\theta}^2 - \sigma^2 + \omega^2 - R_{\alpha\beta}\xi^\alpha \xi^\beta \end{equation}
where $\sigma^2=\sigma^{\alpha\beta}\sigma_{\alpha\beta}$ is the shear and $\omega^2=\omega^{\alpha\beta}\omega_{\alpha\beta}$ is the twist\footnote{See \cite{kar_raych} for a geometrical discussion of these variables, as they do indeed represent the local expansion, shear and twist of a geodesic congruence.}.

Equation (\ref{eqn:raychaudhuri_early}) is important as it is frequently used in the analysis of spacetimes to prove that they are singular, i.e. geodesically incomplete, via the singularity theorems devised by Hawking, Ellis, Penrose and Wald.  One can similarly derive equations for the evolution of the shear and twist, which will be discussed later.

\section{Notation and Conventions}
To establish a common ground for communication let us lay out the general conventions and notations that are used throughout the \emph{remainder} of this thesis.

In GR $N=4$ (i.e. we have four dimensions), and the the geometry is pseudo-Riemannian\footnote{The signature of the metric is indeterminate; either positive or negative depending on the sign convention chosen.}.

We use Greek indices (i.e. $\{\alpha,\beta\}$) to indicate 4-dimensional coordinate indices.

We use Latin indices (i.e. $\{i,j\}$) to represent the 3-spatial coordinate indices, for example $\{\eta,\theta,\varphi\}$.

The use of Einstein summation notation is implied with repeated tensor indices, i.e. 
$$g^{ai}R_{ib} \equiv \displaystyle\sum\limits_{i=1}^3 g^{ai}R_{ib} \;\; \mathrm{or} \;\; g^{\alpha\beta}V_\alpha V_\beta = \sum\limits_{\alpha=1}^4 \sum\limits_{\beta=1}^4 g^{\alpha\beta}V_\alpha V_\beta$$

The use of subscripts with coordinate names (and with or without commas) represents a partial derivatives with respect to those coordinate variables. e.g. $f_{\eta}  \equiv f_{,\eta} \equiv \partial_\eta f \equiv \frac{\partial f}{\partial\eta}$

When writing mixed second rank tensor quantities, i.e. $K^i_j$, we mean that the first covariant index has been raised, i.e. $K^i_j \equiv K^{i}_{\ j}=g^{ia}K_{aj}$

\section{Einstein's Field Equations}
Einstein's general 4-dimensional gravitational equations that couple space-time curvature to the matter and non-gravitational fields are
\begin{equation} G_{\alpha\beta}=\frac{8\pi G}{c^4}T_{\alpha\beta} \end{equation}
where $G$ is the universal/Newtonian gravitational constant and $c$ is the speed of light.  Setting $G=c=1$ we find\footnote{This implies that $1s \equiv 299 792 458m = 1.8016 \times 10^{15} kg$, so once we choose one scale (length, time or mass) we can calculate the other scalings.  This lack of intrinsic scale implies that we are looking at subatomic and galactic scales simultaneously.}:
\begin{equation}\label{eqn:einstein}
G_{\alpha\beta}=8 \pi T_{\alpha\beta}
\end{equation}

The left hand side of equation (\ref{eqn:einstein}) contains purely geometric tensors, and the right hand side contains the energy/momentum density, pressures, etc. associated with matter and non-gravitational fields.  For a discussion of some physical energy momentum tensors, their formulations and meaning see for example \cite{d'inverno}, $\S 12$ or \cite{alcubierre:3p1num}.  As we are working in a vacuum with no matter or non-gravitational fields, we will be setting all the components of the energy-momentum tensor to zero ($T_{\alpha\beta}=0$).

The greatest difficulty in studying (\ref{eqn:einstein}) is that the equations have no general closed-form solution and are, in general, a series of coupled, quasi-linear\footnote{In this context meaning linear in the highest order derivatives and non-linear in lower order derivatives.}, second-order PDEs in the metric quantities.

There are generally three methods that one can use to find a particular solution to these equations:

(a) specify the physical distribution of matter in the spacetime via the Energy-Momentum Tensor ($T_{\alpha\beta}$) and solve the non-linear PDEs for the metric and extrinsic curvature quantities in (\ref{eqn:einstein}).  This is the method we adopt in this thesis, choosing a vacuum ($T_{\alpha\beta}=0$) spacetime.

(b) specify a metric and compute the components of the Einstein tensor, which then determines the components of the energy-momentum tensor $T_{\alpha\beta}$

(c) a hybrid of the first two methods depending on the components of the left or right hand side that we specify \emph{a priori}

These three methods vary in utility and applicability, as the first one involves solving difficult PDEs that can usually only be solved numerically, whereas the second involves a hit-and-miss approach that will generally produce a non-physical energy-momentum tensor.

We will now discuss two of the main physical features that can appear in a vacuum solution to (\ref{eqn:einstein}): black holes and gravitational waves.

\subsection{Black Holes}
Black Holes are characterised by the existence of an event horizon, which is a boundary in the spacetime that marks the divide between the interior region of the black hole and the exterior region.
For a \emph{stationary} black hole the interior region is incapable of transmitting matter or light beyond the event horizon\footnote{Ex-postulated quantum mechanical effects like Hawking radiation.}, hence it is causally disconnected from the exterior region.  The event horizon acts like a one-way membrane, allowing infalling light/matter from the exterior region to penetrate into the interior, but not the other way around.  More specifically, the event horizon marks the boundary in a spacetime between the region where light rays do and do not reach future null infinity\footnote{See \cite{Wald} and \cite{d'inverno} for a discussion on various horizon theories and examples.}.

The simplest black hole solution is the Schwarzschild solution, which is spherically symmetric and has the metric:
\begin{equation}\label{eqn:schw_metric}ds^2=-\left(1-\frac{2m}{r}\right)dt^2 +\left(1-\frac{2m}{r}\right)^{-1}dr^2 + r^2d\theta^2 + r^2\sin^2\theta d\phi^2\end{equation}
where $m$ is the mass of the black hole.

For dynamic black holes it is possible for the boundary that is the event horizon to evolve (however Hawking has shown that the area cannot decrease), so the calculation of exactly where the interior region is requires the entire space-time solution to allow determination of the region from which light rays do not reach future null infinity.  This can complicate determination of the exact location of the event horizon as we require the \emph{whole} spacetime solution.

Ashtekar \cite{ashtekar} describes some alternate methods of classifying and measuring isolated and dynamic horizons, which are not global but rather \emph{local} properties of black hole spacetimes.  Two results that are applicable to this thesis are (1) the quasi-local mass-energy of an isolated horizon can be calculated from the horizon area $A_{AH}$ via
\begin{equation}\label{eqn:mqlah}M_{QL}=\sqrt{\frac{A_{AH}}{16\pi}}\end{equation}
(where we have dropped the black hole angular momentum term $J$ as we consider non-rotating spacetimes in this thesis) and (2) the area of an apparent horizon can be used to place a lower bound on the eventual steady-state black hole mass via equation (\ref{eqn:mqlah}).

There is a small class of known exact, static black hole solutions\footnote{We exclude consideration of the various solutions to, for example, the Schwarzschild solution in the plethora of coordinate systems that people have explored.  By the principle of general covariance they are all physically equivalent.} to (\ref{eqn:einstein}), including the solutions of Schwarzschild (spherically symmetric vacuum), Reissner-Nordstr\"{o}m (charged, static, spherically symmetric), Kerr (rotating vacuum) and Kerr-Newman (rotating vacuum with charge) which are usually used as test bed computations for more complex numerical codes.
One could say that numerical relativity is like experimental GR to some people, and theoretical GR to others.

\subsection{Gravitational Waves in General Relativity}\label{subsec:gravwave}
The Einstein equations can be written as a set of hyperbolic PDEs\footnote{The general question of the hyperbolicity of Einstein's equations is discussed in \cite{friedrich} or \cite{reula}.  In specific cases they can be shown to be hyperbolic or hyperbolic/elliptic.  Alcubierre \cite{alcubierre:3p1num} gives a detailed examination and classification of ``hyperbolicity'' of a few select formulations (whose details differ from our formulation).} which allow for ``wave-like'' motion; i.e. any displacement that is present in the initial data will propagate with finite speed along characteristic curves.  These disturbances are therefore felt at large distances only after the passage of a non-zero time interval.

In the case of the Einstein equations this is manifested in all of the physical variables (i.e. the metric variables $g_{\alpha\beta}$, curvature, gauge variables, etc.).  Analogously to the time dependent Maxwell's equations, (\ref{eqn:einstein}) admit wavelike solutions that propagate at the speed of light.

In the linearized form of (\ref{eqn:einstein}), there is wave-like behaviour exhibited by the metric and one can find solutions for plane-polarized waves\footnote{See for example \cite{d'inverno}, chapters 20 and 21 for a discussion of the formulation, results and limitations of this theory.}.  Because these waves are plane-polarised they require two spatial dimensions for propagation, leading us to 2+1\footnote{By this we mean a two spatial plus one temporal coordinate slicing of the general 4D spacetime.} or higher formalisms in order to investigate their behaviour.

A gravitational monopole cannot radiate due to Birkhoff's theorem, which states that any spherically symmetric solution of the \emph{vacuum} Einstein equations must be \emph{stationary} and \emph{asymptotically flat}, i.e. the Schwarzschild solution.  A gravitational dipole cannot radiate due to the conservation of momentum.  Therefore when considering any multipole expansion in the asymptotic radiation zone we would expect to need quadrupole or higher moments to be present in order to characterise the waves.  

The fully non-linear form of (\ref{eqn:einstein}) also admits some exact solutions of plane and cylindrical waves, solitons or other wave-like behaviour, however these spacetimes are generally not asymptotically flat.

Another interesting question to investigate is whether pure gravitational waves can ``collapse'' to form a black hole - i.e. can one create what is traditionally thought of as the final state of a matter collapse from a pure vacuum wave with no matter or other fields present?  This is one of the questions that we aim to investigate.

As gravitational radiation will require one or two transverse degrees of freedom for propagation, we have to move beyond spherical symmetry to examine the physical/numerical characteristics of asymptotically flat gravitational wave spacetimes with higher spatial dimensions.

\subsection{Energy in GR}
The general covariance of General Relativity allows one to transform to alternate coordinate systems freely, so it is impossible to have an invariant \emph{local} energy measure\footnote{\cite{MTW} $\S 20.4$ has an interesting discussion/analogy as to why this is the case.} and one must use global measures of energy instead.  Globally energy is not necessarily conserved, however there are some conservative measures for asymptotically flat solutions that have been developed.

One idea is to consider the gravitational effect of an isolated system on a test particle in the asymptotically flat region far away from any strongly non-linear regions.  If we push the point of measurement out to spatial infinity then as any gravitational waves propagate at a finite speed and never reach spatial infinity we know that this measured mass must remain constant, and is the ADM mass (see for example \cite{MTW,bernstein,Wald,alcubierre:3p1num}).

Another measure is to compare one's metric values in the asymptotic region to the Schwarzschild solution metric variables in equation (\ref{eqn:schw_metric}) and thereby calculate an equivalent mass, which is called the mass aspect.

We will eventually make use of the ADM mass inside our code to investigate convergence and critical behaviour\footnote{There are various other measures of mass/energy (i.e. Bondi \cite{bernstein,Zhang}, Brill \cite{Brill}, Hawking \cite{hawkingmass}, Penrose \cite{penrosemass}) that we do not make use of but may incorporate as additional checks at some point in the future.}.

\section{Why Numerical GR?}
There often comes a point in the study of any sufficiently complex problem where one cannot proceed any further due to an integral that doesn't have a closed form expression, a differential equation with no (known) closed-form solution, a set of coupled equations that exhibit complex chaotic behaviour, etc.  At that point numerical analysis and computational methods are needed to investigate the problem any further\footnote{Unless perturbation methods are appropriate to the problem being studied, however they are not applicable in general.}.

While numerical techniques introduce aspects of imprecision (to be discussed later), they can motivate analytical or experimental work through the results they produce.

In general the Riemann curvature tensor is a non-linear combination of metric variables and their derivatives (see equation \ref{eqn:riemann}), as are the Ricci tensor and Ricci scalar.  Because of this one cannot solve the resulting \emph{general} Einstein equations analytically.  So investigation of time evolutions of realistic spacetimes requires numerical modeling in the majority of cases.

To this end we will discuss a Cauchy evolution formulation of GR in the next section that permits a construction of dynamical solutions to the Einstein equations.

\section{Cauchy IVP formulation of GR}\label{sec:cauchyivp}
Analogously to the IVP (initial value problem) formulation of Maxwell's equations (or any set of hyperbolic PDEs), we look for a formulation of Einstein's field equations that involves the definition of an initial set of data defined on a 3D spatial hypersurface, and a set of evolution equations to evolve that data off of the initial hypersurface onto subsequent hypersurfaces either into the future or the past \cite{ADM,Lichnerowitz}.

The method of taking a 4-D spacetime and creating a slicing that most closely mimics our human experience is to take a 3-D spatial hypersurface as our Cauchy surface and use time to thread Cauchy surfaces into an evolution scenario (also called 3+1).  Much like humans experience the world in three spatial dimensions and evolve through time, we wish to formulate our Cauchy problem in this manner.

There are many slicing methods that do not use time as their level-surface coordinate, and there are benefits and problems with doing so.  Sometimes choosing our coordinates to be, for example, null coordinates will simplify the analysis of a particular physical situation.  It might also be the case that a difficult differential equation will turn into a solvable one in an alternate coordinate formulation\footnote{See for example Walsh \cite{walsh} or Rinne \cite{RinneCQG}.}.  Whatever the reason, choosing an appropriate coordinate system is very important to the solvability of any problem, and not all coordinate systems allow for an easy space/time separation\footnote{For a discussion on these and many other considerations see for example Alcubierre \cite{alcubierre:3p1num}.}.

This choice is also difficult to change once one has started coding - generally changing slicings or coordinate systems (which are also associated with gauge choices) will mean recoding your entire program and re-deriving all of your equations, and these often introduce a new set of numerical problems.

In this thesis we proceed with the choice of a 3-spatial Cauchy surface and time-like normal vector.
A time evolution method allows for the use of some well developed singularity-avoiding techniques assuming that such singularities arise in the future of the Cauchy initial data.

This 3(spatial)+1(time) splitting also gives a ``natural'' method of searching for horizon formation (i.e. \cite{Thornburg:AH,evans,bernstein}) and extracting gravitational wave information.  These methods have also been used extensively in the past, so their properties are better-known than some other formulations and we can utilise some of the results from the broad base of literature in the field.

Much like in electrodynamics one can use a Hamiltonian approach to derive the appropriate equations of motion (and constraints) from Einstein's field equations.  In performing this split we do need to make the distinction between a general 4-D tensor quantity and a 3-D \emph{spatial} tensor quantity.  If it is unclear we will prefix 4-D quantities only with a ${}^{(4)}$.

Two major methods of splitting the Einstein equations into a space+time problem that we consider\footnote{See also Alcubierre \cite{alcubierre:3p1num}.} are called \emph{ADM}\footnote{Short for \emph{Arnowitt, Deser and Misner} \cite{ADM}.} and \emph{BSSN}\footnote{Short for \emph{Baumgarte, Shapiro, Shibata and Nakamura} \cite{BSSN,BSSN2}.}.  We will discuss the ADM formulation now, and leave consideration of BSSN until a later time.

\subsection{ADM formulation overview}\label{subsec:admoverview}
\begin{figure}
\centering
\includegraphics{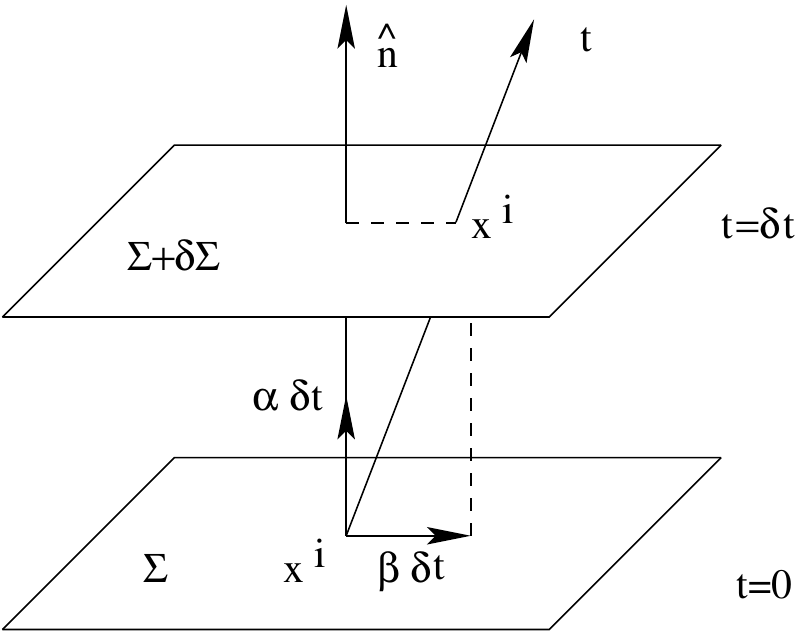}
\caption[Spacetime 3+1 splitting]{A schematic diagram of the splitting of the space-time using the 3+1 ADM formalism.  $\hat{n}$ is the normal vector to the surface at each point, $\alpha$ gives the distance, in time, between points on successive sheets and $\beta$ represents the spatial shift of the coordinates from one sheet to the next.  $\delta\tau=\alpha\,\delta t$, where $\tau$ is the proper time}
\label{fig:foliation}
\end{figure}

For the majority of this thesis, we use the 3+1 ADM\footnote{See for example \cite{MTW} section $21.4$, \cite{cookphd} Chapter 2, \cite{alcubierre:3p1num} or \cite{bernstein}.} formulation of the general 4-dimensional Einstein equations.  
As mentioned earlier, the goal is to construct the spacetime by slicing it along the ``time'' coordinate into 3-D spatial hypersurfaces.

The technical formulation of the 3+1 equations and the mathematical difficulties that arise from it are covered in many different texts including \cite{ADM,paul_thesis,MTW,frontiers,hawking,bernstein,alcubierre:3p1num}, so the exposition that follows will be brief.
Our goal is to reduce the 4-D equations in (\ref{eqn:einstein}) to a 3-D spatial tensor format and to this end we can describe the 4-metric as
\begin{equation}\label{eqn:genmetric}
g_{\mu\nu} = \left( \begin{array}{cc}
-\alpha^2+\beta^{a}\beta_{a}&\beta_a\\
\beta_a&\gamma_{ab}\\
\end{array} \right),\end{equation}
where $\alpha$ represents the scalar \emph{lapse} function and $\beta_a$ represents the \emph{shift} vector function, using the ADM formulation of 3+1 spacetime (see figure \ref{fig:foliation}).  Note that we denote the 3-spatial metric as $\gamma_{ab}$ (i.e. the metric on a spatial hypersurface $\Sigma$).

The lapse, $\alpha(x^{\xi})$, is a scalar function of time and space that represents orthogonal proper time progression at each point on a $t=$constant spatial hypersurface via 
$$\delta\tau=\alpha\,\delta t$$
$\delta\tau$ measures the increment of proper time, i.e. the amount of time that a comoving observer measures.  The quantity $\delta t$ measures the increment of coordinate time between adjacent spacelike hypersurfaces and is usually chosen to be a constant as we iterate through the code\footnote{Unless we wish to slow the evolution in a particular area to examine horizon formation or other critical behaviour.  In GR, critical behaviour typically spans a very small portion of the overall evolution so sometimes we need a way to ``zoom'' into a particular time region.}.
If $\alpha \rightarrow 0$, the proper time increments measured by $\alpha\,\delta t$ vanish, which can allow the ``evolution'' to progress for infinitely long in \emph{coordinate} time while not progressing in proper time.
Thus $\alpha$ provides us with a tool to cause the proper time evolution to progress at different rates at different points in the space-time and prevent coordinate points from ``running into'' areas of large curvature (or numerical singularities like at $r=0$ and $\theta=0$ in our case).
This can also be used to help prevent numerical errors in one region of the grid from destabilising the evolution in another and allows for much longer coordinate time evolutions (i.e. raw number of iterations).

The shift vector, $\beta_a(x^{\xi})$, is a vector function of time and space that represents the change of the spatial coordinates from one time slice to the next\footnote{If we are employing a numerical grid as described in chapter \ref{chap:nummethod}, the shift vector can be employed to provide optimal grid resolution in areas of numerically unstable curvature, while minimizing grid point calculations in low-curvature areas.
This can allow for a variable spatial increments (grid ``width''), and the system can be allowed to respond dynamically to the need for grid points, instead of having a static linear or logarithmic scaling forced upon it (with the added complication that the grid points can be moved around by other parts of the dynamics).}.
The problem, however, is that introducing a non-zero shift can complicate the evolution equations\footnote{Non-zero shift vectors sometimes introduce numerical regularisation problems when employing numerical methods. Alcubierre \cite{alcubierre:3p1num}, however, argues that a static, vanishing shift vector (i.e. $\beta_i=0$) is unstable for ADM numerical codes.}, depending on the other gauge and slicing conditions that are chosen.

From these considerations we can see that the distance the metric must measure from the lower to the upper surface is\footnote{Or, as Misner, Thorne and Wheeler \cite{MTW} put it, the structure of the ribbons of steel that bind the hypersurfaces together.}.
\begin{eqnarray}
{ds}^2 & = & ( \textrm{proper distance on spatial hypersurface} )^2 - ( \textrm{proper time} )^2 \nonumber \mbox{} \\
& = & \gamma_{ij}(dx^i+\beta^i dt)(dx^j+\beta^j dt) -(\alpha dt)^2
\end{eqnarray}
where $\gamma_{ij}$ is the spatial metric tensor formed from $g_{\alpha\beta}$, i.e.
$$\gamma_{ij}=g_{ij} \;;\; i,j \in \{1,2,3\}$$
This gives us the metric in equation (\ref{eqn:genmetric}).

\subsection{Mathematical Treatment of the 3+1 ADM decomposition}\label{sec:ADM3plus1}
Recalling that the Einstein tensor $G_{\alpha\beta}$ is symmetric (\ref{eqn:einsteintsymm}), it provides us with 10 equations in 4-D to describe the spacetime.  Because we are splitting spacetime into spatial hypersurfaces and time evolution, we consider the 3-D spatial portion of the Einstein field equations separately from the time (0-index) quantities.  So we will end up with six evolution equations for the spatial metric quantities and four constraint equations on each hypersurface.

\begin{figure}
\centering
\includegraphics{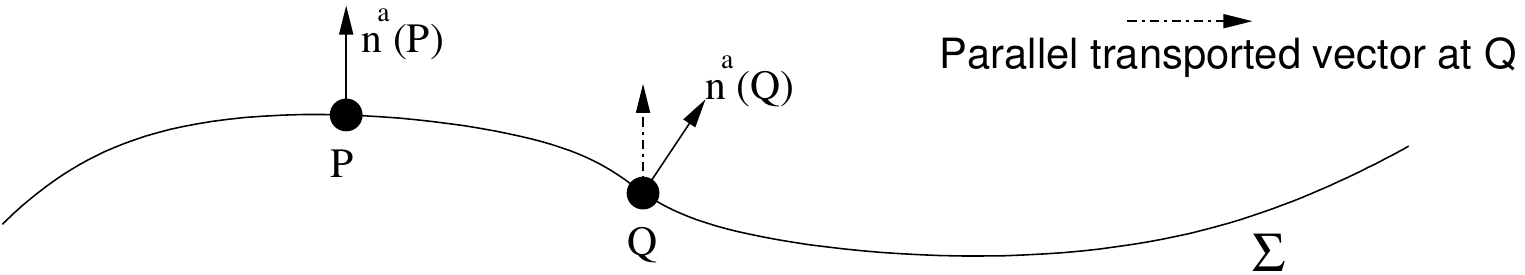}
\caption[Parallel Transport on a curved manifold]{Parallel transport of the normal vector $n^{a}$ from a point $P$ to the point $Q$ on a curved manifold $\Sigma$.  The difference between the normal vector that is ``parallel transported'' and the non-transported normal vector gives the curvature of the manifold in an embedding manifold, i.e. the extrinsic curvature.}
\label{fig:parallel_transport}
\end{figure}

The extrinsic curvature is defined by measuring the deviation of a parallel transported normal vector between points on the 3-manifold $\Sigma$ as shown in figure (\ref{fig:parallel_transport})
and is given by \cite{alcubierre:3p1num}
$$K_{\mu\nu}=-(\nabla_\mu n_\nu+n_\mu n^{\delta} \nabla_\delta n_\nu)$$
where $n^\mu$ is the time-like unit normal to our spatial hypersurface $\Sigma$.

There are many ways to arrive at the full set of equations.  One can define a Hamiltonian (or Lagrangian) density and calculate the resultant equations using the Hamilton (or Euler-Lagrange) equations of motion.  In this formalism one treats the lapse scalar and shift vector as Lagrange multipliers\footnote{The original ADM paper \cite{ADM} takes this approach, or for example \cite{MTW} pp. 505-538}.  The extrinsic curvature ($K_{ij}$) relates to the conjugate momenta in this formulation, leaving us with $\gamma_{ij}$, $K_{ij}$, $\alpha$ and $\beta_i$ as our variables that are necessary to describe the physics of the spacetime.

Alternately, Evans \cite{evans} and Bernstein \cite{bernstein} use projections of the 4-D Riemann tensor onto the 3-D spatial surface to derive the constraints and use Lie derivatives to transport the fundamental forms onto subsequent hypersurfaces and derive the appropriate time evolution equations.  The intrinsic curvature is measured by the ${}^{(3)}$Riemann tensor which is a function of the metric variables.

Evans \cite{evans} gives a very relevant derivation of the equations particular to the mixed form of the extrinsic curvature variables that we use.

We can make a comparison from the structure of Maxwell's 3D equations in Gaussian units to those of the 3+1 split GR equations.  Writing Maxwell's equations for the electric field ($\vec{E}$) and magnetic field ($\vec{B}$) with charge density $\rho$ and current density $\vec{j}$ we have the well known result:
\begin{eqnarray}
\label{eqn:maxwellcon} \nabla \cdot \vec{E} = 4 \pi \rho & ; & \nabla \cdot \vec{B} = 0 \\
\label{eqn:maxwelldyn} \partial_t \vec{E} = \nabla \times \vec{B} - 4 \pi \vec{j} & ; & \partial_t \vec{B} = - \nabla \times \vec{E}
\end{eqnarray}
We can identify two constraint equations (\ref{eqn:maxwellcon}) which are time independent and six evolution equations for the vector components of the fields (\ref{eqn:maxwelldyn}).  One can then generate a solution to Maxwell's equations by first solving the IVP (from equations (\ref{eqn:maxwellcon}) at $t=0$) and then employing equations (\ref{eqn:maxwelldyn}) to evolve the field quantities off the $t=0$ surface.

One finds a similar structure when performing the ADM 3+1 splitting, which leads to the following identifications in GR\footnote{further discussion is presented in \cite{d'inverno} $\S 13.5$}:
\begin{itemize}
\item $G^{00}=T^{00}$ as the ``Hamiltonian constraint'' (time independent)
\item $G^{0b}=T^{0b}$ are the ``Momentum constraints'' (time independent)
\item $G^{ab}=T^{ab}$ are the evolution equations for the $K_{ij}$
\end{itemize}
For example, following the conventions of Evans who defines a one-form $\omega=dt$, with $t$ the local coordinate time, we have 
$$g^{\mu\nu} \omega_{\mu} \omega_{\nu} = -\alpha^{-2}$$
where $\alpha$ is strictly positive and the sign convention is chosen to ensure that our quantities are space-like.  This leads to a normalised one-form
$$\Omega_{\mu}=\alpha \omega_{\mu}$$
(i.e. $\alpha$, the lapse, is the scaling factor between coordinate time $\omega_{\mu}$ and proper time $\Omega_{\mu}$) and leads to our unit normal vector being
$$n^{\mu}=-g^{\mu\nu}\Omega_{\nu}$$
(see figure \ref{fig:foliation} for a visualisation of the normal vector $n$).
The Lie derivative is therefore defined along the general vector field
$$t^{\mu}=\alpha n^{\mu} + \beta^{\mu}$$
(see figure \ref{fig:foliation} for a schematic of the motion of the coordinates, $x^i$), where we have decomposed it into orthogonal components (time-like and space-like).  To fulfill the requirement that
$$\omega_\mu t^{\mu}=1$$
we require
$$\beta_{\mu} n^{\mu} = 0$$
i.e. $\beta_{\mu}$ is a purely spatial quantity, which we refer to as the shift vector.

Choosing our coordinate basis accordingly (i.e. 3 spatial basis vectors that are tangent to each time slice) we find that
$$t^{\mu}=(1,0,0,0)$$
which implies that
$$\pounds_t = \partial_t$$
and
$$n^{\mu}=(\alpha^{-1},-\alpha^{-1}\beta^i) \; ; \; n_{\mu}=(-\alpha,0,0,0)$$

From here, one can split the 4-tensor equations into 3-spatial tensor evolution equations plus some constraints as mentioned above, by using a projection of the 4-D Riemann tensor onto a 3-D sub-manifold.

The equations that result from the projection and contraction of the 4-Riemann Tensor are referred to as the Gauss-Codazzi-Ricci equations (which are 0 order, 1st order and 2nd order contractions of ${}^{(4)}R_{\mu\nu\gamma\delta}$ projected onto the 3-surface).\footnote{see \cite{hawking} for a formal derivation}  The results that are relevant to the 3+1 ADM formalism are:

The general ADM Hamiltonian Constraint is:
\begin{equation}R+({\rm Tr}K)^2-K^{ab}K_{ab}=2\rho,\label{eqn:3p1:ham3}\end{equation}
where $\rho=n^c n^d T_{cd}$ is the energy density, and ${\rm Tr}K = \gamma^{ij}K_{ij}=K^j_j$ is the trace of the Extrinsic Curvature tensor, $R_{ij}$ is the 3-spatial Ricci curvature, and $R=\gamma^{ij}R_{ij}$.  We will also use the convention that 3-spatial covariant derivatives are denoted by $D_a$ compared to 4D covariant derivatives $\nabla_\alpha$.

The general ADM Momentum Constraints are:
\begin{equation}p_b=D_aK^{a}_{b}-D_b{\rm Tr}K=S_{b}\end{equation} 
or
\begin{equation}p^a=D_b (K^{ab}-\gamma^{ab}{\rm Tr}K) = S^a\label{eqn:3p1:mom3}\end{equation} 
where $S_b=-\gamma^c_b n^d T_{cd}$ is the momentum density.  One interesting feature of these constraints is that they are independent of the variables $\alpha$ and $\beta^i$.

The general evolution equations for covariant spatial metric quantities $\gamma_{ab}$ (aka First Fundamental Form) are:
\begin{equation}\partial_t \gamma_{ab}=-2\alpha K_{ab}+D_a\beta_b+D_b\beta_a.
\label{eqn:3p1:gammadot}\end{equation}

From this we can alternately define the extrinsic curvature tensor via
$$K_{ab}=-\frac{1}{2}\pounds_n \gamma_{ab}$$
where $n^a$ is our unit normal time vector, and this gives the interpretation of the extrinsic curvature as the ``velocity'' of the 3-metric.  Note also that (\ref{eqn:3p1:gammadot}) contains no $T_{\alpha\beta}$ terms and arises purely from geometric considerations, i.e. it is derived independently of Einstein's field equations.

The general evolution equations for the extrinsic curvature in mixed ($K^a_b$) and covariant ($K_{ab}$) forms are:
\begin{eqnarray}\label{eqn:genmixkevol}
\partial_t K^a_b & = & -D^a D_b \alpha + \alpha [ R^a_b + K K^a_b - S^a_b - \frac{1}{2}\delta^a_b(\rho - S)] \nonumber \\
& & + \beta^lD_l K^a_b + K^a_lD_b\beta^l - K^l_bD_l\beta^a. \label{eqn:3p1:Kevol3}
\end{eqnarray}
\begin{eqnarray}\label{eqn:gencovkevol}
\partial_t K_{ab}&=&-D_aD_b\alpha+\alpha[R_{ab}-2K_{ac}K^c_b+({\rm Tr}K) K_{ab}-S_{ab}-\frac{1}{2}\gamma_{ab}(\rho-S)]\nonumber\\
& &+\beta^c D_c K_{ab}+K_{ac} D_b \beta^c+K_{cb}D_a \beta^c.
\end{eqnarray}
These equations contain the source terms from the right hand side of Einstein's field equations.

\subsection{Contracted Christoffel Symbols}
One can create the contracted Christoffel symbols in the following manner:
$$\Gamma^\alpha=g^{\beta\delta}\Gamma^\alpha_{\beta\delta}$$
The general idea behind this is to change the numerical nature of the evolution equations by creating (or eliminating) auxiliary variables to capture various non-linear terms of the Ricci curvature  (see for example \cite{harmonic:var,lindblom:harmonic} for ``Harmonic coordinate'' formulations, and \cite{BSSN} for a BSSN formulation of the field equations).

In the simplest form the Harmonic condition on coordinates
\begin{equation}\Gamma^\alpha=0 \label{eqn:harmonicchristoff}\end{equation}
is used to eliminate all second-order derivative terms in the Ricci tensor (\ref{eqn:riccitensor}), and therefore the Einstein tensor (\ref{eqn:einsteintensor}), \emph{except} for a wave-like operator.  For example, the vacuum Einstein equations $R_{\alpha\beta}=0$ turn into \cite{lindblom:harmonic}:
$$g^{\delta\epsilon}\partial_\delta \partial_\epsilon g_{\alpha\beta}=2 g^{\delta\epsilon}g^{\rho\sigma}(\partial_{\rho} g_{\delta\alpha} \partial_{\sigma}g_{\epsilon\beta} - \Gamma_{\alpha\delta\rho}\Gamma_{\beta\epsilon\sigma} )$$
where $\Gamma_{\alpha\beta\delta}$ are the Christoffel symbols of the first kind, that can be calculated by lowering the contravariant index of the Christoffel symbols of the second kind.
$$\Gamma_{\alpha\beta\delta}=\frac{1}{2}(g_{\alpha\delta,\beta}+g_{\beta\delta,\alpha}-g_{\alpha\beta,\delta})$$
One can formulate various alternatives to the harmonic condition (\ref{eqn:harmonicchristoff}) to arrive at a harmonic or BSSN formulation, which provide alternate methods of solving the Cauchy IVP and computing the evolution of the appropriate dynamic variables.

\section{The Bianchi Identities}
One of the basic properties of the Riemann tensor is that it satisfies the Bianchi identities, which are differential identities that take the form:
\begin{equation} \nabla_{\epsilon} R^{\alpha}_{\beta\gamma\delta} + \nabla_{\delta} R^{\alpha}_{\beta\epsilon\gamma} + \nabla_{\gamma} R^{\alpha}_{\beta \delta \epsilon} \equiv 0
\label{eqn:bianchi}\end{equation}
These identities arise simply from a consideration of the underlying geometry that defines the Riemann tensor, and as such they are independent of any other conditions we put upon the space-time.

Equation (\ref{eqn:bianchi}) can also be reworked into the contracted Bianchi identities, which are
\begin{equation} \nabla_\alpha G_\beta^\alpha \equiv 0 \label{eqn:contractedbianchi}\end{equation}
using (\ref{eqn:einstein}) this is equivalent to
\begin{equation}\label{eqn:tabcon} \nabla_\alpha T_\beta^\alpha \equiv 0 \end{equation}
From d'Inverno\footnote{Reference \cite{d'inverno} $\S 13.5$} (\ref{eqn:contractedbianchi}) is equivalent to
$${G_\alpha^0}_{,0}= C^{\beta a}_{\alpha}{G^0_\beta}_{,a} + D^\beta_\alpha G_\beta^0$$
where $C^{\beta a}_{\alpha}$ and $D^\beta_\alpha$ are solely functions of the metric and its first derivatives, and the comma notation indicates a partial derivative.  Therefore the system of equations above has only one solution for $G_\alpha^0$, which is chosen to be uniquely zero on our initial Cauchy surface (i.e. our ``Hamiltonian'' and ``momentum'' constraints are satisfied on the initial slice).

This shows that if the constraints are satisfied on the initial slice, then they are consistent with the evolution equations at all future times, i.e. the evolution equations propagate the constraints onto future time slices.\footnote{In theory only, of course.  In numerical simulations the constraints can be used as a test of regularity, convergence, accuracy, etc.}

Therefore equations (\ref{eqn:3p1:ham3}), (\ref{eqn:3p1:mom3}), (\ref{eqn:3p1:gammadot}) and (\ref{eqn:3p1:Kevol3}) provide a complete method for solving the Cauchy problem in a 3+1 ADM spacetime.

One other note regarding the Bianchi Identities we wish to make is that they provide a level of internal self-consistency to GR not present in Maxwell's Equations.  As Maxwell's equations do not account for the motions of the \emph{sources}, one must employ Newton's Second Law (or some other physical law) to do so.  Then, when trying to account for the motion of source particles in classical E\&M one ends up with the radiation-reaction/damping problem\footnote{See \cite{marion_em,jackson_em} or others.} where the motion of the source creates a field that in turn interacts with the source via independent equations that create a recursive relationship.

In GR, the motion of the sources is instead built into the field equations.  Considering $\beta=0$ in (\ref{eqn:tabcon}) for example we see that
\begin{equation} \nabla_\alpha T_0^\alpha = \nabla_0 T_0^0 + \nabla_aT_0^a = 0 \end{equation}
which says that the time derivative of the energy density is (-) the spatial ``divergence'' of the momentum density, which is a 4D tensor version of the continuity equation.  Letting $\beta=b$ (spatial indices) we find that
\begin{equation} \nabla_\alpha T_b^\alpha = \nabla_0 T_b^0 + \nabla_aT_b^a = 0 \end{equation}
which says that the time derivative of the momentum density must be (-) the ``gradient'' of the momentum flux - which is the 4D tensor version of Newton's second law.  Hence the equations of motion for the sources are built into the field equations.

  So we do not encounter the same consistency problems in GR that we do in classical E\&M.

\section{The Brill Wave Criteria}\label{subsec:brillformalism}
The first positive energy result for vacuum gravitational waves came from Brill \cite{Brill}, following in the footsteps of Bondi's work.
Brill described one set of conditions to ensure that we have a physical situation to study, which are\footnote{In Brill's words \cite{Brill}: \emph{...every time-symmetric axially-symmetric gravitational wave which has an asymptotically Schwarzschildian character necessarily has a positive definite mass.}}:
\emph{In an asymptotically flat, axi-symmetric spacetime, the mass of the time-symmetric initial hyper-surface is non-negative (i.e. physically meaningful and well-defined) provided:}
\begin{itemize}
\item \emph{The line element of 3-space at a fixed moment in time $(t=0)$ is chosen to have the form (in spherical polar coordinates)}\footnote{Brill originally devised his proof in cylindrical coordinates.}:
\begin{equation}\label{eqn:brillmetricchap1} dl^2 = \psi^4 \left[ e^{2q}(dr^2 + r^2 d\theta^2) + r^2 \sin^2\theta d\phi^2 \right]
\end{equation}
\emph{This presents a ``conformal decomposition'' of the metric (where the $\psi^4$ term has been factored out).  This is so that our initial time-symmetric slice satisfies $R=0$ with an appropriate choice of $\psi$.}
\end{itemize}
Conditions on the metric (\ref{eqn:brillmetricchap1}) that ensure asymptotic flatness and positive energy are:
\begin{itemize}
\item The functions $q=q(r,\theta)$, $\psi=\psi(r,\theta)$
\item The functions $\psi$ and $q$ must be symmetric across the plane $z=0$ (where $z=r\cos\theta$ and $z=0 \rightarrow \theta=\pi/2$)
\item The function $q$ obeys the condition $q(r,0)=0$
\item The function $q$ falls off faster\footnote{Some authors require integer falloff powers in $1/r$, however that is not required for Brill's analysis to hold.} than $\frac{1}{r}$ asymptotically:  $q(r,\theta) = O(r^{-1-\delta}) \;;\; \delta>0$
\item We use time-symmetric initial data so $\partial_t g_{ij}=0$ initially.
\end{itemize}
This places a set of limitations on the metric that will guide some of our choices in future sections.  It turns out in Brill's analysis that the global mass measurement is solely dependent on the volume integral of the conformal factor $\psi$ over the entire spacetime\footnote{Note the presence of a $\log$ term, which serves as one of many motivations for the choice of an exponential variable to replace $\psi$ in this thesis.} \cite{Brill}.
$$M=\frac{1}{2\pi}\int_{\mathbf{all\ space}} (\nabla \log \psi)^2 dV \ge 0$$
which is a positive definite quantity (a volume integral over a positive definite quantity), and represents the first positive mass theorem for vacuum gravitational waves.

\section{Critical Phenomena}\label{sec:critical}
One of the more recent contributions of numerical relativity to the study of gravitational collapse problems has been critical collapse simulations.
Choptuik \cite{choptuik2, choptuik1} and others have performed investigations into the critical nature of gravitational collapses from an asymptotically flat, \emph{non-vacuum}, set of smooth initial data. (for a more detailed overview of work in the field, see Gundlach \cite{critical})
The surprising trend that is common in all of these situations is that the mass of the black hole formed from the collapse is governed by a scaling law,
$$ M = C(p-p_{\ast})^{\gamma}$$
where $p$ is some parameter that is indicative of the initial ``strength'' of the data set and $\gamma \simeq 0.37$ (and is independent of what $p$ is).
These systems thus seem to exhibit universal scaling of the black hole mass, of two types:

Type I - finite mass formation at $p=p_\ast$

Type II - infinitesimal mass formation at $p=p_\ast$ - leads to naked singularities (but needs infinite fine tuning)

In the case of the axisymmetric vacuum system we are studying, let us consider a general Gaussian-type wave of the form:
$$q(r,\theta)=A r^2 e^{-\frac{r^2}{\sigma^2}} \sin^4\theta$$
We can characterise (at least) two parameters $p$ that define the initial data set:
\begin{enumerate}
\item The initial amplitude $A$ of the wave.  Larger amplitude gravitational waves have more energy associated with them.
\item The initial width or spread, $\sigma$ of the wave.  Waves with a smaller $\sigma$ will have a higher group velocity and a higher energy associated with them.  We are currently using a Gaussian-type wave for the initial profile, so this is true.  Other wave types would require different analysis.
\end{enumerate}

One goal of creating this evolution framework is to study if this critical / self-similar phenomena in phase space is present in vacuum Brill wave evolutions.

\section{Brill Gravitational Collapse}\label{sec:brillcollapse_historic}
The initial purpose of this thesis was to numerically discover (a) if pure gravitational curvature (with a \emph{vanishing} energy-momentum tensor) can collapse to form a black hole? (b) if there are critical parameters that govern the collapse of a Brill gravitational wave to form a black hole (c) if so, what those parameters are and (d) what the potential critical behaviour is in the parameter space.

During the course of setting up the evolution equations and code to investigate this problem, however, it became apparent that the regularisation of the Brill gravitational wave evolution problem in spherical polar coordinates was the most important piece of the puzzle.  Researchers have laboured for 40+ years to devise a stable numerical scheme for looking at this problem, but the results have not been definitive.  Current and historical evolutions of the GR zero mass wave collapse problem fall into a few classes:
\begin{enumerate}
\item Perturbation techniques on spherical (Schwarzschild) symmetry or superposition of a preexisting black hole+Brill wave to remove the problematic region near the origin \cite{evans,bernstein,Abrahams}.  While these are interesting studies of vacuum spacetimes they pre-suppose the existence of a black hole in conjunction with the wave, an assumption we wish to do away with.
\item Non-Brill waves, i.e. linearised gravitational waves \cite{pfeifferIVP} superimposed on a black hole.  Once again we wish to analyze spacetimes with no black hole present \emph{a priori}.
\item Cylindrical coordinates with a massive scalar field ($T_{\mu\nu} \neq 0$) \cite{choptuik1}\cite{RinnePHD}\cite{Rinne}\cite{Brown}\cite{Murchadha}\cite{gentle} \emph{et al}, which has become a de facto standard after Choptuik \emph{et al}'s success.  The imposition of a non-zero energy-momentum tensor has significant implications for the structure of the Einstein equations, so while they are interesting they are not pure vacuum evolutions. Numerical dissipation is used heavily in these schemes.

\item Mixed-success results using Cartesian coordinate\footnote{The intent of using Cartesian coordinates is to remove \emph{coordinate} singularities that exist in cylindrical and polar coordinates.} evolutions, which suffer from enormous complexity and implementation problems (e.g. \cite{Alcubierre,Thornburg:cartesian}).  There are insufficient details available in the literature to make an honest appraisal of the work done here, and with the switch to black hole inspiral codes by a large number of GR groups these projects seem to have been abandoned by the community.
\item 3-D Cactus code\footnote{Cactus is a numerical/computational framework that allows the use of modular ``thorns'' depending on the problem being solved.  It was first developed with the GR community and has since branched out to other scientific and engineering uses.} vacuum evolution in cylindrical coordinates \cite{alcubierre:3dbrill} with large\footnote{$\sim 10$ orders of magnitude larger than ours.} Hamiltonian violations.

\item Mixed Cartesian/cylindrical coordinates that have incomplete results and potential issues around momentum constraint formulation \cite{miyama}.  The formulation here seems to assume that a solution to $\nabla F=0$ is the same as the solution to $F=0$, without consideration of the arbitrary functions that can be added to $F$ in the first case.

\item Vacuum Brill wave evolutions in cylindrical coordinates with few results, large amounts of numerical dissipation and different gauge and variable choices \cite{RinneCQG}.\footnote{These results mostly relate to fixing the Conformal Thin Sandwich formulation's uniqueness problem.}
\item Spherical polar coordinates with very few ($3$) time steps \cite{eppley} most likely due to the difficulties discussed later around gauge conditions on the lapse $\alpha$.
\item Harmonic gauge vacuum evolution with ``constraint damping'' and various strongly dissipative numerical schemes in conformally compactified cylindrical coordinates \cite{sorkin,sorkin:code}.  Compactified coordinates present their own brand of difficulties as the time required to move from one grid point to another in the heavily compactified asymptotic region grows and grows, requiring smaller and smaller time steps to model properly.  The \emph{heavy} use of numerical dissipation is also a cause for concern.
\item Cylindrical compactified coordinates with a different gauge and extrinsic curvature variables, maximal slicing, and a short evolution due to outer boundary condition problems \cite{garfinkle}.
\end{enumerate}
A summary of these results is presented in table \ref{tbl:litreview}.  While most of the work is interesting in its own right, we wish to study (i) a vacuum spacetime, (ii) with no superimposed black hole present initially and (iii) a \emph{lack} of large amounts of artificial numerical dissipation.  This will give us the most representative exploration of \emph{pure} gravitational radiation's physical characteristics.  Eppley's results \cite{eppley} come the closest to what we wish to study, however he was only able to complete 3 time steps\footnote{Which is the same problem the author encountered for years before realising that certain regularity conditions were required for a stable evolution.}.

\begin{table}\begin{center}
\begin{tabular}{|c|c|} \hline
Ref & Issues/Differences \\ \hline
\cite{evans,bernstein,Abrahams} & BH \\
\cite{pfeifferIVP} & BH, NBW \\
\cite{choptuik1} \emph{et al} & NV,ND,Cyl \\
\cite{Alcubierre,Thornburg:cartesian} & IR,Car \\
\cite{alcubierre:3dbrill} & IR,Car \\
\cite{miyama} & see text \\
\cite{RinneCQG} & ND, Cyl \\
\cite{eppley} & Max \\
\cite{sorkin,sorkin:code} & ND, Cyl, CC, NADM \\
\cite{garfinkle} & Cyl, CC \\
\hline
\end{tabular}
\caption[Summary of major papers in ``Brill wave evolutions'']{Summary of major papers in ``Brill wave evolutions''. See text for more details. BH - superimposed Black Hole present \emph{a priori}; NBW - Non-Brill Wave used; NV - Non-Vacuum system, i.e. $T_{\alpha\beta} \neq 0$; ND - large amounts of Numerical Dissipation used; Cyl - Cylindrical coordinates used; CC - Compactified Coordinates used; Max - Maximal slicing with spherical polar coordinates used; IR - Insufficient Results presented in literature to draw meaningful conclusions; Car - Cartesian 3+1 splitting is used which solves some numerical issues and adds others; NADM - Non-ADM splitting used}
\label{tbl:litreview}
\end{center}\end{table}

As a general philosophy, this author objects to the use of dissipative or viscous numerical methods in strongly non-linear \emph{vacuum} gravitational simulations if for no other reason than the fact that one cannot know the conditioning of the system \emph{a priori} to pick the ``right'' solution scheme\footnote{For example, the heat diffusion equation is stable under some numerical schemes, and unstable for others \cite{burden} $\S 12.2$, and \cite{recktenwald}}.  One can ask the question: if one adds a dissipative scheme to the code, and the curvature builds up then dissipates, is it a physical or numerical result?

While it is reasonable to add artificial numerical viscosity to a hydrodynamical simulation to account for diffusion and/or dissipation, it makes no sense in a vacuum with only gravitational waves present.  These dissipative schemes have, in essence, added extra terms to the field equations which act like a non-zero $T_{\mu\nu}$.  These would possibly mimic a matter distribution present in the spacetime which indicates that Brill gravitational waves need matter present to prevent collapse.

Frequently, a lack of understanding of the conditioning of the underlying problem was at the heart of numerical issues, and correcting any misconceptions was what yielded the best results.

\fancyhead[RO,LE]{\thepage}
\fancyfoot{} 
\chapter{Numerical Methods and Discretisation}\label{chap:nummethod}
\bigskip
Until this point we have discussed some methods to mathematically model GR systems using the language of continuous functions, derivatives and integrals.  We will now discuss how to take these continuous systems and approximate them by discretisation for use in a computer.  The goal is to calculate approximate numerical solutions for the variables in physical systems whose equations (a) have no (known) closed-form analytic solution and (b) we cannot simply create in a laboratory environment to measure their behaviour.

Numerical simulations of continuous systems has enjoyed a rich history of application in many areas of study including GR\footnote{For a general discussion of Numerical GR see for example Alcubierre \cite{alcubierre:3p1num}.}, financial markets, engineering, physical (i.e. E\&M, heat transfer, fluid dynamics) and astrophysical (i.e. stellar collapse, supernovas, neutron star formation) systems, chemical modeling, and many more\footnote{Although sometimes the continuous equations used for modeling in these realms of study are actually approximations of discrete underlying behaviour themselves...}.

\section{Gridding}
To begin discretising a continuous system, it is necessary to choose discrete coordinate nodes (called a grid) on which to evaluate the variables in question.  For example, we can evaluate a one-dimensional function $f(x)$ at various coordinate nodes $x_i$ (where $i$ is an integer) which we denote
$$f_i=f(x_i)$$
where the $x_i$ can be determined in many different manners (i.e. constant or non-constant spacing between the nodes).

Another important factor in discretising a continuous system is to choose coordinates and their boundaries dependent on the situation being modeled.  For example, we might choose two-dimensional Cartesian coordinates and confine ourselves to the interior of a square such that
\begin{equation}\label{eqn:square_aa_ex} 0 \leq x \leq a \;;\; 0 \leq y \leq a\end{equation}
Boundaries should be chosen carefully as one is required to provide boundary conditions on variables at those points for most problems.

One requires a finite interval as computers are only capable of handling finite numbers (except in overflow situations).  This can be partially overcome by compactifying any domain that extends to infinity with a coordinate transformation, for example
$$0 \leq r < \infty \mapsto \left\{ 0 < \bar{r} \leq 1 \;;\; \bar{r}=\frac{r}{r+a}\right\}$$
where $a>0$ is some real number.

This compactification method has its own problems as a small change in $\bar{r}$ can correspond to a large or small change in $r$, depending on where you are in the interval and it may cause physical problems such as causality violation in hyperbolic systems.

\begin{figure}[h]
\centering
%\psfrag{D}{$\Delta$}
\includegraphics{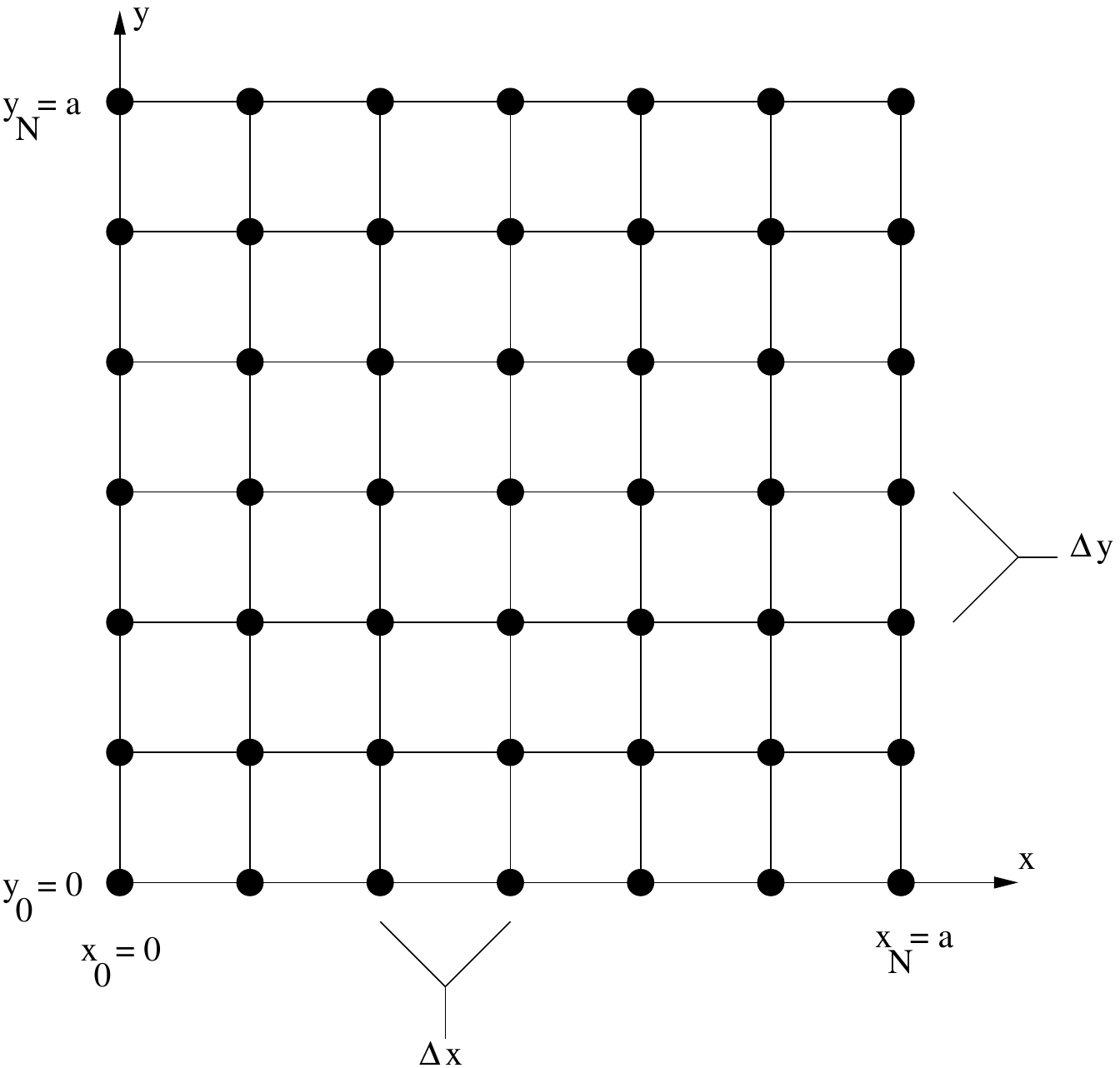}
\caption[Example square grid]{A schematic of the square grid discretisation example defined on $0 \le x \le a \;;\; 0 \le y \le a$}
\label{fig:square_grid}
\end{figure}

We then divide the coordinate intervals in question into finite sized intervals (called grid zones) depending on the problem at hand.  We will restrict our attention to a ``Finite Difference'' scheme, as Finite Element and Finite Volume (and generalized Galerkin methods) add large layers of complexity that are beyond the scope of the current discussion\footnote{And one of their main strengths lies in adaptation to arbitrary grid or mesh configurations which we are not concerned with here.}.
The simplest method to create grid zones is to split each domain into equally sized zone intervals by deciding on the number of grid zones required in each coordinate direction.  For the square region above if we want $N$ grid zones in both the $x$ and $y$ directions, we would set our zone size to $a/N$ and have $N+1$ discrete grid points in each coordinate direction ($(N+1)^2$ total)
$$x_j=\frac{aj}{N} \;;\; y_k=\frac{ak}{N} \;;\; j,k \in {0,1,2 \ldots N}$$
We define $\Delta x$ as the distance between neighbouring grid points in the $x$ direction, which in this case is a constant given our constant zone size:
$$\Delta x = x_{j+1}-x_j = \frac{a}{N}$$
and similarly in the $y$ direction:
$$\Delta y = y_{j+1}-y_j = \frac{a}{N}$$
See figure \ref{fig:square_grid} for a visualisation of this grid.

In the case of a time evolution that we wish to discretise we can alternately specify the initial time value (usually $t_0=0$) and a time interval $\Delta t$ without an upper boundary and let the code evolve until certain conditions are met.

We can now discretise our variables by using the grid shown in figure \ref{fig:square_grid}.  If $F(x,y)$ is a function defined on our square region, its discretisation is calculated by evaluating the function at the nodes/grid points defined above.
$$F_{ij}=F(x_i,y_j)$$

In practice the grid and boundaries that one chooses depend largely on the problem being studied, and sometimes require re-formulation after working with the problem for a while (i.e. when new boundary conditions, or a larger grid, or a coordinate basis change are required).

\section{Discretisation of Derivatives (Finite Differences)}\label{sec:discderiv}
The next step is to discretise our derivative operators.  When taking derivatives of a continuous function $F(x,y)$ we use the standard expression:
$$\frac{\partial F}{\partial x}=\lim_{\Delta x \rightarrow 0}\frac{F(x+\Delta x,y)-F(x,y)}{\Delta x}$$
if we are taking a derivative along the $x$ direction.  Since we cannot take the limit as the grid zone interval goes to zero on a finite computing device, we must approximate our derivatives and set a bound on the errors we encounter.

Taylor's theorem for approximating the value of a function $F(x,y)$ at an adjacent point ($F(a+\Delta x,b)$) to a known point ($F(a,b)$) is
\begin{eqnarray}\label{eqn:taylor}
F(a+\Delta x,b) & = & F(a,b) + \frac{\partial F(a,b)}{\partial x} (\Delta x) + \frac{\partial^2F(a,b)}{\partial x^2} \frac{(\Delta x)^2}{2!} + \ldots \nonumber \\ \mbox{} & &
 + \frac{\partial^nF(a,b)}{\partial x^n} \frac{(\Delta x)^n}{n!} + R_n \\
R_n & = & \frac{\partial^{n+1}F(\xi,b)}{\partial x^{n+1}} \frac{(\Delta x)^{n+1}}{(n+1)!} = o((\Delta x)^n) \;;\; a \leq \xi \leq a+\Delta x
\end{eqnarray}
We note that $R_n$ is called the remainder or error term and it is ``little o'' (i.e. of order) of $(\Delta x)^n$, which means that it goes to zero \emph{faster} than $(\Delta x)^n$.  So when we say that a numerical scheme is ``$n$th order correct'', we mean that it is $o((\Delta x)^n)$ as the error converges faster than $n$th order.  We will now use $h = \Delta x$ as a finite approximation on our grid spacing, and using equation (\ref{eqn:taylor}) we can define discretised derivatives.

There are three commonly used difference schemes, which are based on the forward difference ($D_f$), backward difference ($D_b$) and central difference ($D_c$).  Given a differentiable function $F=F(x,y)$ and grid spacing $h$ the simplest differencing equations are given by:
$$D_f[F] = F(x+h,y) - F(x,y) $$
$$D_b[F] = F(x,y) - F(x-h,y)$$
$$D_c[F] = F(x+h,y) - F(x-h,y)$$
And the errors in these differencing schemes are given by simple application of equation (\ref{eqn:taylor}):
$$\lim_{h \rightarrow 0}\frac{D_f[F]}{h} - F_x = o(h)$$
$$\lim_{h \rightarrow 0}\frac{D_b[F]}{h} - F_x = o(h)$$
$$\lim_{h \rightarrow 0}\frac{D_c[F]}{2h} - F_x = o(h^2)$$
One can see how the central differencing method might be preferable from an error perspective.
We can therefore define higher order correct derivatives by application of (\ref{eqn:taylor}) at multiple grid points, and choose the differencing operator that is appropriate for our situation based on boundary conditions and/or a perturbation analysis of the structure of the variables' differential equations\footnote{Frequently referred to as the ``conditioning'' of the problem.}.

\subsection{Second Order Derivatives}\label{sec:2ndordderiv}
\begin{figure}
\centering
\includegraphics{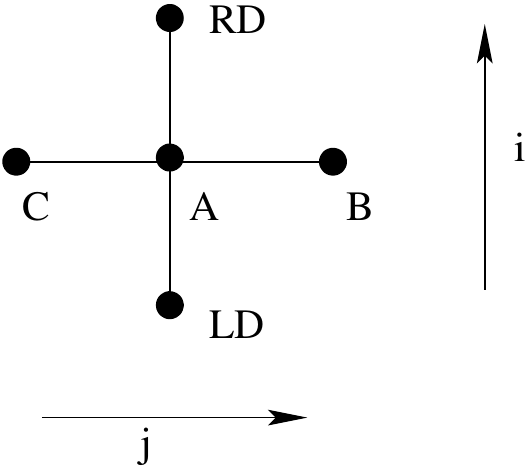}
\caption[2nd order 2-D spatial stencil]{Stencil used in a 2nd-order correct 2-D discretisation.}
\label{fig:2ndordstencil}
\end{figure}

For calculating 2nd order derivatives, we use the method of fitting a single variable quadratic function to three points equally spaced around a central node.
Our stencil for 2nd-order correct evolutions is given in figure (\ref{fig:2ndordstencil}).
The general fitting function in one direction is
$$ f(x)=A x^2 + B x + C $$
with 3 known values\footnote{Note that because the \emph{derivative} of a function does not change if we perform horizontal or vertical translations, we can transform our general point $x=x_0$ to $x=0$ when calculating derivative terms to simplify the equations we have to solve.}
$$ f(0)= C $$
$$ f(\Delta x)=A (\Delta x)^2 + B (\Delta x) + C $$
$$ f(-\Delta x)=A (\Delta x)^2 - B (\Delta x) + C $$
which yields the familiar 2nd order correct centered difference equations with equal spacing between nodes:
\begin{equation}\label{eqn:2ordderiv1} f_{x (i,j)}=\frac{f_{i+1,j}-f_{i-i,j}}{2 \Delta x}\end{equation}
\begin{equation}\label{eqn:2ordderiv2} f_{xx (i,j)}=\frac{f_{i+1,j}-2 f_{i,j} + f_{i-1,j}}{(\Delta x)^2}\end{equation}

\subsection{Fourth order correct derivatives}\label{sec:4thordderiv}
\begin{figure}
\centering
\includegraphics{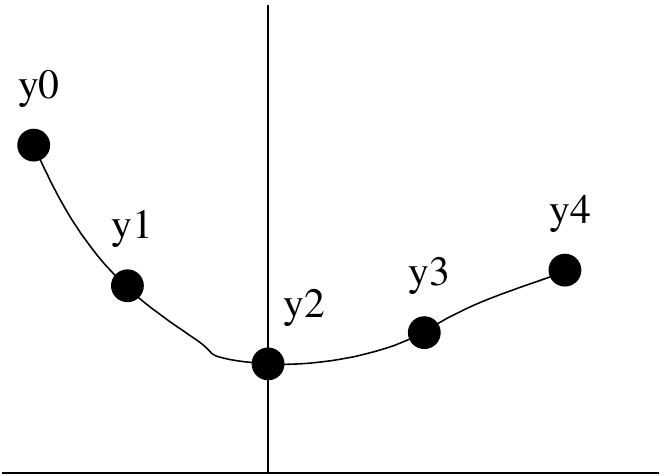}
\caption[Fitting a 4th order polynomial to 5 evenly spaced points]{conceptual diagram of fitting a 4th order polynomial to 5 evenly spaced points}\label{fig:4ordderiv}
\end{figure}

\begin{figure}
\centering
\includegraphics{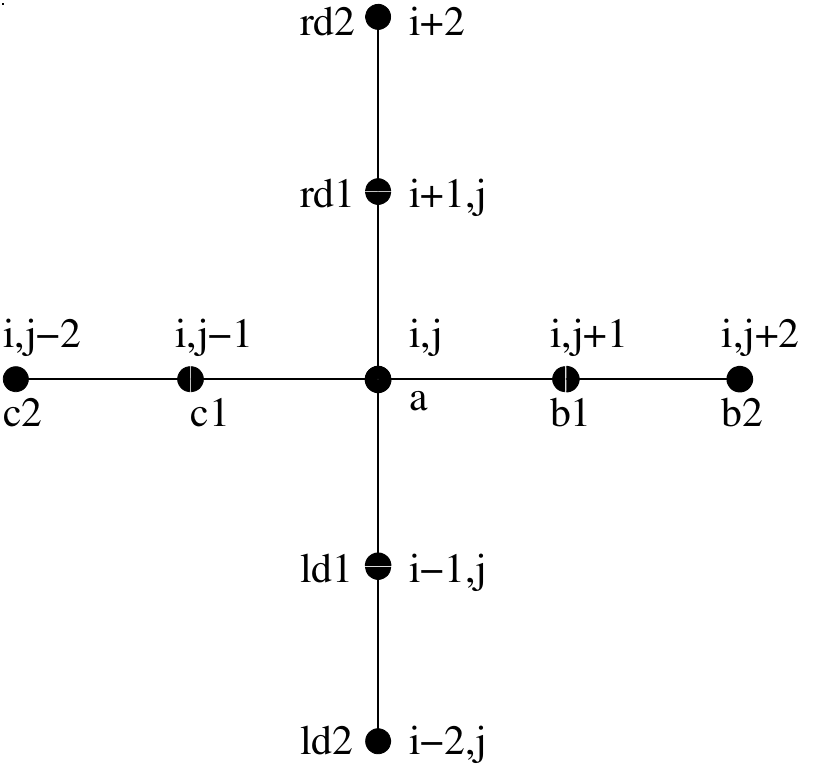}
\caption[4th order 2-D spatial stencil]{The 4th order 2-D stencil with corresponding labels used inside the code to identify stencil points}
\label{fig:4thordstencil}
\end{figure}

We will need the equations for 4th order correct derivatives, so we use the same general method as 2nd order, with an expanded grid.
To this end we fit a 4th order polynomial to 5 points centered and equally spaced around the central node\footnote{This is equivalent to using Taylor's Theorem to find a particular fitting - it is not a unique solution, but it is \emph{symmetric}.} (see figure \ref{fig:4ordderiv}).
In order to have sufficient grid points to have a higher order correct method, instead of 3 points in each coordinate direction we use 5 as in figure (\ref{fig:4thordstencil}).
Our general fitting function is:
$$ f(x)=A x^4 + B x^3 + C x^2 + D x + E $$
with 5 known values:
$$ y_2=f(0)= E $$
$$ y_3=f(\Delta x)=A (\Delta x)^4 + B (\Delta x)^3 + C (\Delta x)^2 + D (\Delta x) + E $$
$$ y_1=f(-\Delta x)=A (\Delta x)^4 - B (\Delta x)^3 + C (\Delta x)^2 - D (\Delta x) + E $$
$$ y_4=f(2\Delta x)=16 A (\Delta x)^4 + 8 B (\Delta x)^3 + 4 C (\Delta x)^2 + 2 D (\Delta x) + E $$
$$ y_0=f(-2\Delta x)=16 A (\Delta x)^4 - 8 B (\Delta x)^3 + 4 C (\Delta x)^2 - 2 D (\Delta x) + E $$
where once again we note that we can translate vertically or horizontally (i.e. $x_0 \rightarrow 0$ to simplify the equations) and the \emph{derivative} terms will be the same.  Solving the above for the coefficients (i.e. 4th order derivatives) yields:
$$A=\frac{f^{(4)}_{x}(i,j)}{4!}=\frac{1}{24(\Delta x)^4}[y_0-4y_1+6y_2-4y_3+y_4] $$
$$B=\frac{f^{(3)}_{x}(i,j)}{3!}=\frac{1}{12(\Delta x)^3}[-y_0+2y_1-2y_3+y_4] $$
$$C=\frac{f_{xx}(i,j)}{2!}=\frac{1}{24(\Delta x)^2}[-y_0+16y_1-30y_2+16y_3-y_4] $$
$$D=f_{x}(i,j)=\frac{1}{12(\Delta x)}[y_0-8y_1+8y_3-y_4] $$
$$E=f(i,j)=y_2$$
the terms we use most frequently are:
\begin{equation}\label{eqn:4ordderiv1} f_{x} (i,j)=\frac{f_{i-2,j}-8 f_{i-1,j}+8 f_{i+1,j}-f_{i+2,j}}{12 \Delta x} \end{equation}
\begin{equation}\label{eqn:4ordderiv2} f_{x x} (i,j)=\frac{-f_{i-2,j}+16 f_{i-1,j}-30 f_{i,j}+16 f_{i+1,j}-f_{i+2,j}}{12 (\Delta x)^2} \end{equation}
which leads to the 2-D stencil in figure (\ref{fig:4thordstencil}).

Higher order derivatives can be found with
\begin{equation}\label{eqn:4ordderiv3} f^{(3)}_{x}=\frac{1}{2(\Delta x)^3}[-y_0+2y_1-2y_3+y_4] \end{equation}
\begin{equation}\label{eqn:4ordderiv4} f^{(4)}_{x}=\frac{1}{(\Delta x)^4}[y_0-4y_1+6y_2-4y_3+y_4] \end{equation}

Solving for mixed second derivatives yields:
\begin{eqnarray}\label{eqn:4ordmixedbad}
f_{xy}(i,j)&=&\frac{1}{144~\Delta x~\Delta y}(f_{i-2,j-2} - 8 f_{i-1,j-2} + 8 f_{i+1,j-2} - f_{i+2,j-2} \nonumber \\ \mbox{}
& & - 8 f_{i-2,j-1} + 64 f_{i-1,j-1} - 64 f_{i+1,j-1} + 8 f_{i+2,j-1} \nonumber \\ \mbox{}
& & + 8 f_{i-2,j+1} - 64 f_{i-1,j+1} + 64 f_{i+1,j+1} - 8 f_{i+2,j+1}  \nonumber \\ \mbox{}
& & - f_{i-2,j+2} + 8 f_{i-1,j+2} - 8 f_{i+1,j+2} + f_{i+2,j+2})
\end{eqnarray}
Which is improved upon in section \ref{sec:mixed4thord}.

\subsubsection{Mixed 4th order derivatives}\label{sec:mixed4thord}
When taking mixed derivatives in both the ``$x$'' and ``$y$'' directions, the resulting stencil has a strange coupling that leads to unstable numerical results.

This was especially evident when attempting to create a potential formulation of the momentum constraints using equations (\ref{eqn:3p1:mom3}); many possible expressions had only first order derivative terms and the mixed partial derivative $\partial_{xy}$.  As the first order derivatives don't involve any terms containing the value at the points in question\footnote{See equation (\ref{eqn:2ordderiv1}) for 2nd order or (\ref{eqn:4ordderiv1}) for 4th order - neither have any $f_{i,j}$ terms.}, the mixed partial derivatives don't either - so the coefficient matrix for the differential operator has no diagonal terms.  The lack of coupling becomes evident if we write the coefficients in equation (\ref{eqn:4ordmixedbad}) in a more visual form:

$$\begin{array}{cc|c|cc}
1 & -8 & 0 & 8 & -1 \\
-8 & 64 & 0 & -64 & 8 \\ \hline
0 & 0 & 0 & 0 & 0 \\ \hline
8 & -64 & 0 & 64 & -8 \\
-1 & 8 & 0 & -8 & 1 \\
\end{array} \times\frac{1}{144} $$

This leads to the coefficient matrix being weighted on the bands, and there is no diagonal dominance in the resulting matrix equation to help push the solution towards stability.  To fix this we created a stencil that has non-zero entries along the major axis and is symmetrical to avoid any numerical artefacts that tend to show up in asymmetric stencils.  As there are any number of stencils that will calculate the 4th order derivatives at a point, we need only put some constraints on the resulting set of linear equations in order to generate our stencil.

To wit, we use the Taylor expansion of a function in two coordinates centered about the point $(x_0,y_0)$, with neighbouring points on our grid labeled via
$$(x_k,y_l)=(x_0+k\Delta x,y_0+l\Delta y)~;~k=-2...2~;~l=-2...2$$
which gives the Taylor expansion
\begin{eqnarray}f(x_k,y_l) & = & f(x_0,y_0) + \sum^{n_{\rm max}}_{n=1}\sum_{i=1}^{n+1} \frac{k^{i-1}l^{n+1-i}}{(i-1)!(n+1-i)!}\frac{\partial^{i-1}}{{\partial x}^{i-1}}\frac{\partial^{n+1-i}}{{\partial y}^{n+1-i}}f(x_0,y_0)~ (\Delta x)^{i-1} (\Delta y)^{n+1-i} \nonumber \\ \mbox{}
& &  + O(\Delta x^{n_{\rm max}+1})
\end{eqnarray}

This allow us to express any derivative term as:
$$\frac{\partial^i}{\partial x^i}\frac{\partial^j}{\partial y^j}f(x_0,y_0) = \sum_{k=-2}^2 \sum_{l=-2}^2 c^{(ij)}_{kl}f(x_k,y_l)$$
where the $c^{(ij)}_{kl}$ represent the required finite difference coefficients.

As we are using the $25$ stencil points around $(x_0,y_0)$, we have $25$ equations in 
$$\frac{(n_{\rm max}+1)(n_{\rm max}+2)}{2}-1$$
 unknowns (the derivatives themselves). To 4th order we have an overdetermined system, so we are allowed a large amount of freedom in choosing our stencil coefficients.  Therefore we choose some constraints such that the stencil is as symmetric as possible to reflections across our coordinate axis, as we don't wish to introduce an artificial numerical directional preference.  We also want non-zero entries on the coordinate axis (as all other coupling is in those directions), and a non-zero entry for $(x_0,y_0)$.

One possible solution\footnote{Many thanks to D. Hobill for working out the details.}, that we use in this thesis, is:

$$\mathrm{coeff}(i,j)=-\frac{1}{48\Delta  x \Delta y}\left[\begin{array}{ccccc}
0 & -1 & -1 & 1 & 1 \\
-3 & 18 & 4 & -18 & -1 \\
3 & 0 & -6 & 0 & 3 \\
-1 & -18 & 4 & 18 & -3 \\
1 & 1 & -1 & -1 & 0 \\
\end{array}\right]$$

and the mixed derivative at any point $(k,l)$ is therefore:

$$f_{xy}(k,l)=\sum_{i,j=1}^{5} f(k+3-i,l-3+j) \times  \mathrm{coeff}(i,j)$$

\subsection{Boundary Conditions}\label{subsec:bc_gen}
Another set of important considerations when calculating finite representations of derivatives on a discretised grid are the boundaries.  Since the finite difference operators all involve some combination of different grid points, calculating derivatives along boundaries must be treated differently than derivative calculations at the interior points depending on whether we employ Dirichlet or Neumann boundary conditions\footnote{Or some combination thereof, or at least a local functional conditioning, something we introduce later in this thesis.}.  A Dirichlet boundary condition specifies the value of a function along a boundary, whereas a Neumann boundary condition specifies the derivative along the boundary.  Using our square domain example from above in equation (\ref{eqn:square_aa_ex}) we could specify Dirichlet conditions along $x_0=0,x_N=a$ by
$$F(0,y)=F(a,y)=0$$
We could also specify Neumann boundary conditions along $y_0=0,y_N=a$ by
$$F_y(x,0)=F_y(x,a)=0$$
The first $x$ derivative using the central difference operator at $i=1,j=1$ is given by:
$$F_x(x_1,y_1)=\frac{F(x_2,y_1)-F(x_0,y_1)}{2\Delta x}=\frac{F(x_2,y_1)}{2\Delta x}$$
To calculate the first $x$ derivative at $i=0,j=1$ requires that we either (a)(i) know something about the symmetry properties across the boundary or (ii) some method of extrapolating values (like a local polynomial approximation) to create a ``phantom grid point'' at $i=-1,j=1$, (b) use the simplest forward differencing operator (which has lower order accuracy than its equivalent centered difference version) to calculate the derivative, or (c) use a higher order forward differencing operator and include more grid points.

Any of these methods will work, and each have their own merits and faults depending on the equations being discretised.  We use methods (a)(i) and (a)(ii) exclusively in this thesis to ensure consistent conditioning of our differencing operators throughout the grid.

The first $y$ derivative using the central difference operator at $i=1,j=1$ is given by
$$F_y(x_1,y_1)=\frac{F(x_1,y_2)-F(x_1,y_0)}{2\Delta y}$$
The first $y$ derivative at $i=1,j=0$ is known from our boundary conditions to be zero.  Calculating the second $y$ derivative at $i=1,j=0$ one encounters the same problem as above, namely that the central difference operator requires a grid point outside our grid zone. This can be dealt with using any of the same methods as before.

\section{Discretisation and the Evaluation of Integrals}\label{sec:discintegral}
Recalling that the geometric interpretation of a single variable integral is to calculate an area under a curve, we can divide up a continuous curve into discrete grid zones and approximate the function using a constant, linear, quadratic, etc. fitting function\footnote{This is frequently the motivation for the definition of the Riemann integral in first year calculus.} then calculate the area under that known, simple, fitting function.

\begin{figure}[h]
\centering
\psfrag{D}{$\Delta$}
\includegraphics{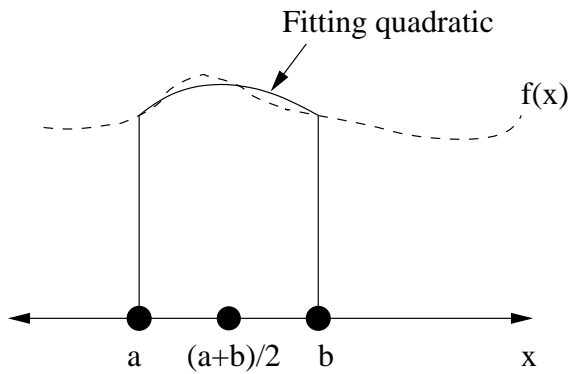}
\caption[Simpson's Rule schematic]{Schematic of fitting a quadratic function on the interval $[a,b]$ for use in Simpson's rule to calculate numerical integrals.}
\label{fig:simpson_fit}
\end{figure}
As an example, we can approximate the function $f(x)$ on the interval $[a,b]$ with a quadratic function as in figure \ref{fig:simpson_fit}.  As we need three points to define a parabola uniquely we will choose the points $x\in\{a,(a+b)/2,b\}$ so as to fit onto a grid with constant zone intervals $h=\frac{b-a}{2}$.  Simpson's Rule then states that
$$\int_a^bf(x)dx = \frac{b-a}{6}\left[f(a) + 4f\left(\frac{a+b}{2}\right)+f(b)\right] + R(h)$$
where $R(h)$ is once again our error term and is given by
$$R(h)=\frac{(b-a)^5}{2880}\frac{d^4f(\xi)}{dx^4} \;;\; a \leq \xi \leq b$$

We can then compute the integral over a larger interval by taking Simpson's rule on each subinterval and adding them together\footnote{See any elementary calculus book for a formal derivation of this, for example \cite{anton} $\S 9.7$}.

The basic methodology of dividing up a continuous interval into subintervals and calculating areas with approximating functions is fairly straight forward and easy to program, provided we choose the appropriate numerical method to match the location and number of grid zones while also satisfying a preset limitation on the size of acceptable errors.

\section{Numerical Errors}\label{sec:numerrors}
The two notable types of errors we must keep in mind when performing numerical calculations are roundoff and truncation errors.  We offer a brief introduction to each in this thesis, however Goldberg \cite{goldberg} offers an in-depth discussion.

\subsection{Truncation Errors}\label{subsec:truncerror}
When defining a finite difference operator, we must decide to what ``order''  we wish to truncate our Taylor series approximation when deriving the formulas for calculating numerical derivatives.  At this point, we are neglecting all terms of higher order in the grid spacing $\Delta x$, which means that there is an error in the finite difference representation of a derivative. This is called a ``truncation error''.

There are two methods to mitigate truncation errors:
\begin{enumerate}
\item Use higher order correct schemes which require more grid points to approximate a derivative. This in turn requires more CPU computational cycles and memory.  There also can come a point where the data is ``overfitted'' if the fitting polynomial is too high an order. Theoretically if one has $N+1$ total grid points along a dimension (as in our square region example) you can't fit the numerical data with a polynomial of order higher than $N$th order (although this is rarely a concern for the number of grid points vastly exceeds the order of accuracy of the numerical method in most cases).
\item Use smaller grid spacing $\Delta x$.  In equation (\ref{eqn:taylor}) the remainder term is dependent on $(\Delta x)^{n+1}$, so decreasing $\Delta x$ will make the remainder term smaller.  This also means that one would need more grid points to cover the same bounded region.  If, for example, $\Delta x \rightarrow \frac{\Delta x}{2}$ then for our example square region above we need almost twice as many grid points $(N+1 \rightarrow 2N+1)$ along the $x$ coordinate to cover the same region.  On a 2D grid this would then quadruple the CPU cycle and memory requirements, so it is not a decision one makes lightly, with only limited resources available.
\end{enumerate}

\subsection{Rounding Errors}\label{subsec:rounderror}
Rounding errors arise because computers are finite devices and must therefore store numbers as finite length binary numbers.  This means that there is a minimum precision that can be stored that depends on hardware architecture\footnote{i.e. x86, x64, SPARC, IBM mainframes.} and the software compiler.  Most modern hardware/software systems are based on $64$-bit architectures\footnote{There are a few modern true $128$ bit implementations, but they are only on IBM mainframe or z-series machines.} and the vast majority use IEEE 754 as an implementation standard.  This means that the $64$th bit, being either $1$ or $0$, determines the roundoff error in every calculation for double precision floating point algebra.

IEEE 754 specifies that of the $64$ bits used in storing a double precision floating point number, one bit is used for the sign $S$ of the number, 11 bits are used for the exponent $E$ and the remaining 52 bits are used for the fractional part.  Assuming that the bits representing the fractional part are
$$b_0,b_1,b_2,\ldots,b_{51} \;;\; b_i\in\{0,1\}$$
the conversion from binary to decimal representation uses the formula
$$(-1)^S \left(1+\sum_{i=1}^{52}b_{52-i}2^{-i}\right)2^{E-1023}$$
From this it is evident that the last bit being $0$ or $1$ gives us a maximum rounding error \emph{relative} to our numerical value of one part in $2^{-53}\sim 1.11\times 10^{-16}$, or approximately $16$ decimal places of precision in base $10$.

One interesting property of this method of storing decimal numbers as binary digits in a computer is that some very simple numbers, like $0.1$, do not have a finite \emph{exact} representation in IEEE 754.

While $2^{-53}$ may seem like a ``small'' relative error, when roundoff error is compounded over thousands and millions of calculations in an ill-conditioned problem it can cause measurable/large errors.  This was more of an issue when the maximum floating point size was $16$ or $32$ bits in older architectures, as relative rounding errors are larger and can swamp high-precision codes quickly.  There are two methods of mitigating these rounding errors
\begin{itemize}
\item Move to a larger number of bits for storing floating point numbers.  This is not currently practical given the specialised hardware required for 128-bit implementations.
\item Careful re-arrangement of operands can decrease the total number of roundoff errors that are performed at each calculation step.  Expressions that are analytically equivalent are not always equivalent numerically due to roundoff errors.
\end{itemize}

Another form of roundoff error that can occur is when we add two numbers and must discard a portion of one number or the other.  For example, if we have a computer that stores two decimal digits, when we add the two numbers $12$ and $1.1$ we must round $12+1.1=13$ losing $0.1$ along the way.

While the re-arrangement of operands can mitigate this problem, it does significantly complicate and slow down even simple summation processes.

\section{Solving Partial Differential Equations via Discretisation}
Now that we have a mathematical framework with which to analyze discretised approximations to differential equations, numerical algorithms for dealing with different types of partial differential equations need to be introduced. In 3+1 ADM GR we encounter both elliptic and hyperbolic PDEs.

\subsection[Fourth Order Correct Runge-Kutta Algorithm]{Fourth Order Correct Runge-Kutta Algorithm - A Numerical Method For Differential Equations Depending On a Single Dimension}
Throughout the course of attempting to solve various differential equations in this thesis, there were times when it was appropriate to attempt the problem at hand with a Runge-Kutta algorithm.

The fourth order Runge-Kutta algorithm RK4 (which is a particular member of the Runge-Kutta family of solvers) is used to solve differential equations in one independent variable.  For a single equation in one variable one specifies an initial value at one point on the grid, and proceeds to fill in the values for the solution as you move to the other end of the grid.

Consider, for example, the equation
\begin{equation}\label{eqn:rk1}A~\alpha_{rr}+B~\alpha_r+C~\alpha=0\end{equation}
where $\alpha=\alpha(r)$ on the interval $[r_0,r_{max}]$.  We then define an auxiliary variable
\begin{equation}\label{eqn:rkaux}\omega=\alpha_r=f(\omega)\end{equation}
which allows us to rewrite (\ref{eqn:rk1}) as a system of two first-order ODE's.  The first equation is given by \ref{eqn:rkaux}, and the second is:
\begin{equation}\omega_r=-\frac{B}{A}\omega-\frac{C}{A}\alpha=g(\omega,\alpha,A,B,C)\end{equation}
Let us discretise the $r$ interval $[r_0,r_{max}]$ into $n_{max}$ finite intervals of equal size $h$, so that
$$r_n=r_0+nh$$
Once we have specified the boundary (or initial) conditions
$$\alpha_0=\alpha(r_0), \omega_0=\omega(r_0)$$
we can proceed with computing successive values of $\alpha_n$ and $\omega_n$ using the following algorithm.\footnote{The derivation of this algorithm can be found in many books on numerical methods, including \cite{burden}. See also \cite{ross_ode,mathphys} for discussions.}
\begin{eqnarray}
\alpha_{n+1} & = & \alpha_n + K \\ \nonumber \mbox{}
K & = & \frac{1}{6}(k_1+2k_2+2k_3+k_4) \\ \nonumber \mbox{}
k_1 & = & h\times f(\omega_n) \\ \nonumber \mbox{}
k_2 & = & h\times f(\omega_n+\frac{m_1}{2}) \\ \nonumber \mbox{}
k_3 & = & h\times f(\omega_n+\frac{m_2}{2}) \\ \nonumber \mbox{}
k_4 & = & h\times f(\omega_n+m_3) \\
\omega_{n+1} & = & \omega_n + M \\ \nonumber \mbox{}
M & = & \frac{1}{6}(m_1+2m_2+2m_3+m_4) \\ \nonumber \mbox{}
m_1 & = & h\times g(\alpha_n,\omega_n,A_n,B_n,C_n) \\ \nonumber \mbox{}
m_2 & = & h\times g(\alpha_n+\frac{k_1}{2},\omega_n+\frac{m_1}{2},A_{n+\frac{1}{2}},B_{n+\frac{1}{2}},C_{n+\frac{1}{2}}) \\ \nonumber \mbox{}
m_3 & = & h\times g(\alpha_n+\frac{k_2}{2},\omega_n+\frac{m_2}{2},A_{n+\frac{1}{2}},B_{n+\frac{1}{2}},C_{n+\frac{1}{2}}) \\ \nonumber \mbox{}
m_2 & = & h\times g(\alpha_n+k_3,\omega_n+m_3,A_{n+1},B_{n+1},C_{n+1}) \\ \nonumber
\end{eqnarray}

While this method provides $O(h^4)$ precision, higher order RK methods as well as techniques that employ an adaptive step size are also well known.

This method was used several times while studying the problems discussed in Appendix \ref{chap:testgridcoord} where many of the PDE's of the 3+1 ADM decomposition reduce to ODEs (i.e. the elliptic PDEs reduce to two-point boundary value problems in one spatial dimension).  One can then use the RK method in conjunction with a ``shooting'' method to find a solution across the entire grid.

In the above case, as we have a two point boundary problem (at $r_0$ and $r_{max}$) and a second order PDE we need to specify two boundary values (for example $\alpha$ and $\omega$) on each of the boundaries.  We can then integrate from one boundary to the other\footnote{Or from both ends and match the solutions in the middle somewhere.} (i.e. ``shoot'') using the RK algorithm to see if the values on the boundary obtained using the RK algorithm match the imposed boundary values to within a specified error.  If they do not, we can perturb/interpolate/extrapolate our boundary values to provide a new set of boundary values and integrate again until the error in the solution falls within a specified tolerance.

For example, suppose the conditions $\alpha(r_{max})=1, \; \omega(r_{max})=0$ are imposed on the outer boundary. We can guess an inner boundary value for $\alpha(r_0)=\alpha_0$, use the symmetry condition $\omega(r_0)=0$ and proceed to fill in the rest of the grid using the algorithm above.  We then calculate the errors in our outer boundary values at $r_{max}$ relative to the desired values, and use them to adjust our guess for the inner boundary condition $\alpha_0$ (using a binary search or some other appropriate method).

This method has also been used very successfully in the study of geodesic motion in a given spacetime (e.g. around Kerr black holes and other gravitational objects).

\subsection[Solving elliptic PDEs to second order]{Solving Multi-Dimensional Second Order Linear Elliptic PDEs With Second Order Discretisation}\label{sec:2orpde2ord}
Assuming the general form of a 2nd order elliptic equation we wish to solve numerically is of the form\footnote{We omit consideration of the cross-term $\psi_{xy}$ for the simple reason that we never need it.}
$$ A \;\psi_{xx} + B \;\psi_{yy} +  C \; \psi_{x} + D \; \psi_{y} + E \; \psi = G $$
and using the second order accurate finite difference discretisation for the first and second order partial derivatives of $\psi$ (given in equations (\ref{eqn:2ordderiv1}) and (\ref{eqn:2ordderiv2})), then collecting terms, we find that
\begin{eqnarray}
\left(\frac{A}{(\Delta x)^2}+\frac{C}{2 \Delta x}\right) \psi_{i+1,j}
+\left(\frac{A}{(\Delta x)^2}-\frac{C}{2 \Delta x}\right) \psi_{i-1,j}
+\left(-\frac{2 A}{(\Delta x)^2}-\frac{2 B}{(\Delta y)^2}+E\right) \psi_{i,j}
& & \nonumber \\ \mbox{}
+\left(\frac{B}{(\Delta y)^2}+\frac{D}{2 \Delta y}\right) \psi_{i,j+1}
+\left(\frac{B}{(\Delta y)^2}-\frac{D}{2 \Delta y}\right) \psi_{i,j-1} = G_{i,j}\;\;\; & & {}
\end{eqnarray}
This can be written schematically as
$$rd_{i,j} \psi_{i+1,j} + ld_{i,j} \psi_{i-1,j} + a_{i,j} \psi_{i,j} + b_{i,j} \psi_{i,j+1} + c_{i,j} \psi_{i,j-1} = G_{i,j} $$
using the stencil in figure (\ref{fig:2ndordstencil}).  If our subscript indices $i$ and $j$ are integers given by
$$1 \le i \le m \;;\; 1 \le j \le n$$
(i.e. our discretised grid has $m \times n$ nodes), we can create the vector $\Psi$ such that
$$\Psi_k=\Psi_{(i-1)n+j}=\psi_{i,j} \;;\; 1 \le k \le mn$$
and similarly we can construct the vector $B$ such that
$$B_k=B_{(i-1)n+j}=G_{i,j} \;;\; 1 \le k \le mn$$
This leads to the following matrix equation:

\pagebreak

\rotatebox{270}{ $ 
\left[\begin{array}{cccccccccccc}
a_{1,1} & b_{1,1} & 0 & ... & rd_{1,1} & 0 & \cdots & & & & & 0 \\
c_{1,2} & a_{1,2} & b_{1,2} & & & rd_{1,2} & & & & & & \\
0 & c_{1,3} & a_{1,3} & b_{1,3} & & & rd_{1,3} & & & & & \\
\vdots & & \ddots & \ddots & \ddots & & & \ddots & & & & \\
ld_{2,1} & & & c_{2,1} & a_{2,1} & b_{2,1} & & & rd_{2,1} & & & \\
0 & ld_{2,2} & & & c_{2,2} & a_{2,2} & b_{2,2} & & & rd_{2,2} & & \\
\vdots & & \ddots & & & & \ddots & & & &\ddots & \\
 & & & ld_{m-1,n} & & & c_{m-1,n} & a_{m-1,n} & b_{m-1,n} & & & rd_{m-1,n} \\
 & & & & \ddots & & & & \ddots & & & \\
 & & & & & \ddots & & & & \ddots & & 0 \\
 & & & & & & ld_{m,n-1} & & & c_{m,n-1} & a_{m,n-1} & b_{m,n-1} \\
0 & \cdots & & & & & 0 & ld_{m,n} & ... & & c_{m,n} & a_{m,n} \\
\end{array} \right]
\left[\begin{array}{c} \Psi_{1} \\ \Psi_{2} \\ \Psi_{3} \\ \vdots \\ \Psi_{n+1} \\ \vdots 
\\  \\ \Psi_{(m-1)n} \\ \vdots \\   \\ \Psi_{mn-1} \\ \Psi_{mn} \\ \end{array} \right]
=\left[\begin{array}{c} B_{1} \\ B_{2} \\ B_{3} \\ \vdots \\ B_{n+1} \\ \vdots 
\\  \\ B_{(m-1)n} \\ \vdots \\   \\ B_{mn-1} \\ B_{mn} \\ \end{array} \right]
$ }

\pagebreak
Where we have used $rd$ to represent the ``right diagonal'' terms and $ld$ to represent the ``left diagonal'' coefficients.

This is a sparse, $(mn) \times (mn)$, banded matrix equation of the form $Ax=b$ that we wish to solve for $x$, and can be solved using a variety of methods.

\subsection[Solving elliptic PDEs to 4th order]{Solving Multi-Dimensional Second Order Linear Elliptic PDEs With Fourth Order Discretisation}\label{sec:2orpde4ord}
Analogously to the previous section, if we assume that our general 2nd order mixed PDE that we want to solve to 4th order in $\Delta x$ and $\Delta y$ is of the form\footnote{Once again we omit consideration of the $\psi_{xy}$ terms.}
\begin{equation} A~ \psi_{xx} + B~ \psi_{yy} + C~ \psi_{x} + D~ \psi_{y} + E~ \psi = G
\end{equation}
we find that after substitution of the fourth order correct discretised derivative operators our matrix coefficients become
\begin{eqnarray}
a & = & -30A{(\Delta y)}^2-30B(\Delta x)^2+12E(\Delta x)^2{(\Delta y)}^2 \nonumber \\ \mbox{}
ld2 & = & (-A+C\Delta x){(\Delta y)}^2 \nonumber \\ \mbox{}
ld1 & = & (16A - 8C\Delta x){(\Delta y)}^2 \nonumber \\ \mbox{}
rd1 & = & (16A + 8C\Delta x){(\Delta y)}^2 \nonumber \\ \mbox{}
rd2 & = & (-A - C\Delta x){(\Delta y)}^2 \nonumber \\ \mbox{}
c2 & = & (-B + D\Delta y)(\Delta x)^2 \nonumber \\ \mbox{}
c1 & = & (16B - 8D\Delta y)(\Delta x)^2 \nonumber \\ \mbox{}
b1 & = & (16B + 8D\Delta y)(\Delta x)^2 \nonumber \\ \mbox{}
b2 & = & (-B - D\Delta y)(\Delta x)^2 \nonumber \\ \mbox{}
r & = & 12G(\Delta x)^2{(\Delta y)}^2 \label{eqn:4thordmatrix}
\end{eqnarray}
which yields a matrix problem similar to that in section (\ref{sec:2orpde2ord}), with wider bands now that we have four additional off-diagonal terms\footnote{We again use ``ld'' to signify left (lower) diagonal banded terms and ``rd'' to signify right (upper) diagonal banded terms, see also Figure (\ref{fig:4thordstencil}).}.
\begin{eqnarray}\label{eqn:4ordstencilexpand}
{a}_{i,j} \, \psi_{i,j} + {b1}_{i,j} \, \psi_{i,j+1} + {b2}_{i,j} \, \psi_{i,j+2} + {c1}_{i,j} \, \psi_{i,j-1} + {c2}_{i,j} \, \psi_{i,j-2} & & \nonumber \\ \mbox{}
+ {rd1}_{i,j} \, \psi_{i+1,j} + {rd2}_{i,j} \, \psi_{i+2,j} + {ld1}_{i,j} \, \psi_{i-1,j} + {ld2}_{i,j} \, \psi_{i-2,j} & = & {r}_{i,j}
\end{eqnarray}

\subsection{Solving Matrix Equations}\label{subsec:matrixeqns}
There are many methods for solving matrix equations of the form $Ax=b$ for the vector $x$, each suited to the particularities of the problem at hand.

For smaller, well-conditioned systems one can use a Gaussian elimination method which closely mimics the traditional method one first learns of solving linear equations: add and subtract equations (or rows in a matrix) in multiples until a matrix with only upper or lower triangular elements is obtained.  Back-substitution is then used to arrive at a solution.

These methods can be slow and numerically error prone\footnote{See, for example, \cite{burden} for a discussion.}. Although improvements can be introduced (e.g. partial pivoting and scaling), the Gaussian elimination class of solvers are not well suited to large, sparse matrix problems. If $A$ is a matrix that is mostly ``empty'' (composed of a large number of zero values) one can instead use a number of well-designed algorithms that take advantage of the sparseness of the coefficient matrix.

For a more in-depth discussion of various advanced techniques for solving matrix equations we refer the reader to \cite{golub,saad_sparse}.  One class of algorithms well suited to sparse matrix problems begins with an approximate solution and iterates until a solution with an acceptable tolerance is reached.  The methods are called ``relaxation techniques'' and rely on approximating an elliptic equation with a diffusion process.

Another class of algorithms calculates gradients of steepest descent for the remainder term $b-Ax$ to converge to a solution, and are called conjugate gradient methods.  We will briefly discuss relaxation techniques and leave discussion of an appropriate conjugate gradient technique until later.

\subsubsection{Relaxation Techniques}\label{subsubsec:relaxmatrix}
One form of sparse matrix solver that was initially implemented (and abandoned) was the Gauss-Seidel/Jacobi family of iterative solvers.

Firstly we discuss the Jacobi method which is derived by taking our discretised differential equation (for example equation (\ref{eqn:4ordstencilexpand})), and solving for the $\psi(i,j)$ term iteratively where $n$ represents the $n$th iteration:
\begin{eqnarray}\label{eqn:4ordrelax}
\psi^{n+1}_{i,j} & = & {r}_{i,j} - \frac{1}{{a}_{i,j}}\left[ {b1}_{i,j} \, \psi^n_{i,j+1} + {b2}_{i,j} \, \psi^n_{i,j+2} + {c1}_{i,j} \, \psi^n_{i,j-1} + {c2}_{i,j} \, \psi^n_{i,j-2} \right. \nonumber \\ \mbox{} & & \left.
+ {rd1}_{i,j} \, \psi^n_{i+1,j} + {rd2}_{i+2,j} \, \psi^n_{i,j} + {ld1}_{i,j} \, \psi^n_{i-1,j} + {ld2}_{i,j} \, \psi^n_{i-2,j}\right]
\end{eqnarray}

The first difficulty with this method is the requirement that all $a_{i,j}$ terms are non-zero, which cannot be guaranteed in the general case\footnote{One class of problematic situations is mentioned in section \ref{sec:mixed4thord}.}.  The largest difficulty, however, is that we must specify a good initial guess at $n=0$ (preconditioning) to ensure convergence\footnote{Saad \cite{saad_sparse} states: ``In general, the reliability of iterative techniques, when dealing with various applications, depends much more on the quality of the preconditioner than on the particular Krylov subspace accelerators used''.}.

We can speed convergence in some cases by using the already-updated values of $\psi$, called a Gauss-Seidel method.  Assuming we start at $i=2, j=2$ (see figure \ref{fig:2dgrid} for an idea of what this means), and iterate angularly first, then radially, we find:
\begin{eqnarray}\label{eqn:4ordrelax2}
\psi^{n+1}_{i,j} & = & {r}_{i,j} - \frac{1}{{a}_{i,j}}\left[ {b1}_{i,j} \, \psi^n_{i,j+1} + {b2}_{i,j} \, \psi^n_{i,j+2} + {c1}_{i,j} \, \psi^{n+1}_{i,j-1} + {c2}_{i,j} \, \psi^{n+1}_{i,j-2} \right. \nonumber \\ \mbox{} & & \left. 
+ {rd1}_{i,j} \, \psi^n_{i+1,j} + {rd2}_{i+2,j} \, \psi^n_{i,j} + {ld1}_{i,j} \, \psi^{n+1}_{i-1,j} + {ld2}_{i,j} \, \psi^{n+1}_{i-2,j}\right]
\end{eqnarray}

We then iterate through
$$\psi^{n+1}=g(\psi^n,\psi^{n+1})$$
until a convergence criteria is satisfied or we exceed a pre-determined number of iterations.

This second method sometimes yields better results. However it can also cause problems if the difference $\psi^n_{i,j}-\psi^{n+1}_{i,j}$ is too large to allow a smooth ``diffusion'' process at points in the grid.

In either case, the speed of convergence is less than one tenth what is possible by using a conjugate gradient method. In some cases this method is incapable of converging in a reasonable number of iterations if the initial guess for the solution is far from the actual solution.

Furthermore, the ``direction'' of iteration seems to have a huge effect on convergence - if one starts at $(i_{max}+2,j_{max})$ (the outer radial boundary on the equator) instead of $(2,2)$ (inner radial boundary on the axis) one gets significantly different characteristics of convergence, stability, and accuracy.  Conjugate gradient algorithms do not suffer from such problems.

\subsection{Explicit Time Evolution}\label{subsec:explicit_time}

In the 3+1 ADM formulation of GR all of the evolution equations with a time derivative component are of the form
\begin{equation}\label{eqn:gentimevol}\frac{\partial \vec{\xi}(i,j)}{\partial t}=\vec{F}(\vec{\xi})\end{equation}
where $\vec{\xi}$ is a vector whose components are the dynamic variables (i.e. metric and extrinsic curvature variables) and $\vec{F}$ is a (quasi-linear) differential operator in the first and second \emph{spatial} derivatives of the components of $\vec{\xi}$.

One method to numerically evolve these equations is to use the 4th order correct centered first derivative in equation (\ref{eqn:4ordderiv1}) but now in the temporal coordinate $t$, and solve explicitly for the current time step.  If $t=k \Delta t$ where $k$ is our time step counter, this leads to the schematic:
\begin{equation}\label{eqn:4ordtime} \vec{\xi}_{k}(i,j) = -12 (\Delta t) \vec{F}_{k-2}(i,j) + \vec{\xi}_{k-4}(i,j) - 8 \vec{\xi}_{k-3}(i,j) + 8 \vec{\xi}_{k-1}(i,j)\end{equation}
So the values $\vec{\xi}_k$ that we are solving for are explicitly calculated using variables that are already known from previous time steps.

To discuss stability, we note that in its simplest one-dimensional form the diffusion equation has the form:
\begin{equation}\label{eqn:diffusion}\frac{\partial u}{\partial t} = \zeta \frac{\partial^2 u}{\partial x^2}\end{equation}
From this we find that explicit second order forward finite discretisation \cite{burden} is conditionally stable as long as
$$\Delta t \le \frac{(\Delta x)^2}{2 \zeta}$$
This means that time steps $\Delta t$ must be quadratically small relative to our spatial steps $\Delta x$ to ensure that numerical errors do not swamp the code.

As the 3+1 ADM evolution equations have many first order time derivatives on the LHS and second order spatial derivatives on the RHS, we can use this result as a conservative\footnote{The wave equation in its simplest form leads to the condition $\Delta t \sim \Delta x$ for many finite difference operators, so this condition is stricter.}, rough guideline or starting point when attempting to determine stability conditions\footnote{Although $\zeta$ can assume a large range of values across the grid points on a hypersurface, let alone over coordinate time, in our simulations.}.

\subsection{Implicit Crank-Nicholson Time Evolution}\label{subsec:cn}
Another method of solving first order hyperbolic time evolution equations is via the Crank-Nicholson method, which is an implicit method.  Recall that our evolution equations with time derivatives have the form of equation (\ref{eqn:gentimevol}).
With $i,j$ our spatial counters in the $x$ and $y$ directions and $k$ our time counter as before, the C-N method can be written as:
\begin{equation}\label{eqn:gen_cn}\frac{\vec{\xi}_{ij}^{k+1}-\vec{\xi}_{ij}^k}{\Delta t} = \frac{1}{2}\left[\vec{F}_{ij}^{k+1}(\vec{\xi}^{k+1})+\vec{F}_{ij}^{k}(\vec{\xi}^k) \right]\end{equation}
where we are trying to solve for $\vec{\xi}^{k+1}$ which appears on both the left and right hand sides, at both the future and current time steps, and across various spatial grid points.

Examining stability once more, when applied to the diffusion equation (\ref{eqn:diffusion}) the Crank-Nicholson algorithm is unconditionally stable, and we find that $\Delta t \sim \Delta x$ for proper error propagation \cite{burden}, which is much more desirable than the explicit method above.  The Crank-Nicholson method\footnote{Or any other implicit method for that matter.} has the added complication of requiring simultaneous solutions to all variables on future time steps, a difficulty we will revisit later.

\chapter{$2+1$ Coordinate, Metric and Gauge Choices}\label{chap:coordgauge}
\bigskip
Now that we have laid down a basic mathematical foundation for GR and numerical analysis, let us proceed with a description of the various coordinate, metric and gauge choices that were made to arrive at a set of equations to evolve for the $2+1$ Brill wave problem.

All symbolic calculations were performed via Maxima\footnote{Maxima is the open source version the computer algebra package Macsyma (circa 1982), which was originally developed at MIT in the 60s.  Macsyma was then later commercialised with limited success and faces an unsure future - hence the split of the code base.} (or hand), and Appendix \ref{appendix:maxima} gives scripts and methodologies for reproducing these results.  As we move into higher dimensions the use of symbolic calculation programs to generate the respective equations becomes more important.

The majority of the numerical work was performed in FORTRAN, with some auxiliary C programs used for quick data conversions.  Matlab\footnote{Matlab is a 4th generation programming language (as opposed to third generation ones like FORTRAN and C) and numerical computation software package.  We forego the computational capabilities of Matlab for the most part as they are too slow for our purposes, except in some simple cases, and instead utilise its graphing capabilities.} was used for visualisations, some small test beds and visual spot checks on boundary conditions (which are very important, given the wide variety of ways that boundary conditions can fail).

\section{The Axisymmetric Formulation}\label{sec:aximetric}
The use of the term \emph{axisymmetric} in this thesis means the following: we use spherical polar coordinates $\{\eta,\theta,\varphi\}$ where $\eta$ is a radial coordinate, $\theta$ is the polar angle and $\varphi$ is the azimuthal angle\footnote{We have the unfortunate notation conflict that our conformal factor is $\phi$ and the azimuthal angle is generally denoted $\phi$, so we will instead label our azimuthal angle as $\varphi$ for the purpose of this thesis.}, and our system is axially symmetric around the $z$ axis ($\theta=0$).\footnote{i.e. $\frac{\partial}{\partial\varphi}$ is a Killing vector.}

We add reflection symmetry across the $z=0$ plane (i.e. $\theta=\frac{\pi}{2}$) so our dynamic (or static) variables are symmetric (or anti-symmetric) across $\theta=0$, $\theta=\frac{\pi}{2}$, $\eta=0$, and also to comply with the Brill conditions (see section \ref{subsec:brillformalism}).

As discussed in Chapter \ref{chap:mathrel}, we use the standard ADM methodology of splitting our metric into spatial and time components, and our general line element takes the form given by equation (\ref{eqn:genmetric})
\begin{equation}{ds}^2=(-\alpha^2+\beta^i\beta_i){dt}^2 + 2\beta_i{dx}^idt + \gamma_{ij}{dx}^i{dx}^j\label{eqn:genmetricaxi}\end{equation}
As we are in an axisymmetric formalism the $\gamma_{13}$ and $\gamma_{23}$ spatial components of the 3-metric vanish (i.e. no $\varphi$ dependence), as well as $\beta^3=\beta^{\varphi}=0$.  This is because under the transformation $d\varphi \rightarrow -d\varphi$ we expect $ds^2$ to be unchanged, so any terms in (\ref{eqn:genmetricaxi}) that have a linear $d\varphi$ dependence (i.e. $\gamma_{13}\;d\eta \; d\varphi$, $\gamma_{23}\;d\theta \;d\varphi$ and $\beta_{3}\;d\varphi\; dt$) must vanish (i.e. the metric and all other quantities are independent of $\varphi$).

\subsection{Gauge Choices}\label{sec:gauge}
For the metric we have $10$ dynamical variables in a $4D$-spacetime\footnote{i.e. there are $10$ possibly unique entries in a symmetric $4\times 4$ matrix}, however $3$ are removed by axisymmetry, which leaves us with $7$ dynamical variables in order to fully describe the system of differential equations.  We are provided with 3 constraints (the Hamiltonian and two momentum) which leaves us with 4 gauge choices.  Some examples include $\beta_i=0$ (vanishing shift), $\alpha=1$ (coordinate time=proper time) or $\alpha=f(r,\theta)$ (static lapse).  For a further detailed discussion of different slicing and gauge choices, see \cite{bardeen}.

As discussed in section \ref{sec:ADM3plus1}, we expect 6 metric and 6 extrinsic curvature evolution equations in the Cauchy formulation\footnote{i.e. $6$ potential unique variables in a symmetric $3\times 3$ matrix}.  With axi-symmetry we drop two of those (the $G^{13}$ and $G^{23}$ equations), leaving us with 4 equations and therefore four metric and four extrinsic curvature variables.  Because we have 3 constraints as well (the Hamiltonian and two momentum constraints since the $G^{03}$ constraint is trivially satisfied) we have an overdetermined system and must determine what level of free vs. constrained evolution to follow inside our code.

Gauge choice \#1) To simplify the metric evolution equations, add stability\footnote{We wish to avoid prescribing the shift vectors \emph{a priori} for, as Alcubierre \cite{alcubierre:3p1num} shows, such a system is not a well-posed hyperbolic formulation.} \cite{physrevd5008} and to agree with the form of the Brill metric (\ref{eqn:brillmetricchap1}), we use the condition that\footnote{This leads to constraints on the shift vectors, $\beta^{i}$ - see section \ref{sec:shiftvec}.} $\gamma_{12}=0$.  The spatial part of our line element then takes the form
$$dl^2=a {dr}^2 + b r^2 {d\theta}^2 + d r^2 \sin^2\theta{d\phi}^2$$
with
$$a=a(t,r,\theta),b=b(t,r,\theta),d=d(t,r,\theta)$$
Making the coordinate transformation
\begin{equation}r=f(\eta)\label{eqn:rfeta}\end{equation}
to redefine the radial coordinate, we find that
$$dr = f_{\eta}d\eta$$
and we arrive at the line element
$$dl^2=a f_{\eta}^2{d\eta}^2 + b f^2 {d\theta}^2 + d f^2 \sin^2\theta{d\phi}^2$$
The freedom to make this coordinate transformation was built into the code to allow us the choice of various radial functions.  It is not mathematically necessary, but allows easy reformulation in the event a different radial function is desired, and also ensures that in the outer grid regions that the grid zones are approximately ``square''.  Normally in spherical polar coordinates a section of our grid has area
$$(\Delta r) (r \Delta\theta)$$
for a fixed $\Delta r$ and $\Delta\theta$, which becomes increasingly rectangular as we move into the asymptotic region of the grid ($r \rightarrow \infty,r \Delta\theta >> \Delta r$).  If we instead employ (\ref{eqn:rfeta}) we find that a section of our grid has area
$$(f_\eta \Delta \eta)(f \Delta \theta) $$
which for $f=\sinh\eta$, $f_\eta=\cosh\eta$ gives an approximately square grid in the asymptotic region as
$$\sinh\eta \sim \cosh\eta$$
for large $\eta$ which implies that
$$f_\eta \Delta \eta \sim f \Delta \theta $$
if
$$\Delta \eta \sim \Delta \theta$$

Gauge choice \#2) We employ the isothermal gauge, $a=b$, which has been shown in the past \cite{evans} to stabilise some aspects of the evolution, and is also required to satisfy the Brill criteria on the metric given in equation (\ref{eqn:brillmetricchap1}).

Gauge choice \#3) $d=1$  When employing a conformal decomposition on the metric, this will move the evolution into the conformal factor.  A conformal decomposition is required as seen in equation (\ref{eqn:brillmetricchap1}), for the Brill formalism employs a conformal factor, $\psi=e^{\phi}$, that is ``factored out'' of the metric components.

Gauge choice \#4) Static or Maximal slicing.  If one wants to avoid areas of large curvature, the basic idea is to construct spatial hyper-surfaces with maximal enclosed \emph{volume}, as high curvature areas increase the surface area to volume ratio.  The condition that creates hyper-surfaces of maximal volume is that the trace of the extrinsic curvature vanishes, i.e.
\begin{equation}\label{eqn:maxslic}K = \mathrm{Tr}K = K^i_{\;i} = 0
\end{equation}
The Lie derivative of the trace of the extrinsic curvature yields
\begin{equation}\label{eqn:liemaxslice}
\pounds_t({\rm Tr}K) = -D^a D_a \alpha + \alpha [R + ({\rm Tr} K)^2] + \pounds_{\beta} ({\rm Tr} K)
\end{equation}
which yields the ``maximal slicing equation'' to solve for $\alpha$ if ${\rm Tr} K=0$ at all times.  This ensures correct propagation of the gauge conditions to all future time steps.

Maximal slicing has the property that it has certain singularity avoiding capabilities \cite{Lichnerowitz,bernstein}, as it compensates in regions of large curvature by collapsing the lapse function $\alpha$ to zero via the maximal slicing equation. This essentially halts the evolution of \emph{proper} time in that area of the grid (see section \ref{subsec:admoverview}).  Singularities without regions of large curvature will not be avoided, this only helps to avoid so called ``crushing singularities'' where the curvature diverges.

For the axi-symmetric Brill wave evolution code, in order to preserve regularity we instead employ a static lapse function of the form
\begin{equation}\label{eqn:staticlapse}\alpha=\tanh^n(\eta)\sin^l(\theta) \;;\; n=4, l=2\end{equation}
which maintains regularity near singular points ($r=0$ and $\theta=0$)\footnote{See section (\ref{subsec:alphareg}) for more details.}.  It may be possible to merge this lapse with a more singularity avoiding one (like maximal slicing) in the future.

We do not implement the polar slicing condition\footnote{$K^2_2+K^3_3=0$, for reasons discussed in \cite{evans}, Chapter IV, Section C - namely that polar slicing introduces irregularity into the mixed tensor quantities.}, and we briefly discuss some alternate slicings in future chapters and their advantages and disadvantages.

As mentioned above we also choose to conformally decompose the metric to aid in the measurement of the mass in section \ref{subsec:massmeasure} as well as to follow the strict form of the Brill formalism in section \ref{subsec:brillformalism}.  See also \cite{mastersonmsc, paul_thesis,evans} for a discussion of how this relates to mass measurements.

The sum of all these gauge choices leads to the 3-metric being given by:

\begin{eqnarray}
\gamma_{ij} & = & \psi^4 \left[\begin{array}{ccc}
a f_{\eta}^2 & 0 & 0 \\
 0 & a f^2 & 0 \\
 0 & 0 & f^2 \sin^2\theta \end{array} \right] \nonumber
\end{eqnarray}

Furthermore introducing the functions $q(t,\eta,\theta)$ and $\phi(t,\eta,\theta)$ such that:
\begin{equation}\label{eqn:etoq} a = e^q \end{equation}
and
\begin{equation}\label{eqn:etophi} \psi = e^{\phi} \end{equation}
one obtains simplifications for the expressions involving logarithmic derivatives that appear in connection coefficients and curvature expressions.  These functions also provide greater numerical stability (to be discussed in section (\ref{sec:christoffel})).

As further motivation for this change of variables, Deadman \cite{deadman} in his mathematical treatment of outer boundary conditions notes the presence of numerous logarithmic terms that arise from using traditional variables, noting how this will likely cause issues with a numerical implementation.

These variable changes lead to the final form of our 3-spatial metric:
\begin{eqnarray}\gamma_{ij} & = & e^{4\phi} \left[\begin{array}{ccc}
e^q f_{\eta}^2 & 0 & 0 \\
 0 & e^q f^2 & 0 \\
 0 & 0 & f^2 \sin^2\theta \end{array} \right] \label{eqn:3dmetric}
\end{eqnarray}
and thusly
\begin{eqnarray}\gamma^{ij} & = & \left\{\begin{array}{ll} \frac{1}{\gamma_{ij}} & i=j \\  0 & i \ne j \\ \end{array} \right. \nonumber \end{eqnarray}
Our 3-line element has the final form:
\begin{equation}\label{eqn:3metricfinal}
dl^2=e^{4\phi}\left[e^qf_{\eta}^2 d\eta^2 + e^qf^2d\theta^2 + f^2\sin^2\theta d\phi^2  \right]
\end{equation}

We can write the 4-metric in matrix notation following the conventions of Bernstein \cite{bernstein} and our gauge, which helps with visualisation and variable transformations we employ later:
\begin{eqnarray}\label{eqn:4dgmunu}
g_{\mu\nu}&=&\left[ \begin{array}{cccc}
-\alpha^2+\beta^i\beta_i & \beta_\eta & \beta_\theta & 0 \\
\beta_\eta & A e^{4\phi} & 0 & 0 \\
\beta_\theta & 0 & B e^{4\phi} & 0 \\
0 & 0 & 0 & D \sin^2\theta e^{4\phi} \\
\end{array} \right]
\end{eqnarray}
given that
$$A=e^qf_{\eta}^2,B=e^qf^2,D=f^2$$

\subsection{Extrinsic Curvature}
The covariant form of our extrinsic curvature is chosen to be in the form
\begin{eqnarray}\label{eqn:origcovkijtensor}
K_{ij} & = & \left[\begin{array}{ccc}
f_{\eta}^2 e^{q+4\phi} H_a & f_{\eta}^2 e^{q+4\phi} H_c & 0 \\
f_{\eta}^2 e^{q+4\phi} H_c & f^2 e^{q+4\phi} H_b  & 0 \\
 0 & 0 & f^2 e^{4\phi} H_d \sin^2\theta \end{array} \right] \nonumber
\end{eqnarray}
Where
$$H_a=H_a(t,\eta,\theta) \;;\; H_b=H_b(t,\eta,\theta) \;;\; H_c=H_c(t,\eta,\theta) \;;\; H_d=H_d(t,\eta,\theta)$$
Knowing that our spatial metric is diagonal yields
$$K^i_j=\gamma^{ia}K_{aj}$$
for the mixed extrinsic curvature components.  This gives\footnote{We use the mixed tensor in our evolution equations for the reasons that (1) it leads to more desirable asymptotic spatial behaviour for the extrinsic curvature variables (hence more numerical stability) and (2) it simplifies the imposition of $K=K^i_i=0$ (Maximal Slicing), as the $K^i_i$ are our dynamic variables giving a simple algebraic equation.}
\begin{eqnarray}\label{eqn:mixkijtensor}
K^i_j & = & \left[\begin{array}{ccc}
H_a & H_c & 0 \\
\frac{f_{\eta}^2}{f^2} H_c & H_b  & 0 \\
 0 & 0 & H_d \end{array} \right]
\end{eqnarray}
for the final form of the mixed extrinsic curvature components that we will use in the evolution equations.  This also yields a constraint on our dynamic variables via the first maximal slicing equation (\ref{eqn:maxslic})
\begin{equation}\label{eqn:hahbhdcon}
H_a + H_b + H_d = 0
\end{equation}
Which we can make use of several times to remove $H_d$ from equations (when employing maximal slicing).

\subsection{Christoffel Symbols}\label{sec:christoffel}
The general mixed Christoffel symbols that we use are given by the spatial components of (\ref{eqn:christoffel}).
Because of our diagonal metric, the mixed Christoffel symbols are given by (dropping Einstein summation notation for this one equation)
$$\Gamma^a_{bc}=\frac{1}{2 \gamma_{aa}}(\partial_c\gamma_{ab}+\partial_b\gamma_{ac}-\partial_a\gamma_{bc})$$
using (\ref{eqn:3dmetric}) we get
$$\begin{array}{ccc} \gamma_{11}=e^{q+4\phi}f_{\eta}^2; & \gamma_{22}=e^{q+4\phi}f^2; & \gamma_{33}=e^{4\phi}f^2\sin^2\theta \end{array} \\ $$
which for $f,f_{\eta} > 0 $ gives\footnote{i.e. our radial function is monotonically increasing and non-negative.  $r=0$ poses, as always, difficulties that must be dealt with separately.}
\begin{eqnarray}
\ln(\gamma_{11}) & = & q + 4\phi + 2\ln(f_{\eta}) \nonumber \\
\ln(\gamma_{22}) & = & q + 4\phi + 2\ln(f) \nonumber \\
\ln(\gamma_{33}) & = & 4\phi + 2\ln(f) + 2\ln|\sin(\theta)|
\end{eqnarray}
Knowing that our coordinates are defined on $0 \leq \theta \leq \frac{\pi}{2}$ we obtain\footnote{The Christoffel symbols are written in the more suggestive manner $\partial\ln(g_{ab})$ to indicate that our choice of exponential variables in the metric (\ref{eqn:3dmetric}) has a very solid reason - it simplifies the expressions and is more numerically stable.}
$$\begin{array}{ccccc}
\Gamma^i_{jk} & = & \Gamma^i_{kj} & & \nonumber \\
\Gamma^1_{11} & = & \frac{1}{2}\partial_{\eta}\ln(g_{11}) & = & \frac{1}{2}\left[q_{\eta}+4\phi_{\eta}+2\frac{f_{\eta\eta}}{f_{\eta}}\right] \nonumber \\
\Gamma^1_{12} & = & \frac{1}{2}\partial_{\theta}\ln(g_{11}) & = & \frac{1}{2}[q_{\theta}+4\phi_{\theta}] \nonumber \\
\Gamma^1_{13} & = & 0 & & \nonumber \\
\Gamma^1_{22} & = & \frac{-f^2}{2f_{\eta}^2}\partial_{\eta}\ln(g_{22}) & = & \frac{-f^2}{2f_{\eta}^2}\left[q_{\eta}+4\phi_{\eta}+2\frac{f_{\eta}}{f}\right] \nonumber \\
\Gamma^1_{23} & = & 0 & & \nonumber \\
\Gamma^1_{33} & = & \frac{-1}{2g_{11}}\partial_{\eta}(g_{33}) & = & \frac{-\sin^2\theta}{e^q}\left[2\phi_{\eta}\left(\frac{f}{f_{\eta}}\right)^2+\frac{f}{f_{\eta}}\right] \nonumber \\
\Gamma^2_{11} & = & \frac{-f_{\eta}^2}{2f^2}\partial_{\theta}\ln(g_{11}) & = & \frac{-f_{\eta}^2}{2f^2}[q_{\theta}+4\phi_{\theta}] \nonumber \\
\Gamma^2_{12} & = & \frac{1}{2}\partial_{\eta}\ln(g_{22}) & = & \frac{1}{2}\left[q_{\eta}+4\phi_{\eta}+2\frac{f_{\eta}}{f}\right] \nonumber \\
\Gamma^2_{13} & = & 0 & & \nonumber \\
\Gamma^2_{22} & = & \frac{1}{2}\partial_{\theta}\ln(g_{22}) & = & \frac{1}{2}\left[q_{\theta}+4\phi_{\theta}\right]\nonumber \\
\Gamma^2_{23} & = & 0 & & \nonumber \\
\Gamma^2_{33} & = & \frac{-1}{2g_{22}}\partial_{\theta}g_{33} & = & \frac{-\sin^2\theta}{e^q}(2\phi_{\theta} + \cot\theta )\nonumber \\
\Gamma^3_{11} & = & \Gamma^3_{12} & = & 0 \nonumber \\
\Gamma^3_{13} & = & \frac{1}{2}\partial_{\eta}\ln(g_{33}) & = & \left[2\phi_{\eta}+\frac{f_\eta}{f}\right] \nonumber \\
\Gamma^3_{22} & = & 0 & & \nonumber \\
\Gamma^3_{23} & = & \frac{1}{2}\partial_{\theta}\ln(g_{33}) & = & \left[2\phi_{\theta} + \cot\theta \right] \nonumber \\
\Gamma^3_{33} & = & 0 & & \\
\end{array}$$\begin{eqnarray}\label{eqns:mcs}\end{eqnarray}

\subsection{Ricci Curvature Variables}\label{sec:extrinsiccurvature}
From the Christoffel symbols we can derive the covariant form of the Ricci Tensor
\begin{equation}\label{eqn:ricciten} R_{ij} = R^k_{ikj} = \partial_k \Gamma^{k}_{ij} - \partial_j \Gamma^{k}_{ik} +  \Gamma^{l}_{ij}\Gamma^{k}_{lk} - \Gamma^{l}_{ik}\Gamma^{k}_{lj}\end{equation}
Numerical note: While it may be tempting to \emph{numerically calculate} the $\Gamma^i_{jk}$'s and then \emph{numerically calculate} the Ricci Tensor (and the evolution equations) from these relations, the fact that the Riemann tensor measures the non-commutativity of covariant differentiation on the manifold\footnote{i.e. the deviation from Euclidean flat space.} means that there should be a number of terms that cancel exactly.  While this will happen analytically, numerically things are different when it comes to cancellation.  For the same reasons that we expound upon in section (\ref{subsec:addterms}), performing naive or poorly ordered summation of these terms can cause significant numerical errors.

Thus it is preferable to treat all terms separately to avoid numerically clumping together curvature terms that arise from different parts of the covariant differentiation operator.

This is especially important as certain spacetime regions can contain large non-linear quantities and others have an asymptotically Schwarzschild-like solution, so the nature of commutation of curvature terms will be different throughout the grid.  Rather than trying to guess the ``correct'' order to numerically piece the curvature terms together in, we'll let the computer do it instead.

For example, using (\ref{eqn:ricciten}) we find
$$R_{11} = \partial_{\eta} \Gamma^1_{11} + \partial_\theta \Gamma^2_{11} - \partial_{\eta} (\Gamma^1_{11} + \Gamma^2_{12} + \Gamma^3_{13}) + \Gamma^{1}_{11}\Gamma^{k}_{1k} + \Gamma^{2}_{11}\Gamma^{k}_{2k} - \Gamma^{l}_{1k}\Gamma^{k}_{l1}$$
which can be written explicitly as
\begin{equation}\label{eqn:r11_gam}R_{11} = \partial_{\theta}\Gamma^{2}_{11} - \partial_{\eta}\Gamma^{2}_{12} - \partial_{\eta}\Gamma^{3}_{13} + \Gamma^{1}_{11}(\Gamma^{2}_{12}+\Gamma^{3}_{13}) + \Gamma^{2}_{11}(\Gamma^{2}_{22}+\Gamma^{3}_{23}-\Gamma^{1}_{22}) - \left[\Gamma^{2}_{12}\right]^2 - \left[\Gamma^{3}_{13}\right]^2\end{equation}
In theory one could numerically evaluate the $\Gamma$'s and perform the computation of $R_{11}$ in equation (\ref{eqn:r11_gam}).  However the analytic expressions given in equations (\ref{eqns:mcs}) lead to exact cancellations that may not be properly accounted for in a naive numerical implementation.

Furthermore, because of the $\cot\theta$ terms in $\Gamma^{2}_{33}$ and $\Gamma^{3}_{23}$ we need to be able to apply l'Hopital's rule to those terms at $\theta=0$, so breaking down some of the $\Gamma$'s into their constituent parts is necessary.  Using
$$R^i_{\;j} = \gamma^{ia}R_{aj}$$
we find that

\begin{eqnarray}\label{eqn:r11}
R^1_{\;1} & = & \frac{1}{f^2 e^{(q+4\phi)}}\left[-\frac{q_{\theta}\cot\theta}{2}
-2\phi_{\theta}\cot\theta
-\frac{q_{\theta\theta}}{2}
-\phi_{\theta}q_{\theta}
-\frac{q_{\eta\eta} f^2}{2f_{\eta}^2}
+\frac{\phi_{\eta}q_{\eta} f^2}{f_\eta^2}
+\frac{q_{\eta} f^2 f_{\eta\eta}}{2f_{\eta}^3}
-2\phi_{\theta\theta} \right.
\nonumber \\ \mbox{}
 & & \left. -4(\phi_{\theta})^2
-\frac{4\phi_{\eta\eta} f^2}{f_{\eta}^2}
+\frac{4\phi_{\eta} f^2 f_{\eta\eta}}{f_{\eta}^3}
-\frac{4\phi_{\eta} f}{f_\eta} \right]
\end{eqnarray}

This is much more stable and less prone to numerical noise than the version without exponential functions (due to fewer truncation and roundoff errors):

\begin{eqnarray}
R^1_{\;1}&=&-\frac{2 \psi_{\theta} \cos\theta}{a f^2 \psi^5 \sin\theta}
-\frac{ a_{\theta} \cos\theta}{2 a^2 f^2 \psi^4 \sin\theta}
-\frac{2 \psi_{\theta\theta}}{a f^2 \psi^5}
-\frac{2 \psi_{\theta}^2}{a f^2 \psi^6}
-\frac{a_{\theta} \psi_{\theta}}{a^2 f^2 \psi^5}
-\frac{4 \psi_{\eta\eta}}{a {f_{\eta}}^2 \psi^5}
\nonumber \\ & & \mbox{}
+\frac{4 {\psi_{\eta}}^2}{a {f_{\eta}}^2 \psi^6}
+\frac{4 f_{\eta\eta} \psi_{\eta}}{a {f_{\eta}}^3 \psi^5}
-\frac{4 \psi_{\eta}}{a f f_{\eta} \psi^5}
+\frac{a_{\eta} \psi_{\eta}}{a^2 {f_{\eta}}^2 \psi^5}
+\frac{a_{\eta} f_{\eta\eta}}{2 a^2 {f_{\eta}}^3 \psi^4}
-\frac{a_{\eta\eta}}{2 a^2 {f_{\eta}}^2 \psi^4}
\nonumber \\ & & \mbox{}
+\frac{a_{\eta\eta}}{2 a^3 {f_{\eta}}^2 \psi^4}
-\frac{a_{\theta\theta}}{2 a^2 f^2 \psi^4}
+\frac{a_{\theta\theta}^2}{2 a^3 f^2 \psi^4}
\label{eqn:nonexpr11}\end{eqnarray}

The other mixed Ricci tensor terms are:

\begin{eqnarray}\label{eqn:r12}
R^1_{\;2} & = & \frac{1}{f_\eta^2 e^{(q+4\phi)}}\left[\frac{q_{\eta}\cot\theta}{2}
+\phi_{\eta}q_{\theta}
+\frac{q_{\theta} f_\eta}{2f}
+\phi_{\theta}q_{\eta}
+4\phi_{\eta}\phi_{\theta}
+\frac{2\phi_{\theta}f_{\eta}}{f}
-2\phi_{\eta\theta} \right]
\end{eqnarray}

\begin{eqnarray}
R^2_{\;1} & = & \frac{f_{\eta}^2}{f^2} R^1_2
\end{eqnarray}

\begin{eqnarray}\label{eqn:r22}
R^2_{\;2} & = & \frac{1}{f^2 e^{(q+4\phi)}}\left[\frac{q_{\theta}\cot\theta}{2}
-2 \phi_{\theta} \cot\theta
-\frac{q_{\theta\theta}}{2}
+\phi_{\theta}q_{\theta}
-\frac{q_{\eta\eta} f^2}{2f_{\eta}^2}
-\frac{\phi_{\eta}q_{\eta} f^2}{f_{\eta}^2}
+\frac{q_{\eta} f^2 f_{\eta\eta}}{2f_{\eta}^3}
-\frac{q_{\eta} f}{f_{\eta}} \right.
\nonumber \\ \mbox{}
 & & \left. -4 \phi_{\theta\theta}
-\frac{2 \phi_{\eta\eta} f^2}{f_{\eta}^2}
-\frac{4 \phi_{\eta}^2 f^2}{f_{\eta}^2}
+\frac{2 \phi_{\eta} f^2 f_{\eta\eta}}{f_{\eta}^3}
-\frac{6 \phi_{\eta} f}{f_{\eta}} \right]
\end{eqnarray}

and

\begin{eqnarray}\label{eqn:r33}
R^3_{\;3} & = & \frac{1}{f^2 e^{(q+4\phi)}}\left[-4 \phi_{\theta}\cot\theta
-2 \phi_{\theta\theta}
-4 \phi_{\theta}^2
-\frac{2 \phi_{\eta\eta} f^2}{f_{\eta}^2}
-\frac{4 \phi_{\eta}^2 f^2}{f_{\eta}^2} \right.
\nonumber \\ \mbox{}
& & \left. +\frac{2 \phi_{\eta} f^2 f_{\eta\eta}}{f_{\eta}^3}
-\frac{6 \phi_{\eta} f}{f_{\eta}} \right]
\end{eqnarray}

\subsection{The Radial Function, f}\label{sec:fnx}
\begin{figure}[h]
\centering
\includegraphics{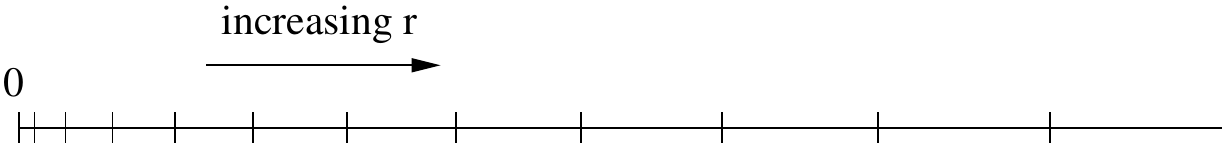}
\caption[Radial grid spacing]{A schematic diagram of the grid point spacing in the radial direction using our radial function $f(\eta)$}
\label{fig:exponential_coords}
\end{figure}

We use a radial function, $f$, that gives better grid coverage near the origin where the dynamics are strongly nonlinear, while allowing for large values of $r$ to extract radiation terms in the asymptotic zone:
\begin{equation}\label{eqn:fdefn} f(\eta)=\sinh(\eta)\end{equation}
This is a numerical consideration which comes into play when we discretise the spacetime with a grid - this will naturally place more of our finite number of grid points near the origin and less further away (see figure \ref{fig:exponential_coords}).

\begin{figure}[h]
\centering
\includegraphics{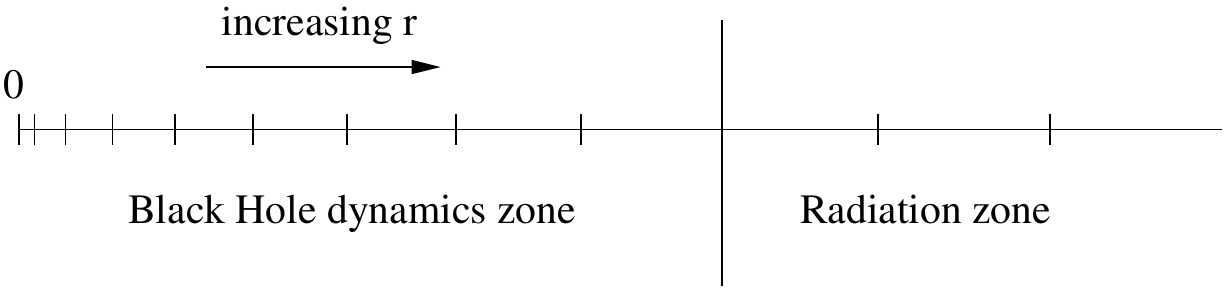}
\caption[Radial grid spacing splitting]{A schematic diagram of the splitting of a coordinate grid spacing into the black hole zone and the radiative zone}
\label{fig:grid_zones}
\end{figure}

While some formulations create separate grid areas for black hole and wave zones to allow for a different treatment of the highly non-linear zone and the radiative zone (see figure \ref{fig:grid_zones}), we will dispense with the inherent complications in matching the \emph{boundaries} between different grid zones\footnote{See for example \cite{Thornburg:patches}.} and attempt to adapt our radial function to perform the same purpose.

Some identities that will come in useful to rearrange our equations into a friendlier numerical form are given below:

\begin{equation}1-\frac{f f_{\eta\eta}}{f_{\eta}^2}=\frac{d}{d\eta}\left(\frac{f}{f_\eta}\right)\label{eqn:dxfnfx}\end{equation}

$$-\left(\frac{f_\eta}{f}\right)^2\frac{d}{d\eta}\left(\frac{f}{f_\eta}\right)=\frac{d}{d\eta}\left(\frac{f_\eta}{f}\right)$$

These can be used to simplify the output generated by Maxima and collect terms to generate more precise numerical algorithms.  For example, if we define $f=\sinh(\eta)$ then we find
$$\frac{d}{d\eta}\left(\frac{f}{f_\eta}\right)=\frac{1}{\cosh^2(\eta)}$$
which can be numerically calculated more precisely than the LHS of equation (\ref{eqn:dxfnfx}).

\subsection{The Scalar Curvature ($R$)}\label{sec:scurv}
The spatial scalar curvature is defined by (\ref{eqn:ricciscalar})
$$R=\gamma^{ij}R_{ij}$$
or
\begin{eqnarray}
R & = & \frac{1}{f^2 e^{q+4\phi}}\left(-8\phi_{\theta}\cot\theta
-q_{\theta\theta}
-\frac{q_{\eta\eta} f^2}{f_{\eta}^2}
+\frac{{q}_{\eta} f^2 f_{\eta\eta}}{f_{\eta}^{3}}
-\frac{q_{\eta} f}{f_{\eta}}
-8\phi_{\theta\theta}
-8\phi_{\theta}^{2} \right.
\nonumber \\ \mbox{}
& & \left. -\frac{8\phi_{\eta\eta} f^2}{f_{\eta}^2}
-\frac{8\phi_{\eta}^2 f^2}{f_{\eta}^2}
+\frac{8\phi_{\eta} f^2 f_{\eta\eta}}{f_{\eta}^3}
-\frac{16\phi_{\eta} f}{f_{\eta}}\right)
\label{eqn:scurvunmodified}\end{eqnarray}

\subsection{Metric Evolution Equations for $q, \phi$}\label{sec:metricevol}
The covariant metric evolution equations are given by (\ref{eqn:3p1:gammadot}) \cite{MTW,evans}:
\begin{equation} \partial_t \gamma_{ij} = -2\alpha K_{ij} + \partial_i\beta_j
-\Gamma^l_{ji}\beta_l+\partial_j\beta_i-\Gamma^l_{ij}\beta_l
\end{equation}
From the gauge conditions given in section (\ref{sec:aximetric}), the dynamic metric variables are chosen to be $q$, $\phi$. In addition two constraints on the shift vector components $v_1$ and $v_2$ are introduced, where we define:
$$\beta_i=( v_1 , v_2 , 0 )$$
The non-zero metric evolution equations become:
\begin{equation}
\frac{\dot{\gamma}_{11}}{f_{\eta}^2e^{q+4\phi}} = \dot{q}+4\dot{\phi} = v_2q_{\theta} + 4\phi_{\theta}v_2 +2{v_1}_{,\eta} + q_{\eta}v_1 + 4\phi_{\eta}v_1 + 2\frac{f_{\eta\eta}}{f_{\eta}}v_1 - 2\alpha H_a
\label{eqn:gam11dot}\end{equation}
\begin{equation}
\frac{\dot{\gamma}_{22}}{f^2e^{q+4\phi}} = \dot{q}+4\dot{\phi} = 2{v_2}_{,\theta} + v_2q_{\theta} + 4\phi_{\theta}v_2 + q_{\eta}v_1 + 4\phi_{\eta}v_1 + 2\frac{f_{\eta}}{f}v_1 - 2\alpha H_b
\label{eqn:gam22dot}\end{equation}
\begin{equation}
\frac{\dot{\gamma}_{12}}{f_{\eta}^2} = \frac{\dot{\gamma}_{21}}{f_{\eta}^2} = 0 = \left(\frac{f}{f_\eta}\right)^2{v_2}_{,\eta} + {v_1}_{,\theta} - 2 \alpha H_c
\label{eqn:gam12dot}\end{equation}
\begin{equation}
\frac{\dot{\gamma}_{33}}{f^2e^{4\phi}\sin^2\theta} = 4 \dot{\phi} = 4 \phi_{\theta}v_2 + 4 \phi_{\eta}v_1 + 2 \frac{f_{\eta}}{f}v_1 - 2\alpha H_d + 2 v_2 \cot\theta
\label{eqn:gam33dot}\end{equation}
Note: We can divide through by $\sin^2\theta$ in (\ref{eqn:gam33dot}) because we know that when employing l'Hopital's rule we require ${v_2}_{\theta}=0$ at $\theta=0$, i.e. $v_2$ is symmetric across $\theta=0$.  We impose this symmetry because we don't wish to have a coordinate shift in the $\theta$ direction at $\theta=0$, otherwise we would have coordinate points ``sliding'' across the axis and violating our symmetry conditions at $\theta=0$.

We can choose to use equation (\ref{eqn:gam33dot}) to evolve $\phi$, or we can use the Hamiltonian constraint to solve for it.  As second order elliptic PDEs are generally more stable than first order hyperbolic equations we choose to constrain $\phi$ instead. This also ensures that energy conservation is coupled into the evolution.

The evolution equation for the metric variable $q$ can be found from either of equations (\ref{eqn:gam11dot}) or (\ref{eqn:gam22dot}) in conjunction with (\ref{eqn:gam33dot}).  This yields
\begin{equation}\label{eqn:qdot}
\dot{q} = 2{v_2}_{,\theta} + v_2 q_{\theta} + q_{\eta}v_1 + 2\alpha(H_d-H_b) - 2v_2\cot\theta
\end{equation}
or
\begin{equation}\label{eqn:qdotdontuse}
\dot{q} = v_2q_{\theta} + 2 {v_1}_{,\eta} + q_{\eta}v_1 + 2v_1\left(\frac{f_{\eta\eta}}{f_{\eta}}-\frac{f_{\eta}}{f}\right) + 2\alpha(H_d-H_a) - 2v_2\cot\theta\end{equation}
which are equivalent provided we solve the series of equations for the shift vector components. We choose to use equation (\ref{eqn:qdot}) for its relative numerical simplicity.  At $\theta=0$ use of l'Hopital's rule and the fact that
$${\left.{\frac{d(\tan\theta)}{d\theta}}\right|}_{\theta=0} = {\left.{\frac{1}{\cos^2\theta}}\right|}_{\theta=0}=1$$
leads to the boundary condition:
$${\left.{\dot{q}}\right|}_{\theta=0} = v_2 q_{\theta} + q_{\eta}v_1 + 2\alpha(H_d-H_b)$$

\subsection{Solving for Shift Vector Components ($v_1, v_2$)}\label{sec:shiftvec}
Using equations (\ref{eqn:gam11dot}) and (\ref{eqn:gam22dot}) we can eliminate time dependent quantities and get one constraint on the shift vectors.  Equation (\ref{eqn:gam12dot}) yields the other constraint on $v_1$ and $v_2$.  These are
\begin{eqnarray}
f^2{v_2}_{,\eta} + f_{\eta}^2{v_1}_{,\theta} & = & 2 f_{\eta}^2 H_c \alpha
\nonumber \\ \mbox{}
2{v_1}_{,\eta} - 2{v_2}_{,\theta} - 2v_1\left(\frac{f_{\eta}}{f}-\frac{f_{\eta\eta}}{f_{\eta}}\right) & = & 2 \alpha(H_a-H_b) \nonumber
\end{eqnarray}
Which can be rewritten in the form
\begin{eqnarray}\label{eqn:shiftvcoupled}
\left(\frac{f}{f_{\eta}}\right)\partial_{\eta}\left(\frac{f_{\eta}}{f}v_1\right) - \partial_{\theta}(v_2) & = & \alpha(H_a - H_b) \nonumber \\ \mbox{}
\partial_{\eta}(v_2) + \partial_{\theta}\left(\left(\frac{f_{\eta}}{f}\right)^2 v_1\right) & = & 2 \alpha \left(\frac{f_\eta}{f}\right)^2 H_c
\end{eqnarray}
We define two scalar functions (called \emph{shift vector potentials}) $\Phi(\eta,\theta)$ and $\chi(\eta,\theta)$ such that\footnote{There are two reasons for this.  Firstly we can decouple the shift equations, and secondly first-order elliptic PDEs are inherently unstable when solved with a finite differencing operator that isn't coupled to the central point, so we wish to solve for the potentials instead.}
\begin{eqnarray}\label{eqn:shiftvpotdef}
\left(\frac{f_{\eta}}{f}\right)v_1 & = & \left(\frac{f}{f_{\eta}}\right) \chi_{\eta} + \Phi_{\theta} \\ \mbox{}
v_2 & = & \left(\frac{f}{f_{\eta}}\right)\Phi_{\eta} - \chi_{\theta}
\end{eqnarray}
Which leads to two decoupled, spatial partial differential equations for the shift vector potentials:
\begin{eqnarray}\label{eqn:decshiftvecs}
\left(\frac{f}{f_{\eta}}\right)^2 \chi_{\eta\eta} + \chi_{\theta\theta} + \left(\frac{f}{f_{\eta}}\right)\partial_{\eta}\left(\frac{f}{f_{\eta}}\right)\chi_{\eta} & = & \alpha(H_a - H_b) \nonumber \\ \mbox{}
\left(\frac{f}{f_{\eta}}\right)^3\Phi_{\eta\eta} + \left(\frac{f}{f_{\eta}}\right) \Phi_{\theta\theta} + \left(\frac{f}{f_{\eta}}\right)^2\partial_{\eta}\left(\frac{f}{f_{\eta}}\right)\Phi_{\eta} & = & 2 \alpha H_c 
\end{eqnarray}
These are 2nd order elliptic PDEs for the shift vector potentials $\Phi(\eta,\theta)$ and $\chi(\eta,\theta)$ that can be solved using the method described in section \ref{sec:4thordderiv}.

\subsection{Extrinsic Curvature Evolution Equations for $H_a, H_b, H_c, H_d$}\label{sec:extrincurvevol}
The general equation for the evolution of the mixed extrinsic curvature components, $K^i_{\;j}$ is derived in \cite{evans}\footnote{pg. 26} and is given by:
\begin{eqnarray}\label{eqn:mixkevolvacuum} \partial_t K^i_{\;j}&=&-\gamma^{id}\left[\partial_d\partial_j\alpha-\Gamma^e_{\;jd}\partial_e\alpha\right]
+\alpha\left[R^i_{\;j}+K^i_{\;j} K\right]
+\beta^c\left[\partial_c K^i_{\;j} - \Gamma^d_{\;jc} K^i_{\;d} 
+ \Gamma^i_{\;dc} K^d_{\;j}\right]
\nonumber \\ & & \mbox{}
+K^i_{\;c}\left[\partial_j \beta^c + \Gamma^c_{\;ej} \beta^e\right]
-K^c_{\;j}\left[\partial_c \beta^i + \Gamma^i_{\;ec} \beta^e\right]
\end{eqnarray}
using the conventions of this thesis.  
In terms of our metric and extrinsic curvature variables one obtains four evolution equations and one ``deficit'' equation:
\begin{eqnarray}\label{eqn:haevol}
\frac{\partial H_a}{\partial t} & = & \frac{1}{f^2 \, e^{q+4\phi}}\left(
-\frac{\alpha\, q_{\theta} \,\cot\theta }{2}
-2\,\alpha\, \phi_{\theta} \,\cot\theta
-\frac{\alpha\, q_{\theta\theta} }{2}
-\alpha\, \phi_{\theta} \, q_{\theta}
-\frac{ {\alpha}_{\theta} \, q_{\theta} }{2} \right. \nonumber \\ \mbox{} & & \left.
-\frac{\alpha\, q_{\eta\eta} f^2 }{2f_\eta^2}
+\frac{\alpha\, \phi_{\eta} \, q_{\eta} f^2}{f_\eta^2}
+\frac{{\alpha}_{\eta} \, q_{\eta} f^2}{2 f_\eta^2}
+\frac{ \alpha\, q_{\eta} \,f^2 \, f_{\eta\eta}}{2 f_\eta^3}
-2\,\alpha\, \phi_{\theta\theta} \right. \nonumber \\ \mbox{} & & \left.
-4\,\alpha\,{\phi_{\theta}}^2
-2\, {\alpha}_{\theta} \, \phi_{\theta}
-\frac{4\,\alpha\, \phi_{\eta\eta} f^2}{f_\eta^2}
+\frac{2\, {\alpha}_{\eta} \, \phi_{\eta} f^2}{f_\eta^2}
+\frac{4\, \alpha\, \phi_{\eta} \, f^2 \, f_{\eta\eta}}{f_\eta^3} \right. \nonumber \\ \mbox{} & & \left.
-\frac{4\,\alpha\, \phi_{\eta} f}{f_\eta}
-\frac{{\alpha}_{\eta\eta} f^2}{f_\eta^2}
+\frac{{\alpha}_{\eta} \, f^2 \, f_{\eta\eta}}{f_\eta^3} \right) \nonumber \\ \mbox{} & &
+ H_c {v_2}_{,\eta} + {H_a}_{,\theta} v_2 - \frac{H_c {v_1}_{,\theta}f_\eta^2}{f^2} + {H_a}_{,\eta} v_1
\nonumber \\ \mbox{} & & + [\alpha H_a(H_d + H_b + H_a)]
\end{eqnarray}
\begin{eqnarray}\label{eqn:hcevol}
\frac{\partial H_c}{\partial t} & = & \frac{1}{f_{\eta}^2 \, e^{q+4\,\phi}}\left(
\frac{\alpha\, q_{\eta} \,\cot\theta }{2 }
+\alpha\, \phi_{\eta} \, q_{\theta}
+\frac{ {\alpha}_{\eta} \, q_{\theta} }{2}
+\frac{\alpha\, q_{\theta} f_\eta}{2\,f}
+\alpha\, \phi_{\theta} \, q_{\eta} \right. \nonumber \\ \mbox{} & & \left.
+\frac{ {\alpha}_{\theta} \, q_{\eta} }{2}
+4\,\alpha\, \phi_{\eta} \, \phi_{\theta}
+2\, {\alpha}_{\eta} \, \phi_{\theta}
+\frac{2\,\alpha\, \phi_{\theta} f_\eta}{f}
-2\,\alpha\, \phi_{\eta\theta} \right. \nonumber \\ \mbox{} & & \left.
+2\, {\alpha}_{\theta} \, \phi_{\eta}
+\frac{ {\alpha}_{\theta} f_\eta}{f }
-{\alpha}_{\eta\theta} \right) \nonumber \\ \mbox{} & &
+H_c\, {v_2}_{,\theta} + {H_c}_{,\theta} \,v_2 - H_b\, {v_1}_{,\theta} + H_a\, {v_1}_{,\theta} - H_c\, {v_1}_{,\eta} + {H_c}_{,\eta} \,v_1
\nonumber \\ \mbox{} & & + [\alpha H_c(H_d + H_b + H_a)]
\end{eqnarray}
\begin{eqnarray}\label{eqn:h12h21diffmix}
\frac{\partial\left(K^1_{\;2}\right)}{\partial t}-\left(\frac{f}{f_\eta}\right)^2\frac{\partial \left(K^2_{\;1}\right)}{\partial t}
 & = &(H_b-H_a) \left[\left(\frac{f}{f_\eta}\right)^2  {v_2}_{,\eta}
+  {v_1}_{,\theta}\right] \nonumber \\ \mbox{} & &
+2 H_c \left[{v_1}_{,\eta} - {v_2}_{,\theta} - v_1 \left(\frac{f_\eta}{f}\right) \partial_\eta\left(\frac{f}{f_\eta}\right)\right] \nonumber \\ \mbox{}
 & = & 0
\end{eqnarray}
\begin{eqnarray}\label{eqn:hbevol}
\frac{\partial H_b}{\partial t} & = & \frac{1}{f^2 \, e^{q+4\phi}}\left(
\frac{\alpha\, q_{\theta} \cot\theta }{2}
-2\,\alpha\, \phi_{\theta} \cot\theta
-\frac{\alpha\, q_{\theta\theta} }{2}
+\alpha\, \phi_{\theta} \, q_{\theta}
+\frac{ {\alpha}_{\theta} \, q_{\theta} }{2} \right. \nonumber \\ \mbox{} & & \left.
-\frac{\alpha\, q_{\eta\eta}  \, f^2}{2 f_\eta^2}
-\frac{\alpha\, \phi_{\eta} \, q_{\eta} \, f^2}{f_\eta^2}
-\frac{ {\alpha}_{\eta} \, q_{\eta} \, f^2 }{2 f_\eta^2}
+\frac{ \alpha\, q_{\eta}  \, f^2 \, f_{\eta\eta}}{2 f_\eta^3}
-\frac{\alpha\, q_{\eta} \, f}{f_\eta} \right. \nonumber \\ \mbox{} & & \left.
-4\,\alpha\, \phi_{\theta\theta}
+2\, {\alpha}_{\theta} \, \phi_{\theta}
-\frac{2\,\alpha\, \phi_{\eta\eta} \, f^2}{f_\eta^2}
-\frac{4\,\alpha\, \phi_{\eta}^2 \, f^2}{f_\eta^2}
-\frac{2\, {\alpha}_{\eta} \, \phi_{\eta} \, f^2}{f_\eta^2} \right. \nonumber \\ \mbox{} & & \left.
+\frac{2\, \alpha\, \phi_{\eta} \, f^2 \, f_{\eta\eta}}{f_\eta^3}
-\frac{6\,\alpha\, \phi_{\eta} \, f}{f_\eta}
-{\alpha}_{\theta\theta}
-\frac{{\alpha}_{\eta} \, f}{f_\eta} \right) \nonumber \\ \mbox{} & &
-H_c\, {v_2}_{,\eta} + {H_b}_{,\theta} \,v_2 + \frac{H_c\, {v_1}_{,\theta} f_\eta^2}{f^2}+ {H_b}_{,\eta} \,v_1
\nonumber \\ \mbox{} & & + [\alpha H_b(H_d + H_b + H_a)]
\end{eqnarray}

\begin{eqnarray}\label{eqn:hdevol} \frac{\partial H_d}{\partial t} & = &
\frac{1}{f^2 e^{q+4\phi}}\left[ -4 \alpha \phi_{\theta} \cot\theta
-\alpha_{\theta} \cot\theta
-2 \alpha \phi_{\theta\theta}
-4 \alpha \phi_{\theta}^2
-2 \alpha_{\theta} \phi_{\theta}
-\frac{2 \alpha \phi_{\eta\eta} f^2}{f_\eta^2}
 \right. \nonumber \\ \mbox{} & & \left.
-\frac{4 \alpha \phi_{\eta}^2 f^2}{f_\eta^2}
-\frac{2 \alpha_{\eta} \phi_{\eta} f^2}{f_\eta^2}
+\frac{2 \alpha \phi_{\eta} f_{\eta\eta} f^2}{f_\eta^3}
-\frac{6 \alpha \phi_{\eta} f}{f_\eta}
\right. \nonumber \\ \mbox{} & & \left.
-\frac{\alpha_{\eta} f}{f_\eta} \right]
+{H_d}_{,\theta} v_2 + {H_d}_{,\eta} v_1 
\nonumber \\ \mbox{} & &+ [\alpha H_d(H_d + H_b + H_a)]
\end{eqnarray}
Where the last term in each evolution equation vanishes if we choose maximal slicing ($TrK=(H_d + H_b + H_a)=0$).

These are first order time evolution equations that can solved using the finite differencing methods discussed in chapter \ref{chap:nummethod}.

\subsection{Hamiltonian Constraint}\label{sec:hamcon}
The general vacuum\footnote{$T^{ab}=0$, therefore $\rho=0$} Hamiltonian Constraint can be found from (\ref{eqn:3p1:ham3}):
\begin{equation}\label{eqn:hamconfullvaccum}R + (TrK)^2 - K_{ij}K^{ij} = 0\end{equation}
If we employ the maximal slicing gauge, $Tr(K)=0$, this constraint becomes
\begin{equation}R-K_{ij}K^{ij}=0\label{eqn:hamcontensor}\end{equation}
However equation (\ref{eqn:hamconfullvaccum}) becomes, under our coordinate conditions,
\begin{eqnarray}\label{eqn:hamcon}
 -\frac{8\, {\phi}_{\theta} \,\cot\theta }{f^2}
-\frac{q_{\theta\theta} }{f^2}
-\frac{q_{\eta\eta} }{f_{\eta}^2}
+\frac{ f_{\eta\eta} \,q_{\eta} }{f_{\eta}^3}
-\frac{q_{\eta} }{f\, f_{\eta} } & & \nonumber \\ \mbox{}
-\frac{8\, {\phi}_{\theta\theta}}{f^2}
-\frac{8\,{\phi}_{\theta}^2}{f^2}
-\frac{8\, {\phi}_{\eta\eta}}{{ f_{\eta} }^2}
-\frac{8\,{\phi}_{\eta}^2}{f_{\eta}^2}
+\frac{8\, f_{\eta\eta} \, {\phi}_{\eta}}{f_{\eta}^{3}}
-\frac{16\, {\phi}_{\eta} \,}{f\, f_{\eta} } & & \nonumber \\ \mbox{}
+{e^{q+4\phi}}\left(2 H_b H_d + 2 H_a H_d + 2 H_a H_b -2\,{H_c}^2 \frac{f_\eta^2}{f^2}\right) & = & 0
\end{eqnarray}
which can be re-arranged to give the following:
\begin{eqnarray} & & \left(\frac{f}{f_{\eta}}\right)^2\phi_{\eta\eta} + \phi_{\theta\theta} +  \frac{f}{f_{\eta}}\left[1+\partial_{\eta}\left(\frac{f}{f_{\eta}}\right)\right] \phi_{\eta} + \cot(\theta) \phi_{\theta} + \left(\frac{f}{f_{\eta}}\right)^2 \phi_{\eta}^2 + \phi_{\theta}^2 \nonumber \\ \mbox{} & &
 = \frac{1}{8}e^q[f^2(2 H_b H_d + 2 H_a H_d + 2 H_a H_b)-2f_{\eta}^2H_c^2] e^{4\phi} \nonumber \\ \mbox{} & &
- \frac{1}{8}\left[q_{\eta\eta}\left(\frac{f}{f_{\eta}}\right)^2+q_{\eta}\left(\frac{f}{f_{\eta}}\right)\partial_{\eta}\left(\frac{f}{f_{\eta}}\right)+q_{\theta\theta}\right] \label{eqn:hamconphiearly}
\end{eqnarray}
which we will use to solve the time symmetric initial value problem for $\phi$ given the function $q(\eta,\theta)$.

\subsection{Momentum Constraints}\label{sec:momcon}
The general momentum vacuum constraint equations in the ADM formalism are given by equations (\ref{eqn:3p1:mom3}).  In a vacuum\footnote{$T_{ab}=0$, so $S_b=0$} these reduce to
$$p^i=\partial_j K^{ij}+\Gamma^i_{\;lj} K^{lj}+\Gamma^j_{\;lj} K^{il}-\gamma^{ij}\partial_j(TrK)=0$$
which become\footnote{After factoring out $\left(f_\eta^2e^{q+4\phi}\right)$ and $\left(f^2e^{q+4\phi}\right)$ respectively.}
\begin{eqnarray}\label{eqn:momcons}
\frac{f_{\eta}^2\,H_c\,\cot\theta }{f^2}
+\frac{f_{\eta}^2\,H_c\, q_{\theta} }{f^2}
-\frac{H_b\, q_{\eta} }{2}
+\frac{H_a\, q_{\eta} }{2}
+\frac{6\,f_{\eta}^2\,H_c\, {\phi}_{\theta} }{f^2} & & \nonumber \\ \mbox{}
-2\,H_d\, {\phi}_{\eta} 
-2\,H_b\, {\phi}_{\eta} 
+4\,H_a\, {\phi}_{\eta} 
-\frac{ f_{\eta} \,H_d}{f}
+\frac{f_{\eta}^2\, {H_c}_{\theta} }{f^2} & & \nonumber \\ \mbox{}
-\frac{ f_{\eta} \,H_b}{f}
-{H_d}_{\eta}-{H_b}_{\eta}
+\frac{2\, f_{\eta} \,H_a}{f} & = & 0 \nonumber \\ \mbox{}
-H_d\,\cot\theta
+H_b\,\cot\theta
+\frac{H_b\, q_{\theta} }{2}
-\frac{H_a\, q_{\theta} }{2}
+H_c\, q_{\eta} & & \nonumber \\ \mbox{}
-2\,H_d\, {\phi}_{\theta} 
+4\,H_b\, {\phi}_{\theta} 
-2\,H_a\, {\phi}_{\theta} 
+6\,H_c\, {\phi}_{\eta} 
+{H_c}_{\eta} & & \nonumber \\ \mbox{}
+\frac{ f_{\eta\eta} \,H_c}{f_{\eta}}
+\frac{2\, f_{\eta} \,H_c}{f}
-{H_d}_{\theta}-{H_a}_{\theta} & = & 0
\end{eqnarray}
And while one may be tempted to use the maximal slicing condition (\ref{eqn:hahbhdcon}) to remove $H_d$ from these equations and use them to solve for some of the extrinsic curvature variables ($H$'s), this yields many problems which we discuss in section \ref{sec:hahbcon}.

In general these equations should only be used as a check on the code as they are first order and degenerate\footnote{i.e. all the terms multiplying extrinsic curvature quantities vanish identically at the $\theta=0$ boundary, so the extrinsic curvature values can literally be anything; something that numerical solvers sniff out and therefore fail to converge.} at $\theta=0$, making them poor choices for obtaining numerical solutions.  Comparing to the more general equations in \cite{physrevd5008}, we find that this is due to our gauge choice of diagonalising the metric by setting $\gamma_{12}=0$, which removes the non-degenerate terms in the momentum constraints.\footnote{Although the authors of \cite{physrevd5008} were still unable to use the momentum constraints for evolution, and found the diagonal metric to be the most stable.}

\subsection{Maximal Slicing Equation ($\alpha$)}\label{sec:maxslice}
If we employ the gauge choice that $TrK=0$, and equation (\ref{eqn:liemaxslice}), we can calculate the \emph{Maximal Slicing constraint}, which ensures that if $TrK=0$ on the initial time slice, it remains so on all future time slices.\footnote{see section \ref{sec:ADM3plus1} and equation (\ref{eqn:liemaxslice}) for more details}
\begin{equation}\label{eqn:maxslicegauge} \gamma^{ad}\left[\partial_d\partial_a\alpha - \Gamma^e_{\;ad}\partial_e\alpha\right] = \alpha R \\ \end{equation}
which becomes
\begin{eqnarray}\label{eqn:maxslice}
& & \frac{ \alpha_{\theta} \,\cot\theta}{f^2}
+\frac{2\, \alpha_{\theta} \, {\phi}_{\theta} }{f^2}
+\frac{2\, \alpha_{\eta} \, {\phi}_{\eta} }{f_{\eta}^2}
+\frac{ \alpha_{\theta\theta} }{f^2}
+\frac{ \alpha_{\eta\eta} }{f_{\eta}^2} \nonumber \\ \mbox{} & &
-\frac{ f_{\eta\eta} \, \alpha_{\eta} }{f_{\eta}^3}
+\frac{2\, \alpha_{\eta} }{f\, f_{\eta} } \nonumber \\ \mbox{}
& = & \alpha\,\left(
-\frac{8\, {\phi}_{\theta} \,\cot\theta}{f^2}
-\frac{q_{\theta\theta} }{f^2}
-\frac{q_{\eta\eta} }{f_{\eta}^2}
+\frac{ f_{\eta\eta} \, q_{\eta} }{f_{\eta}^3}
-\frac{q_{\eta} }{f\, f_{\eta} }  \right. \nonumber \\ \mbox{} & & \left.
-\frac{8\, {\phi}_{\theta\theta} }{f^2}
-\frac{8\,{ {\phi}_{\theta} }^2}{f^2}
-\frac{8\, {\phi}_{\eta\eta} }{f_{\eta}^2}
-\frac{8\,{ {\phi}_{\eta} }^2}{f_{\eta}^2}
+\frac{8\, f_{\eta\eta} \, {\phi}_{\eta} }{f_{\eta}^3}
-\frac{16\, {\phi}_{\eta} }{f\, f_{\eta} } \right)
\end{eqnarray}

This can be re-arranged to give:

\begin{eqnarray}
\left(\frac{f}{f_\eta}\right)^2 \alpha_{\eta\eta}
+\alpha_{\theta\theta}
+\left[2 \phi_{\eta} \left(\frac{f}{f_\eta}\right)^2 - \left(\frac{f_{\eta\eta}}{f_\eta}\right)\left(\frac{f}{f_\eta}\right)^2 + 2\left(\frac{f}{f_\eta}\right)\right] \alpha_{\eta}
& & \nonumber \\ \mbox{}
+\left[\cot\theta + 2 \phi_{\theta}\right] \alpha_{\theta}
- \left[-8\, {\phi}_{\theta} \,\cot\theta
-q_{\theta\theta}
-\left(\frac{f}{f_\eta}\right)^2 q_{\eta\eta} \right. & & \nonumber \\ \mbox{} \left.
+\left(\frac{f_{\eta\eta}}{f_\eta}\right)\left(\frac{f}{f_\eta}\right)^2 \, q_{\eta}
-\left(\frac{f}{f_\eta}\right) q_{\eta}
-8\, {\phi}_{\theta\theta}
-8\,{ {\phi}_{\theta} }^2
-8\, {\phi}_{\eta\eta} \left(\frac{f}{f_\eta}\right)^2 \right. & & \nonumber \\ \mbox{} \left.
-8\,{ {\phi}_{\eta} }^2\left(\frac{f}{f_\eta}\right)^2
+8\, \left(\frac{f_{\eta\eta}}{f_\eta}\right) \left(\frac{f}{f_\eta}\right)^2 \, {\phi}_{\eta}
-16\, \left(\frac{f}{f_\eta}\right) {\phi}_{\eta} \right]\alpha
 & = & 0
\end{eqnarray}
which can be further consolidated using the fact that
$$\frac{d}{d\eta}\left(\frac{f}{f_{\eta}}\right)=1-\frac{f f_{\eta\eta}}{f_{\eta}^2}$$
to give
\begin{eqnarray}\label{eqn:maxslicealpha}
\left(\frac{f}{f_\eta}\right)^2 \alpha_{\eta\eta}
+\alpha_{\theta\theta}
+\left[2 \phi_{\eta} \left(\frac{f}{f_{\eta}}\right)^2 + \frac{f}{f_{\eta}}\left[1+\frac{d}{d\eta}\left(\frac{f}{f_{\eta}}\right)\right]\right] \alpha_{\eta}
& & \nonumber \\ \mbox{}
+\left[\cot\theta + 2 \phi_{\theta}\right] \alpha_{\theta}
- \left[-8\, {\phi}_{\theta} \,\cot\theta
-q_{\theta\theta}
-\left(\frac{f}{f_\eta}\right)^2 q_{\eta\eta} \right. & & \nonumber \\ \mbox{} \left.
-\left(\frac{f}{f_\eta}\right)\frac{d}{d\eta}\left(\frac{f}{f_{\eta}}\right) q_{\eta}
-8\, {\phi}_{\theta\theta}
-8\,{ {\phi}_{\theta} }^2
-8\, {\phi}_{\eta\eta} \left(\frac{f}{f_\eta}\right)^2 \right. & & \nonumber \\ \mbox{} \left.
-8\,{ {\phi}_{\eta} }^2\left(\frac{f}{f_\eta}\right)^2
-8\, \left(\frac{f}{f_\eta}\right) \left[1+\frac{d}{d\eta}\left(\frac{f}{f_\eta}\right)\right] \, {\phi}_{\eta} \right]\alpha
 & = & 0
\end{eqnarray}
This 2nd order elliptic PDE for $\alpha=\alpha(\eta,\theta,t)$ can be solved numerically using the method described in section \ref{sec:4thordderiv}, and as discussed previously allows us to keep our coordinates from evolving into regions of large curvature.

\section[Weyl Curvature and 4D Invariants]{Weyl Curvature, The Newman Penrose Formalism and Curvature Invariants}\label{sec:weyl_np}
One major issue with analysing the results of any numerical relativity simulation which introduces a particular foliation of a generic 4D spacetime in a generally covariant formulation is to relate quantities that are coordinate or slicing dependent to quantities that are independent of the choice of coordinates.  To this end, we can calculate scalar invariants of the 4D Riemann curvature that will tell us what the spacetime is doing in a manner that can be compared across formalisms.

The full four-dimensional Riemann tensor can be decomposed into the trace (Ricci tensor) and traceless (Weyl tensor) parts via\footnote{We use the usual notation that ${}_{[\;]}$ represents the antisymmetric portion of the tensor.}
$${}^{(4)}R_{\mu\nu\alpha\beta} = C_{\mu\nu\alpha\beta} -(g_{\nu\lbrack\alpha}{}^{(4)}R_{\beta\rbrack\mu} - g_{\mu\lbrack\alpha}{}^{(4)}R_{\beta\rbrack\nu}) - \frac{1}{3}{}^{(4)}R \;g_{\mu\lbrack\alpha}g_{\beta\rbrack\nu} $$
where the superscript ${}^{(4)}$ is used to indicate the full four-dimensional tensor and $C_{\mu\nu\alpha\beta}$ is the Weyl tensor.

In a vacuum ($T_{\alpha\beta}=0$) the Einstein equations imply the vanishing of the Ricci Tensor portions ($R_{\alpha\beta}={}^{(4)}R=0$), and we find that 
\begin{equation}\label{eqn:ricciweyl}{}^{(4)}R_{\mu\nu\alpha\beta} = C_{\mu\nu\alpha\beta}\end{equation}
i.e. pure vacuum gravitational radiation is solely expressed in terms of Weyl curvature.  We will therefore use the Weyl curvature to explore the full structure of the 3+1 solution.

In the 1960s Newman, Penrose and others presented a spinor formulation of GR \cite{newmanpenrose}, which has some attractive features in helping to describe and understand the structure of the Einstein field equations.  Specific to our interests here, the mathematical description of an asymptotically flat vacuum spacetime via null tetrads gives a direct correlation between the Weyl scalars and the amplitude of plane polarised gravitational waves.

Various derivations of the Weyl Scalars ($\Psi_0$, $\Psi_1$, $\Psi_2$, $\Psi_3$ and $\Psi_4$) from a null tetrad formalism can be found in \cite{alcubierre:3p1num,bernstein,newmanpenrose,newmantod,Wald}.
The Weyl scalars are the five complex-valued independent components of the Weyl tensor\footnote{Because of the reduced dimensionality in axisymmetry, the Weyl scalars and null tetrad values are all real-valued instead of complex valued, decreasing the number of variables by half; so we have five instead of ten.} expressed in a particular tetrad, and therefore completely describe the Riemann tensor in a vacuum.  In the tetrad used by \cite{newmanpenrose} we find:

\begin{equation} \Psi_0=-C_{(1)(3)(1)(3)} \;;\; \Psi_1=-C_{(1)(2)(1)(3)} \;;\;\Psi_2=-\frac{1}{2}(C_{(1)(2)(1)(2)}+C_{(1)(2)(3)(0)})\nonumber\end{equation}
\begin{equation}\Psi_3=C_{(1)(2)(2)(0)} \;;\; \Psi_4=-C_{(2)(0)(2)(0)}=C_{(0)(2)(2)(0)}\nonumber\end{equation}

where the subscript notation ${}_{(i)}$ indicates projection of that index along the tetrad.
A complete expansion of the Weyl scalars in terms of dynamic spacetime variables in a comparable metric/extrinsic curvature formulation (and the tetrad $\{t,r,\theta,\varphi\}$ which differs from above) can be found in Bernstein \cite{bernstein} (Appendix G), and we omit their derivation as it is outside the scope of this thesis.  We present the equations and necessary coordinate transformations in section \ref{sec:PSINeqns} to calculate the Weyl Scalars in Bernstein's tetrad with our metric/extrinsic curvature variables.

In a vacuum we can also construct two real-valued\footnote{They are complex in general, however once again our symmetries cause them to be real-valued.} non-vanishing 4D scalar invariants $I$ and $J$ of the Riemann curvature from the Weyl Scalars. We note\footnote{The two sources have a slightly different scale factor.} \cite{alcubierre:3p1num,bernstein} that the Riemann curvature invariants $I$ and $J$ have the form:
\begin{equation}\label{eqn:4dIinvariant}
I=8\left[\Psi_0 \Psi_4 - 4 \Psi_1 \Psi_3 + 3 (\Psi_2)^2\right]
\end{equation}
\begin{equation}\label{eqn:4dJinvariant}
J=\left| \begin{array}{ccc}
\Psi_4 & \Psi_3 & \Psi_2 \\
\Psi_3 & \Psi_2 & \Psi_1 \\
\Psi_2 & \Psi_1 & \Psi_0 \\
\end{array}\right| = \Psi_4[\Psi_2\Psi_0-(\Psi_1)^2] - \Psi_3[\Psi_3\Psi_0-\Psi_1\Psi_2] + \Psi_2[\Psi_3\Psi_1-(\Psi_2)^2]
\end{equation}
As $I$ and $J$ are scalar invariants they provide formulation-independent information about what the full 4D spacetime curvature is doing and can be used to compare results from different formulations.

Alcubierre \cite{alcubierre:3p1num} demonstrates that the second time derivative of the amplitude of plane polarised linearised gravitational waves\footnote{In axisymmetry we have only $+$ polarised waves present, and the $\times$ polarisation mode is precluded.} in the asymptotic wave zone is related to $\Psi_0$ and $\Psi_4$, namely that radially \emph{outgoing} waves satisfy
\begin{equation}\label{eqn:outgravwavepsi4}{}_{\rightarrow}\ddot{h}=-\Psi_4\end{equation}
where $h$ is the amplitude of the plane gravitational wave and the overdot represents differentiation with respect to proper time. Radially \emph{ingoing} waves satisfy
\begin{equation}\label{eqn:ingravwavepsi0}{}_{\leftarrow}\ddot{h}=-\Psi_0\end{equation}
so by studying the behaviour of the Weyl scalar quantities in the exterior region of our numerical grid one can determine what gravitational radiation is passing through the region. Another property of interest is that the Weyl Scalars asymptotically $(r\rightarrow \infty)$ obey the ``peeling theorem'' \cite{newmanpenrose,alcubierre:3p1num} in the wave zone of an asymptotically flat spacetime
\begin{equation}\label{eqn:peelweyl}\Psi_n = O(r^{n-5})\end{equation}
in the Newman Penrose tetrad\footnote{See \cite{newmanpenrose} for a formal definition of $O(r)$ in this case and the limitations of this theory.}.

\section{Summary}
Using the ADM formulation of General Relativity in axisymmetry, we have arrived at a set of evolution equations for the metric and extrinsic curvature components, and 3 constraints on these equations (Hamiltonian and 2 momentum).

Through our gauge conditions we have arrived at $2$ dynamic metric variables ($q,\phi$) plus two gauge variables ($v_1,v_2$) that all depend on $(\eta,\theta,t)$ and can evolve via the metric evolution equations.  Using auxiliary variables $\Phi$ and $\chi$ we can decouple the equations to be solved for the shift vector quantities $v_1,v_2$.  We have also derived $4$ evolution equations for the extrinsic curvature variables, and presented some conditions on the lapse functions.

As we have $3$ additional constraints we can use, the overdetermined nature of these equations allows us a certain amount of freedom to do a maximally constrained evolution (3 constraints + 6 evolution equations), a free evolution (9 evolution equations) or something in the middle.

In this thesis we choose\footnote{Many different combinations were attempted, and this was the one that yielded the best results.} to do a mostly free evolution by using the 3 of the 4 metric evolution equations to solve for $q$ and create constraints on the shift vector components $v_1$ and $v_2$, leaving the fourth evolution equation as a numerical check.  We employ the Hamiltonian constraint to solve for $\phi$, and leave the momentum constraints for use as numerical checks.

We use the 4 evolution equations for the extrinsic curvature variables $H_a$, $H_b$, $H_c$ and $H_d$.  If we employ maximal slicing we instead use 3 extrinsic curvature evolution equations and a constraint to solve for $H_d$.

In total, the equations presented here give a complete framework for solving the evolution of the spacetime, which are further refined in Chapter \ref{chap:numcode}.

\fancyhead[RO,LE]{\thepage}
\fancyfoot{} 
\chapter{Numerical Methods and Lessons Learned}\label{chap:changes}
\bigskip

In this chapter we will discuss (1) a variety of numerical methods that were needed to tackle the large project of creating a numerical Brill wave simulation computer code and (2) a number of problems (and their solutions) that were encountered in the process of attempting to understand the origin of certain numerical irregularities.  A discussion of two tests that were performed in 1+1 dimensions (e.g. an alternate formalism (BSSN\footnote{Baumgarte-Shapiro-Shibata-Nakamura \cite{BSSN,BSSN2}.}) and an alternate gridding method) is presented in Appendix \ref{chap:testgridcoord}.

The most difficult part of writing an axi-symmetric Brill gravitational wave evolution code in spherical-polar coordinates is to regularise all of the equations that in theory can be solved using different coordinate systems, gauges, numerical splittings, etc.  Historical approaches have proven insufficient or unsuccessful, so this project necessitated a careful re-evaluation of these methods.
More specifically: in the course of creating a numerical code to analyse the set of differential equations presented in chapter \ref{chap:mathrel}, one must make decisions regarding what one chooses to use for coordinates, variables, gridding, finite difference operators, gauge conditions, convergence techniques, boundary conditions, constrained versus free evolution and numerical solvers to name the main ones.

In some cases, certain methods may seem self-evident or obvious. However given the lack of literature discussing the difficulties that arise from implementing such a numerical code in GR, this chapter represents both a review and an introduction to old and new techniques.

\section{Gridding}\label{sec:grid}
As discussed in chapter \ref{chap:nummethod}, throughout this thesis we use a finite difference approach to solving the PDEs that arise from the equations that are derived from equation (\ref{eqn:einstein}).  For the axisymmetric problem this involves discretising radial, angular and time coordinates into a grid of discrete positions and times, and using $n$-th order approximations to compute partial derivatives at each point along the grid.

\begin{figure} \centering
%\psfrag{P}{$\pi$}
\includegraphics{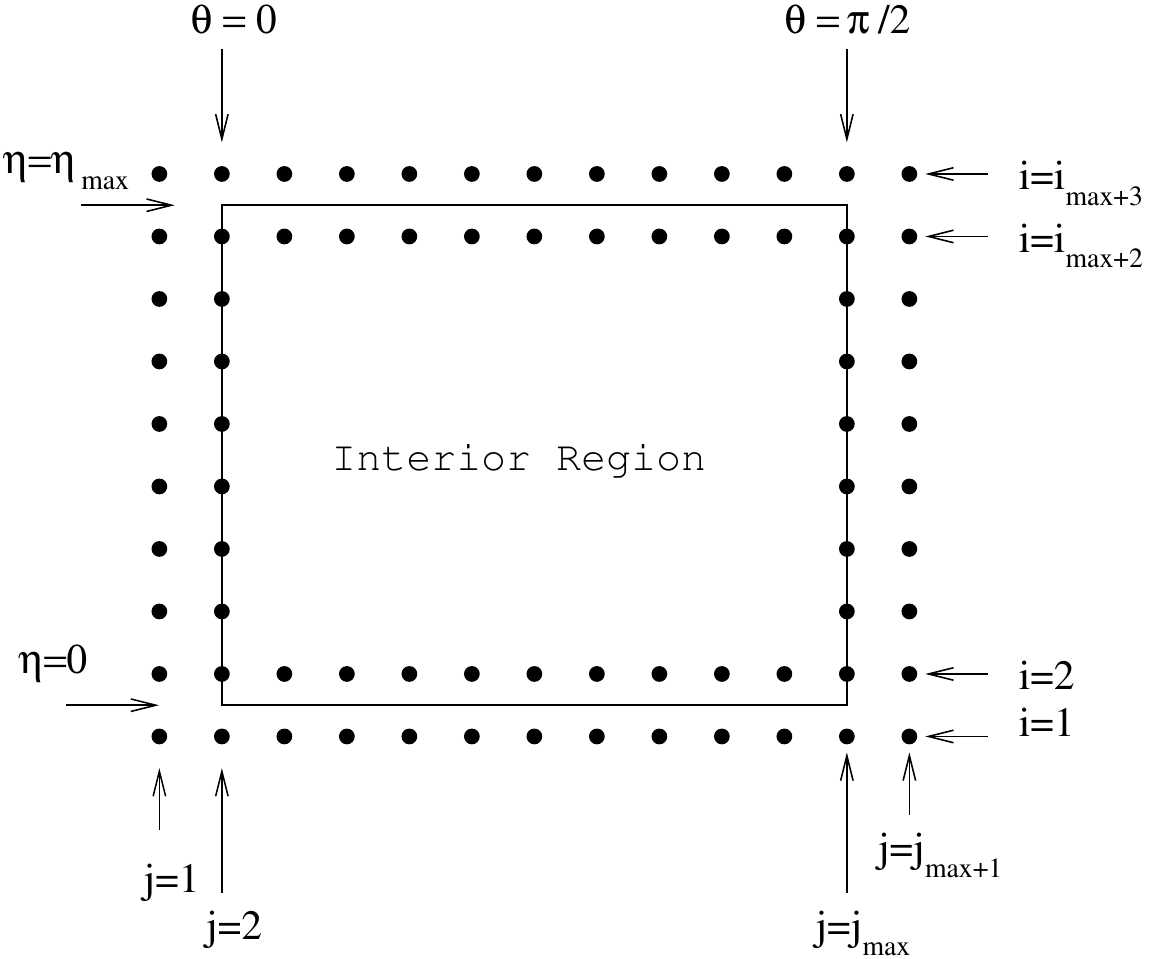}
\caption[Numerical Grid]{Schematic of the grid positioning in $\eta$ and $\theta$.  The square box represents the physical boundary of the coordinate region $\eta=\left[0 ,\eta_{\mathbf{max}}\right] , \theta=\left[0 ,\frac{\pi}{2}\right]$}\label{fig:2dgrid}
\end{figure}

See figure \ref{fig:2dgrid} for a depiction of the spatial gridding, and table \ref{tbl:gridcoords} for a list of the coordinates and their gridding.  We generally keep the radial, angular and time step sizes consistent\footnote{In exploring some late-time evolutions we have altered $\Delta t$ such that $t_k=t_{k-1}+\Delta t_k$, with $\Delta t_k$ not constant.} and define our time counter integer $k \geq 0$ such that $t=k \Delta t$.  This gridding also means that we can relate our physical coordinate values to grid points via:
$$\eta=\left(i-\frac{3}{2}\right)\Delta\eta \;;\; \theta=(j-2)\Delta\theta$$

\begin{table}\begin{center}
\begin{tabular}{|c|c|c|c|}\hline
variable & grid spacing & inner region & inner region \\
 & & (start) & (end)\\ \hline
$r$ or $\eta$ & $\Delta\eta$ & $i=2$ & $i=imax+2$ \\
$\theta$ & $0.9\Delta\eta\leq\Delta\theta\leq 1.1\Delta\eta$& $j=2$ & $j=jmax$ \\
$t$ & $\Delta t<\Delta\eta \;(0.1\Delta\eta-0.5\Delta\eta)$ & $t=0$,$k=0$ & ?? \\ \hline
\end{tabular}\caption[Relation between grid points and coordinates]{Relation between grid points and coordinates (see also figure \ref{fig:2dgrid}). } \label{tbl:gridcoords}
\end{center} \end{table}

So-called ``phantom'' grid points are employed outside the fixed spatial boundaries, as discussed in section \ref{subsec:bc_gen}, which are necessary for defining partial derivatives on or near the boundaries.

\subsection{Staggered or Non-Centered Differencing}
As discussed in section \ref{sec:CTCSSLFtests}, after an investigation of a non-centered vs centered in time finite differencing schemes, we found that the centered scheme performed better.

We also expect in a system with evolving, non-linear behaviour that we should implement spatial finite difference operators that are directionally symmetric.  There are cases when choosing a non-centered finite difference operator is advantageous (e.g. when one has a preferred spatial direction in the advection operator). However in the case of non-linear wave behaviour that involves self-interactions and backscattering there is no clear, preferred direction of propagation\footnote{Except perhaps near the outer radial regions of the grid where one expects gravitational waves to propagate radially outwards.}.

\section{Roundoff Errors and Summation}\label{subsec:addterms}
In the process of adding together the terms on the RHS of equations such as (\ref{eqn:haevol}), we can have a large number of terms that need to be summed.  As we only have a finite amount of precision available to us\footnote{With 64-bit double precision floating point numbers, we get about 16 decimal places of precision under the IEEE 754 specification.}, it happens that we need to be more careful about the order of summation than just summing terms in the order in which they are listed (recall the discussion around rounding errors in general that was presented in section \ref{subsec:rounderror} - what follows is a demonstration of this phenomena in action and how to mitigate it).

The smaller terms will get dwarfed by the larger ones, as we can have many orders of magnitude difference between the values of operands.  As such, the terms need to be sorted then summed from smallest to largest to ensure no more decimal places get lost than are unavoidable.  Some of our equations have upwards of 17 terms to sum together, so we need to employ a rapid but necessary reordering of the summation.

For example, let us consider taking the mixed spatial second derivative of one of our functions, $\Phi(\eta,\theta)$.  The equation for taking a fourth order correct mixed derivative is given by
\begin{eqnarray}\label{eqn:phibadderiv}
\Phi_{\eta\theta}(i,j)&=&\frac{1}{144~\Delta\eta~\Delta\theta}(\Phi_{i-2,j-2} - 8 \Phi_{i-1,j-2} + 8 \Phi_{i+1,j-2} - \Phi_{i+2,j-2} \nonumber \\ \mbox{}
& & - 8 \Phi_{i-2,j-1} + 64 \Phi_{i-1,j-1} - 64 \Phi_{i+1,j-1} + 8 \Phi_{i+2,j-1} \nonumber \\ \mbox{}
& & + 8 \Phi_{i-2,j+1} - 64 \Phi_{i-1,j+1} + 64 \Phi_{i+1,j+1} - 8 \Phi_{i+2,j+1}  \nonumber \\ \mbox{}
& & - \Phi_{i-2,j+2} + 8 \Phi_{i-1,j+2} - 8 \Phi_{i+1,j+2} + \Phi_{i+2,j+2})
\end{eqnarray}
where $i$ and $j$ represent radial and angular grid counters respectively.  We will break this calculation down into two steps: (i) summing together 16 ``terms'' and (ii) dividing by the scaling factor $({144~\Delta\eta~\Delta\theta})$.

Let us take equation (\ref{eqn:phibadderiv}) and write it as such:
\begin{equation}\label{eqn:phiderivarrange}\Phi_{\eta\theta}(i,j)=\frac{1}{144~\Delta\eta~\Delta\theta}\left(\sum_{i=1}^{16}s_i(i,j)\right)\end{equation}
where
$$s_1=\Phi_{i-2,j-2}$$
$$s_2=-8\Phi_{i-1,j-2}$$
etc.  We now compare the difference between a ``direct'' calculation of $\Phi_{\eta\theta}$ which involves performing the sum $\sum_{i=1}^{16}s_i$ in the order that the terms appear in equation (\ref{eqn:phibadderiv}):
$$\sum_{i=1}^{16}s_i=s_1 + s_2 + s_3 + s_4 + s_5 + \ldots $$
versus re-ordering the summation order such that we have a one-to-one mapping
$$u_j=s_i \; ; \; i,j=\{1,2,3 \ldots 16\}$$
where
$$|u_1| \le |u_2| \le |u_3| \le |u_4| \le |u_5| \ldots \le |u_{16}| $$
and performing the mathematically equivalent sum
\begin{equation}\label{eqn:rearrange_add}\sum_{i=1}^{16}s_i=\ldots ((((u_1 + u_2) + u_3) + u_4) + u_5) + \ldots\end{equation}
instead when computing values via equation (\ref{eqn:phiderivarrange}).\footnote{In reality we add all terms $u_i$ of the same order $n$ together first, where $n=floor(log(|u_i|))$, starting with the smallest $n$ first.}

Table \ref{tbl:phiderivarrange} demonstrates the different results obtained by using the two different methods, at some chosen radial and angular grid points.
\begin{table}\begin{centering}
\begin{tabular}{|c|c|c|c|} \hline
$i$ & $j$ & $\Phi_{\eta\theta}$ (direct) & $\Phi_{\eta\theta}$ (re-ordered addition order)  \\ \hline
 $50$ & $27$ & $0.19178074794456\mathbf{801}$     & $0.19178074794456\mathbf{795}$ \\
 $50$ & $28$ & $0.344900061903187\mathbf{50}$     & $0.344900061903187\mathbf{34}$   \\  
 $50$ & $29$ & $0.51236262322511184$      & $0.51236262322511184$ \\ 
 $50$ & $30$ & $0.765964939485532\mathbf{24}$      & $0.765964939485532\mathbf{01}$ \\  
 $50$ & $31$ & $\mathbf{-1.82372661856680216\times 10^{-016}}$  & $\mathbf{0.0000000000000000}$ \\   
 $51$ & $2$ & $\mathbf{1.55683979633751402\times 10^{-016}}$  & $\mathbf{0.0000000000000000}$ \\    
 $51$ & $3$ & $2.052669220767\mathbf{50048}\times 10^{-2}$ & $2.052669220767\mathbf{48834}\times 10^{-2}$ \\
 $51$ & $4$ & $2.631945962671\mathbf{81174}\times 10^{-3}$ & $2.631945962671\mathbf{64304}\times 10^{-3}$ \\
 $51$ & $5$ & $-2.9546433347372\mathbf{0742}\times 10^{-3}$ & $-2.9546433347372\mathbf{6554}\times 10^{-3}$ \\ \hline
\end{tabular}\caption[$\Phi_{\eta\theta}$ calculated using direct or re-ordered addition]{Comparative analysis of the fourth order correct mixed partial derivative $\Phi_{\eta\theta}$ using a direct and re-ordered addition order for numerical calculation. Bolded digits indicate a difference between the two calculation methods.}\label{tbl:phiderivarrange}\end{centering}\end{table}

Note how the derivatives at $j=31$ ($\theta=\frac{\pi}{2}$) and $j=2$ ($\theta=0)$, which should be $0$ due to symmetric boundary conditions across those boundaries (that are implemented analytically in the code), are non-zero if we do not re-order the addition of terms in equation (\ref{eqn:phibadderiv}).

This is a demonstration of how the non-commutativity of addition in finite precision algebra affects the implementation of our code.  For a discussion of algorithms and coding for implementing this procedure in general see section \ref{sec:algorithmdesign}.

The use of a 4th order finite differencing scheme highlighted the need for this change; as we become more demanding of the accuracy of the differencing operators other numerical methods must be adjusted as well.

\begin{figure} \centering
\includegraphics{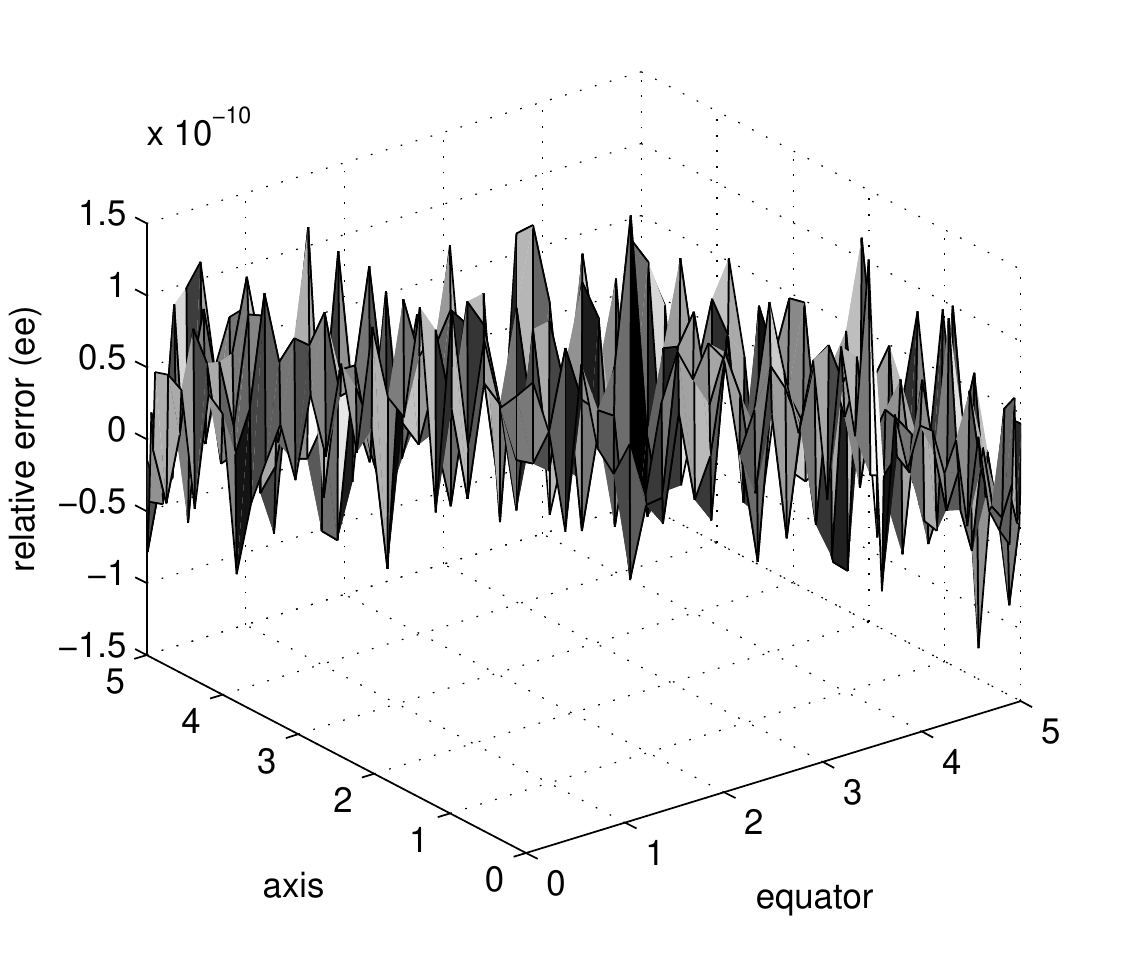}
\caption[Relative difference of the coefficient of $\alpha$]{Relative difference in the coefficient that multiplies $\alpha$ when trying to solve the maximal slicing equation at $t=\Delta t$ using two different methods.  See equation (\ref{eqn:eediff}).}\label{fig:alphaee_reldiff}
\end{figure}

As another example, consider the elliptic equation for the lapse, $\alpha$, given by (\ref{eqn:maxslicealpha}).  The coefficient for the term that multiplies $\alpha$ is given by:
\begin{eqnarray}
e_e & = & - \left[-8\, {\phi}_{\theta} \,\cot\theta
-q_{\theta\theta}
-\left(\frac{f}{f_\eta}\right)^2 q_{\eta\eta} \right. \nonumber \\ \mbox{} & & \left.
-\left(\frac{f}{f_\eta}\right)\frac{d}{d\eta}\left(\frac{f}{f_{\eta}}\right) q_{\eta}
-8\, {\phi}_{\theta\theta}
-8\,{ {\phi}_{\theta} }^2
-8\, {\phi}_{\eta\eta} \left(\frac{f}{f_\eta}\right)^2 \right. \nonumber \\ \mbox{} & & \left.
-8\,{ {\phi}_{\eta} }^2\left(\frac{f}{f_\eta}\right)^2
-8\, \left(\frac{f}{f_\eta}\right) \left[1+\frac{d}{d\eta}\left(\frac{f}{f_\eta}\right)\right] \, {\phi}_{\eta} \right]
\end{eqnarray}
This coefficient has $9$ terms in the sum.  Figure \ref{fig:alphaee_reldiff} shows the relative difference
\begin{equation}\label{eqn:eediff}\frac{\tilde{e}_e-e_e}{e_e}\end{equation}
at the outer boundary ($5$ outer-most radial grid points) between the coefficient $e_e$ calculated using summation of terms in the order listed above versus $\tilde{e}_e$ calculated using re-ordering of operands.  The relative error is much larger than numerical precision ($6$ orders of magnitude larger than $\sim 10^{-16}$) and very ``choppy''.

\begin{figure} \centering
\includegraphics{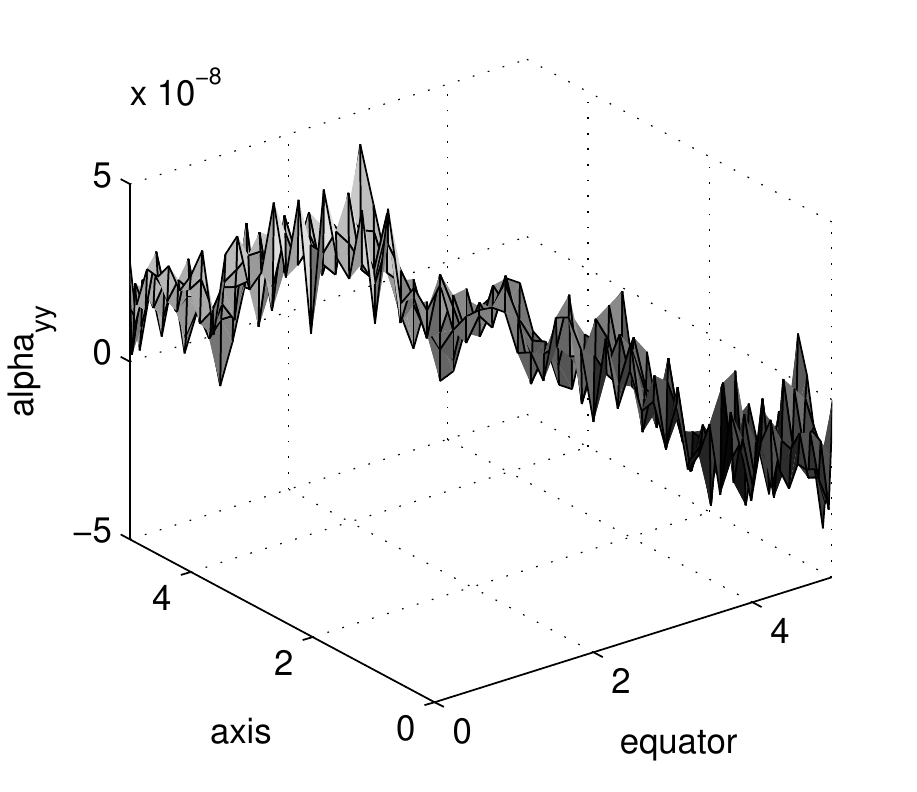}
\caption[$\alpha_{\theta\theta}$ without coefficient operand rearrangement]{Second angular derivative of the lapse, $\alpha_{\theta\theta}$, near the outer boundary calculated from the maximal slicing equation without rearrangement of coefficient operands.}\label{fig:alphayy_OB_norearrange}
\end{figure}

\begin{figure} \centering
\includegraphics{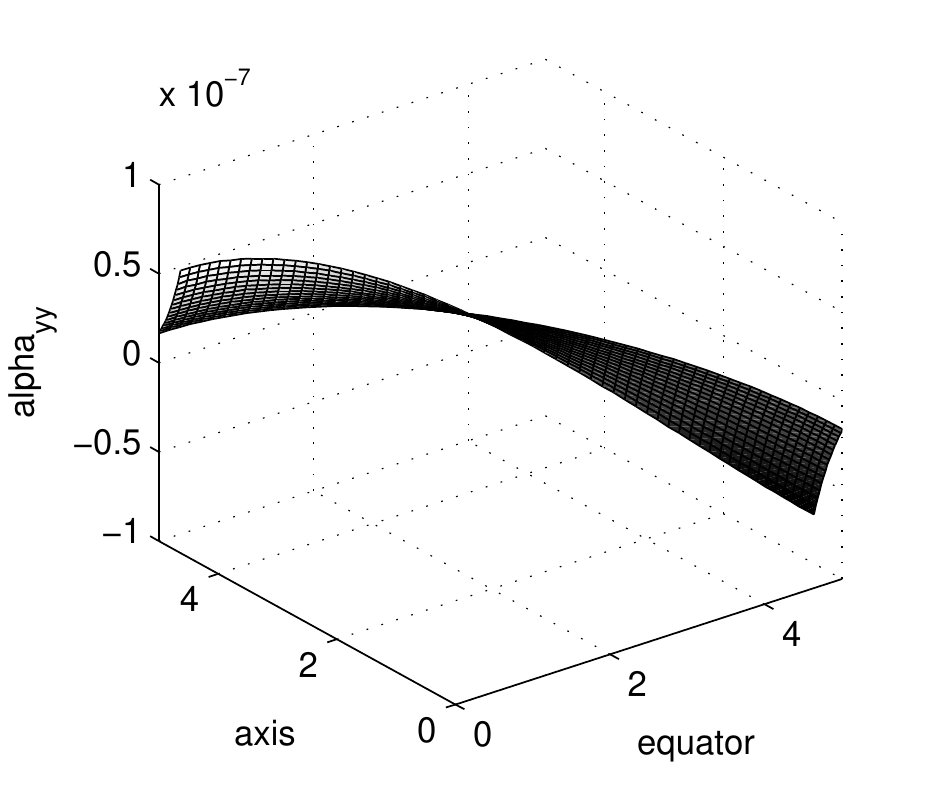}
\caption[$\alpha_{\theta\theta}$ \emph{with} coefficient operand rearrangement]{Second angular derivative of the lapse, $\alpha_{\theta\theta}$, near the outer boundary calculated from the maximal slicing equation \emph{with} rearrangement of coefficient operands.  The only difference between this method and figure \ref{fig:alphayy_OB_norearrange} is that we use the algorithm described in equation (\ref{eqn:rearrange_add}) to add operands before solving (\ref{eqn:maxslicealpha}) for $\alpha$.}\label{fig:alphayy_OB_yesrearrange}
\end{figure}

We then show the second angular derivative of the solution that is calculated for $\alpha$ at these grid points generated in the first (non-rearranged) and second (with rearrangement) cases; see figures \ref{fig:alphayy_OB_norearrange} and \ref{fig:alphayy_OB_yesrearrange} respectively.

From this we see that the ``choppiness'' in calculating our coefficients propagates into the calculation of the solution to our elliptic equation, and can have a significant impact on our numerical solution. With the re-arrangement of operands we achieve a smooth solution to equation (\ref{eqn:maxslicealpha}) instead of a choppy one.

This demonstrates once again the non-commutativity of certain numerical operations and that a naive application of even the summation process can have a profound effect on the subsequent results.

\section{Finite Differencing Considerations}

\subsection{Issues with 2nd order accurate spatial derivatives}\label{sec:4thorder}
To examine the behaviour of a second order correct finite differencing method, consider the function $F(x)$ at the point $x_2$, and defining $x_1=x_2-\Delta x$ we know that
\begin{equation}\label{eqn:taylorx1} F(x_1)=F(x_2) + \frac{\partial F}{\partial x}(-\Delta x) + \frac{\partial^2 F}{\partial x^2}\frac{(-\Delta x)^2}{2} + \frac{\partial^3 F}{\partial x^3}\frac{(-\Delta x)^3}{3!} + \frac{\partial^4 F(\xi_1)}{\partial x^4}\frac{(-\Delta x)^4}{4!} \;;\; \xi_1 \in [x_1,x_2] \end{equation}
using Taylor's theorem (\ref{eqn:taylor}), where the derivative terms are evaluated at $x_2$.  Similarly defining $x_3=x_2+\Delta x$ Taylor's theorem tells us that
$$F(x_3)=F(x_2) + \frac{\partial F}{\partial x}(\Delta x) + \frac{\partial^2 F}{\partial x^2}\frac{(\Delta x)^2}{2} + \frac{\partial^3 F}{\partial x^3}\frac{(\Delta x)^3}{3!} + \frac{\partial^4 F(\xi_2)}{\partial x^4}\frac{(\Delta x)^4}{4!} \;;\; \xi_2 \in [x_2,x_3] $$
Combining these we know that:
$$\left.\frac{\partial^2F}{\partial x^2}\right|_{x_2} = \frac{F(x_1) - 2 F(x_2) + F(x_3)}{(\Delta x)^2} + E_2$$
where $E_2$ is our ``second order'' error term and is given by:
$$E_2=\left[\frac{\partial^4 F(\xi_1)}{\partial x^4}-\frac{\partial^4 F(\xi_2)}{\partial x^4}\right]\frac{(\Delta x)^2}{4!} \;;\; x_1 \leq \xi_1 \leq x_2 \leq \xi_2 \leq x_3$$
Note also that in (\ref{eqn:taylorx1}) that our lowest order error term from all sources must be $\sim(\Delta x)^4$ in order for the above equations to hold.

For the sake of this discussion we now define the error term order $\varepsilon$ such that
$$\varepsilon(E_2)\sim(\Delta x)^2$$
in our example above and $\varepsilon$ represents the lowest order dependence on grid spacing of a function's error term.
Thus any second derivative term has an error term $\varepsilon(F_{xx})\sim (\Delta x)^2$ for a second order discretisation method.  This error dependence can obviously cause problems if we substitute this value into an equation for some other variable and try to take too many derivatives.  To wit, if a functional
$$C=C(F,F_x,F_{xx})$$
and possibly depends on other variables, then $C$ will have an error term $\varepsilon(C)\sim (\Delta x)^2$, and $C_{xx}$ will have an error term of order $\varepsilon(C_{xx})\sim 1$, which is ill-conditioned.

\subsubsection[Motivation for fourth order spatial derivatives]{Motivation for moving to fourth order correct spatial derivatives}
More specific to the problem at hand, if we examine the extrinsic curvature evolution equations in section \ref{sec:extrincurvevol} we see that they all depend on the second spatial derivatives of $q$ and $\phi$, and on the first spatial derivatives of each other.  Using a second order discrete derivative method implies then that each extrinsic curvature variable has an error term of
$$\varepsilon(H_i)\sim \left[(\Delta \eta)^2 + (\Delta \theta)^2\right](\Delta t)$$
from their dependence on various derivatives of $q$ and $\phi$, and thusly an error term of order $\sim\left[(\Delta \eta) + (\Delta \theta)\right](\Delta t)$ when substituted into each other, or $\sim(\Delta \eta)^2$ assuming
$$\Delta \eta \simeq \Delta \theta \simeq \Delta t$$

Further, examining the evolution equation for $q$ (\ref{eqn:qdot}) we see that it depends linearly on the extrinsic curvature variables, so if the extrinsic curvature variables have error terms of order $\varepsilon(H_i)\sim(\Delta \eta)^2$ then the second derivative of $q$ will fail to be regular when propagated onto future time slices\footnote{From above, the second derivative is two orders lower in correctness than the values used, so $(2-2=0)$ which means we have errors of order $(\Delta \eta)^0=1$}.  So while $q$ may have a regular second derivative on one time slice, when we propagate forward to the next time slice we lose two orders of correctness.

There are two methods to help correct this numerical problem and \emph{both} had to be employed in the 2+1 code to achieve stability:
\begin{enumerate}
\item Employing finite difference expressions that are correct to 4th order in the discretisation to provide second derivatives that are well-defined on future time steps.  For example, instead of the error term for $\varepsilon(\phi_{\eta\eta})\sim (\Delta \eta)^2$ in a second order method we find $\varepsilon(\phi_{\eta\eta})\sim (\Delta \eta)^4$ in a fourth order method.  This means that our extrinsic curvature variables have errors $\varepsilon(H_i)\sim (\Delta \eta)^3$ and similarly for $q$, so its second derivative $q_{\eta\eta}$ is regular.  This allows you to ``start'' evolving variables onto the next time slice in a well-posed manner, although the lowest order error is still degrading (i.e. $O((\Delta \eta)^4) \rightarrow O((\Delta \eta)^2)$).  It is important to note that this degradation of order prevents the Taylor series expansion from having the desired form (\ref{eqn:taylorx1}), and therefore our numerical methods will fail.

To demonstrate this, if we set the relationship between temporal and spatial discretisation to be $\Delta t = (\Delta \eta)^3 < (\Delta \eta)^2$ with 2nd order finite differencing we find that the code still fails to evolve properly in the explicit time evolution case (see section \ref{subsec:explicit_time}).  For example, the code crashes in $8$ time steps for the initial data set with Amplitude and ``width'' $(A=-1 \times 10^{-10}, s_0=3)$ simply due to numerical errors across the grid.  

Setting $\Delta t = (\Delta \eta)^3$ according to CLF stability criteria for linear hyperbolic and parabolic PDEs should provide a stable evolution using 2nd order finite difference approximations.  This clearly does not occur for the equations being evolved here.  Instead, what we observe is that the values along the outer boundary are dominated by numerical error within very few time steps (see figure \ref{fig:qy_a-1e-10_s03_explicit_t_evol_t7} for an example of what happens).\footnote{Even setting the time step several order of magnitude smaller has no effect - the author was unable to create a stable explicit time evolution.}

\begin{figure} \centering
\includegraphics{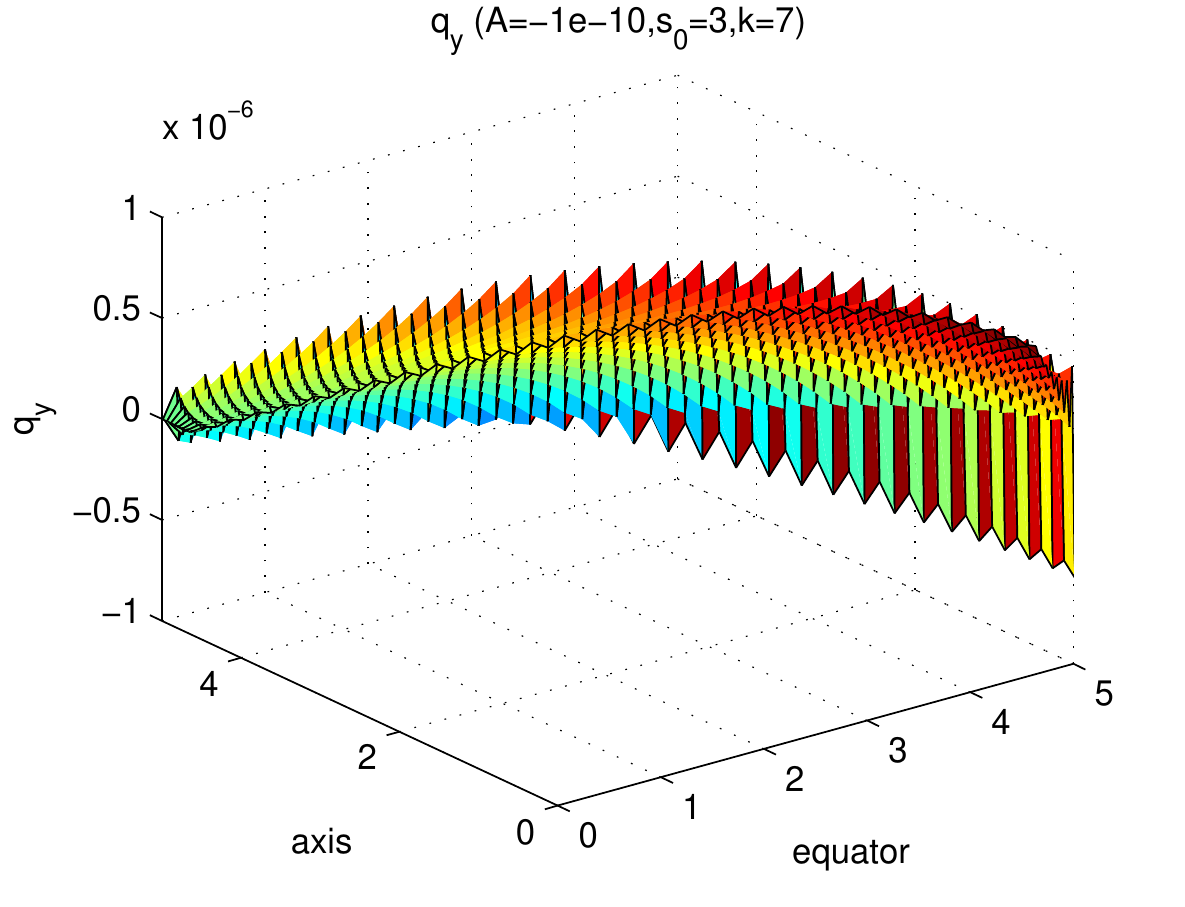}
\caption[$q_{y}$ at outer boundary for explicit time evolution]{First angular derivative of $q$ near the outer radial boundary for $A=-1 \times 10^{-10}$, $s_0=3$, $t=7 \Delta t (k=7)$ using an explicit time evolution.  The code crashes on the next time step $k=8$.  The errors are on the scale of the numerical grid and typical of poorly chosen numerical methods.}\label{fig:qy_a-1e-10_s03_explicit_t_evol_t7}
\end{figure}

\item An implicit time evolution method must be used to couple all variables on neighbouring time steps to each other, in order to keep finite differencing orders from degrading.  If we are using fourth order finite differences we need $\varepsilon(q)\sim(\Delta \eta)^5$, however as we saw in the previous point when solving the equations together in an explicit manner we find that $\varepsilon(q)\sim(\Delta \eta)^3$ on future time slices.  By coupling the current and future time slices together (i.e via an implicit time evolution technique such as a Crank-Nicholson method) we can recover $\varepsilon(q)\sim(\Delta \eta)^5$ so that our Taylor series is well-defined.
\end{enumerate}

Furthermore, gravitational waves need at least quadrupole terms to propagate in the asymptotic radiative zone\footnote{See section \ref{subsec:gravwave}} - so 2nd order correct methods won't allow radiative terms to propagate properly and will instead cause unnecessary dissipation.  To see this, we examine the typical quadrupole Legendre polynomial
$$Q(r,\theta)=\frac{P_2(\cos\theta)}{r^3}=\frac{3\cos^2\theta-1}{2r^3}$$
Consider the scenario where $\theta$ is near the axis (i.e.  $\theta=k \Delta\theta$, $k$ some small integer) then
$$Q \sim \frac{1}{r^3}\left(1 + \frac{3k^2}{4}(\Delta\theta)^2\right) $$
to first non-vanishing order in our spatial gridding in order to capture angular information.  This means that any numerical scheme that isn't correct to at least $(\Delta\theta)^3 $ will not be able to properly capture the angular component of any quadrupole waves.

As additional motivation to employ higher order discretisation, in a collaboration of multiple numerical relativity groups working on gravitational wave predictions from Black Hole Binary inspirals (called the NINJA project \cite{NINJA}), all of the groups were using at least fourth order spatial discretisation, with some moving more recently to 8th order \cite{Lousto}.

The transition from a second order spatial finite differencing method to a fourth order one required a major recoding effort, but probably yielded one of the largest benefits of any of the changes mentioned here.  Most regularity problems near the origin and axis that remained after the implementation of other numerical regularisation techniques described in this chapter disappeared with the implementation of a 4th order spatial correct numerical scheme.\footnote{Also important to note, as mentioned in section \ref{sec:momcon}, is that the inability to use the momentum constraints as Choptuik \emph{et. al.} \cite{choptuik1,choptuik2} do in their cylindrical symmetry codes is possibly what sets this coordinate system (spherical polar) apart in needing 4th order to propagate.}

\subsection{Time evolution of hyperbolic equations in our formalism}\label{sec:4ordtime}
The evolution equations for the metric and extrinsic curvature we encounter are first order in time.  For reasons discussed in section \ref{sec:4thorder} we must use an implicit method to couple future time slices to the current time slice.  So we abandon the use of the explicit centered in time method discussed in \ref{subsec:explicit_time} and instead employ one of the family of Crank-Nicholson methods discussed in section \ref{subsec:cn}.

\subsubsection[Doubly Iterative, Implicit Crank-Nicholson]{Doubly Iterative, Modified Crank-Nicholson, Implicit Time Evolution}\label{subsubsec:cntimeevol}
For the problem at hand we introduce an implicit method based on a Crank-Nicholson method that we use iteratively.  Let us write equation (\ref{eqn:qdot}) schematically as
\begin{equation}\label{eqn:cn-q}\frac{\partial q}{\partial t}=F(q,v_2,H_d,H_b,\alpha)\end{equation}
which leads to
\begin{eqnarray}\frac{q^{k+1,(n,m)}-q^{k}}{\Delta t} &=&\frac{1}{2}\left[F^{k+1}\left(q^{(n-1,m-1)},v_2^{(n-1)},H_d^{(n-1)},H_b^{(n-1)},\alpha^{(n-1)}\right)\right. \nonumber \\ \mbox{}
& & \left. +F^{k}(q,v_2,H_d,H_b,\alpha)\right]
\end{eqnarray}
Where we evaluate at $(n,m)$ unless stated otherwise, and
\begin{itemize}
\item $k+1$ is the time step we are solving for
\item $n$ is a counter that represents what iteration through the global iterative Crank-Nicholson algorithm we are on (iterating through all equations on the time step)
\item $m$ represents what iteration through the individual variable's iterative scheme we are on\footnote{Hence ``doubly iterative'' - we have to iterate on each variable for convergence then on the global variable set for a second convergence.}.
\end{itemize}

We then solve for $q^{k+1,(n,m)}$ and iterate through the variable's Crank-Nicholson scheme until we reach a specified tolerance
$$\left|\frac{q^{k+1,(n,m)}-q^{k+1,(n,m-1)}}{q^{k+1,(n,m)}}\right| \le 10^{-16}$$
where $10^{-16}$ is chosen as this solver converges very quickly to within machine precision (usually $4-5$ iterations).

We further note that while equation (\ref{eqn:cn-q}) for $q$ involves only linear combinations of variables so it would be possible to concurrently solve for all variables using a matrix technique, the equations for the extrinsic curvature terms (e.g. (\ref{eqn:haevol})) have many non-linear terms in them necessitating a global iterative scheme.

Even with the possibility of using auxiliary variables the complication (and thusly potential for complicated numerical error, typographical mistakes, etc.) inherent in putting together an implicit solver that couples all of the neighbouring (in space and time) grid points across all $\sim 11$ equations makes the author believe that the effort is not warranted at this point\footnote{The quasi-linear equation for $\phi$ and the equations for the shift vector potentials $(\chi,\Phi)$ would further complicate this endeavour. \cite{allen} created an implicit scheme for the linearised equations, however the fully non-linear equations require some form of iteration.}.

Once we have solved for a variable, we proceed on to the next one.  As the variables are all coupled there is no preferred order in which they are evaluated\footnote{It may be possible to speed convergence by altering the order that variables are solved for; a possible future extension.}.  We then look for a global convergence criteria on all variables
\begin{equation}\label{eqn:relcnconv}\frac{\xi^{(k+1,n)}-\xi^{(k+1,n-1)}}{\xi^{(k+1,n)}} \le 10^{-4}\end{equation}
where $\xi$ represents any of $(q,\phi,\alpha,H_a,H_b,H_b,H_d,v_1,v_2)$, this check is performed if $2 \le n \le 10$ and the value $10^{-4}$ is a maximum relative error that is based on experimentation/observation\footnote{In general this algorithm cannot converge to better than one part in $\sim 10^4$ across the entire grid for all variables.}.  This algorithm suffers from some of the same difficulties discussed in section \ref{subsubsec:bicgimprove} where some variables are of vastly differing orders of magnitude in various regions of the grid; the variables may possess strongly non-linear behaviour in the interior region and demonstrate wave behaviour in the exterior region.  This variability in behaviour of the variables causes difficulties in defining a \emph{global} convergence measure.

It is possible that a better convergence condition than equation (\ref{eqn:relcnconv}) is required, however in practice the algorithm seems to perform well, only failing to converge in $10$ passes through the global loop when we have:
\begin{itemize}
\item strongly non-linear behaviour in some of the variables in the interior zone and 
\item the values of the variables in the outer (large radius) regions of the grid are having difficulty settling down to some small value (so the relative errors are larger).
\end{itemize}

Perhaps a relaxation technique or linear interpolation/extrapolation would speed convergence, however this would be an area for future research.

\section{Numerical Methods for solving multi-dimensional elliptic PDEs}

\subsection{The Stabilised Bi-Conjugate Gradient Algorithm (BiCGStab)}\label{subsec:bicgstab}
As discussed in section \ref{subsec:matrixeqns}, there are many ways to solve the matrix equation $Ax=b$ depending on the problem at hand.  We will now proceed with a discussion of a conjugate gradient method that is appropriate to large, sparse systems such as those we encounter when solving elliptic PDEs as presented in section \ref{sec:2orpde4ord}.
For a mathematical development of the algorithm (including stabilisation considerations compared to a typical Conjugate Gradient algorithm) see \cite{bicgstab}.

The implementation of this algorithm by the first generation of coders\footnote{From the code:

\emph{c This code was first written by Peter Anninos, David Bernstein, David Hobill,}

\emph{c Edward Seidel, Larry Smarr, and John Towns at NCSA, with support from}

\emph{c the University of Illinois, the State of Illinois, the National Science}

\emph{c Foundation, and other federal agencies.}
} provided a number of advantages.  Thankfully their work has been preserved so that it could be incorporated into this thesis.  I will present the theoretical algorithm as implemented in the code, and it has been tested extensively (both by myself and previous generations of coders) for accuracy.

It was modified and retested for this thesis to allow for 4th order, 9-point stencils due to the requirements mentioned in section \ref{sec:4thorder}, however the basic framework remains intact.  This BiCG algorithm only performs matrix multiplications, and does not perform a direct inversion to solve the matrix problem $Ax=b$. The technique is to calculate multi-directional paths of ``steepest descent'' in order to obtain a stable solution.

As such, the speed of the method is largely dependent on the ability of the user to define an efficient matrix multiplication algorithm, and is therefore well suited to large, sparse systems.  The matrix problems encountered are generally defined on a spatial grid with the total number of nodes given by:
$$i_{max}~ \times ~j_{max} \simeq 200 \times 60 = m \times n$$
(i.e. $200$ radial grid zones and $60$ angular grid zones).  The corresponding matrix problem we need to solve in (\ref{eqn:4thordmatrix}) is a $12,000 \times 12,000$ matrix, with only 9 entries on most rows - i.e it is highly sparse.  This is obviously poorly suited to direct inversion techniques such as Gaussian elimination, partial pivoting, etc. and is an obvious candidate for iterative methods.

One last advantage of this routine over the relaxation techniques discussed in section \ref{subsubsec:relaxmatrix} is that conjugate gradient methods are directionally agnostic - they don't start at one point in the grid and spread from there - they simultaneously shift the entire solution across the entire grid at each iteration.

The input required to initialize the algorithm consists of:
\begin{itemize}
\item $A$ (coefficient matrix, $(mn)\times (mn)$, obtained from the form of the elliptic equation to be solved)
\item $x$ (initial guess, return value is solution, $(mn)\times 1$ vector)
\item $b$ (RHS of matrix equation $Ax=b$, $(mn)\times 1$ vector)
\item $tolerance$, the maximum tolerable residual $|b-Ax|^2$
\item Maximum number of iterations is set at $2 i_{max} j_{max}$
\end{itemize}

\verb+BEGIN INITIALIZATION+

%Initialization stage of algorithm:
Diagonally scale the matrix problem if possible, then:
\begin{eqnarray}
r & = & b \nonumber \\ \nonumber \mbox{}
a_p & = & Ax \\ \nonumber \mbox{}
r & = & r-a_p \\ \nonumber \mbox{}
p & = & r \\ \nonumber \mbox{}
\Delta & = & \sum_{l}r_{l} \\ \nonumber \mbox{}
a_p & = & Ap \\ \nonumber \mbox{}
\phi & = & \frac{\sum_{l}a_p}{\Delta} \\ \nonumber \mbox{}
rnorm & = & \sqrt{\sum_{l}r_{l}^2}
\end{eqnarray}

\verb+END INITIALIZATION+

\verb+BEGIN MAIN LOOP+
%Loop part of algorithm:

\begin{eqnarray}
\Omega & = & \frac{1}{\phi} \nonumber \\ \nonumber \mbox{}
w & = & r-\Omega a_p \\ \nonumber \mbox{}
a_s & = & Aw \\ \nonumber \mbox{}
\chi_1 & = & \sum_{l}\left[(a_sw)_{l}\right] \\ \nonumber \mbox{}
\chi_2 & = & \sum_{l}(a_s)_{l}^2 \\ \nonumber \mbox{}
\chi & = & \frac{\chi_1}{\chi_2} \\ \mbox{}
r & = & w-\chi a_s \; \rightarrow \; r=b-A x \label{eqn:bicgshiftr} \\ \nonumber \mbox{}
x & = & x+\Omega p + \chi w \\ \nonumber \mbox{}
\Delta_p & = & \Delta \\ \nonumber \mbox{}
\Delta & = & \sum_{l}r_{l} \\ \nonumber \mbox{}
p & = & r+(p-\chi a_p)\frac{\Omega \Delta}{\chi \Delta_p} \\ \nonumber \mbox{}
a_p & = & Ap \\ \nonumber \mbox{}
\phi & = & \frac{\sum_{l}a_p}{\Delta} \\ \nonumber \mbox{}
rnorm & = & \sqrt{\sum_{l}r_{l}^2}
\end{eqnarray}

\verb+END MAIN LOOP+

Where $1 \le l \le i_{\mathbf{max}} \times j_{\mathbf{max}}$.  The code originally checked to see if $rnorm > tolerance$ at each iteration (which is improved on below), but still returns an error if the maximum number of iterations is reached before achieving the desired tolerance.  The code then returns its best guess for $x$.

It is possible to parallelise this algorithm when implementing it in a code, however parallelisation in this case is not terribly efficient.  As the calculation of the ``gradient'' variables $\Delta$, $\chi$, $\Omega$ on each iteration depends on the values from \emph{every} $i$ and $j$, the best one can do is parallelise the small sub-loops.

Unfortunately the overhead from parallelising these small loops is larger than the gains of splitting $200\times 60=12000$ calculations over multiple processors given the speed of modern processors.  With the algorithm as written it is better to overclock a single processor designed for single-threaded calculations as much as possible.  At the time of writing an overclocked $5$GHz processor was used, so one can see why splitting up $12000$ relatively simple calculations is not beneficial.

\subsubsection[Improvements on BiCGStab]{Improvements on the Stabilised Bi-Conjugate Gradient Algorithm (BiCGStab)}\label{subsubsec:bicgimprove}
Note that because the algorithm ``shifts''/updates the RHS of the matrix equation on each iteration (see equation \ref{eqn:bicgshiftr}), virtually arbitrary tolerances can be achieved in theory.

In reality, one reaches the limits of numerical precision from accumulated error rather rapidly and while the algorithm may calculate that the ``residual'' is less than the specified tolerance, and that the $r_{l}$ terms are getting smaller, the reality is that you can still be quite far from the actual solution.  To fix this, we replace the calculation of the residual term in (\ref{eqn:bicgshiftr}) with
$$r=b-A x$$
where $b$ is the original right hand side of the matrix equation we are trying to solve.  This change alone caused a significant increase of the ability of the solver to converge to within a specified tolerance.

The second major improvement we make to the algorithm is related to equation (\ref{eqn:rearrange_add}) and the discussion around summation of a large number of terms of varying magnitude in section \ref{subsec:addterms}.

The previous discussion demonstrates the \emph{numerical} difference between summing $16$ terms together in various manners and how re-arrangement benefits the calculation of the actual sums.  In the case of calculating, for example, $\Delta$ in this algorithm we must add together $m \times n$ terms which for a standard grid is $200\times 60=12000$ terms!  To this end, a function was developed\footnote{See section \ref{sec:algorithmdesign} for a discussion.} to add together all the elements of a matrix via reordering and the differences were very noticeable.  Before the implementation of a rearrangement algorithm $\Delta$ was $0.3$ in one case (that was unable to converge to a solution) and afterwards it was $0.7$ (and the algorithm was able to converge).

This change in summation methodology reduced the speed of the solver due to the addition of a sorting routine\footnote{i.e. the move from dumb addition (fast) to a sorting algorithm (slow).}.  While more work could probably be done to develop a more efficient sorting algorithm, its introduction ensured that the BICG solver could converge for a much wider range of problems.

The third and last improvement is in terms of convergence criteria.  Once again, accumulated error in calculating the value of variables over the grid causes a calculation of $rnorm$ to be suspect, and it is also unreasonable to expect the convergence in a region where $x_{ij} \sim 10^{-2}$ to be the same as a region where $x_{ij} \sim 10^{-300}$.  Furthermore, comparing the \emph{absolute} differences (via $rnorm$) in these regions is meaningless.

An attempt was made to compare \emph{relative} errors, however let us instead take the approach of looking at the leading order behaviour of our ``gradient'' variables $\chi$, $\Omega$ and $\Delta$.

To very rough leading order
$$\chi = \frac{\chi_1}{\chi_2} \sim \frac{\max(A)w^2}{\max(A)^2w^2} \sim \frac{1}{A_{ij}(max)}$$
$$\Omega = \frac{\Delta}{a_p} = \frac{\sum{r_{l}}}{\sum{Ap}} \sim \frac{1}{A_{ij}(max)}$$
$$\Delta=\sum_{l}{r_{l}}=\sum_{l}{(Ax-b)_{l}}=\sum_{l}{\left(\sum_{p}{({Ax})_{lp}}-b_l\right)}$$
where $p$ is an index that iterates through the $5$ to $9$ non-zero entries in the sparse matrix multiplication of one row of $A$ with the solution vector $x$.

This analysis tells us that $\chi$ and $\Omega$ are largely unaffected by anything but the maximum value of the coefficient matrix, which doesn't change from iteration to iteration\footnote{This is verified by monitoring the values as the solver iterates through thousands of iterations; they vary by about two orders of magnitude but are always around $1\rightarrow 100$ with a diagonally scaled problem.}.

The value of $\Delta$ decreases significantly the closer we get to our ``actual'' solution\footnote{In theory it should $\rightarrow 0$ as we converge on the correct solution, so it can vary over 15+ orders of magnitude depending on the initial guess.}.  The best we can ask of $\Delta$ is that it is less than (machine numerical precision) $\times$ (the maximum value of an individual term $(Ax)_{lp}$ over \emph{all} $(l,p)$).  Any values less than this will be insignificant compared to numerical noise.\footnote{Runs were tried with $10^{-15}$ and $10^{-16}$ as the multiplying factor, but the extra order of magnitude requirement for precision just slowed the code down significantly for no benefit as the end results were virtually indistinguishable.  So we use $10^{-15}$.}
\begin{equation}\Delta \le 10^{-15} \times \max_{l,p}|(Ax)_{lp}| \end{equation}
This change in convergence criteria allowed the algorithm to be used (and converge) for a wide variety of problems, from small perturbative waves to large amplitude cases\footnote{This sort of convergence analysis was lacking from previous implementations, and had a significant impact on the utility of the solver.}.

\subsection{Third Party Software for Numerical Relativity}
Some attempt was made to use the built-in numerical solvers of programs like Maxima, Maple or Matlab, however these engines are much higher-level programming languages and suffer from the ``abstraction penalty'' - the inability of easily programmable languages to offer efficient and highly customisable code.  While optimally everything could be coded in assembly language for fastest execution, the barrier to entry for non-specialists is large and the time to code/debug is prohibitive\footnote{If one were to spend time on such an initiative, the largest gains would probably be in coding the matrix multiplication routine for the elliptic solver in assembly language.}.  Conversely, programs that offer a large array of pre-canned algorithms (like Matlab) hide the fine details from the user and make it impossible to perform some of the required numerical customisations that we have implemented.  All attempts to use high-level languages resulted in prohibitively long calculation times or non-convergence, so their use was limited to things like calculating matrix condition numbers instead of spending the effort to code a condition number solver.

Therefore as a compromise, a ``mid-level'' language like C or Fortran is perfectly suited to this sort of problem.

Another class of pre-canned methods for tackling the complexities of numerical relativity code lies in initiatives like the CACTUS code and associated ``thorns'', which some members of the relativity community have been putting great effort into (e.g. \cite{etoolkit}).  This suite of code offers some pre-coded adaptive mesh refinement algorithms, wave information extraction routines, parallel computation implementation and other numerical tools instead of having to re-invent the wheel for every simulation.

There is, however, still a barrier to entry and it seems that groups still have to expend significant energy defining custom ``thorns'' to suit their particular situation.  It is possible that if this code needed an adaptive mesh refinement implementation or higher order derivatives that it would be less work to port to a CACTUS/thorn model than develop them from scratch, however this is debatable.

\section{Condition Numbers}\label{sec:condition}
One measure of the well-posedness of a matrix problem
$$Ax=b$$
is the condition number of the coefficient matrix $A$, which indicates its propensity to amplify or remove error from calculations when it multiplies a solution vector $x$.

This is especially important in iterative schemes, as each pass through the matrix multiplier can potentially cause unbounded numerical noise if the coefficient matrix is ill-conditioned.

We can define the $n$-condition number $K_n$ in the following manner after \cite{burden}: if $E_r$ is the relative error between our approximate solution $\tilde{x}$ and the actual solution $x$ defined by
$$E_r=\frac{||x-\tilde{x}||_n}{||x||_n}$$
and $r$ is the residual represented by
$$r=b-A\tilde{x}$$
and $||.||_n$ is the $n$-norm, then the condition number $K_n$ of the coefficient matrix $A$ satisfies the following inequality:
$$E_r \leq K_n\frac{||r||_n}{||b||_n}$$
and is defined exactly by
$$K_n=||A||_n||A^{-1}||_n$$
It is not easy to compute condition numbers for $12,000 \times 12,000$ matrices. Instead, we modify the grid size and obtain the conditions number for smaller coefficient matrices using MATLAB (with advances in computing power over the last $5-10$ years we can now accomplish in 10 minutes on an overclocked $5$ GHz processor what was not possible in the past).  In the case of the elliptic equation for $\phi$ (\ref{eqn:hamconphi}) we present the condition numbers of the coefficient matrix for various grid sizes (including the $200 \times 60$ size used in various simulations) in table \ref{tbl:condition}.

\begin{table}\begin{center}\begin{tabular}{|c|c|c|c|}\hline
Grid Size & Norm Order $n$ & Condition Number $K_n$ & $K_n/\#_{points}$\\ \hline
$28 \times 10$ & $1$ & $4680$ & $16.7$ \\
$28 \times 10$ & $2$ & $2119$ & $7.6$ \\
$28 \times 10$ & $\infty$ & $4025$ & $14.4$\\
$57 \times 20$ & $1$ & $25 503$ & $22.3$ \\
$57 \times 20$ & $2$ & $9516$ & $8.3$\\
$57 \times 20$ & $\infty$ & $25 020$ & $21.9$ \\
$200 \times 60$ & $1$ & $2.8215\times 10^6$ & $235$ \\
$200 \times 60$ & $2$ & $1.9835\times 10^5$ & $165$ \\
$200 \times 60$ & $\infty$ & $2.6415\times 10^5$  & $220$ \\\hline
\end{tabular}\end{center}\caption[Condition numbers $K_n$]{Condition numbers $K_n$ to solve equation (\ref{eqn:hamconphi})}\label{tbl:condition}\end{table}

A condition number near ``$1$'' means that any numerical error is approximately carried through each matrix multiplication with no increase or decrease in amplitude (less than ``$1$'' is preferred, as it can cause contraction).  The condition numbers listed above, however, are typical of ``ill-conditioned'' matrices ($10^3-10^6$), which create a host of issues for large scale numerical solutions and are very sensitive to perturbations and numerical methodology.

Also included is a rough measure of how fast the condition numbers blow up as the grid size increases (condition number per grid point).  This has no firm/real meaning for measures like the $\infty$ norm, which is measuring the maximum error, but it demonstrates that ``more grid points'' is not necessarily better.  While more grid points may reduce the amount of error in calculating a derivative for example, it may cause a blow-up of the error in the attendant matrix problems (such as those encountered in the Hamiltonian constraint and shift vector equations).

There are some cases \cite{bergamaschi,dijkstra} where one can show analytically that when solving Laplace's equation as a discretised matrix problem the condition number depends on the \emph{inverse of the grid spacing}
$$K_n \sim \frac{1}{(\Delta \eta)^p} \;;\; p\ge 1$$
so this result is not entirely unexpected.

This also helps to account, however, for the large degree of difficulty in obtaining a numerical solution to this particular problem for the last 40+ years...

\section{Regularity of tensors near $r=0$}\label{sec:r0reg}
As discussed in Evans \cite{evans} (Chapter 4, Section C), we expect a ``regular''\footnote{Regular tensors are defined as tensors whose ``Cartesian components are analytic, i.e. expandable in a Taylor series in the neighbourhood of each point''.} second rank tensor $U_{ab}$ quantity to have the form

$$U_{rr} = k_1 \sin^2\theta + k_2 \cos^2 \theta + k_3 r^2 \sin^4\theta + 2k_4 r^2\cos^2\theta\sin^2\theta$$
$$U_{\theta\theta} = r^2(k_1 \cos^2\theta + k_2 \sin^2 \theta) + (k_3-2k_4) r^4 \cos^2\theta\sin^2\theta $$
$$U_{\phi\phi} = k_1 r^2\sin^2\theta$$
$$U_{r\theta} = [k_1 - k_2 + k_3r^2\sin^2\theta + k_4r^2\cos(2\theta)]r\cos\theta\sin\theta$$
near the origin, where $k_i=k_i(\rho^2,z^2)$, and similarly for the $h_i$ and $f_i$ used below.

We will now investigate the regularity conditions of the metric near $r=0$. Making the identification 
$$g_{11}=U_{rr} \;;\; g_{22}=U_{\theta\theta} \;;\; g_{33}=U_{\phi\phi}$$
and remembering that our metric is given by (\ref{eqn:3dmetric}), we find that the $U_{\phi\phi}$ condition becomes
$$e^{4\phi}f^2\sin^2\theta=k_1 r^2\sin^2\theta$$
which immediately allows us to say that
\begin{equation}\label{eqn:regk1}e^{4\phi}=k_1\end{equation}
and therefore we expect our conformal factor, $\phi$, to be even and single valued as it approaches $r=0$.

Evaluating the other two conditions and noting that for our metric because of our isothermal gauge
$$\frac{f^2}{f_\eta^2} g_{11}=g_{22}=f^2e^{q+4\phi}$$
we see that
$$r^2 U_{rr} = U_{\theta\theta}$$
which allows for the identification
$$k_1 \sin^2\theta + k_2 \cos^2 \theta=k_1 \cos^2\theta + k_2 \sin^2 \theta \rightarrow k_1=k_2$$
and
$$r^2\sin^2\theta(k_3\sin^2\theta+2k_4\cos^2\theta)=r^2\sin^2\theta(k_3\cos^2\theta-2k_4\cos^2\theta)$$
which can be uniquely solved to eliminate either $k_3$ or $k_4$.  Therefore we can write our last regularity condition as
$$e^{4\phi}e^q=k_1\left[1+r^2\frac{k_4}{k_1} l(\theta)\right]$$
Recalling (\ref{eqn:regk1}), and noting that $e^q=1+q+\ldots$ we find that
\begin{equation}\label{eqn:qr2regmetric}q\sim r^2\end{equation}
and therefore $q$ must be ``small'' near the origin to ensure that the exponential series converges quickly, the regularity conditions are met and therefore the metric is regular.  Previous incarnations of this code (and some others in the literature\footnote{However codes that have a black hole with a Brill wave superimposed do not necessarily have $r=0$ in the computational domain.}) did not enforce this condition on the initial slice when setting up the initial data for $q$.  Nor did they derive/impose appropriate boundary conditions during the evolution, however we will change our prescribed initial function to ensure that it falls off as $r^2$ as we approach the origin and set our boundary conditions accordingly.

We therefore expect both $q$ and $\phi$ to be single-valued and even as they approach $r=0$, with an additional requirement on $q$ from equation (\ref{eqn:qr2regmetric}).

The behaviour of the second rank mixed tensors is the following:
$$K^r_{\;r} = h_1\sin^2\theta + h_2\cos^2\theta + h_3 r^2\sin^4\theta + 2 h_4 r^2 \cos^2\theta \sin^2\theta$$
$$K^{\theta}_{\;\theta} = h_1\cos^2\theta + h_2\sin^2\theta + (h_3-2h_4) r^2\cos^2\theta\sin^2\theta$$
$$K^{\varphi}_{\;\varphi} = h_1T^2$$
$$K^r_{\;\theta}=[h_1(T^2+2) + 2h_3r^2\sin^2\theta + h_4 r^2\cos(2\theta)]r\cos\theta\sin\theta$$
where
$$T=1+k_5 r^2 \sin^2\theta$$
Recalling how we defined our mixed extrinsic curvature variables in (\ref{eqn:mixkijtensor}), we can expect $K^r_{\;r}$-like quantities (i.e. $H_a$) to be multivalued as we approach the origin in a regular solution.  Similarly $T^{\theta}_{\;\theta}$-like terms (i.e. $H_b$) should be multivalued as we approach the origin, as with $T^{\varphi}_{\;\varphi}$ ($H_d$) and our $T^r_{\;\theta}$ ($H_c$) terms should be single-valued\footnote{This explains the ``kink'' seen in figures \ref{fig:h11} and \ref{fig:h11_orig}}.

Our rank one tensors (i.e. the shift vectors) have the following behaviour
$$W^r = r(\sin^2\theta f_1 + \cos^2\theta f_2)$$
and
$$W^{\theta} = \sin\theta\cos\theta(f_1-f_2)$$
near the origin, meaning that we would expect $\beta_r$ to be single-valued at the origin, and $\beta^{\theta}$ to be multi-valued.

\section{Coordinate conditions at the axis and equator}\label{sec:regconditions}
In attempts to avoid the coordinate singularity at $\theta=0$ a computational grid that straddled that region with grid points at $-\frac{\Delta\theta}{2}$ and  $\frac{\Delta\theta}{2}$ was utilized in a number of axi-symmetric codes.  There are some constraints, however, that must be enforced on the axis ($\theta=0$) and equator ($\theta=\frac{\pi}{2}$) for stability/regularity.  We can determine these additional constraints by examining the curvature terms in our current coordinate system, as well as any evolution equations that we will be using.  Any ill-behaved quantities must be investigated.

\begin{enumerate}
\item The ``$\phi_{\theta}\cot\theta$'' terms present in $R^1_{\;1}$, $R^2_{\;2}$ and $R^3_{\;3}$ (equations \ref{eqn:r11}, \ref{eqn:r22} and \ref{eqn:r33}) mean that $\phi$ must be symmetric across $\theta=\left\{0,\frac{\pi}{2}\right\}$.  Enforcing this with explicit definitions helps ensure regularity along the axis.
\item The ``$q_\eta\cot\theta$'' term in $R^1_{\;2}$ (equation \ref{eqn:r12}) means that $q$ must be anti-symmetric across $\theta=0$
\item The ``$q_\theta\cot\theta$'' term in $R^2_{\;2}$ (equation \ref{eqn:r22}) means that $q$ must be symmetric across $\theta=\left\{0,\frac{\pi}{2}\right\}$.
\item The combination of these two conditions means that $q$ and its derivatives must vanish across $\theta=0$ (the only way to be both symmetric and anti-symmetric).  This is also consistent with the Brill conditions presented in section \ref{subsec:brillformalism}.
\item The evolution equations for the extrinsic curvature variables are derived from the mixed curvature variables, so not surprisingly they yield the same conditions on $\phi$ and $q$ as above.
\item In devising equations (\ref{eqn:gam33dot}) and (\ref{eqn:qdot}) we require that $v_2$ is symmetric across $\theta=\left\{0,\frac{\pi}{2}\right\}$ for regularity.
\item Consideration of the \emph{first} momentum constraint in equations (\ref{eqn:momcons}) show us that $H_c=0$ is a necessary algebraic constraint at both $\theta=0$ and $\theta=\frac{\pi}{2}$ in order to maintain regularity.  Numerical noise seems to be a larger problem than this algebraic constraint and so this constraint is imposed in the code.
\item Consideration of the \emph{second} momentum constraint in equations (\ref{eqn:momcons}) shows us that for regularity along the axis the quantity $(-H_d+H_b)|_{\theta=0}=0$.   This is a numerically unstable algebraic constraint in the \emph{maximal slicing gauge}, so it should be used as a check on the code, as imposing it seems to cause many numerical problems.  Evans \cite{evans} uses this as an impetus for defining a new variable in the \emph{maximal slicing gauge} to mimic this quantity and ease the imposition of the constraints, however at this point I do not see a value in recoding the simulation for this variable shift.  It may be that in the future such a shift is necessary, dependent on the behaviour of this quantity which should be closely monitored\footnote{It turns out to not be an issue for this code, see section \ref{subsec:momcon-hd+hb}}.
\item The maximal slicing equation (\ref{eqn:maxslicealpha}) adds the requirement that $\alpha$ is symmetric across $\theta=\left\{0,\frac{\pi}{2}\right\}$ due to the presence of the ``$\alpha_{\theta}\cot\theta$'' term.
\item Due to the degenerate nature of the \emph{second} momentum constraint in equations (\ref{eqn:momcons}) we require that $\partial_r H_c=0$ at $\theta=0$ to ensure consistency.  As we already require $H_c=0$ it is a redundant requirement.
\end{enumerate}
We will use these conditions to define some boundary conditions on our variables in a future section.

\subsection{Regularity conditions for the lapse function $\alpha$}\label{subsec:alphareg}
The use of spherical polar coordinates necessitates a careful consideration of the regularity of the lapse function $\alpha$.

Firstly it should be noted that the $\frac{1}{f^2}$ coefficient multiplying many of the extrinsic curvature evolution terms in section (\ref{sec:extrincurvevol}) means that all variables multiplied by this term in the region $f=0$ must have at \emph{lowest order} an $f^2$ radial dependence.  This can be accomplished by noting the following in the region near $r=0$:

Let us consider the
\begin{equation}\label{eqn:qreq1}\frac{dH_b}{dt} \sim \frac{\alpha q_{\theta}}{f^2}\end{equation}
term in the evolution equation for $H_b$ (\ref{eqn:hbevol}) and the
\begin{equation}\label{eqn:qreq2}\frac{dq}{dt} \sim \alpha H_b\end{equation}
term in the evolution equation for $q$ (\ref{eqn:qdot}).

Let us assume that 
\begin{equation}\label{eqn:qreq3}\alpha \sim c_0 + c_2 f^2 + c_4 f^4 + O(f^6)\end{equation}
to satisfy the symmetric boundary condition $\alpha(\Delta\eta)=\alpha(-\Delta\eta)$ at the origin, and $c_i=c_i(t,\theta)$ in the most general case.

As we require that $q\sim f^2$ near the origin\footnote{See section \ref{sec:r0reg}.}, (\ref{eqn:qreq1}) tells us that to ``lowest'' order in $f$
$$H_b \sim \alpha$$
When we proceed to the next time step (or use, for example, an implicit time evolution scheme) substituting this dependence into (\ref{eqn:qreq2}) we find that to lowest order
$$q \sim \alpha^2$$
Recalling (\ref{eqn:qreq3}) and the requirement that $q \sim f^2$, we see that $c_0=0$ and
\begin{equation}\label{eqn:qreq4}\alpha \sim c_2 f^2 + c_4 f^4 + O(f^6)\end{equation}

Next we note that a term that frequently appears in the non-rearranged field equations is
$$\left(\frac{f_\eta}{f}\right)^2 H_c$$
which implies that $H_c$ needs at lowest order an $f^2$ dependence to ensure regularity near $f=0$.  Due to the presence of the $\alpha_{\eta\theta}$ and $\frac{\alpha}{f}$ terms in the evolution equation for $H_c$ (\ref{eqn:hcevol}), we now have an additional requirement that
\begin{equation}\label{eqn:hcorigreg} H_c \sim f^2 \rightarrow \alpha \sim f^3\end{equation}
to lowest order.  This, combined with (\ref{eqn:qreq4}), means that $c_2=0$ and our condition on $\alpha$ is now
\begin{equation}\label{eqn:alpharadorig}\alpha \sim c_4 f^4 + O(f^6)\end{equation}

Similarly in the angular direction, we see that the presence of the
$$\frac{dH_d}{dt} \sim \alpha_\theta \cot\theta$$
term in the evolution equation for $H_d$ (\ref{eqn:hdevol}) and the symmetry condition $\alpha(\Delta\theta)=\alpha(-\Delta\theta)$ necessitates
\begin{equation}\label{eqn:alphaangorig}\alpha \sim d_2(t,\eta) \sin^2\theta + O(\theta^4)\end{equation}
near $\theta=0$.

Combining (\ref{eqn:alpharadorig}) and (\ref{eqn:alphaangorig}) we then find that
\begin{equation}\label{eqn:alphaorig}\alpha \sim f^4 \sin^2\theta\end{equation}
To lowest order near the origin and axis.

The equations (\ref{eqn:alpharadorig}) and (\ref{eqn:alphaangorig}) constitute a major result in defining the regularity conditions for the Brill wave \emph{evolution} problem, as one treatment\footnote{Brill \cite{Brill}, while he does not discuss the evolution, does set $\alpha=1$ on the initial slice.} is to set $\alpha=1$ on the entire initial slice, and most others use maximal slicing on the initial slice.  These evolution schemas then proceed to use maximal, polar, etc. slicing on subsequent slices.  This analysis shows that you cannot do this and expect the evolution to proceed in a regular manner, and explains why other attempts (i.e. \cite{eppley}) have failed to evolve past $\sim 3$ time steps (which is what this author faced for many many years, until finally realising the source of the problem).

This also has many other implications, including

- the solution for $q$ cannot evolve on the axis via the portion of the evolution equation (\ref{eqn:qdot}) that contains the lapse (although the portions containing the shift vector can still evolve)

- variables will ``evolve'' at vastly different rates in different regions of the grid.  This is normal in maximal slicing which has the required propery that it slows the evolution in areas of large curvature, however this lapse is not curvature dependent.

- if an apparent horizon (i.e. a trapped surface that encloses a region from $\theta=0 \rightarrow \theta=\frac{\pi}{2}$) is not present on the initial slice it will probably not appear later in the evolution for purely numerical reasons, as we likely cannot fully explore the spacetime (as $\alpha \rightarrow 0$ along the axis and at the origin).

One particular expression which satisfies (\ref{eqn:alphaorig}) and asymptotically $\rightarrow 1$ off the axis is
\begin{equation}\label{eqn:alphastatic}\alpha=\tanh^4(\eta)\sin^2(\theta)\end{equation}

\subsection{General constraints on the lapse in axisymmetry}\label{subsec:genlapsecon}
In \cite{physrevd5008}, the general axisymmetric equations in $(\eta,\theta)$ coordinates, without our gauge assumptions, are derived and presented.  The only assumption is that the metric and extrinsic curvature variables are conformally decomposed in the form:
\begin{equation}\gamma_{ij}=\psi^4\hat{\gamma}_{ij}=\left[\begin{array}{ccc}
a \psi^4 & c \psi^4 & 0 \\
c \psi^4 & b \psi^4 & 0 \\
0 & 0 & d \psi^4 \sin^2\theta \\
\end{array}\right]
\end{equation}

\begin{equation}K_{ij}=\psi^4\hat{K}_{ij}=\left[\begin{array}{ccc}
H_a \psi^4 & H_c \psi^4 & 0 \\
H_c \psi^4 & H_b \psi^4 & 0 \\
0 & 0 & H_d \psi^4 \sin^2\theta \\
\end{array}\right]
\end{equation}
The variable $\delta$, which appears in the denominator of the general terms quite frequently, is defined by:
$$\delta=ab-c^2$$
and is used to simplify expressions.  The same analysis presented above applies to the general equations, as we recall that\footnote{See section \ref{sec:r0reg}.}
$$a \sim O(1) \;;\; b \sim r^2 \;;\; c \sim r$$
so to leading order
$$\frac{1}{\delta} \sim \frac{1}{r^2} $$
As before, we define the general form of $\alpha$ as
\begin{equation}\label{eqn:alpreggen1}\alpha \sim c_0 + c_2 f^2 + c_4 f^4 + O(f^6)\end{equation}
So similarly to what we saw in section \ref{subsec:alphareg}, the $R_{\theta\theta}$ terms in \cite{physrevd5008} are of lowest leading order $1$ near the origin\footnote{For example, anything with the terms $\frac{b}{\delta} \sim 1$ or $\frac{c^2}{\delta}\sim 1$.}, so
$$\dot{H}_b \sim \frac{\alpha R_{\theta\theta}}{\psi^4} \rightarrow H_b \sim \alpha$$
this means that when we plug this back into the evolution equation for $b$ we find that
$$\dot{b} \sim \alpha H_b \rightarrow b \sim \alpha^2$$
and the requirement that $b\sim r^2$ once again restricts $\alpha$ via (\ref{eqn:alpreggen1}) to
\begin{equation}\label{eqn:alpreggen2}\alpha \sim c_2 f^2 + c_4 f^4 + O(f^6)\end{equation}
Noting once again the presence of terms of the form
$$\frac{H_c}{\delta} \times O(1)$$
throughout the evolution and momentum constraint equations \cite{physrevd5008}, we once again require\footnote{Due to the antisymmetry of $H_c$ across $r=0$ this can be further restricted to $H_c \sim f^3$.}
$$H_c \sim f^2 \rightarrow \alpha \sim f^3$$
due to the presence of the $\alpha_{\eta\theta}$ term in the evolution equation for $H_c$.  This once again restricts our lapse function to the form
\begin{equation}\label{eqn:alpreggen3}\alpha \sim c_4 f^4 + O(f^6)\end{equation}
near the origin.

In the angular direction, we note the presence of the term in the evolution equation for $H_d$:
$$\frac{a d \alpha_{\theta} \cot\theta}{\delta\psi^4}$$
The angular components of the general expressions for regularity of the tensors near the origin require that
$$a \sim 1 \;;\; b \sim 1 \;;\; c \sim \sin^2\theta \;;\; d \sin^2\theta \sim \sin^2\theta \rightarrow d \sim 1$$
so
$$\dot{H}_d \sim \alpha_{\theta} \cot\theta \rightarrow \alpha \sim \sin^2\theta$$
as before.

This is a general result that is independent of our gauge choices, Brill waves, etc. and only requires an axisymmetric ADM spacetime in spherical polar coordinates.

\section{Outer boundary conditions}\label{sec:OBcond}
When one is specifying outer boundary conditions, one must be careful to match them to a physically relevant condition.  A ``flat'' outer boundary condition such as a Neumann condition
$$\left.\frac{\partial\alpha}{\partial\eta}\right|_{\eta_{max}}=0 \rightarrow \alpha(\eta_{max}+\Delta\eta)=\alpha(\eta_{max})$$
or a Dirichlet condition
\begin{equation}\label{eqn:dirichletob}H_a(\eta_{max})=H_a(\eta_{max}+\Delta\eta)=0\end{equation}
can result in a non-physical boundary condition with the variables we are examining in this thesis.  While, for example, our extrinsic curvature variables will vanish on the initial slice, they should have non-zero values and derivatives at the outer boundaries for all other times.

\begin{figure}\centering
\includegraphics{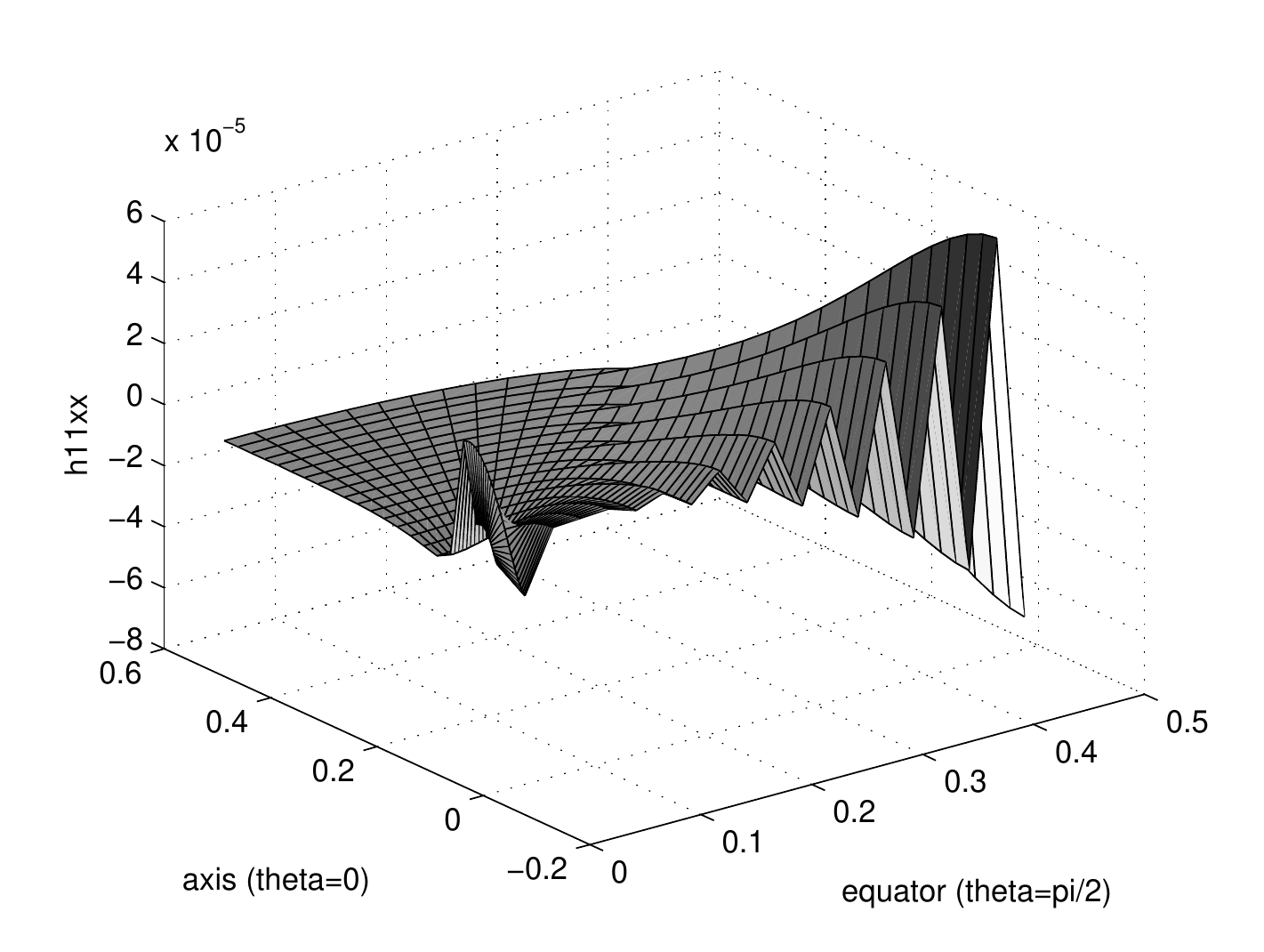}
\caption[Non-convergent solution for $H_a$]{Example of Bi-Conjugate Gradient solver's non-convergent (i.e. the solver reached the maximum number of iterations without reaching a solution that fell within the specified tolerance) ``best guess'' when trying to fit flat (\ref{eqn:dirichletob}) exterior boundary conditions to $H_a$}
\label{fig:unstableOB}
\end{figure}

If one tries to enforce these sorts of ``flat'' boundary conditions onto the solution of an elliptic PDE, the numerical solver can be trying to match an unphysical condition at the outer boundary\footnote{As a Brill wave is present in the entire spacetime our outer boundary values (for all variables) should generally be non-zero on all slices after the initial slice.} and will generally be unable to converge.  As such, the BiConjugate Gradient solver experienced \emph{large} oscillations in the interior region at times to try to match to flat outer boundary conditions.  See figure \ref{fig:unstableOB} for an example of the difficulties experienced by the solver.

If one is trying to enforce these sorts of boundary conditions on an evolution equation like (\ref{eqn:haevol}), one will generally end up with sharp spikes in the radial derivative terms close to the outer boundaries, which will in turn cause other solvers to fail that use those derived terms.

Further, our metric terms should have wave-like behaviour at the outer boundary and settle to Schwarzschild (i.e. non-flat) behaviour in the asymptotic limit.  Therefore any boundary conditions that have non-radiative terms (like the flat conditions above) at the outer radial boundary will interfere with proper wave propagation towards the edge of the grid and are inconsistent with the interior region - so those boundary conditions are bound to cause problems with numerical solvers and should be avoided.

\subsection{Robin outer boundary conditions}\label{sec:robinbc}
Historical treatments \cite{evans,bernstein,paul_thesis} of outer boundary conditions for the conformal factor $\psi$ in a Brill evolution in spherical polar coordinates have employed the Robin boundary condition.  This condition is given by
$$\psi_{\rho} + \frac{\psi-1}{\rho} = 0$$
where $\rho$ is the isotropic radial coordinate commonly used in analysing the Schwarzschild solution and is related to the areal coordinate $r$ by:
$$r=\rho\left(1+\frac{M}{2\rho}\right)^2$$
This boundary condition gives a way to map the conformal factor in the asymptotic region into the expected Schwarzschild one.  The largest difficulty with this assumption is that it neglects any angular information that may be present in the conformal factor, and forces a purely radial solution.  

The initial wave profile used for $q$ in most implementations has an angular dependence on the order of $\sin^{2n}\theta$, so forcing $\psi$ to have no angular dependence is inconsistent with the coupled nature of the equations.  Further, the author's experience with the solutions for $\psi$ indicate that it contains non-trivial angular information at the outer edge of the grid.

Another problem with the Robin condition is that it is only third order correct and is not easily implemented in non-isotropic coordinates\footnote{Bernstein \cite{bernstein} pp 286-7 discusses the order of correctness of the Robin condition, and it varies between 3rd order in an ideal implementation to 2nd order in his implementation (which isn't in isotropic coordinates).}.  For reasons mentioned in section \ref{sec:4thorder} we need to have a 4th order correct finite differencing scheme throughout the grid, which precludes a second or third order correct boundary condition.

The ability to extract gravitational radiation information as it passes the outer boundary (a future consideration) also requires a higher order approach to ensure that quadrupole moments are present in the spacetime.  As such, I needed to re-write the traditional spherical-polar ADM boundary conditions because Robin boundary conditions are unsuitable.  To this end we will investigate spherical harmonics and multipole expansions along the outer boundary.

\subsection{Multipole Expansions on the outer boundary}\label{sec:spherpolob}
As gravitational radiation requires at lowest order a time dependent quadrupole moment to be present for its propagation, we abandon the historical 2nd order method at the outer boundaries which only allows for $l=0$ and $l=1$ modes to be present. Therefore we assume the outer boundary condition for the variables in our axisymmetric formulation can be expanded as\footnote{As our symmetry conditions require that the azimuthal angle is a Killing Vector, we set $m=0$ in the traditional $P^l_m(\cos\theta)$ formulation of spherical harmonics.}
\begin{equation}\label{eqn:spherpolOBdecomp}\xi(t,r,\theta)=\sum_{l}\frac{A_l(t)}{r^{l+1}}P_l(\cos\theta)\end{equation}
as we have a quasilinear wave propagation system in the far wave region of the grid, and we assume a separable solution\footnote{See \cite{deadman} for a current, in-depth look at radiative boundary conditions in numerical relativity}.

So let us examine the imposition of 4th order multipole expansion boundary conditions at the outer boundary.  All of our functions are either symmetrical across $\theta=0$ and $\theta=\frac{\pi}{2}$, or anti-symmetrical across those boundaries.

This method proved unsuccessful for many reasons discussed below, however we include the implementation as the framework proves useful in the future.

\subsubsection{Symmetrical Functions}\label{subsec:spherpolsym}
\begin{figure}
\centering
\includegraphics{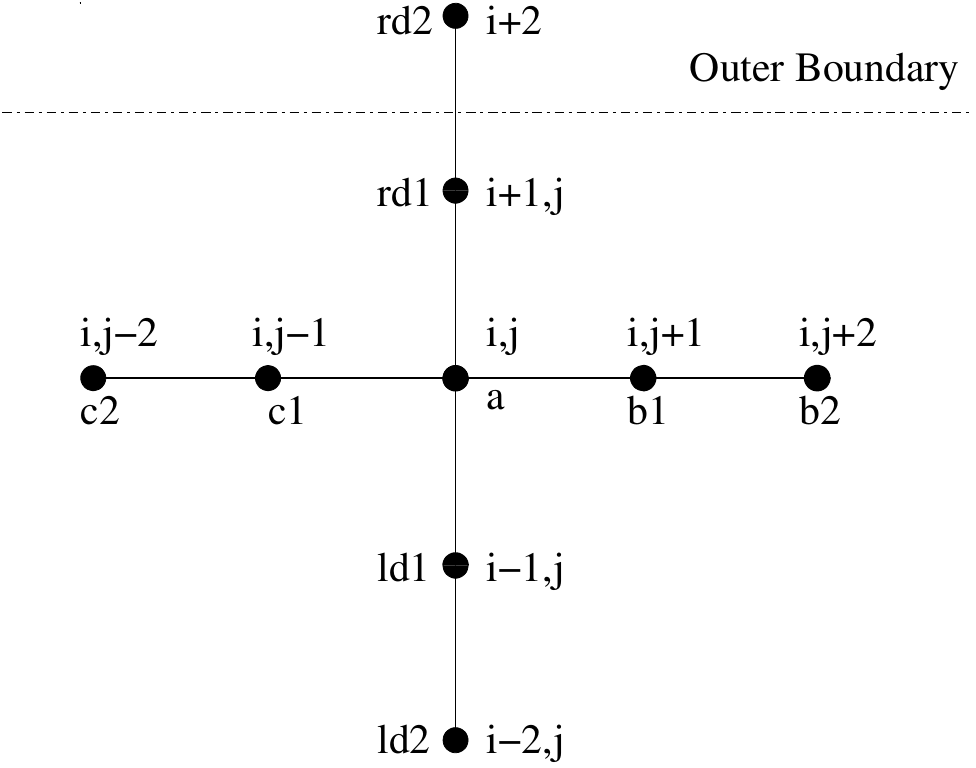}
\caption{Stencil at $i=imax+1$}
\label{fig:4ordstencilnmax-1}
\end{figure}

\begin{figure}
\centering
\includegraphics{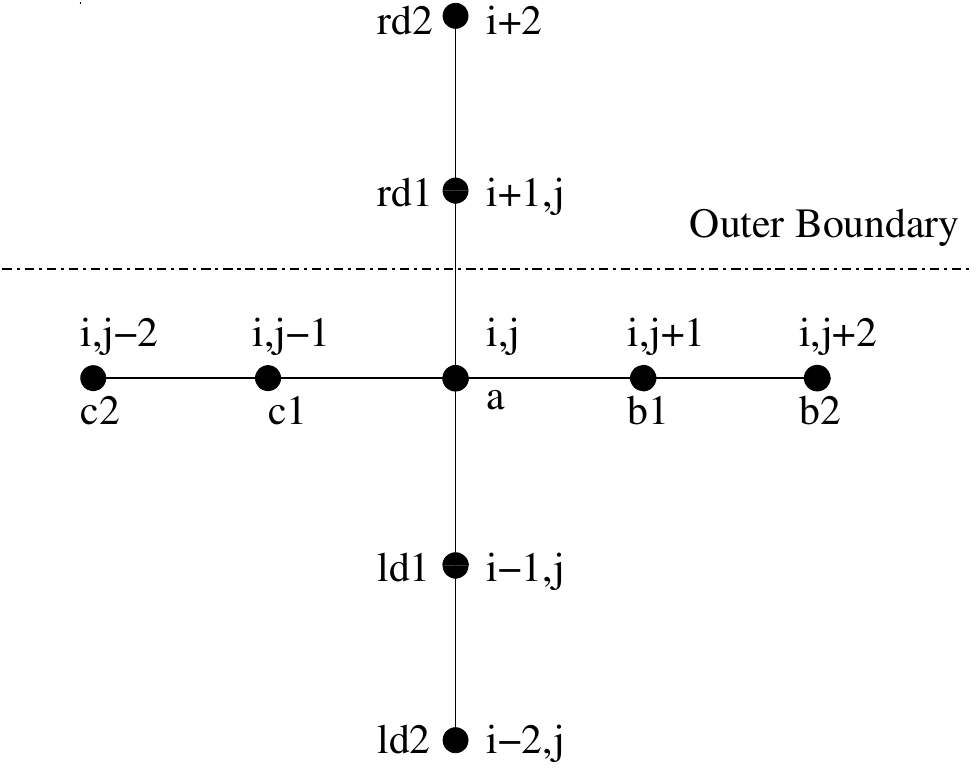}
\caption{Stencil at $i=imax+2$}
\label{fig:4ordstencilnmax}
\end{figure}

For functions that are symmetrical\footnote{In this context symmetrical means symmetrical across $\theta=0$ and $\theta=\frac{\pi}{2}$, and means that $f(\Delta\theta)=f(-\Delta\theta)$ and $f(\frac{\pi}{2}+\Delta\theta)=f(\frac{\pi}{2}-\Delta\theta)$ to 2nd order, with the addition that $f(2\Delta\theta)=f(-2\Delta\theta)$ and $f(\frac{\pi}{2}+2\Delta\theta)=f(\frac{\pi}{2}-2\Delta\theta)$ for 4th order.}, we know then that the $l=1$ and $l=3$ (and higher order odd) modes will vanish in equation (\ref{eqn:spherpolOBdecomp}), leaving terms that look like
$$\frac{A_0}{r}$$
and
$$A_2\left(\frac{3\cos^2(\theta)-1}{2r^3}\right)$$
i.e. we can write the terms $\xi_{i,j}(r_i,\theta_j)$ at grid points in the asymptotic region as:
\begin{equation}\label{eqn:k1k2local}\xi_{i,j}=\frac{\tilde{M}}{r_i} + \frac{\tilde{Q}(3\cos^2\theta_j-1)}{2r_i^3}\end{equation}
where $\tilde{M}$ (the mass aspect) and $\tilde{Q}$ (the quadrupole aspect) are constants for all points in the local stencil region.  It is not possible\footnote{At least not without a great deal of effort and/or completely changing numerical methods.} to devise a scenario that will use $\tilde{M}$ and $\tilde{Q}$ as the same values across the entire outer boundary while making it possible to remove grid points that are outside the computational domain and put those values back into the interior grid points that are only locally coupled.
We therefore assume that $\tilde{M}$ and $\tilde{Q}$ are local parameters\footnote{i.e. $\tilde{M} = \tilde{M}(i,j)$ and $\tilde{Q} = \tilde{Q}(i,j)$}, and our goal then is to absorb the coefficient at $i_{max}+1$ for the $i_{max}+3$ grid point (i.e. $RD2$ in figure \ref{fig:4ordstencilnmax-1}), and at $i_{max}+2$ to absorb the $i_{max}+3$ and $i_{max}+4$ grid point coefficients ($RD1$ and $RD2$ respectively in figure \ref{fig:4ordstencilnmax}).

We can then later calculate the value of $\tilde{M}$ and $\tilde{Q}$ at each point along the outer grid as a check of the consistency of our assumption.

Schematically we want to solve
$$ \left[ \begin{array}{c} \xi_{(1)}(r_1,\theta_1) \\ \xi_{(2)}(r_2,\theta_2) \\ \end{array}\right] =
\left[\begin{array}{cc} \frac{1}{r_1} & \frac{3\cos^2\theta_1-1}{2r_1^3} \\
\frac{1}{r_2} & \frac{3cos^2\theta_2-1}{2r_2^3} \\ \end{array}\right]
\left[\begin{array}{c} \tilde{M} \\ \tilde{Q} \end{array} \right] $$
for $\tilde{M}$ and $\tilde{Q}$ on the interior of the grid and then use those values in the exterior region.  We choose $\xi_{(1)}$ and $\xi_{(2)}$ to be points on the stencil that are on the interior of the problem (i.e. $a$ and $b1$ or $c1$ in figure \ref{fig:4ordstencilnmax} depending on where on the outer boundary you are).

In practice it is even more complicated than that, as local values for the function can cause coupling or cancellation precisely where they are not needed, leaving it as more of an art than science as to which stencil points to choose to absorb the values into.  One has 5-6 stencil points that can be used to solve two equations, so the system is overdetermined and one has to choose which points to use.  I had more success using points at the same radial value\footnote{as $\eta >> \Delta \eta$ along the outer boundary, so we get large rounding errors if $i_1 \ne i_2$} and trying to make sure that the central point ($a$ in figure \ref{fig:4ordstencilnmax}) was included. See table (\ref{tbl:spherpolcoef}) for the stencil points used.
\begin{table} \begin{center}
\begin{tabular}{|c|c|c|c|} \hline
symmetry & $j=2,3$ & $j=maxj$ & other j \\ \hline
yes & $b1$,$b2$ & $c1$,$c2$ & $a$,$c1$ \\
no & $b1$,$b2$ & $c1$,$c2$ & $b1$,$c1$ \\ \hline
\end{tabular}\caption[OB coupled stencil points]{Stencil points used for coupling at outer boundary to solve overdetermined coefficient system (see figure \ref{fig:4ordstencilnmax})} \label{tbl:spherpolcoef}
\end{center} \end{table}

Solving for $\tilde{M}$ and $\tilde{Q}$ yields:
\begin{equation}\left[ \begin{array}{c} \tilde{M} \\ \tilde{Q} \\ \end{array}\right] =
\frac{1}{\kappa_{11}\kappa_{22}-\kappa_{12}\kappa_{21}}\left[\begin{array}{cc} \kappa_{22} & -\kappa_{12} \\ -\kappa_{21} & \kappa_{11} \end{array}\right] \left[\begin{array}{c} \xi_{(1)} \\ \xi_{(2)} \\ \end{array}\right]\label{eqn:sphpolinter}\end{equation}

where the $\kappa_{ij}$ terms are either the multipole expansion coefficients of the symmetric problem noted above or the antisymmetric ones derived below.

Once we have solved for $\tilde{M}$ and $\tilde{Q}$ we then know that at $i_{max+1}$ we can substitute $\xi_{imax+3,j}$ with
\begin{equation}\label{eqn:psiextsherpolob}\xi_{imax+3,j}=\frac{\tilde{M}}{r_{imax+3}}+\frac{\tilde{Q}(3\cos^2(\theta_j)-1)}{2r_{imax+3}^3}\end{equation}
recalling that we don't actually know the values of $\xi_{(1)}$ and $\xi_{(2)}$ above in equation (\ref{eqn:sphpolinter}), rather we know $\tilde{M}=\tilde{M}(\xi_{(1)},\xi_{(2)})$ and $\tilde{Q}=\tilde{Q}(\xi_{(1)},\xi_{(2)})$.  In our stencil equation (\ref{eqn:4ordstencilexpand}) we can now replace the term
$$RD2(i,j)\xi_{i+2,j}$$
at $i=i_{max}+1$ with
$$RD2(i,j)\xi_{imax+3,j}=RD2(i,j)\tilde{\kappa_1}\xi_{(1)} + RD2(i,j)\tilde{\kappa_2}\xi_{(2)}$$
and absorb these terms into their owners (i.e. $c1$, $b1$, etc. from table (\ref{tbl:spherpolcoef}) )

And then we can do the same thing for the
$$RD1(i,j)\xi_{imax+3,j} \;;\; RD2(i,j)\xi_{imax+4,j}$$
terms at $i=i_{max+2}$.  This presents a complete method for defining the problem on the interior region that can then be solved via some matrix solver.

\subsubsection{Anti-Symmetric Functions Across Angular Boundaries}\label{subsec:antisphharm}
We wish to construct orthonormal basis functions to allow decomposition of antisymmetric\footnote{In this context anti-symmetric means anti-symmetric across $\theta=0$ and $\theta=\frac{\pi}{2}$, and means that $f(\Delta\theta)=-f(-\Delta\theta)$ and $f(\frac{\pi}{2}+\Delta\theta)=-f(\frac{\pi}{2}-\Delta\theta)$ to second order, with the addition that $f(2\Delta\theta)=-f(-2\Delta\theta)$ and $f(\frac{\pi}{2}+2\Delta\theta)=-f(\frac{\pi}{2}-2\Delta\theta)$ for 4th order correctness.} functions in the asymptotic wave zone as the multipole expansions above only apply to symmetric functions.

To this end we choose the angular portion of our antisymmetric multipole expansion basis functions to be:
$$\zeta_{(1)} = A\sin(2\theta)$$
$$\zeta_{(2)} = B\sin(2\theta)+C\sin^3(2\theta)$$
and perform a Gram-Schmidt orthogonalization on these basis functions.  Normality on $\zeta_{(1)}$ requires that
$$\int_{0}^{\frac{\pi}{2}}A^2\sin^2(2\theta)d\theta=1$$
which gives
$$A=\frac{2}{\sqrt{\pi}}$$
Orthogonality between $\zeta_{(1)}$ and $\zeta_{(2)}$ requires that
$$2\sqrt{\pi}\int_{0}^{\frac{\pi}{2}}B\sin^2(2\theta)+C\sin^4(2\theta)d\theta=0$$
which gives
$$B=-\frac{3C}{4}$$
and requiring normality on $\zeta_{(2)}$ via
$$\int_{0}^{\frac{\pi}{2}}\zeta_{(2)}^2 d\theta=1$$
yields
$$B = -\frac{6}{\sqrt{\pi}}$$
$$C = \frac{8}{\sqrt{\pi}}$$
Giving our final antisymmetric orthonormal basis functions as:
\begin{eqnarray}
\xi_{(1)}(r,\theta)& = & \left(\frac{2}{\sqrt{\pi}}\right)\frac{\sin(2\theta)}{r^2} \nonumber \\
\xi_{(2)}(r,\theta)& = & \left(-\frac{6}{\sqrt{\pi}}\right)\left(\frac{1}{r^4}\right)\left[\sin(2\theta)-\left(\frac{4}{3}\right)\sin^3(2\theta)\right]
\label{eqn:quasisphpol}\end{eqnarray}
We can then use the general framework established above in section (\ref{subsec:spherpolsym}) and put our coefficients from equation (\ref{eqn:quasisphpol}) into equation (\ref{eqn:sphpolinter}), using the same methodology to absorb the coefficients back into the stencil equation. This also presents a complete method for defining the problem on the interior region that can then be solved via some matrix solver.

\subsubsection{Extracting wave coefficients}
As we have designed our Legendre-like polynomials to form an orthogonal basis, we easily pick off wave coefficients by using the fact that, for example,
$$\int_{0}^{\frac{\pi}{2}}\xi_{(1)} F(\eta,\theta) d\theta = \frac{4}{\pi} \tilde{M}$$
if $F(\eta,\theta)$ is an odd function across $\theta=0$ and uses the basis functions of section \ref{subsec:antisphharm} for decomposition in the wave region of the grid.

\subsubsection{How well do the multipole aspect functions work?}\label{sec:testsphharm}
We will now investigate the merits of this methodology in the context of solving the Hamiltonian constraint equation (\ref{eqn:hamconphiearly}) for $\phi$.

In section \ref{subsec:spherpolsym} we discussed the use of local spherical harmonic terms along the outer boundary.  As we are choosing two of the $7-9$ possible stencil points\footnote{see figures \ref{fig:4ordstencilnmax-1}, \ref{fig:4ordstencilnmax} and equation (\ref{eqn:k1k2local})} to use as our points in the inverse matrix problem (equation \ref{eqn:sphpolinter}), we have an easily invertible matrix problem - a $2 \times 2$ system is easy to invert analytically, which is what we need as we have to specify functions \emph{a priori}.

The disadvantage of this method is that we are not using the full stencil at each \{$i,j$\}, so the spherical harmonics on the outer boundary will not be fully coupled with the solution of the derivative equation when solving, for example, for $\phi$ (see equation \ref{eqn:hamconphi}), $\alpha$ (see section \ref{sec:maxslice} for maximal slicing) or the shift vector potentials (see section \ref{sec:shiftvec}).

Furthermore, the wave coefficients should be global properties of the solution at the outer boundary in a fully separable PDE solution, not local functions\footnote{i.e. we would expect at the outer boundary that the spherical harmonics are a global property of the solution, and $\tilde{M} \neq \tilde{M}(i,j)$ and $\tilde{Q} \neq \tilde{Q}(i,j)$.  With the nonlinearities present in GR separability is not guaranteed, however.  So this can also be thought of as a measure of non-separability of the solution at the outer boundary.} as in equation \ref{eqn:k1k2local}.

So let us examine how ``good'' this approximation is, keeping in mind the discussion from section \ref{sec:4thorder}, i.e. that we need 4th order correct spatial derivatives for proper propagation of the evolution equations.

\begin{figure} \centering
\includegraphics{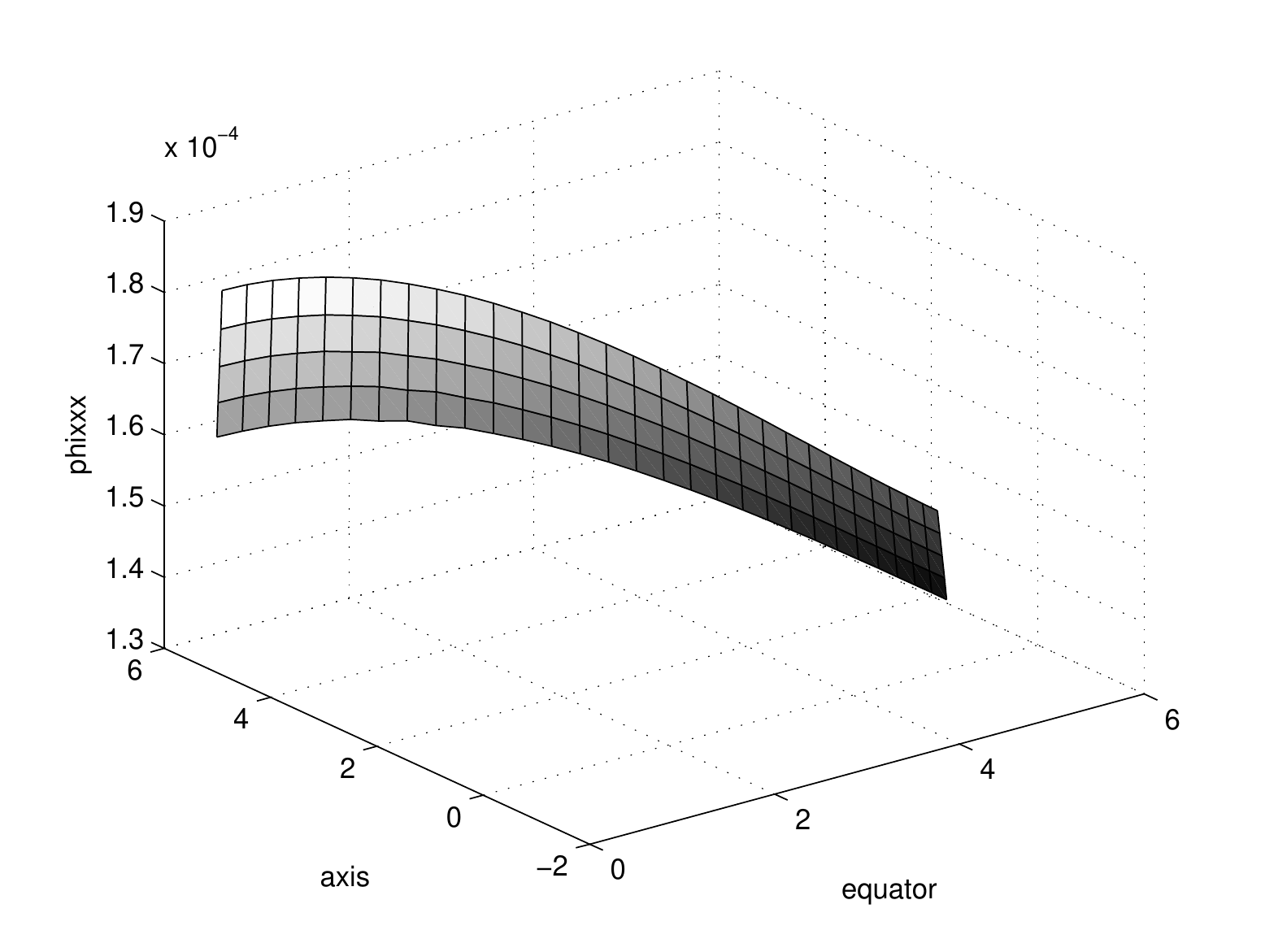}
\caption{Third radial derivative of $\phi$ at the 5 outermost boundary points.}\label{fig:phixxx}
\end{figure}

\begin{figure} \centering
\includegraphics{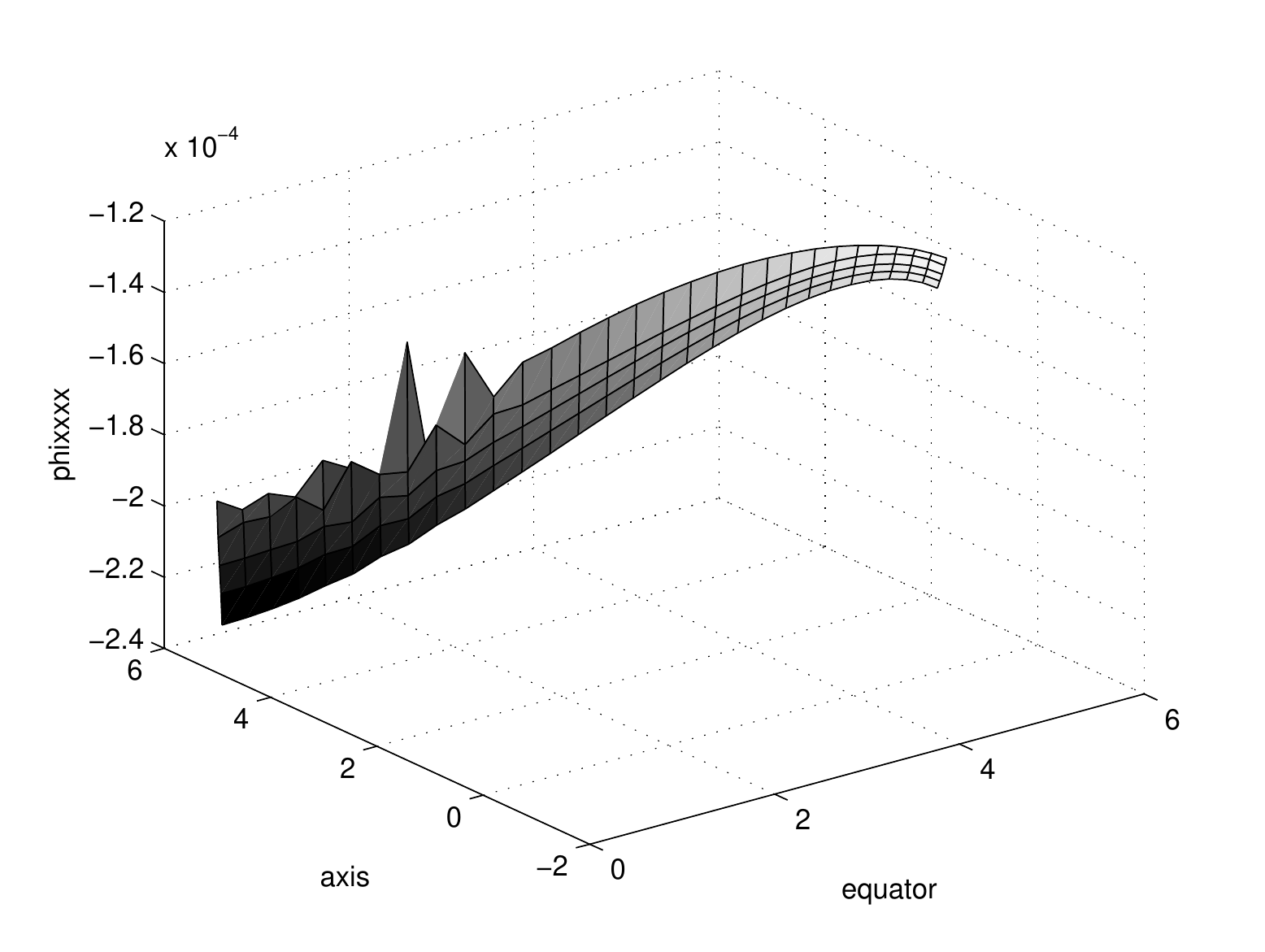}
\caption{Fourth radial derivative of $\phi$ at the 5 outermost boundary points.}\label{fig:phixxxx}
\end{figure}

\begin{figure} \centering
\includegraphics{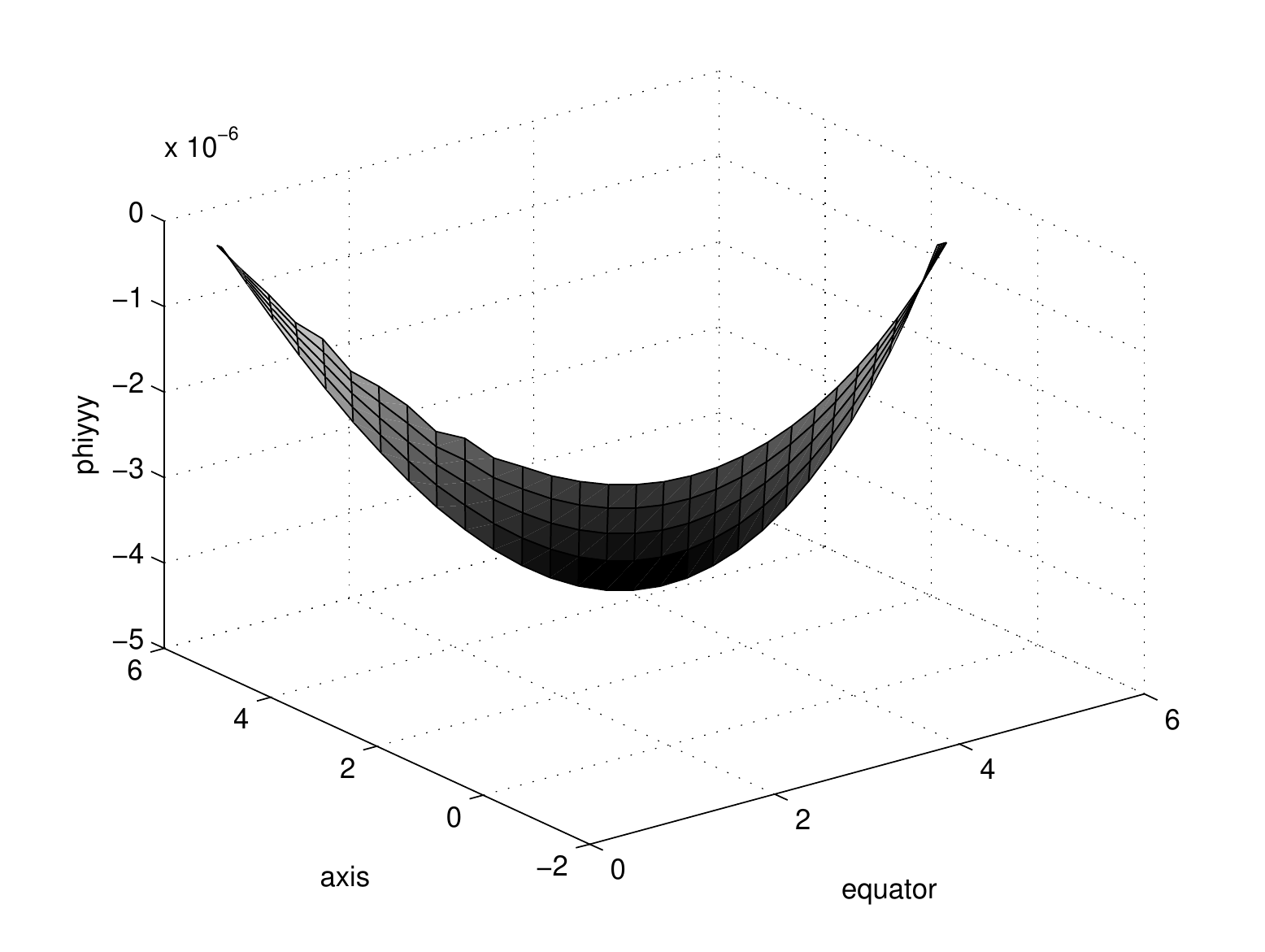}
\caption[Third angular derivative of $\phi$ at the 5 outermost boundary points.]{Third angular derivative of $\phi$ at the 5 outermost boundary points. (note the jagged solution)}\label{fig:phiyyy}
\end{figure}

\begin{figure} \centering
\includegraphics{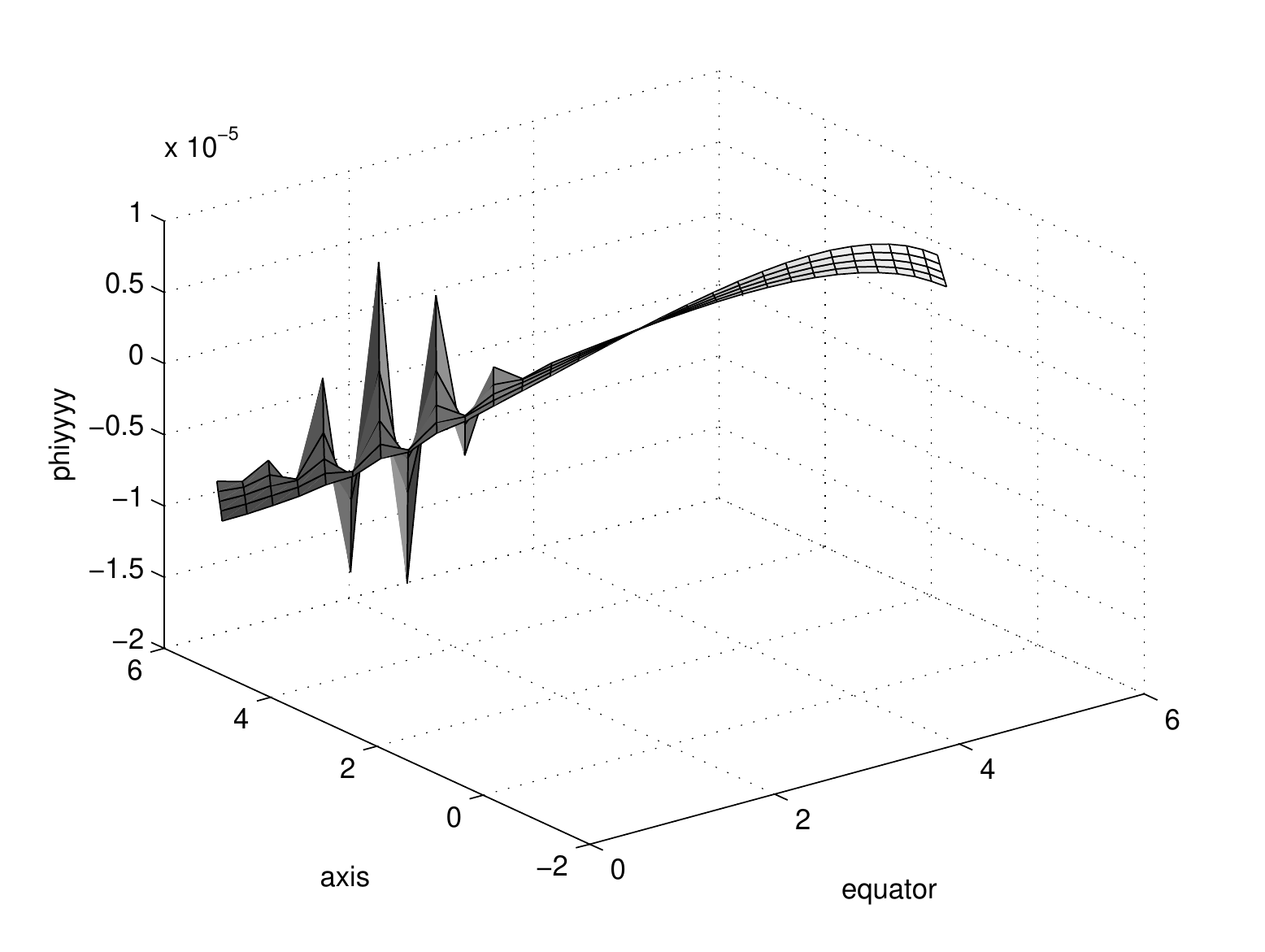}
\caption{Fourth angular derivative of $\phi$ at the 5 outermost boundary points.}\label{fig:phiyyyy}
\end{figure}

\begin{figure} \centering
\includegraphics{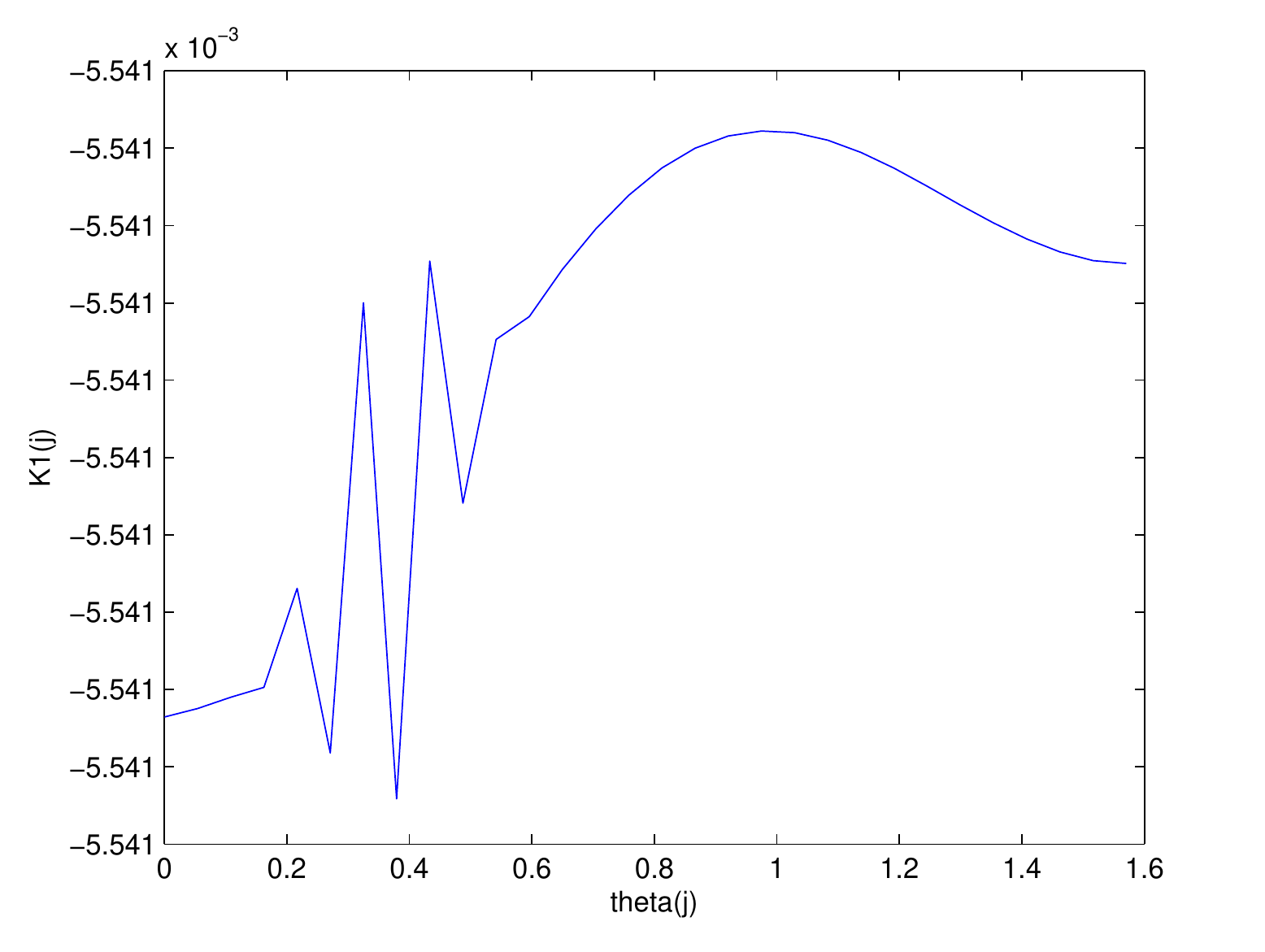}
\caption[$\tilde{M}$ along the outer boundary]{Local spherical harmonic coefficient, $\tilde{M}$, calculated along the outer boundary ($i=i_{max+2}$).}\label{fig:K1OBsphpolphi}
\end{figure}

\begin{figure} \centering
\includegraphics{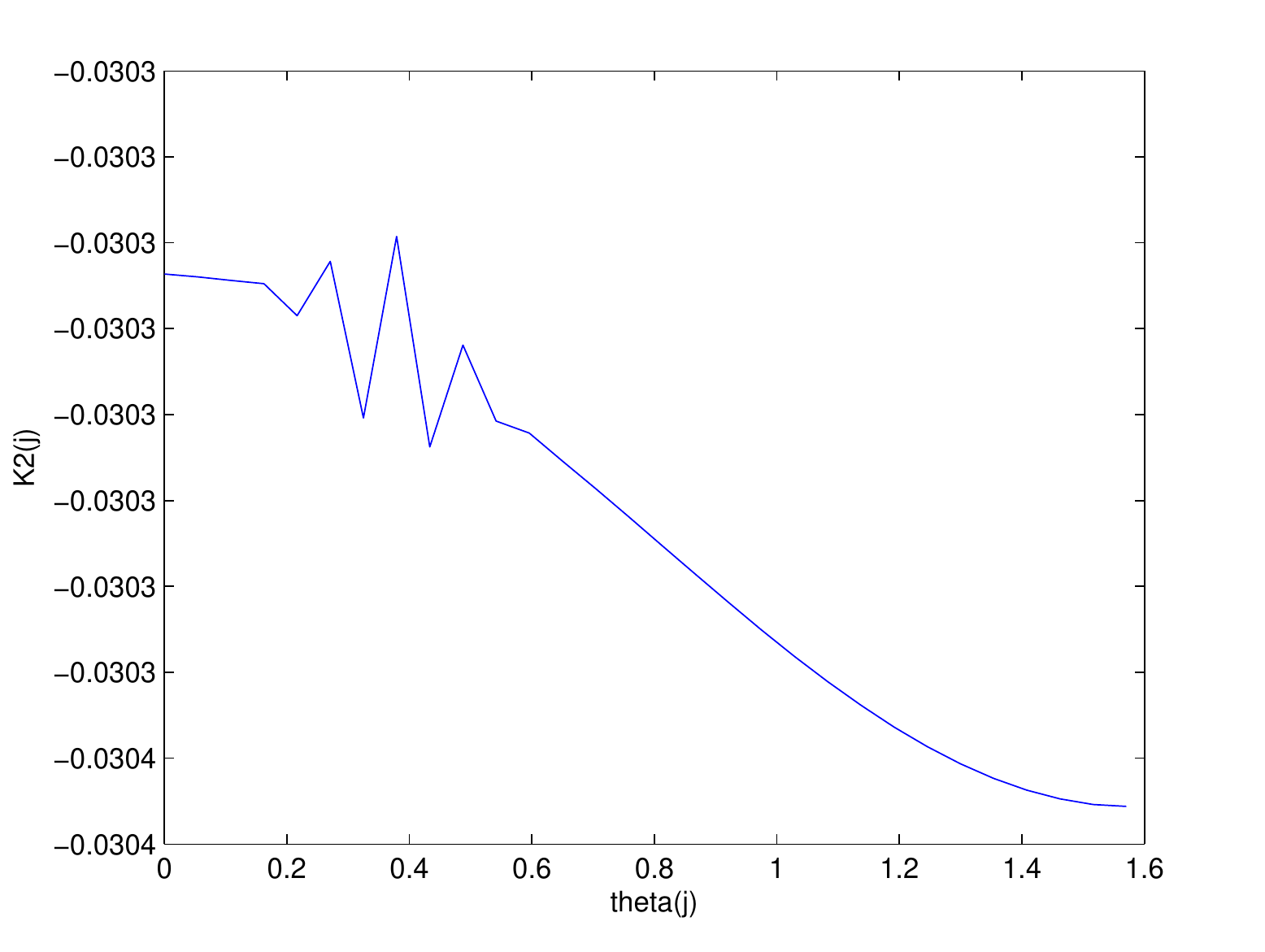}
\caption[$\tilde{Q}$ along the outer boundary]{Local spherical harmonic coefficient, $\tilde{Q}$, calculated along the outer boundary ($i=i_{max+2}$).}\label{fig:K2OBsphpolphi}
\end{figure}

To this end, we require that $\phi$ is solved to 4th spatial order correctness, provided we supply the correct boundary conditions and solve equation \ref{eqn:hamconphi} using the procedure in section \ref{sec:4thordderiv}.  We need to investigate two questions:

\begin{enumerate}
\item We need to ensure that $\phi_{\eta\eta\eta}$, $\phi_{\eta\eta\eta\eta}$, $\phi_{\theta\theta\theta}$ and $\phi_{\theta\theta\theta\theta}$ are smooth and well-behaved to ensure that our solution for $\phi$ will support the rest of the evolution.\footnote{Also their mixed counterparts, but these suffice for now to demonstrate the point.}  Equations (\ref{eqn:4ordderiv3}) and (\ref{eqn:4ordderiv4}) give the method for calculating these higher order derivatives.

See figures \ref{fig:phixxx}, \ref{fig:phixxxx}, \ref{fig:phiyyy} and \ref{fig:phiyyyy} for visualisations of third and fourth spatial derivatives at the outer boundary.  As we can see from these figures, the coupling of the spherical harmonics to the outer boundary conditions is not strong enough and instead yields irregular (non-smooth) fourth derivatives.\footnote{The second derivatives are regular, however, as are the third derivatives for the most part.}

\item How good is the approximation that $\tilde{M} = \tilde{M}(i,j)$ and $\tilde{Q} = \tilde{Q}(i,j)$, i.e. how much variance is there in the coefficients along the outer boundary?  Table \ref{tbl:k1k2sphpolOB} shows the extracted coefficients along the outer boundary, and shows that $\tilde{M}$ values vary by one part in $10^6$ along the outer boundary (see also figure \ref{fig:K1OBsphpolphi}).  However $\tilde{Q}$ is constant to one part in $10^3$ and shows larger variation (see also figure \ref{fig:K2OBsphpolphi}).  It is expected that the higher order harmonics would not be matched as well, however this is a rather large variation.

In both cases, the assumption that we can approximate the global functions $\tilde{M}$ and $\tilde{Q}$ with local ones does not fall within the required level of numerical accuracy.  This is most likely because we are only coupling 2 of the 9 stencil points at the outer boundary into the spherical harmonics and the non-linearities that are present in GR.  Experience has dictated that this sort of approach with other PDEs in numerical relativity also fails, so it is not terribly surprising that this one also falls short of the mark.
\end{enumerate}

\begin{table}\begin{center}
\begin{tabular}{|c|c|c|}\hline
$j$ & $\tilde{M}(j)(\times 10^{-3})$ & $\tilde{Q}(j)(\times 10^{-2})$ \\ \hline
2 & -5.54104470918756475 & -3.03036549374837705 \\
3 & -5.54104449052453263 & -3.03039642981275392 \\
4 & -5.54104419812330440 & -3.03043854478914909 \\
5 & -5.54104394706438075 & -3.03047655710599884 \\
6 & -5.54104138386493639 & -3.03084775599700151 \\
7 & -5.54104563951778040 & -3.03021692208886861 \\
8 & -5.54103399490221693 & -3.03203931908342383 \\
9 & -5.54104682661786876 & -3.02992579479024471 \\
10 & -5.54103292145803443 & -3.03237713698756295 \\
11 & -5.54103917956504880 & -3.03119090752499432 \\
12 & -5.54103494842806722 & -3.03207445742081916 \\
13 & -5.54103435789345643 & -3.03221430756259736 \\
14 & -5.54103313037471006 & -3.03254028234860584 \\
15 & -5.54103209291324783 & -3.03286167757832459 \\
16 & -5.54103121437770491 & -3.03319018729411027 \\
17 & -5.54103051279596255 & -3.03352292596485995 \\
18 & -5.54103000109385391 & -3.03385574420573789 \\
19 & -5.54102968267741920 & -3.03418464650633268 \\
20 & -5.54102955193004447 & -3.03450570891996563 \\
21 & -5.54102959477601169 & -3.03481510692122924 \\
22 & -5.54102978959466336 & -3.03510915707576601 \\
23 & -5.54103010847064106 & -3.03538435677666030 \\
24 & -5.54103051871594580 & -3.03563742004371306 \\
25 & -5.54103098458534962 & -3.03586530836525981 \\
26 & -5.54103146908814168 & -3.03606525212813949 \\
27 & -5.54103193578662362 & -3.03623475654148933 \\
28 & -5.54103235043341275 & -3.03637157667253632 \\
29 & -5.54103268228220425 & -3.03647363923289504 \\
30 & -5.54103290459252718 & -3.03653879988416975 \\
31 & -5.54103298225992746 & -3.03656112236012632 \\ \hline
\end{tabular}\caption[Local spherical harmonic coefficients]{Values of local spherical harmonic coefficients along the outer boundary, as calculated using the method in section \ref{sec:spherpolob}.  Ideally the values in one column should all be equal to each other.  In reality there is $\sim 1$ part in $10^6$ variation in $\tilde{M}$ and $\sim 1$ part in $10^3$ variation in $\tilde{Q}$ } \label{tbl:k1k2sphpolOB}
\end{center} \end{table}

Possible solutions to this problem involve:
\begin{enumerate}
\item \emph{Expansion of equation (\ref{eqn:k1k2local}) to higher order}.  The problem is that by the time we expand out to 8 stencil points (and therefore 8 even $l$ coefficients) we will be at $\frac{1}{r^{15}}$, and those terms will most likely get lost in the numerical noise\footnote{$r^4$-like differences are near the numerical precision of 64-bit doubles for this grid ($\sinh^4(10) \sim 1$ part in $10^{16}$), so $14$ orders of $r$ difference will definitely be lost}.  We would also have to analytically invert an $8 \times 8$ matrix, and we still won't have solved the problem that $\tilde{M}=\tilde{M}(i,j)$.
\item \emph{Predictor-corrector algorithm}.  The development of a predictor-corrector algorithm to include all exterior boundary points and solve for global $\tilde{M}$'s would involve a lot of effort, and while possible is beyond the scope of this thesis.  A basic template could look like: (a) solve for local $\tilde{M}(i,j)$'s using the above approximation (b) calculate a ``mean'' global $\tilde{M}$ (c) calculate the deviation of the 4th derivatives of $\phi$ from smooth\footnote{There are various methods to do this, however experience dictates that the measure would take different forms for various amplitudes of the variables being measured, and would also vary in time.} (d) perturb the global $\tilde{M}$'s and re-measure the deviation of the 4th derivatives (e) attempt to determine an approximately constant global $\tilde{M}$ using some interpolation/extrapolation based on deviations from smoothness for the 4th derivatives.
\end{enumerate}

These barriers comprise large enough hurdles that it is better left for future projects.  An initial attempt at developing a predictor-corrector algorithm leaves the author to believe that it will not be a simple task to unravel the numerical issues.  Let us instead tackle the problem from an alternate angle.

\subsection{The Solution: Separable Spherical Functions}\label{subsec:sepspherharm}
Similar to the methodology in section \ref{sec:spherpolob} we assume a solution that has radial and angular components, however we drop any assumptions about the nature of the functions themselves and instead assume that: (1) a function $G(\eta,\theta)$ that we are trying to solve for along the outer boundary can be written as the product of a function of $\eta$ and a function of $\theta$
$$G(\eta,\theta,t)=N(\eta,t)T(\theta,t)$$
near the outer boundary, (2) if $N(\eta,t) \rightarrow 0$  as $\eta \rightarrow \infty$ we therefore assume an asymptotic falloff that goes as:
\begin{equation}\label{eqn:NsepOB} N = \frac{c_n(t)}{f^n} + \frac{c_{n+1}(t)}{f^{n+1}} + \frac{c_{n+2}(t)}{f^{n+2}} + \ldots \end{equation}
and (3) we evaluate all values of $G$ at the \emph{same} angular coordinate (i.e. the points $a$, $ld1$ and $ld2$ in figure (\ref{fig:4ordstencilnmax})) so that the angular function $T(\theta)$ is the same for our inverse problem\footnote{This significantly simplifies the problem as compared to the presentation of the previous section, which introduced a number of complications with it by having both radial and angular dependence.}.

We can therefore solve for the coefficients $c_n,c_{n+1},c_{n+2}$ at the outer boundary given that we know the values $G_1(\eta_{max}-2\Delta\eta,\theta),G_2(\eta_{max}-\Delta\eta,\theta),G_3(\eta_{max},\theta)$ via a ``simple'' matrix problem.
\begin{eqnarray}
\left[\begin{tabular}{c}$G_1$ \\ $G_2$ \\ $G_3$
\end{tabular}\right] & = &
\left[\begin{tabular}{ccc} $\left(\frac{1}{f(\eta_{max}-2\Delta\eta)}\right)^{n}$ & $\left(\frac{1}{f(\eta_{max}-2\Delta\eta)}\right)^{n+1}$ & $\left(\frac{1}{f(\eta_{max}-2\Delta\eta)}\right)^{n+2}$ \\
$\left(\frac{1}{f(\eta_{max}-\Delta\eta)}\right)^{n}$ & $\left(\frac{1}{f(\eta_{max}-\Delta\eta)}\right)^{n+1}$ & $\left(\frac{1}{f(\eta_{max}-\Delta\eta)}\right)^{n+2}$ \\
$\left(\frac{1}{f(\eta_{max})}\right)^{n}$ & $\left(\frac{1}{f(\eta_{max})}\right)^{n+1}$ & $\left(\frac{1}{f(\eta_{max})}\right)^{n+2}$ \\
\end{tabular}\right]
\left[\begin{tabular}{c}
$c_n$ \\ $c_{n+1}$ \\ $c_{n+2}$ \\
\end{tabular}\right] \nonumber \\ & &
\end{eqnarray}

In the case of a variable that we solve for directly (like $H_a$ via its evolution equation), we can simply calculate the values of $c_i$ and use them to extrapolate to the phantom grid points outside the computational region of the grid.

For variables that we are solving for via an elliptic equation solver, we need to use the same methodology as in section \ref{subsec:spherpolsym} to move the coefficients for the matrix problem into the interior region of the grid.

In practice this method works very well, and is vastly simpler to implement than the use of spherical harmonics.  The downfall would be if we wanted to match higher order terms as we only have $3$ stencil points that are at the same angular value.

\subsubsection{Dynamic outer boundary falloffs}\label{subsec:DynamicOB}
The last unanswered question in all of these methods however is: what is $n$? i.e. what is the leading order fall-off behaviour in our various dynamic variables?  And how far out do we perform the expansion? ($n+1$? $n+2$?)

As for the ``order'' of expansion, it is generally the case that using two terms in equation (\ref{eqn:NsepOB}) provides the best results.  Trying to match to higher orders seems to cause ``phantom'' harmonics to appear and give the elliptic solvers great difficulty.

Also note that
$$\partial_\eta\left(\frac{1}{f}\right) = -\frac{f_\eta}{f^2} \sim -\frac{1}{f} \;;\;\eta\rightarrow\infty$$
if we use $f=\sinh\eta$, so higher order derivative terms do not generally have faster dropoff using our radial coordinates(!).

We further expect that as radiation (or other effects, like the doubly iterative C-N solver settling down to an iterative solution for a variable) influence the outer edge of the grid, we might see other leading order behaviour, or that we may be wrong about our assumptions of what those values of $n$ should be.
To account for this, we allow the code to calculate an estimated value of $n$ across the outer boundary via
$$n_j=\frac{\log\frac{G_3}{G_2}}{\log\frac{f(\eta_{max}-\Delta\eta)}{f(\eta_{max})}}$$
and calculate
$$l={min}_j(n_j)$$
We then set a minimum falloff order of $1$ and maximum\footnote{Sometimes very small values near the edge of the grid for some variables (like $q$ initially) can cause wild results if $n$ isn't capped.} of $6$
$$n=min(max(floor(l*1.1),1),6)$$

The \emph{starting/guessed} values of $n$ ($n_1,n_2$) that we use for each variable\footnote{Note that these are for the mixed extrinsic curvature variables, the covariant extrinsic curvature formulation will differ because of equations (\ref{eqn:Hcovtransform}).} are given in table \ref{tbl:OBradn}.   We also present some observed values at $t=\Delta t \; (k=1)$ for various ``wave strengths'', i.e. strong wave ($A=9,s_0=1$), moderate wave strength ($A=1,s_0=1$) and weak perturbative wave ($A=1e-10,s_0=3$)

Generally most variables dynamically settle into $1$ or $2$ as their leading order falloff\footnote{$\phi$ fails to converge if we allow dynamic determination of its order $n$, so we fix it.  All other variables work well, however.}. This could provide a rich area for future investigation given that (i) we are calculating actual falloffs of various metric/curvature variables and (ii) there is a division in opinion on actual outer boundary falloff rates in the theoretical gravitational wave literature.  This is due to the nonlinear behaviour of gravitational waves.

\begin{table}\begin{center}\begin{tabular}{|c|c|c|c|c|c|}\hline
Variable & $n_1$ & $n_2$ & $n$ Strong Wave $(9,1)$ & $n$ Mid Wave $(1,1)$ & $n$ Weak Wave $(10^{-10},3)$ \\ \hline
$q$ & $2$ & $4$ & $(-11,42,1.72)$(*) & $(2.56,2.49,2.49)$ & $(2.48,2.53,2.51)$ \\
$\phi$ & $1$ & $2$ & $(0.917,0.921,0.919)$ & $(0.988,1.00,0.997)$ & $(-31,36,N/A)(**)$  \\
$H_a$ & $1$ & $2$ & $(1.95,1.96,1.95)$ & $(2.58,2.68,2.59)$ & $(3.77,3.77,3.77)$  \\
$H_b$ & $1$ & $2$ & $(-1.86,2.4,1.31)$ & $(1.99,2.00,1.99)$ & $(1.98,2.01,2.00)$  \\
$H_c$ & $1$ & $2$ & $(0.875,0.877,0.876)$ & $(1.99,1.99,1.99)$ & $(2.00,2.00,2.00)$  \\
$H_d$ & $1$ & $2$ & $(-7.81,8.63,1.37)$ & $(1.75,3.33,2.03)$ & $(2.00,4.35,2.07)$  \\
$v_1$ & $1$ & $2$ & N/A & N/A & N/A \\
$v_2$ & $1$ & $2$ & N/A & N/A & N/A \\
$\chi$ & $1$ & $2$ & $(0.589,0.642,0.612)$ & $(0.66,0.66,0.66)$ & $(0.66,0.67,0.66)$  \\
$\Phi$ & $1$ & $2$ & $(0.45,0.53,0.48)$ & $(1.53,1.61,1.57)$ & $(1.54,1.62,1.58)$  \\
$\alpha$ & $1$ & $2$ & N/A & N/A & N/A \\ \hline
\end{tabular}\end{center}\caption[OB dropoff for variables]{Dynamic variables and their leading order dropoff in $\frac{1}{f^n}$ at the outer boundary via equation (\ref{eqn:NsepOB}).  Wave parameters in row headers are in the form $(A,s_0$), and measured values are in the form $(n_{\mathrm{min}},n_{\mathrm{max}},n_{\mathrm{average}})$. (*) Boundary terms at $\theta=\left\{0,\frac{\pi}{2}\right\}$ tend to skew the max and min measures significantly.  Further, $e^{-r^2}$ falls of so fast that it is zero numerically by $\eta=5$. (**) IVP values are too small to get a meaningful measure.}
\label{tbl:OBradn}\end{table}

\section{Rearrangement/factorisation of operands}
Related to the discussion around re-ordering the way in which numerical computations are performed in section \ref{subsec:addterms} we present another set of numerical tools that are required in ill-conditioned numerical situations.

When numerically computing the value of a function
$$F(r)=\frac{a}{r^2}+\frac{b}{r^2}$$
it is helpful to rearrange our equation computationally as
$$F(r)=\frac{(a+b)}{r^2}$$

Firstly, from a speed perspective, division is a computationally expensive operation so we should strive to minimize the number of operations overall in our right-hand-side terms.

Secondly, each operation we perform involves some rounding error, so we should minimize the number of computations to minimize rounding errors in our code.  As a very simple example, let us calculate the value of
$$F=\frac{1}{3}+\frac{1}{3}$$
to 4 decimal places.  Calculating in the order presented we find\footnote{In reality the conversion of floating point numbers from their binary representation is more complicated than this and is implementation dependent, see for example IEEE 754 and \cite{goldberg}.}
$$F=0.3333 + 0.3333=0.6666$$
rearranging the terms instead we find
$$F=\frac{2}{3}=0.6667$$
While one can ask whether we really care about values that are 16 decimal places down inside the computation\footnote{Recall that double precision reals in IEEE 754 have $\frac{1}{2^{52}} \sim 1.1\times 10^{-16}$ decimal places that can be represented.} the answer is that when one compounds all of these rounding errors over thousands and millions of computations, they do matter.

\begin{figure} \centering
\includegraphics{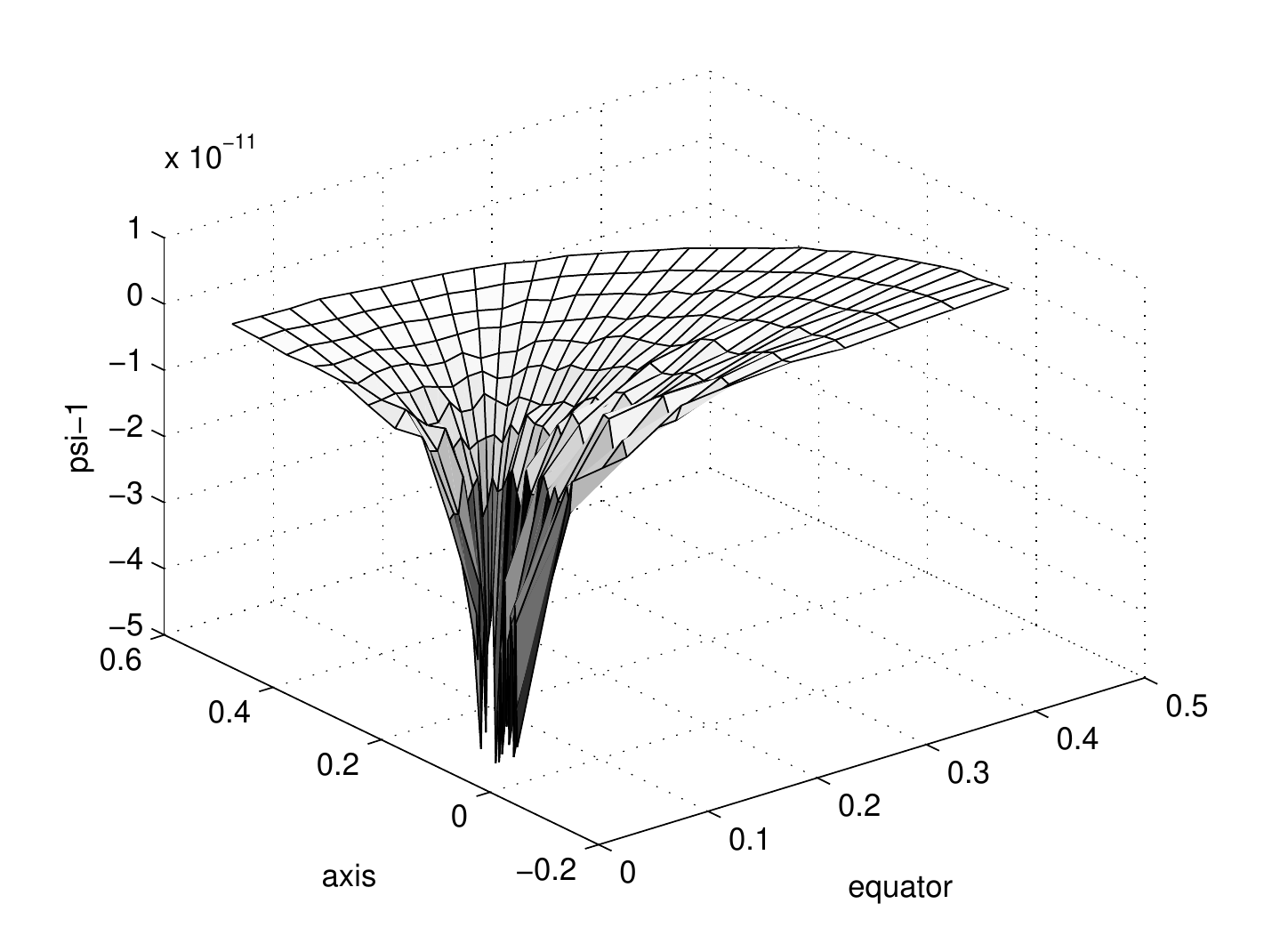}
\caption[Solution for $\psi-1$ near $\eta=0$ \emph{without} rearranging]{Solution for $\psi-1$ near $\eta=0$ \emph{without} rearranging the $\frac{1}{r^2}$ and $\frac{1}{\sin^2\theta}$ terms in the Hamiltonian constraint.}\label{fig:psinoarrange}
\end{figure}

\begin{figure} \centering
\includegraphics{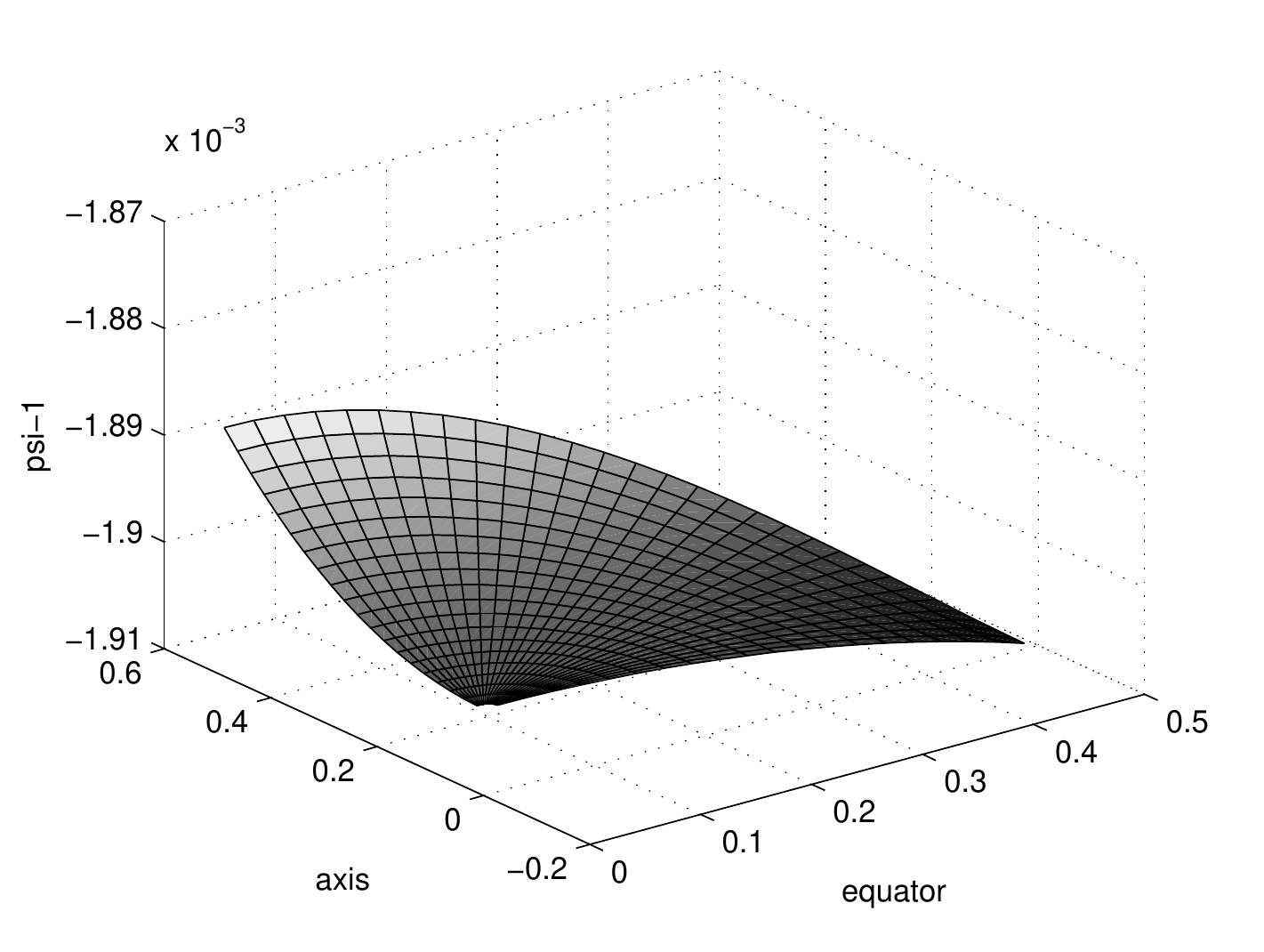}
\caption[Solution for $\psi-1$ near $\eta=0$ \emph{with} rearranging]{Solution for $\psi-1$ near $\eta=0$ \emph{with} rearranging $\frac{1}{r^2}$ and $\frac{1}{\sin^2\theta}$ terms in the Hamiltonian constraint.}\label{fig:psiarrange}
\end{figure}

We now present two examples of numerical situations in which this reordering was employed with noticeable effect.
\begin{enumerate}
\item When re-creating the values of a function along the outer boundary that is using the spherical polar harmonics described in Section (\ref{sec:spherpolob}), it is advantageous to choose radial points such that they are the same and can be factored out when doing the calculations.

To wit, if one uses the $\frac{1}{r^4}$ terms as-is that arise in the coefficients $\tilde{M}$ and $\tilde{Q}$ in equation (\ref{eqn:psiextsherpolob}), one ends up with the results in table \ref{tbl:norfactor} along the outer boundary.  Factoring these values out yields the results in \ref{tbl:yesrfactor} (This shows the difference if we factor out the radial values when performing matrix inversion calculations, then put them back in at the end).

\begin{table}\begin{center}
\begin{tabular}{|c|c|c|c|} \hline
$j$ & $F(imax+2)$ & $F(imax+3)$ (extrapolated) &  $F(imax+4)$ (extrapolated) \\ \hline
$2$ & $8.03172809074999814$ &  $7.8338056988601\mathbf{3892}$ & $7.6407428382314\mathbf{5653}$ \\
$3$ & $8.03176453414853795$ & $7.83383855315038\mathbf{756}$ & $7.6407715251495\mathbf{1998}$ \\
$4$ & $8.03187473684172852$ & $7.833942151458\mathbf{39052}$ & $7.6408694304605\mathbf{0947}$ \\
$5$ & $8.03205611714215609$ & $7.83410858610498\mathbf{030}$ & $7.64102204492032\mathbf{150}$ \\
$6$ & $8.03231155513592300$ & $7.83435239200943\mathbf{852}$ & $7.6412548907830\mathbf{4162}$ \\
$7$ & $8.03262244711370967$ & $7.83462642800181\mathbf{316}$ & $7.6414950895573\mathbf{5946}$ \\
\hline
\end{tabular}\caption[Sample outer boundary harmonics extrapolation with $r$ \emph{not} factored out]{Sample outer boundary harmonics extrapolation with $r$ \emph{not} factored out. (all values $\times 10^{-4}$)} \label{tbl:norfactor}
\end{center} \end{table}

\begin{table}\begin{center}
\begin{tabular}{|c|c|c|c|} \hline
$j$ & $F(imax+2)$ & $F(imax+3)$ (extrapolated) &  $F(imax+4)$ (extrapolated) \\ \hline
$2$ & $8.03172809074999814$ & $7.8338056988601\mathbf{6386}$ & $7.6407428382314\mathbf{7930}$ \\
$3$ & $8.03176453414853795$ & $7.83383855315038\mathbf{864}$ & $7.6407715251495\mathbf{2215}$ \\
$4$ & $8.03187473684172852$ & $7.833942151458\mathbf{41546}$ & $7.6408694304605\mathbf{3116}$ \\
$5$ & $8.03205611714215609$ & $7.83410858610498\mathbf{139}$ & $7.64102204492032\mathbf{476}$ \\
$6$ & $8.03231155513592300$ & $7.83435239200943\mathbf{202}$ & $7.6412548907830\mathbf{3403}$ \\
$7$ & $8.03262244711370967$ & $7.83462642800181\mathbf{641}$ & $7.6414950895573\mathbf{6380}$ \\ \hline
\end{tabular}\caption[Sample outer boundary harmonics extrapolation \emph{with} $r$ factored out.]{Sample outer boundary harmonics extrapolation \emph{with} $r$ factored out. (all values $\times 10^{-4}$)} \label{tbl:yesrfactor}
\end{center} \end{table}

\item The elliptic equation solvers encounter problems and are possibly non-convergent if we do not condition our matrix problems appropriately.  For example, we can take the following equation
$$\frac{1}{\eta^2} F_{\eta\eta}+\cot^2\theta F_{\theta\theta} + \ldots = (RHS)$$
and rearrange it in the form
$$\sin^2\theta F_{\eta\eta}+\eta^2\cos^2\theta F_{\theta\theta} + \ldots = \eta^2 \sin^2\theta (RHS)$$
at all points in the interior region excluding the boundaries (which one deals with separately).  With this sort of re-arrangement, we can take a non-convergent matrix problem and turn it into a convergent one.

Figures \ref{fig:psinoarrange} and \ref{fig:psiarrange} demonstrate a particular example of the solution we obtain with and without the arrangement above. This demonstrates how the ill-conditioned nature of our matrix problems (see section \ref{sec:condition}) makes the code sensitive to these numerical nuances, as the first solution is not a valid solution, whereas the second one is.
\end{enumerate}

\subsection{Re-arrangement of operands, II}
While the expansion of the Ricci curvature in terms of metric variables does result in many helpful (and physical) analytic cancellations\footnote{see section \ref{sec:extrinsiccurvature}}, there are some expansions in the equations that come from Maxima that need to be re-grouped for numerical reasons.  Operands containing terms like
$$1-\frac{f f_{\eta\eta}}{f_{\eta}^2}$$
are problematic numerically, and can be improved upon by using the analytic derivatives of the radial function given in equation (\ref{eqn:dxfnfx}).

\begin{figure} \centering
\includegraphics{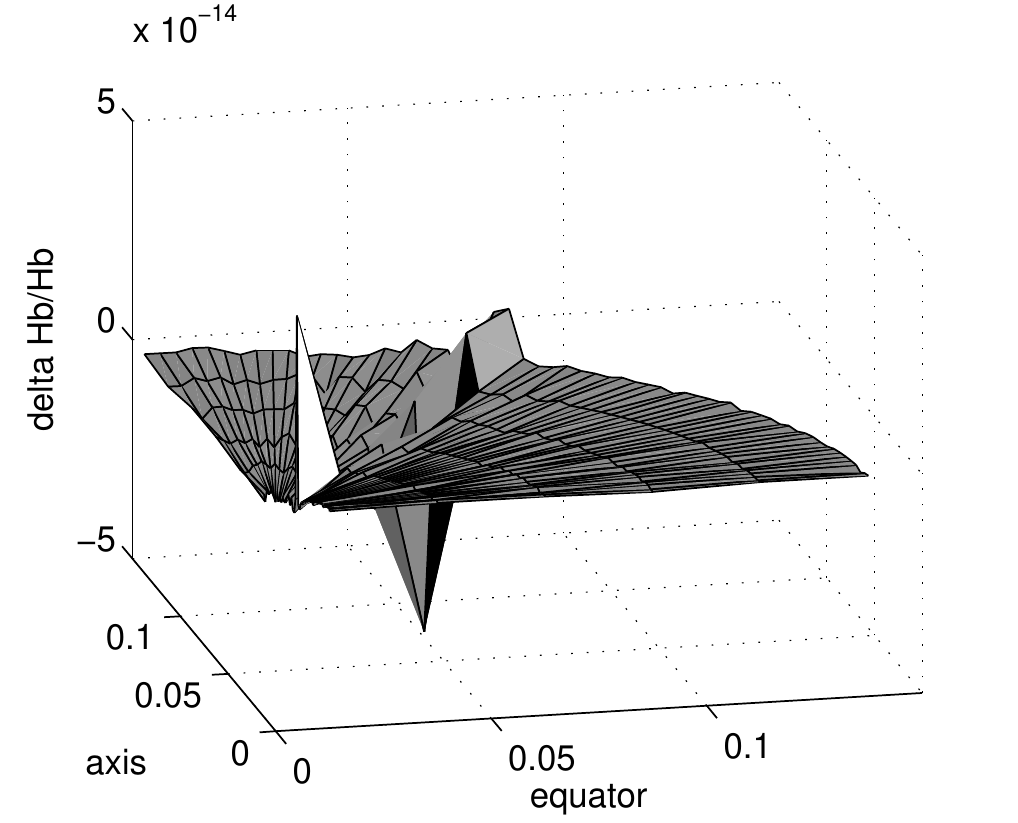}
\caption[Relative difference in $H_b$]{Relative difference in $H_b$ when calculated from two different methods at $t=\Delta t$ near the origin.  See equation (\ref{eqn:relhbdot}).}\label{fig:relhbdot}
\end{figure}

For example, if we take equation (\ref{eqn:hbevol}):
\begin{eqnarray}\label{eqn:hbevolrepeat}
\frac{\partial H_b}{\partial t} & = & \frac{1}{f^2 \, e^{q+4\phi}}\left(
\frac{\alpha\, q_{\theta} \cot\theta }{2}
-2\,\alpha\, \phi_{\theta} \cot\theta
-\frac{\alpha\, q_{\theta\theta} }{2}
+\alpha\, \phi_{\theta} \, q_{\theta}
+\frac{ {\alpha}_{\theta} \, q_{\theta} }{2} \right. \nonumber \\ \mbox{} & & \left.
-\frac{\alpha\, q_{\eta\eta}  \, f^2}{2 f_\eta^2}
-\frac{\alpha\, \phi_{\eta} \, q_{\eta} \, f^2}{f_\eta^2}
-\frac{ {\alpha}_{\eta} \, q_{\eta} \, f^2 }{2 f_\eta^2}
+\frac{ \alpha\, q_{\eta}  \, f^2 \, f_{\eta\eta}}{2 f_\eta^3}
-\frac{\alpha\, q_{\eta} \, f}{f_\eta} \right. \nonumber \\ \mbox{} & & \left.
-4\,\alpha\, \phi_{\theta\theta}
+2\, {\alpha}_{\theta} \, \phi_{\theta}
-\frac{2\,\alpha\, \phi_{\eta\eta} \, f^2}{f_\eta^2}
-\frac{4\,\alpha\, \phi_{\eta}^2 \, f^2}{f_\eta^2}
-\frac{2\, {\alpha}_{\eta} \, \phi_{\eta} \, f^2}{f_\eta^2} \right. \nonumber \\ \mbox{} & & \left.
+\frac{2\, \alpha\, \phi_{\eta} \, f^2 \, f_{\eta\eta}}{f_\eta^3}
-\frac{6\,\alpha\, \phi_{\eta} \, f}{f_\eta}
-{\alpha}_{\theta\theta}
-\frac{{\alpha}_{\eta} \, f}{f_\eta} \right) \nonumber \\ \mbox{} & &
-H_c\, {v_2}_{\eta} + {H_b}_{\theta} \,v_2 + \frac{H_c\, {v_1}_{\theta} f_\eta^2}{f^2}+ {H_b}_{\eta} \,v_1
\nonumber \\ \mbox{} & & + [\alpha H_b(H_d + H_b + H_a)]
\end{eqnarray}

in the maximal slicing gauge ($TrK=0$) and rearrange it we arrive at
\begin{eqnarray}\label{eqn:hbevolarrange}
\frac{\partial H_b}{\partial t} & = & \left\{
\frac{1}{f^2}\left[\frac{\alpha\, q_{\theta} \cot\theta }{2}
-2\,\alpha\, \phi_{\theta} \cot\theta
-\frac{\alpha\, q_{\theta\theta} }{2}
+\alpha\, \phi_{\theta} \, q_{\theta}
+\frac{ {\alpha}_{\theta} \, q_{\theta} }{2} 
-4\,\alpha\, \phi_{\theta\theta}
+2\, {\alpha}_{\theta} \, \phi_{\theta}
-{\alpha}_{\theta\theta}\right]
\right. \nonumber \\ \mbox{} & & \left.
+\left(\frac{1}{f f_\eta}\right)\left[
-\frac{\alpha\, q_{\eta}}{2}\left(1+\frac{d}{d\eta}\left(\frac{f}{f_\eta}\right) \right)
-2 \, \alpha\, \phi_{\eta} \, \left(2+\frac{d}{d\eta}\left(\frac{f}{f_\eta}\right) \right)
-{\alpha}_{\eta} \right]
\right. \nonumber \\ \mbox{} & & \left.
+\left(\frac{1}{f_\eta^2}\right) \left[ -\frac{\alpha\, q_{\eta\eta} } {2}
-\alpha\, \phi_{\eta} \, q_{\eta}
-\frac{ {\alpha}_{\eta} \, q_{\eta} }{2}
-2\,\alpha\, \phi_{\eta\eta}
-4\,\alpha\, \phi_{\eta}^2
-2\, {\alpha}_{\eta} \, \phi_{\eta} \right]
 \right\} \frac{1}{e^{q+4\phi}}\nonumber \\ \mbox{} & &
-H_c\, {v_2}_{\eta} + {H_b}_{\theta} \,v_2 + H_c\, {v_1}_{\theta}\left(\frac{f_\eta}{f}\right)^2+ {H_b}_{\eta} \,v_1
\end{eqnarray}
If we add terms in the square brackets first using the algorithm in (\ref{eqn:rearrange_add}), then perform the resulting sum, and calculate the relative difference
\begin{equation}\label{eqn:relhbdot}\left.\frac{\tilde{H_b}-H_b}{\tilde{H_b}}\right|_{k=1}\end{equation}
where $\tilde{H_b}$ is the value calculated using the arrangement above and $H_b$ is the value calculated via equation (\ref{eqn:hbevolrepeat}), we find that relative differences are up to $10^{-13}$. Recalling that machine precision is $\sim 10^{-16}$ this is a non-trivial improvement in our numerical method as it is $3$ orders of magnitude larger than roundoff errors in places.  For a graph showing the difference between the use of  equation (\ref{eqn:hbevolrepeat}) and equation (\ref{eqn:hbevolarrange}) near the origin see figure \ref{fig:relhbdot}.

\section{Interpolation/Extrapolation techniques and smoothing}

In the course of trying to deal with regularisation issues at the boundaries of our grid, several interpolation and extrapolation techniques were implemented and tested.

We present some mathematical formulations, and will follow with a discussion on their usefulness in tackling this code.

If we know the values of a function $y(x)$ at three points $x_1$, $x_2$ and $x_3$, and wish to find the value at a fourth point $x_4$ using a quadratic fitting extrapolation and assuming all points are evenly spaced 
$$x_n=x_1+(n-1)\Delta x \;;\; n \in \{2,3,4\}$$
we find that
$$y(x_4)=y(x_1) -3 y(x_2) + 3 y(x_3)$$
Moving now to a fourth order correct extrapolation method, if we know the values
$$y(x_i) \;;\; i \in \{1,2,3,4,5\}$$
and
$$x_n=x_1+(n-1)\Delta x \;;\; n \in \{2,3,4,5,6\}$$
we find that
\begin{equation}\label{eqn:4ordextrap}y(x_6)=y(x_1) -5 y(x_2) + 10 y(x_3) - 10 y(x_4) + 5 y(x_5)\end{equation}

In general, using a fourth order polynomial fitting of the above points we can extrapolate to the point
$$y(x_{k+3})=y(x_3+k\Delta x)$$
using the following algorithm\footnote{we use the notation $y(x_i)=y_i$}
\begin{eqnarray}
a & = & \frac{1}{24}\left(y_1 - 4 y_2 + 6 y_3 - 4 y_4 + y_5 \right) \nonumber \\ \mbox{}
b & = & \frac{1}{12}\left(-y_1 + 2 y_2 - 2 y_4 + y_5 \right) \nonumber \\ \mbox{}
c & = & \frac{1}{24}\left(-y_1 + 16 y_2 - 30 y_3 + 16 y_4 - y_5 \right) \nonumber \\ \mbox{}
d & = & \frac{1}{12}\left(y_1 - 8 y_2 + 8 y_4 - y_5 \right) \nonumber \\ \mbox{}
e & = & y_3 \nonumber \\ \mbox{}
y(x_{k+3}) & = &  a k^4 + b k^3 + c k^2 + dk + e\nonumber \\ \mbox{}
y_x(x_{k+3}) & = &  \frac{4ak^3 + 3bk^2 + 2ck + d}{\Delta x} \nonumber \\ \mbox{}
y_{xx}(x_{k+3}) & = & \frac{12ak^2 + 6bk + 2c}{(\Delta x)^2}
\end{eqnarray}
where the special case $k=3$ yields the same results as in equation (\ref{eqn:4ordextrap}).

We can also apply a Fourier transform, spline, Pad\'{e} approximant, or various other numerical techniques to give an extrapolation condition.

These techniques were originally employed when difficulties were encountered at various boundaries, for example $r=0$ or the outer boundary.  Numerically irregular regions are usually indicative of other numerical problems, however, and given the highly non-linear nature of the equations using any linear smoothing algorithm\footnote{Or algorithm like a Fourier transform, which will generally have non-trivial contributions from many higher-order modes.} creates more problems than it solves.  These algorithms also assume a single coordinate direction (e.g. $\eta$ or $\theta$) extrapolation method for determining function values, which is nearly impossible to ``knit'' together to create a consistent multi-spatial directional extrapolation technique.  In all cases it was possible to trace down the \emph{source} of the regularisation problem instead of trying to smooth it out.

For example, attempting to smooth variables at the origin using various fitting curves (i.e. polynomial interpolation, polynomial extrapolation, splines, Fourier decomposition, \emph{et. al.}) was masking the need to move to 4th order correct derivatives from 2nd order (see section \ref{sec:4thorder}) and fix the regularity of our initial prescription of $q$ (see section \ref{sec:r0reg}).

Furthermore, as discussed in section \ref{sec:condition}, these matrix problems are highly susceptible to numerical error/noise due to their large condition numbers, meaning that any ``smoothing'' algorithm is bound to cause amplified (numerical error) perturbations that make convergence impossible.

\section{Choice of 3-metric functions}

\subsection{Representation of the Conformal Factor}\label{sec:psiphiconform}

Logarithmic derivatives of the form
\begin{equation}\label{eqn:logderiv}\frac{\psi_{\eta}}{\psi}=\partial_{\eta}(\ln\psi)\end{equation}
are present in the equations for the Christoffel symbols and the Ricci curvature (i.e. \ref{eqn:nonexpr11}) which are difficult to finite difference properly.

Using a 2nd order correct differencing approximation, we know that the error involved in calculating the first $\eta$ derivative of the function $\Psi$ at $\eta_i$ will be given by\footnote{See for example \cite{burden} pg. 640.}
\begin{equation}\label{eqn:eps2ord1stderiv}\epsilon=\frac{(\Delta\eta)^2}{6} \frac{d^3 [\Psi(\xi_i)]}{d \eta^3}  \; ;\; \xi_i \in [\eta_{i-1},\eta_{i+1}]\end{equation}
Letting
$$\Psi=\ln\psi$$
we can see that the error terms for (\ref{eqn:logderiv}) will look like
$$\epsilon \sim \frac{\psi_{\eta\eta\eta}}{\psi}-3\frac{\psi_{\eta}\psi_{\eta\eta}}{\psi^2}+2\frac{\psi_{\eta}^3}{\psi^3}$$
For small values of $\psi$ or in regions with large non-linearities, this error term will not be well-behaved (as is expected of numerically differentiating a logarithmic function).  Further, increasing the order of the finite differencing scheme will have similar looking terms which can actually increase the error in those regions. 

A second issue is that the traditional conformal factor $\psi$ should asymptotically approach $1$ at the outer boundary to match an asymptotically flat solution, plus leading order behaviour that is typically $\frac{1}{r}$ depending on whether one is matching to Schwarzschild, etc.  With a typical outer boundary for our code set at $r=\sinh(10) \sim 10^4$, as $\psi$ is approaching the outer boundary it is possible to lose some important precision because one is tracking 4-5 extra decimal places that are unnecessary.  While Webster \cite{paul_thesis} attempted sidestep this problem by introducing a shifted conformal factor $\tilde{\psi}=\psi-1$, it creates other problems in defining boundary values while solving the Hamiltonian constraint.

The solution to these problems is to introduce an auxiliary variable $\phi$, such that $\psi=e^{\phi}$.  This removes the growth of error terms noted above and simplifies many equations, benefiting us in both reduced computation time and more manageable errors.

This was a major recoding, as all of our equations had to be re-derived, coded, tested, etc. but was well worth the effort and solved many regularity problems.  The move to a 4th order finite differencing scheme highlighted the need for this change as errors in the operands of the RHS of equations, that were previously masked by the limitations of the finite differencing scheme, came to light then.  There has also been mention of other numerical groups using a similar transformation while employing a BSSN formalism \cite{NINJA,Pollney}.

\subsection{Choice of metric variable $a$}
Analogously to the problems mentioned in section \ref{sec:psiphiconform} regarding the use of $\psi$, we choose to implement an auxiliary variable $a=e^q.$\footnote{Traditional Brill metrics use $e^{2q}$, but we drop the 2 to avoid extra computational steps throughout the code, and instead absorb it into the amplitude to allow for comparison to other codes.}

As before, we note that this was a major coding change but was well worth the effort, and once again the move to a 4th order finite differencing scheme highlighted the need for this change, although it was noticeable in the 2nd order implementation as well.

\section{Shift vector potentials as alternative gauge variables}

The first order equations for the shift vectors (\ref{eqn:shiftvcoupled}) are unstable as they do not couple the two spatial directions for the shift vector quantities, $v_1$ and $v_2$.  By this, we mean that the first equation couples the radial derivative of $v_1$ with the angular derivative of $v_2$, and vice-versa in the second equation.  What this leads to is a ``ridging'' effect in our solution of either variable, as we only require variable and derivative consistency in one of the spatial directions, not both, when solving any particular equation.

Furthermore, the shift vector quantities are present in each equation so that if we try to solve this system we need to either find some way of separating the equations for the variables or developing an iterative convergence scheme.

For example, one could present an initial ``guess'' for $v_2$, solve for $v_1$ using the first equation, then use that solution for $v_1$ as a guess in the second equation and solve for $v_2$.  One can then iterate on this scheme and test for convergence of the solutions.

The problem arises that the derivative information in the two spatial directions is not coupled in any one equation so while, for example, the angular derivatives may be smooth when solving one equation, the radial derivatives tend not to be.  The creates large problems for any iterative scheme as the non-smooth (i.e. radial) derivatives cause pathological failures when you try to use that solution as a ``guess'' in the second equation.

The solution to this problem is to use potentials instead which allows us to do two things: (i) decouple the variables into their own respective equations and (ii) to generate PDEs for that contain derivative information in both spatial directions.  This stabilizes and simplifies the solution methods, at the cost of an extra required order of finite differencing precision. See section \ref{sec:shiftvec} for more details.

\section{Constrained Evolution vs. Free Evolution}
Due to the overdetermined nature of Einstein's equations, we have a certain degree of freedom in choosing the degree of constrained versus free evolution that we wish to implement numerically.  We will investigate a few of the choices available to us.
\subsection{Momentum Constraints}\label{sec:hahbcon}
Solving equations (\ref{eqn:momcons}) for $H_a$ and $H_b$ in the maximal slicing gauge yields the following equations\footnote{A similar analysis holds if we avoid maximal slicing, so this result is not a gauge pathology.}:

First momentum constraint solved for $H_{a}$
\begin{eqnarray}\label{eqn:momcon1nouse}
\lefteqn{H_{a \eta} + \left(\frac{6 \frac{d\psi}{d\eta}}{\psi}+\frac{3 f_{\eta}}{f}+\frac{\frac{da}{d\eta}}{2 a}\right)H_a} \nonumber \\
& = & H_b \left(\frac{a_\eta}{2 a}\right)
-H_c \left(\frac{{f_{\eta}}^2}{f^2}\cot\theta
+\frac{6 {f_{\eta}}^2 \psi_{\theta}}{f^2 \psi}
+\frac{a_{\theta} {f_{\eta}}^2}{a f}\right)
-\frac{{f_{\eta}}^2}{f^2} H_{c \theta}
\end{eqnarray}
Second momentum constraint solved for $H_{b}$
\begin{eqnarray}\label{eqn:momcon2nouse}
\lefteqn{H_{b \theta} + \left(2 \cot\theta+\frac{6 \frac{d\psi}{d\theta}}{\psi}+\frac{\frac{d a}{d\theta}}{2 a}\right)H_b} \nonumber \\
& = & -H_a\left(\cot\theta-\frac{\frac{d a}{d\theta}}{2 a}\right)
-H_c\left(\frac{6 \frac{d\psi}{d\eta}}{\psi}+\frac{f_{\eta\eta}}{f_{\eta}}+\frac{2 f_{\eta}}{f}+\frac{\frac{da}{d\eta}}{a}\right)
-H_{c \eta}
\end{eqnarray}
The first thing to note is that these equations are coupled, so any attempt to find a solution to $H_b$ from these constraints will require a solution for $H_a$ which is either derived from (1) a guess from the other constraint equation with some form of attempted iterative convergence, or (2) the evolution equation (\ref{eqn:haevol}).  The author had no success at mixing momentum constraints and extrinsic curvature evolution equations, and the degeneracy of the second equation is problematic numerically.

The degeneracy of these equations arises from a consideration of boundary conditions along $\theta=0$. As we require symmetry across $\theta=0$ for most variables and from inspection of equation (\ref{eqn:momcon2nouse}), we see that most of the terms in equation (\ref{eqn:momcon2nouse}) vanish identically along the axis, leaving the following:
$$2\cot\theta H_b = -\cot\theta H_a$$
which implies either that 
$$2 H_b + H_a = 0$$
along the axis, which is a numerically unstable algebraic condition\footnote{This was observed in attempted implementations.} or that $H_a$ and $H_b$ are symmetric across $\theta=0$ and can be \emph{anything}.  Allowing either of these makes any attempted solution unstable.

\begin{figure}\centering
\includegraphics{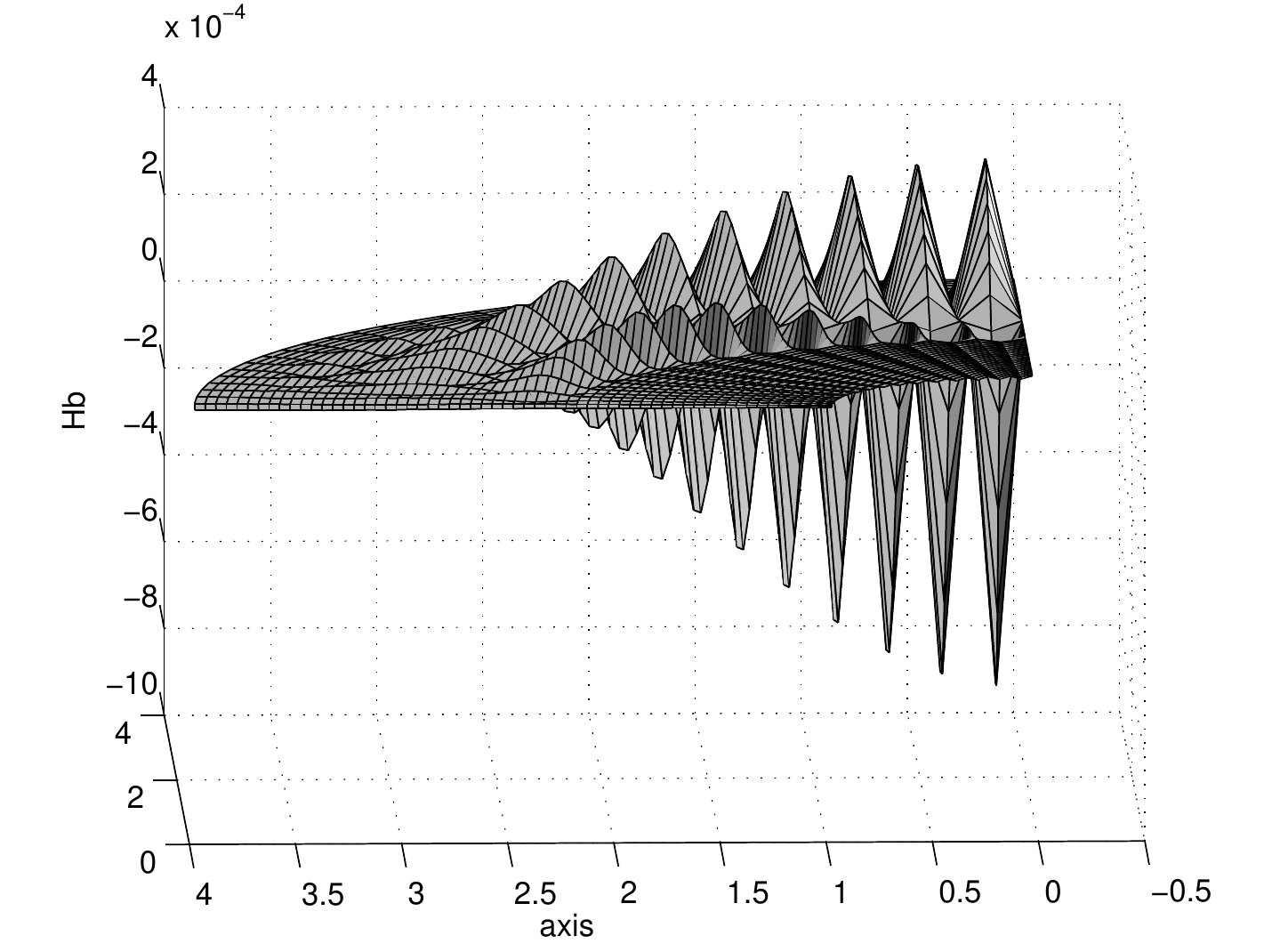}
\caption[Non-convergent solution for $H_b$]{Example of solver's non-convergent ``best guess'' when trying to solve equation \ref{eqn:momcon2nouse} for $H_b$}
\label{fig:unstablemomcon2}
\end{figure}

This degeneracy at the axis caused any numerical solver I attempted to be unable to converge, until I finally realised the source of its problems and abandoned the attempt.  See figure \ref{fig:unstablemomcon2} for an example of why to avoid these equations.

The second issue that arises is that these equations only contain first spatial derivatives of the extrinsic curvature variables - and when one is finite differencing the first derivative terms\footnote{See equation (\ref{eqn:4ordderiv1}).} at $(i,j)$ they do not contain any coupling of the value of the function to the point itself at $(i,j)$ - this can lead to undesirable numerical coupling\footnote{e.g. in a second order scheme this decouples the even and odd indexed grid points.}.  One could attempt to circumvent this by devising a potential formulation if the axis instability mentioned above wasn't present, however there is no simple method to decouple the equations as happens with the shift vectors and their potentials.

Further, any attempt to use potentials to couple the $\eta$ and $\theta$ directions in those equations will involve mixed first partial derivatives and first order spatial derivatives, which in the traditional finite difference formulation contains no coupling to the $(i,j)$ point - hence the need for an alternate numerical method to calculate the mixed partial derivative for example (see section \ref{sec:mixed4thord}).

So in the end, we choose to use these constraints as a check on the consistency and accuracy of the code and use the evolution equations for $H_a$ and $H_b$.

One feature to note is that while Choptuik \cite{choptuik1,choptuik2} in his now-famous cylindrical coordinate code can make use of the momentum constraints, we cannot due to their degeneracy in spherical polar coordinates.

\section{BSSN formalism and the Brill Wave formulation}
As discussed in Appendix \ref{chap:testgridcoord}, there does not seem to be a compelling argument for switching to one (of the many) alternate Cauchy formulations to ADM.  It would be impossible to enumerate and compare all of the alternate formulations (as there are theoretically infinitely many of them), however BSSN has been employed for a number of reasons\footnote{There has also been some success in numerical work using various Harmonic coordinates which arise from variations of equation (\ref{eqn:harmonicchristoff}).}, including increased stability in certain scenarios\footnote{Alcubierre \cite{alcubierre:3p1num} also presents a discussion of the well-posed hyperbolicity of BSSNOK.} due to its use of the auxiliary variable $\Gamma^i$.

For reasons discussed in section \ref{sec:extrinsiccurvature} I wish to avoid the use of ``consolidated'' curvature terms and prefer the natural cancellation present in ``expansion'' of the Christoffel symbols.  This runs counter to the BSSN formalism which seeks to turn the Christoffel coefficients into dynamic variables.

It was attempted, at one point, to roll the ADM formalism into more of a ``consolidated curvature terms'' evolution, with unstable numerical results.  One can calculate the mixed Christoffel symbols explicitly from equations (\ref{eqns:mcs}), then calculate the Ricci Tensor using
\begin{equation}\label{eqn:riccimcs}
R_{ab}=\partial_c\Gamma^c_{ab} - \partial_b \Gamma^d_{ad} + \Gamma^c_{ab}\Gamma^d_{cd} - \Gamma^d_{ac}\Gamma^c_{bd}
\end{equation}
and form the remainder of the terms in equation (\ref{eqn:genmixkevol}) explicitly instead of expanding them as in section \ref{sec:extrincurvevol}.  For example, one can calculate $D^iD_j \alpha$, $\beta^lD_l K^a_b$, $K^a_lD_b\beta^l$ and $K^l_bD_l\beta^a$ then sum them afterwards to give the RHS terms for the evolution of the extrinsic curvature.

After implementing this method and tracking down several numerical problems, they all seemed to stem from forcing a numerical grouping that did not allow natural cancellation - i.e. the terms in a grouping (i.e. $K^a_lD_b\beta^l$) would be of vastly differing orders of magnitude\footnote{This also directly relates to the discussion in section \ref{subsec:addterms}.}, so one would lose decimal places of precision when summing them together, which in turn would propagate into lost precision when calculating the total RHS of equation (\ref{eqn:genmixkevol}).

Further, the BSSN formalism also requires a conformally flat metric which is not compatible with the Brill conditions/metric in spherical polar coordinates\footnote{See (\ref{eqn:brillmetricchap1}) for the required form of the metric.}, and while it may be possible to work around that limitation with a ``modified BSSN'', it would entail a reworking that is not merited by the other considerations.\footnote{As demonstrated in section \ref{sec:bssnvsadm}, the Schwarzschild code worked better with the ADM scheme than BSSN.}

\section{General remarks}
In the course of performing mathematical analysis, debugging, coding and tracking down numerical issues, some important general considerations came to light.  While these are not new concepts, they bear repeating in the context of performing a project like this one:
\begin{itemize}
\item Ensure code is properly commented.  This is general good coding practice, but it is probably the single most ignored aspect of coding.  The author finds it useful to assume you will forget what you have done when revisiting your code 4 years (or six months) from now.
\item Use generic functions/subroutines wherever possible to avoid having to update (or forget to update) calculation methods in several different places that perform the same task.  Never assume that you are only going to have to calculate derivatives in only one place.
\item Use \emph{implicit-none} to ensure that all variables are properly declared and typed in FORTRAN to prevent type mismatches.  Similarly in C ensure that type-mismatches are caught by the compiler and resolved.
\item Do not use common blocks or other specialised quirks of a particular compiler. A good code only survives by being portable across platforms and easily debuggable by someone not familiar with your compiler.  This code has been migrated to no less than 7 different compilers over the years including f77, xlf, hf77, g77, gcc, mpif77 and currently resides on gfortran.  It has also run on various operating systems including various AIX and linux distributions and versions, as well as on 16-bit, 32-bit and 64-bit architectures.
\item Use simple situations to test numerical methods, gauge conditions, coordinate regularity, etc. before implementing complex situations (i.e. Schwarzschild before Brill)
\item Always check the conditioning of your equations around critical points like $r=0$, $\theta=0, \frac{\pi}{2}$, outer edge of your grid, etc.  Write programs to automate checking these regions and always perform visual sanity checks (e.g. graphical and tabular analysis of the data produced by the codes) before believing your results.  Matlab was indispensable for performing visual graphical checks on the various boundaries as it is virtually impossible to catch all of the regularity problems that can arise at various boundaries with coded regularisation checks.  The human brain can find some things that numerical algorithms cannot (as the brain is a highly efficient visual processing machine).
\item Equations must be translated from Maxima or other symbolic output programs to give (a) more efficient methods and (b) more precise methods.  See equation (\ref{eqn:hamaxima}) for an example of non-rearranged symbolic output that is numerically ill-behaved.  While this is useful for forcing analytic calculations into the equations instead of using the unexpanded forms, one must take care to recollect terms afterwards and not blindly apply the code it spits out.
\end{itemize}

An example of the symbolic output from Maxima when calculating the evolution equations from Einstein's equations with our gauge and coordinate conditions:
\begin{eqnarray}\label{eqn:hamaxima}
{H_a}_t & = & -\alpha e^{-q-4 p} q_y cos(y)/(f^2 sin(y))/2.0-2 \alpha p_y e^{-q-4 p} cos(y)/(f^2 sin(y)) \\ \nonumber \mbox{} & &
+{H_c} {v_2}_x+{H_a}_y {v_2}-f_x^2 {H_c} {v_1}_y/f^2+{H_a}_x {v_1}-\alpha e^{-q-4 p} q_{yy}/f^2/2.0 \\ \nonumber \mbox{} & &
-\alpha p_y e^{-q-4 p} q_y/f^2-\alpha_y e^{-q-4 p} q_y/f^2/2.0-\alpha e^{-q-4 p} q_{xx}/f_x^2/2.0 \\ \nonumber \mbox{} & &
+\alpha p_x e^{-q-4 p} q_x/f_x^2+\alpha_x e^{-q-4 p} q_x/f_x^2/2.0+f_{xx} \alpha e^{-q-4 p} q_x/f_x^3/2.0-2 \alpha p_{yy} e^{-q-4 p}/f^2 \\ \nonumber \mbox{} & &
-4 \alpha (p_y)^2 e^{-q-4 p}/f^2-2 \alpha_y p_y e^{-q-4 p}/f^2-4 \alpha p_{xx} e^{-q-4 p}/f_x^2 \\ \nonumber \mbox{} & &
+2 \alpha_x p_x e^{-q-4 p}/f_x^2+4 f_{xx} \alpha p_x e^{-q-4 p}/f_x^3 \\ \nonumber \mbox{} & &
-4 \alpha p_x e^{-q-4 p}/(f f_x)-\alpha_{xx} e^{-q-4 p}/f_x^2+f_{xx} \alpha_x e^{-q-4 p}/f_x^3
\end{eqnarray}

\section{Chapter Summary}
In this chapter a number of physically, mathematically and numerically motivated methodologies have been presented and employed to regularise the various equations that must be solved in the course of creating a Brill wave evolution code.

Careful consideration of tensor regularity and matrix conditioning allowed more robust determination of which of many available numerical methods to employ, and also shaped some of our gauge and variable choices.

Development of appropriate outer boundary conditions is an area of ongoing research in numerical relativity, and presented in this chapter is a methodology that was successfully employed in the Brill wave evolution code.

Various miscellaneous methods were also discussed that, when combined together, give more robust numerical algorithms.  The ill-conditioning of the matrix problems means that much attention must be paid to even seemingly minor numerical errors lest they become amplified and dominate the numerical solution.

\fancyhead[RO,LE]{\thepage}
\fancyfoot{} 
\chapter{Development and Structure of The Numerical Code}\label{chap:numcode}
\bigskip
In this chapter we will step through all the major functional pieces that this Brill wave evolution code requires in order to compute all of the necessary gauge and dynamical variables on a time step.  Each section will require its own special considerations, algorithms, boundary conditions, etc. however we will cover the major issues encountered in each section of the code.

We store our global constants in \verb+param.h+ at compile time. We then read the initial wave parameters from \verb+waveparam.h+ at run time, including ``amplitude'' ($A$) and ``width'' ($s_0$) of the initial (Gaussian) wave packet.

\section{Storage requirements}
Assuming a 2+1 ADM grid with spatial grid dimensions $(200 \times 60)$, $64$ bit double precision floating point numbers, and $\sim 10$ dynamic variables, we require $\sim 7.68$ Mb ($960$KB) of storage space per time step.  At one point the ability to reduce this number was an important consideration due to physical memory limitations, but even low-end modern consumer-grade computers can easily handle these memory requirements.

We also choose to store variables in memory that are derived from the primary dynamic variables (e.g. $\phi_\eta$) instead of computing them each time they are required in a specific calculation. It is computationally cheaper to store them in RAM than to recalculate or page them in and out constantly if you have the available RAM.  This can expand the number of stored variables per time step to over $100$, however it significantly speeds up the BiCGStab solver.  This increase in variables, plus saving $5$ time steps of information for a fourth order time scheme, gives us a fifty-fold increase in memory requirements ($48$MB), making it conceivable that we could increase the resolution by a factor of $6$ in both spatial directions ($48$MB$\times 6^2=1.7$GB) and still stay within the limits of most commercially available 32-bit platforms (let alone supercomputers or 64-bit platforms)\footnote{Running, for example, a ``normal'' $200\times 60$ simulation consumes $\sim 63$MB of memory and causes usage on a single processor to remain constant at $100\%$.  Increasing the grid resolution to $800\times 120$ consumes $\sim 375$MB of memory.}.

\subsection{Computational Hardware}
The original simulations were run many years ago on the UofC ACS cluster.  The code was then ported in 2009 to a high-end retail system ($\sim 8$ years later) with much improved results, then in 2013 it was ported again on to a new custom-built retail system with a fourfold speed increase again.  What used to run overnight now takes a matter of minutes to perform.  Where we used to be constrained primarily by the amount of RAM available and the speed of hard disks, we are now constrained primarily by the speed of an individual CPU\footnote{CPU-bound} more than ever.  This CPU-bound nature of computational bottlenecks has caused a shift in focus to parallelisation efforts and multi-core architectures in many computational realms.

\section{Code Flow Chart}
The following is a basic flow chart of how the code runs:
\pagebreak
\begin{tabbing}
PAR\=AME\=TER\=  \ S\=EARCH \\
\> Iterate through wave parameter phase space and look for \\ \>  horizon formation by calling the Evolution program below for each value \\
EVOLUTION \\
\> INITIAL DATA SETUP \+ \+ \\
(I1) - wave profile ($q$) formed based on parameter space search \\
(I2) - time symmetric initial data for everything else \\
(I3) - solve for $\phi$ (**) (using the Hamiltonian constraint) \\
(I4) - perform a horizon search for black hole formation \- \\
MAIN LOOP \+ \\
Move variables in memory back one time bin; $\xi(t_0) \rightarrow \xi(t_{-1})$ \\
Evolve $q$ (*) \\
IF EVOLVEPHI THEN perform $\phi$ evolution (*)\\
IF EVOLVEHC THEN perform $H_c$ evolution (*) \\
IF EVOLVEHA AND EVOLVEHB THEN perform $H_a$ and $H_b$ evolution (*) \\
IF EVOLVEHD THEN perform $H_d$ evolution (*) \\
IF NOT EVOLVEHA AND NOT EVOLVEHB THEN constrain $H_a$ and $H_b$ \\ \> from the momentum constraints and potentials \\ \> (not recommended for this thesis) \\
IF NOT EVOLVEHC THEN constrain $H_c$ - only if evolving $H_a$ or $H_b$ \\
IF NOT EVOLVEHD THEN update $H_d$ from constraints if using maximal slicing \\
IF NOT EVOLVEPHI THEN constrain $\phi$ (using the Hamiltonian constraint) (**) \\
IF MAXIMAL SLICING THEN constrain lapse, $\alpha$ \\
Constrain shift vector potentials ($\Phi$, $\chi$) \\
Compute shift vectors from potentials ($v_1$,$v_2$ from $\Phi$ and $\chi$) \\
Calculate scalar curvature $R$  \\
Check for convergence of coupled constraints and reiterate if necessary \\ \> for convergence \\
Calculate constraints (e.q. momentum) \\
Perform a horizon search for black hole formation \- \- \\
END OF EVOLUTION \\
\end{tabbing}

(*) - requires a Crank-Nicholson iterative convergence on this variable (see section \ref{subsubsec:cntimeevol})

(**) - requires iterative convergence on a non-linear PDE

\pagebreak

\section{Coding considerations}
The code used for the evolution of the spacetime was originally made \cite{orig_code} as a general axisymmetric evolution code, and was coded in Fortran 77.
It was then used and modified by Paul Webster in his thesis \cite{paul_thesis} to study various black hole systems, and was then further modified to study the above scenario.
Due to the large number of changes needed to study the Brill axisymmetric system, however, only the biconjugate gradient matrix solver routine survived (relatively) unchanged.  For all intents and purposes the program has been re-written from its various incarnations to allow for the improvements listed in chapter (\ref{chap:changes}), including numerical, programmatic and physical considerations that dictate certain choices.

\section{Dynamic Variables}
Inside the program, our dynamic variables are:
\begin{itemize}
\item The 3-metric variables $\phi=\ln\psi$ (logarithm of the conformal factor) and $q$ (free wave data)
\item The extrinsic curvature variables $H_a$, $H_b$, $H_c$, $H_d$
\item The gauge variables $\alpha$ (lapse), $v_1=v_1(\chi_{\eta},\Phi_{\theta})$, $v_2=v_2(\chi_{\theta},\Phi_{\eta})$ (shift vectors) with auxiliary variables $\chi$ and $\Phi$ (shift potentials).
\end{itemize}

We choose to evolve the following variables (those which contain $\frac{\partial}{\partial t}$ time evolution): $q$, $H_a$, $H_b$, $H_c$, $H_d$

The elliptic (constrained) variables are: $v_1$ (gauge), $v_2$ (gauge), $\phi$ (Hamiltonian) and the auxiliary variables $\chi$ and $\Phi$.

Variables which are currently constant in time: $\alpha$

A discussion of the details of each section of the code used for the implementation above follows.

\section{Initial Value Problem}
\subsection{Choice of radial function $f$}
We use a radial function $f=f(\eta)$ that increases monotonically in $\eta$, is zero at the origin of our coordinate system, and grows exponentially as we move into the asymptotic region of our spacetime (see section \ref{sec:fnx}).

We define the derivatives (and combinations like $\frac{f}{f_\eta}$) of $f$ analytically, which was an important regularisation condition. 

\subsection{(I1) Initial Wave Profile}\label{sec:initwave}
The initial wave profile for $q$ is freely chosen subject to the restrictions given in section \ref{subsec:brillformalism}.
Other gravitational wave initial value formulations (for example \cite{evans,paul_thesis,oliveira}) have used wave profiles of the following form:
\begin{equation}\label{eqn:qinitgauss}q(\eta,\theta,t=0)= A \left(e^{-\frac{(f(\eta)-x_0)^2}{s_0^2}}-e^{-\frac{(f(\eta)+x_0)^2}{s_0^2}}\right) \sin^2(\theta)\end{equation}

where
\begin{itemize}
\item $A$ is the Brill wave ``amplitude''
\item $f=x_0$ is the radial location of the peak of the wave and
\item $s_0$ is the $\frac{1}{2}$ ``width'' of the initial wave
\end{itemize}
It is important to note that this form of $q$ does not go as $r^2$ at the origin so as to satisfy the regularity conditions discussed in section \ref{sec:regconditions}.  This wave profile causes critical regularity problems in the code with a pure vacuum gravitational wave and should be avoided.

It is also important to note that our function for $q$ must be symmetric about $\eta=0$ ($r=0$) to ensure that the metric obeys the symmetry conditions, so we cannot have functions that have translations that introduce a $q \sim e^{(f-x_0)^2}$ dependence as they do not satisfy
$$q(\eta) = q(-\eta)$$
hence we must use a function of the form $q \sim e^{(f/s_0)^2}$ if we wish to translate the peak of the wave in or out radially, in combination with an appropriate amplitude.

In \cite{mastersonmsc} we used an initial wave profile of the form 
$$q(\eta,\theta,t=0)=A h(\eta) \sin^n\theta$$
where $h(\eta)$ is a Lorentzian function of the form
$$h(\eta)=\frac{[f(\eta)]^k}{\left(1+\frac{f(\eta)}{\eta_0}\right)^l}$$
where $k,l$ are integers and $A$ is again our amplitude. The radial part of this waveform $h(\eta)$ is very difficult to differentiate numerically and is prone to regularisation problems.  While we can specify the derivatives analytically on the initial slice, all future time slices would suffer from the inability to calculate good numerical derivatives of this waveform as it evolves.

Another waveform that was tried was a continuous piece-wise cubic function with $\frac{1}{\eta^2}$ falloff at the outer edge of the grid.  We start by defining $h_{\eta\eta}$ so that the function is smooth to second order and integrate to determine $h$.

\begin{figure}[h!]  \begin{center}
\includegraphics{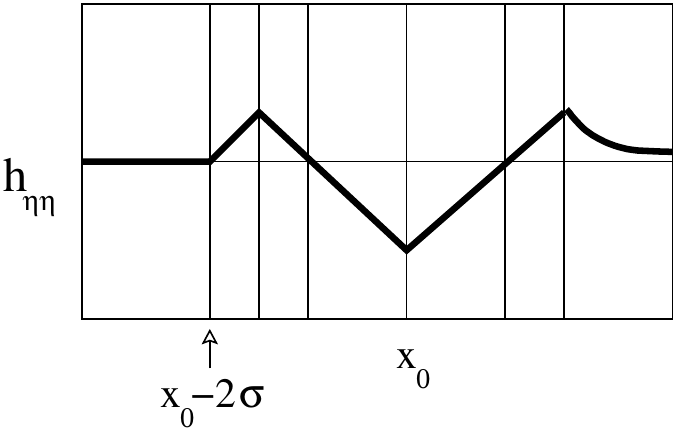}
\caption[$h_{\eta\eta}$ for determining a piecewise cubic $h$]{Schematic of $h_{\eta\eta}$ used to generate a piecewise $C_2$ cubic function for $h$ with $\frac{1}{\eta^2}$ falloff at the outer edge of the grid}\end{center}  \end{figure}

\begin{center}\begin{tabular}{|c|c|c|} \hline
$h_{\eta\eta}(\eta)$ & $\eta$ & $h(\eta)$ \\ \hline
$0$ & $\eta \leq \eta_0-2\sigma$ & $0$ \\ \hline
$\frac{2A}{\sigma}\left(\eta-\eta_0+2\sigma\right)$ & $\eta_o-2\sigma \leq \eta \leq \eta_0-\frac{3}{2}\sigma$ & $\frac{A}{3\sigma}\left(\eta-\eta_0+2\sigma\right)^3$ \\ \hline
$\frac{-2A}{\sigma}\left(\eta-\eta_0+\sigma\right)$ & $\eta_0-\frac{3}{2}\sigma \leq \eta \leq \eta_0 - \sigma$ & $\frac{A\sigma}{2}\left(\eta-\eta_0+\frac{3}{2}\sigma\right)-\frac{A}{3\sigma}\left(\eta-\eta_0+\sigma\right)^3$ \\ \hline
$\frac{-A}{\sigma}\left(\eta-\eta_0+\sigma\right)$ & $\eta_0-\sigma\leq \eta\leq \eta_0$ & $\frac{A\sigma^2}{4} + \frac{A\sigma}{2}\left(\eta-\eta_0+\sigma\right) - \frac{A}{6\sigma}\left(\eta-\eta_0+\sigma\right)^3$ \\ \hline
$\frac{A}{\sigma}\left(\eta-\eta_0-\sigma\right)$ & $\eta_0\leq \eta \leq \eta_0+\sigma$ & $\frac{3A\sigma^2}{4} - \frac{A\sigma}{2}\left(\eta-\eta_0\right) + \frac{A}{6\sigma}\left(\eta-\eta_0-\sigma\right)^3$ \\ \hline
$m\left(\eta-\eta_0-\sigma\right)$ & $\eta_0+\sigma \leq \eta \leq \eta_0 + \frac{3}{2}\sigma$ & $\frac{A\sigma^2}{4}-\frac{A\sigma}{2}\left(\eta-\eta_0-\sigma\right)+\frac{m}{6}\left(\eta-\eta_0-\sigma\right)^3$ \\ \hline
$\frac{\alpha}{(\eta+c)^4}$ & $\eta \geq \eta_0 + \frac{3}{2}\sigma$ & $\frac{\alpha}{6\left(\eta+c\right)^2}$ \\ \hline
\end{tabular}\end{center}

Where $\eta_0$ is the center of the wave, and to ensure $C_0$, $C_1$ and $C_2$ at the boundary $\eta=\eta_0+ \frac{3}{2}\sigma$ between the cubic and $\frac{1}{\eta^2}$ function we need
$$c = -\eta_0 - \sigma$$
$$m = \frac{12 A}{5 \sigma}$$
and
$$\alpha=\frac{3 A \sigma^4}{40}$$
The maximum value of $h(\eta)$ can be found at $\eta = \eta_0$ and is
$$h_{max}(\eta_0)=\frac{7 A \sigma^2}{12}$$

While in principle there is nothing wrong with this prescription for the radial expression $h$ of our initial waveform\footnote{And it resolved all of the regularity problems observed in the Lorentzian IVP wave formulation.}, it is only second order correct (i.e. it fails to be differentiable to any higher order than two) and is rather tedious to piece together.  The move to a fourth order correct numerical scheme would have necessitated an even larger amount of manual work to generate a piecewise fifth-order polynomial function, and at that point it was decided to find an alternate way to prescribe $h$.  With, for example, a Gaussian profile for $h$, we have infinite differentiability so if we decide to go to a sixth or eighth order correct numerical scheme in the future we don't have to re-derive what $h$ is, unlike the case where we use piece-wise polynomials.

Taking all of these considerations together we choose our wave profile to be of the form:
\begin{equation}\label{eqn:qinit}q(\eta,\theta,t=0)=2 A f^4e^{-\left(\frac{f}{s_0}\right)^2}\sin^2\theta\end{equation}
Where the $f^4$ term is used to ensure that the IVP formulation of $q$ has a radial dependence of $r^2$ or higher at the origin.  The factor of $2$ in equation (\ref{eqn:qinit}) is added to account for the fact that we use $q$ in our metric instead of $2q$ like most other groups, so this will allow an easy comparison of amplitude phase space characterisation to other work in the field.

\subsection{(I2) Time-Symmetric Initial Data}
\begin{enumerate}
\item $\alpha$ is defined by equation (\ref{eqn:alphastatic}) for reasons explained in section \ref{subsec:alphareg}.  This means that the $D^aD_b\alpha$ terms will not vanish in the extrinsic curvature evolution equations (\ref{eqn:genmixkevol}), so while we may have the time derivative of the \emph{metric} vanishing, the extrinsic curvature variables will not have a vanishing time derivative initially\footnote{$R_{ab}$ will also be non-zero in general, and this term is the usual driver of the evolution off the initial slice.}.  This can be thought of in terms of the velocity of the metric vanishing, but its acceleration is non-zero.  Most prescriptions set $\alpha=1$ on the initial slice for simplicity, however this is not a well-posed problem for reasons described in section \ref{subsec:alphareg}.
\item We employ the general gauge choices listed in section \ref{sec:gauge}, namely
\subitem $b=a$
\subitem $d=1$
\subitem $\gamma_{12}=c=0$
\item  As discussed in the section on the Brill criteria (\ref{subsec:brillformalism}) we require that our initial time-symmetric hypersurface satisfies $R=0$.  From equation (\ref{eqn:3p1:gammadot}) we see that if
\begin{enumerate}
\item our shift vectors vanish on the initial slice ($v_1=v_2=0$)
\item $K_{ab}=0$
\end{enumerate}
then we have a moment of time symmetry (so all time derivatives of the metric vanish).  This necessitates $K_{ab}=0$ i.e. the extrinsic curvature vanishes on the initial slice\footnote{This means that the initial hypersurface has only intrinsic curvature and no extrinsic curvature in the embedding spacetime, a very special condition.}, represented in our variables by $H_{a}=H_{b}=H_{c}=H_{d}=0$.  This is also consistent with the Hamiltonian constraint condition above which requires that (\ref{eqn:hamconfullvaccum}) reduce to $R=0$ on the initial slice.
\item Set the shift vector potentials $\chi=\Phi=0$ everywhere on the initial slice.
\end{enumerate}

\subsubsection{Miscellaneous IVP Conditions}
\begin{itemize}
\item 4th order correct derivatives are calculated using the methods in section \ref{sec:4thordderiv} for all variables.
\item We must define the stencil we are using for the mixed 4th order derivatives - see section \ref{sec:mixed4thord}.
\item Define $\alpha$ (for now) via equation (\ref{eqn:alphastatic})
\end{itemize}

\subsubsection{Time symmetry}

To complete the IVP, we need to provide all variables values and their spatial derivatives at ($t=-\Delta t,-2\Delta t,-3\Delta t$) if we are using a 4th order correct in time schema.  So we set
$$F_{(t=0)}=F_{(t=-\Delta t)}=F_{(t=-2\Delta t)}=F_{(t=-3\Delta t)}$$
where $F$ can represent any of our dynamic variables.  While we have moved to the time evolution schema described in section \ref{subsubsec:cntimeevol}, which will only require the previous time step, higher order C-N schemes or comparative will require these values.  It is also possible to use lower order correct time evolution schemes to start the evolution, however there is very little cost to defining these values initially to prevent possible coding errors in the future.

\subsection{(I3) - Calculation of the conformal factor $\phi$}\label{sec:constrainconform}
The most difficult portion of the IVP (and subsequent evolution) is solving the Hamiltonian constraint (\ref{eqn:hamconphiearly}) for $\phi$ due to the $e^{4\phi}$, $\phi_{\eta}^2$ and $\phi_{\theta}^2$ terms on the RHS of the equation.

While it is true that we can use equation (\ref{eqn:gam33dot}) to get a simple time evolution equation for $\phi$, we still need to set up the Initial Value Problem for $\phi$.  The method of solving the elliptic equation for $\phi$ on the initial $t=0$ hypersurface will be extended to all times during the evolution.

The numerical techniques developed to deal with this particular problem also carry over into solving for the lapse, $\alpha$, in equation (\ref{eqn:maxslice}) and the shift vectors $v_1$ and $v_2$ in equations (\ref{eqn:decshiftvecs}) as the formulation of a complete problem in the interior region is the same in all cases.

The Hamiltonian constraint with gauge conditions applied as computed by Maxima can be found in equation (\ref{eqn:hamcon}).  Using $\psi=e^{\phi}$ and equation (\ref{eqn:hamcon}), we find that the equation to be solved for $\phi$ is:
\begin{eqnarray} & & \left(\frac{f}{f_{\eta}}\right)^2\phi_{\eta\eta} + \phi_{\theta\theta} +  \frac{f}{f_{\eta}}\left[1+\partial_{\eta}\left(\frac{f}{f_{\eta}}\right)\right] \phi_{\eta} + \cot(\theta) \phi_{\theta} + \left(\frac{f}{f_{\eta}}\right)^2 \phi_{\eta}^2 + \phi_{\theta}^2 \nonumber \\ \mbox{} & &
 = \frac{1}{4}e^q[f^2(H_b H_d + H_a H_d + H_a H_b)-f_{\eta}^2H_c^2] e^{4\phi} \nonumber \\ \mbox{} & &
- \frac{1}{8}\left[q_{\eta\eta}\left(\frac{f}{f_{\eta}}\right)^2+q_{\eta}\left(\frac{f}{f_{\eta}}\right)\partial_{\eta}\left(\frac{f}{f_{\eta}}\right)+q_{\theta\theta}\right] \label{eqn:hamconphi}
\end{eqnarray}
which has the form:

$$A \phi_{\eta\eta} + B \phi_{\theta\theta} + C \phi_{\eta} + D \phi_{\theta}  = G e^{4\phi} + E -  \left(\frac{f}{f_{\eta}}\right)^2 \phi_{\eta}^2 -\phi_{\theta}^2$$
This is a quasi-linear equation, however the equation using $\psi$ is also quasi-linear with a $\psi^5$ non-linearity.  While the equation for $\phi$ is somewhat harder to solve than the non-linear equation involving $\psi$, the benefits are accrued throughout the rest of the code by not using exponential variables.
In avoiding the use of $\psi$ in the remainder of the code, one pays the price by having to solve a PDE with quadratic first derivatives and an exponential term.

For $\theta \neq 0$ we multiply by $\sin\theta$ to remove division by small numbers near $\theta=0$ to obtain the coefficients\footnote{Note: using the non-expanded and analytical form of terms like $\partial_{\eta}\left(\frac{f}{f_{\eta}}\right)$ has a noticeable effect on the stability of the solver, as the numerical error in computing $\tanh(\eta)$ for large values of $\eta$ can cause the elliptic solver to be non-convergent otherwise.}:
\begin{eqnarray}
A&=&\left(\frac{f}{f_{\eta}}\right)^2*\sin(\theta) \nonumber \\
B&=&1*\sin(\theta) \nonumber \\
C&=&\frac{f}{f_{\eta}}\left[1+\partial_{\eta}\left(\frac{f}{f_{\eta}}\right)\right]*\sin(\theta) \nonumber \\
D&=&\cos(\theta) \nonumber \\
E&=&-\frac{1}{8}\left[q_{\eta\eta}\left(\frac{f}{f_{\eta}}\right)^2+q_{\eta}\left(\frac{f}{f_{\eta}}\right)\partial_{\eta}\left(\frac{f}{f_{\eta}}\right)+q_{\theta\theta}\right]*\sin(\theta) \nonumber \\
G&=&\frac{1}{4}e^q[f^2(H_b H_d + H_a H_d + H_a H_b)-f_{\eta}^2H_c^2]*\sin(\theta) \nonumber \\
\end{eqnarray}

for $\theta=0$, we use the fact that\footnote{This falls out of l'H\^{o}pital's rule and the fact that $\phi$ is symmetric across $\theta=0$}
$$\lim_{\theta\rightarrow 0} (\cot(\theta)\phi_{\theta}) = \lim_{\theta\rightarrow 0} \left(\frac{\phi_{\theta}}{\tan\theta}\right) = \phi_{\theta\theta}$$
which yields similar coefficients for $\theta=0$ to above:
\begin{eqnarray}
A&=&\left(\frac{f}{f_{\eta}}\right)^2*\sin(\Delta\theta) \nonumber \\
B&=&2*\sin(\Delta\theta) \nonumber \\
C&=&\frac{f}{f_{\eta}}\left[1+\partial_{\eta}\left(\frac{f}{f_{\eta}}\right)\right]*\sin(\Delta\theta) \nonumber \\
D&=&0 \nonumber \\
E&=&-\frac{1}{8}\left[q_{\eta\eta}\left(\frac{f}{f_{\eta}}\right)^2+q_{\eta}\left(\frac{f}{f_{\eta}}\right)\partial_{\eta}\left(\frac{f}{f_{\eta}}\right)+q_{\theta\theta}\right]*\sin(\Delta\theta) \nonumber \\
G&=&\frac{1}{4}e^q[f^2(H_b H_d + H_a H_d + H_a H_b)-f_{\eta}^2H_c^2]*\sin(\Delta\theta) \nonumber \\
\end{eqnarray}
Where we have scaled all terms at $\theta=0$ by $\sin(\Delta\theta)$ to keep the scaling consistent for our matrix problem\footnote{This is desirable as any iterative method which calculates the matrix multiplication $A_{(m\times n)}x_{(n\times 1)}$ multiple times can be dominated by a non-scaled row.}.  To discretise equation (\ref{eqn:hamconphi}) for a solvable matrix problem we use 4th order finite differencing as described in section (\ref{sec:4thordderiv}) and iterative convergence (see below).

\subsubsection{Boundary Conditions for $\phi$}\label{sec:boundcond}
Since we are using a fourth order correct numerical scheme, and previous attempts to compute the Brill evolution were only second order correct, the imposition of boundary conditions must be revisited.  We maintain even symmetry across $\eta=0$, and $\theta=0,\frac{\pi}{2}$ with the addition of extra phantom grid points.  The outer boundary cannot use Robin conditions as they are not fourth order correct, and we instead use separable functions along the outer boundary (see section (\ref{sec:OBcond}) for details).

When we use separable harmonics at the outer boundary, we can determine an upper limit on the falloff that our numerical methods can resolve.  If our lowest order falloff term at the outer boundary goes like $1/r^n$, we know that the $1/r^{n+4}$ term will differ from the dominant term by $4$ powers of $r$, which for a grid size with
$$\eta_{\mathbf{max}}=10 \rightarrow r \sim 10^{4} \rightarrow r^4 \sim 1.5\times 10^{16}$$
With a numerical precision of $\sim 1$ part in $10^{16}$ using $64$ bit double precision reals this will be close to our numerical resolution, and if present would be difficult to extract.  So we can expect at most $5$ terms in the asymptotic expansion to be resolvable (which fits nicely with the Weyl Scalars discussed in equation (\ref{eqn:peelweyl})).

\subsubsection{Iterative Convergence Scheme}\label{subsec:iterconv}

Given the highly non-linear nature of equation \ref{eqn:hamconphi}, we cannot simply calculate our grid coefficients and solve the matrix equation as in section \ref{sec:2orpde4ord}.  We must instead use an iterative convergence scheme, with $\phi(i,j,t_0)=0$ as our starting guess on the initial time slice and $\phi(i,j,t_m)=\phi(i,j,t_{m-1})$ as our starting guess on all future time slices.  We then recursively solve (\ref{eqn:hamconphi}), which can be represented schematically as:
$$\tilde{U}(\phi_{n+1}(i,j,t_m))=\tilde{V}(\phi_{n}(i,j,t_m))$$
where the operator $\tilde{U}$ is the elliptic LHS, the operator $\tilde{V}$ is the non-linear RHS and $n$ represents the looping/iteration counter through our solver subroutine.

We then calculate the maximum relative difference between $\phi_n(i,j)$ and $\phi_{n+1}(i,j)$ over the interior region
\begin{equation}\label{eqn:errmaxphi}
\epsilon_{max}={max_{i,j}}\left|\frac{\phi_{n+1}-\phi_n}{\phi_n}\right| \;;\; 2 \leq i \leq i_{max+2} \;;\; 2 \leq j \leq j_{max} \;;\; \phi_n \neq 0
\end{equation}
and use this as one of our convergence measures.  See tables \ref{tbl:errmaxphit0} and \ref{tbl:errmaxphit1} for an example of the use of these convergence measures.  Note that we stop the convergence attempt once we get an increase in our error measure or it becomes smaller than the numerical precision of the solver (see section \ref{subsubsec:bicgimprove}).

It is possible that we are just in a local minima and need to let the solver go further to find the global minima, however experience dictates that instead we end up in a position where numerical precision causes the solution to oscillate between several very close values.

We can also test for the absolute or ``total'' distance between solutions
\begin{equation}\label{eqn:errtotphi}
\epsilon_{tot}=\sum_{i,j}\left|\phi_{n+1}(i,j)-\phi_n(i,j)\right|
\end{equation}
and find that it provides a similarly good measure (see tables \ref{tbl:errtotphit0} and \ref{tbl:errtotphit1}).

This iterative solver algorithm has had excellent success using both the total and maximum errors together. Requiring that \emph{both} errors fail to decrease any further before proceeding on with the evolution simply means that there is nothing to be gained by further iterations or refinements as we have reached the numerical limit of the algorithm to converge.

\begin{table} \begin{center}
\begin{tabular}{|c|c|c|c|c|c|}\hline
$n$ & $i_m$ & $j_m$ & relative error & $\phi_{n}(i,j)$ & $\phi_{(n-1)}(i,j)$ \\ \hline
2 & 80 & 30 & 9.85241261122664939E-011 & -3.37570447865094278 & -3.37570447831835445 \\
3 & 80 & 30 & 4.39967504423875535E-011 & -3.37570447879946281 & -3.37570447865094278 \\
4 & 80 & 30 & 1.95217537384831400E-010 & -3.37570447814046609 & -3.37570447879946281 \\
\hline \end{tabular}\caption[Relative error while solving for $\phi$ at $t=0$]{$i_m$ and $j_m$ are the grid points where maximal relative error occurs (see equation \ref{eqn:errmaxphi}).  $\phi$ values are $\times 10^{-6}$.  These values are for $t=0$ (IVP).} \label{tbl:errmaxphit0}
\end{center} \end{table}

\begin{table} \begin{center}
\begin{tabular}{|c|c|}\hline
n & total error \\ \hline
2 & 1.93014671882499973E-012 \\
3 & 1.25420811016786391E-012 \\
4 & 4.08221975611801260E-012 \\ \hline
\end{tabular}\caption[Total error while solving for $\phi$ at $t=0$]{Total error calculated using equation \ref{eqn:errtotphi} at each iteration trying to solve for $\phi$.  These values are for $t=0$ (IVP).}
\label{tbl:errtotphit0} \end{center} \end{table}

\begin{table} \begin{center}
\begin{tabular}{|c|c|c|c|c|c|}\hline
$n$ & $i_m$ & $j_m$ & relative error & $\phi_{n}(i,j)$ & $\phi_{(n-i)}(i,j)$ \\ \hline
2 & 102 & 4 & 2.2534925514629114 & -1.45254986884100770 & -4.72586017890264755 \\
3 & 102 & 4 & 8.66404713220494949E-002 & -1.33673455680679596 & -1.45254986884100770 \\
4 & 102 & 4 & 3.07512171191832388E-003 & -1.33263653725698136 & -1.33673455680679596 \\
5 & 102 & 4 & 1.08822818815247440E-004 & -1.33249153177244644 & -1.33263653725698136 \\
6 & 102 & 4 & 3.85050775893508223E-006 & -1.33248640102322062 & -1.33249153177244644 \\
7 & 102 & 4 & 1.36206501834367435E-007 & -1.33248621952993392 & -1.33248640102322062 \\
8 & 102 & 4 & 4.91632639808671202E-009 & -1.33248621297899678 & -1.33248621952993392 \\
9 & 102 & 4 & 3.65289519134709348E-011 & -1.33248621293032245 & -1.33248621297899678 \\
10 & 102 & 4 & 1.07533931769264338E-010 & -1.33248621278703497 & -1.33248621293032245 \\
\hline \end{tabular}\caption[Relative error while solving for $\phi$ at $t=\Delta t$]{$i_m$ and $j_m$ are the grid points where maximal relative error occurs (see equation \ref{eqn:errmaxphi}).  $\phi$ values are $\times 10^{-6}$.  These values are for $t=\Delta t$.}
\label{tbl:errmaxphit1} \end{center} \end{table}

\begin{table} \begin{center}
\begin{tabular}{|c|c|}\hline
n & total error \\ \hline
2 & 2.90920253965643687E-002 \\
3 & 1.02935917801231028E-003 \\
4 & 3.64230199062157343E-005 \\
5 & 1.28880282345998550E-006 \\
6 & 4.56022069727286064E-008 \\
7 & 1.61270623316132493E-009 \\
8 & 5.81934699729203116E-011 \\
9 & 4.50470258866336477E-013 \\
10 & 1.62156363843805017E-012 \\ \hline
\end{tabular}\caption[Total error while solving for $\phi$ at $t=\Delta t$]{Total error calculated using equation \ref{eqn:errtotphi} at each iteration trying to solve for $\phi$.  These values are for $t=\Delta t$.}
\label{tbl:errtotphit1} \end{center} \end{table}

\subsubsection{BiCG Solver Convergence}\label{sec:bicghamcon}
Lastly, we can ask how good the solution to the total problem is, and we employ a similar technique to testing the convergence of the BICG solver in section \ref{subsubsec:bicgimprove} where we look at the error compared to the largest term in the operands, i.e. we calculate
$$\log_{10}\left|\frac{(LHS)_{i,j}-(RHS)_{i,j}}{max_{k=1,9}(S_k)}\right|$$
where
$$S_1 = a4(i,j)*\phi(i,j) \;;\; S_2=ld2(i,j)*\phi(i-2,j) \ldots$$
using the notation of section \ref{sec:2orpde4ord} and
$$(LHS)_{i,j}=\sum_{k=1}^9 S_k$$
This tells us how well the equation is solved (i.e. $LHS-RHS=0$) relative to the largest term, as the best we can hope for in optimal circumstances is (largest term)$\times$(machine precision) as before.

\begin{figure}[h!]  \begin{center}
\includegraphics{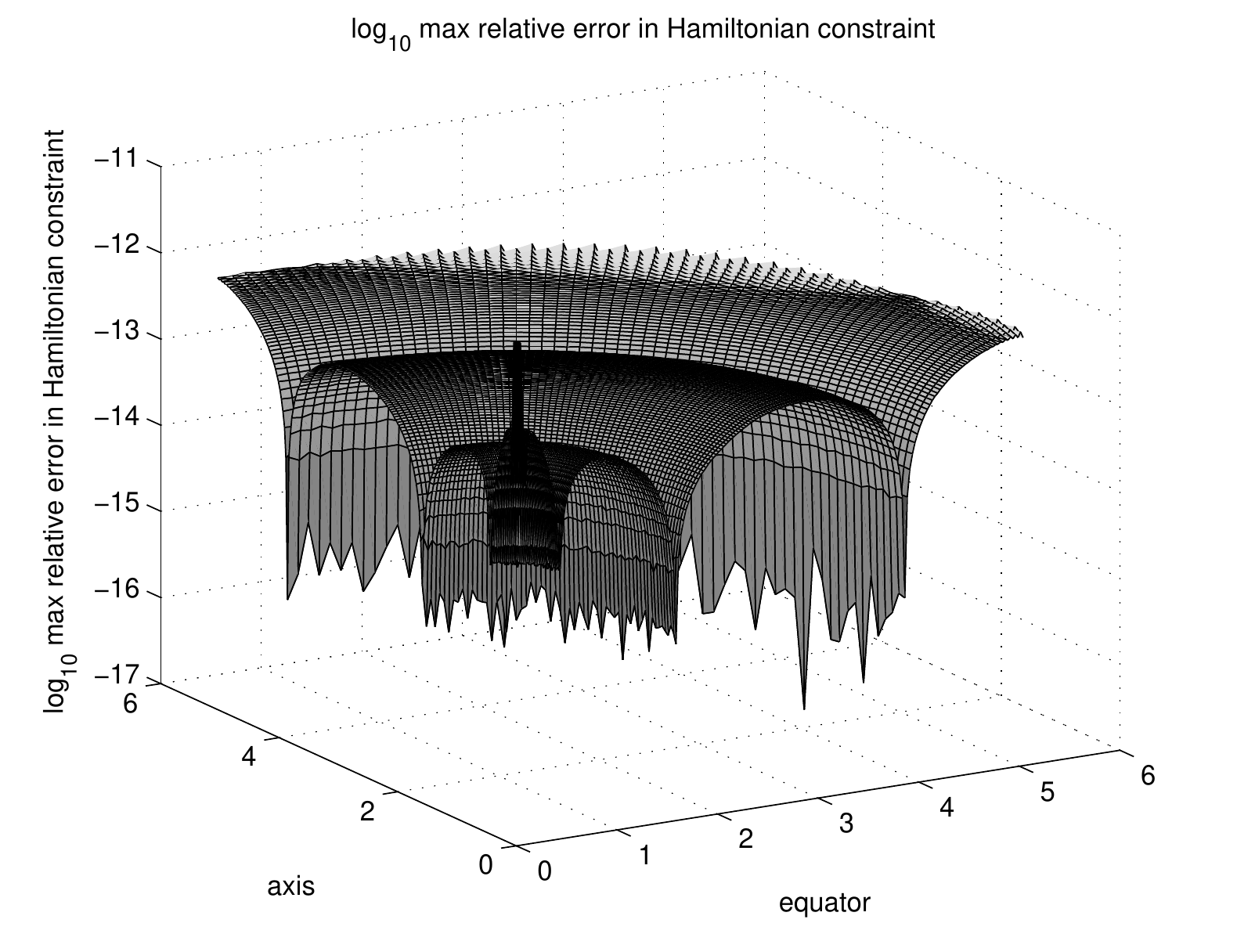}
\caption[$\log_{10}$ Max Relative Error in Hamcon]{$\log_{10}$ of the maximum relative error in the Hamiltonian constraint when solving for $\phi$ as discussed in section \ref{sec:bicghamcon}}\end{center} \label{fig:hamconrelerror} \end{figure}

Figure \ref{fig:hamconrelerror} shows this measure over the entire grid in the late stage of evolution ($k=243$) of a strong wave scenario $(A=-4,s_0=1)$, where one would expect the violations to be largest.  The BiCG solver, however, still performs as expected yielding a solution that is good to one part in $\sim 10^{12}$ despite the presence of two enclosed trapped surfaces and large curvatures.  A similar analysis applies to all the evolutions analysed to date (i.e. the results are consistent across various wave strengths and types of spacetimes), so we can conclude that we are consistently solving the Hamiltonian constraint.

\subsection{(I4) Apparent Horizon Search}\label{sec:hznsearch}
The eventual goal of this code is to look for horizon formation so we can study the dynamics of a spacetime that has a black hole forming in it. In addition we can attempt to characterise the initial value parameter space according to whether or not the initial data contains or does not contain a black hole.

There are several conditions that we can check for
\begin{enumerate}
\item \emph{Look for apparent horizons}.  The presence of an apparent horizon necessitates the presence of an event horizon (and consequently a black hole).  Outermost marginally trapped surfaces\footnote{Which coincide with apparent horizons} are local conditions that are found by locating areas where outgoing orthogonal null geodesics have zero convergence, and these conditions can be tested numerically.

Event horizons are global conditions that represent causal disconnectedness between regions of the spacetime, and can only be located when the entire spacetime solution is known.  They either coincide with or are located outside the apparent horizon.  The two horizons coincide precisely when the spacetime is stationary (i.e. no longer evolving).  As we will most likely never have the complete solution in hand, we prefer to check for local conditions, which also speeds up the numerical search.
\item \emph{Check for negative quasi-local ADM mass}.  Negative quasi-local mass could indicate that the apparent horizon is outside the exterior region of our grid (and our spacetime is thusly not asymptotically flat), so we need to decrease the search parameters to move the horizon into the interior region of our grid.  This criteria will be revisited later and we will find that it is unreliable given the slow rate of radial convergence of the ADM mass.
\end{enumerate}

\subsubsection{Locating Trapped Surfaces and Apparent Horizons}\label{subsubsec:trapsurf}
To search for apparent horizons, we follow the work of Cook\footnote{\cite{cookphd}, Chapter 11} and Bernstein\footnote{\cite{bernstein}, Appendix D, who follows in the footsteps of Cook, Bishop and Cadez (see \cite{bernstein} for references).  For an excellent survey of methodologies across ADM formulations, see Thornburg \cite{Thornburg:AH}.} and we look for the ``trapped'' 2-surface $S$ which is orientable and compact, and whose outward pointing spatial unit normal $s^a$ satisfies
\begin{equation}\label{eqn:trappedsurf}D_a s^a + K_{ab}s^a s^b - trK = 0 \end{equation}
given a 3-hypersurface which has a metric $\gamma_{ab}$ and extrinsic curvature $K_{ab}$ defined on it.  The outermost of all such \textbf{closed} trapped surfaces will form an apparent horizon that guarantees event horizon creation within a finite time whose ``mass'' $\ge$ the ``mass'' of the apparent horizon.

Note that this is a \emph{local} condition at every point on the hypersurface, which leads to a global-in-time result if the trapped surface \emph{encloses} a region of the hypersurface.  It is possible to have trapped surfaces that are not closed. This would allow a light ray to move into a different region of the space-time and escape; the only way to check if a non-closed trapped surface is a global-in-time property would be to have the full time solution and trace light cones that originate ``inside'' such a surface to see if they escape or not\footnote{Or some other algorithm as detailed in Thornburg \cite{Thornburg:AH}.}.

In general $S$ can be parametrised on the axisymmetric hypersurface as
$$S=[\eta(l),\theta(l)],$$
where $l$ is a parametrisation variable, however we follow Cook and Bernstein in assuming that the surface $S$ does not fold over on itself and assume that it can instead be parametrised in the less general format
\begin{equation}\label{eqn:sparam}S=[h(\theta),\theta]\end{equation}
where $h(\theta)$ is a single-valued function of the angle $\theta$.
Given previous results that show a tendency of the horizons to be perturbations on spherical geometry, and the fact that a bending back of the horizon to give multi-valued $\eta$ values would make it extremely difficult to satisfy equation (\ref{eqn:trappedsurf}) we feel this is a safe assumption.  Further, as we are finding trapped surface topology instead of \emph{only} closed trapped surfaces, we should see an indication of a trapped surface that is doubling back on itself.  We have not seen any such surfaces to date (except in extreme circumstances), which indicates that our assumptions are mostly valid.

We do not, however, assume like Bernstein that there is a ``complete'' horizon present in the spacetime on any slice, i.e. one that starts at $\theta=0$ and goes all the way to $\theta=\frac{\pi}{2}$.  We do not make this assumption for several reasons, including:
\begin{enumerate}
\item The fact that our lapse function $\alpha$ is not $1$ everywhere means that we cannot expect the evolution to be at the same proper time everywhere on each spatial hypersurface, so horizon formation may occur at different coordinate times.
\item Partial horizon creation is possible due to the slicing conditions imposed on the spacetime.
\item The topology of partially formed trapped surfaces \emph{in and of itself} is an interesting aspect of the spacetime for study which can give us clues to horizon formation, being inside a black hole, etc.  If we were to limit ourselves to only closed trapped surfaces we would be missing out on some interesting physics.
\end{enumerate}

Using equation (\ref{eqn:sparam}), the tangent to our surface $S$ is given by
$$t^a=(h_\theta,1,0)$$
and we assume a unit normal of the form
$$s^a=(s^{(1)},s^{(2)},s^{(3)}) \; ; \; s_a=\gamma_{ab}s^b$$
Writing our metric (\ref{eqn:3dmetric}) in the form\footnote{Using the metric formulation of Bernstein, which translates to $\psi=e^\phi$, $A=e^q f_\eta^2$, $B=e^q f^2$ and $D=f^2$ in this thesis's conventions.}
\begin{eqnarray}
\gamma_{ab} & = & \psi^4 \left[\begin{array}{ccc}
 A & 0 & 0 \\
 0 & B & 0 \\
 0 & 0 & D \sin^2\theta \end{array} \right] \nonumber
\end{eqnarray}
and using the coordinate symmetry condition $\partial_{\varphi}=0$ from axisymmetry (i.e. all functions are independent of $\varphi$) we find
$$s^a=(s^{(1)},s^{(2)},0) \; ; \; s_a=\gamma_{ab}s^b=(A \psi^4 k \;, B \psi^4 l \;, 0)$$
Now using the conditions that the normal is perpendicular to the tangent vector, i.e.
$$t^a s_a=0$$
and that the normal has unit length
$$s^a s_a = 1$$
we find that
$$s^a=\frac{1}{\psi^2 \sqrt{A} \sqrt{h_\theta^2\left(\frac{A}{B}\right)+1}}\left[1,-h_\theta\left(\frac{A}{B}\right),0\right]$$
which in our metric formulation is:
\begin{equation}\label{eqn:savector}
s^a = \frac{1}{f_\eta e^{\left(\frac{q}{2}+2\phi\right)}\sqrt{h_\theta^2\left(\frac{f_\eta}{f}\right)^2+1}} \left[1,-h_\theta\left(\frac{f_\eta}{f}\right)^2 ,0\right] \end{equation}
Plugging equation (\ref{eqn:savector}) into equation (\ref{eqn:trappedsurf}) yields the equation we need to solve for our radial function, $h(\theta)$:
\begin{eqnarray}\label{eqn:trapsurffull}
h_{\theta\theta} + \left[\cot\theta+\frac{q_\theta}{2}+4\phi_\theta\right] h_\theta & & \nonumber \\ 
- \left[\frac{q_\eta}{2} + 4\phi_\eta + \left(\frac{f_\eta}{f}\right)\left(2+\partial_\eta\left(\frac{f}{f_\eta}\right)\right)\right] h_\theta^2  & &  \nonumber \\
+ \left(\frac{f_\eta}{f}\right)^2\left[\cot\theta + \frac{q_\theta}{2}+4\phi_\theta\right] h_\theta^3 & & \nonumber \\
+ \frac{f^2e^{2\phi-q}}{f_\eta^4}[H_d+H_b+H_a] \delta^3 & & \nonumber \\
+ e^{2\phi} \left[-H_b h_\theta^2+2H_ch_\theta-\left(\frac{f}{f_\eta}\right)^2H_a\right] \delta & & \nonumber \\
-\left(\frac{f}{f_\eta}\right)^2\left[\frac{q_\eta}{2}+4\phi_\eta\right]-2\left(\frac{f}{f_\eta}\right) & = & 0
\end{eqnarray}
where we have defined
$$\delta=\delta(h(\theta))=f_\eta e^{\frac{q}{2}}\sqrt{\left(\frac{f_\eta}{f}\right)^2h_\theta^2+1}$$
This is a quasi-linear equation in the form
\begin{equation}\label{eqn:apphorschematic}A \;h_{\theta\theta} + B \;h_{\theta} + C \;\left(h_{\theta}\right)^2 + D \;\left(h_{\theta}\right)^3 + E \;\delta^3 + F \;\delta + G = 0\end{equation}
and the lowest order behaviour of $\delta \sim h_\theta$, so it is a non-negligible portion of the solution.

We then take second order correct finite difference approximations to the derivatives appearing in equation (\ref{eqn:apphorschematic}) using equations (\ref{eqn:2ordderiv1}) and (\ref{eqn:2ordderiv2}) to find:
\begin{eqnarray}\label{eqn:apphorcubic}
\left(\frac{-D}{8 (\Delta\theta)^3}\right)h_{j-1}^3 & & \nonumber \\
+ \left(\frac{3Dh_{j+1}}{8(\Delta\theta)^3}+\frac{C}{4(\Delta\theta)^2}\right)h_{j-1}^2 & & \nonumber \\
+ \left(\frac{A}{(\Delta\theta)^2}-\frac{B}{2\Delta\theta}-\frac{2h_{j+1}C}{4(\Delta\theta)^2}-\frac{3 h_{j+1}^2D}{8(\Delta\theta)^3}\right)h_{j-1}& & \nonumber \\
+ \left(\frac{A \left(h_{j+1}-2h_j\right)}{(\Delta\theta)^2} + \frac{B h_{j+1}}{2\Delta\theta} + \frac{C h_{j+1}^2}{4(\Delta\theta)^2} + \frac{D h_{j+1}^3}{8(\Delta\theta)^3}+E \;\delta^3(h_{j-1}) + F  \;\delta(h_{j-1}) + G \right) & = & 0 \nonumber \\
\end{eqnarray}
where we have written in the explicit dependence $\delta=\delta(h_{j-1})$ as a reminder that the ``constant'' in the cubic equation (\ref{eqn:apphorcubic}) for $h_{j-1}$ is actually dependent on the variable $h_{j-1}$ and will require an iterative convergence technique to solve fully\footnote{In general this converges very quickly, within 2-4 iterations.}.

We then start at each radial grid point on the equator ($j=jmax$), knowing that $h_\theta=0$ there due to symmetry, and calculate the ``next'' value $h_{jmax-1}(\eta)$ at $j=jmax-1$ using the above equation\footnote{We are hiding some details here, like the need to interpolate the values of all the quantities in equation (\ref{eqn:trapsurffull}) to the actual radial point $h_j(\eta)$, which will generally not fall right on a grid point to give exact numerical values.} and the algorithm described in Section \ref{subsec:cubicroots}.  We choose the equator to start our search at because it advances through time the fastest due to our choice of $\alpha$, and has the largest portion of the initial wave profile from $q$, so horizons are most likely to start forming there.

It is possible, however, to miss trapped surfaces because we only start on the equator and miss any partially formed surfaces that do not touch the equator (or sit in between grid points).

Others, like Bernstein, view this as a tri-diagonal matrix problem to solve across the entire region assuming it touches both the axis and equator (and are therefore \emph{only} looking for apparent horizons), however for the reasons stated above we instead treat this as a ``shooting'' problem.  Further, we get a glimpse into the complicated trapped surface topology and its evolution in this manner, which yields some interesting physical insights.

Returning to the trapped surface algorithm, we then continue to iterate through angular points towards the axis until one of several conditions is met and we stop searching at that radial value:
\begin{enumerate}
\item no suitable value for $h_{j-1}(\eta)$ is found (i.e. all computed roots are outside the computational domain, i.e. greater than $\eta_{\mathbf{max}}$ or less than $0$)
\item $|h_{j-1}(\eta)-h_j(\eta)| > h_{tol}$, where $h_{tol}=2\Delta\eta$ based on experience with the solver jumping around wildly from one trapped surface to another.  In other words, the surface must change smoothly from one angular point to another.
\item We have hit the axis and have a complete closed apparent horizon
\item The non-linear solver was unable to converge within $200$ iterations (never observed)
\end{enumerate}

This then provides a snapshot of the trapped surface structure within that slice of spacetime that has been found by starting at the equator and iterating angularly.

\subsubsection{Trapped Surface Interpretations}\label{subsubsec:trapsurfinterpr}
\begin{figure} \centering
\includegraphics{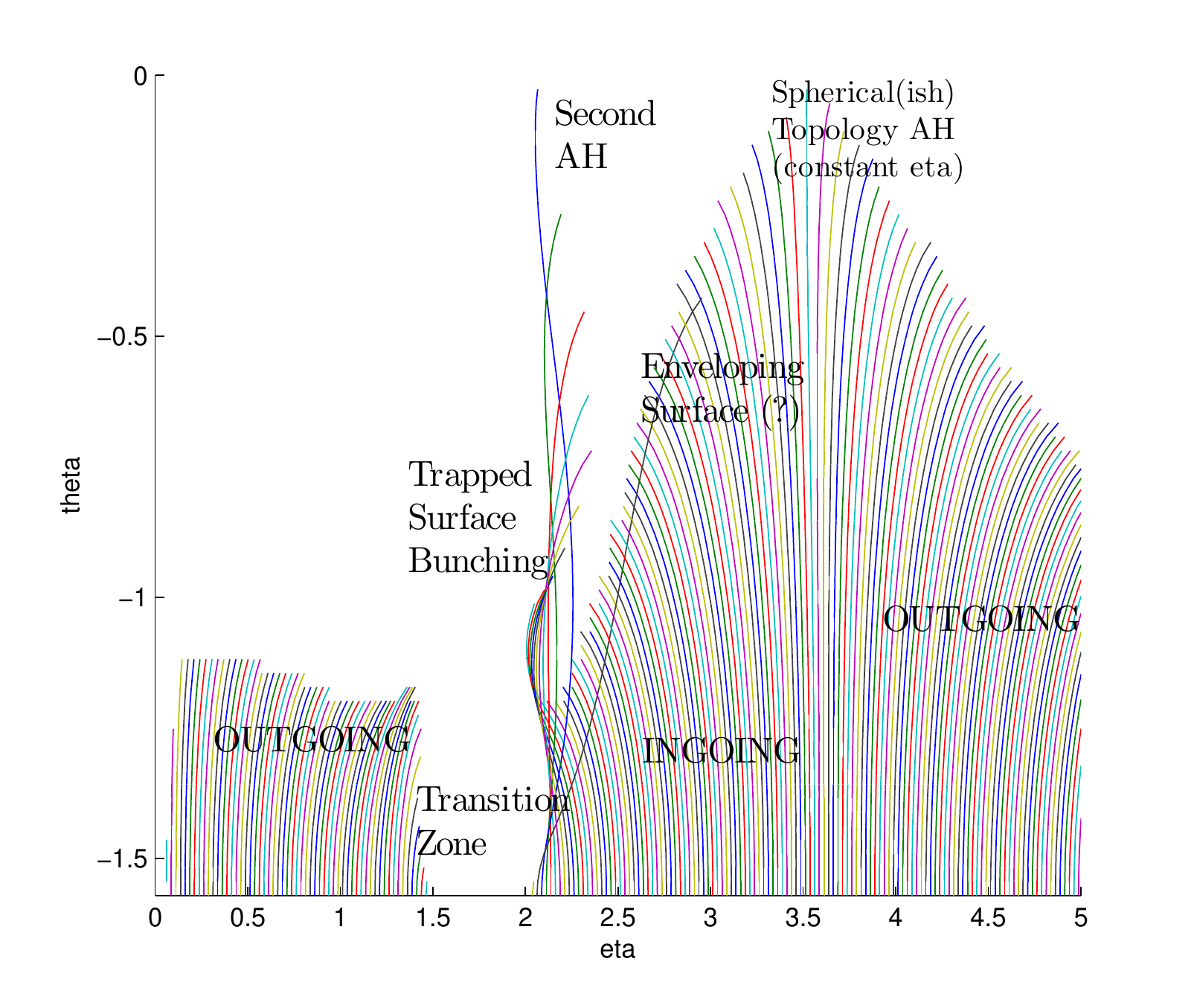}
\caption[Trapped Surfaces for $A=9$]{Trapped Surface structure at $k=36$ for a strong initial wave ($A=9$,$s_0=1$) as discussed in section \ref{subsubsec:trapsurfinterpr}.}
\label{fig:trapsurft0a9annotated}
\end{figure}

\begin{figure} \centering
\includegraphics{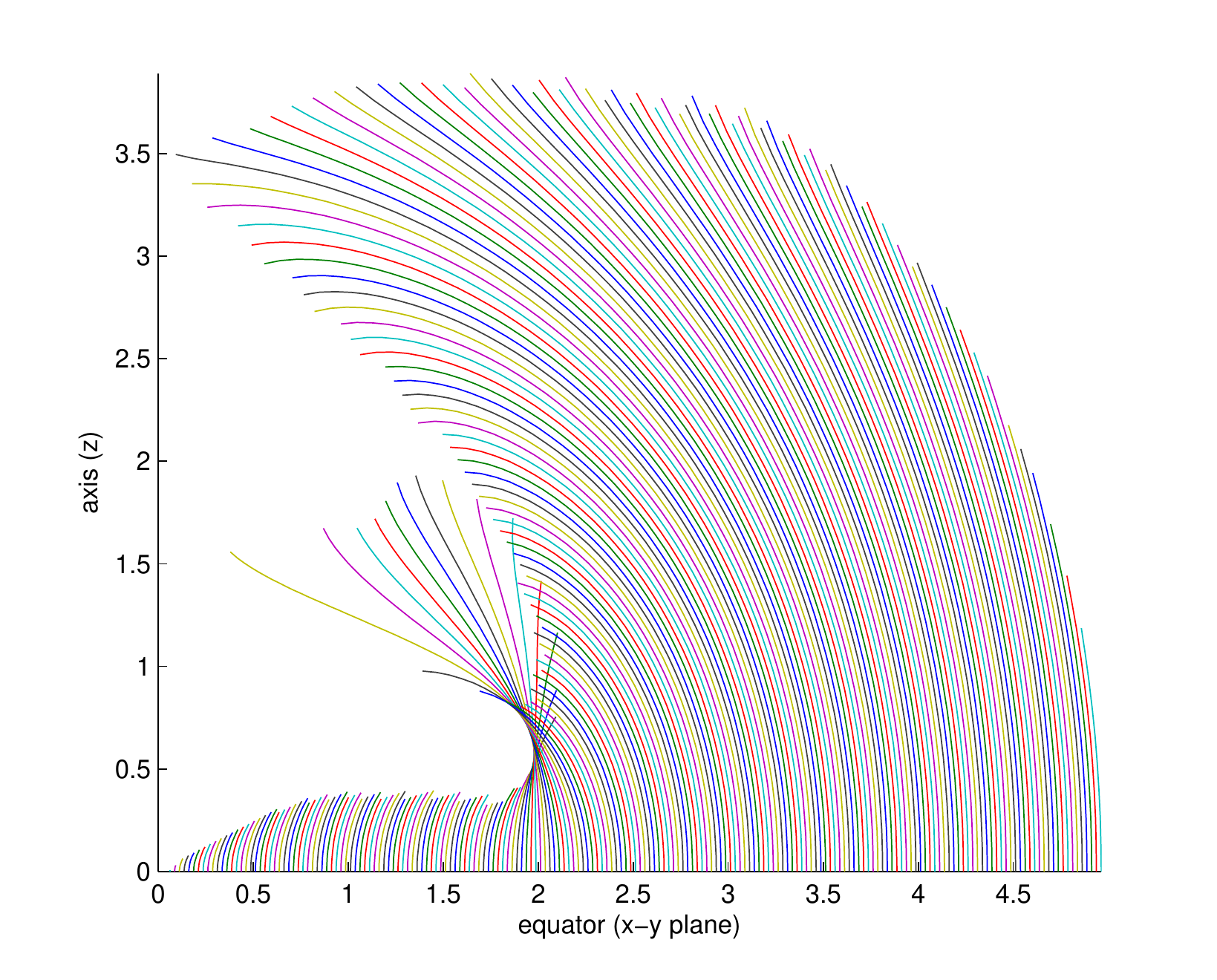}
\caption[Initial Trapped Surfaces for $(A=9,s_0=1)$]{Initial trapped surfaces for a $200 \times 60$ grid with initial wave amplitude $(A=9,s_0=1)$, calculated using the algorithm described in section \ref{subsubsec:trapsurf}}
\label{fig:trapsurft0a9}
\end{figure}

One mathematical note regarding the trapped surface equations; if we consider the Initial Value Problem with $H_i=0$, and flat space so that $q=\phi=0$, then we find along the axis ($h_\theta=0$) that
$$h_{\theta\theta}=2\left(\frac{f}{f_\eta}\right)$$
Which implies that
\begin{eqnarray}
h_\theta&=&2\left(\frac{f}{f_\eta}\right)\theta+C \rightarrow C=0 \; @ \; \theta=0 \nonumber \\ \mbox{}
h&=&\left(\frac{f}{f_\eta}\right)\theta^2+D \rightarrow D=\eta_0 \nonumber \\ \mbox{}
h&=&\left(\frac{f}{f_\eta}\right)\theta^2+\eta_0 \nonumber
\end{eqnarray}
(recalling that $h=h(\theta)$) and a similar argument applies to the equator.  i.e. in flat space we would expect to see ``trapped'' surfaces that curve away from the origin as we leave the equator.  We do indeed see this in the case of ``small'' amplitude IVP waves, with the surfaces curving towards the outer edge of the grid and eventually leaving the computational domain.  In stronger wave regions, however, we see the surfaces curving along constant $\eta$ or towards the origin, so there must be a transition zone where we see curves intersecting (in figures \ref{fig:trapsurft0a9annotated} and \ref{fig:trapsurft0a9} this occurs in the interior of the black hole/apparent horizon).

More generally, equation (\ref{eqn:trappedsurf}) becomes
$$\partial_\eta s^1 = -\partial_\theta s^2$$
in flat space\footnote{Where the 3-spatial metric $\gamma_{ij}=\mathrm{diag}(1,r^2,r^2\sin^2\theta)$, the Christoffel symbols vanish and covariant derivatives become partial derivatives.} and on a flat embedding hypersurface ($K_{ij}=0$).  Starting on the equator and requiring symmetry we see that the outgoing normal must be purely radial at this point, i.e. $s^2=0$.

This is a purely geometric requirement that is satisfied by having outward bending surfaces so that flat space should have ``trapped surfaces'' that start on the equator and are ``outgoing'', as we see in the asymptotic region of our trapped surface topologies.  Then as seen in figure \ref{fig:trapsurft0a9annotated} the surfaces ``tip up'' and we get approximately spherical trapped surfaces in $\eta$ coordinates (which is not necessarily the case in radial proper distances), and a transition to ``ingoing'' trapped surfaces (i.e. near $\eta=3.6$).

Then as the surfaces reach another transition zone where they go from ``ingoing'' to ``outgoing'' again ($\eta=2 \rightarrow 1.5$) we see the surfaces bumping into the large curvature region and bunching up/converging towards each other around $\eta=2$, with another closed trapped surface formed.  This multiple horizon structure has been seen in other Brill IVP analysis, however the trapped surface topology gives some context around which to frame what is happening inside the apparent horizon (i.e. the outermost trapped surface/black hole).

The trapped surface bunching and transition zone analysis also help us understand what is happening in the weak wave cases where we cannot evolve to have a fully closed trapped surface (i.e. apparent horizon) form, but we do see this same structure forming in the trapped surfaces - i.e. it is indicative of the interior region of a black hole forming.  We also see the trapped surfaces ``tipping up'' in the weak wave cases, to form a more spherical topology in $\eta$ coordinates, and presumably the surfaces would eventually move to a ``ingoing'' topology at some point in the hypersurface with a full apparent horizon forming.

\begin{figure} \centering
\includegraphics{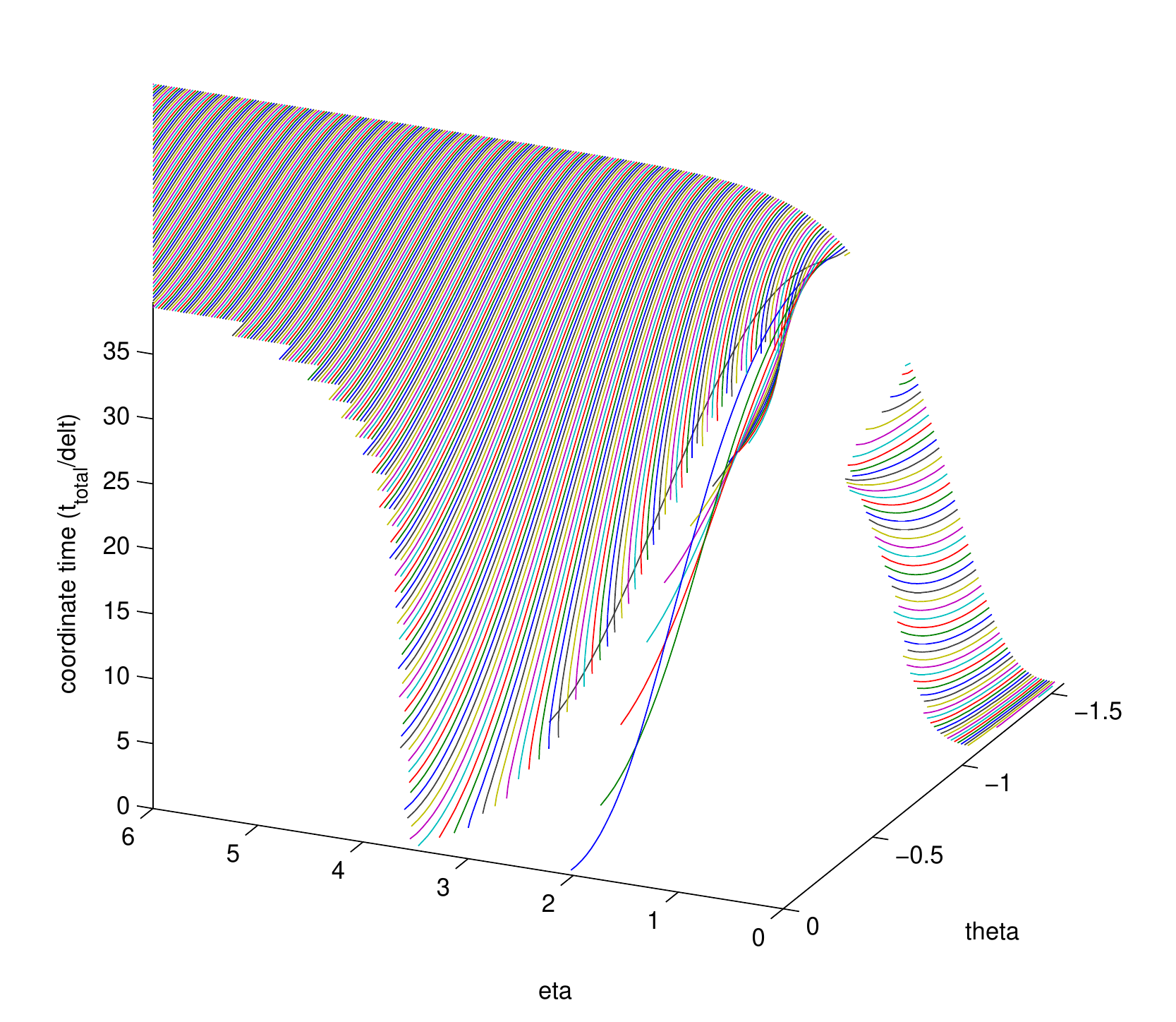}
\caption[Evolved Trapped Surfaces for $A=9$]{Trapped surfaces at $k=36$ for a $400 \times 60$ grid ($\eta_{\mathbf{max}}=10$) with a strong initial wave ($A=9$,$s_0=1$), calculated using the algorithm described in section \ref{subsubsec:trapsurf}.  Note the presence of a second apparent horizon in the interior region near $\eta=2$, where the surfaces change from ``ingoing'' to ``outgoing''.  Vertical axis is $t_k\times \alpha(\eta,\theta)$ at each point to give a perspective of evolved time at each point}
\label{fig:trapsurft0a9nmax10}
\end{figure}

See figure \ref{fig:trapsurft0a9} for an example of the trapped surfaces that we find with a \emph{large} amplitude initial wave ($A=9$) where we would expect an apparent horizon to be present (or almost fully present) on the initial slice.  Figure \ref{fig:trapsurft0a9nmax10} shows what the evolved trapped surface topology looks like just before the code reaches a singularity and halts for the same initial conditions\footnote{In $(\eta,\theta,t)$ coordinates, with the equator in the background and axis in the foreground.  This perspective helps to visualise the progression of the evolution and why for weaker initial waves a \textbf{fully} enclosed trapped surface (i.e. apparent horizon) may not form.}.

I suspect this sort of analysis would be useful in binary black hole collisions, etc. where trapped surface topology could be used to analyse/direct horizon topology efforts (similar to one of the methodologies mentioned in Thornburg \cite{Thornburg:AH}).

Returning to the larger code schematic, if we find a black hole (i.e. an apparent horizon) in the spacetime, we can return to the part of the code that does the parameter search and adjust our critical parameters accordingly.  For example, we could decrease the amplitude of the initial Brill Wave in equation (\ref{eqn:qinit}) and re-run the evolution.

If instead we evolve for a sufficiently long time without the formation of apparent horizons\footnote{Determining what a ``sufficiently long time'' is may prove to be a difficult task.} then we can go back into the parameter searching portion of the code and, for example, increase the amplitude in equation (\ref{eqn:qinit}).  In this manner we can perform a phase space search for our critical parameters and, in theory, tune them to arbitrary precision to study near-critical solutions.  We have \textbf{not}, however, seen any situations in which the macro-indicators of a black hole are not present\footnote{Considering what our slicing conditions permit.}.  i.e. we seem to be able to form black holes with all initial conditions.  While in the large amplitude cases the proper radius\footnote{$R(\theta)=\int_0^{h(\theta)}\sqrt{g_{11}}d\eta$} of the bunched trapped surface region may be on the order of $200+$, in the small wave cases the proper radius is on the order of $1-2$, giving a wide range in observed physical phenomena; small black holes for small waves and large black holes for large waves.

\subsection{Mass Measures and Calculations}\label{subsec:massmeasure}
At this point in the development of the code, we take a somewhat simplistic approach to the calculation of mass measures.  Bernstein \cite{bernstein} has a detailed discussion of many different mass parameters and their use, and Masterson \cite{mastersonmsc} discusses some of the results previously obtained on Brill waves.

We define the (quasi-local) ADM mass\footnote{The ADM mass is only properly defined at spatial infinity, i.e. in the limit where $r\rightarrow\infty$, hence this is a quasi-local measure.} at the outer edge of our computational grid $\eta=\eta_{\mathbf{max}}$ via
\begin{eqnarray}\label{eqn:admmass}
m_{ADM} & = & -\frac{1}{2\pi} \int^{2\pi}_0 \int^\pi_0 r_m^2 \frac{\partial \psi}{\partial r} \sin\theta d\theta d\phi \nonumber \\ \mbox{}
& = & -r_m^2 \int^\pi_0 \frac{\partial \psi}{\partial r} \sin\theta d\theta \nonumber \\ \mbox{}
& = & -\frac{f^2}{f_\eta} \int^\pi_0 \psi_\eta \sin\theta d\theta \nonumber \\ \mbox{}
& = & -\frac{f^2}{f_\eta} \int^\pi_0 \phi_\eta(\eta_{\mathbf{max}},\theta) e^{\phi(\eta_{\mathbf{max}},\theta)} \sin\theta d\theta
\end{eqnarray}
Where $r=f(\eta)$, $r_m=r(\eta_{\mathbf{max}})$ and $f,f_\eta$ are evaluated at $\eta=\eta_{\mathbf{max}}$.  It is important to note that this result assumes that the metric is asymptotically conformally \emph{flat}, or at least has the asymptotic form of Schwarzschild space plus perturbations that fall off faster \cite{alcubierre:3p1num} than $\frac{1}{r}$.  In terms of our variables this means that
\begin{eqnarray}\gamma_{ab} & \sim & e^{4\phi}\left[\begin{array}{ccc} f_\eta^2 & 0 & 0 \\
0 & f^2  & 0 \\
0 & 0 & f^2 \sin^2\theta
\end{array}\right]\end{eqnarray}
which implies that perturbations away from flat space look like (comparing to equation (\ref{eqn:3dmetric}))
$$e^q \sim 1+q+\ldots$$
and therefore $q$ must fall off faster than $\frac{1}{r}$ at the edge of the grid to ensure that the mass information is only contained in the conformal factor $\phi$.  While our initial value formulation and the Brill criteria are consistent with this, we are not guaranteed that this will be \emph{numerically} true, nor that it is true \emph{for all time}.

We also note that if $\psi\sim\frac{1}{r^2}$ or faster at the outer edge of the grid, the ADM mass will be 0.

The mass aspect $\tilde{M}$ is found by equating the metric to a Schwarzschild metric at the outer boundary, and will only be valid in the case of spherical symmetry or as we approach radial infinity.  We have a wave present in the entire spacetime with non-zero $\theta$ dependence, therefore we do not expect a uniform measure across the entire outer boundary.  But it can be used as a check in low amplitude cases or as a measure of deviation from spherical symmetry (see also section \ref{sec:spherpolob}).

\subsection{Boundary Conditions for all dynamic variables}

\begin{table} \begin{center}
\begin{tabular}{|c|c|c|c|c|} \hline
\multicolumn{5}{|c|}{Variables and their symmetry properties} \\
\multicolumn{5}{|c|}{across boundaries in axisymmetry} \\ \hline
variable  & $\eta=0$ & $\theta=0$ & $\theta=\frac{\pi}{2}$ & $\eta=\eta_{\mathbf{max}}$ \\ \hline
$q$ & SP(2) & S & S & RDFSV \\
$\phi$ & SP(0) & S & S & RDFSV \\
$H_a$ & SP(0) & S & S & RDFSV \\
$H_c$ & ASP(3) & AS & AS & RDFSV \\
$H_b$ & SP(0) & S & S & RDFSV \\
$H_d$ & SP(0) & S & S & RDFSV \\
$\alpha$ & SP(2) & S & S & RDFSV \\
$v_1$ & AS & S & S & RDFSV \\
$v_2$ & S & AS & AS & RDFSV \\
$\chi$ & SP(0) & S & S & RDFSV \\
$\Phi$ & SP(0) & AS & AS & RDFSV \\ \hline
\end{tabular}
\end{center}
\caption[Dynamic Variables and their Boundary Conditions]{Dynamic variables and their boundary conditions at the four boundaries of the finite computational region given in figure \ref{fig:2dgrid}.  S = Symmetric, AS = Anti-symmetric, SP(N) = Symmetric Polynomial of order $N$, ASP(N) = Anti-Symmetric Polynomial of order $N$, RDFSV = Radial Dynamic Falloff via Separation of Variables (see section \ref{subsec:sepspherharm})} \label{tbl:boundvar}
\end{table}

In order to properly define the numerical problem for all of our variables in a finite region on a discretised grid, we must provide appropriate boundary conditions at all of the boundaries given in figure \ref{fig:2dgrid}.  This allows calculation of derivatives on or near the boundaries as discussed in section \ref{subsec:bc_gen}.  The conditions are given in table \ref{tbl:boundvar}, and let us now discuss these boundary conditions in some detail.  In order to preserve axisymmetry we require that under the coordinate transformation $\theta \rightarrow -\theta$ our invariant $ds^2$ in equation (\ref{eqn:3metricfinal}) is preserved, i.e.
$$\left.\frac{\partial \gamma_{ij}}{\partial \theta}\right|_{\theta=0}=0 \;;\; i=j$$
and similarly to preserve equatorial plane symmetry we require:
$$\left.\frac{\partial \gamma_{ij}}{\partial \theta}\right|_{\theta=\frac{\pi}{2}}=0 \;;\; i=j$$
these conditions both immediately lead to the identification that $q$ and $\phi$ must satisfy the same conditions.

To similarly preserve $ds^2$ under the transformation $\eta \rightarrow -\eta$ we see that
$$\left.\frac{\partial \gamma_{ij}}{\partial \eta}\right|_{\eta=0}=0 \;;\; i=j$$
so $q$ and $\phi$ must also be symmetric at $\eta=0$.  Following the logic of section \ref{sec:r0reg} we see that $q \sim r^2$ near the origin to lowest order, and from the above it must be symmetric, so we define a local polynomial fit near the origin for $q$ as
$$q(\eta,\theta)=a_2(\theta) f^2 + a_4(\theta) f^4 + a_6(\theta) f^6 \;;\; f \rightarrow 0$$
We choose to expand to three terms as we have a $9$-point stencil that is offset at the origin, so putting our central stencil point at $\left(i=2;\eta=\frac{\Delta\eta}{2}\right)$ allows us to use the three points $i=2,3,4$ to solve for the coefficients $a_2,a_4,a_6$ and then fit the values of a variable for the phantom grid points $$\eta=-\frac{\Delta\eta}{2},-\frac{3\Delta\eta}{2} \;;\; i=1,0$$
In general, we can define a symmetric polynomial fit for a function around $f=0 \;(r=0)$ as
$$SP(N)=a_N(\theta)f^N + a_{N+2}(\theta)f^{N+2} + a_{N+4}(\theta)f^{N+4} \;;\; f \rightarrow 0$$ where $N$ is even.  An antisymmetric polynomial fitting function has the same form however N must be odd.

We have defined the ``inner'' 3 boundary conditions for $q$ and $\phi$ (axis, equator and origin), so let us now examine the inner boundary conditions for our other dynamic variables.

The inner boundary conditions for the lapse, $\alpha$ can be determined as follows: knowing that $q$ is symmetric at the 3 inner boundaries, and analysing the evolution equation for $q$ (\ref{eqn:qdot}), we see that $\alpha$ must match the symmetry sign of $H_a$ and $H_b$.  We do not in \emph{general} want the evolution to stop at the origin (anti-symmetry at $\eta=0 \rightarrow \alpha|_{\eta=0} = 0$), so we choose the symmetry sign of $\alpha$ to be positive.  Further, from the extrinsic curvature evolution equations (\ref{eqn:mixkevolvacuum}) we see that $\phi$ and $\alpha$ must have matching symmetry signs, also implying that 
$$\left.\frac{\partial \alpha}{\partial \theta}\right|_{\theta=0}=0 \;;\; \left.\frac{\partial \alpha}{\partial \theta}\right|_{\theta=\frac{\pi}{2}}=0 \;;\; \left.\frac{\partial \alpha}{\partial \eta}\right|_{\eta=0}=0$$
Then following the logic of section \ref{subsec:alphareg} we can define a symmetric polynomial fit of order $2$ near the origin for $\alpha$ if we are not using a static lapse.

Using the fact that $\alpha$, $H_a$ and $H_b$ must have the same symmetry signs, we find that
$$\left.\frac{\partial H_a}{\partial \theta}\right|_{\theta=0}=0 \;;\; \left.\frac{\partial H_a}{\partial \theta}\right|_{\theta=\frac{\pi}{2}}=0 \;;\; \left.\frac{\partial H_a}{\partial \eta}\right|_{\eta=0}=0$$
$$\left.\frac{\partial H_b}{\partial \theta}\right|_{\theta=0}=0 \;;\; \left.\frac{\partial H_b}{\partial \theta}\right|_{\theta=\frac{\pi}{2}}=0 \;;\; \left.\frac{\partial H_b}{\partial \eta}\right|_{\eta=0}=0$$

The maximal slicing condition (\ref{eqn:hahbhdcon}) implies that $H_d$ must have the same symmetry sign as $H_a$ and $H_b$, i.e.
$$\left.\frac{\partial H_d}{\partial \theta}\right|_{\theta=0}=0 \;;\; \left.\frac{\partial H_d}{\partial \theta}\right|_{\theta=\frac{\pi}{2}}=0 \;;\; \left.\frac{\partial H_d}{\partial \eta}\right|_{\eta=0}=0$$
or it falls out of the evolution equation for $H_d$ if we are not employing maximal slicing.

$H_c$ requires somewhat stricter boundary conditions near the origin as presented in equation (\ref{eqn:hcorigreg}), namely that it is antisymmetric and has lowest order $f^3$ behaviour, so we use an antisymmetric polynomial fit of order $3$ there.  Along the axis and equator $H_c$ must be antisymmetric, i.e.
$$\left. H_c \right|_{\theta=0}=0 \;;\; \left. H_c \right|_{\theta=\frac{\pi}{2}}=0$$
which can be imposed during the evolution to aid in regularity\footnote{\emph{Not} explicitly setting $H_c$ to zero along the axis and equator does cause numerical noise to propagate noticeably through the code.}.

The symmetry requirements for the shift vectors can be derived from the metric evolution equation for $q$ (\ref{eqn:qdot}) and the definition of $f$ (\ref{eqn:fdefn}), and it follows that\footnote{Use of the shift vector potentials implies that $v_2$ won't have $0$ order (constant) terms present at the origin and it will be lowest order $f^2$ to preserve conditions on $q$.  We may need to use SP(2) conditions at some point, however.}
$$\left.\frac{\partial v_1}{\partial \theta}\right|_{\theta=0}=0 \;;\; \left.\frac{\partial v_1}{\partial \theta}\right|_{\theta=\frac{\pi}{2}}=0 \;;\; \left.v_1\right|_{\eta=0}=0$$
and
$$\left.v_2\right|_{\theta=0}=0 \;;\; \left.v_2\right|_{\theta=\frac{\pi}{2}}=0 \;;\; \left.\frac{\partial v_2}{\partial \eta}\right|_{\eta=0}=0$$
These equations also imply that (1) coordinate values are not shifted radially at the origin (as $v_1=0 \; @ \; \eta=0$) and (2) that coordinates are not shifted angularly at the axis and equator (as $v_2=0 \;@ \;\theta=\left\{0,\frac{\pi}{2}\right\}$)

The boundary conditions for $\Phi$ and $\chi$ fall directly out of their definitions (\ref{eqn:shiftvpotdef}), and are
$$\left.\frac{\partial \chi}{\partial \theta}\right|_{\theta=0}=0 \;;\; \left.\frac{\partial \chi}{\partial \theta}\right|_{\theta=\frac{\pi}{2}}=0 \;;\; \left.\frac{\partial \chi}{\partial \eta}\right|_{\eta=0}=0$$
$$\left.\Phi\right|_{\theta=0}=0 \;;\; \left.\Phi\right|_{\theta=\frac{\pi}{2}}=0 \;;\; \left.\frac{\partial \Phi}{\partial \eta}\right|_{\eta=0}=0$$

Our outer boundary conditions follow the separation of variables method using dynamic radial fall-off functions discussed in section (\ref{subsec:sepspherharm}).

\subsubsection{Implementing 4th order boundary conditions}\label{sec:do4ordbound}
Table \ref{tbl:boundvar} has a listing of the various symmetry conditions for our dynamic variables.  To implement these boundary conditions in a numerical code we must consider two scenarios:

\begin{enumerate}
\item Problems where we do not know the values in the interior region \emph{a priori} and must merge the boundary conditions into the matrix problem to be solved.  e.g. Elliptic equations that are to be solved via a matrix solver, which involves creating phantom grid points and mapping the variable values into the interior region somehow - see section \ref{subsec:bc_gen}.
\item Calculations where we do know the values of the function in the interior region of our computational grid.  e.g. Hyperbolic evolution equations, or calculating the value of a variable at phantom grid points in the exterior region \emph{after} having solved an elliptic matrix equation.
\end{enumerate}

In the first scenario, we do not know the value of the variable at each grid point, but generally must merge the exterior region conditions into the interior region to have a complete problem.

\begin{figure} \centering
%\psfrag{P}{$\pi$}
\includegraphics{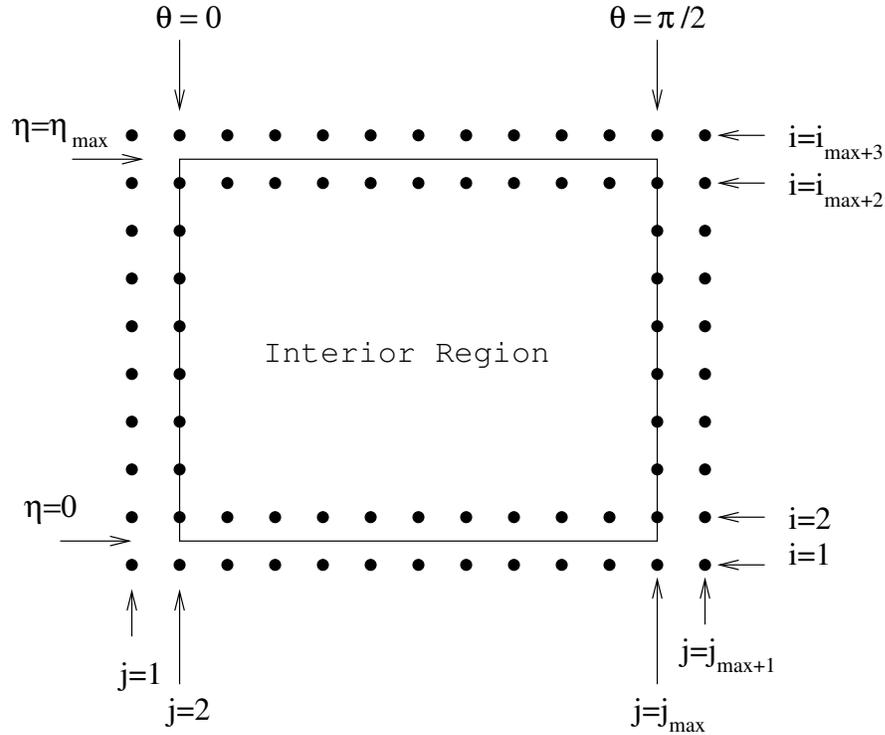}
\caption[Numerical Grid]{Schematic of the grid positioning in $\eta$ and $\theta$.  The square box represents the physical boundary of the coordinate region $\eta=\left[0 ,\eta_{\mathbf{max}}\right] , \theta=\left[0 ,\frac{\pi}{2}\right]$}\label{fig:2dgrid_reprint}
\end{figure}

Recall that our grid is configured as in figure \ref{fig:2dgrid_reprint} and that

\begin{itemize}
\item $j=2$ corresponds to $\theta=0$
\item $j=jmax$ corresponds to $\theta=\frac{\pi}{2}$
\item $i=2$ corresponds to $r=\frac{\Delta\eta}{2}$ (see section \ref{sec:grid})
\end{itemize}
and also that we have the stencil defined in equation (\ref{eqn:4ordstencilexpand}).  Let us now describe how to implement a simple boundary condition for $\phi$.

We know that $\phi_{i,1}=\phi_{i,3}$ as $\phi_{i,j}$ is symmetric across $\theta=0$ ($j=2$).  This tells us that in our stencil equation (\ref{eqn:4ordstencilexpand}) we can replace all instances of $\phi_{i,1}$ with $\phi_{i,3}$.  Specifically we see that the expression
$${b1}_{i,j} \, \phi_{i,j+1} + {c1}_{i,j} \, \phi_{i,j-1}$$
for $j=2$ becomes
$${b1}_{i,2} \, \phi_{i,3} + {c1}_{i,2} \, \phi_{i,1}$$
Then applying the symmetry condition $\phi_{i,1}=\phi_{i,3}$
\begin{eqnarray}
{b1}_{i,2} \, \phi_{i,3} + {c1}_{i,2} \, \phi_{i,3} & \rightarrow & \nonumber \\
({b1}_{i,2} + {c1}_{i,2}) \, \phi_{i,3} & \rightarrow & \nonumber \\
\tilde{b1}_{i,2} \, \phi_{i,3} & &
\end{eqnarray}
From this methodology we can then produce the following FORTRAN code for a function that is symmetric on all 3 interior\footnote{On the axis ($\theta=0$), equator ($\theta=\frac{\pi}{2}$) and at the origin ($\eta=0$).} boundaries:

\begin{verbatim}
do j=2,jmax
  i=2
  a4ord(i,j)=a4ord(i,j)+ld1(i,j); ld1(i,j)=0.d0
  rd1(i,j)=rd1(i,j)+ld2(i,j); ld2(i,j)=0.d0
  i=3
  ld1(i,j)=ld1(i,j)+ld2(i,j); ld2(i,j)=0.d0
enddo

do i=2,imax+2
  j=2
  b1(i,j)=b1(i,j)+c1(i,j); c1(i,j)=0.d0
  b2(i,j)=b2(i,j)+c2(i,j); c2(i,j)=0.d0
  j=3
  a4ord(i,j)=a4ord(i,j)+c2(i,j); c2(i,j)=0.d0
  j=jmax
  c1(i,j)=c1(i,j)+b1(i,j); b1(i,j)=0.d0
  c2(i,j)=c2(i,j)+b2(i,j); b2(i,j)=0.d0
  j=jmax-1
  a4ord(i,j)=a4ord(i,j)+b2(i,j); b2(i,j)=0.d0
enddo
\end{verbatim}
This effectively folds the boundary conditions into the interior region by setting the coefficients of the variables in the stencil equation at points \emph{outside} the computational grid to zero.  If we did not do this, we would have a stencil equation that included coupling to points outside the computation domain which is ill-posed.

For the second scenario where we know the value of the function in the interior region and merely need to populate grid points in the exterior region to calculate derivative information near the boundary, the symmetry conditions easily translate into values for the exterior region.\footnote{In practice we try to avoid explicitly defining all these points and instead prefer to pass a boundary condition parameter to the derivative calculation routine.}

i.e. $g(i,1)=g(i,3)$ for a function $g$ that is symmetric across $\theta=0$ ($j=2$).

\section{Main Loop portion of code}
We will now proceed with a discussion of the time evolution portion of the code as we have completely specified the Initial Value Problem.  This discussion provides the steps necessary to compute all variable values on a single time step, and the procedure can be iterated upon to evolve the spacetime.

\subsection{Archive historical time step information}
As we only store the values of variables on the current time step $k$ and the previous $n$ time steps to save on memory consumption while the code is running, we need to migrate the values of variables to a historical storage memory location (and clear the $(k-n-1)$th value out as it is no longer needed) once a new time step is started.

This portion of the code does a simple iteration through all variables and pushes historical values back one slot to make room for the current time step. i.e. if $\phi(i,j,n)$ is the value of $\phi$ at all grid points $n$ time steps in the past, we perform
$$\phi(i,j,n-1) \rightarrow \phi(i,j,n)$$
$$\phi(i,j,n-2) \rightarrow \phi(i,j,n-1)$$
etc. and $\phi(i,j,0)$ becomes the storage location for the new time step $k$.

\subsection{Evolution of $q$}
At this point we are ready to start the evolution on this time step, and our first step is to evolve the metric parameter $q$ via equation (\ref{eqn:qdot}).

We utilise commutative summation (see section \ref{subsec:addterms}), 4th order correct spatial derivatives (see section \ref{sec:4thordderiv}), and the boundary conditions in table \ref{tbl:boundvar}.  See section \ref{sec:do4ordbound} for more details on boundary conditions.  See section \ref{sec:4ordtime} for information about the time evolution.

We also monitor the values of the variables as we evolve (looking for NANQs, for example), and one advantage of evolving $q$ instead of $a$ is that we will never end up in the unphysical situation\footnote{As sometimes happened before we switched into exponential variables...} where $g_{11}=a e^{4\phi}=f_\eta^2 e^{q+4\phi}<0$.

\subsection{(Option) Evolution of $\phi$}

We can choose to evolve $\phi$ using either equation \ref{eqn:gam33dot}, or the Hamiltonian formulation in section \ref{sec:constrainconform}.  Once the Hamiltonian formulation is fully functional, that becomes the preferred method.  The Hamiltonian formulation maintains energy conservation and helps stabilise the overall evolution\footnote{As fully free evolutions will generally have large constraint violations.}.

As such we avoid the use of the evolution equation for $\phi$ (except as a numerical check on the code later).

\subsection{Evolution of $H_a$, $H_b$, $H_c$ and $H_d$}\label{sec:codeevol}
We evolve the extrinsic curvature variables $H_a$ and $H_b$ instead of constraining them, for the reasons mentioned in section \ref{sec:hahbcon}.

The evolution equations for $H_a$, $H_b$, $H_c$ and $H_d$ are (\ref{eqn:haevol}), (\ref{eqn:hbevol}), (\ref{eqn:hcevol}) and (\ref{eqn:hdevol}).

We utilise commutative summation (see section \ref{subsec:addterms}), 4th order correct spatial derivatives (see section \ref{sec:4thordderiv}), and the boundary conditions in table \ref{tbl:boundvar}.  See section \ref{sec:do4ordbound} for more details on boundary conditions.  See section \ref{sec:4ordtime} for information about the time evolution.

The introduction of the exponential variables $\phi$ and $q$, where $\psi=e^{\phi}$ and $a=e^q$, helped in the regularization of these equations, especially near the origin ($\eta=0$).  In addition the use of analytical derivatives terms for our radial function $f$ (see section \ref{sec:fnx}) and imposition of  regularity conditions on $\alpha$ (see section \ref{subsec:alphareg}) led to improvements in stability of the evolution equations.

\begin{figure} \centering
\includegraphics{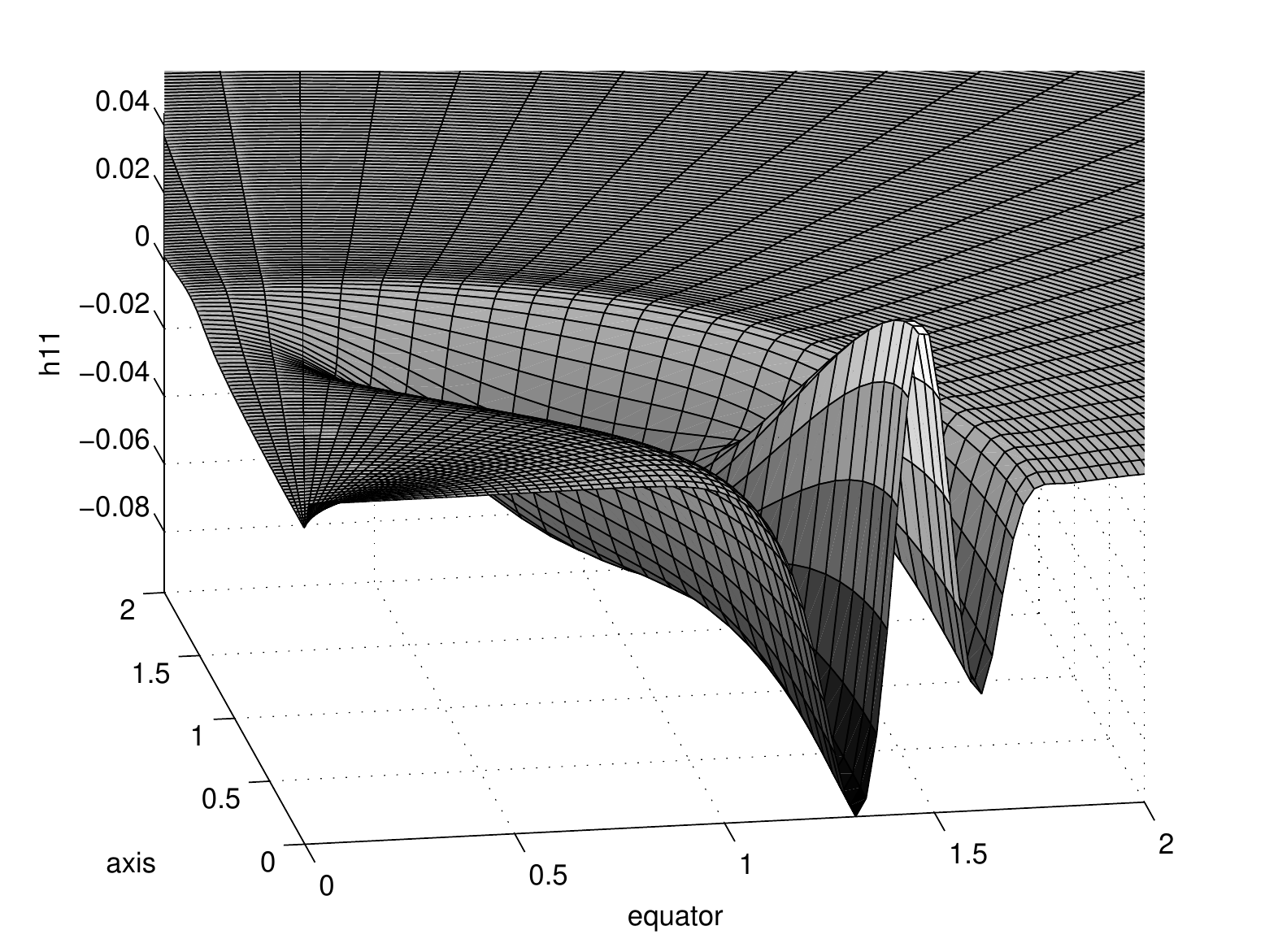}
\caption[An example of a solution for $H_a$ at $t=\Delta t$]{An example of a solution for the first extrinsic curvature variable, $H_a$ at $t=\Delta t$.  The majority of the nonlinear wave is present near the equator in the region $\eta \sim 1 \rightarrow 1.5$, and at this visual scale the outer radial region appears flat.}\label{fig:h11}
\end{figure}

\begin{figure} \centering
\includegraphics{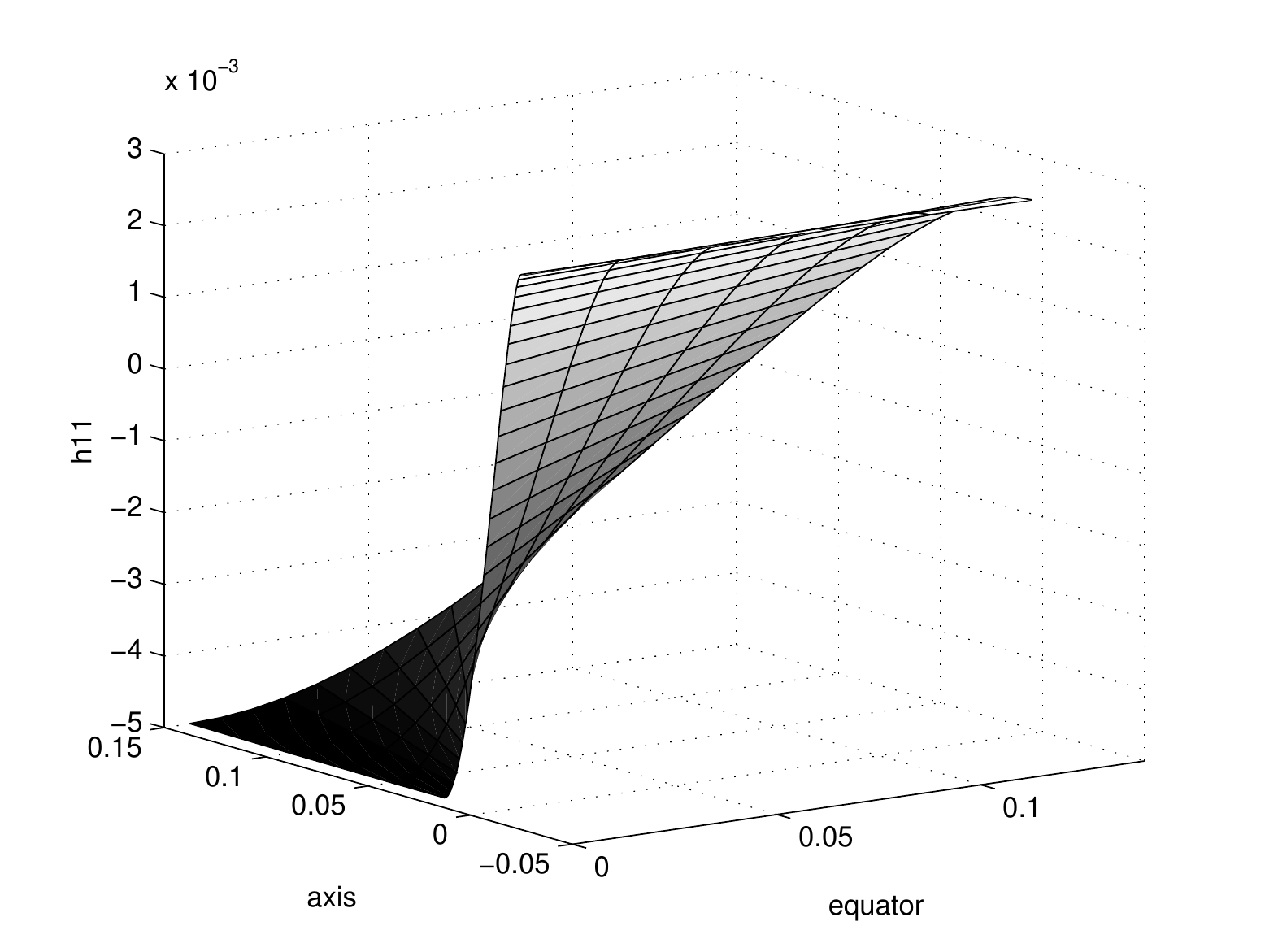}
\caption[Solution for $H_a$ near origin]{An example of a solution for $H_a$ in the first six grid points out from $r=0$ at $t=\Delta t$.  Close-up of figure \ref{fig:h11}.  Many variables exhibit this sort of behaviour at the origin causing a small, localised ``kink'' in the values (see also section \ref{sec:r0reg}) that is expected.  Regularising solutions in the area of this ``kink'' (near $r=0$) is one of the major challenges that leads to many of the choices discussed through chapter \ref{chap:changes}.}\label{fig:h11_orig}
\end{figure}

\begin{figure} \centering
\includegraphics{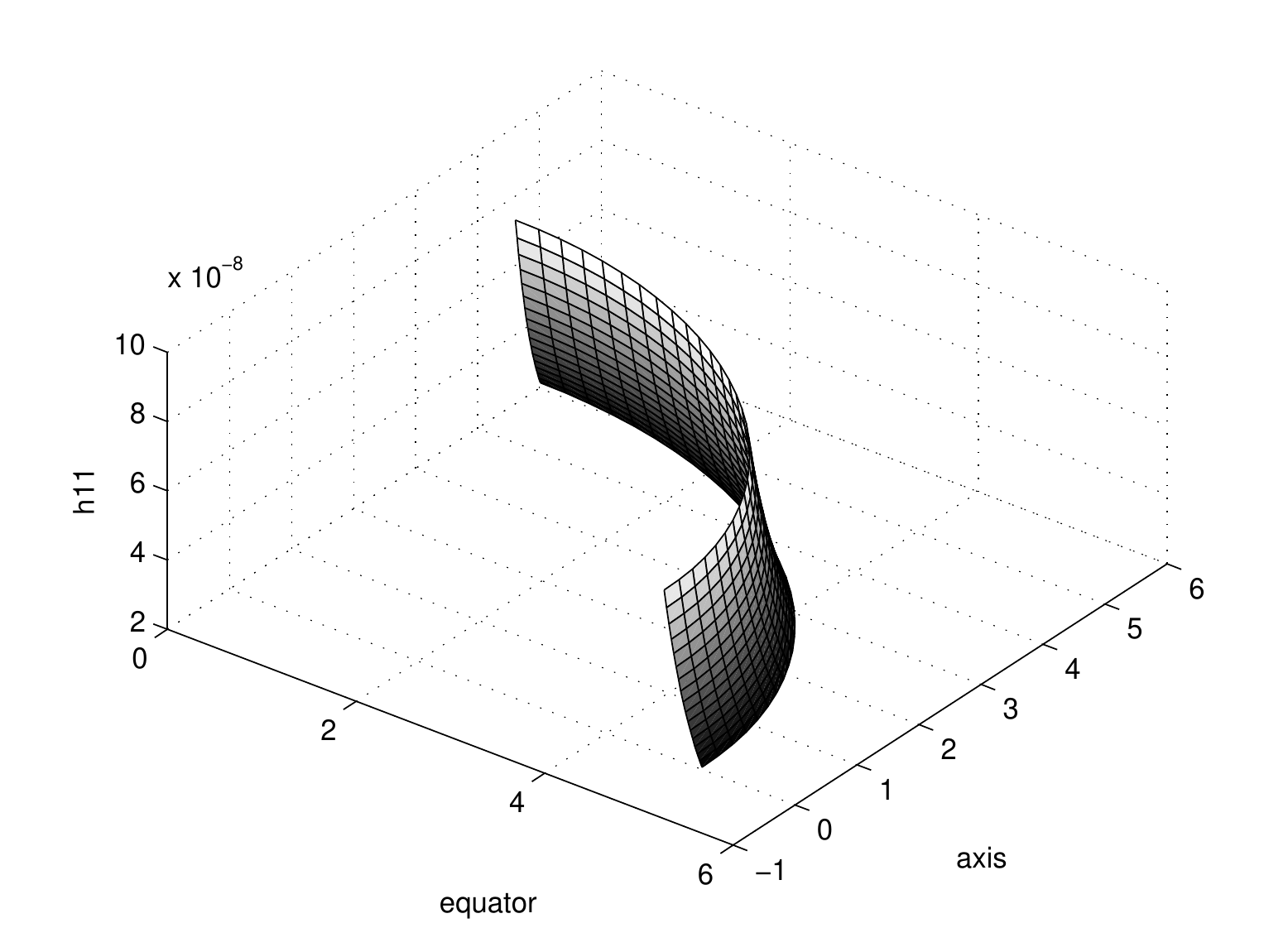}
\caption[Solution for $H_a$ near outer boundary]{A cutaway of a solution for $H_a$ near the outer boundary at $t=\Delta t$.  While the values are close to zero relative to the nonlinear interior zone, they are in fact still non-zero and require appropriate boundary conditions at the outer radial edge of the grid.  This is the exterior radial boundary of figure \ref{fig:h11}.}\label{fig:h11_ob}
\end{figure}

A sample solution for $H_a$ can be found in figures \ref{fig:h11}, \ref{fig:h11_orig} and \ref{fig:h11_ob}.  As the majority of the IVP wave is present near the equator and generally in the region $\eta \sim 1 \rightarrow 1.5$ we see that largest derivative terms and function values are present in this region.  The values become very close to zero in the radiative zone (in this case $\eta > \sim 2$) relative to the values in the nonlinear near zone.

\begin{figure} \centering
\includegraphics{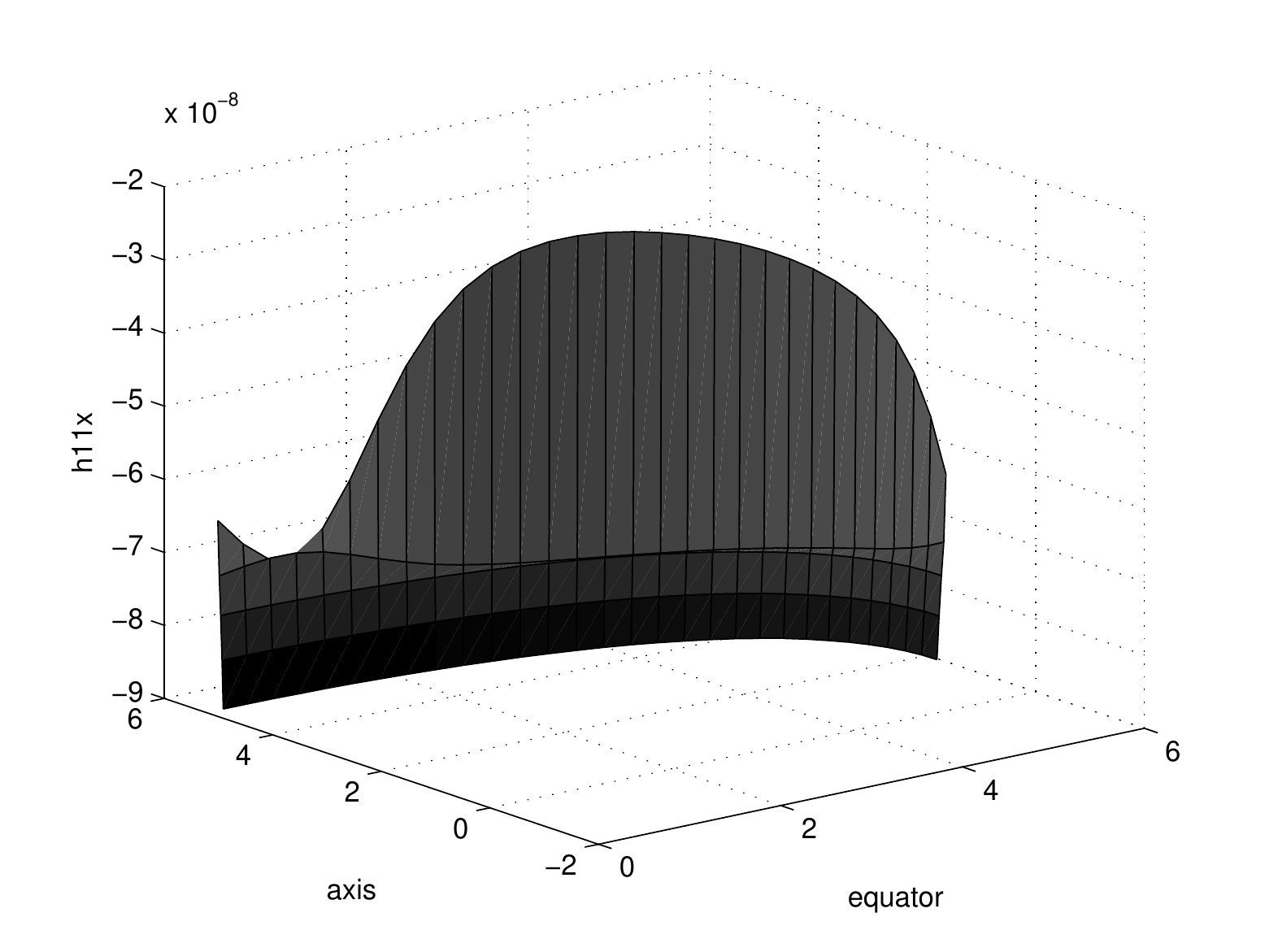}
\caption[Solution for $\frac{\partial H_a}{\partial\eta}$ near outer boundary]{A cutaway of a solution for $\frac{\partial H_a}{\partial \eta}$ in the outer grid points at $t=\Delta t$ demonstrating the propagation of poor higher order derivative behaviour for $\phi$ into the rest of the code if we don't use fourth order correct derivatives.}\label{fig:h11_bad}
\end{figure}

A demonstration of why the errors present in the higher order derivatives of $\phi$ can cause grief through the rest of the code if second order correct discretization methods are used can be seen in figure \ref{fig:h11_bad} (hence the need for fourth order correct discretisation discussed in section \ref{sec:4thorder}).

See section \ref{sec:r0reg} for a discussion of the ``kinks'' in some variables near the origin.

\subsection{Calculation of $H_d$ from constraints (Maximal Slicing \emph{only})}
We have the option of either evolving $H_d$ by using equation (\ref{eqn:hdevol}), or by using the condition that $H_a + H_b + H_d = 0$ if we are employing maximal slicing.

As we make use of the algebraic condition
\begin{equation}\label{eqn:hd_maximal}H_d = -H_a - H_b \end{equation}
several times through the code to simplify equations when using maximal slicing, it makes sense to continue its use.  At some point in the future it may be found that the stability of the code is adversely affected by the choice to substitute the condition in equation (\ref{eqn:hd_maximal}) into multiple other equations, but for now it will suffice \emph{if} we use maximal slicing.

\subsection{Computation of $\phi$ using the Hamiltonian Constraint}
We use the same procedure to calculate $\phi$ on future time steps as for the Initial Value Problem, as discussed in section \ref{sec:constrainconform}, with the notable exception that our extrinsic curvature variables are no longer guaranteed to be identically zero\footnote{Numerically, this equates to another term added to the RHS of the matrix equation to solve, which has no impact on the matrix solver algorithm used.}.  Future stability considerations may indicate that we use the hyperbolic evolution equation (\ref{eqn:gam33dot}) instead of the elliptic Hamiltonian constraint (\ref{eqn:hamconphi}), which is a simple change.  The question of whether it is better to evolve or better to constrain in numerical relativity is still an open one and seems to depend on the problem being studied.

\subsection{Computation of $\alpha$ (if using maximal slicing)}
Using our general framework that is in place for solving elliptic, 2nd order PDEs, we can solve for $\alpha$ from equation (\ref{eqn:maxslicealpha}).

\begin{figure} \centering
\includegraphics{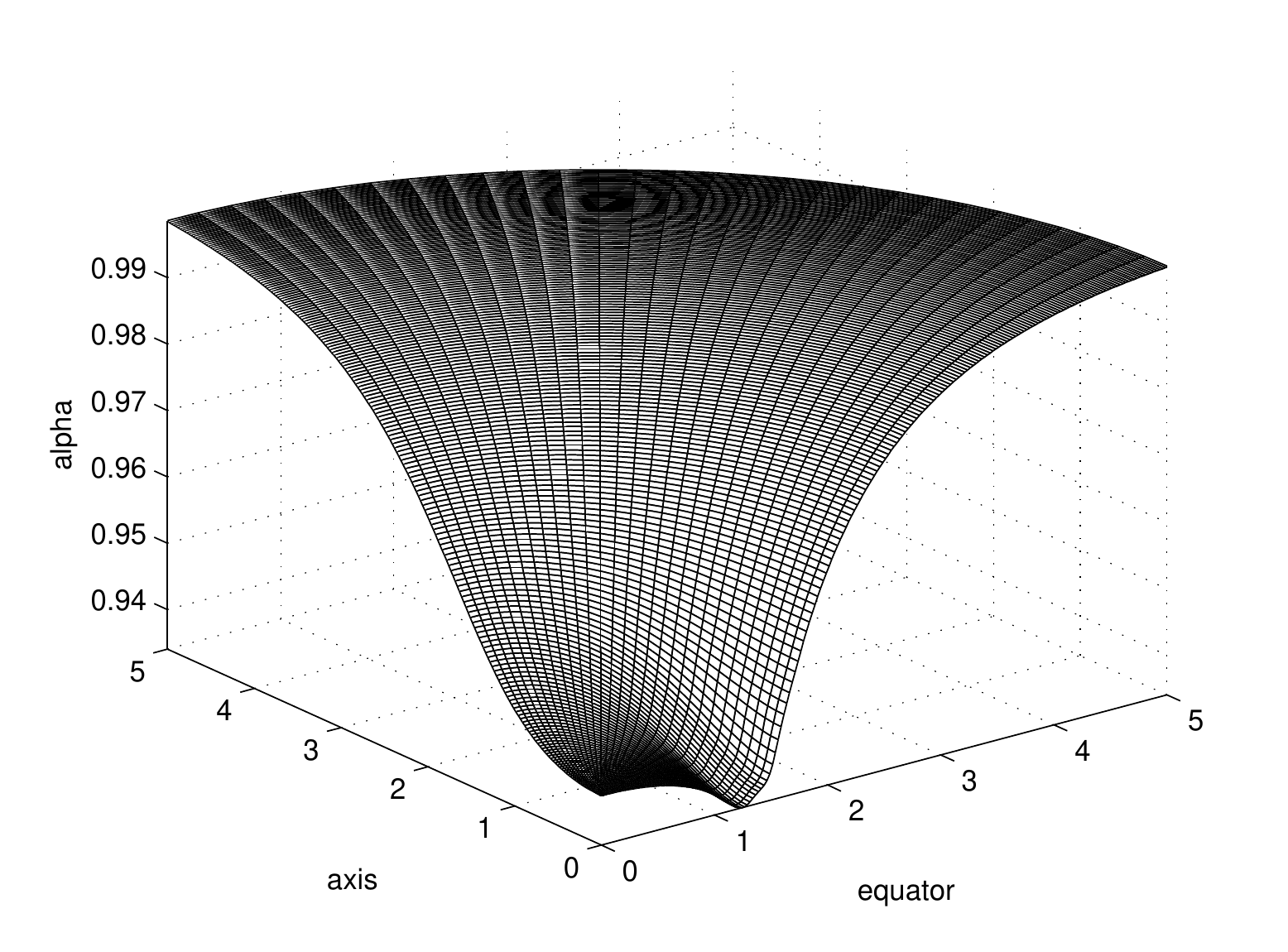}
\caption[Sample solution for $\alpha$]{An example of a solution for the lapse function, $\alpha$ at $t=\Delta t$ when employing maximal slicing.  Note how this is \emph{not} consistent with $\alpha\sim r^2 \sin^2\theta$ near $r=0$.}\label{fig:alpha}
\end{figure}

Boundary conditions are listed in table \ref{tbl:boundvar}, implementation considerations are listed in sections \ref{sec:4thordderiv}, \ref{sec:do4ordbound} and \ref{sec:spherpolob}.  See figure \ref{fig:alpha} for an example of a solution for $\alpha$.  Note that in the outer radial region $\alpha \rightarrow 1$, and across the whole grid $0 < \alpha < 1$ (time marches forward).

With the move away from maximal slicing for reasons discussed in section \ref{subsec:alphareg}, this portion of the code is kept (but not used) in case a merge between static algebraic and dynamic lapses is needed at some point in the future.

\subsection{Computation of Shift Vector Potentials $\chi$ and $\Phi$}
As we managed to decouple the two shift vector constraints in equations (\ref{eqn:decshiftvecs}) that arise from our gauge choices by the definition of potentials in equations (\ref{eqn:shiftvpotdef}), we essentially have two more second order elliptic PDEs to solve\footnote{Without this decoupling a rather complicated iterative convergence technique had to be introduced.  This increased computation time, and in the end the method was unable to converge to a satisfactory answer.}.

\begin{figure} \centering
\includegraphics{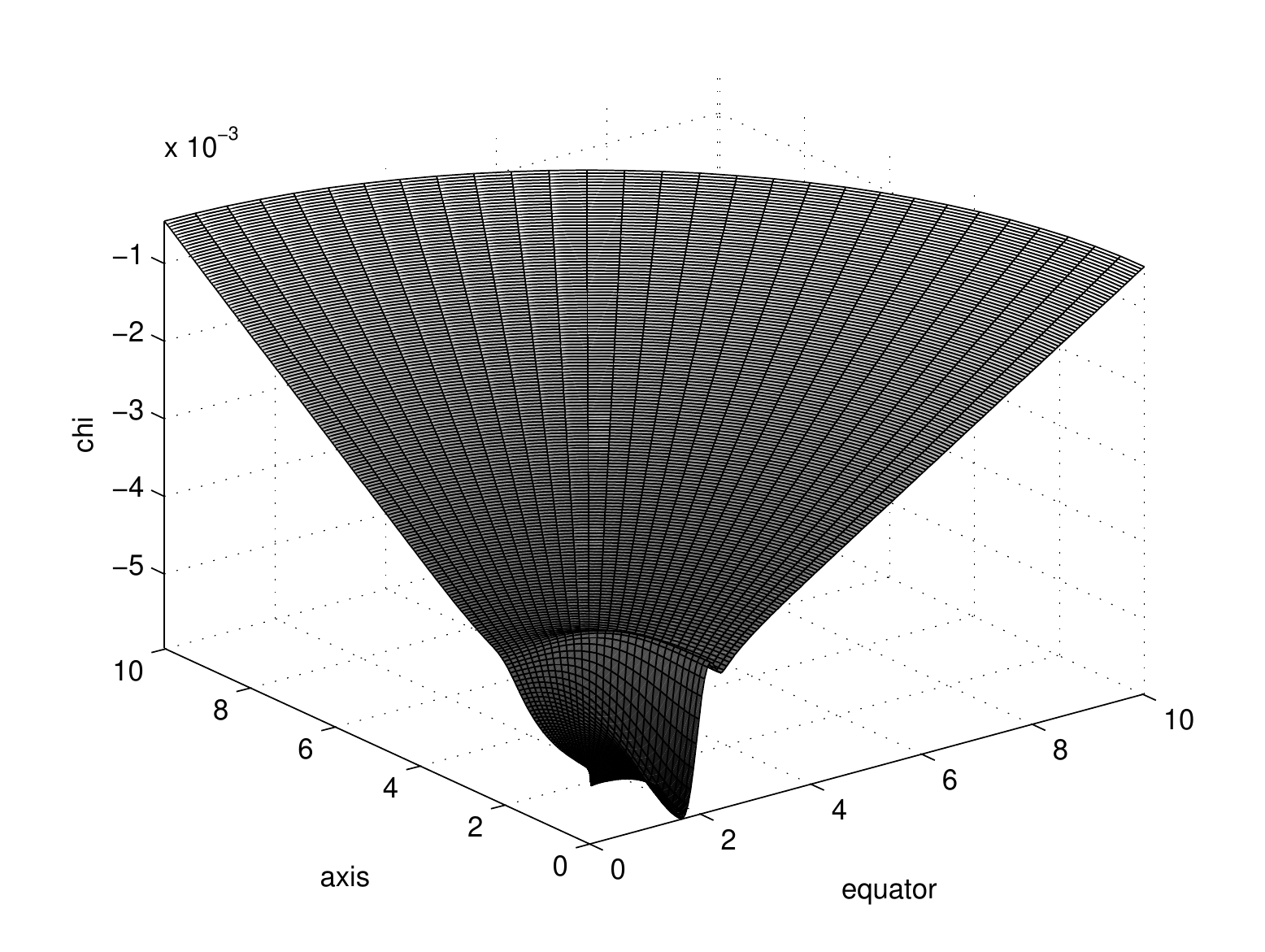}
\caption[An example of a solution for $\chi$ at $t=\Delta t$.]{An example of a solution for $\chi$ at $t=\Delta t$.  Note the ``kink'' near the origin discussed in figure \ref{fig:h11_orig} and section \ref{sec:r0reg}.}\label{fig:chi}
\end{figure}

\begin{figure} \centering
\includegraphics{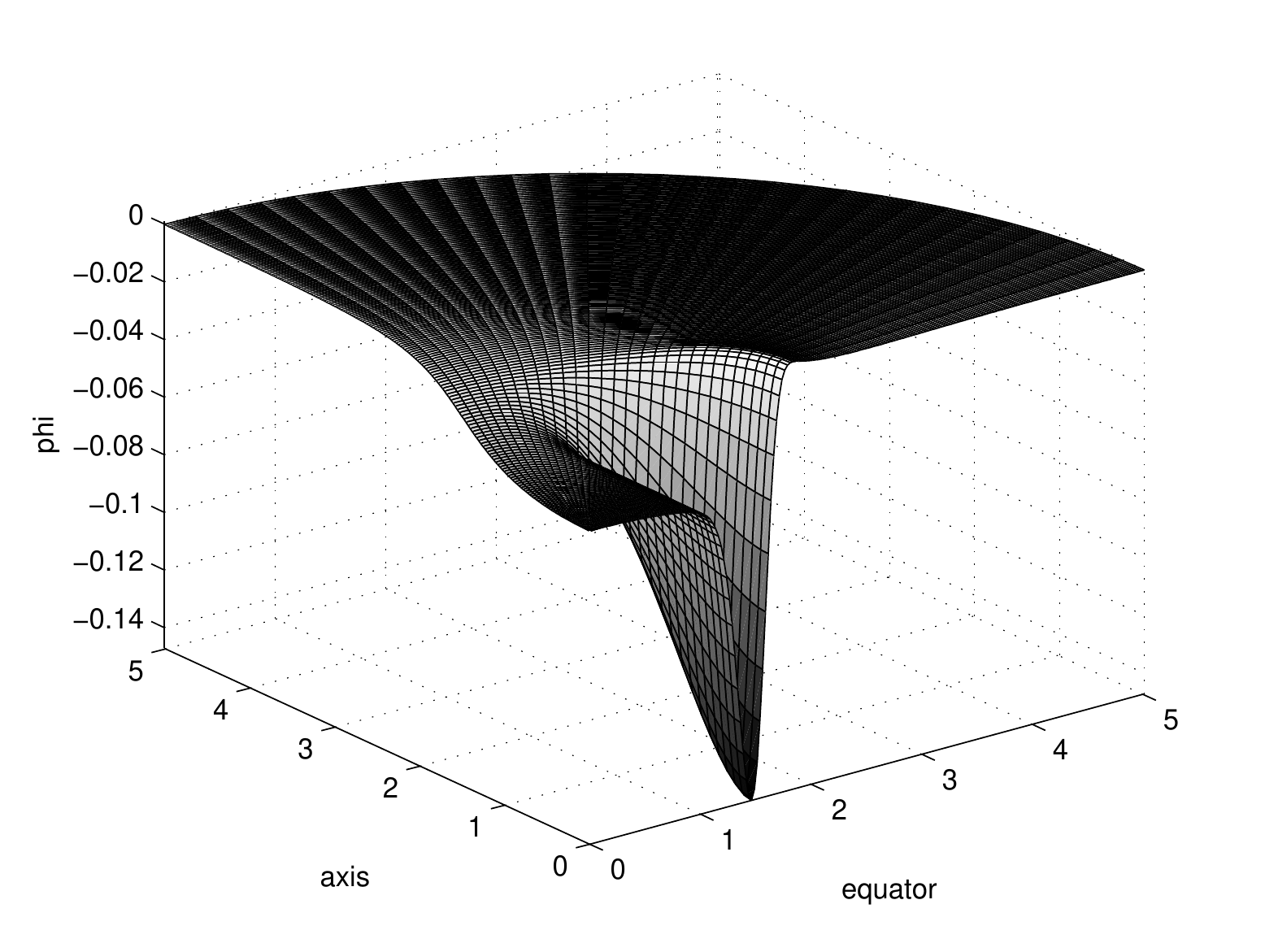}
\caption[An example of a solution for $\Phi$ at $t=\Delta t$]{An example of a solution for $\Phi$ at $t=\Delta t$.  Note the large variable and derivative values near the peak of the non-linear wave (close to the equator and $\eta\sim 1.5$).}\label{fig:phi}
\end{figure}

Boundary conditions are listed in table \ref{tbl:boundvar}, implementation considerations are listed in sections \ref{sec:4thordderiv}, \ref{sec:do4ordbound} and \ref{sec:spherpolob} or \ref{subsec:antisphharm}.  See figure \ref{fig:chi} for an example of a solution for $\chi$, and figure \ref{fig:phi} for an example of $\Phi$.  In both cases we again see that the large non-linear wave near the equator around $\eta \sim 1.5$ translates into larger derivative terms in that region.  We also see that the functions are single or multi-valued near the origin as expected from regularity conditions.

\subsection{Construction of the shift vectors from their potentials}
This is a simple algebraic problem based on equation (\ref{eqn:shiftvpotdef}), then boundary conditions given in table \ref{tbl:boundvar} are applied.

This portion of the overall numerical algorithm is generally where the ill-conditioning of second order correct differencing methods discussed in section \ref{sec:4thorder} starts to manifest noticeably in the outer boundary conditions.  It took considerable effort to determine the root cause of the ill-conditioning, however visual graphical sanity checks on the boundary conditions for all variables were indispensable in guiding the search.

\subsection{Calculation of the Scalar Curvature, $R$}
This is a simple calculation via equation (\ref{eqn:scurvunmodified}) as it is derived from our dynamic variables.

\subsection{Search for Apparent Horizon formation}
See section \ref{sec:hznsearch} for a discussion.

This is the end of the main loop in the code, and presents a complete method for evolving Brill Gravitational Waves in spherical polar coordinates.

\section{Code Halting/Crashing}\label{sec:crashhalt}
The methodology described in this chapter has been successfully employed to create a computer code that allows for the study of a wide variety of initial value problems and their evolutions.  Depending on the choice of initial conditions the code will run for a very large or small number of time steps, e.g. the code performed a non-trivial evolution over $\sim 9$ months before halting\footnote{The longest run achieved was over a period of $268$ days with a $5$GHz processor that was $100\%$ utilized for the entire duration of the run; which equates to quadrillions of calculations.}, or it can hang while trying to determine a solution to the Initial Value Problem.  A detailed analysis will be provided in the next chapter however these results indicate an important \emph{physical}, not numerical, phenomenon at play.

It is therefore important to note a few halting conditions on the code as all numerical simulations have an endpoint\footnote{Barring numerically static codes or limit cycle attractors, neither of which are anticipated or encountered here.}.  There are a few checks for NANQ values inside the code which will cause a halt or return to the parameter space searching portion of the code.  These are rarely if ever triggered now after the implementation of exponential metric variables and having fixed the numerical regularity problems on various parts of the grid.  The existence of NANQs now generally indicates that there is a coding or numerical algorithm problem.

Generally the code ``hangs'' or ``halts'' during attempts to solve the Hamiltonian constraint equation for $\phi$.  As the code iterates through the non-linear scheme described in section \ref{subsec:iterconv} at a fixed time step to attempt to converge to a solution, we see the values of $\phi$ grow during each iteration until they overflow and the routine cannot converge, at which point the code hangs.  Attempts to catch this overflow before the code hangs have had mixed results\footnote{Generally the checks involve slowing the BICGStab routine down substantially for the net effect of halting instead of hanging... which presents no numerical advantage apart from not having to check manually every now and then to see what the code is doing.}, and is an area for future research.

If, however, we are not solving the Hamiltonian constraint for $\phi$ we typically observe crashing occurs because the values of a variable blow up during the ``local'' iterative Crank-Nicholson iteration on that variable at a fixed time step.  The values grow without bound in a local region of the grid and eventually grow larger than HUGE\footnote{The largest floating point number that is representable on that compiler.} and turn into NANQs or INFs depending on how it blows up.

Physically, this can represent a few things:

(1) The curvature has grown in an unbounded way which is usually associated with singularity/horizon formation.  There are some clever ways to create spacetimes that have pathological curvature in localised areas, however we do not have any such exotic spacetimes here.  Wald \cite{Wald} calls singularities that are associated with unbounded curvature growth ``scalar curvature singularities''\footnote{Wald \cite{Wald} $\S 9.1$ presents a discussion on different types of singularities, their meanings and definitions.}.  This will cause the code to halt as computers can only hold finite sized values, so at some point the variables will overflow.

(2) We have a geodesically incomplete spacetime.  Wald \cite{Wald} $\S 9.5$ shows that under certain conditions (that are satisfied here\footnote{We trivially satisfy the strong, weak and dominant energy conditions as $T_{\alpha\beta}=0$}) the presence of an apparent horizon necessitates geodesic incompleteness, i.e. that there is at least one future directed null geodesic that starts at the trapped surface, is inextensible and terminates in a finite affine length\footnote{i.e. if the distance along the geodesic is parametrised via $\lambda$, $\lambda$ has an upper bound and the geodesic terminates.}.  This will cause the code to halt as the geodesics that some of the coordinates are following will fail to exist and the solver will be unable to converge.  This is where singularity avoiding slicing techniques are handy to avoid halting codes.

(3) We have specified a problem for which there is no solution, generally because there is an apparent horizon of infinite extent.  As discussed in section \ref{subsec:qampbounds}, there are bounds on the IVP parameters (or shape of $q$) for which a solution exists to the Hamiltonian constraint for $\phi$, and exceeding those values has been shown analytically to constitute an ill-posed problem (to be discussed later).

\section{Chapter Summary}
In this chapter a detailed methodology for constructing a Brill wave evolution code based on various considerations discussed in previous chapters was presented.

A complete method for constructing an initial value problem solution for a Brill wave (that is axisymmetric, in a vacuum and in spherical polar coordinates) has been presented including trapped surface detection and mass measures.

An evolution scheme was then developed to propagate the IVP solution forward through time, including a consideration of the various methods in which we may accomplish this feat.

Finally we discuss halting conditions for the code, and physical interpretations of these conditions.  Let us now proceed to a discussion of the results obtained from this code.

\chapter{2+1 Code Results and Analysis}\label{chap:results}
\bigskip
The code that was developed using the methodologies of this thesis has been used to investigate the time evolution of a vacuum Brill gravitational wave spacetime by providing a few key inputs; most importantly the amplitude $A$ and the ``width'' $s_0$ of the initial wave profile for $q$.  Table \ref{tbl:videoresults} contains results and links to $1000+$ \emph{evolution} videos\footnote{Alternately one can go to \href{http://www.youtube.com/channel/UCGg4GCBzxc8nEuatm5k1P-w/videos}{http://www.youtube.com/channel/UCGg4GCBzxc8nEuatm5k1P-w/videos}, or search for \emph{Andrew Masterson}'s channel on YouTube to view these.} of the Weyl Curvature variables, ${}^{(3)}R$ (Scalar Curvature), metric/extrinsic curvature/gauge variables, trapped surfaces and quasi-local ADM mass as a function of $\eta$ for various combinations of these initial value parameters.

The results in the table provide mostly an exploration of amplitude space, however there are also links to runs with alternate lapse functions, variable spatial/time grid size, backwards in time evolution, and varying exterior boundary location.

Quasi-local ADM masses at the outer boundary are presented for the IVP, first time step and last time step, with data also presented for \emph{last time step-1} in the case of perturbative waves to demonstrate the unstable nature of the final time steps.

The links in the table lead to groups of videos of the evolution of various variables for that particular set of conditions; they are immensely useful in visualising what is happening during the evolution, as it is difficult to present the full suite of results here on a static piece of paper.  The reader is encouraged to investigate some of the videos.

\begingroup
\footnotesize
\begin{center}
\begin{longtable}{|ccccccc|}

\hline
$A$ & $s_0$ & $\eta_{\mathbf{max}}$ & Halt $k\dag$ & $M_{ADM}$(quasi-local) & Notes & YouTube\\[-0.1in]
  &       &             &  $(t=k \Delta t)$    & $t_0,t_1,t_{kmax}$ &    & Video Link\\\hline
\endfirsthead

\multicolumn{7}{c}%
{{\bfseries \tablename\ \thetable{} -- continued from previous page}} \\ \hline
$A$ & $s_0$ & $\eta_{\mathbf{max}}$ & Halt $k\dag$ & $M_{ADM}$(quasi-local) & Notes & Video Link \\ \hline
\endhead

\hline \multicolumn{7}{|r|}{{Continued on next page}} \\ \hline
\endfoot

\hline \hline
\endlastfoot
\multicolumn{7}{|c|}{{Large Negative Amplitude Waves}} \\[-0.1in]
$\le -4.35$ & $1$ & $5$ & N/A & N/A & no IVP & N/A \\[-0.1in]
$-4.3$ & $1$ & $5$ & $13404$ & N/A (BH interior) & AHIVP (?) & ? \\[-0.1in]
$-4$ & $1$ & $5$ & $639$ & $23.34,23.34,(7.3,11.5)$ & AHIVP$\times$2 & \href{http://www.youtube.com/playlist?list=PLIwR6Lx72g5qhSJI_hm31qLgV9KGjTdVK}{Link} \\[-0.1in]
$-3.5$ & $1$ & $5$ & $135$ & $6.21,6.21,5.80$ & AHIVP$\times$2 & \href{http://www.youtube.com/playlist?list=PLIwR6Lx72g5pTWAKBUT-qBaXsr4YWvmh7}{Link} \\[-0.1in]
$-3.5$ & $1$ & $5$ & $>1000$ & $6.21,6.21,7.44$ & AHIVP$\times$2,$\alpha$300 & \href{https://www.youtube.com/playlist?list=PLIwR6Lx72g5rCllJP3wL35pIwU9CPm4iH}{Link} \\[-0.1in]
$-3.4$ & $1$ & $5$ & $114$ & $5.20,5.20,6.01$ & AHF & \href{http://www.youtube.com/playlist?list=PLIwR6Lx72g5rEoLKt9JSJPcWBZ3h_9oU4}{Link} \\[-0.1in]
$-3$ & $1$ & $5$ & $77$ & $2.83,2.83,3.52$ & AHF,CC & \href{http://www.youtube.com/playlist?list=PLIwR6Lx72g5rP5xhgJ4baSPRM0FH7Qb0J}{Link} \\[-0.1in]
$-2$ & $1$ & $5$ & $75$ & $0.752,0.752,2.16$ & CC,TSB & \href{http://www.youtube.com/playlist?list=PLIwR6Lx72g5qr8qeH_-BH2sYrjISCq5uI}{Link} \\[-0.1in]
$-1$ & $1$ & $5$ & $153$ & $0.141,0.141,0.798$ & CC,TSB & \href{http://www.youtube.com/playlist?list=PLIwR6Lx72g5obMAnPVIRci_OcutSi6YjD}{Link} \\[-0.1in]
$-1$ & $1$ & $5$ & $757$ & $0.141,0.141,0.425$ & CC,TSB,$\ddag$ & \href{http://www.youtube.com/playlist?list=PLIwR6Lx72g5qQD2CYcf9zC_OIv3NHS-Jg}{Link} \\
\hline
\pagebreak[2]
\multicolumn{7}{|c|}{{``Perturbative'' Waves}} \\[-0.1in]
& & & & ($t_0,t_1$)($t_{n-1},t_n$) & & \\ [-0.1in]
$-1*10^{-20}$ & $3$ & $5$ & $126$ & $2*10^{-22},2*10^{-6}$ & CC,TSB,$\alpha$300 & \href{https://www.youtube.com/playlist?list=PLIwR6Lx72g5o8Pm74K-NQZylYAb5K_tyW}{Link} \\[-0.1in]
& & & &$(3.5*10^{-2},-1.2*10^{-2})$ & & \\ [-0.1in]
$-1*10^{-10}$ & $3$ & $5$ & $127$ & $-2*10^{-12},2.4*10^{-6}$ & CC,TSB,AHF(*) & \href{http://www.youtube.com/playlist?list=PLIwR6Lx72g5oWaJBalM5YrzQ1Odss4vv_}{Link} \\[-0.1in]
& & & &$(-0.012,24.85)$ & & \\ [-0.1in]
$-1*10^{-10}$ & $3$ & $5$ & $126$ & $-2*10^{-12},2.4*10^{-6}$ & CC,$-dt$ & \href{http://www.youtube.com/playlist?list=PLIwR6Lx72g5pCrX8fDbPR_BqYAM--V8n3}{Link} \\[-0.1in]
& & & &$(0.036,-0.0077)$ & & \\ [-0.1in]
$-1*10^{-10}$ & $3$ & $5$ & $147$ & $-2*10^{-12},9.6*10^{-4}$ & CC,EP & \href{http://www.youtube.com/playlist?list=PLIwR6Lx72g5o0M3aFWmKIeXfaPC5cBp10}{Link} \\[-0.1in]
& & & &$(0.36,0.38)$ & & \\ [-0.1in]
$-1*10^{-10}$ & $3$ & $10$ & $248$ & $-2*10^{-15},2.1*10^{-9}$ & CC,TSB,$\frac{dt}{2}$ & \href{https://www.youtube.com/playlist?list=PLIwR6Lx72g5pjcAdwfvdcT06B1dHGSXVD}{Link} \\[-0.1in]
& & & &$(-2.8*10^{-2},-0.13)$ & $800\times 120$& \\ [-0.1in]
$-1*10^{-10}$ & $3$ & $10$ & $902$ & $-6*10^{-16},1.7*10^{-8}$ & CC,EP,$\alpha$300 & \href{http://www.youtube.com/playlist?list=PLIwR6Lx72g5oN2n0TYkj9pwKdyMil3kCp}{Link} \\[-0.1in]
& & & &$(-0.94,-2.44)$ & & \\ [-0.1in]
$1*10^{-10}$ & $3$ & $5$ & $126$ & $2*10^{-12},2.4*10^{-6}$ & CC,TSB & \href{http://www.youtube.com/playlist?list=PLIwR6Lx72g5r2OJIH6j4u1hbjGPncsEsB}{Link} \\[-0.1in]
& & & &$(0.036,-7.7*10^{-3})$ & & \\ [-0.1in]
$-1*10^{-10}$ & $8$ & $5$ & $126$ & $-1.9*10^{-10},2.4*10^{-6}$ & CC,TSB & \href{http://www.youtube.com/playlist?list=PLIwR6Lx72g5rypmlvNMzA-_QfuOKAm7cJ}{Link} \\[-0.1in]
& & & &$(0.036,-7.7*10^{-3})$ & & \\ [-0.1in]
$-1*10^{-5}$ & $3$ & $5$ & $126$ & $3.5*10^{-8},2.4*10^{-6}$ & CC,TSB & \href{http://www.youtube.com/playlist?list=PLIwR6Lx72g5osv6qF2u6GzG63_eGjrHuZ}{Link} \\[-0.1in]
& & & &$(0.035,-0.011)$ & & \\ [-0.1in]
$-1*10^{-5}$ & $3$ & $10$ & $126$ & $2.3*10^{-7},2.4*10^{-7}$ & CC,TSB & \href{https://www.youtube.com/playlist?list=PLIwR6Lx72g5qOeBpyo7AfWw1PvioG_RQ8}{Link} \\[-0.1in]
& & & &$(6.3*10^{-3},-0.1211)$ & & \\ [-0.1in]
$-1*10^{-5}$ & $3$ & $5$ & $340$ & $3.5*10^{-8},2.5*10^{-6}$ & CC,TSB,$\alpha$40 & \href{http://www.youtube.com/playlist?list=PLIwR6Lx72g5ryNfaVF6bQAQM5EMK2gP9K}{Link} \\[-0.1in]
& & & &$(0.29,-4.13*10^{-2})$ & & \\ [-0.1in]
$-1*10^{-4}$ & $8$ & $5$ & $137$ & $0.166,0.166$ & CC,TSB & \href{http://www.youtube.com/playlist?list=PLIwR6Lx72g5o9IgHZiSz6t1paOQmBCYh3}{Link} \\ [-0.1in]
& & & &$(0.152,-0.48)$ & & \\
\hline
\pagebreak[2]
\multicolumn{7}{|c|}{{Large Positive Amplitude Waves}} \\[-0.1in]
$1$ & $1$ & $5$ & $110$ & $0.104,0.104,0.292$ & CC,TSB & \href{http://www.youtube.com/playlist?list=PLIwR6Lx72g5pRbbYG-MIL97yjAGn9tgfm}{Link} \\[-0.1in]
$2$ & $1$ & $5$ & $45$ & $0.388,0.388,1.785$ & CC,TSB & \href{http://www.youtube.com/playlist?list=PLIwR6Lx72g5pSDYdHcWQVjv3J6g6LKwL9}{Link} \\[-0.1in]
$2$ & $1$ & $5$ & $220$ & $0.388,0.388,1.05$ & CC,TSB,$\frac{dt}{5}$ & \href{http://www.youtube.com/playlist?list=PLIwR6Lx72g5pO9AvGbM3yS9t0dZID8u5Y}{Link} \\[-0.1in]
$2$ & $1$ & $5$ & $238$ & $0.388,0.388,5.01$ & CC,TSB,$\alpha$40 & \href{http://www.youtube.com/playlist?list=PLIwR6Lx72g5pDPpFrG0pO4p_T0t7L6E1B}{Link} \\[-0.1in]
$2$ & $1$ & $5$ & $1083$ & $0.388,0.388,-0.24$ & CC,TSB,$\alpha$300 & \href{https://www.youtube.com/playlist?list=PLIwR6Lx72g5qYXq0H84o8H47w3szZTsDc}{Link} \\[-0.1in]
$2$ & $1$ & $5$ & $316$ & $0.388,0.388,1.207$ & CC,TSB,$\ddag$ & \href{http://www.youtube.com/playlist?list=PLIwR6Lx72g5pdv7ye14O-4mOGdEW2vja1}{Link} \\[-0.1in]
$3$ & $1$ & $5$ & $26$ & $0.843,0.844,1.501$ & CC,TSB & \href{https://www.youtube.com/playlist?list=PLIwR6Lx72g5oAis1E9zCC28mQX-CSP4RH}{Link} \\[-0.1in]
$5$ & $1$ & $5$ & $18$ & $2.427,2.428,3.389$ & CC,TSB & \href{https://www.youtube.com/playlist?list=PLIwR6Lx72g5p64eRJ5m8VhFNwTxX4EdSh}{Link} \\[-0.1in]
$6.6$ & $1$ & $5$ & $18$ & $4.87,4.87,5.85$ & AHF,CC & \href{https://www.youtube.com/playlist?list=PLIwR6Lx72g5pIEwrcaviKTZNRisqKr4PK}{Link}\\[-0.1in]
$6.6$ & $1$ & $5$ & $18$ & $4.87,4.87,5.85$ & AHF,CC,$-dt$ & \href{http://www.youtube.com/playlist?list=PLIwR6Lx72g5rIvY2rwN4a-jCGhej5pEES}{Link}\\[-0.1in]
$6.8$ & $1$ & $5$ & $19$ & $5.305,5.307,6.82$ & AHF,CC & \href{http://www.youtube.com/playlist?list=PLIwR6Lx72g5p8jKxWjRNw8qcCBhiDvK7C}{Link} \\[-0.1in]
$9$ & $1$ & $5$ & $38$ & $15.69,15.70,19.45$ & AHIVP,ETSF & \href{http://www.youtube.com/playlist?list=PLIwR6Lx72g5rqmu0NNVQTWNnvo6jOWFqX}{Link} \\[-0.1in]
$9$ & $1$ & $5$ & $38$ & $15.69,15.70,19.45$ & AHIVP,ETSF,$-dt$ & \href{http://www.youtube.com/playlist?list=PLIwR6Lx72g5qzVTmb-h5kCIwagKrI3PVP}{Link} \\[-0.1in]
$9$ & $1$ & $5$ & $1917$ & $15.69,15.69,88$ & AHIVP,$\alpha 300$ & \href{http://www.youtube.com/playlist?list=PLIwR6Lx72g5rsOfDJRvz-ERbzsCEXUuXe}{Link} \\[-0.1in]
$9$ & $1$ & $10$ & $38$ & $15.72,15.74,19.3$ & AHIVP,ETSF & \href{http://www.youtube.com/playlist?list=PLIwR6Lx72g5rOzXs11wrdjclv6_5aAtrQ}{Link} \\[-0.1in]
$9$ & $1$ & $20$ & $38$ & $15.72,15.73,19.3$ & AHIVP,ETSF & \href{http://www.youtube.com/playlist?list=PLIwR6Lx72g5rmz2flnpkeVDZv4Om3YQ-D}{Link} \\[-0.1in]

$\ge 10.1$ & $1$ & $5$ & N/A & N/A & no IVP & N/A \\
\hline

\caption[Video links and results of time evolutions for various IVPs]{Evolution data for various IVPs, see links for detailed videos of evolution of various parameters.  $\Delta t=\pm 0.0125$ for all $200\times 60$ runs.
(CC) $\rightarrow$ Weyl and Ricci scalar curvatures grow without bound indicating singularity formation.
(TSB) $\rightarrow$ Trapped Surface Bunching is observed instead of apparent horizon formation.
$\alpha N$ $\rightarrow$ $\alpha\sim\tanh^N(\eta)$ (normal is $N=4$).
(no IVP) $\rightarrow$ no Initial Value Problem solution.
(AHF) $\rightarrow$ Apparent Horizon (outermost trapped surface) Formation during the course of the evolution.
(AHIVP) $\rightarrow$ Apparent Horizon present in Initial Value Problem.
(ETSF) $\rightarrow$ Enclosed (non-outermost) Trapped Surface Formation during the evolution (black hole interior topology).
(*) $\rightarrow$ AH forms on last time step.
(?) $\rightarrow$ Topology looks like black hole \emph{interior}.
$-dt$ $\rightarrow$ Time reversed evolution.
$\frac{dt}{N} \rightarrow$ $\Delta t$ is $1/N$th standard size.
EP $\rightarrow$ Evolve $\phi$ instead of constraining (see section \ref{subsec:evolphi}).
$\dag$ $\rightarrow$ Rows with entries of the form ``$>N$'' completed \emph{without} encountering a singularity.
$\ddag \rightarrow$ $\Delta t$ varies over the course of the evolution to study near-critical times
}
\label{tbl:videoresults}
 \\

\end{longtable}
\end{center}
\endgroup

Beyond what is listed in the table, we can also specify the grid resolution, number of free vs. constrained quantities used in the evolution, various solver tolerances, convergence criteria and wave profile shape to name a few.  We will discuss some of these other choices in the context of optimizing how the code runs, verifying our assumptions or even getting the code to run in the first place.

\section{What is $q$, and what does it mean?}\label{sec:meaningofq}
To understand the physical meaning of changing the ``amplitude'' of the variable $q$ at various points in the spacetime, we return to our metric (see equation (\ref{eqn:3metricfinal})) and the notion that metric variables measure spacetime distances.

\begin{figure} \centering
%\psfrag{deta}{$f_\eta d\eta$}
%\psfrag{dtheta}{$f d\theta$}
%\psfrag{dphi}{$f sin\theta d\varphi$}
%\psfrag{*e^q}{$\times e^q$}
\includegraphics{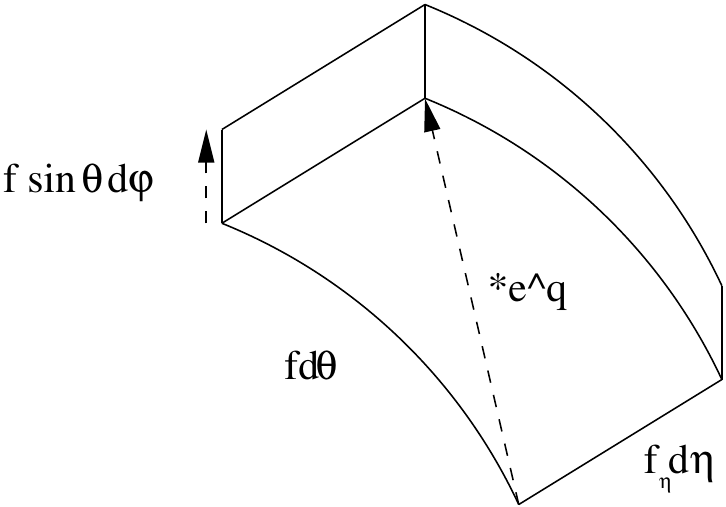}
\caption{Proper radial distances in the $(\eta,\theta)$ plane versus $\varphi$.}\label{fig:qspherepolvolelement}
\end{figure}

On our initial slice with $\beta^i=0$, we can write the spacetime distance as
\begin{equation}\label{eqn:4metricbeta0}ds^2 = -\alpha^2 dt^2 + e^{4\phi}\left(e^q\left[f_\eta^2d\eta^2 + f^2 d\theta^2\right] + f^2 \sin^2\theta d\varphi^2 \right)\end{equation}
For light rays $ds^2=0$, and we have a static\footnote{As this is a gauge condition, the same physical results should apply for various lapse functions.} $\alpha$, so we find that
$$\alpha^2 = e^{4\phi}\left(e^q\left[f_\eta^2\left(\frac{d\eta}{dt}\right)^2 + f^2 \left(\frac{d\theta}{dt}\right)^2\right] + f^2 \sin^2\theta \left(\frac{d\varphi}{dt}\right)^2  \right) = \mathrm{constant}$$
Let us now introduce the notion of $q$ as a \textbf{cost function} for motion of light rays in the $(\eta,\theta)$ plane.  Relative to flat space where $q=0$ and $\phi=0$, $q$ represents the \emph{relative} cost of a light ray moving in the plane versus moving perpendicularly along the $\varphi$ direction (see figure \ref{fig:qspherepolvolelement} for a visualisation).

\begin{figure} \centering
\includegraphics{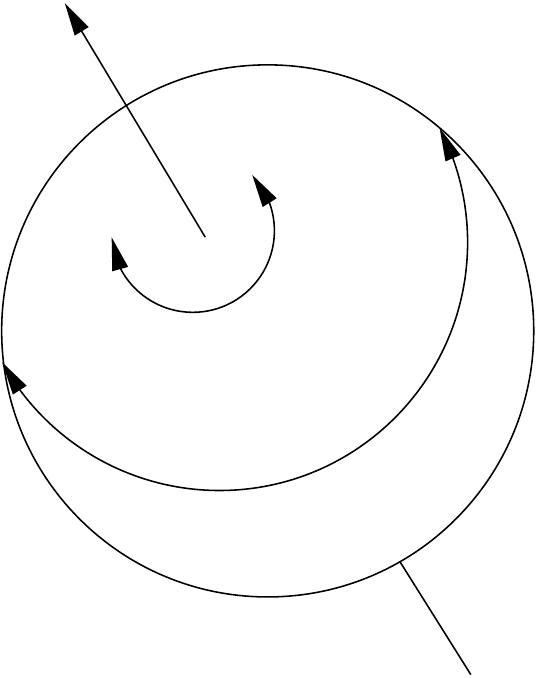}
\caption[Positive $q$ constraining light to latitudes]{A large positive value for $q$ has the effect of constraining light to motion in the $\varphi$ direction.  The more positive $q$ is, the more constrained light is to this motion.}\label{fig:qspherelinelat}
\end{figure}

If $q$ is large and positive, then the cost of moving a small $d\eta$ or $d\theta$ is large as it gets multiplied by $e^q$, relative to a small $d\varphi$.  This can be visualised as light rays that have a trajectory that varies slightly off the $(\eta,\theta)$ plane turning into trajectories along lines of latitude in our $(\eta,\theta,\varphi)$ coordinate system.  See figure \ref{fig:qspherelinelat} for a visualisation of this effect.  One can think of this as large stretching or impedance of the spacetime on the $(\eta,\theta)$ plane relative to $\varphi$.  With the Gaussian shape for $q$ used in the code this can be visualised as a toroid-like shape around the equator $(\theta \sim \frac{\pi}{2}, \eta \sim 1.5)$ where light rays that deviate from the $(\eta,\theta)$ plane get trapped orbiting along lines of latitude when $q$ is large and positive.

This scenario is short-lived, which in this sense is related to the number of time steps the code can evolve for before encountering a singularity.  As the grid points are running into singularities and are terminating due to geodesic incompleteness this is a \emph{physical}, and not numerical, effect.  Brill wave spacetimes seem to preferentially evolve to negative $q$ values and this ``torsion'' of light trajectories unravels\footnote{Perhaps the lack of any coupling terms between $(\eta,\theta)$ and $\varphi$ causes this.  Or, perhaps, because we are in a sense constraining two spatial directions due to large ``curvature impedance'' the system prefers a ``lower energy'' state of folding in one spatial direction ($\varphi$) instead of two - and evolves in that direction.}.

\begin{figure} \centering
%\psfrag{eta}{$\eta$}
%\psfrag{th}{$\theta$}
\includegraphics{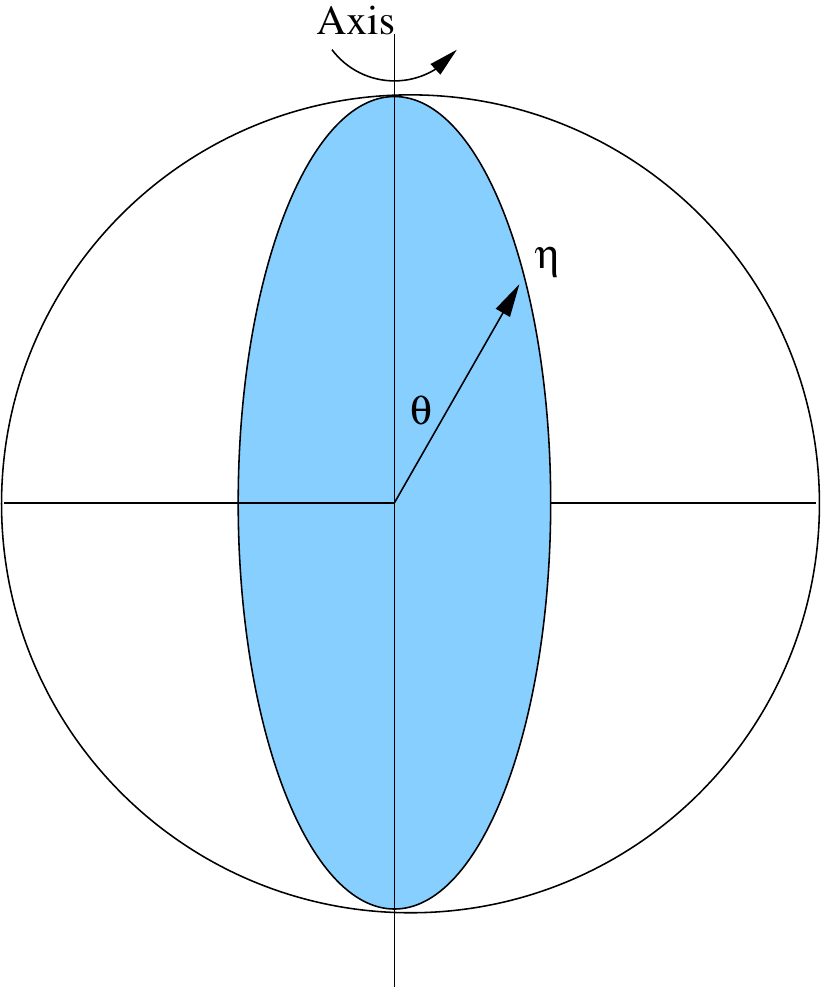}
\caption[Negative $q$ constraining light to planes]{A negative $q$ value effectively constrains light to the $(\eta,\theta)$ plane, which is represented here in blue.  Larger negative values have a more constraining effect.}\label{fig:qspherecrosssec}
\end{figure}

If $q$ is large and negative, then the cost of moving a small $d\eta$ or $d\theta$ is small relative to $d\varphi$, which means that light rays that deviate slightly from motion in the $\varphi$ direction would propagate mostly into the plane (which turns into a cross-section disk once we apply our symmetry conditions around $(\theta=0,\frac{\pi}{2})$).  See figure \ref{fig:qspherecrosssec}.  The spacetime is then ``rigid'' to light rays rotating around the axis in large negative $q$ regions, and has large curvature/impedance in the $\varphi$ direction.  This scenario seems to have longer evolutions before encountering a singularity than the positive $q$ case, and we observe smaller $q$ variations during the evolution with IVP values $A\sim -4,s_0=1$ for our $q$ function.

With the Gaussian shape for $q$ used in this thesis one can visualise this as a toroid-like shape around the equator $(\theta \sim \frac{\pi}{2}, \eta \sim 1.5)$ where most light rays get trapped in the $(\eta,\theta)$ plane when $q$ is large and negative.

After we have solved for $q$, $\phi$ then adjusts accordingly when solving the Hamiltonian constraint to maintain the relationship in equation (\ref{eqn:4metricbeta0}) by scaling the \emph{entire} spatial metric, i.e.
$$\alpha^2 \sim e^{4\phi}[L e^q + P]$$
where $L=L(f,f_\eta,\dot{\eta},\dot{\theta})$ and $P=P(f,\theta,\dot{\varphi})$.
So there is a complex interplay between $q$ and $\phi$ to ensure that the spatial distances are scaled properly. It is possible that the failure of the solver when trying to find a solution for $\phi$ in later stage evolutions is due to the inability to properly scale spatial distances at some points in the hypersurface given the above considerations and trying to balance the perpendicular ``costs'' of moving in the $(\eta,\theta)$ plane and out (along $\varphi$).

\subsection{IVP Amplitude Phase Space for $q$ $(s_0=1)$}
To characterise the Initial Value Problem solutions we first wish to numerically investigate the effect of changing the amplitude $A$ of our initial wave profile on the resulting spacetime trapped surface structure.

\begin{figure}[h] \centering
\includegraphics{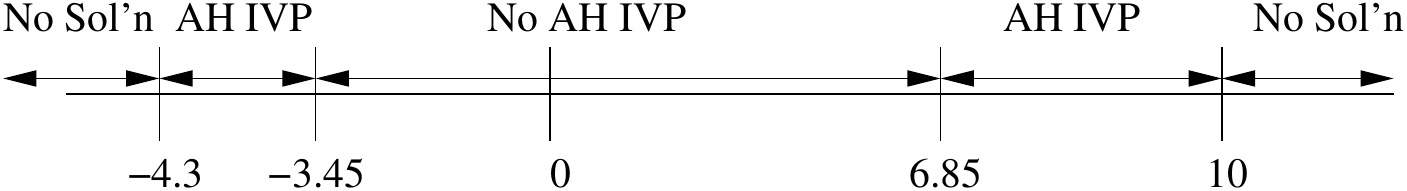}
\caption[IVP phase space (regular $q$)]{IVP phase space in the amplitude $A$ of the metric function $q$ ($s_0=1$).  Regions are divided roughly where the IVP has or does not have Apparent Horizons present, and where the code cannot solve the IVP as we have passed $A_{\pm}$} \label{fig:IVPphaseqa}
\end{figure}

Numerically it was determined that the IVP phase space in the amplitude $A$ of the metric variable\footnote{See equation (\ref{eqn:qinit}) for a reminder of what our initial wave profile looks like.} $q$ (with $s_0=1$) exhibits a maximum ($A_+ \sim 10$) and minimum $(A_- \sim -4.3)$ amplitude of $q$ for which the IVP converges\footnote{See section \ref{subsec:qampbounds} for a further discussion.}. The phase space also contains regions with and without $(-3.45<A<6.8)$ an apparent horizon present for the IVP.  See figure \ref{fig:IVPphaseqa} for a visual mapping of the phase space.

Quasi-local ADM masses at the outer radial edge of the grid on the initial time slice $(t_0)$ are presented in table \ref{tbl:videoresults}, and the masses demonstrate large non-linear growth as we approach the two critical values $A_\pm$.

\subsection{Bounds on $q$}\label{subsec:qampbounds}
Based on the amplitude phase space mapping presented above, we wish to determine why there are bounds on the amplitude of $q$. \'O Murchadha \cite{Murchadha} performs an excellent mathematical analysis of some of the properties of the function $q$.  He discerns that as the amplitude of $q$ is increased (or decreased) to some critical value\footnote{That is dependent on the functional form of $q$ - i.e. it will change depending on what \emph{shape} $q$ has, with nothing more specific available than that.} $A_+$ (or $A_-$) the apparent horizon that is present in the IVP moves outward to spatial infinity.

\begin{figure} \centering
\includegraphics{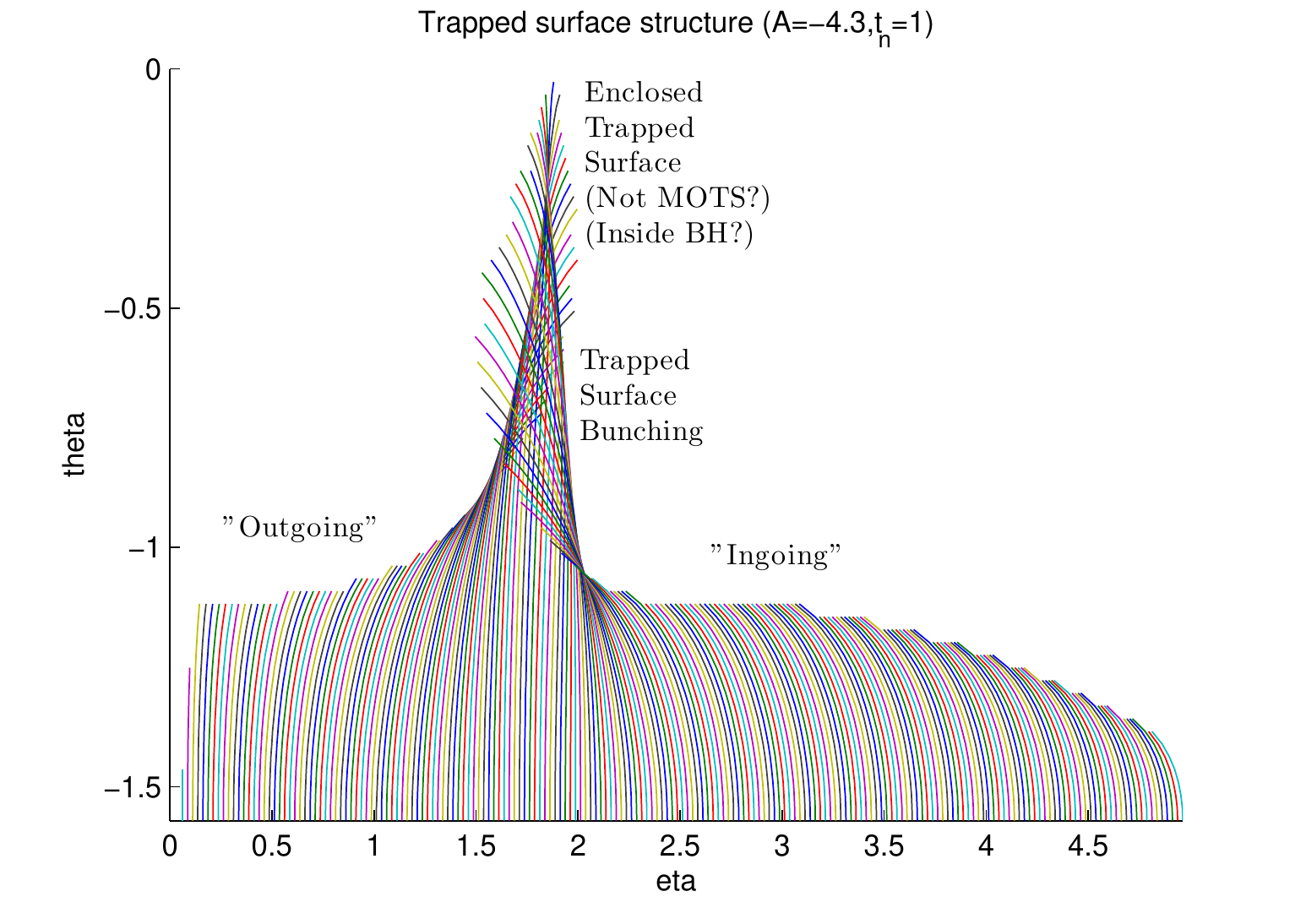}
\caption[Trapped surface topology for $A=-4.3$]{Trapped surface topology for $(A=-4.3,s_0=1,t=\Delta t)$.  Note the ``ingoing'' region at large $\eta$, in stark contrast to the ``outgoing'' topology expected.  This indicates that the whole spacetime is inside a black hole. }
\label{fig:trapsurftop_a_4.3s01t1}
\end{figure}

\begin{figure} \centering
\includegraphics{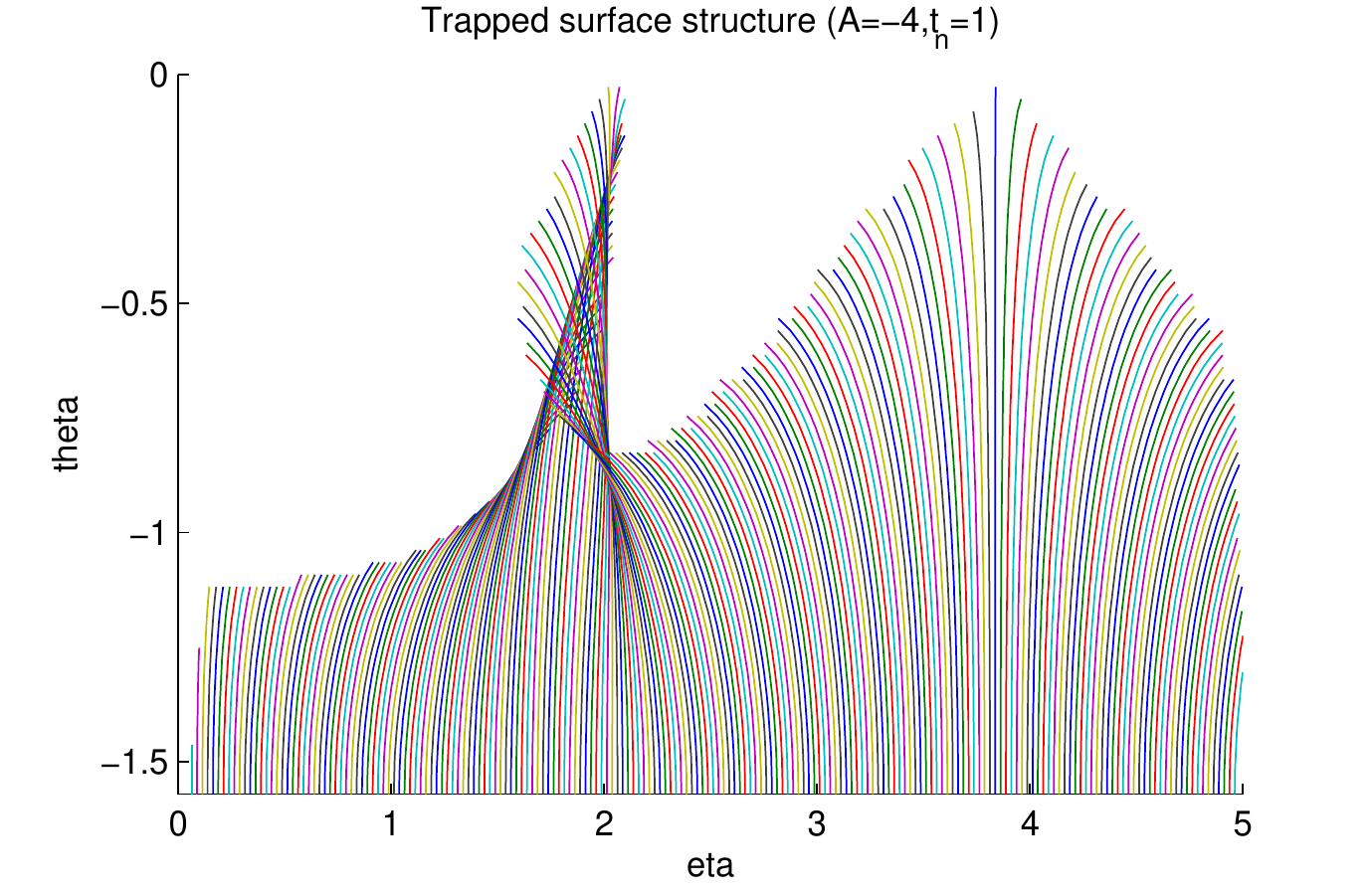}
\caption[Trapped surface topology for $A=-4$]{Trapped surface topology for $(A=-4,s_0=1,t=\Delta t)$.  This is similar to figure \ref{fig:trapsurft0a9nmax10}.}
\label{fig:trapsurftop_a_4s01t1}
\end{figure}

In our case we observe that for $A_-=-4.3$ the outer horizon disappears on our finite grid and we are left only with the ``interior'' region trapped surface topology that is ``outgoing'' in the outer region.  Examining the trapped surface topology of the $(A=-4.3)$ solution in figure \ref{fig:trapsurftop_a_4.3s01t1} we see that it closely resembles the \emph{interior} enclosed trapped surface of a black hole as in the region $\eta=2$ in figure \ref{fig:trapsurft0a9nmax10} or \ref{fig:trapsurftop_a_4s01t1}.  This likely indicates the formation of an apparent horizon near $r=\infty$ as \'O Murchadha details.

To test this we extend the edge of the grid out to $\eta_{\mathbf{max}}=20$, and we find an apparent horizon forming at $\eta \sim 6.6$ and the simulation encounters a singularity in a few time steps.  Conversely, if we evolve the ``interior'' of the black hole ($\eta_{\mathbf{max}}=5$) for the $A=-4.3$ amplitude case, the simulation runs for $>10000$ time steps, indicating that the singularity (due to geodesic incompleteness) predicted by black hole singularity theorems is present outside the computational domain, closer to the apparent horizon, in the region $\eta > 5$.

\begin{figure} \centering
\includegraphics{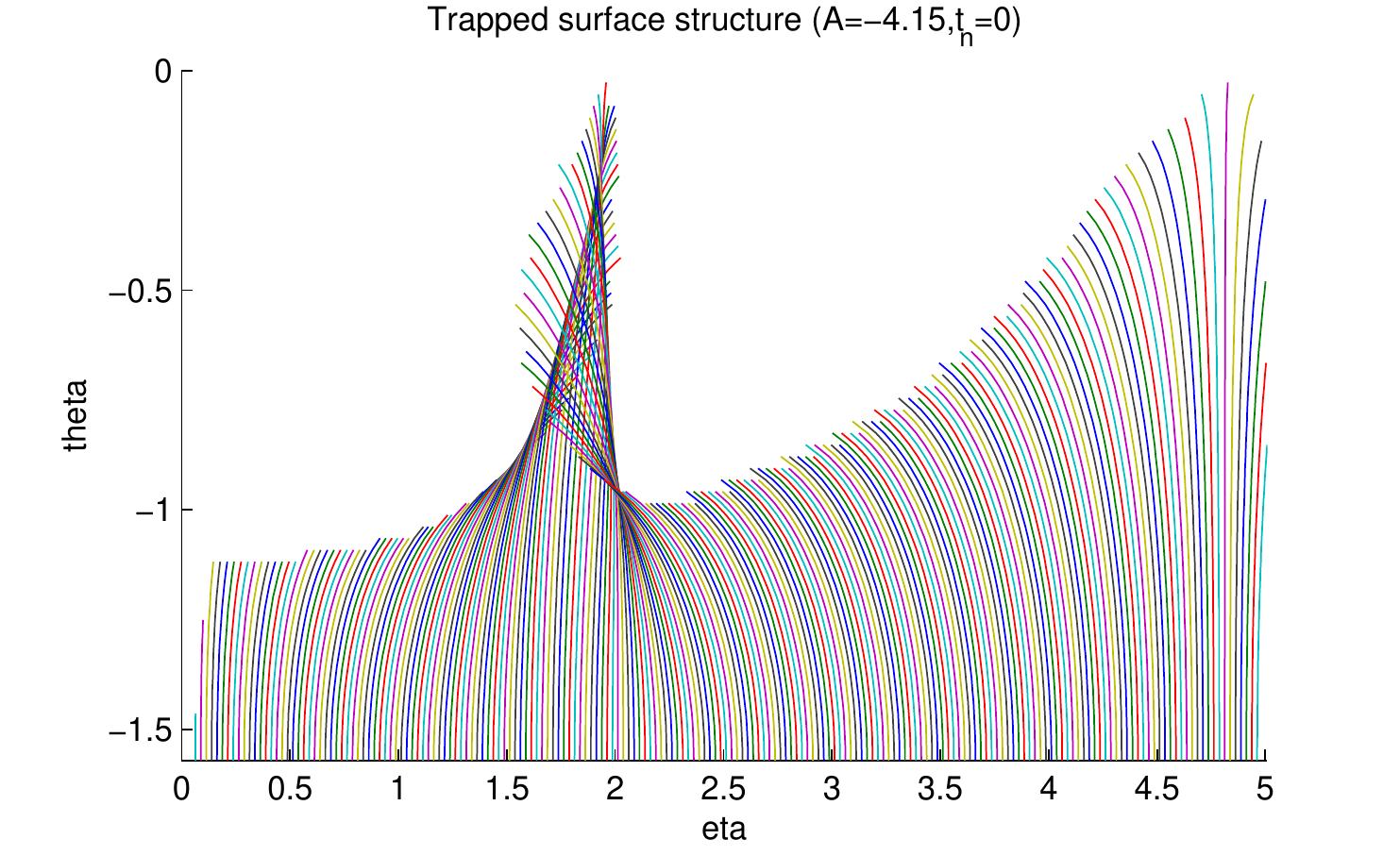}
\caption[Trapped surface topology for $A=-4.15$]{Trapped surface topology for $(A=-4.15,s_0=1,t=0)$.  Note the ``outgoing'' region has almost disappeared near $\eta=5$ and our outermost trapped surface has almost fallen off the edge of the grid. This indicates that almost the whole computational spacetime is (almost) inside a black hole and we are almost at the critical value $A_-$.}
\label{fig:trapsurftop_a_4.15s01t0}
\end{figure}

Figure \ref{fig:trapsurftop_a_4.15s01t0} also captures this effect showing that the AH is located close to the edge of the grid ($A=-4.15$) and would disappear off the edge of the computational domain with a further small increase in $A$.

Near the critical regions of phase space the change in surface area is large for small changes in the value of $A$. e.g. $A=-4,$ Area $\sim 10^3$ vs $A=-4.3,$ Area $\sim 10^{16}$ which indicates some sort of near-critical behaviour (i.e. we are very close to $A_-$).

\'O Murchadha's analysis is consistent with the results of this thesis as seen in table \ref{tbl:videoresults}.

\subsection{Alternate functional forms of $q$}\label{sec:altformq}
While we have chosen to use an exponential function\footnote{In our radial variable $\eta$} for $q$ of the form given in equation (\ref{eqn:qinit}), we can investigate if the same ``critical'' phase space behaviour in the amplitude occurs with other forms for $q$ that satisfy the Brill criteria (as predicted by \'O Murchadha).  We have chosen two other functions to investigate the IVP phase space behaviour for: a modulated version of equation (\ref{eqn:qinit}) and a trigonometric/polynomial function\footnote{See \cite{Abrahams2} for an analysis of a different phase space in cylindrical coordinates, which shows similar properties to what we observe here.}.

\begin{figure} \centering
\includegraphics{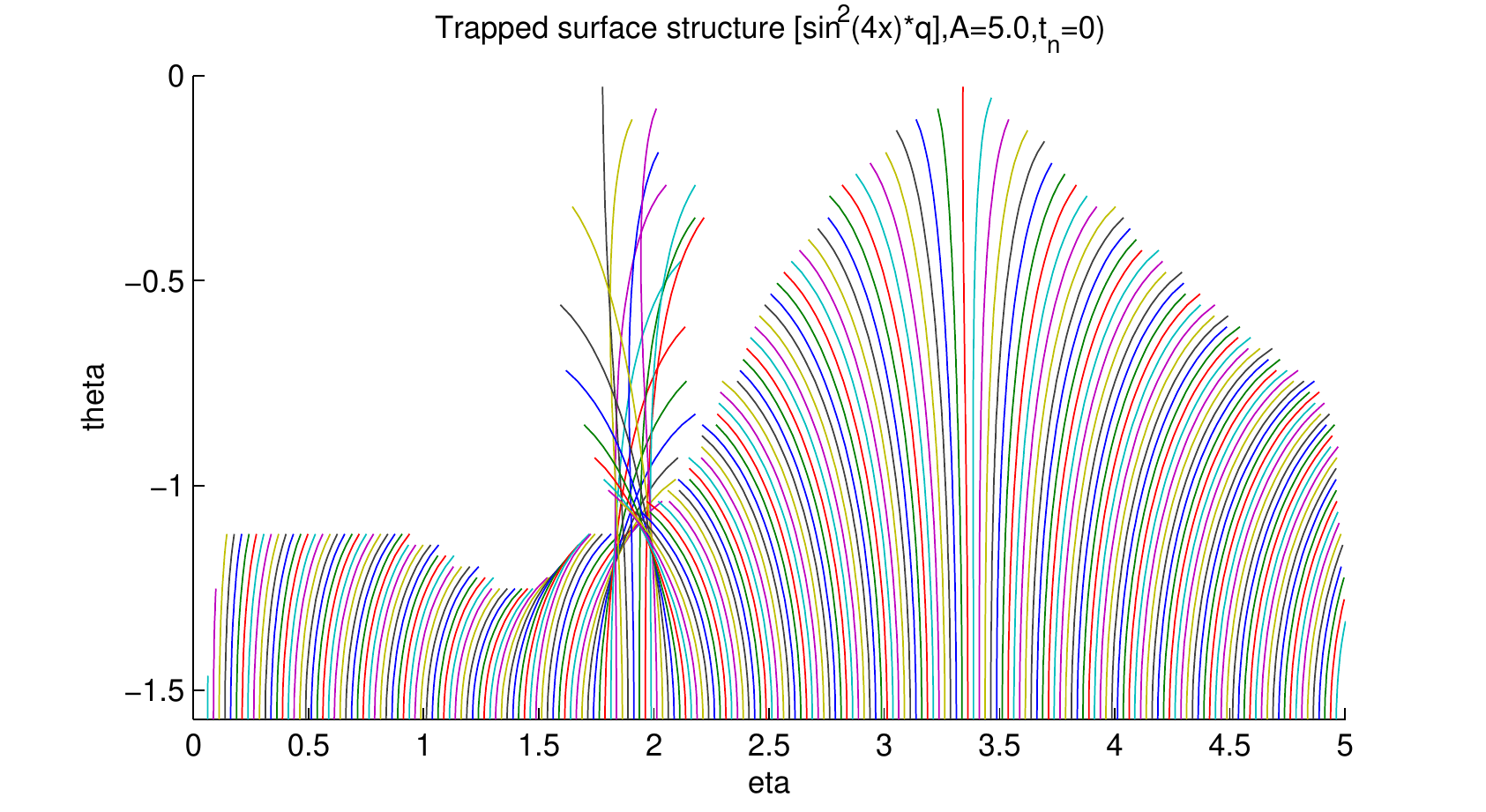}
\caption[Trapped surface topology for modulated $q$, $A=5$]{Trapped surface topology for $(A=5,s_0=1,t=0)$ and the $q$ function given by equation (\ref{eqn:qinitsinmod}).  Note the similarity to the trapped surface topology of the non-modulated $q$ function.}
\label{fig:trapsurftop_qmod_A5s01t0}
\end{figure}

\begin{figure} \centering
\includegraphics{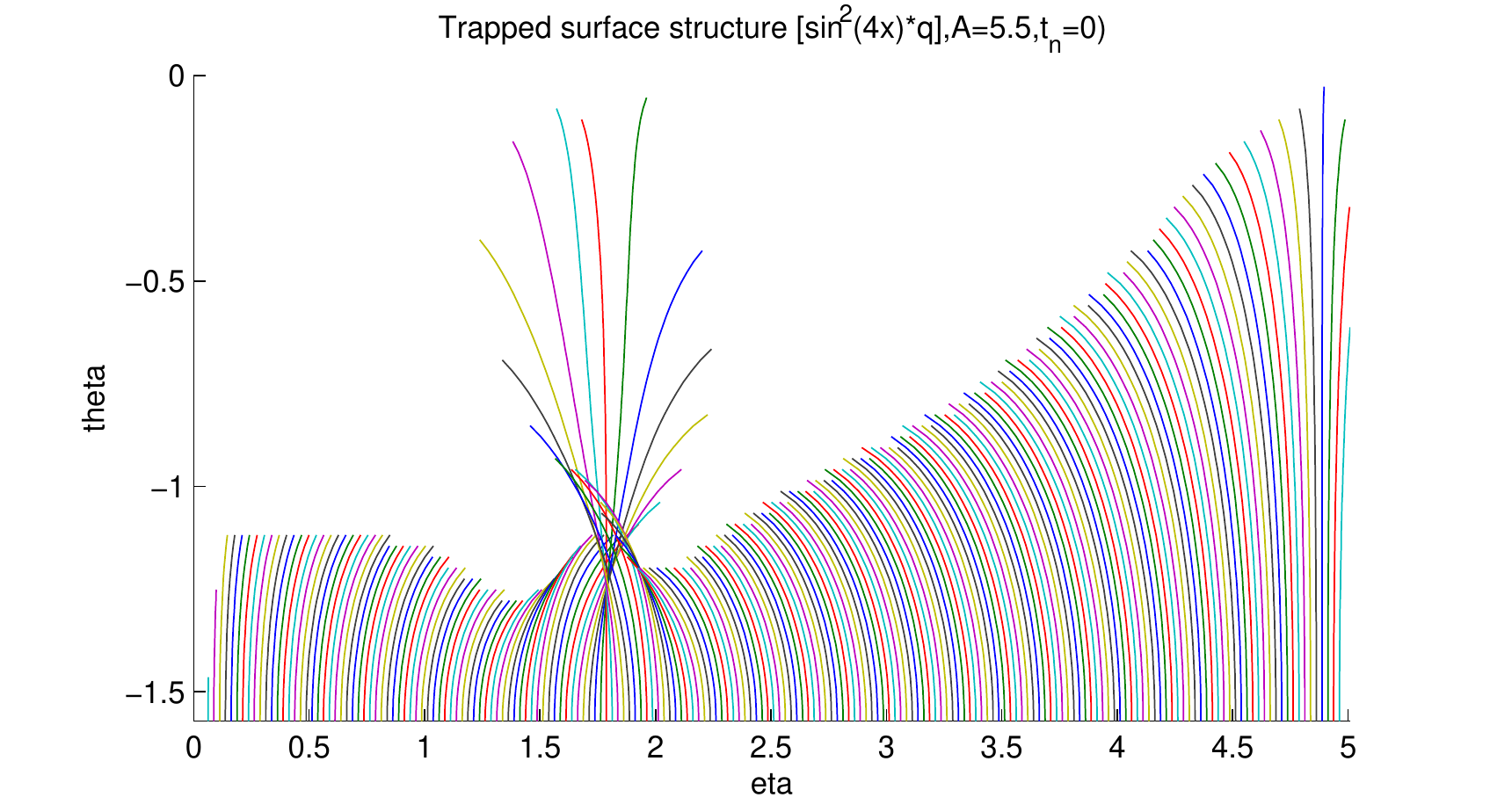}
\caption[Trapped surface topology for modulated $q$, $A=5.5$]{Trapped surface topology for $(A=5.5,s_0=1,t=0)$ and the $q$ function given by equation (\ref{eqn:qinitsinmod}).  Note the similarity to the trapped surface topology of the non-modulated $q$ function and how the outermost trapped surface is leaving the computational grid, in line with \'O Murchadha's predictions and the observations from our other solutions.  Reference to figure \ref{fig:trapsurftop_qmod_A5s01t0}. }
\label{fig:trapsurftop_qmod_A5.5s01t0}
\end{figure}

\begin{figure} \centering
\includegraphics{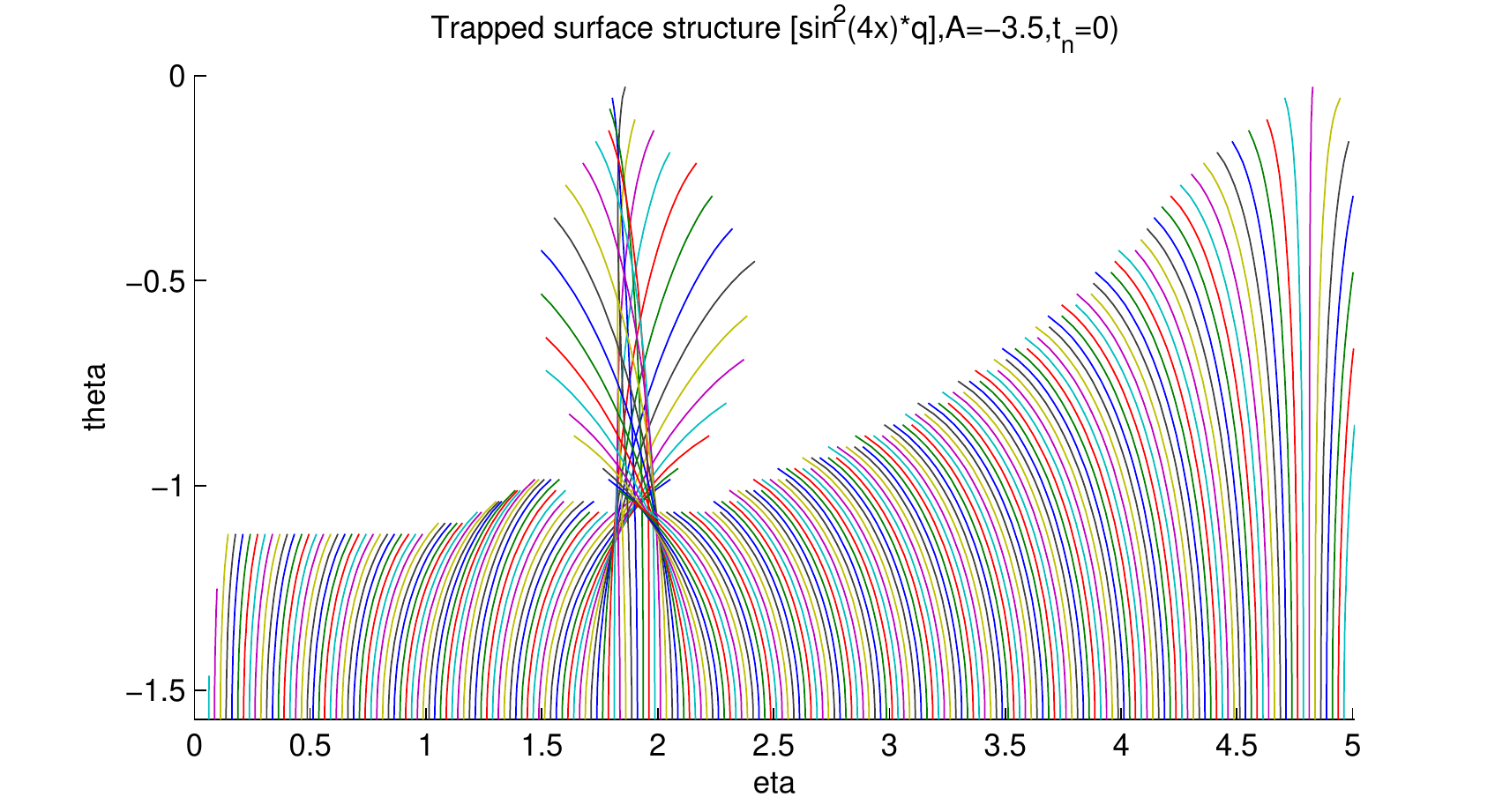}
\caption[Trapped surface topology for modulated $q$, $A=-3.5$]{Trapped surface topology for $(A=-3.5,s_0=1,t=0)$ and the $q$ function given by equation (\ref{eqn:qinitsinmod}).  Note the similarity to the trapped surface topology of the non-modulated $q$ function and how the outermost trapped surface is leaving the computational grid, in line with \'O Murchadha's predictions and the observations from our other solutions.  Reference to figure \ref{fig:trapsurftop_qmod_A5.5s01t0} which shows the same phenomenology with a near-critical positive amplitude. }
\label{fig:trapsurftop_qmod_A-3.5s01t0}
\end{figure}

If we apply a modulation to $q$ via our radial coordinate to give
\begin{equation}\label{eqn:qinitsinmod}q(\eta,\theta,t=0)= \sin^2(4\eta)\left[2 A f^4e^{-\left(\frac{f}{s_0}\right)^2}\sin^2\theta\right]\end{equation}
we can see some of the resulting trapped surface topology in figures \ref{fig:trapsurftop_qmod_A5s01t0} ($A=5$), \ref{fig:trapsurftop_qmod_A5.5s01t0} ($A=5.5$) and \ref{fig:trapsurftop_qmod_A-3.5s01t0} ($A=-3.5$) which mirror the phenomenology we observed with our Gaussian $q$ function.  As we approach $A_-$ or $A_+$ the outermost trapped surface moves further out and eventually leaves the computational domain.

\begin{figure} \centering
\includegraphics{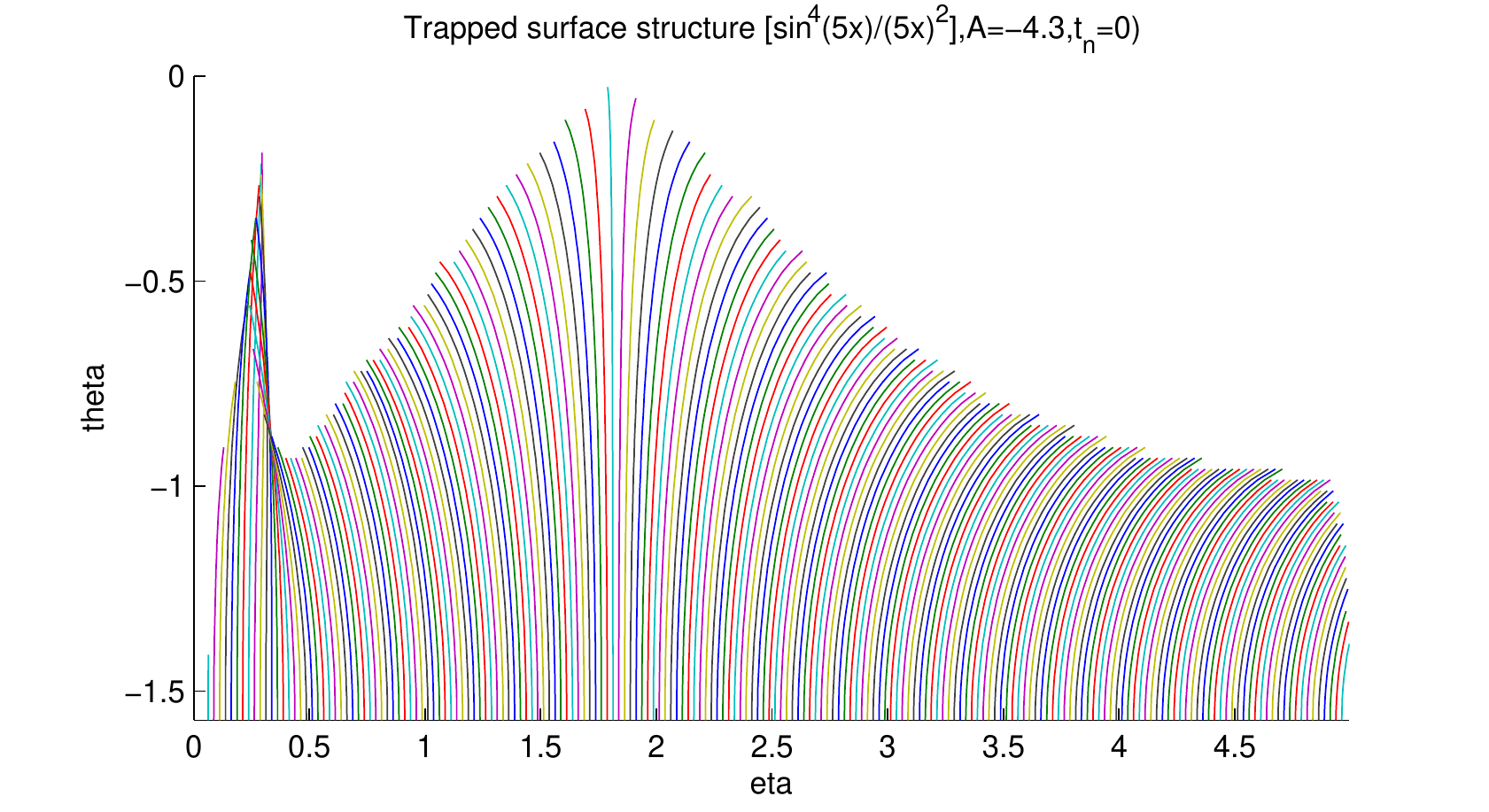}
\caption[Trapped surface topology for trig/poly $q$, $A=-4.3$]{Trapped surface topology for $(A=-4.3,s_0=1,t=0)$ and the $q$ function given by equation (\ref{eqn:qinitsinpoly}).  Note the similarity to the trapped surface topology of the other $q$ functions.}
\label{fig:trapsurftop_qsinpoly_A-4.3s01t0}
\end{figure}

\begin{figure} \centering
\includegraphics{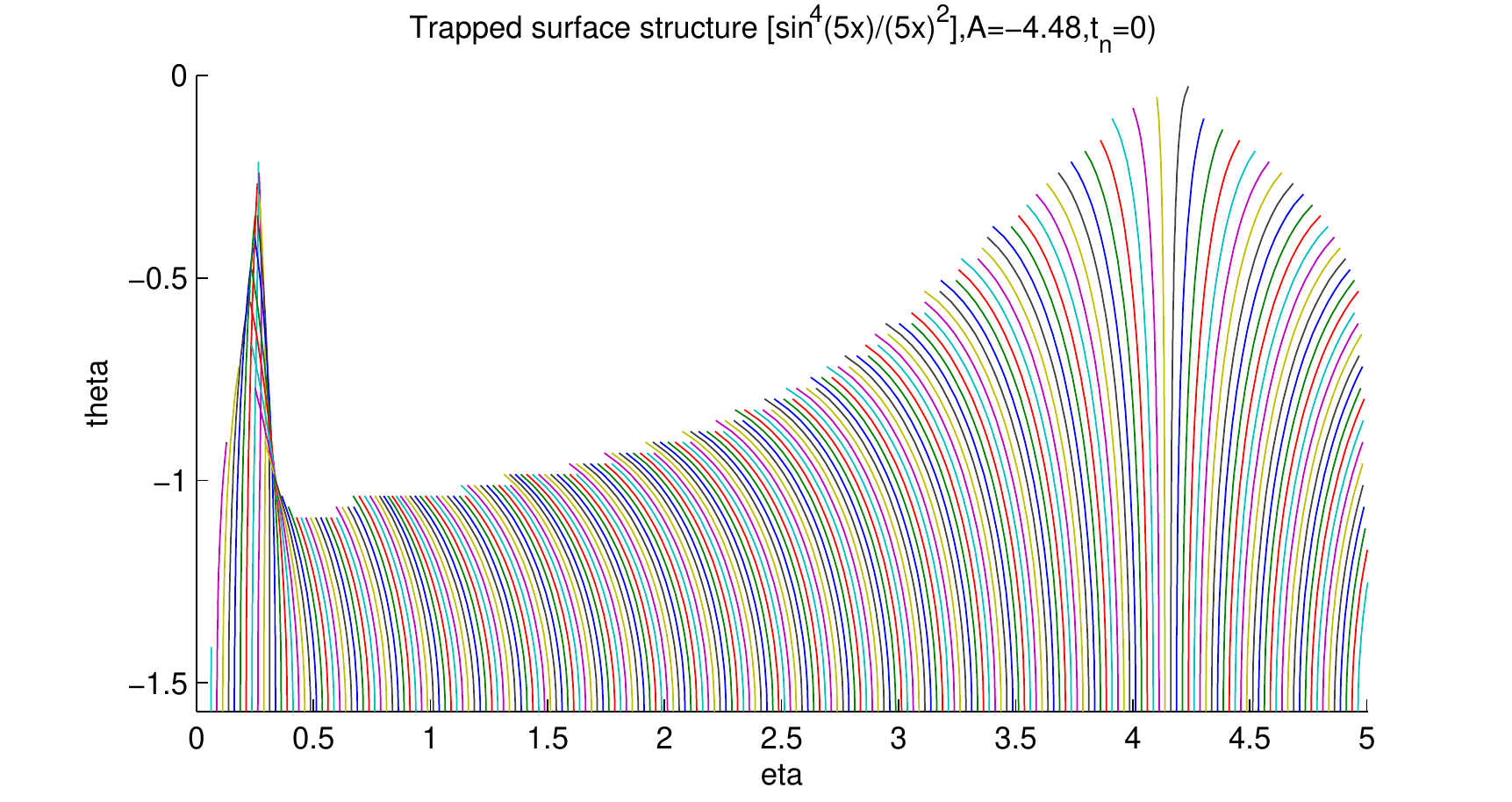}
\caption[Trapped surface topology for trig/poly $q$, $A=-4.48$]{Trapped surface topology for $(A=-4.48,s_0=1,t=0)$ and the $q$ function given by equation (\ref{eqn:qinitsinpoly}).  Note the outer trapped surface moving $\rightarrow \infty$ and off the edge of the grid.  Compare to figure \ref{fig:trapsurftop_qsinpoly_A-4.3s01t0}}
\label{fig:trapsurftop_qsinpoly_A-4.48s01t0}
\end{figure}

\begin{figure} \centering
\includegraphics{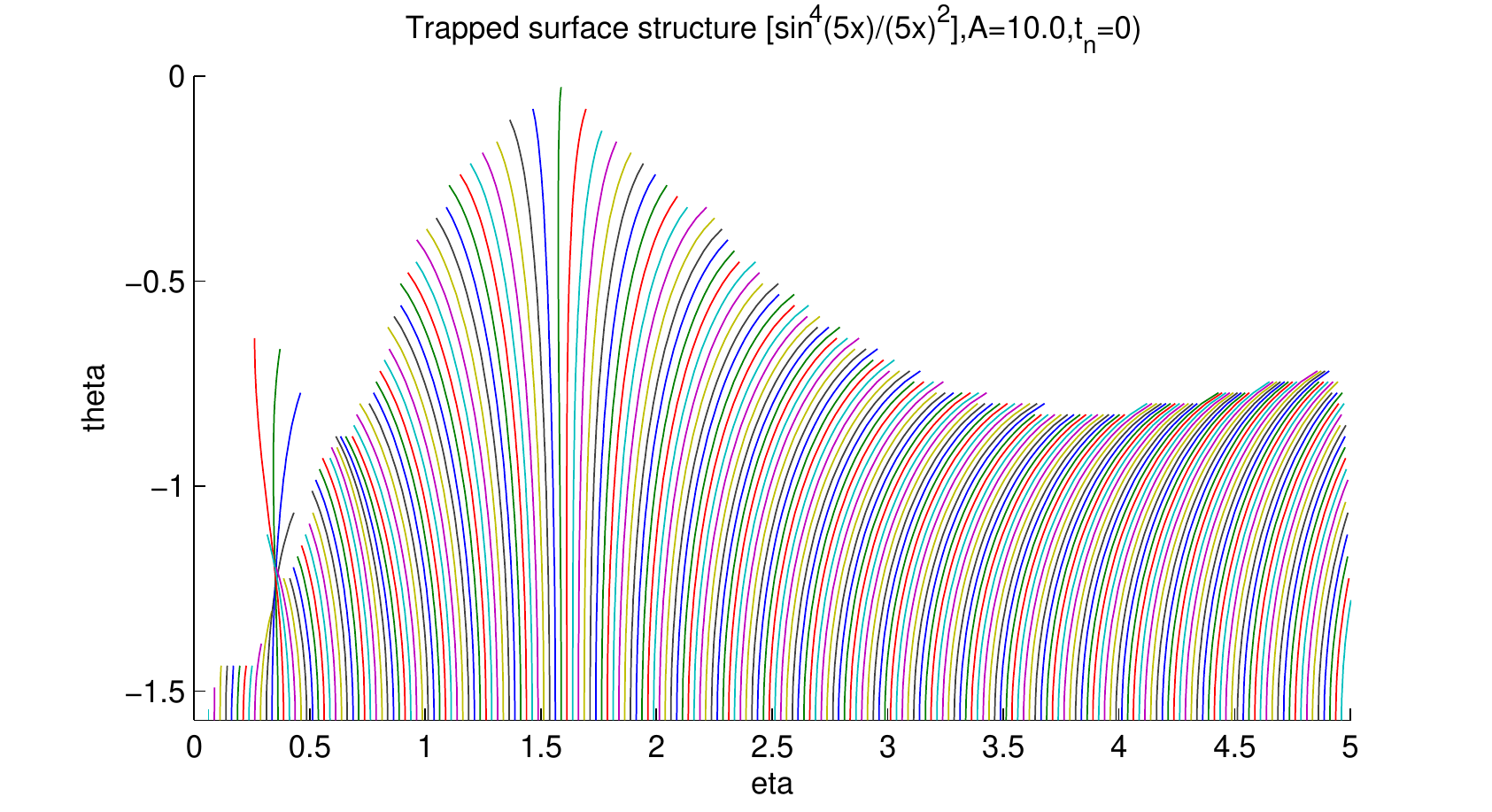}
\caption[Trapped surface topology for trig/poly $q$, $A=10.0$]{Trapped surface topology for $(A=10.0,s_0=1,t=0)$ and the $q$ function given by equation (\ref{eqn:qinitsinpoly}).  Note the similarity to the trapped surface topology of the other $q$ functions.}
\label{fig:trapsurftop_qsinpoly_A10.0s01t0}
\end{figure}

If instead we use a $q$ function with the non-exponential form
\begin{equation}\label{eqn:qinitsinpoly}q(\eta,\theta,t=0)=2 A \left(\frac{\sin^4(5\eta)}{(5\eta)^2}\right)\sin^2\theta\end{equation}
we see the trapped surface topologies show in figures \ref{fig:trapsurftop_qsinpoly_A-4.3s01t0} ($A=-4.3$), \ref{fig:trapsurftop_qsinpoly_A-4.48s01t0} ($A=-4.48$) and \ref{fig:trapsurftop_qsinpoly_A10.0s01t0} ($A=10$).  We see the same phenomenology that was observed with the two previous $q$ functions, indicating that we are observing the global phenomena described by \'O Murchadha.

This demonstrates that various $q$ functions of vastly different shapes, extents and periodic behaviours all exhibit the same amplitude dependent behaviour:
\begin{itemize}
\item critical minimum ($A_-<0$) amplitude for which $A<A_-$ has no solution for the Hamiltonian constraint as the initial apparent horizon $\rightarrow \infty$
\item critical maximum ($A_+>0$) amplitude for which $A>A_+$ has no solution for the Hamiltonian constraint as the initial apparent horizon $\rightarrow \infty$
\item $2$ regions (one positive, one negative) of phase space where an apparent horizon is present in the initial value problem, i.e. $$A_-<A<A_{AH-}<0$$ and $$A_+>A>A_{AH+}>0$$
\item a region of phase space which spans $A=0$ (flat space) with no outermost trapped surface (apparent horizon) present on the initial slice i.e. $$A_{AH-} < A < A_{AH+}$$
\end{itemize}

\begin{figure} \centering
\includegraphics{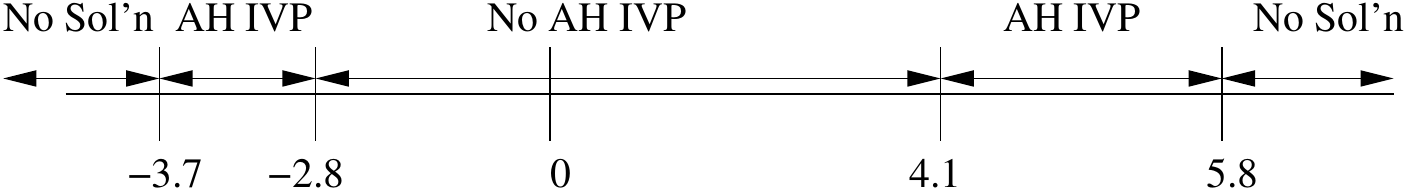}
\caption[IVP phase space (modulated $q$)]{IVP phase space in the amplitude $A$ of the modulated metric function $q$ as given by equation (\ref{eqn:qinitsinmod}).  Regions are divided roughly where the IVP has or does not have Apparent Horizons present, and where the code cannot solve the IVP as we have passed $A_{\pm}$} \label{fig:IVPphaseqa_sinmod}
\end{figure}

\begin{figure} \centering
\includegraphics{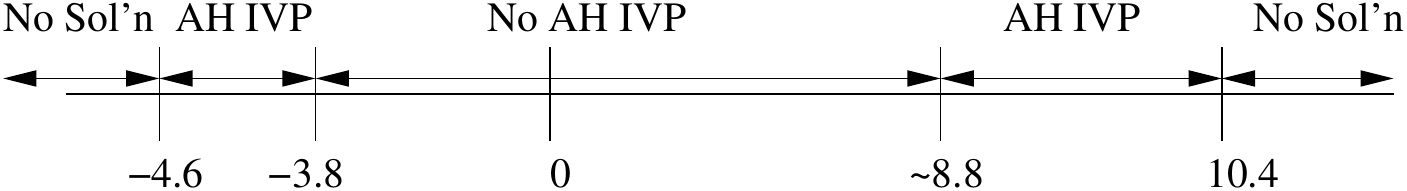}
\caption[IVP phase space (trig/poly $q$)]{IVP phase space in the amplitude $A$ of the trig/poly metric function $q$ as given by equation (\ref{eqn:qinitsinpoly}).  Regions are divided roughly where the IVP has or does not have Apparent Horizons present, and where the code cannot solve the IVP as we have passed $A_{\pm}$} \label{fig:IVPphaseqa_polytrig}
\end{figure}

We summarise our findings of the IVP phase space in figures \ref{fig:IVPphaseqa_sinmod} (modulated $q$ function) and \ref{fig:IVPphaseqa_polytrig} (poly/trig $q$ function).  Compare to figure \ref{fig:IVPphaseqa} for the Gaussian $q$ function.

These results are also consistent with \'O Murchadha \cite{Murchadha} and provide strong confirmation that we are observing physical and not purely numerical effects. It also confirms that the trapped surface solver is performing as expected and that the trapped surface topology in and of itself is useful in understanding the physics in these spacetimes.  They also indicate that black holes are only so ``strong'' - i.e. there is a critical shear\footnote{Between the $(\eta,\theta)$ plane and $\varphi$} in the spacetime curvature caused by $q$ that cannot be made stronger by a bigger black hole/wave amplitude, as the black hole interior already encompasses the entire spacetime.  There seems to be a larger physical result buried in here related to the capacity of an anisotropic function like $q$ to cause stretching of the spacetime, however that is a topic for future investigation.

We do not, however, observe periodicity or other extensions of the amplitude phase space that \'O Murchadha conjectures in his paper, as numerical investigation of the amplitude phase space well beyond $A_\pm$ yielded no solutions to the IVP\footnote{This can also be understood via the \emph{positive} definiteness of the mass as proven by Brill, and no naked singularities were discovered.}.

Performing the time evolution of these alternate forms for $q$ yields results similar to the Gaussian $q$ function given in equation (\ref{eqn:qinit}) and discussed in section \ref{sec:results} (and is consistent with the theoretical framework presented later).

\section{Black Hole and Spacetime Measures}
We will now discuss a few measures of spacetimes that are commonly used to characterise black hole systems and how they apply to the Brill spacetimes presented herein.  These measures will aid in understanding the structure of the spacetime both on the initial slice and as it evolves.

\subsection{Horizon Areas}\label{subsec:horizonarea}
The surface area of a black hole \emph{event} horizon is known to increase or stay constant over time (it cannot decrease), so let us calculate the area of the \emph{apparent} horizon which must be inside the event horizon.  

\begin{figure} \centering
\includegraphics{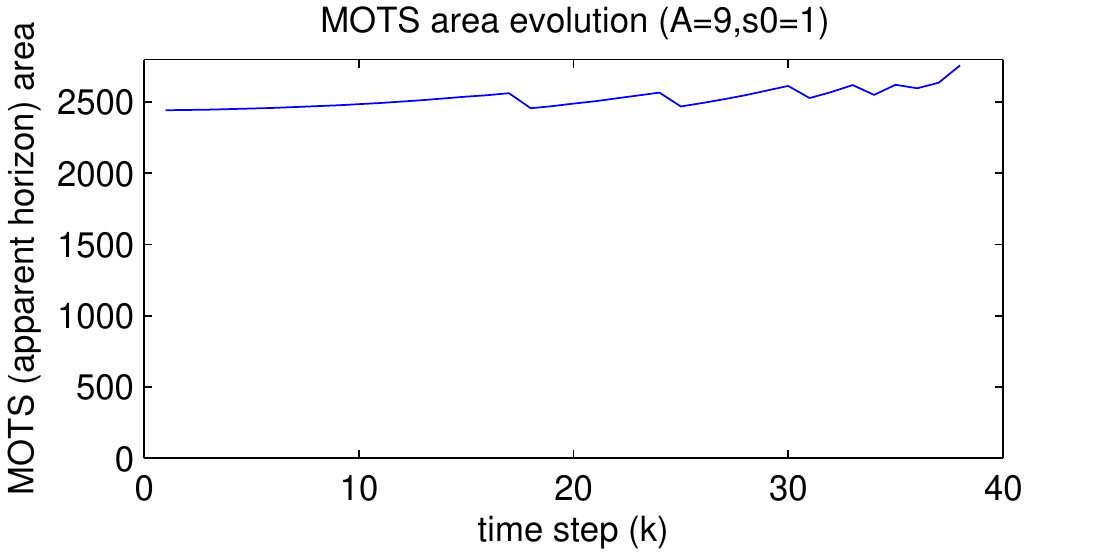}
\caption[Area of MOTS for $A=9$ at various times]{Area of the outermost trapped surface (apparent horizon) for $(A=9,s_0=1,\eta_{\mathrm{max}}=5)$ (see equation \ref{eqn:apphorarea}).}\label{fig:apphorarea_a9s01}
\end{figure}

\begin{figure} \centering
\includegraphics{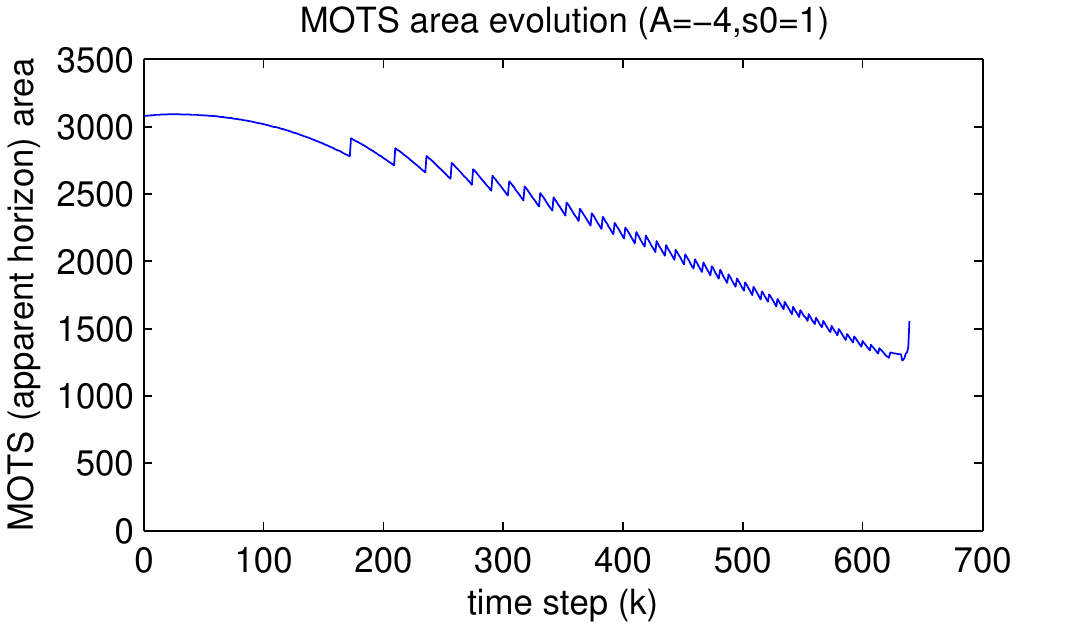}
\caption[Area of MOTS for $A=-4$ at various times]{Area of the outermost trapped surface (apparent horizon) for $(A=-4,s_0=1,\eta_{\mathrm{max}}=5)$ (see equation \ref{eqn:apphorarea}).}\label{fig:apphorarea_a_4s01}
\end{figure}

\begin{figure} \centering
\includegraphics{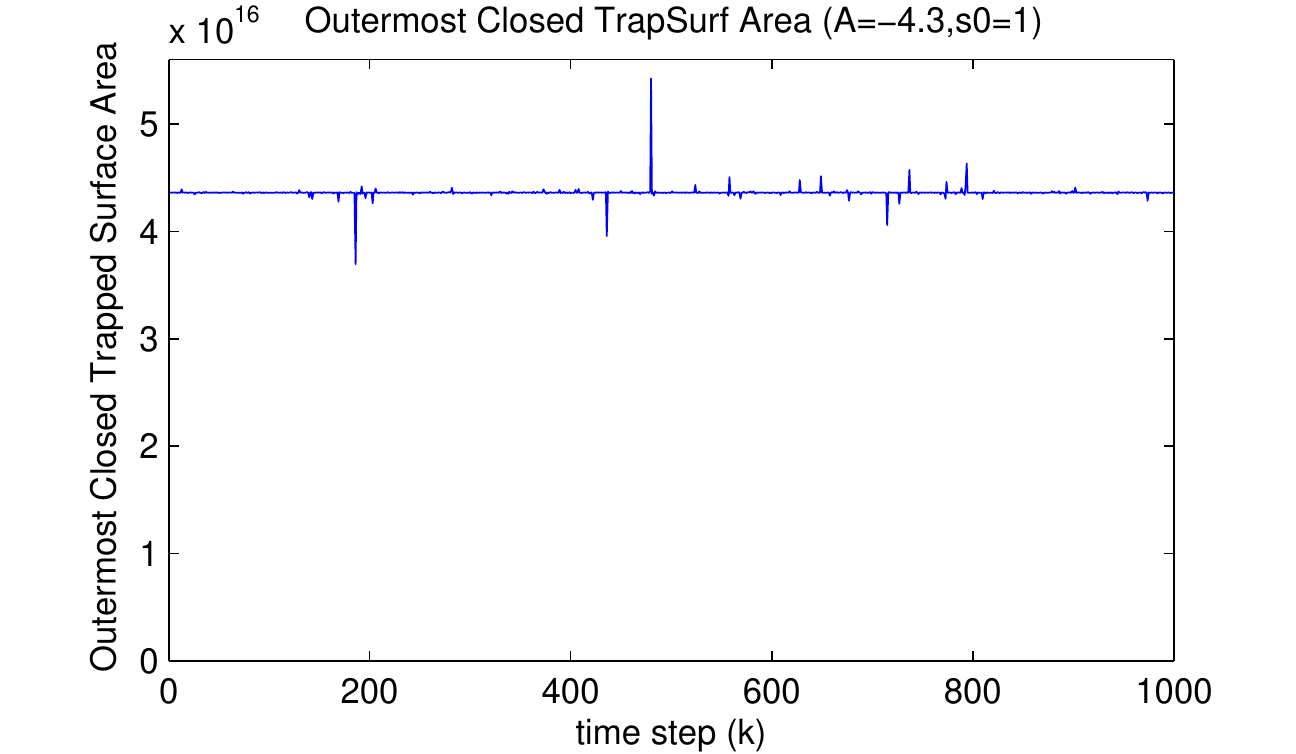}
\caption[Area of MOTS for $A=-4.3$ at various times]{Area of the outermost trapped surface that is present in the computational domain with $\eta_{\mathrm{max}}=5$, which is not the apparent horizon, for $(A=-4.3,s_0=1)$ (see equation (\ref{eqn:apphorarea})).  If we set the outer boundary at $\eta_{\mathbf{max}}=20$ we find a MOTS (which is the apparent horizon) around $\eta \sim 6.6$ and the code forms a singularity in a few time steps.}\label{fig:apphorarea_a_4.3s01}
\end{figure}

For the majority of the evolutions presented the apparent horizons stretch across $\eta \sim \mathrm{constant}$, with a usual deviation of a couple percent, and maximum deviation of $\sim 10 \%$.  We know that if we hold $\eta$ constant that the area of the apparent horizon is given by
\begin{eqnarray}\label{eqn:apphorarea}
\mathcal{A} &=& \int dl_\theta dl_\phi \nonumber \\ \mbox{}
 &=& \int \sqrt{\gamma_{22}}d\theta \;\sqrt{\gamma_{33}}d\phi \nonumber \\ \mbox{}
 & = & \int \left(e^{q/2+2\phi} f d\theta\right) \left(e^{2\phi} f \sin\theta d\phi\right) \nonumber \\ \mbox{}
 & = & f^2 \int^{2\pi}_0 \int_0^{\pi} e^{\frac{q}{2}+4\phi} \sin\theta d\theta d\phi \nonumber \\ \mbox{}
& =& 4 \pi f^2 \int_0^{\frac{\pi}{2}} e^{\frac{q}{2}+4\phi} \sin\theta d\theta
\end{eqnarray}
with our symmetry conditions.  The area of the apparent horizon over time for a large positive amplitude wave is shown in figure \ref{fig:apphorarea_a9s01} and for a large negative amplitude wave in figures \ref{fig:apphorarea_a_4.3s01} and \ref{fig:apphorarea_a_4s01}.  Non-monotonicity of the values of the areas are attributable to the fact that we do not locate the horizon precisely so it will oscillate slightly as the trapped surfaces move around and one eventually comes closer to the axis than any other\footnote{This effect can be observed in any of the videos showing trapped surface evolution - the trapped surfaces look like grass waving in the wind and an adjacent trapped surface will evolve to become the new longest surface as the horizon evolves.}.

The decrease in the apparent horizon (AH) area for the one negative amplitude wave could be due to:
\begin{itemize}
\item an event horizon that isn't decreasing in area but an AH that is as they are not guaranteed to coincide
\item the formation of another AH outside the computational domain
\item the fact that $\eta$ isn't constant across the horizon so our area calculations from equation (\ref{eqn:apphorarea}) aren't quite correct\footnote{These errors are in the $\sim 1\%$ range.}
\item interpolation errors
\end{itemize}
We would need to study the global event horizon structure to meaningfully comment on the reason for this, however the first scenario is the most likely as it explains larger deviations over the entire evolution.

\subsection{Proper Radial Distance Embeddings}\label{subsec:radstdist}
\begin{figure} \centering
\includegraphics{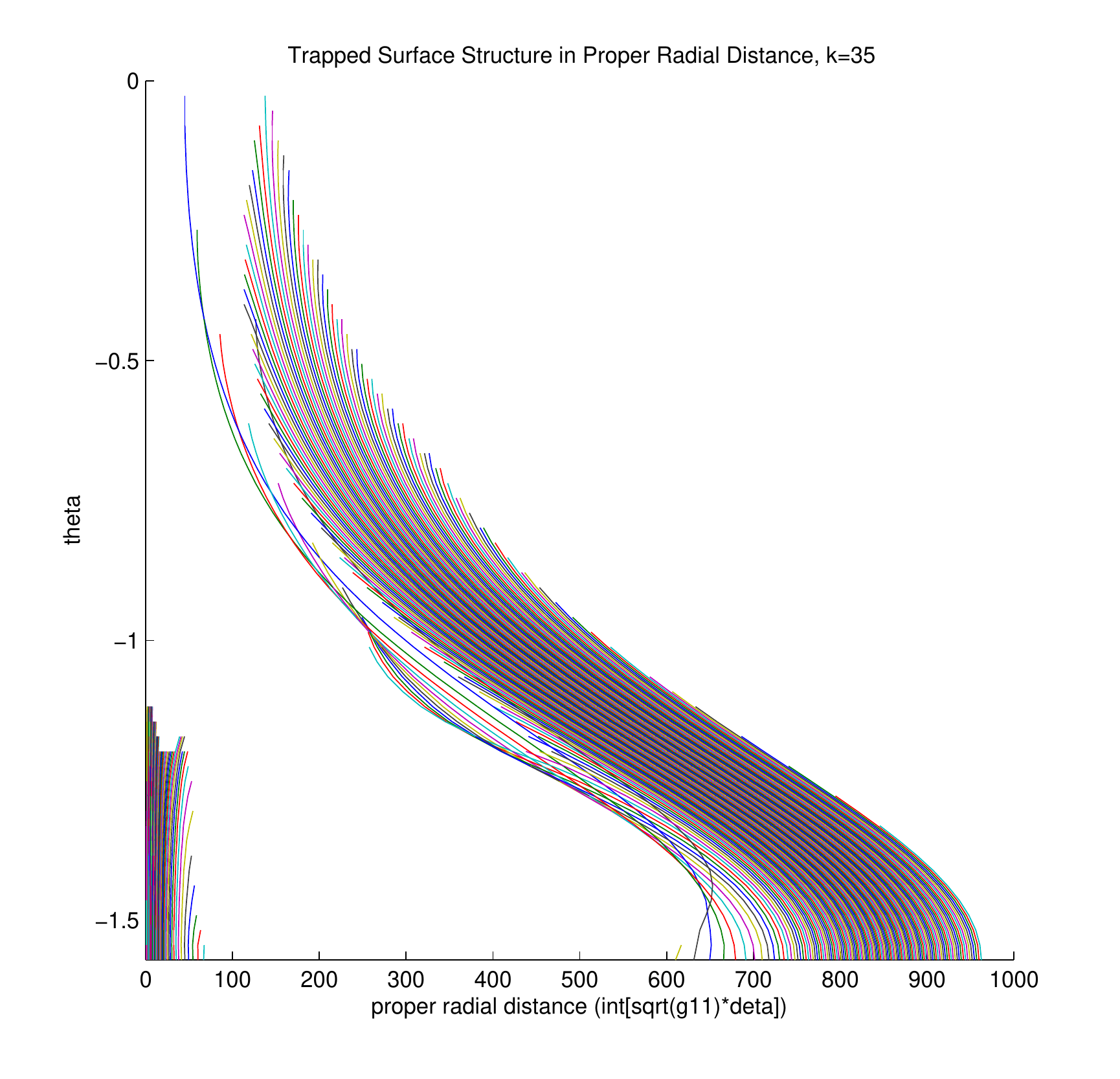}
\caption[Trapped Surfaces for $A=9$ in $\bar{r}$]{Trapped Surface structure at $k=35$ for a strong initial wave ($A=9$,$s_0=1$) transformed into proper radial distances $\bar{r}$ as in equation (\ref{eqn:spacetimerdist}).  $\theta=0$ corresponds to the axis, and $\theta=\pm \frac{\pi}{2}$ corresponds to the equator (which has proportionally larger distances $\bar{r}$ for the surfaces above).}
\label{fig:trapsurft0a9radspacetime}
\end{figure}

Let us consider the mapping of the trapped surface evolution resulting from figure \ref{fig:trapsurft0a9annotated} into a proper radial distance coordinate $\bar{r}$ where we set
\begin{equation}\label{eqn:spacetimerdist}\bar{r}=\int_{0}^{h(\theta)} \sqrt{g_{11}} d\eta\end{equation}
When we plot the resulting graph we see that our $\eta$ coordinates get transformed into $\bar{r}$ coordinates that look drastically different (see figure \ref{fig:trapsurft0a9radspacetime}).

\begin{figure} \centering
\includegraphics{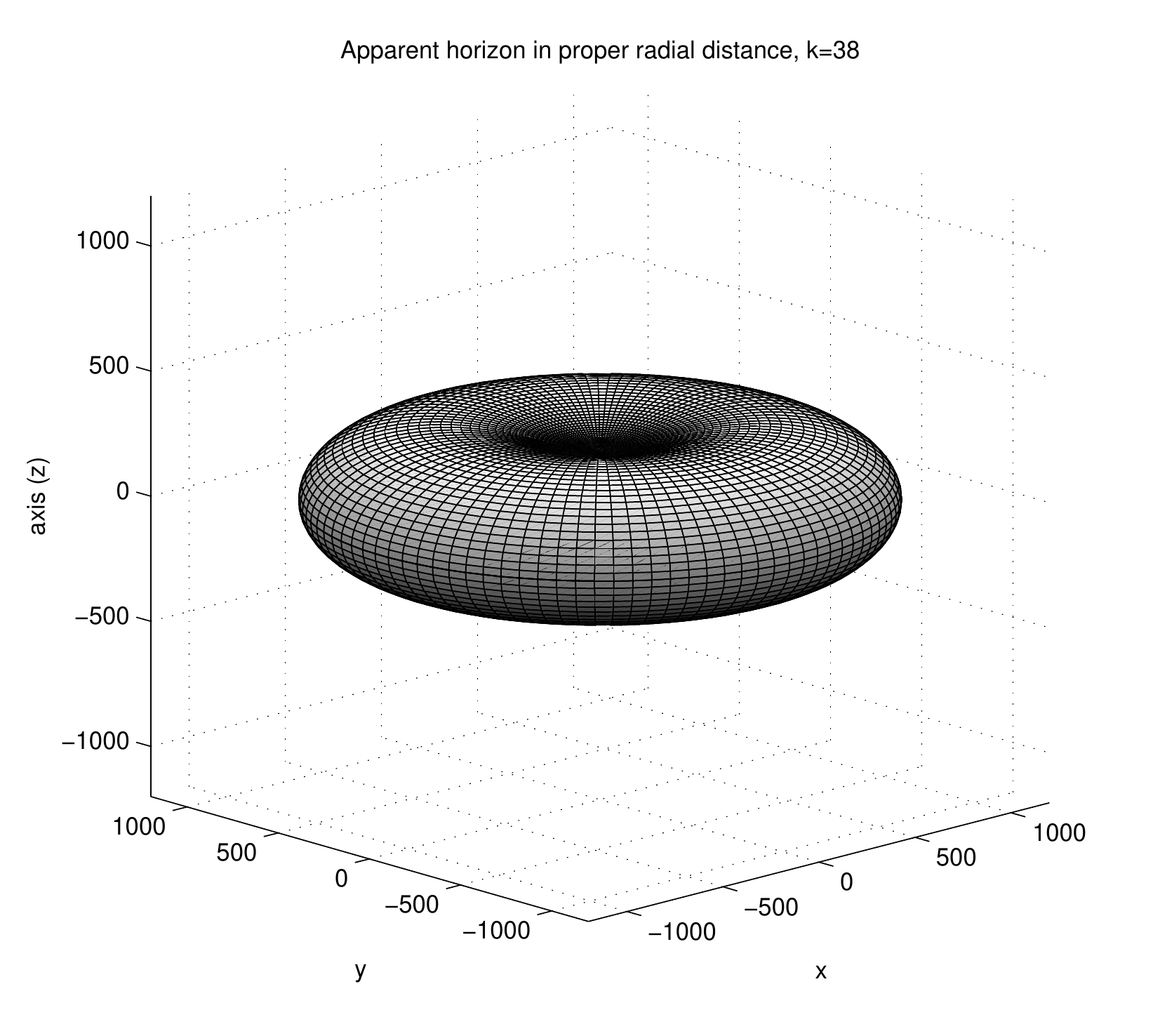}
\caption[Spacetime embedding of MOTS for $A=9$]{Embedding of the outermost trapped surface (apparent horizon) for $(A=9,s_0=1)$ in $\bar{r}$ proper radial distances at $k=38$. (see equation \ref{eqn:spacetimerdist})}\label{fig:apphorradst_a9s01t38}
\end{figure}

If we now use this methodology to map the topology of the outermost \emph{closed} trapped surface in proper radial distances $\bar{r}$ and set
$$x=\bar{r}\sin\theta \;;\; z=\bar{r}\cos\theta$$
then mirror the resulting curve in the $x-y$ plane, and rotate the result around the $z$ axis we end up with the topology show in figure \ref{fig:apphorradst_a9s01t38}.  Note the ``pancake/hemoglobin'' nature of the apparent horizon (which is observed for all large \emph{positive} $q$ cases), and the horizon grows as the evolution progresses.

\begin{figure} \centering
\includegraphics{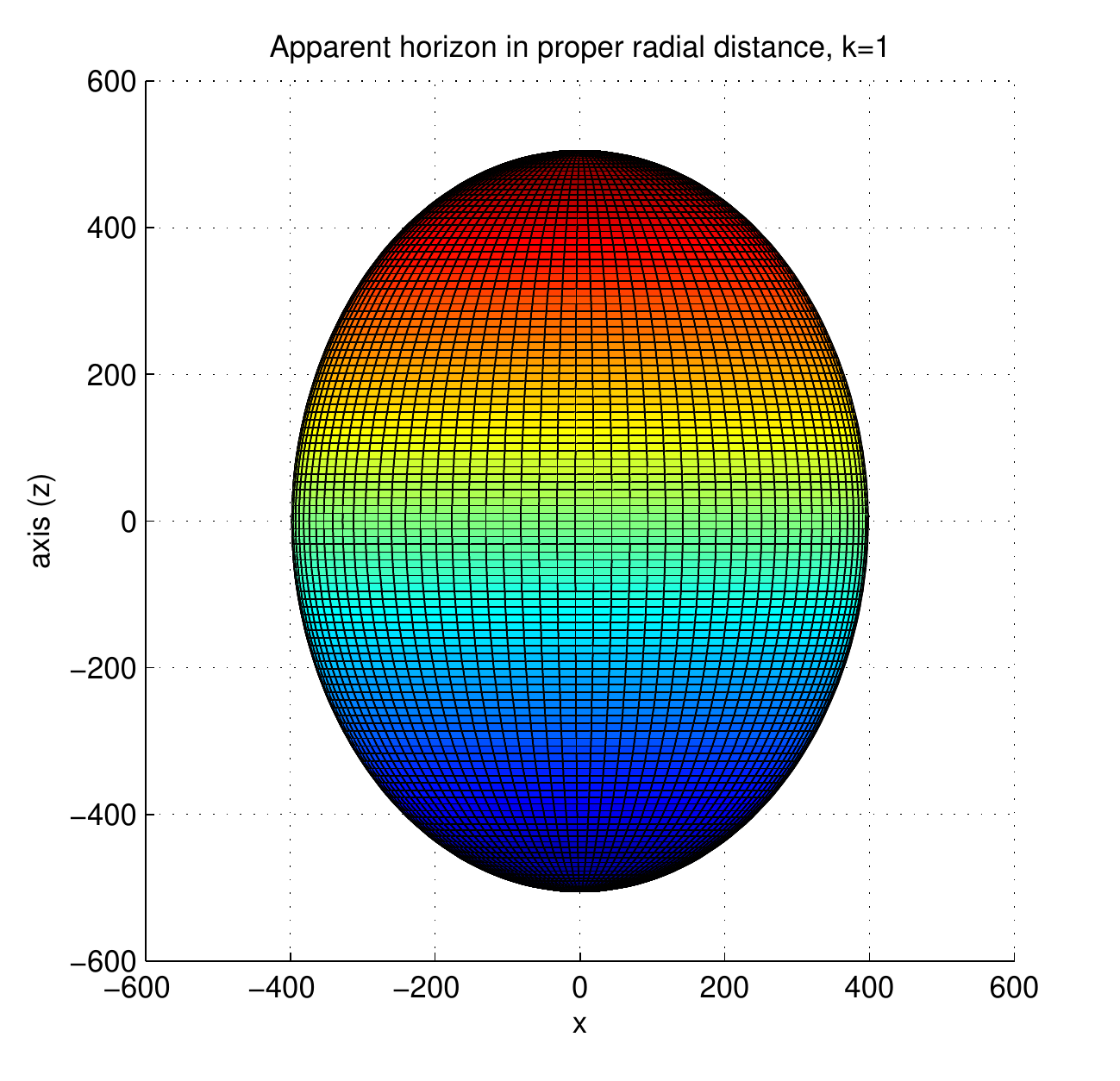}
\caption[Spacetime embedding of MOTS for $A=-4$ at $t=\Delta t$]{Embedding of the outermost trapped surface (apparent horizon) for $(A=-4,s_0=1)$ in $\bar{r}$ proper radial distances at $t=\Delta t (k=1)$. (see equation \ref{eqn:spacetimerdist})}\label{fig:apphorradst_a_4s01t1}
\end{figure}

For large \emph{negative} $q$ values, if we once again map out the topology of the outermost closed trapped surface (apparent horizon) we arrive at a totally different global topology.  See figure \ref{fig:apphorradst_a_4s01t1}, where we have an prolate ellipsoid that shrinks with time.

\subsection{ADM Mass and Mass Aspect}\label{subsec:massresults}
As one measure of the ``energy'' of the system, we calculate the (quasi-local) ADM Mass as described in section \ref{subsec:massmeasure}.  Although as noted there, if the radial fall-off of the conformal factor $\phi$ is given by
$$\phi(\eta\rightarrow\infty) \sim \frac{1}{r^n} \;;\; n>1$$
to leading order at the outer radial edge of the grid there will no calculable contribution to the ADM mass and it will be zero (or ``close'' to zero in the case of a finite grid size).

Recalling that for Brill waves $q$ has the initial asymptotic form
$$q(\eta\rightarrow\infty) \sim \frac{1}{r^n} \;;\; n > 1,t=0$$
we see that it produces no contribution to the ADM mass on the initial slice.

As the spacetime evolves, there will be variable contributions from the extrinsic curvature, etc. to the asymptotic form of $\phi$ that cause the quasi-local ADM mass at the edge of the grid to evolve as we have a non-infinite outer boundary.

\begin{figure} \centering
\includegraphics{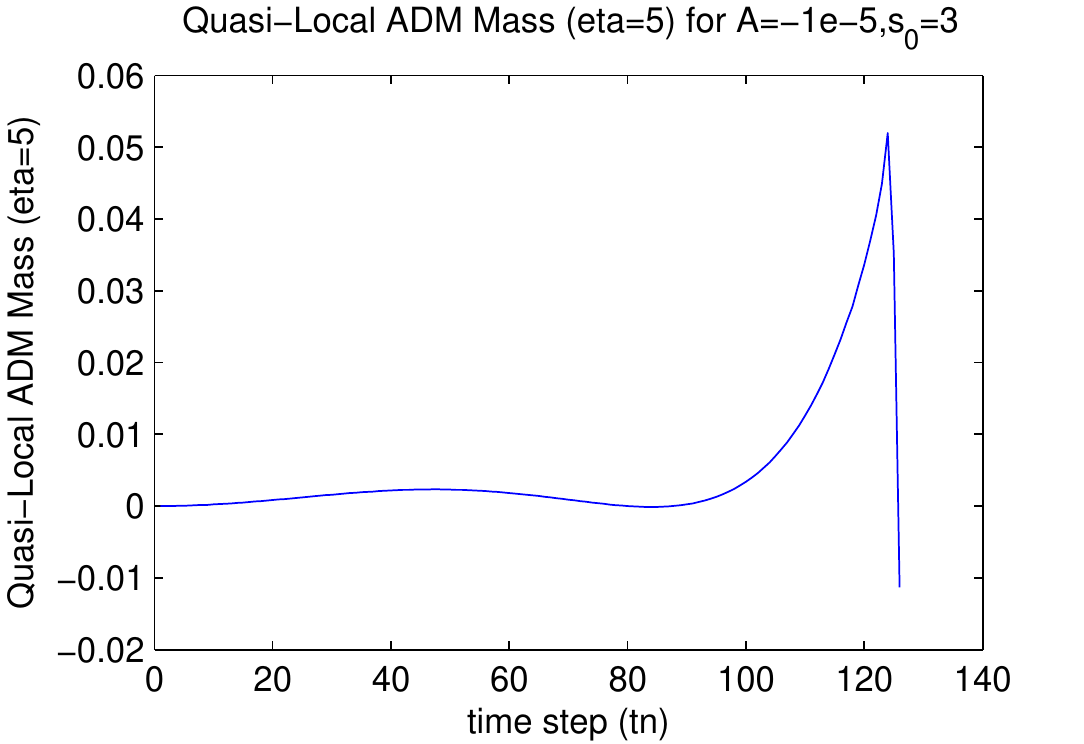}
\caption[QL ADM Mass over time ($A=-1e-5,s_0=3$)]{Quasi-local ADM Mass measured at $\eta=5$ for $(A=-1 \times 10^{-5},s_0=3)$ as discussed in section \ref{subsec:massresults}.  The region around $k=80\rightarrow 90$ has small negative values, and the last two time steps just before singularity formation give the discontinuity at the far right of the graph} \label{fig:admmass_t127_a-1e-5so3}
\end{figure}

A sample plot of the quasi-local ADM mass measured at the radial edge of the computational grid as we progress through time can be seen in figure \ref{fig:admmass_t127_a-1e-5so3}.  The mass does briefly turn negative around $k=80$ as demonstrated in table \ref{tbl:admmass_a-1e-5s03}, then goes positive again and reaches a rapid growth rate around $t\sim 100$.  The very small negative mass is probably due to some portion of the curvature propagating off the grid, and the last bit of the time evolution looks like the contribution from a large wave passing by the ``mass measurement device'' at the edge of the grid.

Further, as constraint violations are generally monotonically increasing during the evolution they would not produce contributions that variably increase and decrease the quasi-local ADM mass over time.   Therefore the violation of the momentum constraints isn't just ``adding energy'' to the system and causing collapse.

Around the point where the code encounters a singularity we see a sharp discontinuity in the quasi-local ADM mass (similar to those seen in the Hamiltonian/momentum constraint evolution as well as in various other measures).

\begin{table}\begin{center}
\begin{tabular}{|c|l|}\hline
Time step $k$ & QL ADM Mass \\ \hline
78 &	0.00015071920818 \\
79 &	8.05035136458513E-05 \\
80 &	1.93751990739007E-05 \\
81 &	-3.13389696646122E-05 \\
82 &	-7.02396803792265E-05 \\
83 &	-9.58476186526466E-05 \\
84 &	-0.00010660669498 \\
85 &	-0.00010087322745 \\
86 &	-7.69178704180812E-05 \\
87 &	-3.29214341187539E-05 \\
88 &	3.30331405863194E-05 \\
89 &	0.0001229795619 \\ \hline
\end{tabular}\caption[Q-L ADM Mass measured at $\eta=5$ over time]{Quasi-Local ADM Mass measured at $\eta=5$ for select time steps as described in section \ref{subsec:massresults} $(A=-1\times 10^{-5})$.} \label{tbl:admmass_a-1e-5s03}
\end{center} \end{table}

\begin{figure} \centering
\includegraphics{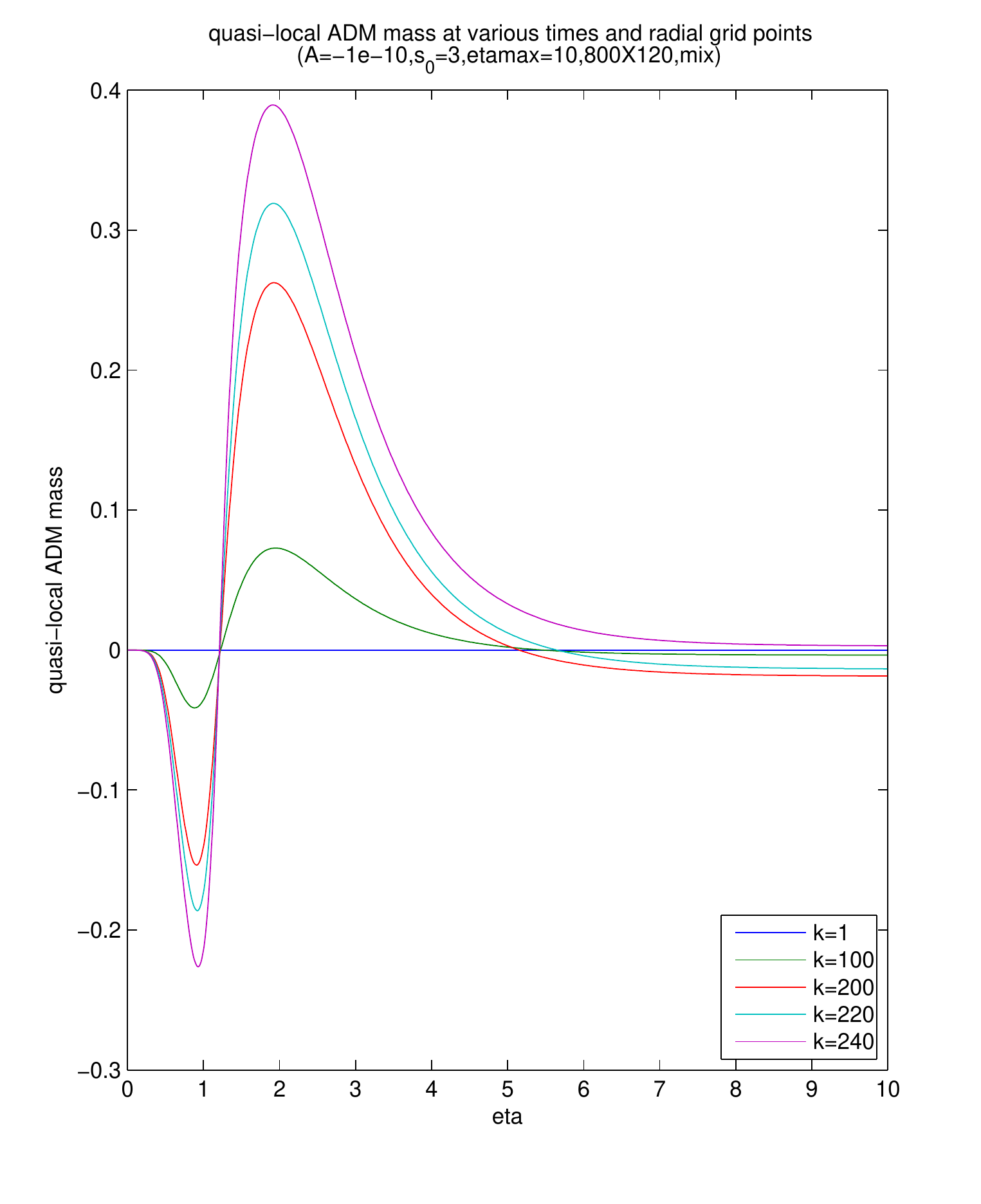}
\caption[QL ADM Mass at various times ($A=-1 \times 10^{-10},s_0=3$)]{Quasi-Local ADM Mass at varying time steps $k$ measured at each radial grid point for $(A=-1 \times 10^{-10},s_0=3)$ as discussed in section \ref{subsec:massresults}.  Note how the asymptotic values are both positive and negative at various time steps and still not converging at $\eta \sim 10 \rightarrow r \sim 11,000$.} \label{fig:admmass_composite_a-1e-10so3}
\end{figure}

A plot of the quasi-local ADM mass for a perturbative wave as a function of $\eta$ for different times throughout the evolution as measured using equation (\ref{eqn:admmass}) can be seen in figure \ref{fig:admmass_composite_a-1e-10so3}.  This shows that as the perturbative wave $(A=-1 \times 10^{-10},s_0=3)$ evolves that even $\eta=10$ is not large enough\footnote{Recalling that $r=\sinh(\eta)$, this is $r \sim 11,000$.} to approximate ``infinity'', and also that the locally measured ADM mass at the edge of the grid oscillates from positive to negative values.  We see similar effects in all simulations, with positive and negative deviations from the IVP ``baseline'' as the evolution progresses.  Some videos showing the full time evolution of Quasi-Local ADM mass measurements for various IVPs are included in the links in table \ref{tbl:videoresults}.  This indicates that the ``near zone'' is the entire computational domain, and we cannot extend the grid out much further so as to model the ``wave zone'' without reaching the limits of $64$ bit numerical precision.

\begin{figure} \centering
\includegraphics{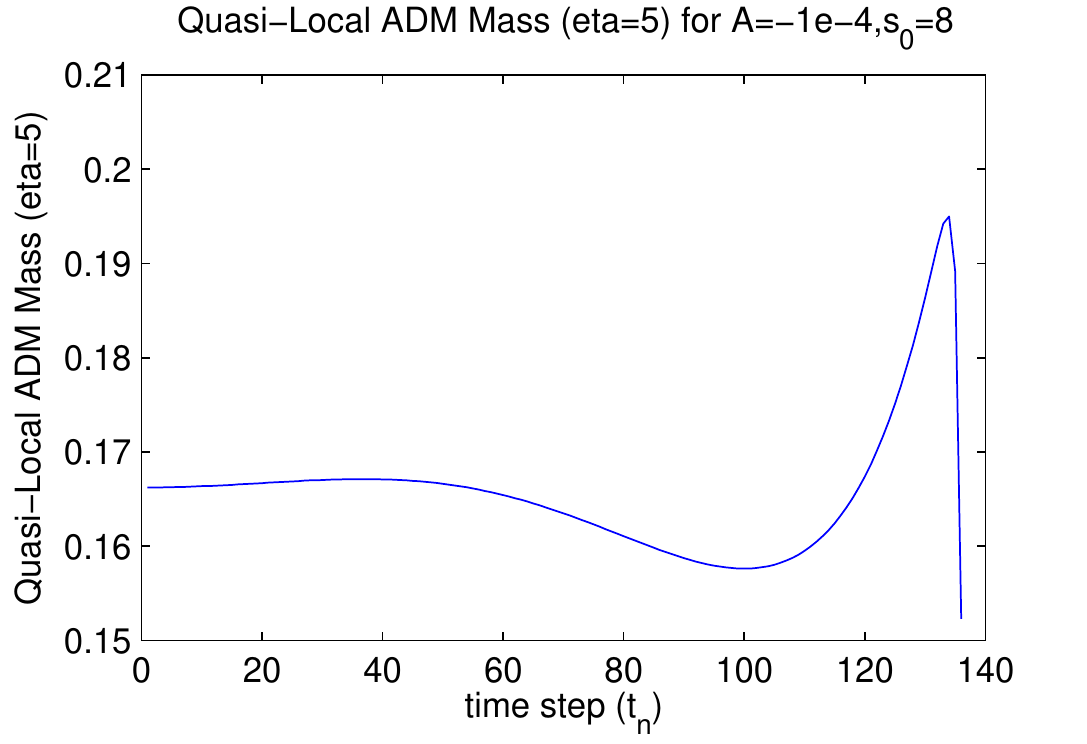}
\caption[QL ADM Mass over time ($A=-1 \times 10^{-4},s_0=8$)]{Quasi-Local ADM Mass measured at $\eta=5$ for $(A=-1\times 10^{-4},s_0=8)$ (wave in far zone) as discussed in section \ref{subsec:massresults}.} \label{fig:admmass_t137_a-1e-4so8}
\end{figure}

Another figure showing a ``moderate'' wave that starts off in the ``far'' zone ($A=-1 \times 10^{-4}$, $s_0=8$) can be seen in figure \ref{fig:admmass_t137_a-1e-4so8}, with slightly different behaviour, as it doesn't just look like a perturbation on flat space due to the non-zero initial quasi-local ADM mass.  The amplitude is only one order of magnitude larger, but the peak of the wave is located at a much larger radius\footnote{Recalling that $\eta$ is an exponential coordinate.}.

\begin{figure} \centering
\includegraphics{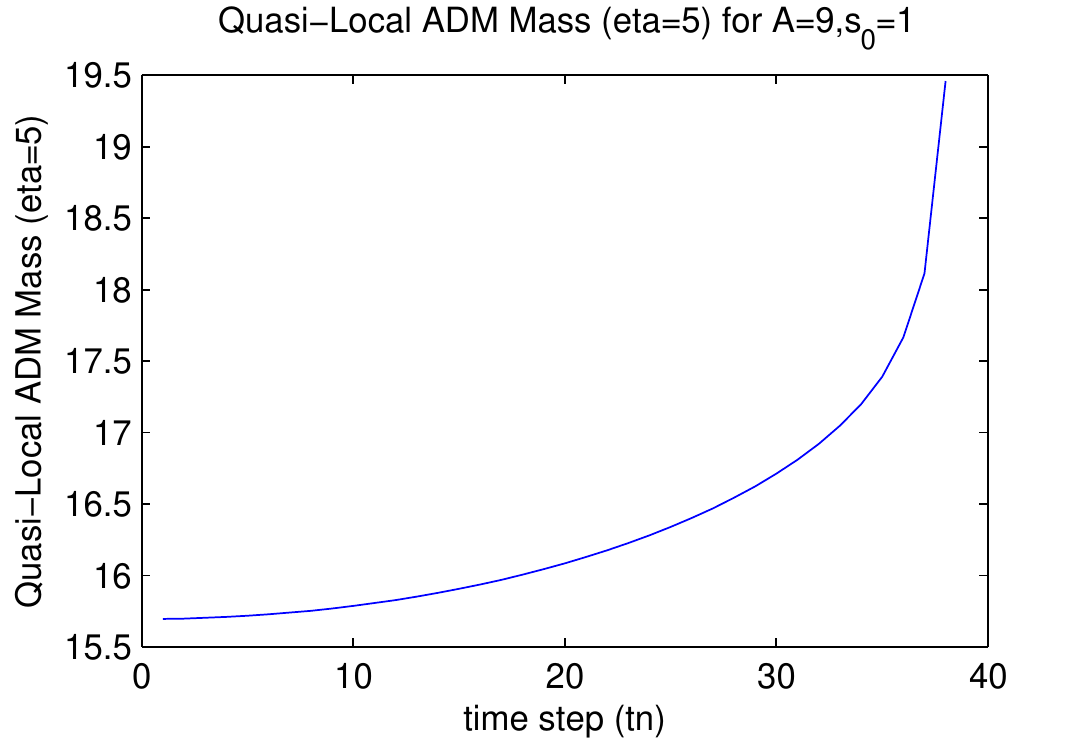}
\caption[QL ADM Mass over time ($A=9,s_0=1$)]{Quasi-Local ADM Mass measured at $\eta=5$ for $(A=9,s_0=1)$ (strong IVP wave) as discussed in section \ref{subsec:massresults}.} \label{fig:admmass_t38_a9so1}
\end{figure}

\begin{figure} \centering
\includegraphics{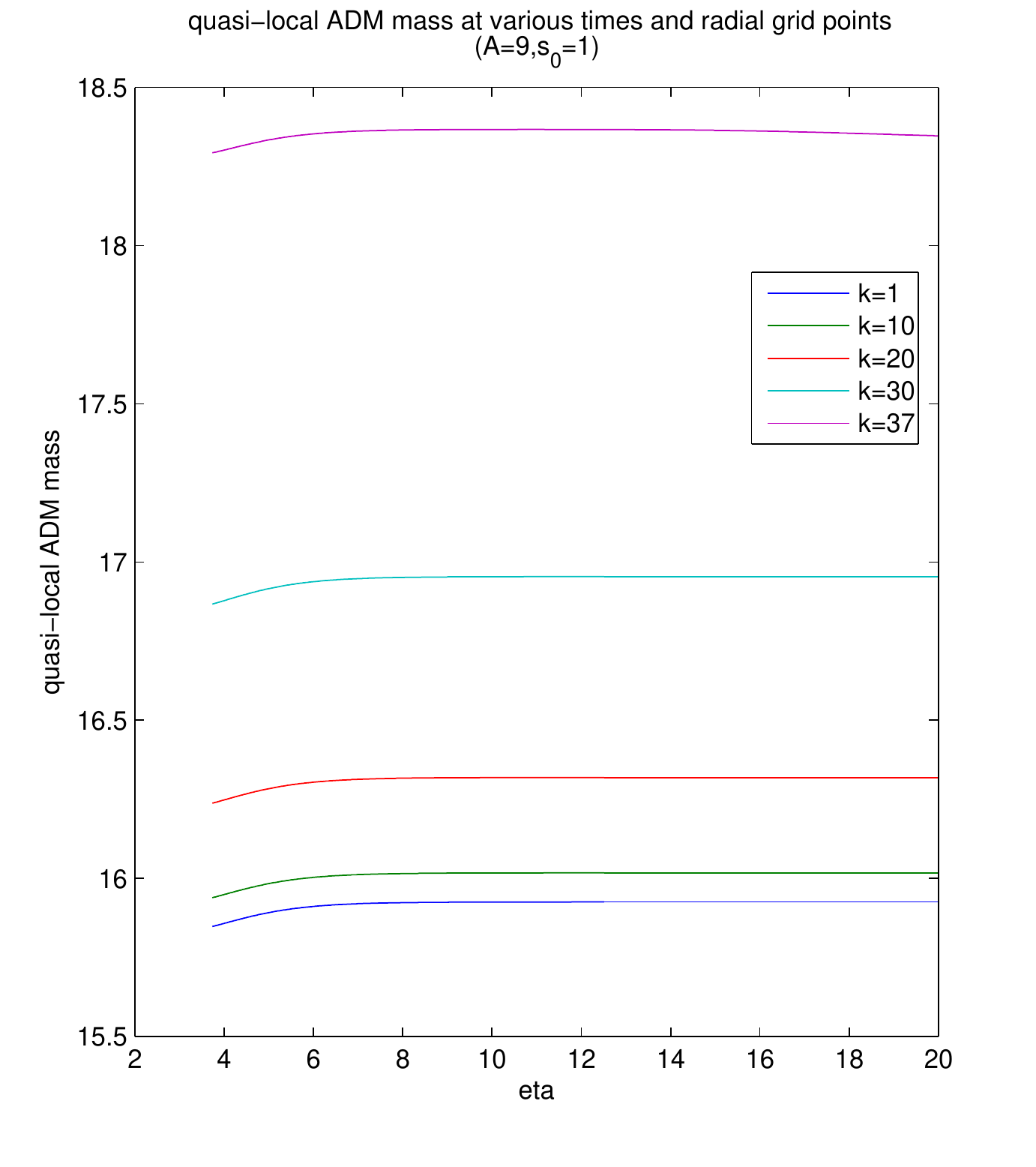}
\caption[QL ADM Mass at various times ($A=9,s_0=1$)]{Quasi-Local ADM Mass at various radial grid points for $(A=9,s_0=1)$ as discussed in section \ref{subsec:massresults}.  Note how the asymptotic values seem to have non-trivial information encoded in them (positive and negative oscillations) at various time steps and are still not converging at $\eta \sim 20 \rightarrow r \sim 2.4 \times 10^8$ (as the values are curling slowly downwards).} \label{fig:admmass_composite_a9s01etamax20}
\end{figure}

A figure showing the quasi-local ADM mass over time for $(A=9,s_0=1)$ (strong IVP wave) can be seen in figure \ref{fig:admmass_t38_a9so1}.  As mentioned above, moving the outer boundary out substantially indicates that a quasi-local measure of the ADM mass is insufficient as the evolution progresses with any reasonably sized grid.  This can be seen in figure \ref{fig:admmass_composite_a9s01etamax20} where we have moved the outer boundary out even further, to $\eta=20$ ($r\sim 2.4\times 10^8$) and we still see non-convergent values of the mass.

The mass aspect $\tilde{M}$ has been previously discussed in section \ref{sec:spherpolob}, and we find that the assumption of spherical symmetry at the outer boundary is correct to approximately one part in $10^6$ (see also table \ref{tbl:k1k2sphpolOB}).

\subsection{$C_p/C_e$ ratio}
As one measure of the deviation from spherical symmetry of the apparent horizons we can calculate the ratio of the polar ($C_p$) to equatorial ($C_e$) circumferences.  The polar circumference is measured by calculating the proper spatial distance along the horizon from $\theta=0$ to $\theta=\frac{\pi}{2}$ at a constant value of $\varphi$.  From our metric we know that
$$dl_{\mathrm{polar}}^2=e^{q+4\phi}f^2 d\theta^2$$
assuming a constant $\eta$ for an apparent horizon, which is approximately correct\footnote{We can add correction terms however they are generally $< 1$\% of the value.}.  From this we find
\begin{equation}\label{eqn:cp}C_p=4 \int_0^{\frac{\pi}{2}}f e^{(\frac{q}{2}+2\phi)}d\theta\end{equation}
The equatorial circumference is measured by taking the horizon at the equator ($\frac{\pi}{2}$) and integrating over $\varphi$ from $0$ to $2\pi$ (i.e. spinning it around the axis of symmetry) to create a circle, then calculating the proper distance around that circle.  From our metric we know that\footnote{$\theta=\frac{\pi}{2} \rightarrow \sin\theta=1$}
$$dl_{\mathrm{equatorial}}^2=f^2e^{4\phi}d\phi^2$$
Therefore
\begin{equation}\label{eqn:ce}C_e=2\pi f e^{2\phi(\eta,\frac{\pi}{2})}\end{equation}

For a sphere in flat space the two circumferences are equal, so $\frac{C_p}{C_e}=1$

For a ``pancake'' which is infinitesimally thin the ratio should be
$$\frac{C_p}{C_e}=\frac{4R}{2 \pi R} = \frac{2}{\pi} \sim 0.637$$

\begin{figure} \centering
\includegraphics{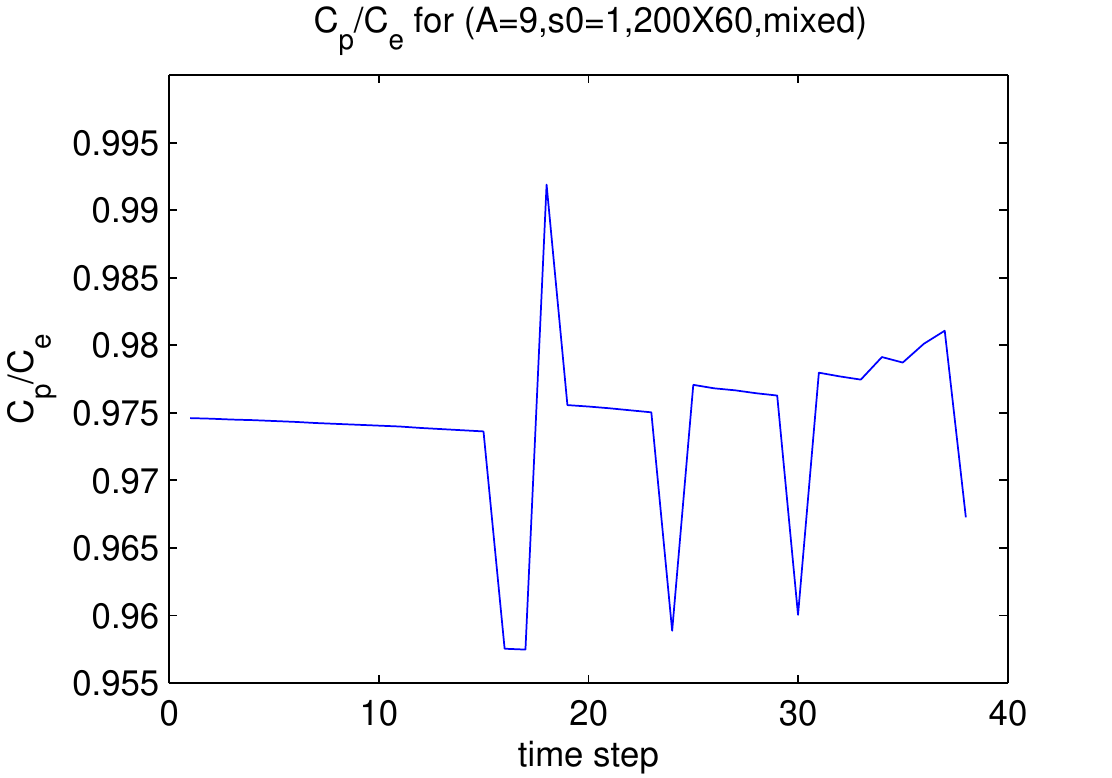}
\caption[$C_p/C_e$ for $(A=9,s_0=1)$]{$C_p/C_e$ at various time steps for the $(A=9,s_0=1)$ oblate ``hemoglobin'' configuration.}\label{fig:cpce_a9_s01}
\end{figure}

In the oblate ``hemoglobin'' shaped strong positive amplitude case $A=9$ we find that the ratio varies from $0.96$ to $0.99$ during the course of the evolution, which indicates a near-spherical horizon embedding.\footnote{Perhaps a ``radius of curvature'' measure of the horizon embedding would serve as a better measure as it would have a positive radius of curvature at the equator and negative curvature at the axis.}  While we do have a very oblate, ``hemoglobin'' topology in proper \emph{radial} distances, the polar distance will actually integrate out larger than flat space due to the fact that (i) there is a non-zero (and very large) factor $q$ in the numerator's integral ($C_p$) via equation (\ref{eqn:cp}) which is not present in the denominator (see equation (\ref{eqn:ce})) and (ii) the dimple at the poles increases the proper distance versus an oblate ellipsoid.  See figure \ref{fig:cpce_a9_s01} for a graph of this ratio over the evolution.

In the strong negative amplitude case $A=-3.5$ (which is prolate in radial proper distance visualisations), we find that the ratio varies from $0.99 \rightarrow 0.86$ during the evolution, indicating that the embedding looks like a ball that is being sat on as the evolution progresses.  Due to the fact that $q$ is large and negative, the value of $C_p$ will (likely) integrate out smaller in equation (\ref{eqn:cp}) than $C_e$, which is independent of $q$.

For $A=-4$ the ratio indicates approximate spherical symmetry $\pm$errors from the fact that we do not locate horizons precisely ($0.98\rightarrow 1.02$).

\section{Evolving the Interior of a Black Hole}\label{subsec:BHInt}
As seen in table \ref{tbl:videoresults}, if we evolve a near-critical case $(A=-4.3,s_0=1)$ by setting the outer computational boundary \emph{inside} the apparent horizon (black hole), then we can evolve for very long times.  In this case we set the outer boundary at $\eta=5$ when the apparent horizon is located around $\eta \sim 6.6$, so we avoid the singularity formation problems detailed in section \ref{sec:crashhalt}, namely condition (2) which states that an apparent horizon generates inextendible geodesics which will cause the code to encounter a singularity.

Further, Wald \cite{Wald} $\S 9.5$ demonstrates that the affine length $\Lambda$ of these particular inextendible, finite geodesics are bounded by the inverse of the maximum expansion $\bar{\theta}$, i.e.
\begin{equation}\label{eqn:affinelength}\Lambda \le \frac{2}{|\bar{\theta}|}\end{equation}
where $\bar{\theta}$ is a (negative) quantity that represents the expansion (contraction) of geodesics.  $\bar{\theta}$ evolves in proper time and depends on the shear $\sigma_{\alpha\beta}$ and itself (in this case) via the Raychaudhuri equation (\ref{eqn:raychaudhuri}) as we have no angular momentum ($\omega_{\alpha\beta}=0$) and we are in a vacuum ($R_{\alpha\beta}=0$).  So a large shear or expansion will cause $\bar{\theta}$ to become more negative very quickly, but the interior of a black hole is very flat.

This can also be understood via the discussion in section \ref{sec:meaningofq} on the meaning of $q$ as a cost function of motion in various coordinate directions.  If $q$ is large (positive or negative) then orthogonal geodesics can have a large convergence due to the ``curvature impedance'' in various directions; i.e. geodesics will prefer motion along lines of latitude or the $(\eta,\theta)$ plane and will converge due to the large cost of motion in perpendicular directions.  This also shortens the total affine length $\Lambda$ before we hit a singularity due to equation (\ref{eqn:affinelength}) and the large negative value of $\bar{\theta}$. Further, this implies that there is a singular structure near the apparent horizon (and not just at the origin as in Schwarzschild topology) as the geodesics do not have time to evolve very far from the horizon before terminating.

This also explains why the evolutions listed in table \ref{tbl:videoresults} encounter a singularity and halt within a small number of time steps once an apparent horizon is present.

By excising the apparent horizon/exterior region and only evolving the black hole interior we see a very flat spacetime (very small curvatures) that is very stable and almost static.  This long-run stability also indicates that the code is performing as required and expected and not amplifying numerical error.  It also demonstrates that the code is evolving without general numerical pathologies due to gauges, slicing, etc.

\section{Common Features of Brill Wave Evolutions}\label{sec:results}
Let us now discuss some of the more general results that are commonly seen with the code being run using various IVPs.

- In general the Weyl scalars\footnote{See section \ref{sec:weyl_np} for a discussion of the Weyl curvature and scalars.} ($\Psi_0$, $\Psi_1$, $\Psi_2$, $\Psi_3$,$\Psi_4$), Riemann Scalars ($I$, $J$) and ${}^3R$ grow without bound near the origin and further out on the equator near $\eta>0.5$ (where curvature grows as the waves evolve) for many scenarios tested that do not have AH's present initially (including ``perturbative'' waves).  This indicates curvature singularities forming (i.e. black holes).

- Weyl curvature terms $\Psi_4$ and $\Psi_0$ show ``outgoing'' and ``ingoing'' curvature wave forms.  This is easily observable in the $(A=\pm 1,s_0=1)$ cases (see table \ref{tbl:videoresults} for links to these videos).

- The Weyl scalars $\Psi_n$ demonstrate non-trivial evolution at the outer boundary, which can be interpreted as ingoing and outgoing plane-polarised gravitational radiation via equations (\ref{eqn:outgravwavepsi4}) and (\ref{eqn:ingravwavepsi0}).\footnote{Remembering that we are not using the NU-tetrad.}

\begin{figure} \centering
\includegraphics{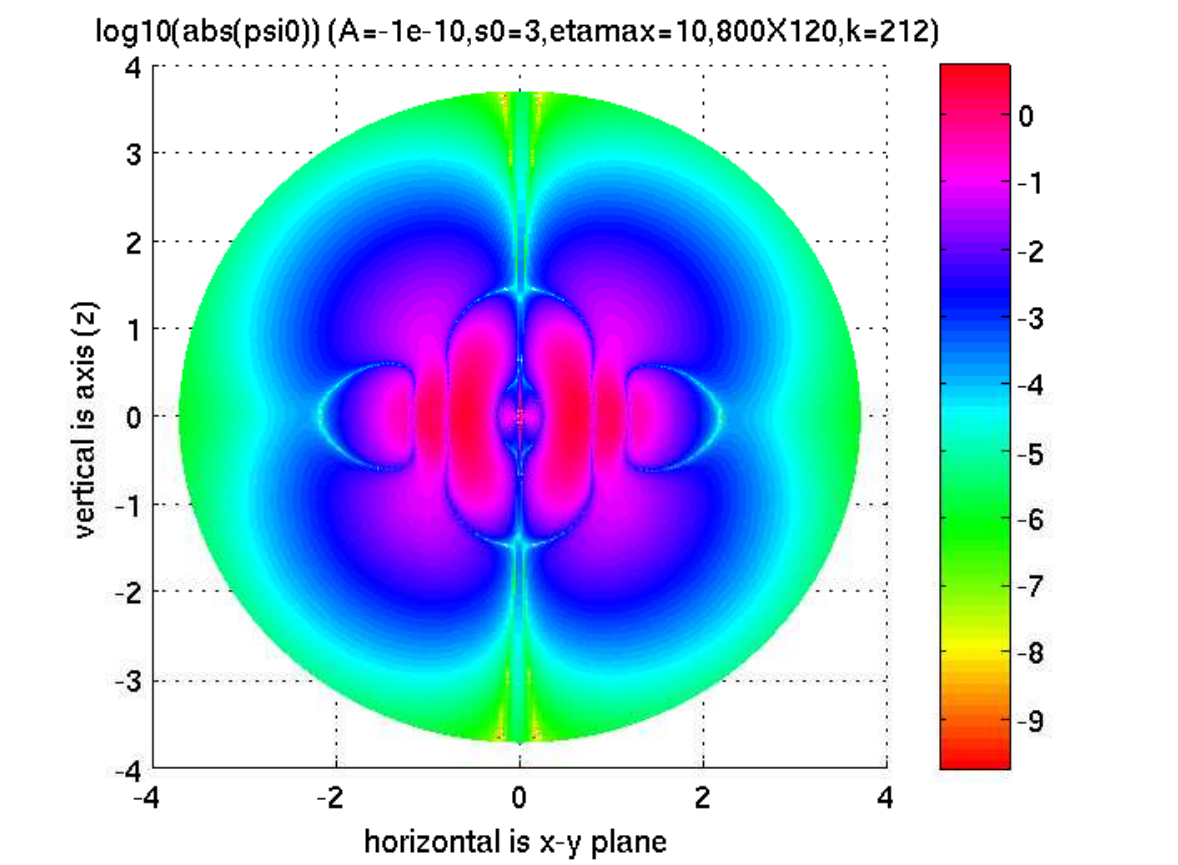}
\caption[Four quadrant example of $log_{10}|\Psi_0|$]{$log_{10}|\Psi_0|$ at $t=212\Delta t$ when reflected across $\theta=0,\frac{\pi}{2}$ into a four quadrant picture (due to the symmetries across those axis), to aid in visualising the spherical harmonics that are present.} \label{fig:log10_abs_psi0_A-1e-10_s03_etamax10_800X120_k212_FOURQUADRANT}
\end{figure}

\begin{figure} \centering
\includegraphics{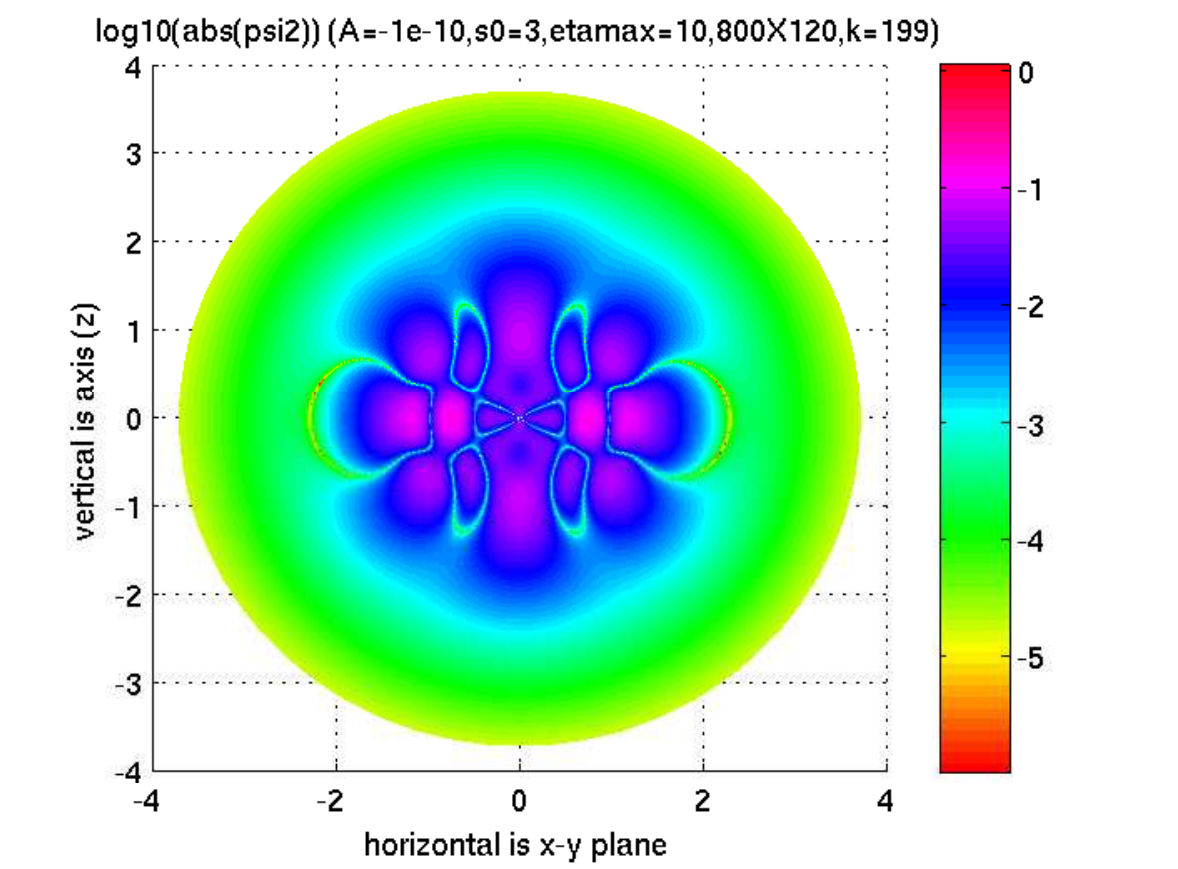}
\caption[Four quadrant example of $log_{10}|\Psi_2|$]{$log_{10}|\Psi_2|$ at $t=199\Delta t$ when reflected across $\theta=0,\frac{\pi}{2}$ into a four quadrant picture, to aid in visualising the spherical harmonics that are present.} \label{fig:log10_abs_psi2_A-1e-10_s03_etamax10_800X120_k199_FOURQUADRANT}
\end{figure}

\begin{figure} \centering
\includegraphics{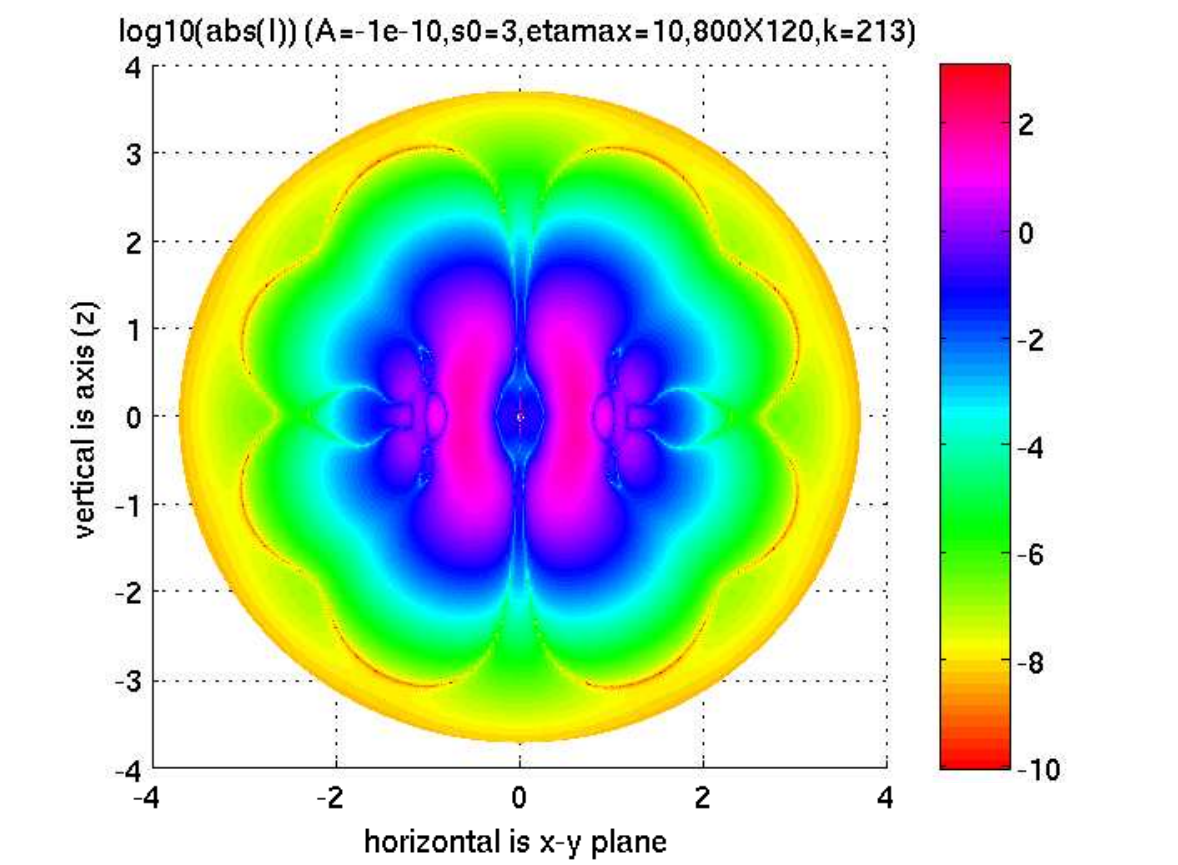}
\caption[Four quadrant example of $log_{10}|I|$]{$log_{10}|I|$ at $t=213 \Delta t$ when reflected across $\theta=0,\frac{\pi}{2}$ into a four quadrant picture, to aid in visualising the spherical harmonics that are present.} \label{fig:log10_abs_I_A-1e-10_s03_etamax10_800X120_k213_FOURQUADRANT}
\end{figure}

\begin{figure} \centering
\includegraphics{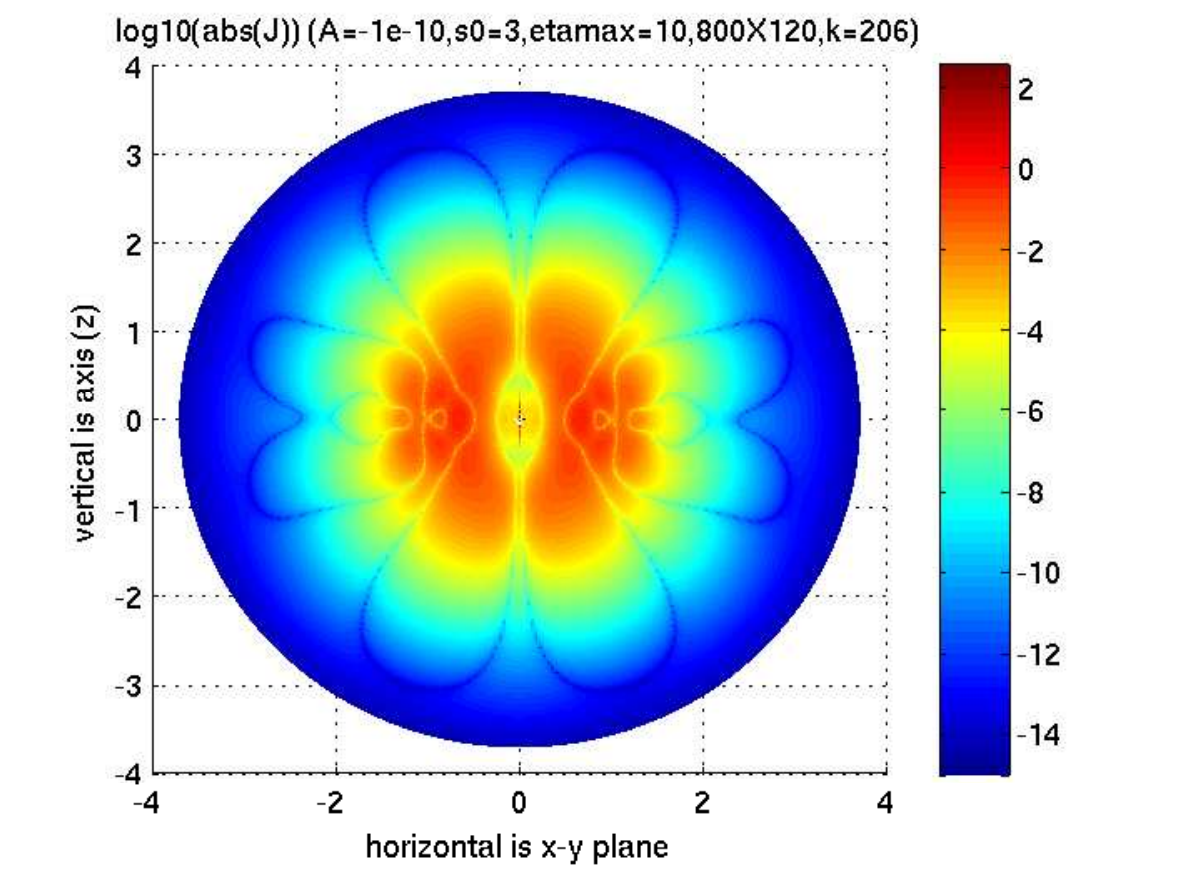}
\caption[Four quadrant example of $log_{10}|J|$]{$log_{10}|J|$ at $t=206 \Delta t$ when reflected across $\theta=0,\frac{\pi}{2}$ into a four quadrant picture, to aid in visualising the spherical harmonics that are present.} \label{fig:log10_abs_J_A-1e-10_s03_etamax10_800X120_k206_FOURQUADRANT}
\end{figure}

- The Weyl scalars all demonstrate complicated spherical harmonic behaviour (see figures \ref{fig:log10_abs_psi0_A-1e-10_s03_etamax10_800X120_k212_FOURQUADRANT}, \ref{fig:log10_abs_psi2_A-1e-10_s03_etamax10_800X120_k199_FOURQUADRANT}, \ref{fig:log10_abs_I_A-1e-10_s03_etamax10_800X120_k213_FOURQUADRANT} and \ref{fig:log10_abs_J_A-1e-10_s03_etamax10_800X120_k206_FOURQUADRANT} for examples).\footnote{What we see is qualitatively similar to the IVP contours explored in \cite{Abrahams2}.}

- All tested shapes for the function $q$ exhibit the same IVP phase space behaviour in the amplitude $A$: (i) a critical maximum (positive) and minimum (negative) value for which the IVP has a solution and (ii) phase space regions with and without an apparent horizon present.

- No scenario that demonstrates a collapse followed by a re-expansion has been observed\footnote{i.e. dispersion of gravitational radiation with no black hole formation.} after the wave ``hits'' the origin (there are no pressure terms, twist or $T_{\mu\nu}$ to cause dispersion so this should be expected).

- No critical mass scaling phenomena exists - all pure vacuum Brill waves collapse to form black holes, whether they are weak perturbative waves or moderate strength waves with no apparent horizon present on the initial surface. Because of our gauge conditions it is difficult to say for certain that an apparent horizon (black hole) has formed as we have variable time evolution across the grid, however the signatures of curvature growth without bound, trapped surface bunching and trapped surfaces evolving from outgoing to ingoing (even in proper radial distance coordinates $\bar{r}$) all point towards black hole formation.  This is in line with analytical results discussed below.

- Most evolutions have shrinking (more negative) $q$ as the spacetime evolves, i.e. $\frac{dq}{dt}<0$.

\section{Why all Brill waves collapse to form black holes}\label{sec:brill_collapse_always}
Motivated by the numerical results presented above demonstrating the collapse of all Brill wave IVP value characterisations to a black hole, let us revisit the theoretical framework of singularities in the same manner as Wald \cite{Wald} and Kar \cite{kar_raych}.

\subsection{The Raychaudhuri Equation(s) and Pseudo-Riemannian Geodesics}
Consider the flow of a bundle of timelike geodesics $\xi^\alpha$ on a Pseudo-Riemannian manifold with metric $g_{\alpha\beta}$ and projection tensor $h_{\alpha\beta}=g_{\alpha\beta}+\xi_{\alpha}\xi_{\beta}$.  Examining the gradient of this flow we can decompose it in $n$ dimensions as:
$$B_{\alpha\beta} \equiv \nabla_\beta\xi_\alpha=\sigma_{\alpha\beta}+\omega_{\alpha\beta}+\frac{1}{n-1}h_{\alpha\beta}\bar{\theta}$$
where

(i) $\bar{\theta}$ is the trace of $B_{\alpha\beta}$ and is the average expansion of infinitesimally close timelike geodesics

(ii) the shear (symmetric traceless part) is given by
\begin{equation}\label{eqn:raych_shear_def}\sigma_{\alpha\beta}=B_{(\alpha\beta)}-\frac{1}{n-1}h_{\alpha\beta}\;\bar{\theta}\end{equation}

(iii) the twist (antisymmetric ``rotation'') is given by
$$\omega_{\alpha\beta}=B_{[\alpha\beta]}$$
and we use the usual notation that the symmetric and antisymmetric parts are given respectively by:
$$a_{(bc)}=\frac{1}{2}\left(a_{bc}+a_{cb}\right)\;,\; a_{[bc]}=\frac{1}{2}\left(a_{bc}-a_{cb}\right)$$
From these definitions one arrives at several equations\footnote{Kar \cite{kar_raych} calls them identities, as they are just geometric properties of pseudo-Riemannian manifolds and are independent of the Einstein equations.} that are very useful for examining the properties of the manifold.  One of the major results that is used extensively in singularity theorems and their proofs is the ``Raychaudhuri equation''\footnote{For non-geodesic motion there is an additional acceleration term $\nabla_\alpha(\xi^\beta\nabla_\beta\xi^\alpha)$ \cite{kar_raych,raychaudhuri2003}. We can ignore this term, however, as we are in a vacuum and free to choose gauges however we wish (and all gauges should produce the same \emph{physical} results).}\footnote{We employ the timelike geodesic congruence version of these equations, see \cite{kar_raych} for a discussion.}
\begin{equation}\label{eqn:raychaudhuri} \frac{d\bar{\theta}}{d\tau} = -\frac{1}{3}\bar{\theta}^2 - \sigma^2 + \omega^2 - R_{\alpha\beta}\xi^\alpha \xi^\beta \end{equation}
where $\tau$ is a scalar affine parameter defined along the geodesic curves, $R_{\alpha\beta}$ is the Ricci tensor and
$$\sigma^2=\sigma^{\alpha\beta}\sigma_{\alpha\beta}$$
$$\omega^2=\omega^{\alpha\beta}\omega_{\alpha\beta}$$

The evolution equations for the shear and twist are
\begin{equation}\label{eqn:raych_shear} \frac{d(\sigma_{\alpha\beta})}{d\tau} = -\frac{2}{3}\bar{\theta}\sigma_{\alpha\beta} - \sigma_{\alpha\delta}\sigma^\delta_{\;\beta} - \omega_{\alpha\delta}\omega^\delta_{\;\beta} \nonumber \\ \mbox{}
+\frac{1}{3}h_{\alpha\beta}\left[\sigma^2-\omega^2\right] + C_{\delta\beta\alpha\epsilon} \xi^{\delta} \xi^{\epsilon} + \frac{1}{2}\bar{R}_{\alpha\beta} \end{equation}
and
\begin{equation}\label{eqn:raych_twist} \frac{d(\omega_{\alpha\beta})}{d\tau} = -\frac{2}{3}\bar{\theta}\omega_{\alpha\beta} - 2\sigma^{\epsilon}_{\left[\beta\right.}\omega_{\left.\alpha\right]\epsilon} \end{equation}
where 
$$\bar{R}_{\alpha\beta}=h_{\alpha\epsilon}h_{\beta\delta}R^{\epsilon\delta}-\frac{1}{3}h_{\alpha\beta}h_{\epsilon\delta}R^{\epsilon\delta}$$
and $C_{\delta\beta\alpha\epsilon}$ is the Weyl tensor (see section \ref{sec:weyl_np}).

If the expansion $\bar{\theta}$ is \emph{strictly} negative for $\tau>\tau_0\ge 0$ across the spacetime, we will encounter singularities as all geodesics intersect in the spacetime and terminate, giving a singular spacetime.  The cosmic censorship conjecture would then require a black hole covering this singularity\footnote{As we are considering \emph{generic} initial data and not finely-tuned naked singularity producing initial data, there is some motivation to consider this approach.}.

\subsection{Brill Waves and the Raychaudhuri Equations}
Let us now investigate a few properties of these equations as they apply to our spacetime by imposing conditions specific to Brill waves.  The first two terms in equation (\ref{eqn:raychaudhuri}) are negative definite.  In axisymmetry with no rotation\footnote{The introduction of rotation would break symmetry across the $z=0$ plane and violate the Brill conditions.}, the twist term $\omega_{\alpha\beta}=0$.  In a vacuum spacetime, as we have here, $R_{\alpha\beta}=0$.

Equation (\ref{eqn:raychaudhuri}) under Brill wave conditions therefore becomes:
\begin{equation}\label{eqn:raych1}\frac{d\bar{\theta}}{d\tau} = -\frac{1}{3}\bar{\theta}^2 - \sigma^2\end{equation}
We therefore deduce that the RHS of equation (\ref{eqn:raych1}) is $\le 0$, meaning that the \emph{change} in expansion $\frac{d\bar{\theta}}{d\tau}$ is non-positive for all proper time, and all points on a spatial hypersurface.

Vanishing twist $(\omega_{\alpha\beta}=0)$ implies that the geodesic congruences are hypersurface orthogonal and $\xi^a$ are in fact the spatial hypersurface normal vectors, which makes $\bot B_{\alpha\beta}=K_{\alpha\beta}$ the extrinsic curvature, where $\bot$ is the projection operator onto the hypersurface.  This also means that the expansion is
\begin{equation}\label{eqn:raych_theta_kij}\bot \bar{\theta}=TrK=K^a_a\end{equation}

From the Brill wave IVP formulation, we know that all extrinsic curvature components are identically zero initially\footnote{This is a result of time symmetry of the metric, a requirement for Brill waves.}, which means that $\left.B_{\alpha\beta}\right|_{t=0}=0$ at that moment and the initial expansion $\bar{\theta}_0=0$ from equation (\ref{eqn:raych_theta_kij}). Using equation (\ref{eqn:raych_shear_def}) we also see that the shear $\sigma^2$ is zero initially.  This means that all terms on the RHS of equation (\ref{eqn:raych1}) are zero initially and the expansion and its ``velocity'' are zero everywhere on the initial time-symmetric hypersurface.

All of the terms in equation (\ref{eqn:raych_twist}) vanish identically as well, so as expected the twist is initially zero and does not evolve to non-zero values.

Examining the evolution equation for the shear (\ref{eqn:raych_shear}), however, we see that all of the terms are zero initially \emph{except} for the $C_{0ji0}$ Weyl tensor term (as the congruences are hypersurface orthogonal so $\xi^\alpha=(\xi^0,0,0,0)$).  This will cause the shear to start evolving once we move off the initial hypersurface:
$$\left.\frac{d(\sigma_{ij})}{d\tau}\right|_{t=0} = C_{0ji0} \xi^0 \xi^0$$
The Brill wave spacetime is an algebraically general, non-flat spacetime, therefore all of the $C_{0ij0}$ terms cannot vanish \cite{VandenBergh:2003yd,Zakhary_elecweyl}.\footnote{Alternately, one can note that (i) the spacetime is Ricci flat, (ii) equation (\ref{eqn:ricciweyl}) holds and (iii) the spacetime is non-flat.}
This evolution of the shear to a non-zero value will then cause the expansion to become negative via equation (\ref{eqn:raych1}) as $\bar{\theta}_0=0$, so any negative term added to it will cause it to become immediately negative.

This is equivalent to $\bar{\theta}$ and $\bar{\theta}$'s ``velocity'' being zero initially, but having non-zero $\bar{\theta}$ ``acceleration''.

This immediately tells us that the spacetime must become singular in a finite time.  To see this, consider $\bar{\theta}_1 < 0,\sigma^2 \ge 0$ for $\tau=\tau_1>0$ where $\tau_0=0$ is the time parameter for the initial hypersurface.  We then know that
\begin{equation}\frac{d\bar{\theta}}{d\tau} \le -\frac{1}{3}\bar{\theta}^2\end{equation}
$$\int_{\bar{\theta}_1}^{\bar{\theta}_2}\frac{-3 d\bar{\theta}}{\bar{\theta}^2} \ge \int_{\tau_1}^{\tau_2}d\tau$$
We define $\Delta\tau=\tau_2-\tau_1$ and simplify to find
$$\bar{\theta}_{2} \le 3\left(\frac{\bar{\theta}_1}{3+\Delta\tau\bar{\theta}_1}\right)$$
Recalling that $\bar{\theta}_1<0$, we see that $\bar{\theta}_2<0$ and $\bar{\theta}_2 \rightarrow -\infty$ as $\Delta\tau \rightarrow \frac{-3}{\bar{\theta}_1}$.\footnote{These are not just caustics due to the global structure.}  As we have a globally hyperbolic spacetime where the expansion becomes negative at all points, various singularity theorems in \S 9.5 from Wald \cite{Wald} apply\footnote{Excising a compact sub-manifold allows us to use other singularity theorems from Wald as well; the spacetime is very singular.}.

This proves that all non-zero Brill wave spacetimes must become singular within a finite time (and therefore form a black hole if we employ the cosmic censorship conjecture).

From this we can state a theorem that incorporates (i) the fact that all non-flat initial configurations of Brill waves\footnote{The only requirements are that (a) the twist $\omega_{\alpha\beta}=0$, (b) the last term in equation (\ref{eqn:raychaudhuri}) is negative and (c) we have a moment where $B_{\alpha\beta}$ vanishes but the Weyl tensor does not, so this result applies to a larger class of spacetimes than just pure Brill waves.} will form a singularity, and (2) they must do so in a finite proper time as $\tau$ has an upper limit.

\emph{Theorem: All non-zero Brill wave spacetimes will form a singularity (and therefore a black hole) within finite proper time.}

and the stronger version;

\emph{Theorem: All non-zero, vacuum spacetimes with ``twist'' $\omega_{\alpha\beta}=0$ that have a Cauchy surface such that (i) the shear and expansion vanish $(\bar{\theta}=\sigma=0)$ with (ii) at least one non-vanishing Weyl curvature component $C_{0ji0}$, will form a singularity (and therefore a black hole) within finite proper time.}

This theorem also implies that flat space is unstable to perturbations of this form as the theorem applies to arbitrarily small Brill (or more general) wave amplitudes.  This differs from \cite{CK-Minkowski-stability} in that we have a slice on which the extrinsic curvature vanishes identically, whereas CK do not \cite{lindblad-minkowski-stability}.

Using the strong energy condition we can state another theorem which is (essentially) a restatement of one of the Hawking/Penrose singularity theorems\footnote{The difference lies in the degeneracy noted above, where \emph{all} terms in the ``Raychaudhuri equation'' are identically zero and we must invoke the shear evolution equation to understand the properties of the evolution.}:

\emph{Theorem: All non-zero spacetimes which obey the strong energy condition $R_{\alpha\beta}\xi^\alpha \xi^\beta \ge 0$ with ``twist'' $\omega_{\alpha\beta}=0$ that have a Cauchy surface such that (i) the shear and expansion vanish $(\bar{\theta}=\sigma=0)$ with (ii) at least one non-vanishing Weyl curvature component $C_{0ji0}$, will form a singularity (and therefore a black hole) within finite proper time.}

The proof of this theorem is the same as above, however the physical interpretation is not as clear; given that matter and light do not necessarily follow geodesics in non-vacuum regions of a spacetime (even if it obeys the Strong Energy Condition), it is not clear that this last result is physically meaningful.  Generally the acceleration term will cause this analysis to change for quantities of physical interest, i.e. non-geodesics\footnote{While cosmic dust, for example, can follow geodesics, there is debate about whether dust represents physical matter given that it is non-interacting.}.

The numerical results presented in this thesis confirm the first theorem and together they give a strong proof that there is no critical black hole collapse behaviour in the IVP phase space parametrisation for Brill waves as they \textbf{all} collapse to form black holes.  It is equally important to note that this result is independent of coordinate systems, gauge choices, slicing conditions, scale (it applies on cosmic and microscopic scales), etc. and relies on some specialised, yet still fairly generic conditions.

The addition of non-vacuum terms ($T_{\alpha\beta}\neq 0$) can possibly alter the dynamics by allowing the last term in equation (\ref{eqn:raychaudhuri}) to become positive, however for physically relevant matter distributions the last term will generally be negative via the strong energy condition.  This implies that there are not likely to be physically relevant situations under which an axisymmetric spacetime with zero twist can have non-collapsing Brill-like waves\footnote{This is likely the source of critical collapse in Choptuik \emph{et al}'s work: the scalar field violates the strong energy condition and adds a positive term to Raychaudhuri's equation.}.

We use the term Brill-like as Brill waves are only defined in a vacuum. Further, the addition of artificial numerical dissipation\footnote{Which seems to have become a de facto methodology...} to numerical vacuum evolution schemes has the net effect of adding non-zero terms to the RHS of the Einstein equations, which acts like a non-zero source term and could violate the strong energy condition.  This could potentially cause schemes with heavy numerical dissipation to observe critical phenomena as they have a ``damping'' mechanism to dissipate the gravitational radiation which is non-physical in the sense of studying the \emph{vacuum}\footnote{This is likely the source of ``critical collapse'' observed in other attempts at pure Brill wave simulations, e.g. Sorkin \cite{sorkin}, and the analysis above demonstrates that their results are inconsistent.}.

This implies that to have critical behaviour in the vacuum we must break symmetry about the $(\eta,\theta)$ plane and allow twist, which violates the Brill conditions and his positive energy theorem.  While this is certainly an interesting question, it is outside the scope of the discussion for Brill waves. 

There are other statements that can be made about similar spacetimes depending on the initial expansion $\bar{\theta}_0$, etc., however this analysis underscores the importance of looking at one's IVP and spacetime from the point of view of the Raychaudhuri equations; it's possible that the answer is already present in analytical form.

\subsection{Black Hole Collapse in Shear-Free Spacetimes}\label{sec:shearfree_bh}
Let us examine what happens if we have no shear $(\sigma_{\alpha\beta}=0)$ in addition to no twist and the strong energy conditions satisfied.  In this case equation (\ref{eqn:raych1}) is altered to
\begin{equation}\label{eqn:raych3}\frac{d\bar{\theta}}{d\tau} \le -\frac{1}{3}\bar{\theta}^2\end{equation}
from which we find
$$\int_{\bar{\theta}_0}^{\bar{\theta}_\tau}\frac{-3 d\bar{\theta}}{\bar{\theta}^2}\ge\int_0^{\tau}d\tau$$
which simplifies to
$$\bar{\theta}_{\tau}\le 3\left(\frac{\bar{\theta}_0}{3+\tau\bar{\theta}_0}\right)$$
If $\bar{\theta}_0>0$ (positive initial expansion at a point in the spacetime) in this case we find that $\bar{\theta}_{\tau} \rightarrow 0$.  If, however, $\bar{\theta}_0<0$ we see that as $\tau \rightarrow \frac{-3}{\bar{\theta}_0}$ that $\bar{\theta}_{\tau} \rightarrow -\infty$.

Therefore as long as there is a Cauchy surface in a shearless, twist-free, strong energy\footnote{The same comments regarding the questionable physical relevance of \emph{geodesic} analysis as above also apply; the acceleration term is important in understanding strong energy astrophysical collapse scenarios like white dwarfs.} spacetime which has a negative initial expansion there will be a singularity which forms, and it will do so in a finite amount of proper time.

\subsection{Black Hole Collapse in Non-Zero Shear Spacetimes}\label{sec:nzshear_bh}
From the positive definiteness of $\sigma^{\alpha\beta}\sigma_{\alpha\beta}$ let us now assume instead that a constant $\sigma>0$ exists such that
$$3 \sigma^{\alpha\beta}\sigma_{\alpha\beta} \ge \sigma^2 \; \forall \tau$$
(i.e. there is a non-zero minimum shear present for a point in the spacetime, as in some cosmological models) which transforms equation (\ref{eqn:raych1}) into
\begin{equation}\label{eqn:raych2} 3\frac{d\bar{\theta}}{d\tau} \le -\bar{\theta}^2 - \sigma^2 \end{equation}
By specifying the initial condition $(\tau=0,\bar{\theta}=\bar{\theta}_0)$ we find:
$$\frac{-3 d\bar{\theta}}{\bar{\theta}^2+\sigma^2} \ge d\tau$$
$$-3 \int_{\bar{\theta}_0}^{\bar{\theta}_{\tau}}\frac{d\bar{\theta}}{\bar{\theta}^2+\sigma^2} \ge \int_0^{\tau}d\tau$$
$$\frac{-3}{\sigma}\left[\tan^{-1}\left(\frac{\bar{\theta}_{\tau}}{\sigma}\right) - \tan^{-1}\left(\frac{\bar{\theta}_{0}}{\sigma}\right)  \right] \ge \tau$$
From which it is evident that as we move forward in $\tau$, $\bar{\theta}$ must become \emph{more} negative, and eventually $\bar{\theta}$ must become negative irrespective of whether it was positive or not initially.  This also demonstrates that $\tau$ has an upper limit defined by
$$\tau_{\mathbf{max}} \le \frac{3}{\sigma}\left[\frac{\pi}{2}+\tan^{-1}\left(\frac{\bar{\theta}_{0}}{\sigma}\right)\right]$$
and larger negative values of the initial expansion $\bar{\theta}_0$ will shorten the length of geodesics, and thusly the length of proper time that the evolution can progress.  Further, larger values of $\sigma$ will shorten the proper time taken for geodesics to terminate (this can be seen heuristically in the numerical results as an accelerating of collapse as the extrinsic curvature values grow).

\section{Chapter Summary}
In this chapter a complete picture has been presented of the Brill wave IVP solutions and subsequent evolutions, from numerical, theoretical and physical standpoints.
\begin{enumerate}
\item A discussion on the physical significance of the metric function $q$ has been presented, as well as a numerical examination of the IVP amplitude phase space and alternate shapes of the IVP function.  \'O Murchadha's theoretical work has been utilised to provide an understanding and verification of various results for critical values of the Brill wave amplitude $A$.
\item Various measures of the computed Brill wave spacetimes were presented in this chapter, providing insight into the quasi-local ADM mass structure and horizon topologies.
\item Overarching numerical results for Brill wave evolutions have been presented, demonstrating \emph{universal} collapse of Brill waves to form black holes.
\item A theoretical framework for understanding the universal collapse of Brill waves has been presented, along with a generalisation of the results to a specific class of vacuum gravitational waves.  An interesting ramification of these results is that Minkowski space is unstable to these forms of perturbations.
\end{enumerate}

\chapter{Error Analysis and Testing Alternate Evolution Schemes}\label{chap:errs}
\section[Error Analysis and Convergence Tests]{Error Analysis: Precision, Accuracy, Convergence and All That}\label{sec:erroranalysis}
One critical portion of any numerical work is an analysis of the results in the context of avoiding ``GIGO'' - Garbage In, Garbage Out (and also good data in, garbage out, which is easier to troubleshoot).  There are generally few simplistic analytic solutions to compare to, as the reason one constructs numerical solutions in the first place is the lack of an analytically solvable system.  It is therefore crucial to verify internal consistency via any constraints or ``checks'' that are available.  In addition analytical solutions and other numerical solutions can act as testbeds for the development of new codes\footnote{e.g. Alcubierre et. al. \cite{alcubierre:testbed}, however they mistakenly suggest using low amplitude Brill wave dissipation as a numerical test.}.

It is in this respect that numerical work is like experimental work - various authors generate results which must be compared to establish some sort of inductive approach to ``truth''.  When asked whether Numerical Relativity is theoretical or experimental physics, the answer is truly ``both''.

As there are few \emph{detailed}\footnote{While, for example, \cite{Thornburg:cartesian,miyama} present some results, there are many details like slicing conditions, incompatible gauge conditions, boundary conditions, equations used, etc. that are missing from the discussions which make it difficult to do a thorough analysis of the results.  \cite{miyama} uses inconsistent gauge conditions to solve the system (i.e. assuming that the solution $f$ to $\nabla F(f)=0$ is the same as the solution to $\nabla \nabla F(f)=0$) so I am unsure what to make of the results.} accounts of formulations and results in pure vacuum axisymmetric spherical polar Brill gravitational wave evolutions, it is difficult to compare to other work in the field.  So let us analyse some internal consistency checks that provide some measure of how the code is performing.

\subsection[Metric evolution equation for $q$]{Convergence Test: Metric evolution equation for $q$}\label{sec:qdoterror}
As discussed in section \ref{sec:metricevol} we end up with two equations which can be used to evolve the metric quantity $q$, namely (\ref{eqn:qdot}) and (\ref{eqn:qdotdontuse}).  If our formulation is consistent we should obtain the same answer for $q$ propagated onto the next time slice from both methods.  So we take a measure of
\begin{equation}\label{eqn:qdoterr}\epsilon_q=\left|\frac{q_1-q_2}{q_1}\right|\end{equation}
where $q_1$ is obtained from the first evolution equation (\ref{eqn:qdot}) and $q_2$ is obtained from equation (\ref{eqn:qdotdontuse}) .  This gives a measure of the relative difference between the two solutions and tells us how well we are solving the whole system of equations.

There are three considerations that must be made when analysing this error; firstly the individual variable $q$ has its own iterative Crank-Nicholson algorithm, secondly there is an iteration through the global Crank-Nicholson algorithm and thirdly there is a value for this error once the solution has ``settled'' down and proceeds to the next time step.

Let us examine the error at the point $\eta\sim=1 (i=40), \theta=\frac{\pi}{2}-\Delta\theta (j=60)$ on a $200\times 60$ grid.  This point was chosen as the evolution frequently ``blows up'' near this point, the momentum constraint violations are largest around this area and because of our slicing the evolution proceeds the fastest near the equator.

Firstly, we examine the effect of the variable's Crank-Nicholson iteration in table \ref{tbl:qdoterrorlocal}.  This particular example is for ($k=4$, global C-N count$=1$, $A=-1\times 10^{-5}$, $s_0=3$ and mixed evolution).  As is evident, there isn't much improvement in the error $\epsilon_q$ between the two equations by performing a local iteration.  This is expected as the values for various portions of the evolution equation for $q$ (including shift vectors, extrinsic curvature, etc.) haven't been ``updated'' yet.

\begin{table}\begin{center}
\begin{tabular}{|c|c|c|}\hline
$q_1$ & $q_2$ & rel error $\epsilon_q$ \\ \hline
-0.04711511791987 &	-0.04711557976532 &	9.802489E-06 \\
-0.04711403518266 &	-0.04711449702811 &	9.802714E-06 \\
-0.04711403504779 &	-0.04711449689324 &	9.802714E-06 \\
-0.0471140350478 &	-0.04711449689325 &	9.802714E-06 \\
-0.0471140350478 &	-0.04711449689325 &	9.802714E-06 \\\hline
\end{tabular}\caption[Relative error in $q$ after local C-N]{Relative error in two values for $q$ after performing local iterative C-N scheme as described in section \ref{sec:qdoterror}} \label{tbl:qdoterrorlocal}
\end{center} \end{table}

Next we examine the performance of this error as we perform iterations through the global convergence algorithm across all variables\footnote{We compare values taken from after the local iterative C-N schema has converged, so essentially the last value in the previous table.} in table \ref{tbl:qdoterrorglobal}.

\begin{table}\begin{center}
\begin{tabular}{|c|c|c|c|}\hline
global C-N & $q_1$ & $q_2$ & rel error $\epsilon_q$ \\ \hline
1 & -0.0471140350478 &	-0.04711449689325 &	9.80271482293374E-06 \\
2 & -0.04710720101077 &	-0.04710720101193 &	2.46645125568013E-11 \\
3 & -0.04710720182801 &	-0.04710720165621 &	3.64690362495247E-09 \\
4 & -0.0471072018315 &	-0.04710720166219 &	3.59419185442902E-09 \\
5 & -0.04710720183298 &	-0.04710720167727 &	3.30556827665655E-09 \\
6 & -0.04710720183296 &	-0.0471072017146 &	2.5126252167864E-09 \\
7 & -0.04710720183408 &	-0.04710720172312 &	2.3552940170334E-09 \\
8 & -0.04710720183468 &	-0.0471072017158 &	2.52362381796691E-09 \\
9 & -0.04710720183209 &	-0.0471072017322 &	2.12045367039491E-09 \\
10 & -0.04710720183369 &	-0.0471072017423 &	1.94000284158628E-09 \\\hline
\end{tabular} \caption[Relative error in $q$ after global iteration]{Relative error in two values for $q$ after performing global iterative convergence scheme as described in section \ref{sec:qdoterror}} \label{tbl:qdoterrorglobal}
\end{center} \end{table}

The relative error $\epsilon_q$ slowly decreases as we converge on a global solution for this time step.

\begin{figure} \centering
\includegraphics{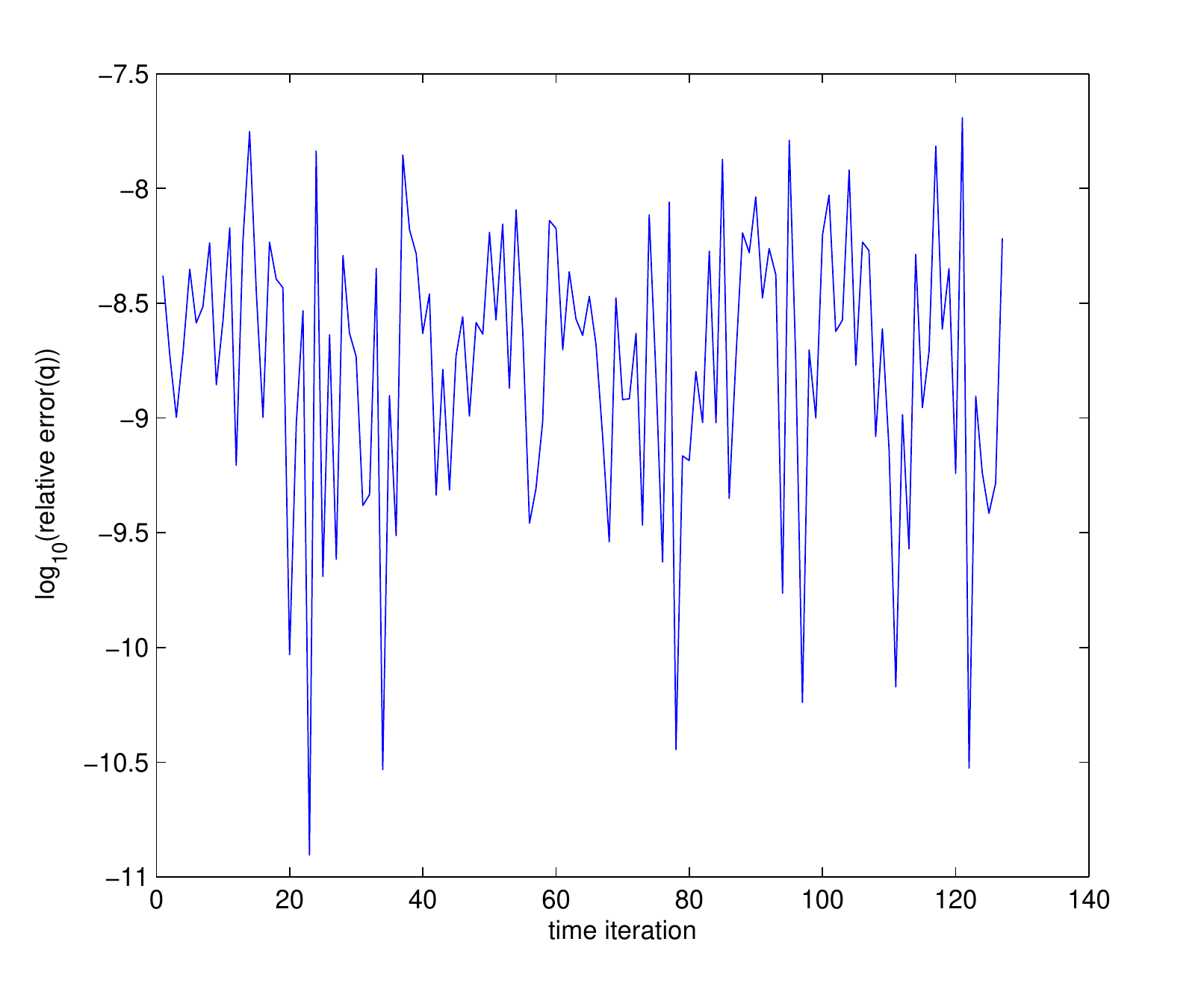}
\caption[$q$ evolution errors in time]{Relative difference in $q$ obtained via two different methods on each time step as discussed in section \ref{sec:qdoterror}}\label{fig:qdot_relerror_time}
\end{figure}

Lastly, we compute $\epsilon_q$ at successive time steps.  Once again, we take the last value from the previous table for each time step, after the global iterative scheme has converged, and plot it in figure \ref{fig:qdot_relerror_time} for the 127 time steps that the code ran for.

Given that we are solving to second order accuracy in time and fourth order accuracy in space, this represents a fairly robust solution (an error of one part in $\sim 10^8$).

\begin{figure} \centering
\includegraphics{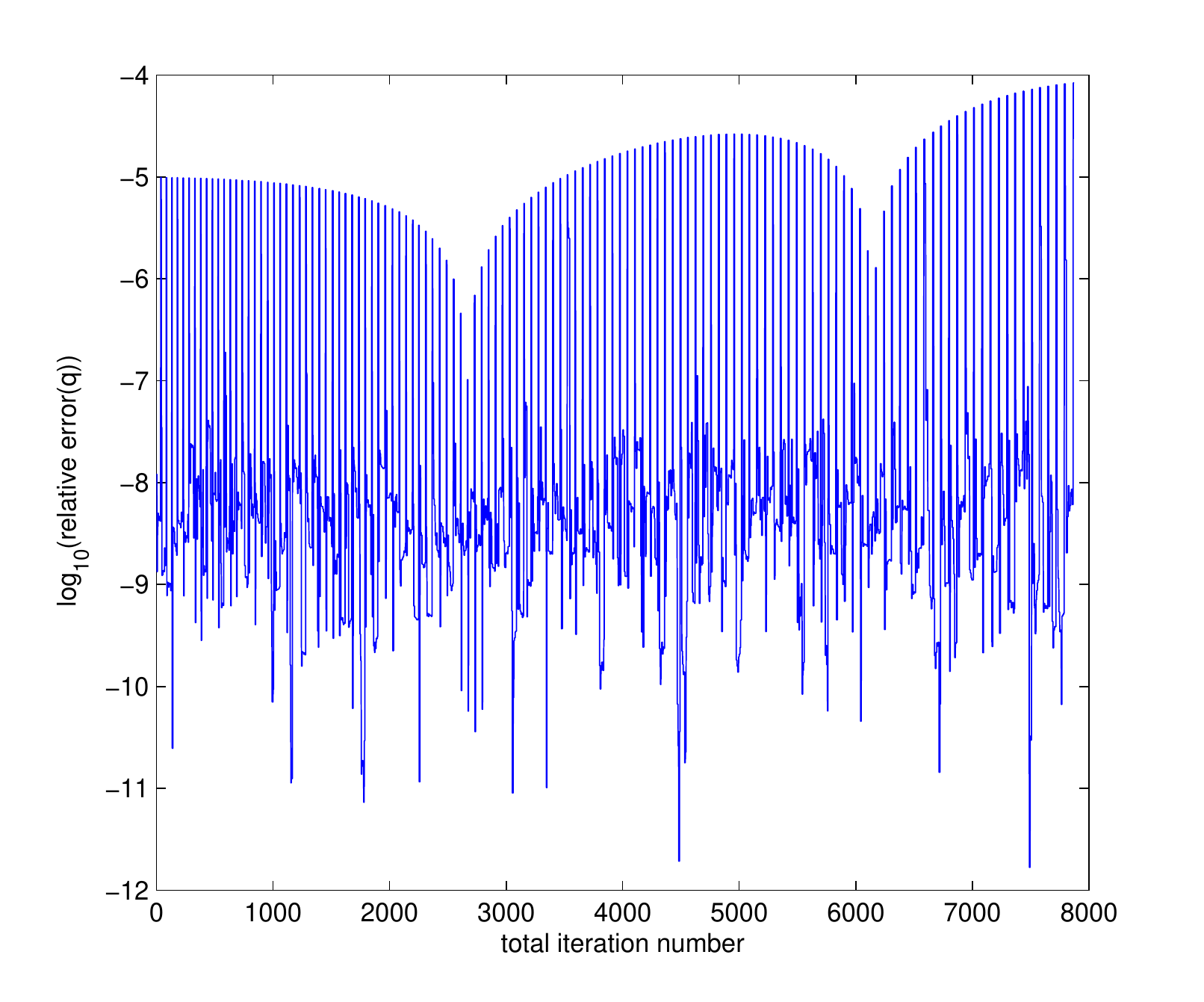}
\caption[$q$ evolution errors (global)]{Relative difference in $q$ ($\epsilon_q$) obtained via two different methods as discussed in section \ref{sec:qdoterror}.  This graph shows $\epsilon_q$ on every pass through the convergence algorithms as we iterate through $127$ time steps (sharply peaked local maxima represent the start of the global iteration on a time step).  The low points correspond to figure \ref{fig:qdot_relerror_time}.}\label{fig:qdot_relerror_all}
\end{figure}

For complete visualisation of the way in which these nested convergence schemes interact, included is a graph of \emph{all} the measures taken including (i) within $q$'s local iterative C-N scheme, (ii) on the global iterative scheme and (iii) finally across all time steps in figure \ref{fig:qdot_relerror_all}.  The sharply peaked local maxima represent the first iteration on the time step, and the answer gets slowly refined as we approach global convergence (i.e. $\epsilon_q$ decreases), only to peak again at the start of the next time step's iterative algorithm.  This indicates that the algorithm is indeed converging as we iterate on each time step.

\subsection[Metric Evolution Equation vs. Hamiltonian Constraint for $\phi$]{Convergence Test: Metric Evolution Equation vs. Hamiltonian Constraint for determining $\phi$}\label{sec:phidoterror}
We have two possible methods to solve for the conformal factor $\phi$, namely the evolution equation (\ref{eqn:gam33dot}) and the Hamiltonian constraint (\ref{eqn:hamconphi}) as discussed in section \ref{sec:constrainconform}.  So once again we have three levels of detail to examine: the convergence on ``local'' iteration through $\phi$'s evolution equation, global C-N iteration and the evolution on each time step.  The difference
\begin{equation}\label{eqn:phierrrel}\epsilon_\phi = \left|\frac{\phi_{evol}-\phi_{Ham}}{\phi_{evol}}\right|\end{equation}
within the local $\phi$ evolution C-N scheme is shown in table \ref{tbl:phidoterrorlocal}.  This indicates that local C-N iteration has little to no effect on the difference between these values.

\begin{table}[h]\begin{center}
\begin{tabular}{|c|c|c|}\hline
$\phi_{evol}$ & $\phi_{Ham}$ & rel error $\epsilon_\phi$ \\ \hline
0.00031146673904 &	0.00029312248531 &	0.05889634887786 \\
0.00031146673974 &	0.00029312248531 &	0.05889635098566 \\
0.00031146673974 &	0.00029312248531 &	0.05889635098591 \\ \hline
\end{tabular}\caption[Relative error in $\phi$ after local C-N]{Relative error $\epsilon_\phi$ in the two values for $\phi$ after performing local iterative C-N scheme as described in section \ref{sec:phidoterror}} \label{tbl:phidoterrorlocal}
\end{center} \end{table}

The effects of converging towards a global iterative solution on a particular time step can be seen in table \ref{tbl:phidoterrorglobal}, and we see that global convergence also has little to no effect on the difference.

\begin{table}\begin{center}
\begin{tabular}{|c|c|c|c|}\hline
global C-N & $\phi_{evol}$ & $\phi_{Ham}$ & rel error $\epsilon_\phi$ \\ \hline
1 &	0.00031146673957 &	0.00029313130636 &	0.05886802948979 \\
2 &	0.0003114667397 &	0.00029312247137 &	0.05889639564506 \\
3 &	0.0003114667397 &	0.00029312248535 &	0.05889635075866 \\
4 &	0.00031146673971 &	0.00029312248537 &	0.0588963507036 \\
5 &	0.00031146673973 &	0.00029312248517 &	0.05889635140073 \\
6 &	0.00031146673973 &	0.00029312248517 &	0.05889635140637 \\
7 &	0.00031146673973 &	0.00029312248527 &	0.05889635108088 \\
8 &	0.00031146673973 &	0.00029312248497 &	0.05889635207034 \\
9 &	0.00031146673974 &	0.00029312248531 &	0.05889635098591 \\ \hline
\end{tabular}\caption[Relative error in $\phi$ after global iteration]{Relative error $\epsilon_\phi$ in the two values for $\phi$ after performing global iterative scheme as described in section \ref{sec:phidoterror}} \label{tbl:phidoterrorglobal}
\end{center} \end{table}

\begin{figure} \centering
\includegraphics{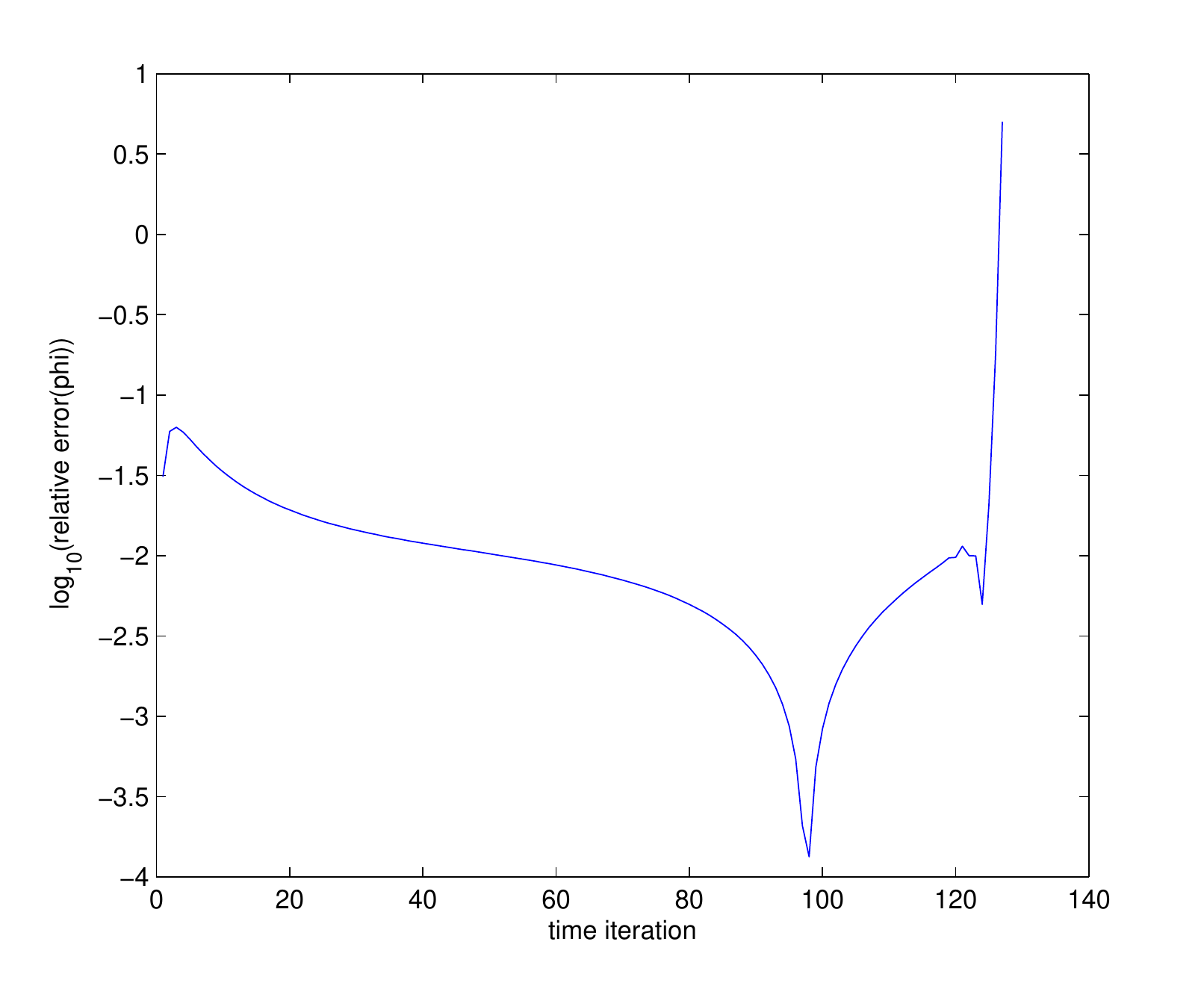}
\caption[$\phi$ evolution errors in time]{Relative difference in $\phi$ ($\epsilon_\phi$) obtained via two different methods on each time step as discussed in section \ref{sec:phidoterror}}\label{fig:phidot_relerror_time}
\end{figure}

The difference on \emph{each time step} once global iterative convergence has been achieved can be seen in figure \ref{fig:phidot_relerror_time}.

\begin{figure} \centering
\includegraphics{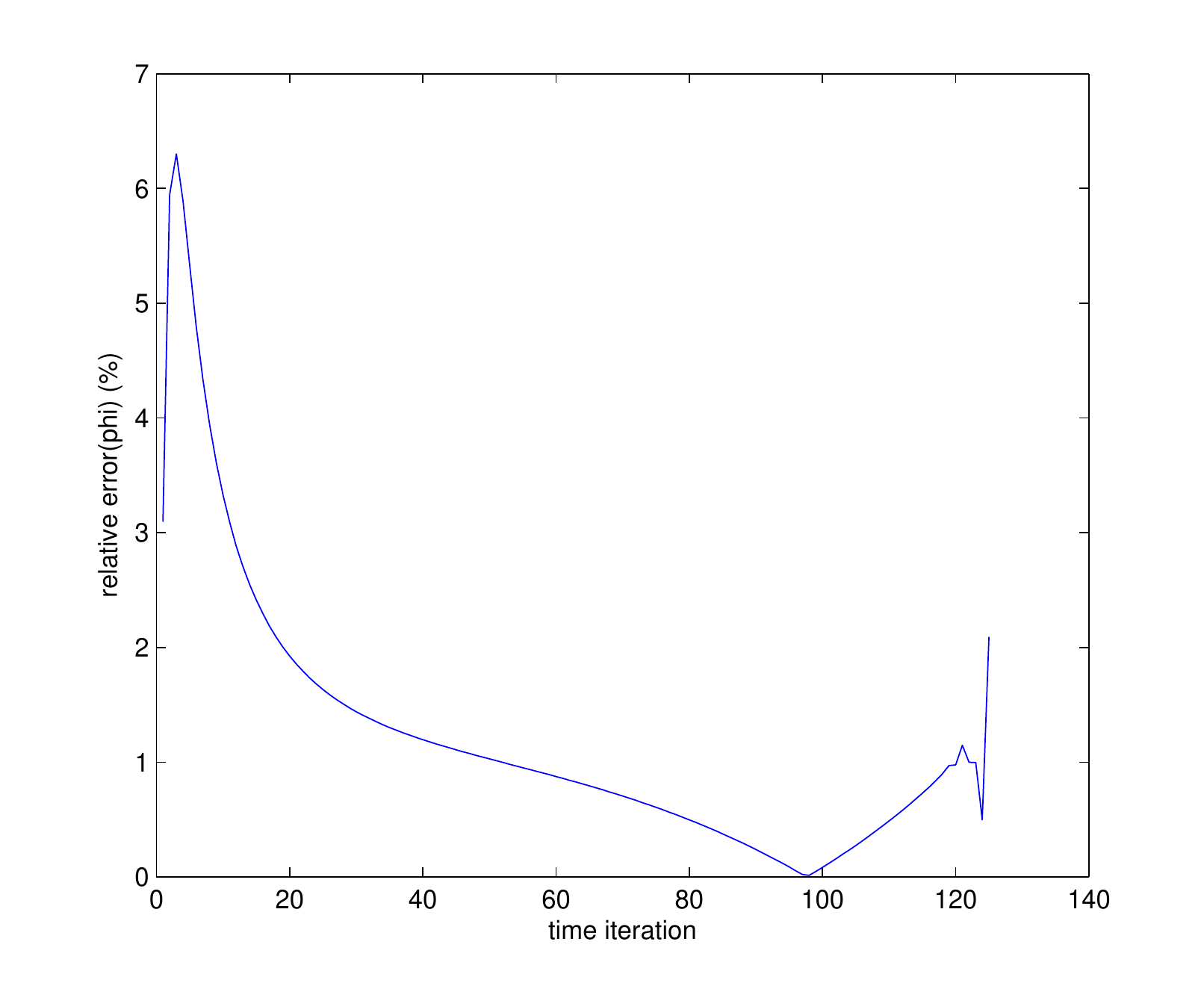}
\caption[$\phi$ evolution errors in time (\%)]{Relative difference (\%) in $\phi$ obtained via two different methods on each time step as discussed in section \ref{sec:phidoterror}}\label{fig:phidot_relerror_time_percent}
\end{figure}

Note the large error that appears on the last time step just before the code reaches a singularity.  To see a graph depicting the percent error (excluding the last time step), see figure \ref{fig:phidot_relerror_time_percent}.

As to the size of the errors, a $5\%$ error or less is not as good as our results from the $q$ equations above (which are orders of magnitude better), but is certainly acceptable when looking at constrained vs. free evolution.  It is especially good when we note that the constraints remain well-behaved as the evolution progresses, even after the formation of apparent horizons and very large values for various curvature measures and their derivatives, that are associated with singularity formation (recall figure \ref{fig:hamconrelerror}).

The entire system is partially constrained in that we use the value of $\phi$ computed from the Hamiltonian constraint in our evolution, in an attempt to conserve the total ``energy'' of the system.  This also relates to the momentum constraints which we will discuss below.

One lens through which to analyze the discrepancy between the two results are the condition numbers of the matrix problem, as discussed in section \ref{sec:condition}.  It is possible that the ill-conditioned nature of the matrix problem for solving the Hamiltonian constraint means that it is impossible to obtain smaller values for $\epsilon_\phi$.

\subsection[Mixed Extrinsic Curvature Evolution Equations Difference]{Convergence Test: Mixed Extrinsic Curvature Evolution Equations Difference}\label{sec:hcdeficit}
When using the mixed form of the extrinsic curvature to evolve the spacetime, we end up with another ``constraint'', or ``difference'' equation that results from computing the difference in equation (\ref{eqn:h12h21diffmix}) which arises from consideration of equation (\ref{eqn:mixkijtensor}):
$$\zeta = max_{(i,j)}\left| \frac{d\left(K^1_2\right)}{dt}-\left(\frac{f}{f_\eta}\right)^2\frac{d\left(K^2_1\right)}{dt}\right| =  0$$

At first it was thought that the existence of this difference identity was an error that arose in the calculation of the evolution equations as none of the previous authors that studied the mixed form of the extrinsic curvature mentioned it, however it is a real phenomenon of the evolution equations.  A calculation of this ``difference'' $\zeta$ as the evolution progresses should also provide a check on the internal consistency of the equations.

\begin{figure} \centering
\includegraphics{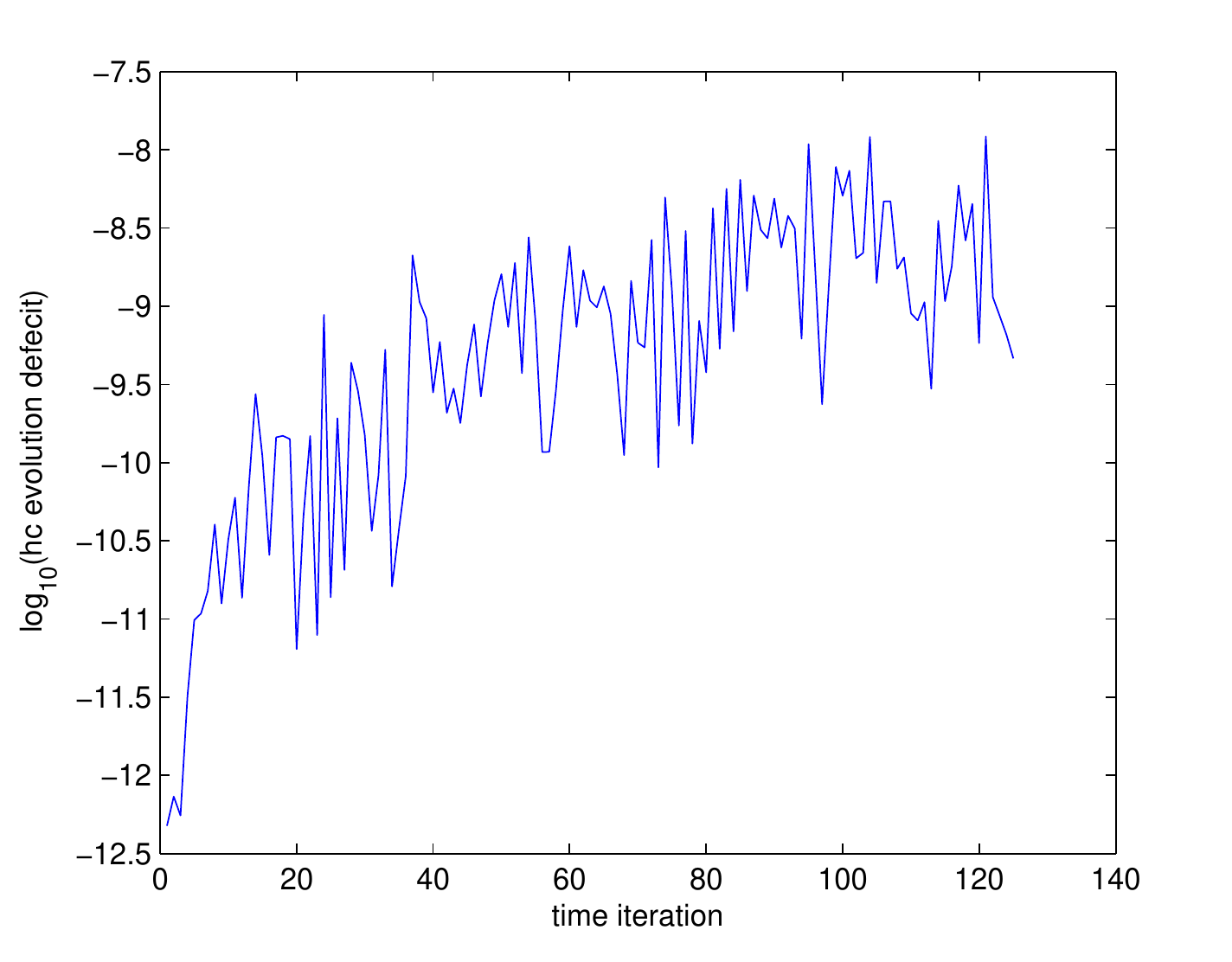}
\caption[$\dot{H_c}$ Difference]{$\log_{10}\zeta$, the difference between the two forms of the extrinsic curvature evolution equation for $H_c$ as discussed in section \ref{sec:hcdeficit}}\label{fig:hcdeficit_time}
\end{figure}

A graph of this measure of internal consistency at each time step is provided in figure \ref{fig:hcdeficit_time}.  As we are using a variety of numerical methods and $max_{i,j}(\dot{H_c}) \sim 1$ on any time step, our results represent an excellent level of internal consistency.
 We find about one part in $\sim 10^8$ error, or about the same as found in section \ref{sec:qdoterror} when comparing the evolution equations for $q$, so we seemingly have consistency between the different evolution equations.

\subsection[Checking Momentum Constraints]{Convergence Test: Checking Momentum Constraints}\label{sec:momconactual}
As discussed in section \ref{sec:momcon} there are two momentum constraints that, because of their degeneracy at $\theta=0$, are unfit for usage in constraining the extrinsic curvature variables. Instead they should be used as checks of the accuracy of the code.  As the Hamiltonian constraint is used to solve for $\phi$, the momentum constraints act as consistency checks on the mixed constrained/free evolution algorithm.

One difficulty in asking ``how well are the momentum constraints obeyed by the evolution'' is that they are supposed to be zero\footnote{Section \ref{sec:CDCSSLFstatictest} also contains a numerical demonstration of deviation from zero.}.  It is important to note that we can rescale equations (\ref{eqn:momcons}) by \emph{any} factor we wish and they should still be zero.  For example, as was mentioned in section \ref{sec:momcon}, we factor out a common
$$f_\eta^2 e^{q+4\phi} \;\; \mathrm{or} \;\; f^2 e^{q+4\phi}$$
when computing the contravariant form of the constraints.  Numerically, however, things are not so clear.  Let us write the momentum constraints schematically as
$$p_l=\sum_{k=1}^{n_l} u_{lk} \;;\; l=1,2$$
where $n_l$ is the number of terms in each of equations (\ref{eqn:momcons}). For example, looking at the radial $(l=1)$ constraint we have
$$u_{11}=\frac{f_{\eta}^2\,H_c\,\cot\theta }{f^2} \;;\; u_{12}=\frac{f_{\eta}^2\,H_c\, q_{\theta} }{f^2} \;;\; \ldots$$

If we were to rescale all of the individual terms $u_{1k}$ by a factor of $10^{-40}$, for example, our constraints would appear to be very well behaved by an absolute error measure when in fact they may not be.  So we are less concerned with the actual value of the total constraint $p_l$ and rather the relative magnitudes of the terms $u_{lk}$ to $p_l$.

The \emph{maximum} values of the \emph{non-scaled} momentum constraints over the entire evolution are generally in the $10^{-5}$ (short run large positive wave) to $10^{-3}$ (long evolution perturbative wave) range, which is orders of magnitude better than other simulations (i.e. \cite{alcubierre:3dbrill,sorkin:code}).

Specifically, a (potentially) more meaningful measure of the momentum constraint violation can be obtained by looking at
$$\xi_l^m = \frac{max_{grid}|p_l|}{||u_{lk}||_m}$$
where $$||\cdot||_m$$ has the usual meaning of an $m$-norm.  The most meaningful measures are probably the $1$ and $\infty$ norms, as the $\infty$-norm represents the best the numerical algorithm can achieve given that we cannot capture values that are less than \emph{(machine precision)*(max value)}, and the $1$-norm represents the absolute worst case for the total value of $p_l$ if we have many terms $u_{lk}$ with large values.

\begin{figure} \centering
\includegraphics{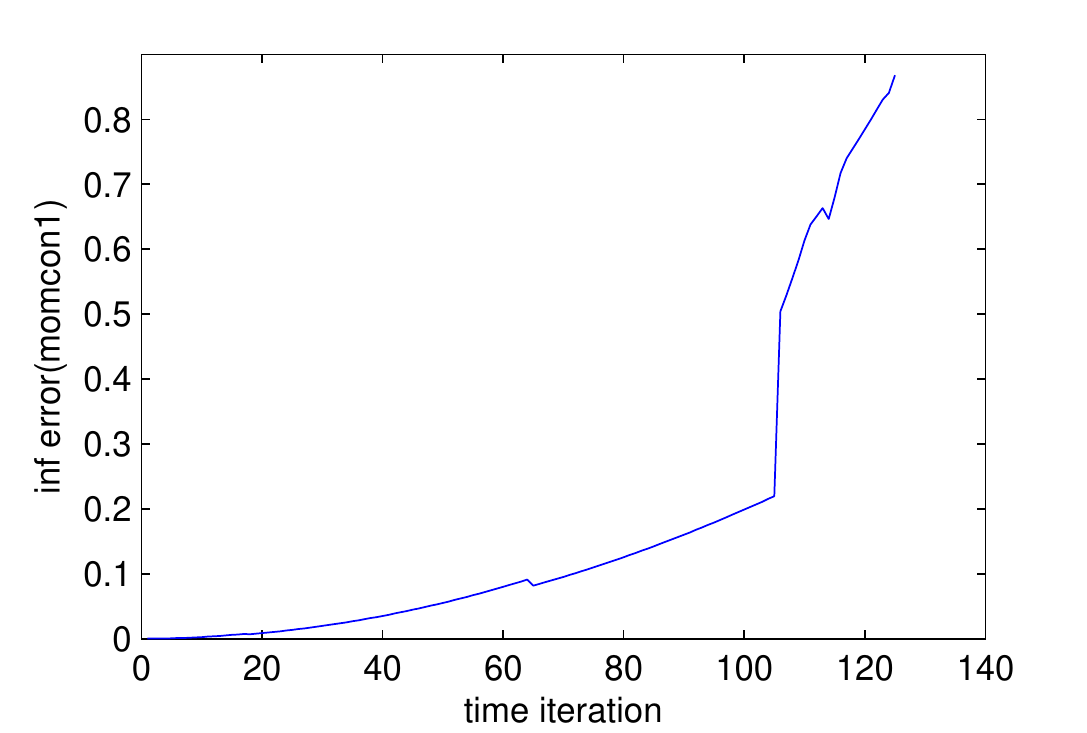}
\caption[Momcon1 Inf Error]{$\xi_1^{\infty}$ ($\infty$-norm error on radial momentum constraint) as discussed in section \ref{sec:momconactual}}\label{fig:momcon1_inf_error}
\end{figure}

\begin{figure} \centering
\includegraphics{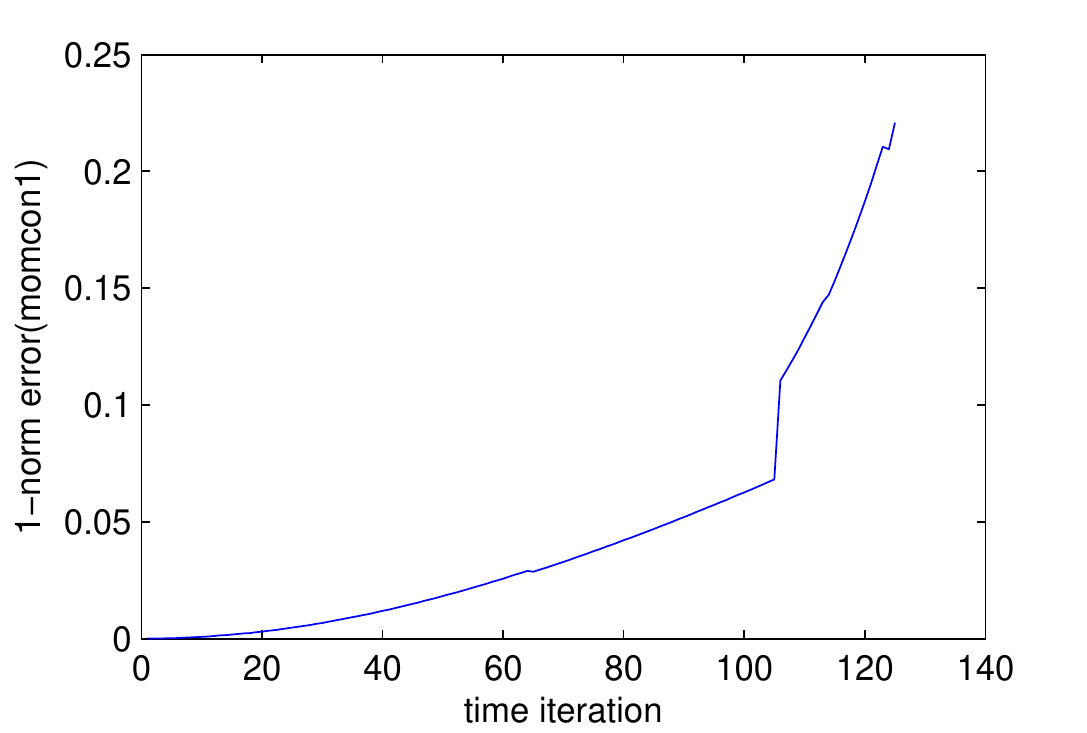}
\caption[Momcon1 $1$-norm Error]{$\xi_1^{1}$ ($1$-norm error on radial momentum constraint) as discussed in section \ref{sec:momconactual}}\label{fig:momcon1_1norm_error}
\end{figure}

Since the maximum values of the constraint over the entire grid are used, this should represent the worst constraint violation, and the ratio removes any issues around rescaling the equations.  Graphs that demonstrate how $\xi^1_1$ and $\xi^\infty_1$ evolve over time\footnote{For the same evolution as described above.} are shown in figures \ref{fig:momcon1_inf_error} and \ref{fig:momcon1_1norm_error}.

We can see that the radial $(l=1)$ momentum constraint starts off very well (both $\xi_1^1$ and $\xi_1^{\infty}$ are $<10^{-4}$), and grows until $t\sim106$ where $\xi_1^{\infty}\sim 22\%$, then it skyrockets in one timestep to $50\%$.  The change in $\xi_1^{1}$ isn't as dramatic as it seems that many $u_{lk}$ values grow on that timestep and cause the relative change to be smaller.  There is no sudden change in any of the variables' evolution that is visible at this time step, but there is definitely a sudden increase in the measure of this constraint.  This is because a single point on the axis or far out in the wave zone suddenly grows to have a larger violation and has relatively little effect on the entire simulation\footnote{As we are finding the maximum value over the entire grid, the actual maximum violation point can vary significantly.}.

By the time the code encounters a singularity $\xi_1^{\infty}\sim 171\%$ and $\xi_1^{1}\sim 36\%$ (we have omitted the very last time step as above to give better scales on the graphs).

\begin{figure} \centering
\includegraphics{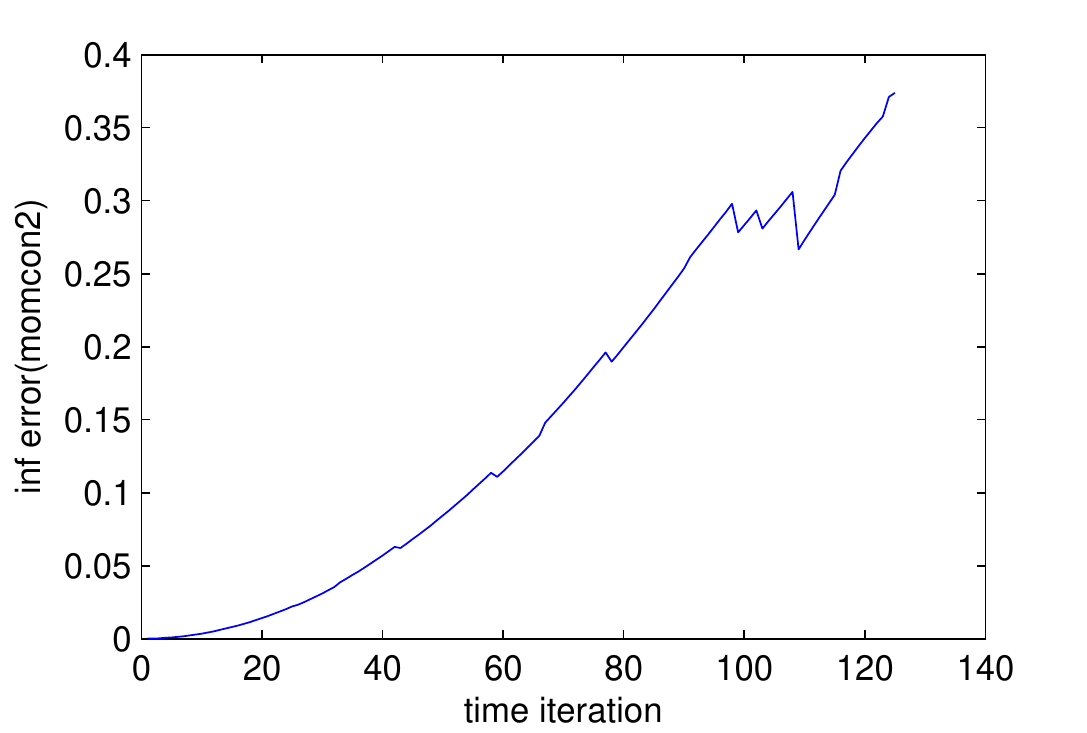}
\caption[Momcon2 Inf Error]{$\xi_2^{\infty}$ ($\infty$-norm error on angular momentum constraint) as discussed in section \ref{sec:momconactual}}\label{fig:momcon2_inf_error}
\end{figure}

\begin{figure} \centering
\includegraphics{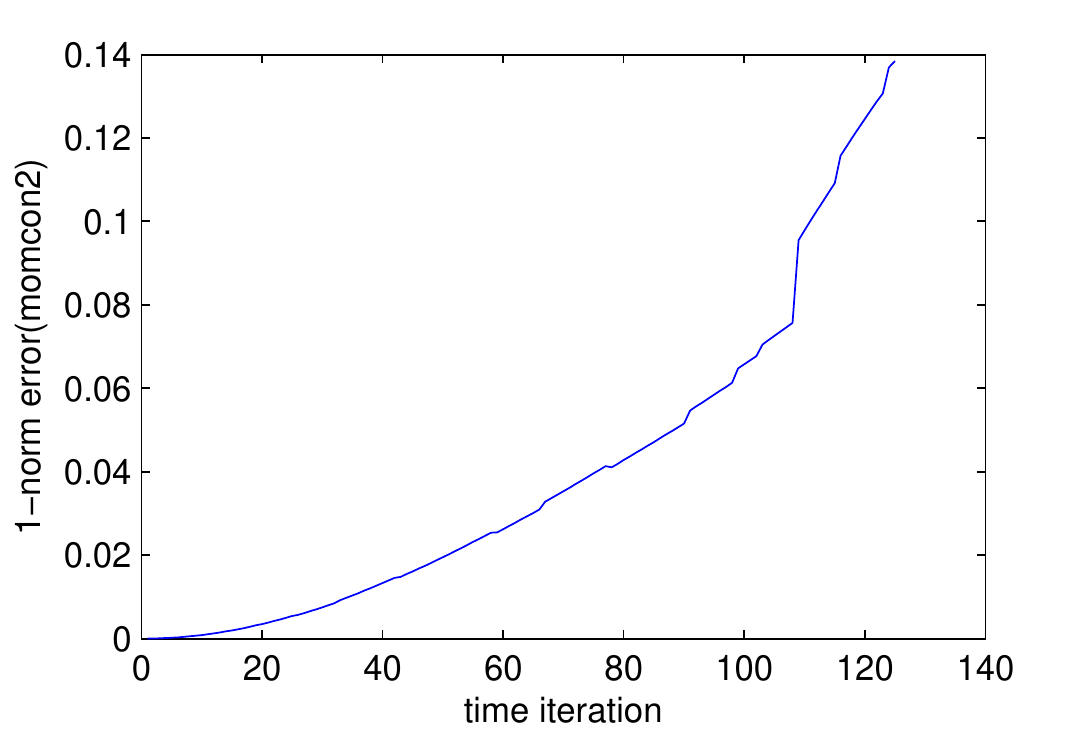}
\caption[Momcon2 $1$-norm Error]{$\xi_2^{1}$ ($1$-norm error on angular momentum constraint) as discussed in section \ref{sec:momconactual}}\label{fig:momcon2_1norm_error}
\end{figure}

The evolution of the angular $(l=2)$ momentum constraint errors are shown in figures \ref{fig:momcon2_inf_error} and \ref{fig:momcon2_1norm_error}.

As above, the angular momentum constraint starts off very well (both $\xi_2^1$ and $\xi_2^{\infty}$ are $<10^{-4}$ as well), and slowly increases as the code progresses.  This constraint is better behaved than the first one, although it still reaches $\xi_2^{\infty}\sim 94\%$ and $\xi_2^{1}\sim 48\%$ by the end of the simulation.

One possible explanation for the growth of these errors is that as the evolution proceeds and the values of the variables from which the momentum constraints are calculated grow, the absolute error grows\footnote{Assuming we have constant $\sim$ one part in $10^8$ precision}.  As this absolute error grows, the relative error near the crossing points of the variables from positive to negative values grows without bound as well.  Due to this it is unclear if $\xi$ is a good measure of error or not, and will require further investigation.

\begin{figure} \centering
\includegraphics{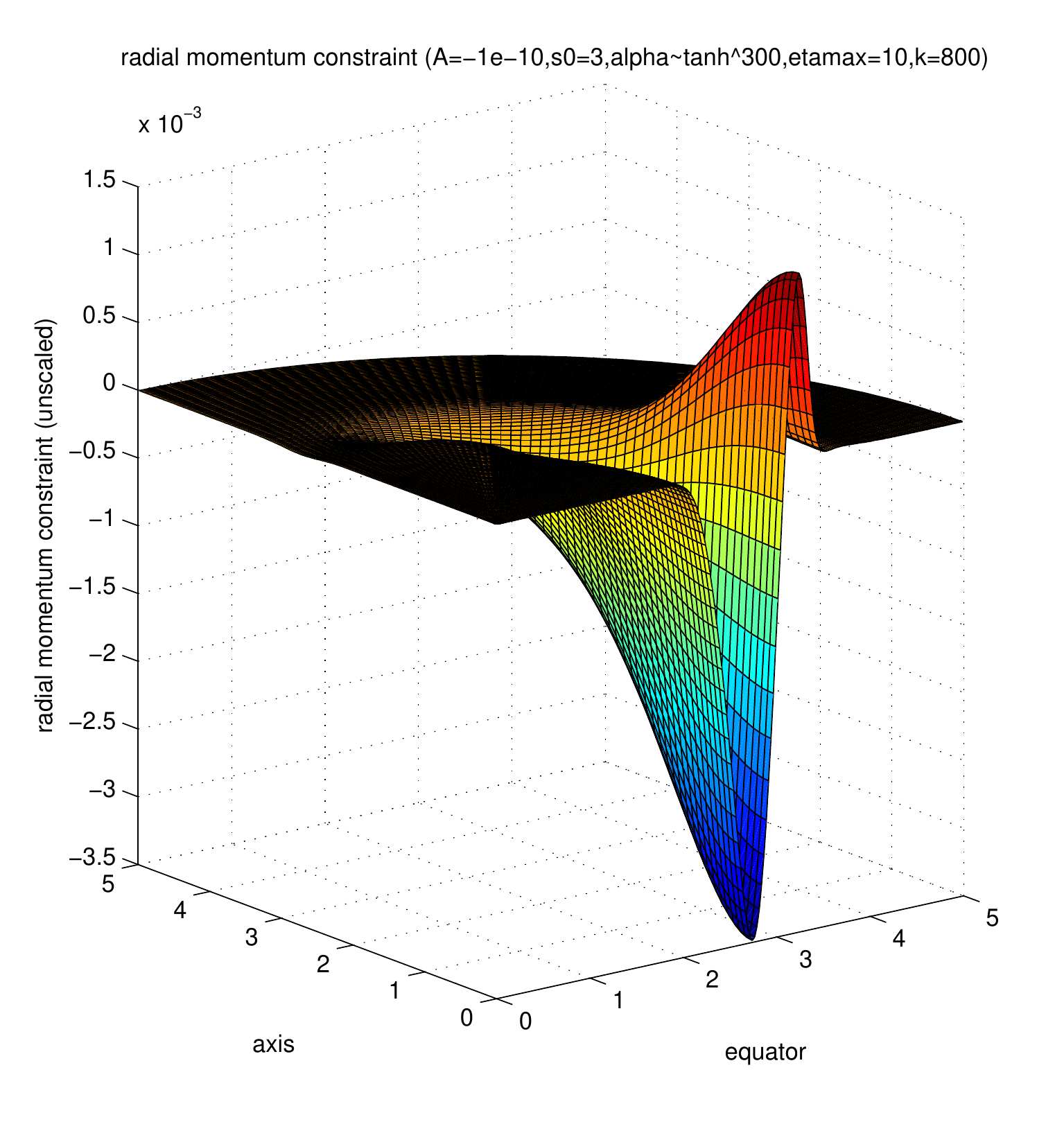}
\caption[Momcon1 value]{First momentum constraint values for interior region at $k=800$ as discussed in section \ref{sec:momconactual}.  Values in the outer zones are too close to zero to distinguish at this scale.}\label{fig:momcon1_val_t800}
\end{figure}

\begin{figure} \centering
\includegraphics{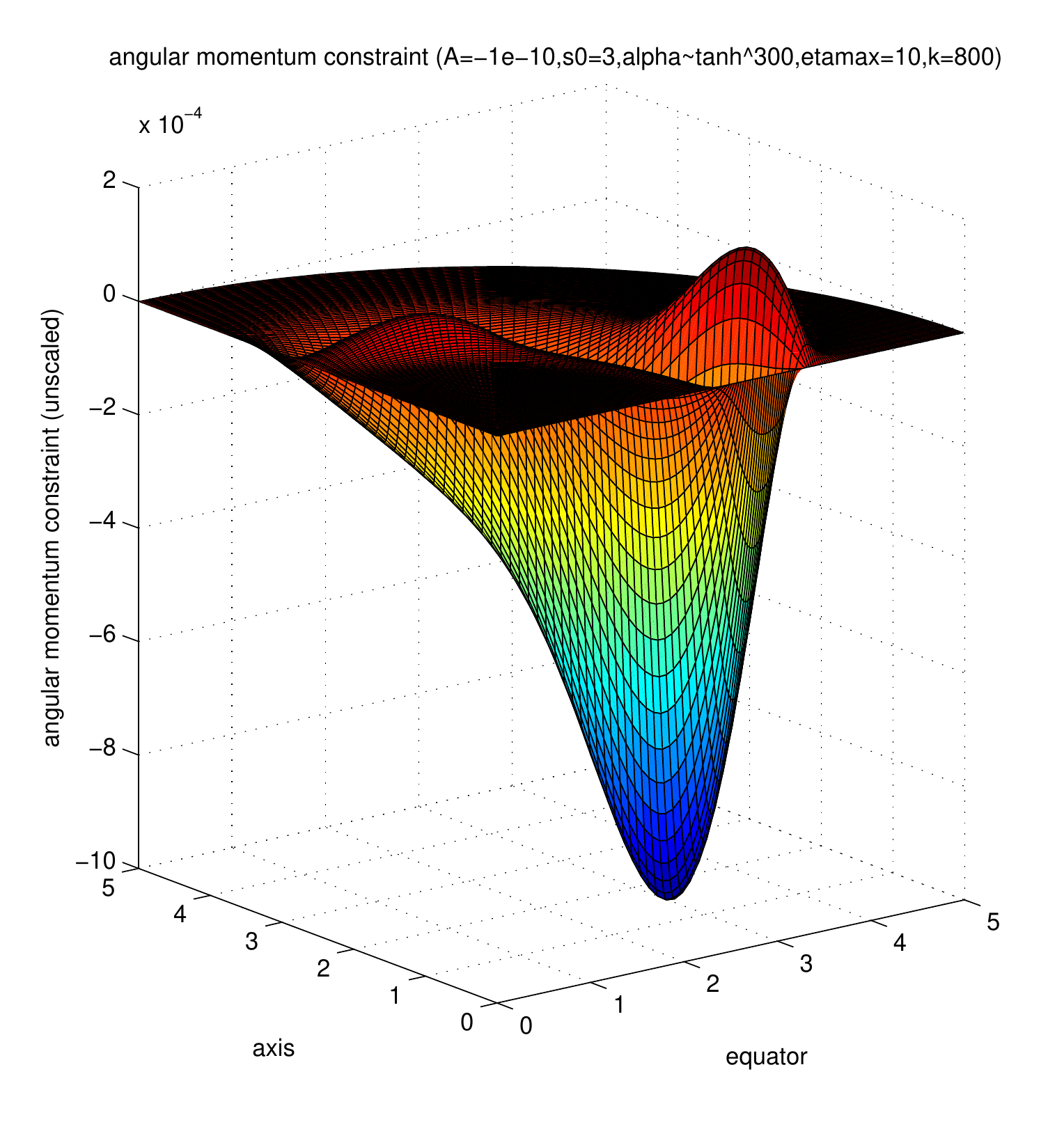}
\caption[Momcon2 value]{Second momentum constraint values for interior region at $k=800$ as discussed in section \ref{sec:momconactual}.  Values in the outer zones are too close to zero to distinguish at this scale.}\label{fig:momcon2_val_t800}
\end{figure}

Some graphs of the actual values of the constraints over the entire grid are shown in figures \ref{fig:momcon1_val_t800} and \ref{fig:momcon2_val_t800}.

\subsection[Checking Momentum Constraint Regularity via $-H_d+H_b=0$]{Convergence Test: Checking Momentum Constraint Regularity via $-H_d+H_b=0$}\label{subsec:momcon-hd+hb}
Recalling that the angular momentum constraint in (\ref{eqn:momcons}) requires the algebraic condition
$$-H_d+H_b=0$$
be satisfied in order that all $\cot\theta$ terms are regular at the axis ($\theta=0$), we will now examine the behaviour of this quantity.  It is observed that this quantity is \emph{exactly} $0$ in the code, which at first glance seems suspicious for a numerically calculated quantity, until we examine the evolution equations themselves.

Recalling that on the axis ($\theta=0$):
\begin{itemize}
\item $q=0$ from the Brill conditions
\item $\alpha=\alpha_{\theta\theta}=0$ from regularity conditions on the lapse function
\item $\phi$ is symmetric (i.e. $\phi_\theta=0$)
\item $H_c=0$ for regularity
\item Other extrinsic curvature quantities ($H_a,H_b,H_d$) are symmetric
\end{itemize}
Therefore the evolution equation for $H_b$ (\ref{eqn:hbevol}) along the axis reduces to:
\begin{eqnarray}\label{eqn:hbevol_axis}
\frac{\partial H_b}{\partial t} & = & \frac{1}{f^2 \, e^{q+4\phi}}\left(
-\frac{2\, {\alpha}_{\eta} \, \phi_{\eta} \, f^2}{f_\eta^2}
-\frac{{\alpha}_{\eta} \, f}{f_\eta} \right)
+ {H_b}_{,\eta} \,v_1
\end{eqnarray}
Similarly, the evolution equation for $H_d$ (\ref{eqn:hdevol}) along the axis reduces to:
\begin{eqnarray}\label{eqn:hdevol_axis}
\frac{\partial H_d}{\partial t} & = & \frac{1}{f^2 \, e^{q+4\phi}}\left(
-\frac{2\, {\alpha}_{\eta} \, \phi_{\eta} \, f^2}{f_\eta^2}
-\frac{{\alpha}_{\eta} \, f}{f_\eta} \right)
+ {H_d}_{,\eta} \,v_1
\end{eqnarray}

\begin{figure} \centering
\includegraphics{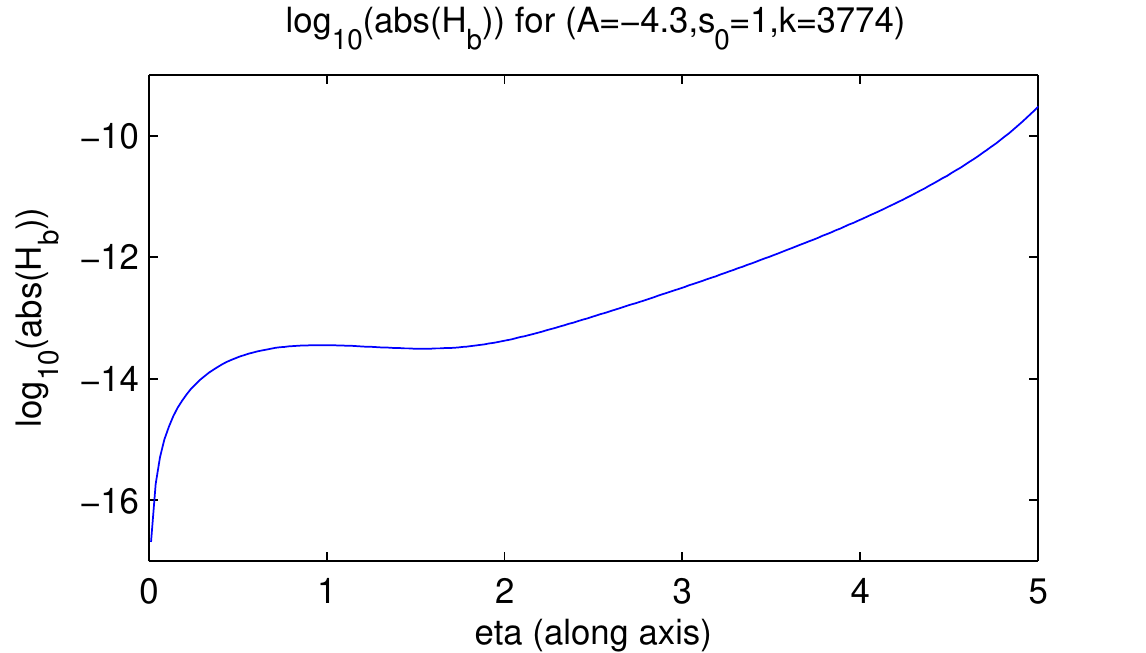}
\caption[Checking $-H_d+H_b=0$]{Plot of $\log_{10}|H_b|$ for $(A=-4.3,s_0=1,k=3774)$ along the axis, showing that the \emph{numerical} cancellation $-H_d+H_b=0$ is not trivial due to their non-zero values.}\label{fig:log10absh22_A_-4.3_s0_1_t3774}
\end{figure}

Thus $H_b$ and $H_d$ should evolve numerically in exactly the same manner along the axis provided we have properly enforced the numerical conditions/symmetries listed above and $H_b=H_d=0$ initially.  So we are guaranteed that $-H_d+H_b=0$ for all time analytically, which is verified numerically.  Figure \ref{fig:log10absh22_A_-4.3_s0_1_t3774} demonstrates both $H_b$ and $H_d$ are non-zero along the axis even after long evolutions ($k=3774$ in this case), so this is not a trivial result.

Since we observe that the other $36$ numerical terms from the extrinsic curvature evolution equations cancel exactly along the axis in the code we have a high degree of confidence that the boundary conditions along the axis and all of the numerical methods discussed in chapter \ref{chap:changes} that have employed to ensure that derivatives, etc. of the dynamical variables are calculated properly have been successful.  This also indicates that we will probably not need to define auxiliary variables as Evans \cite{evans} did in attempts to stabilise the evolution of the dynamic variables along the axis.  Further, this indicates that we do not observe the ``axis instability'' present in many other numerical codes, as we have enforced specific regularity conditions to prevent this instability from arising.

\subsection{Long Duration Runs}
As further evidence that the code does not get overwhelmed with numerical noise or other erroneous calculations or bugs, it is worth noting that the code was capable of running for $\sim 9$ months and performing over $10^{12}$ calculations, on a non-trivial spacetime, before reaching a singularity and halting.  This was achieved by evolving the interior of a black hole (see section \ref{subsec:BHInt}) and there were $130 430$ calculations of the ADM mass performed during this run, all of which were \emph{unique} values.  This indicates that the conformal factor $\phi$ was continuously changing (i) during the evolution from time step to time step and (ii) on each time step's doubly iterative Crank-Nicholson scheme.

With this many non-trivial calculations being performed in a stable manner we have further confidence that the numerical schemes are performing as intended.

\section{Determination of Other Code Parameters}

\subsection{Testing Variable Resolution Evolutions}
Many authors have studied the effect of grid resolution on the values of constraints to check for convergence (see for example \cite{bernstein}).  While there is validity in this method, it is often difficult to determine the exact nature of the factor or factors that limit the precision of the code's evolution.  For example Bernstein \cite{bernstein} observes decreasing errors and or convergence for some numerical methods and or grid sizes and not for others in numerical constructions of the Schwarzschild spacetime.

The code discussed here employs fourth order spatial derivative terms, second order time derivative terms, a quasi-linear equation to solve for $\phi$, nonlinear coupling among all variables whose evolutions require iterative convergence, warped slicing of the spacetime that causes horizons to form differently across the time slice, shifting of coordinate points that distorts distances and exponentially increasing radial coordinates.  We have tried to mitigate these errors as much as possible, to varying degrees of effect.

So let us examine the results of three different grid resolutions and their time evolutions, for the same initial data ($A=2e-5$, $s_0=1$, mixed $K$, $\eta_{\mathbf{max}}=5$, $\Delta t=0.5\Delta\eta$) in table \ref{tbl:compgridresolu}.  Note that as the number of grid points increases, $\Delta t$ decreases due to its tie to $\Delta\eta$ (see table \ref{tbl:gridcoords}); this means that we have a finer grid resolution which will \emph{also} evolve more slowly\footnote{This also makes it difficult to meaningfully comment on a change in grid size at the same time step - as $\Delta t$ will be getting larger and larger relative to $\Delta\eta$, causing causality violations or other effects if we do not adjust $\Delta t$ accordingly}.
\begin{table}\begin{centering}
\begin{tabular}{cccccc} \hline
Grid size & $k_{max}$ & $\Delta\eta$ & $\Delta t$ & total $t_{kmax}$ & $(t,\xi^{\infty}_1,\xi^{\infty}_2,M_{ADM}\times 10^{-2})$ \\ \hline \noalign{\vskip 2mm} 
$200 \times 60$ & $120$ & $0.025$ & $0.0125$ & $1.5$ & $(118,\frac{5.45}{6.89}=0.791,\frac{1.82}{5.21}=0.35,2.84)$ \\ \noalign{\vskip 2mm} 
$300 \times 90$ & $179$ & $0.01\bar{6}$ & $0.008\bar{3}$ & $1.492$ & $(177,\frac{5.44}{6.92}=0.786,\frac{1.83}{5.28}=0.35,2.87)$ \\ \noalign{\vskip 2mm} 
$400 \times 120$ & $238$ & $0.0125$ & $0.00625$ & $1.4875$ & $(236,\frac{5.44}{6.85}=0.794,\frac{1.83}{5.21}=0.35,2.88)$ \\ \noalign{\vskip 2mm} 
\hline
\end{tabular}\caption{Comparison of evolutions at varying grid sizes}\label{tbl:compgridresolu}
\end{centering}\end{table}

What is compelling about these results is that not only do the spacetimes become singular at the same time to within the precision of their $\Delta t$'s, they also land on approximately the same momentum constraint measures $\xi^{\infty}_1$,$\xi^{\infty}_2$ and the same quasi-local ADM mass $M_{ADM}$, despite having radically different grid spacings. In addition the grids at one resolution do not share any grid points with the evolutions at alternate resolutions\footnote{As $\eta=(i-0.5)\Delta\eta$, there are no points on the grids that can align.}.

This indicates that the simulation is already running at or near the limit of its numerical precision with a $200\times 60$ grid.  It also indicates that there is not likely to be an error in the code's equations given the unlikeliness of having two iterative convergence loops operating on different grid points in all three scenarios, at different time step intervals, all converging on the same answer if there was an error.

\subsection{Verifying Lapse Behaviour and an Alternate Lapse Form}\label{subsec:results_constraint}
The results presented for the momentum constraints are not abnormal for evolutions of the Einstein equations, and in fact those computed in this research are well-behaved compared to some of the other results in the literature.  Even Minkowski (flat) and Schwarzschild (static spherically symmetric black hole) numerically generated space-times are ill-behaved in a number of cases (see for example \cite{bernstein,bernuzzi,richter}), and our results are clearly more robust than those obtained by alternative approaches.

\begin{figure} \centering
\includegraphics{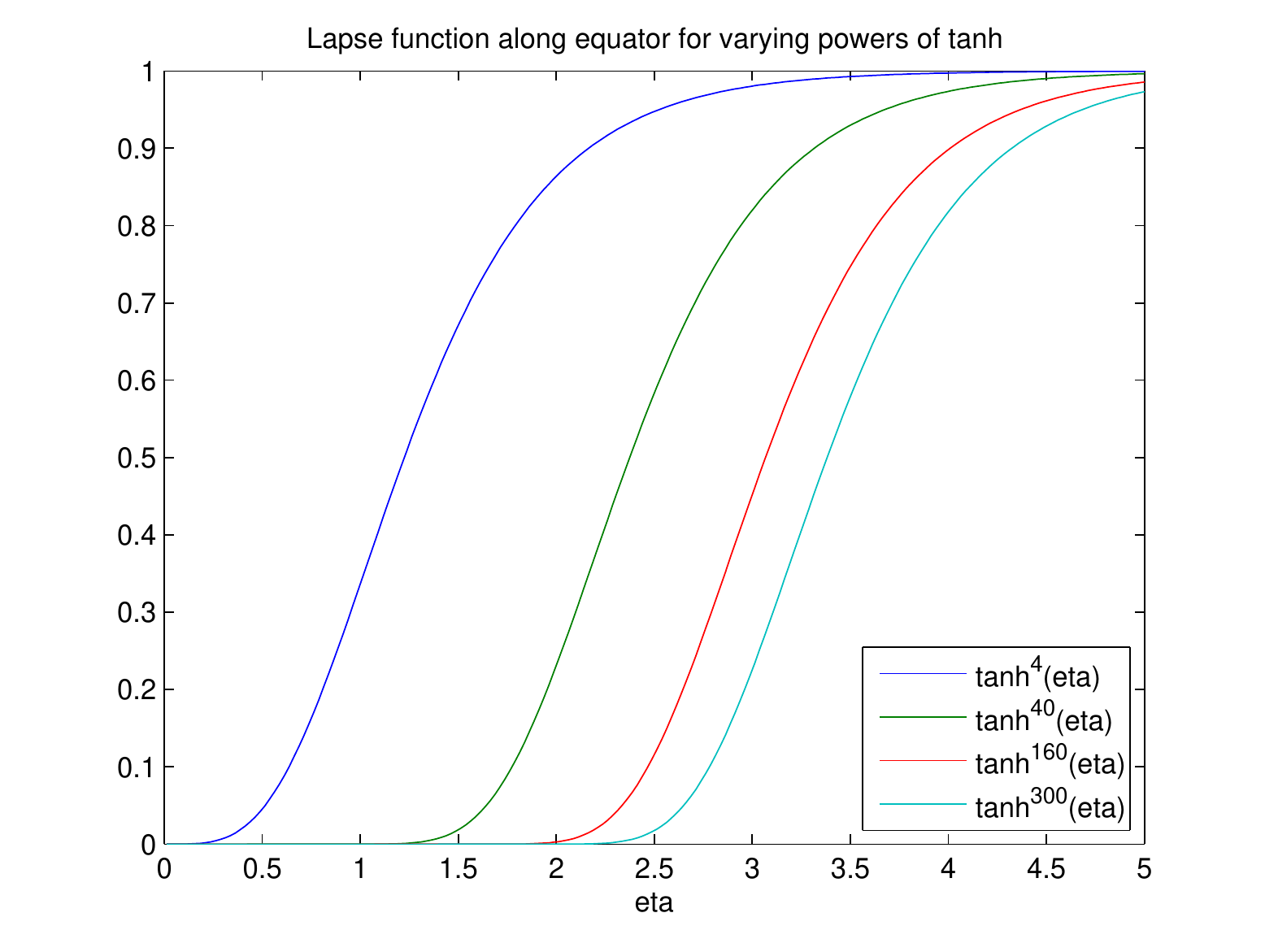}
\caption[Lapse shape for various $\tanh$ powers]{Shape of the lapse function $\alpha$ along the equator ($\theta=\frac{\pi}{2}$) for varying powers of the hyperbolic tangent function (i.e. equation \ref{eqn:alp40}) .} \label{fig:tan_n_eta}
\end{figure}

Part of the difference seems to be that our slicing condition does not prevent areas of large curvature from evolving (whereas other conditions such as maximal slicing do), and therefore the numerical errors can grow quickly in these regions.  By altering the lapse function to the form
\begin{equation}\label{eqn:alp40}\alpha=\tanh^{40}(\eta)\sin^2\theta\end{equation}
we can significantly slow down the evolution in the area of the grid near $\eta=1$, which seems to be where the spacetime frequently ``blows up''\footnote{``blows up'' is used to mean that the values of the scalar curvature $R$ become large (i.e. $<-100$), as well as the fact that the conformal factor $\phi$ is unable to converge when using the Hamiltonian constraint, and the momentum constraints as shown above have large violations which double or more each time step before the code halts; these three things always happen in tandem.}, allowing the evolution to proceed for much longer times.  Figure \ref{fig:tan_n_eta} demonstrates how the lapse function along the equator ($\theta=\frac{\pi}{2}$) depends on various powers of the hyperbolic tangent function.

Using the initial data ($A=-1 \times 10^{-5}$, $s_0=3$, mixed extrinsic curvature) and using the lapse function in equation (\ref{eqn:alp40}), we find that the evolution runs for $340$ time steps before developing singular behaviour.  So the code runs almost $3$ times as long ($t_{max}=4.25$ instead of $t_{max}=1.5$).  It is also apparent that the lapse retards the evolution near the origin, as one would expect, so that the blow-up in the curvature variables, $\Psi_4$, etc. happens further out on the equator where the magnitude of initially small variations (oscillations) grows without bound during the evolution.

\begin{figure} \centering
\includegraphics{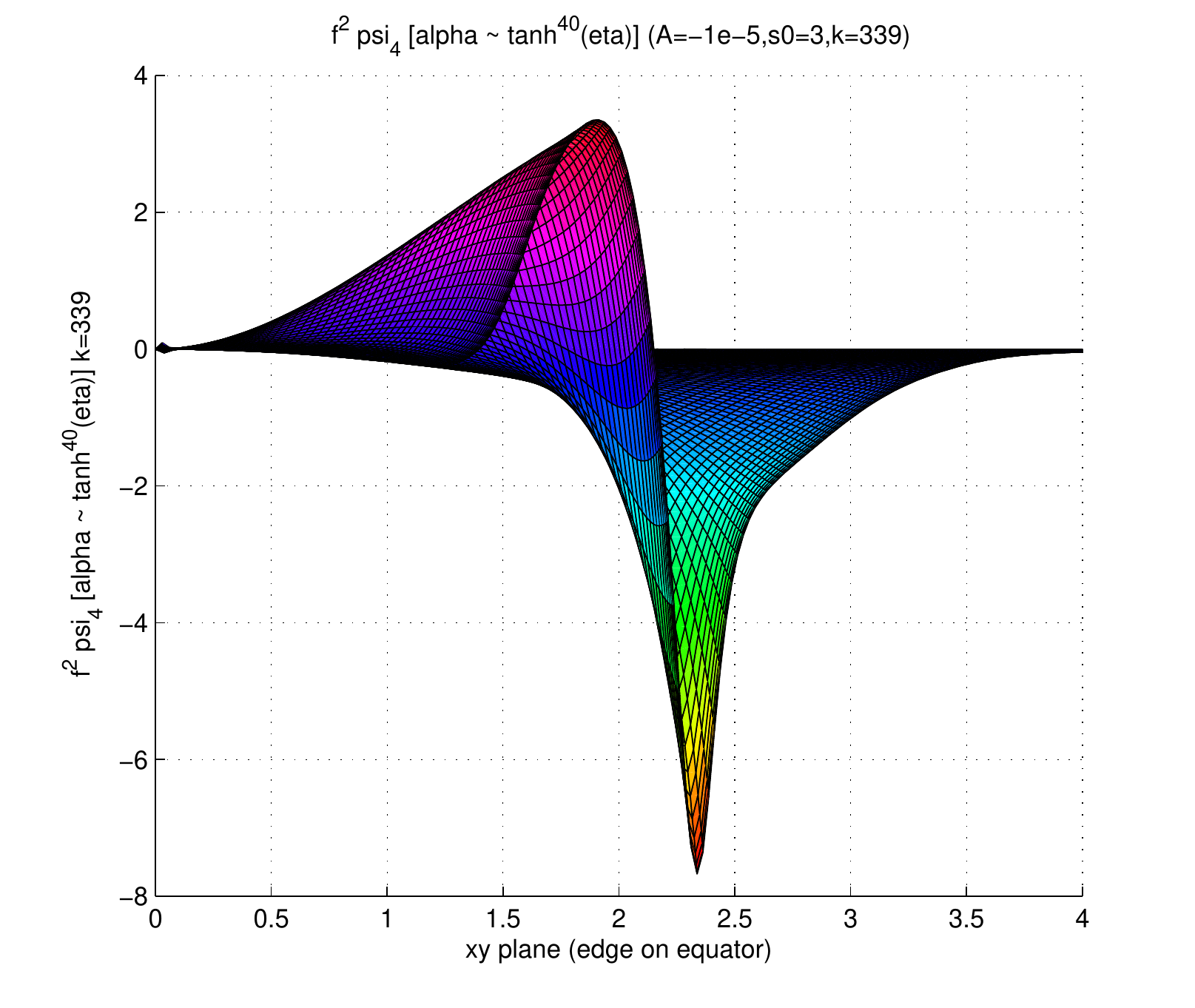}
\caption[$f^2\Psi_4$ for $\alpha\sim\tanh^{40}(\eta)$]{$f^2\Psi_4$ for $\alpha\sim\tanh^{40}(\eta)$ looking edge on from the equator (in the foreground) at the $339$th time step just before singularity formation.  Note the values are growing large (positive) at $\eta\sim 1.8$ and (negative) at $\eta\sim 2.4$ as discussed in section \ref{subsec:results_constraint}.  Also note the sharp peak being formed near $\eta\sim 2.4$.} \label{fig:f2psi4_t339_alp_tanh40}
\end{figure}

Figure \ref{fig:f2psi4_t339_alp_tanh40} shows $f^2 \Psi_4$ with $\alpha\sim\tanh^{40}$ at the $339$th time step, with the first $45$ radial points removed\footnote{$\Psi_4\sim\frac{1}{\alpha}$ in Bernstein's tetrad (see equation (\ref{eqn:psi4bern})), so really small values for the lapse cause an ``artificial'' blow-up at the origin that has no effect on the rest of the evolution, so we remove those points to see the real effect further out on the grid.}.

\begin{figure} \centering
\includegraphics{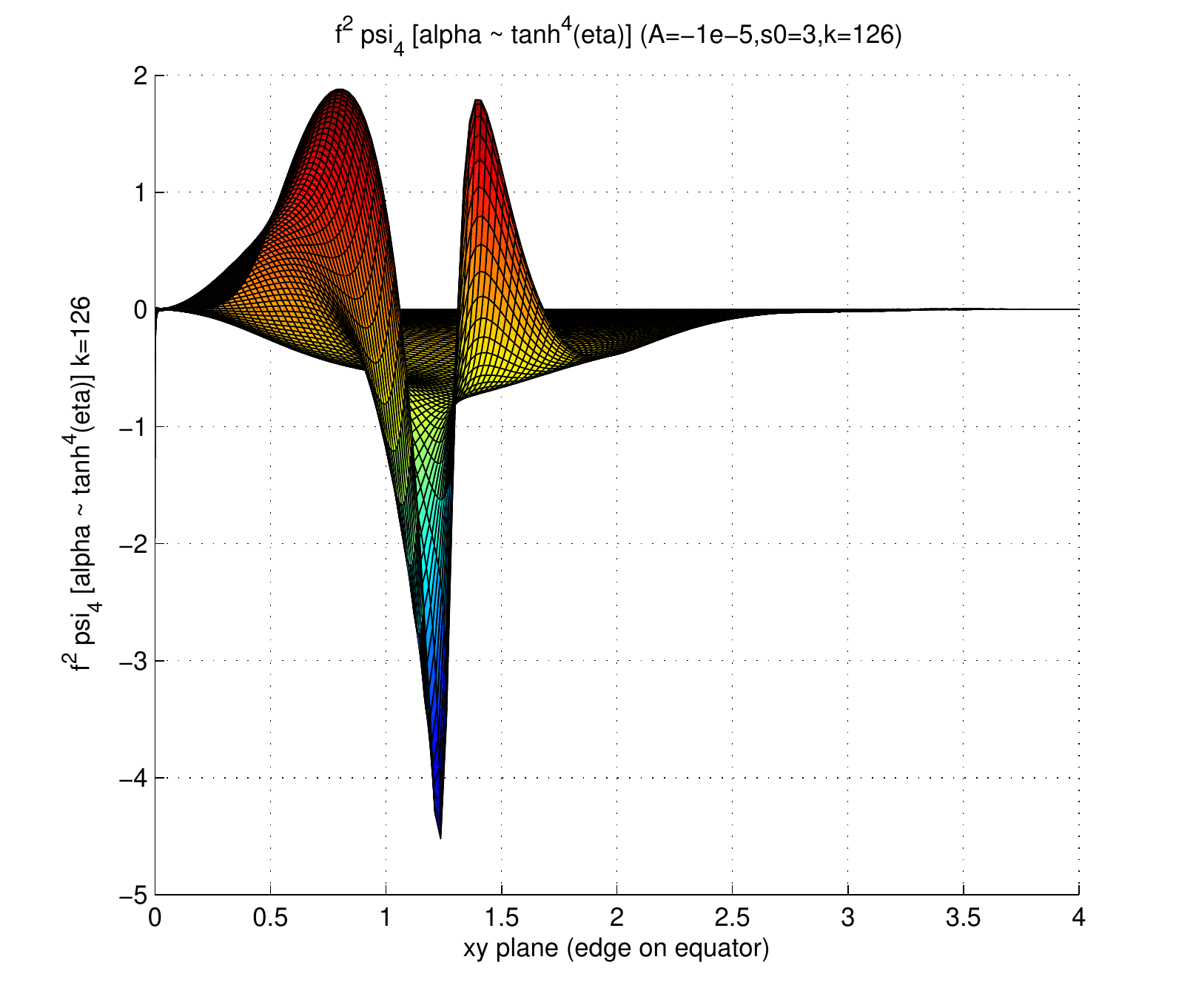}
\caption[$f^2\Psi_4$ for $\alpha\sim\tanh^{4}(\eta)$ near origin]{$f^2\Psi_4$ for $\alpha\sim\tanh^4(\eta)$ and initial data ($A=-1 \times 10^{-5},s_0=3$) looking edge on from the equator (in the foreground) at the $126$th time step just before singularity formation.  Compare the scale to figure \ref{fig:f2psi4_t339_alp_tanh40} as discussed in section \ref{subsec:results_constraint}. Also note the sharp peak being formed near $\eta\sim 1.2$.} \label{fig:f2psi4_t126_alp_tanh4_orig}
\end{figure}

For comparison, see figure
\ref{fig:f2psi4_t126_alp_tanh4_orig} which show the values for $f^2\Psi_4$ in a late stage evolution for $\alpha\sim\tanh^4(\eta)$.

\begin{figure} \centering
\includegraphics{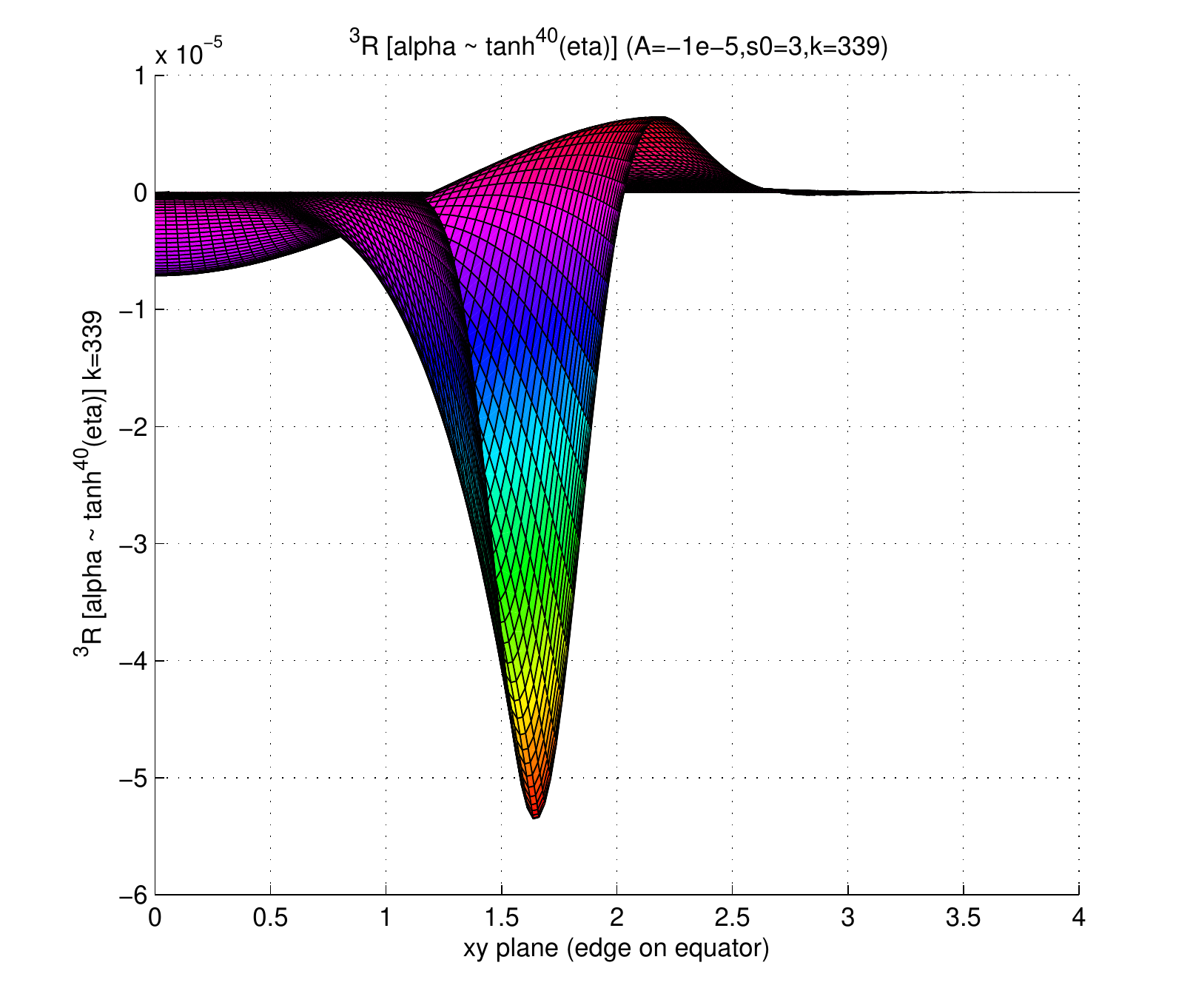}
\caption[${}^{(3)}R$ for $\alpha\sim\tanh^{40}(\eta)$]{${}^{(3)}R$ for $\alpha\sim\tanh^{40}(\eta)$ looking edge on from the equator (in the foreground) at the $339$th time step just before singularity formation.  Note the values are growing large (negative) at $\eta\sim 1.7$ and (positive) at $\eta=2.2$ as discussed in section \ref{subsec:results_constraint}.  The area from $\eta=0\rightarrow\eta=1$ is essentially flat along the equator (in the foreground).} \label{fig:scurv_t339_alp_tanh40}
\end{figure}

\begin{figure} \centering
\includegraphics{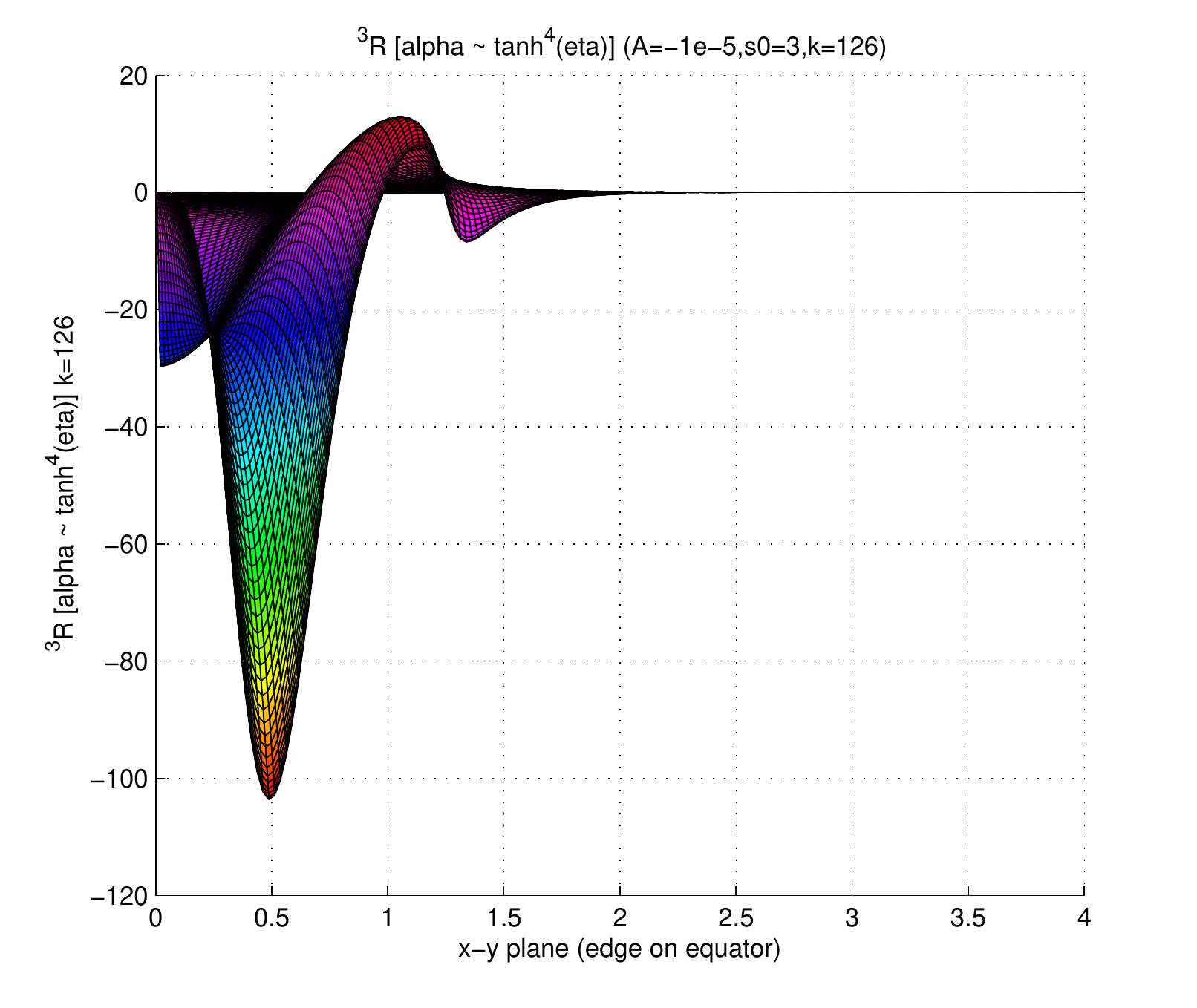}
\caption[${}^{(3)}R$ for $\alpha\sim\tanh^{4}(\eta)$]{${}^{(3)}R$ for $\alpha\sim\tanh^4(\eta)$ looking edge on from the equator (in the foreground) at the $126$th time step (region near origin cut away) just before singularity formation.  Note the oscillating nature that dies out as we move along the equator as discussed in section \ref{subsec:results_constraint}. Compare to figure \ref{fig:scurv_t339_alp_tanh40}.} \label{fig:scurv_t126_alp_tanh4}
\end{figure}

We also perform a similar comparison with the scalar curvature ${}^{(3)}R$ in figures \ref{fig:scurv_t339_alp_tanh40} and \ref{fig:scurv_t126_alp_tanh4}.  The lapse is obviously suppressing the evolution of the scalar curvature near the origin and the oscillations further out on the axis dominate at later times.

What does this tell us?  That the lapse ($\alpha$) is doing its job of slowing the evolution in areas where we define it to be really small (close to zero), and that eventually the curvature terms further out on the equator cause the code to encounter a singularity even if we ``slice'' out the area near the origin from evolving.  This indicates that we aren't just capturing some erroneous $\frac{1}{f}$-type error in the evolution, as we are far enough out on the grid that those terms disappear.  We can also deduce that there are likely multiple singularities that develop in the spacetime.

It also indicates that we are avoiding, for longer coordinate time periods, the incomplete geodesics that originate on the apparent horizon, as the progression of proper time is significantly slowed in the area around the apparent horizon.

Further evidence of this can be seen in the results presented in table \ref{tbl:videoresults} where the use of an alternate lapse function allows the spacetime to evolve for a larger number of iterations with all other variables kept the same.  For example, with a large initial wave $(A=9,s_0=1)$ the code encounters a singularity after $38$ time steps with $\alpha\sim\tanh^4(\eta)$.  By altering the lapse form to $\alpha\sim\tanh^{300}(\eta)$ the code encounters a singularity after $1917$ time steps.

Some hybrid lapse function which combines the need to have $r^4$-like behaviour at the origin and maximal-slicing-area-of-high-curvature-avoiding properties may aid in studying longer evolutions of these spacetimes, although if the trapped surfaces and large curvature values are indeed indicating the formation of a black hole, then we have accomplished our goal of discovering black hole formation.

\subsection{Testing Alternate Outer Boundary Locations}
We generally run the code with an outer radial boundary located at $\eta=5$, which translates to $r\sim 74.2$ with our radial function\footnote{Using $r=\sinh(\eta)$}, however it is possible to move the outer radial computational boundary out further in order to obtain a better approximation to spatial infinity in an asymptotically flat spacetime.

Running weak wave scenarios with $\eta_{\mathbf{max}}=10$ (moving the outer edge of the grid out substantially as this translates to $r\sim 11,000$ from $74.2$) produces the same overall results.

\begin{figure} \centering
\includegraphics{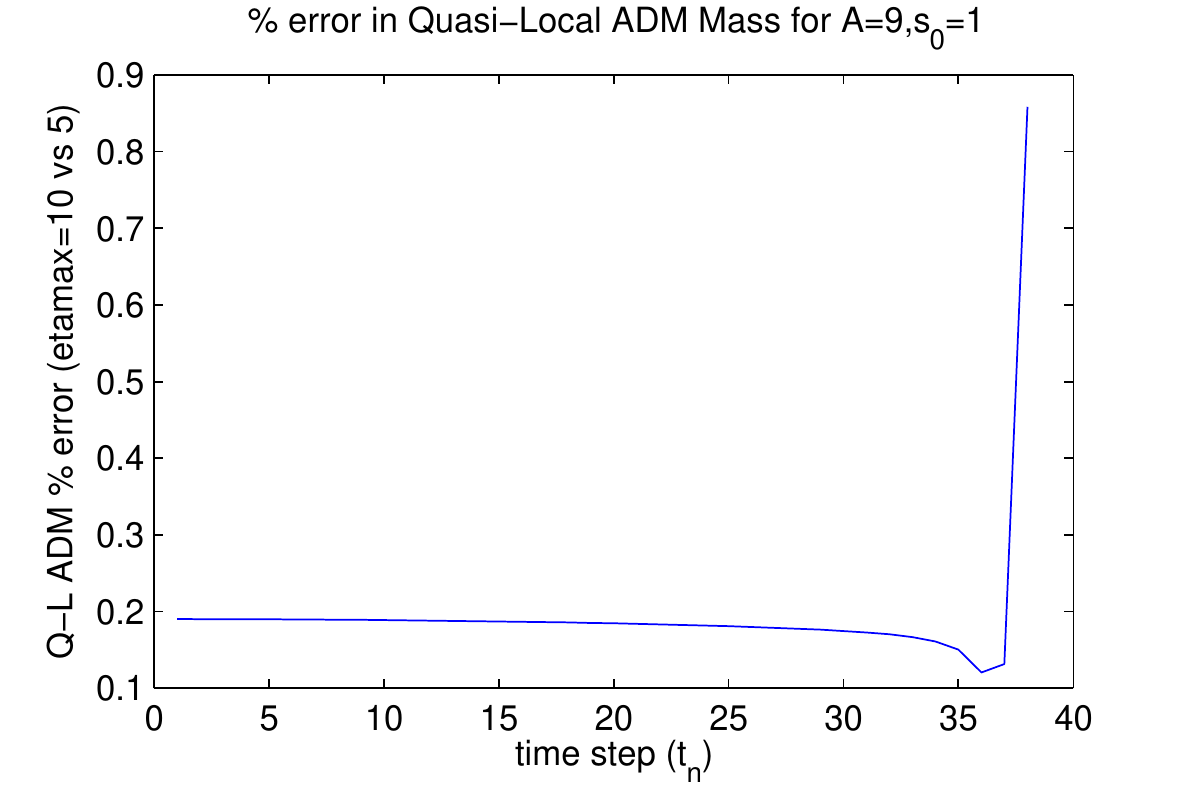}
\caption[Quasi-Local ADM Mass difference for different outer boundaries]{Percent difference in the quasi-local ADM mass measured over time at the outer edge of the grid in the situations $\eta_{\mathbf{max}}=5$ and $\eta_{\mathbf{max}}=10$ for $(A=9,s_0=1)$.  The percent error is less than 0.2\% for most of the evolution until the codes reaches the singularity and halts.} \label{fig:ADMMass_A_9_s0_1_nmax10vs5}
\end{figure}

For the strong wave case $(A=9,s_0=1)$:
\begin{itemize}
\item we see singularity formation in the same amount of time $(k=38)$ if we move the outer boundary out significantly (i.e. to $\eta_{\mathbf{max}}=10$ or $20$)
\item apparent horizons form in the same location as evidenced by figures \ref{fig:trapsurft0a9} and \ref{fig:trapsurft0a9nmax10} which were run with $\eta_{\mathbf{max}}=5$ and $10$ respectively
\item the (quasi-local) ADM mass, which is measured at the outer radial edge of the grid is essentially unchanged.  See figure \ref{fig:ADMMass_A_9_s0_1_nmax10vs5} for a comparison of the masses calculated over time for two different outer boundary locations.  The percent error is less than $0.2\%$ until the code encounters a singularity, which indicates a very slow rate of convergence given that our outer radial boundary is $150$ times further out.
\end{itemize}

\begin{figure} \centering
\includegraphics{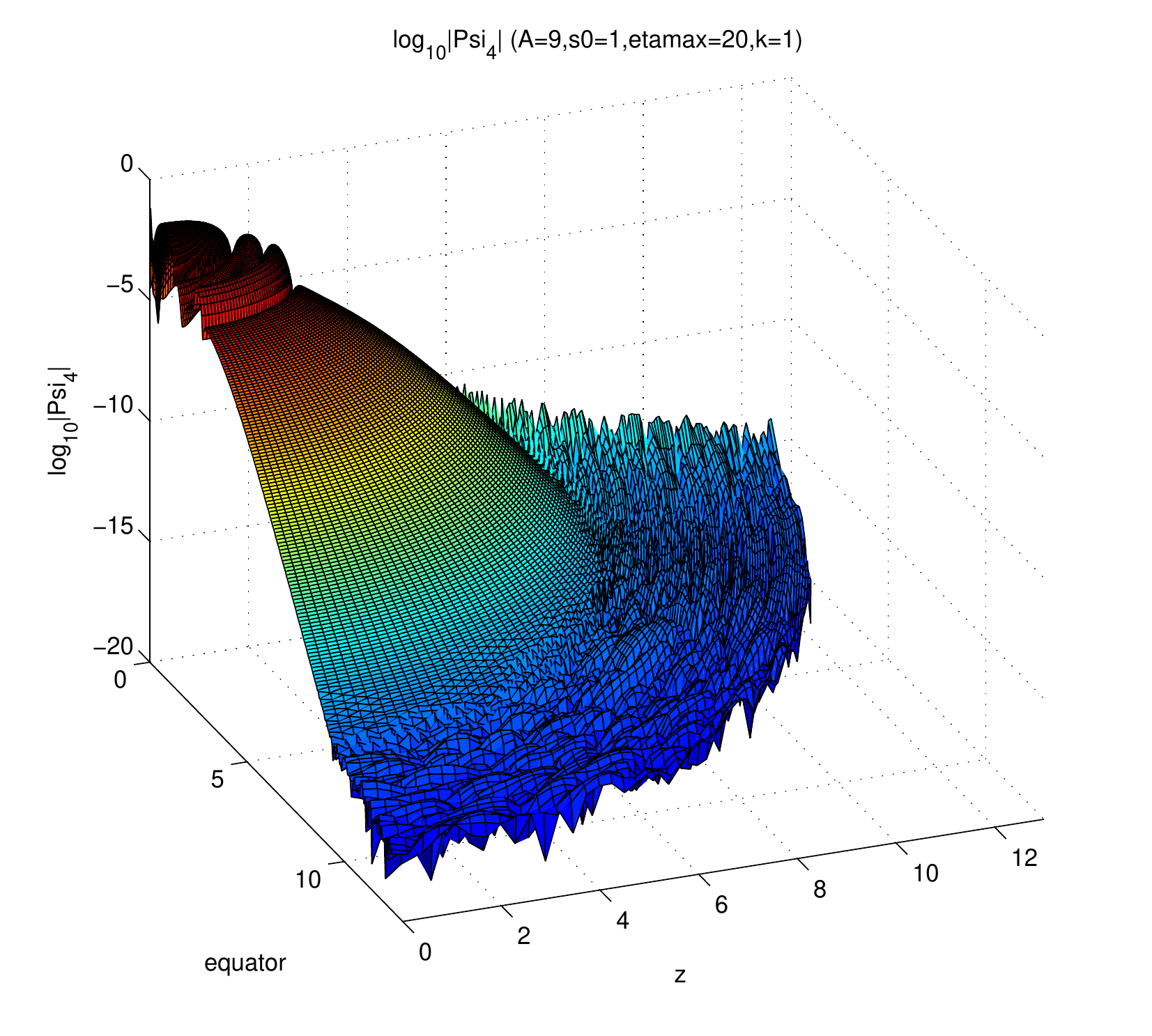}
\caption[Numerical noise far out for $\Psi_4$]{$\log_{10}|\Psi_4|$ for the run $(A=9,s_0=1,\eta_{\mathbf{max}}=20)$ on the first time step $t=\Delta t$.  Further out radially than $\eta\sim 10$ there is only numerical noise present in the values of $\Psi_4$.} \label{fig:log10_abs_psi4_etamax20_a9_s01_k1}
\end{figure}

We also note that moving the boundary out too far causes the calculation of quantities such as the Weyl scalars to encounter numerical limitations near the edge of the grid.  We see, for example, that when running with $\eta_{\mathbf{max}}=20$ in table \ref{tbl:videoresults} that once we pass $\eta\sim 7-10$ we are only seeing numerical error/noise in derived values.  Figure \ref{fig:log10_abs_psi4_etamax20_a9_s01_k1} demonstrates how the smooth falloff in $\Psi_4$ ends as we move out radially and there is only numerical noise present as we go further out on the grid.

Given (i) the large amount of noise present across the majority of the grid in this run due to the greater outer boundary ($r \sim 2 \times 10^8$), (ii) the similar results and (iii) the inability of noise to swamp the code, we are further confident of the code stability and results presented herein.

\subsection{Testing Time Reversal of Evolution}\label{subsec:timereverse}
If we instead alter our time variable to run backwards in time via the transformation
$$\Delta t \rightarrow -\Delta t$$
we can investigate whether the moment of time symmetry we start at is a minimum, maximum or an inflection point in the evolution of the dynamic variables.

In table \ref{tbl:videoresults} I have included a few links to video series for the time reversed evolution ($A\in \{6.6,9,-1\times 10^{-10}\}$).  The time-reversed evolutions demonstrate:
\begin{itemize}
\item the same overall behaviour as their forward in time equivalents discussed in section \ref{sec:results}
\item they encounter singularities in the spacetime in the same number of time steps
\item the same quasi-local ADM masses\footnote{Plus or minus for the perturbative wave case which can be seen as the error in trying to hit near-flat 3-space in a spacetime topology that goes from singularity $\rightarrow$ almost flat space $\rightarrow$ back to a singularity.}
\item negative extrinsic curvatures and shift vectors (as compared to their forward-in-time equivalents)
\item ``white'' hole explosion with ``black'' hole collapse global topology when considering the forward and backwards evolutions together.
\end{itemize}

This matches the global structure discussed in \'O Murchadha \cite{Murchadha_trap}, and leads to the motivation for a possible switch to non-time symmetric initial data for future work.

\section{Alternate Evolution Schemes}

\subsection{Covariant vs. Mixed Extrinsic Curvature Evolutions}
As detailed in section \ref{sec:coveqns}, we can alternatively use the covariant form of the extrinsic curvature tensor in equation (\ref{eqn:covkijtensor}) compared to the mixed form of the extrinsic curvature presented in equation (\ref{eqn:mixkijtensor}).

This provides a somewhat independent check of the numerical algorithms used in this thesis, as it is fairly easy to code the covariant equations beside the mixed equations while using the common code infrastructure\footnote{i.e. subroutines designed to calculate derivatives, interpolations, extrapolations, elliptic BiCG solver, trapped surface finder, etc.}.  One simply sets a switch as to which version of the equations we wish to use.  In theory there should be no difference between the two evolutions in terms of the curvature variables, however there are some numerical differences which can mainly be seen in (\ref{eqn:Hcovtransform}), i.e. there are some $\frac{1}{f^2}$-type factors that appear in or disappear from the extrinsic curvature variable evolution equations.

This translates into different boundary conditions and different behaviour especially at the outer boundaries, where what may previously have been $\frac{1}{f^2}$-type behaviour in the evolution of an extrinsic curvature variable can now become $O(1)$.

The covariant extrinsic curvature evolution equations are completely different as can be seen in equations (\ref{eqn:covkevolvacuum}), and they have their own numerical peculiarities and have different numerical conditioning.

What we see, however, is that despite all these differences that the numerical results in the main are the same.  The codes with the same initial parameters develop large Weyl curvatures, scalar curvatures, quasi-local ADM mass growth, trapped surface bunching, and reach singularities in the same time frame\footnote{The covariant codes encounter singularities about 10 steps sooner, mostly due to poorly implemented outer boundary conditions.  More work would need to be done on the exact outer boundary conditions to give a robust comparison.}.  See table \ref{tbl:covariantresults} for some results.

This gives further confidence, along with the error analysis performed in section \ref{sec:erroranalysis}, that the code is finding real (physical) results.

\begin{table}\begin{center}
\begin{tabular}{ccc}\hline \noalign{\vskip 2mm} 
$K^i_{\;j}$ or $K_{ij}$? & parameters & \# time steps to reach singularity \\ \noalign{\vskip 2mm} \hline
Covariant & $A=1$,$s_0=1$ & $98$ \\
Mixed & & $107$ \\ \hline
Covariant & $A=-1$,$s_0=1$ & $143$ \\
Mixed & & $149$ \\ \hline
Covariant${}^*$ & $A=-1\times 10^{-5}$, $x_0=4$ & $100$ \\
Mixed & & $127$ \\ \hline
\end{tabular}\caption[Covariant vs. Mixed evolutions for various IVPs]{Covariant vs. Mixed evolutions for various IVPs. ${}^*$-this particular run was an erroneous IVP formulation as $q$ is not symmetric across $\eta=0$} \label{tbl:covariantresults}
\end{center} \end{table}

\subsection{Calculating $\phi$ via an Evolution Equation}\label{subsec:evolphi}
As mentioned previously we have the option of using the metric evolution equation (\ref{eqn:gam33dot}) to evolve $\phi$.  Let us examine the behaviour of this type of evolution and how it relates to using the Hamiltonian constraint, keeping in mind the error analysis performed in section \ref{sec:phidoterror}.

If $(A=-1 \times 10^{-10},s_0=3)$ and equation (\ref{eqn:gam33dot}) is used in the iterative Crank-Nicholson algorithm what do we observe?  Table \ref{tbl:videoresults} has a link to the video results for this scenario, and overall some of the results are similar to the results obtained using the Hamiltonian constraint.  

\begin{figure} \centering
\includegraphics{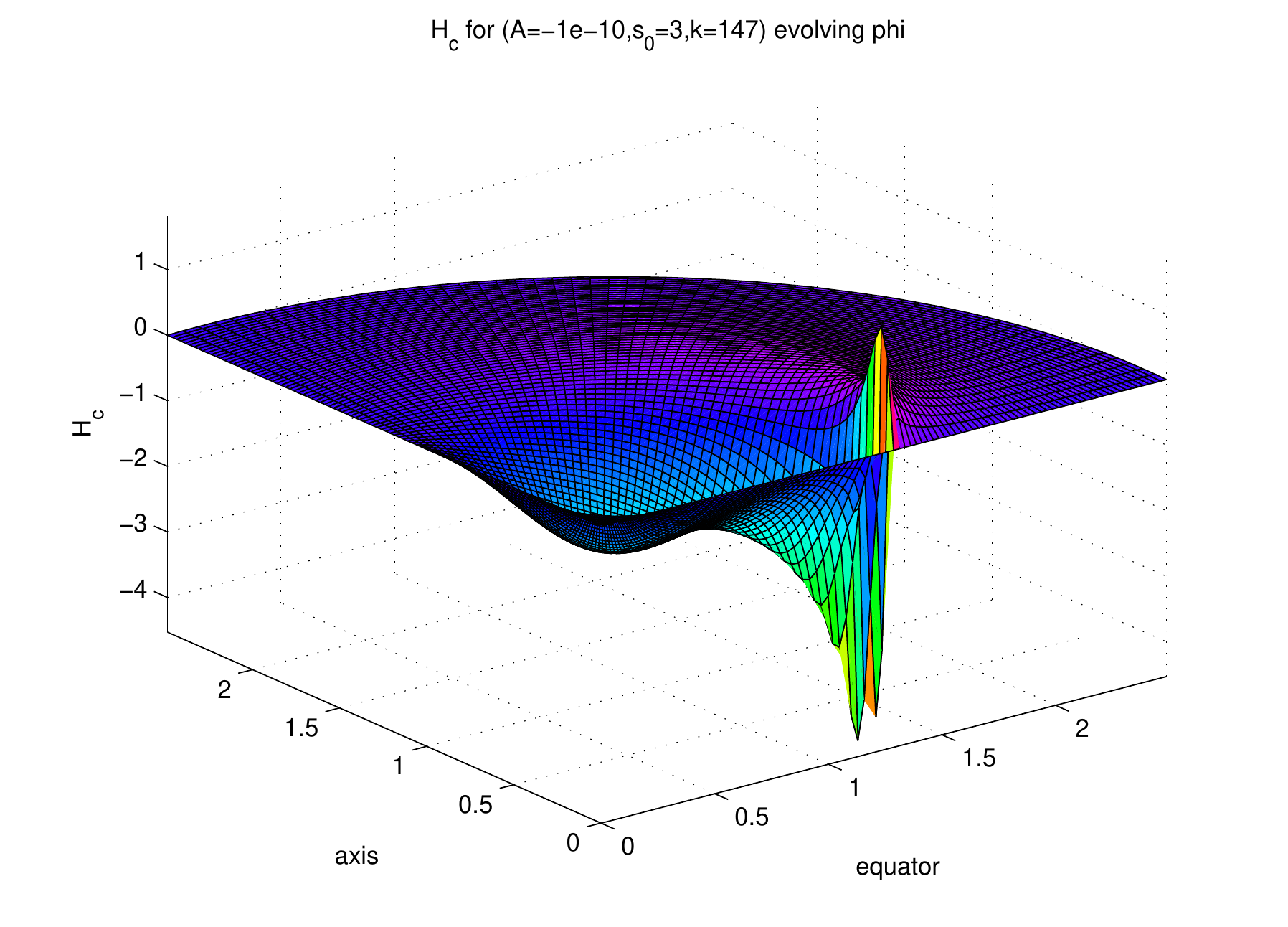}
\caption[$H_c$ singularity formation when evolving $\phi$]{Localised singularity formation in $H_c$ as discussed in section \ref{subsec:evolphi} when evolving $\phi$ instead of using the Hamiltonian constraint equation to calculate a solution. $k=147$ in this figure and the code encounters a singularity on the next time step.}\label{fig:Hc_A-1e-10_s0_3_t147_evolve_phi}
\end{figure}

The evolution method is capable of running for a little bit longer before reaching the singularity ($147$ time steps vs. $127$), since there is no elliptic constraint equation for $\phi$ to solve with all of the smoothness demands that an elliptic solver has.  However the extra time steps are clearly just evolving a pathological spacetime which has localised irregularities in all of the variables on the equator around $(\eta \sim 1 \rightarrow 1.5, \theta=\frac{\pi}{2})$.  As an example of the steep gradients that develop see figure \ref{fig:Hc_A-1e-10_s0_3_t147_evolve_phi} which shows of one of the extrinsic curvature variables ($H_c$) on the time step before the code encounters a singularity.

It is important to note that \emph{all} of the variables (including metric, extrinsic curvature and gauge variables) are ill-behaved in this region, as are all of the curvature measures, and this is the same type of localised singular behaviour as we observe in the constrained case.  When we have a numerical conditioning problem we tend to see numerical error creeping in across a large region of the grid on the order of grid spacing (see for example figures \ref{fig:qy_a-1e-10_s03_explicit_t_evol_t7} or \ref{fig:alphayy_OB_norearrange}), however here we see localised large gradients (with a smooth lead-up to the gradient) and singularities forming in both the constrained and non-constrained cases, which is most likely indicative of black hole horizon formation.

If the Hamiltonian constraint is not enforced then the errors in the constraint become large across the grid.  In addition the quasi-local ADM measures are $10$ times what they are in the constrained case.

\subsection{Employing Momentum Constraints in an Evolution Scheme}
\begin{figure} \centering
\includegraphics{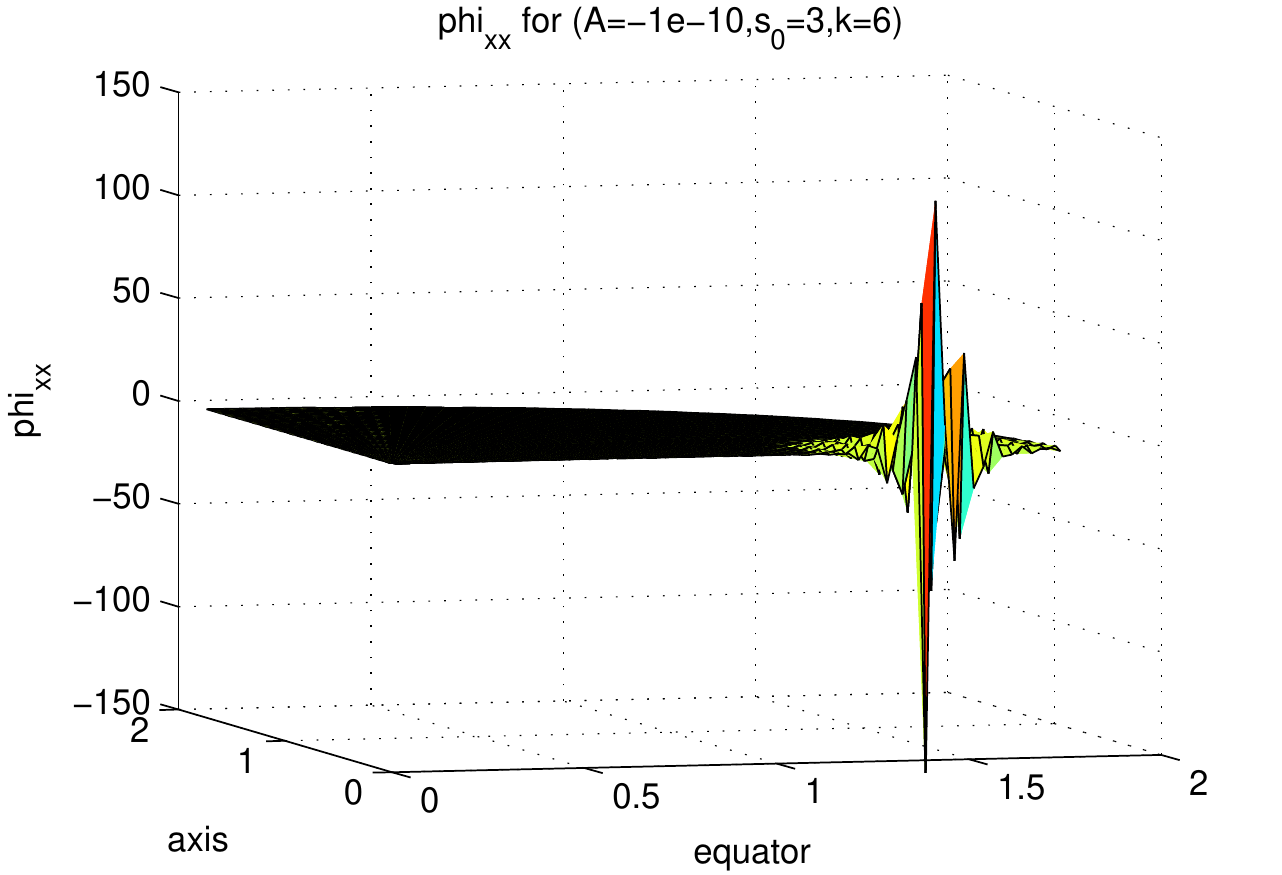}
\caption[$\phi_{\eta\eta}$ using first momentum constraint]{Graph of $\phi_{\eta\eta}$ in the interior region of our grid if we attempt to use equation (\ref{eqn:momcon1_solveha}) to solve for $H_a$ instead of its evolution equation.  The solution is dominated by numerical noise in very few time steps for a small initial wave. $(A=-1 \times 10^{-10},s_0=3,k=6)$ }\label{fig:phixx_a-1e-10_s03_momcon1_ha_t6}
\end{figure}

A naive implementation of an evolution scheme which employs the radial momentum constraint was attempted. Derivative terms for $H_b,H_c$, and $H_d$ are present in equations (\ref{eqn:momcons}), therefore a solution for $H_a$ is easily obtained and it will be used in place of the evolution equation for $H_a$.  This leads us to the algebraic constraint:
\begin{eqnarray}\label{eqn:momcon1_solveha}
H_a & =&\frac{-1 }{\left(4 \phi_\eta + \frac{q_\eta}{2} + \frac{2 f_\eta}{f}\right)}\left[ \frac{f_{\eta}^2\,H_c\,\cot\theta }{f^2}
+\frac{f_{\eta}^2\,H_c\, q_{\theta} }{f^2}
-\frac{H_b\, q_{\eta} }{2}
+\frac{6\,f_{\eta}^2\,H_c\, {\phi}_{\theta} }{f^2}\nonumber \right.\\ \mbox{} & & \left.
-2\,H_d\, {\phi}_{\eta} 
-2\,H_b\, {\phi}_{\eta} 
-\frac{ f_{\eta} \,H_d}{f}
+\frac{f_{\eta}^2\, {H_c}_{,\theta} }{f^2}
-\frac{ f_{\eta} \,H_b}{f}
-{H_d}_{,\eta}-{H_b}_{,\eta} \right]
\end{eqnarray}
If we keep the rest of the code constant (which is well-behaved) and implement this condition, we find that the code crashes very quickly due to amplification of numerical error\footnote{Widespread pathological peaks/valleys that are on the order of grid spacing, as seen before.} near the wave peaks.  For an example of the effect on $\phi_{\eta\eta}$ after $6$ time steps for $(A=-1 \times 10^{-10},s_0=3)$ see figure \ref{fig:phixx_a-1e-10_s03_momcon1_ha_t6}.

\section{Chapter Summary}
In this chapter a variety of code tests and alternate evolution schemes have been presented, which all provide support for the following statements:
\begin{enumerate}
\item The code performs numerically as expected, with errors that are well-behaved (except \emph{perhaps} in some regions for the errors in the momentum constraints).  The large number of tests that all return the same results provides compelling verification that the numerical methods used are sound.
\item Appropriate boundary conditions, boundary locations and grid resolutions (in spatial and temporal coordinates) are being employed in the code.
\item The choices that were made regarding constrained vs. free evolution lead to a numerically well-behaved code, and we can verify the same overall physical spacetime behaviour and singularity formation with alternate evolution schemes.
\end{enumerate}
We now proceed to a discussion of the conclusions derived from the work presented herein.

\fancyhead[RO,LE]{\thepage}
\fancyfoot{} 
\chapter{Conclusions}\label{chap:concl}
\bigskip
In this thesis a numerical evolution code for studying the Brill Gravitational Wave problem in spherical polar axi-symmetry, using ADM 3+1 spacetime slicing has been presented. It has been demonstrated (numerically and analytically) that all Brill wave IVP formulations\footnote{Brill waves in the strictest sense: time symmetric in the metric, vacuum spacetimes that obey Brill's conditions.} eventually lead to the formation of spacetime singularities.

A number of regularisation problems have been identified and resolved in the process of achieving a stable numerical evolution, including the following major paradigm shifts:
\begin{enumerate}
\item The usual ADM framework has been reworked to employ exponential variables for the metric quantities, i.e $a=e^{q}$ and $\psi=e^{\phi}$.  A simple change like this requires recoding and debugging vast amounts of code, but yields great improvements in numerical regularity.
\item All numerical methods were reworked to be fourth order correct.  While this does involve approximately twice the computational cycles\footnote{Going from 5 to 9 stencil points on our stenciled elliptic equations, which is what consumes the vast majority of program time.}, it is necessary for proper propagation of the evolved quantities.
\item Outer boundary conditions that are more appropriate for Brill wave spacetimes have been derived and employed.  The numerical implementation may need more work for dynamic or non-integer falloff powers, but the current implementation yields vastly superior results to any other method tried to date and can be used for insight into theoretical outer boundary work.
\item An additional regularity requirement has been imposed on the metric variable $q$ to ensure that that it goes as $r^2$ near the origin.
\item New regularity conditions on the lapse $\alpha$ have been derived and employed in order that the Brill wave evolution problem is regular in this coordinate system/gauge.  See section \ref{subsec:alphareg} for more details; this is a major result as it explains why previous attempts to solve this problem have failed - the evolution equations are ill-conditioned and guaranteed to blow up in $\sim 3$ time steps unless some rigid constraints are placed on $\alpha$.  This also eliminates the ``axis instability'' problem referenced in many other places\footnote{For example \cite{bernstein} when performing black hole/Brill wave perturbations.}.
\item The regularity conditions on the lapse have been extended to the general axisymmetric case as described in section \ref{subsec:genlapsecon}.  This explains the presence of the axis instability in various axisymmetric spherical polar coordinate evolutions.  The presence of black holes or non-zero $T_{\mu\nu}$ probably masked the regularity problems at the origin in previous evolution codes.
\item An understanding of approximately what amount of constrained versus free evolution is appropriate has been acquired, which is important in any over-determined system.
\item Various other regularisation techniques have been discussed in Chapters \ref{chap:changes} and \ref{chap:numcode} and employed in the code.
\item The numerical methods that have been employed for summation of operands in calculating curvature terms and other variables during the evolution have been reworked for greater robustness and consideration of finite precision computations.  See section \ref{subsec:addterms} for details.  While this may seem like a small nuance it is a shift from previous ``add and go'' models and has a large numerical impact.  This will also be imperative for any other 2+1 or higher order system with evolution equations that contain dozens (or hundreds or thousands\footnote{See for example section \ref{subsubsec:bicgimprove}}) of terms.
\item The matrix equations that arise in solving this problem have been classified as ``ill-conditioned'' $(K_1,K_{\infty} > 10^4)$ and require careful attention to numerical error/noise.  It has further been demonstrated that smaller grid spacing/more grid points is not necessarily better.  See section \ref{sec:condition} for more details.
\item Outer boundary falloff rates have been numerically determined for various dynamic variables.  See section \ref{subsec:DynamicOB}.
\end{enumerate}

There is also some doubt in the back of the author's mind, after the difficulties encountered in section \ref{sec:constrainconform}, as to whether evolving the conformal factor is entirely advisable - it may be necessary to revisit this assumption at some point.

The evolution is stable for ``long'' ($t>125,k>10000$) evolutions depending on the slicing/boundary conditions chosen, and shows excellent internal consistency ($<1$ part in $10^8$) for most measures, with the momentum constraints the most notable exception.  This violation of the momentum constraints is observed in many other numerical implementations in GR.  Sorkin \cite{sorkin,sorkin:code}, for example, is only capable of damping the constraint violations \emph{over time} with large amounts of numerical dissipation.  Choptuik \emph{et al} \cite{choptuik1,choptuik2} only measure the convergence with respect to grid size, and have somewhat inconsistent results.  Most other authors do not include an in-depth error analysis as we have presented here\footnote{Given the over-determined nature of Einstein's equations, there are some natural internal consistency checks that should be implemented in any numerical work.}.  The magnitude of our momentum constraints is in line with or better than other implementations.  Given the self-consistency of all the rest of the results and agreement with theory the results are compelling.

All amplitudes and strengths of pure, vacuum, Brill gravitational waves collapse to form black holes.  We do not see any evidence of collapse/re-expansion at the origin with gravitational waves dispersing to $\infty$, or critical phenomenon where black hole formation turns on at a specific critical IVP phase space value; all Brill gravitational waves head towards one attractor given enough time: singularity (black hole) formation.  This is in contrast to the more commonly studied case of a scalar wave superimposed on a Brill-like wave, but it is fundamentally different as there is no energy-momentum tensor in this thesis.  It also forces any numerical irregularities to be dealt with instead of being masked by the much larger numerical size of the scalar wave or a superimposed black hole.

This also indicates that \emph{time-symmetric, non-rotating, vacuum gravitational waves need matter/non-gravitational fields added to the system or rotation present to prevent collapse}, and our results agree with the theoretical predictions of Raychaudhuri's equations presented herein. This result is proven analytically in section \ref{sec:brill_collapse_always}.

A further implication of this result is that Minkowski space is unstable to the sorts of perturbations described by Brill waves (and a more general class of gravitational waves), and a general theorem with a proof is presented.

One sees a prolate ellipsoidal horizon in the case of large negative $q$ evolutions, and pancake/hemoglobin-shaped equatorially dominated solution in the case of large positive $q$ evolutions.

Over $1000$ videos of the evolution of metric, curvature and horizon variables to help in understanding the structure and breakdown of the evolution have been produced and are available online.  See table \ref{tbl:videoresults} for links to the videos sorted by IVP, giving a fairly robust exploration of the amplitude parameter phase space and visualisations to aid the reader/author in comprehension.

Examination of the initial value parameter phase space in the amplitude $A$ indicates that it has a region with and without an initial apparent horizon present, and that there is an upper (positive) and lower (negative) bound on $A$ for which the IVP has a solution.  These results are consistent with \'O Murchadha's theoretical analysis discussed in section \ref{subsec:qampbounds}, and can be seen in table \ref{tbl:videoresults} and figure \ref{fig:IVPphaseqa}.  An analysis of alternate $q$ functions showing the same IVP phenomenology is presented in section \ref{sec:altformq}, demonstrating the universal phase space features predicted by theory.

Exploration of the time-symmetric evolution demonstrates that the initial IVP state had to arrive from a white hole explosion, after which a black hole was formed (re-collapse) as discussed in section \ref{subsec:timereverse}.

\section{Future Extensions}
Based on the conclusions above and various questions that have arisen during the course of this thesis, presented below are some areas where future work could be directed:

- Implementation of a new lapse function $\alpha$ to incorporate (i) the singularity avoidance aspects of maximal slicing with (ii) the regularity conditions derived herein, which would allow for longer evolutions and more complete spacetime exploration while still maintaining regularity.

- To speed up the convergence of the numerical elliptic equation solver one could (i) develop a new parallelisable algorithm for solving the elliptic PDE's (an ongoing area of research in numerical mathematics) or (ii) use the BiCGStab algorithm as an elliptic preconditioner (requires serial processing) to follow up with a relaxation method (which can make better use of parallel processing).

- To allow for event horizon detection within the code, null ray tracing (or one of the methods mentioned by Thornburg \cite{Thornburg:AH}) could be added to the code.  This could provide the added benefit of precisely locating terminating geodesics and singularities.

- The precision of the apparent horizon detector could be improved to achieve more precise results on horizon locations, areas, etc.  It is unclear if increased grid resolution (which is offset by worse matrix conditioning), better interpolation methods or some other numerical method would be required (like solving the trapped surface equation a second time and searching for only a complete horizon as in other codes, which has its own set of problems to overcome).

- The phase space of IVP values for $A$ and $s_0$ could be mapped out more completely; the results presented here cover a large range of amplitudes but there are other areas of phase space that could be explored.  The expectation is that the results are in line with what is provided in this thesis and preliminary investigation confirms this, however numerical values of various spacetime measures can be mapped out more thoroughly.

- A more in-depth investigation of alternate $q$ function shape evolutions would provide a comparative analysis and additional verification for the observed numerical behaviour, however preliminary results indicate that the overall results are the same as those obtained when using a Gaussian wave profile for $q$.  Other interesting behaviour may be encountered during the evolution, as the trapped surface topology has some differences for alternate functional forms of $q$.

- Due to the nature of the Brill wave problem (and mostly the time-symmetry of the metric on the initial slice) there is no region of IVP phase space which leads to non-singular behaviour. Non-time symmetric initial data has the potential to possess critical regions of phase space, however, and on either side of the critical boundary one could have singular and non-singular behaviour.  The exploration of a non-time symmetric problem is within the realm of possibility for this code, however there would be many theoretical and numerical problems to overcome.

- With the outer boundary conditions as implemented it is possible to perform a more detailed investigation of the outer boundary falloff rate results presented in section \ref{subsec:DynamicOB}.  This could yield insight into the asymptotic structure of gravitational radiation, an open question in GR (provided that difficulties around non-trivial dynamics in the outer grid regions during the Brill wave evolution are accounted for - it may be that we require a $128$ bit compiler to extend the outer boundary out further in a meaningful way).

- Based on the discussion of what $q$ means in section \ref{sec:meaningofq}, an analysis of why there is a maximum amplitude for $q$ and thusly a maximum curvature shear possible between the $(\eta,\theta)$ plane and $\varphi$, despite an infinite black hole extent, should provide interesting insights into the structure of Einstein's field equations.

\appendix
\fancyhead[RO,LE]{\thepage}
\fancyfoot{} 
\chapter{Appendix: Maxima code to generate field equations}\label{appendix:maxima}
\bigskip

The following programs are batch scripts that can be run in Maxima\footnote{Maxima is the open source version the computer algebra package Macsyma (circa 1982), which was originally developed at MIT in the 60s.  Macsyma was then later commercialised with limited success and faces an unsure future - hence the split of the code base.} to symbolically evaluate the ADM 3+1 equations.  Batch the code in section \ref{sec:admk2} first to define the symbolic functions, then batch the code in section \ref{sec:2dexp} to generate the equations for Brill waves in spherical polar coordinates.

At the time of writing Maxima is still freely available in pre-compiled format for many Linux platforms, including Fedora, and represents a powerful (and free) symbolic calculation platform.

\section{admk2.mac}\label{sec:admk2}
This program sets up all the necessary routines to calculate the ADM 3+1 equations.

\begin{verbatim}
/* Dec 1989  adm  D.W. Hobill----NCSA
2000-2013 ADM fix, trapped surface eqns,
NewmanPenrose - A.M. Masterson U Calgary */ 
/* turn off echo during batch loading */
/* eval_when(batch,ttyoff:true)$ */
eval_when(translate,
          define_variable:mode,
	  transcompile:true)$
   
/* set switches */
derivabbrev:true;
ratfac:true$

/* a quick calculation of the contravariant metric   */
gcont():=ug:block([detout:true],ratsimp(lg^^(-1)))$

/*  calculate the mixed components of the extrinsic curvature  */
/* NOTE: first index is raised, second index is lowered */
/* declare(kmix,special)$*/
kmix(dis):=block([],
  for i from 1 thru dim do  
    (for j from 1 thru dim do
      mk[i,j]:
        sum(ug[i,k]*lk[k,j],k,1,dim)),
if dis = true then for i from 1 thru dim do
  (for j:1 thru dim do
    ldisplay(mk[i,j])),
done)$

/* calculate the contravariant extrinsic curvature terms */
/*declare(uk,special)$*/
kcont(dis):=block([],
  for i from 1 thru dim do  
    (for j from 1 thru dim do
      uk[i,j]:
        sum(ug[k,j]*mk[i,k],k,1,dim)),
if dis = true then for i from 1 thru dim do
  (for j:1 thru dim do
    ldisplay(uk[i,j])),
done)$

/* calculate the lower shift vector terms */
/*declare(lv,special)$*/
lowerv(dis):=block([],
  for i from 1 thru dim do
    lv[i]:sum(uv[k]*lg[k,i],k,1,dim),
if dis = true then for i from 1 thru dim do
  ldisplay(lv[i]),
done)$

/*   routine for computing the christoffel symbols */
/* NOTE: first index in mixed christoffel symbols mcs is a raised index */
/* the rest are lowered */
/*declare([lcs,mcs],special)$*/
christof(dis):=block([],
  for b from 1 thru dim do
    (for c from 1 thru dim do
      (for d from 1 thru dim do
        lcs[d,b,c]
          :(diff(lg[d,c],omega[b])
           +diff(lg[d,b],omega[c])
           -diff(lg[b,c],omega[d]))/2)),
  for b from 1 thru dim do
    (for c from 1 thru dim do
      (for a from 1 thru dim do
        (mcs[a,b,c]:expand(ratsimp(
          sum(ug[a,d]*lcs[d,b,c],d,1,dim)))))),
  if dis = mcs then for i thru dim do
    (for j:1 thru dim do
      (for k thru dim do
        ldisplay(mcs[i,j,k]))),
  if dis = lcs then for i thru dim do
    (for j:i thru dim do
      (for k thru dim do
        ldisplay(lcs[i,j,k]))),
done)$

/* covariant components of the ricci tensor */
/*declare(lr,special)$*/
lricci(dis):=block([],
  for i from 1 thru dim do
    (for j from 1 thru dim do
      lr[i,j]:expand(
        sum(diff(mcs[k,i,j],omega[k])
          -diff(mcs[k,i,k],omega[j]),k,1,dim)+
           sum(sum((mcs[l,i,j]*mcs[k,l,k]
          -mcs[l,i,k]*mcs[k,l,j]),l,1,dim),k,1,dim))),
   if dis = true then for i from 1 thru dim do
     (for j:1 thru dim do
       ldisplay(lr[i,j])),
done)$

/* mixed ricci tensor component calculation */
/*declare(mr,special)$*/
mricci(dis):=block([],
  for i from 1 thru dim do
    (for j from 1 thru dim do
      mr[i,j]:expand(sum(ug[i,l]*lr[l,j],l,1,dim))),
  if dis = true then for i from 1 thru dim do
    (for j:1 thru dim do
       ldisplay(mr[i,j])),
done)$

/* computes scalar curvature */
scurvature():=tracer:expand(sum(sum(lr[i,j]*ug[i,j],i,1,dim),j,1,dim))$

/*routine to determine trace k   */
/*declare(trk,special)$*/
tracek():=trk:sum(mk[j,j],j,1,dim)$
/* trk = 0 for maximal slicing */
/*tracek():=trk:0$*/

/* routine for calculating the metric evolution */
/*declare(lgdot,special)$*/
dgdt(dis):=block([],
  for i from 1 thru dim do
    (for j from 1 thru dim do
      lgdot[i,j]:expand(-2*n*lk[i,j]
  +diff(lv[j],omega[i])-sum(mcs[l,j,i]*lv[l],l,1,dim)
  +diff(lv[i],omega[j])-sum(mcs[l,i,j]*lv[l],l,1,dim))),
if dis = true then for i from 1 thru dim do
  (for j:1 thru dim do
    ldisplay(diff(lg[i,j],t)=expand(lgdot[i,j]))),
done)$

/* routine for calculating the extrinsic curvature evolution */
/*declare(lkdot,special)$*/
dkdt(dis):=block([],
  for i from 1 thru dim do
    (for j from 1 thru dim do
      lkdot[i,j]:
        n*(lr[i,j]+trk*lk[i,j]
          -2*sum(lk[i,l]*mk[l,j],l,1,dim))
        -diff(n,omega[i],1,omega[j],1)
        +sum(mcs[k,j,i]*diff(n,omega[k]),k,1,dim)
        +sum(uv[l]*(diff(lk[i,j],omega[l])
	 -sum(mcs[k,l,i]*lk[k,j]
           +mcs[k,l,j]*lk[i,k],k,1,dim))
	 +lk[i,l]*(diff(uv[l],omega[j])
           +sum(mcs[l,j,k]*uv[k],k,1,dim))
	 +lk[l,j]*(diff(uv[l],omega[i])
           +sum(mcs[l,i,k]*uv[k],k,1,dim))
	,l,1,dim)),
if dis = true then for i from 1 thru dim do         
  (for j:1 thru dim do  
    ldisplay(diff(lk[i,j],t)=expand(lkdot[i,j]))),
done)$

/* routine for calculating the mixed extrinsic curvature evolution */
/*declare(mkdot,special)$*/
dmkdt(dis):=block([],
  for i from 1 thru dim do
    (for j from 1 thru dim do
      mkdot[i,j]:
        sum(-ug[i,d]*( diff(diff(n,omega[j]),omega[d])
          -sum(mcs[e,j,d]*diff(n,omega[e]),e,1,dim)),d,1,dim)
        +n*(mr[i,j]+trk*mk[i,j])
/* method 1 */
        +sum(ug[i,d]*
          (sum(uv[l]*(diff(lk[d,j],omega[l])
            -sum(mcs[k,l,d]*lk[k,j]
              +mcs[k,j,l]*lk[d,k],k,1,dim))
           +lk[d,l]*(diff(uv[l],omega[j])
            +sum(mcs[l,k,j]*uv[k],k,1,dim))
	   +lk[l,j]*(diff(uv[l],omega[d])
            +sum(mcs[l,k,d]*uv[k],k,1,dim))
	 ,l,1,dim)),d,1,dim)
         +sum(lk[d,j]*
          sum(
           -ug[l,d]*(diff(uv[i],omega[l])
             +sum(mcs[i,k,l]*uv[k],k,1,dim))
           -ug[i,l]*(diff(uv[d],omega[l])
             +sum(mcs[d,k,l]*uv[k],k,1,dim))
           +uv[l]*(diff(ug[i,d],omega[l])
	    +sum(mcs[i,k,l]*ug[k,d]
              +mcs[d,k,l]*ug[i,k],k,1,dim))
          ,l,1,dim),d,1,dim)
       ),
/* method 2 */
/*+sum(uv[c]*(diff(mk[i,j],omega[c])
   +sum(mcs[i,d,c]*mk[d,j],d,1,dim)
   -sum(mcs[d,j,c]*mk[i,d],d,1,dim)),c,1,dim)
 +sum(mk[i,c]*(diff(uv[c],omega[j])
     +sum(mcs[c,e,j]*uv[e],e,1,dim)),c,1,dim)
 -sum(mk[c,j]*(diff(uv[i],omega[c])
     +sum(mcs[i,e,c]*uv[e],e,1,dim)),c,1,dim) ),*/

if dis = true then for i from 1 thru dim do         
  (for j:1 thru dim do  
    ldisplay(diff(mk[i,j],t)=expand(mkdot[i,j]))),
done)$

/* calculate the Hamiltonian constraint  */
/*declare (hamcon,special)$*/
hamiltonian():=block([],hamcon:tracer+expand(trk*trk-
		sum(sum(lk[i,j]*uk[i,j],j,1,dim),i,1,dim)),
                hamconsimp:expand(-hamcon*lg[2,2]/8),
                ldisplay(hamcon),ldisplay(hamconsimp),
done)$

/* calculate the contravariant components of the momentum constraint  */
/*declare(momcon,special)$*/
momenta(dis):=block([],
  for i from 1 thru dim do
    momcon[i]:expand(sum(diff(uk[i,j],omega[j]),j,1,dim)
     +sum(sum(  mcs[i,l,j]*uk[l,j]+mcs[j,l,j]*uk[i,l]
      ,j,1,dim),l,1,dim)
     -sum(  ug[i,j]*(diff(trk,omega[j]))  ,j,1,dim)
/* should be zero as K is a scalar */
/*     -sum(sum(sum(  ug[i,j]*mcs[l,m,j]*mk[m,l]
  ,j,1,dim),l,1,dim),m,1,dim)
     +sum(sum(sum(  ug[i,j]*mcs[m,l,j]*mk[l,m]
  ,j,1,dim),l,1,dim),m,1,dim)*/
    ),    	
    for i from 1 thru dim do
      momconsimp[i]:expand(momcon[i]*lg[2,2]),
  if dis = true then for i from 1 thru dim do 
    ldisplay(momcon[i]),
  if dis = true then for i from 1 thru dim do 
    ldisplay(momconsimp[i]),
done)$

/* calculate the maximal slicing equation */
/*declare(maximal,special)$*/
maxslice():=block([],
  maximal:expand(sum(sum(ug[a,d]*(diff(diff(n,omega[a]),omega[d])
    -sum(mcs[e,a,d]*diff(n,omega[e]),e,1,dim)),a,1,dim),d,1,dim)),
ldisplay(maximal=n*tracer),done)$

/* calculate nonlinear PDE that determines
radial function "h" (hhor) for apparent horizon solver*/
/* define unit normal vector and defining eqn
(see Bernstein thesis, section Appendix D or */
/* Masterson thesis Chapter 6) */
sahoreqns(dis):=block([],
  saden:(diff(hhor,omega[2])^2*ametric^2/bmetric+ametric)^(1/2)*exp(2*p),
  sahor:[1/saden,-ametric*diff(hhor,omega[2])/(bmetric*saden),0],
  dasahor:expand(expand(
    sum(diff(sahor[a],omega[a]),a,1,dim)
    + sum(sum(mcs[b,a,b]*sahor[a],a,1,dim),b,1,dim)
    + sum(sum(lk[a,b]*sahor[a]*sahor[b],a,1,dim),b,1,dim)
    -trk)*saden^3
  *-f^2/(diff(f,omega[1]))^4/exp(q+4*p)
    ),
  dxdasahor:expand(diff(dasahor,omega[1])),
  ldisplay(saden),
  for i from 1 thru dim do ldisplay(sahor[i]),
    ldisplay(dasahor),
/*        ldisplay(dxdasahor),*/
done)$

/* calculate analytic derivatives of lapse function if appropriate*/
lapsederivs():=block([],
  lapsean[0,0]:tanh(x)^w1*sin(y)^w2,
  lapsean[1,0]:expand(diff(lapsean[0,0],x)),
  lapsean[2,0]:expand(diff(diff(lapsean[0,0],x),x)),
  lapsean[0,1]:expand(diff(lapsean[0,0],y)),
  lapsean[0,2]:expand(diff(diff(lapsean[0,0],y),y)),
  lapsean[1,1]:expand(diff(diff(lapsean[0,0],x),y)),
  for i from 0 thru 2 do
    (for j from 0 thru 2 do ldisplay(lapsean[i,j])),
  done)$
/* xthru combine radcan*/

newmanpenrose():=block([],
  for i from 1 thru dim+1 do(
    kunp[i]:expand((tunp[i]+xunp[i])/sqrt(2)),
    lunp[i]:expand((tunp[i]-xunp[i])/sqrt(2)),
    munp[i]:expand((yunp[i]+%I*zunp[i])/sqrt(2)),
    mbarunp[i]:expand((yunp[i]-%I*zunp[i])/sqrt(2)),
    klnp[i]:expand((tlnp[i]+xlnp[i])/sqrt(2)),
    llnp[i]:expand((tlnp[i]-xlnp[i])/sqrt(2)),
    mlnp[i]:expand((ylnp[i]+%I*zlnp[i])/sqrt(2)),
    mbarlnp[i]:expand((ylnp[i]-%I*zlnp[i])/sqrt(2))
   ),
    for i from 1 thru dim+1 do ldisplay(kunp[i]),
    for i from 1 thru dim+1 do ldisplay(lunp[i]),
    for i from 1 thru dim+1 do ldisplay(munp[i]),
    for i from 1 thru dim+1 do ldisplay(mbarunp[i]),
    for i from 1 thru dim+1 do ldisplay(klnp[i]),
    for i from 1 thru dim+1 do ldisplay(llnp[i]),
    for i from 1 thru dim+1 do ldisplay(mlnp[i]),
    for i from 1 thru dim+1 do ldisplay(mbarlnp[i]),
  done)$

/* END OF SCRIPT */
/* turn echo suppression off  */
eval_when(batch,ttyoff:false)$
\end{verbatim}

\section{2dexp.mac}\label{sec:2dexp}
Routine used to do the calculations for the specific Brill setup described in this thesis.
\begin{verbatim}
dim:3;
omega:[x,y,z];
depends([q,b,d,ha,hb,hc,hd,n,v1,v2,p],[x,y,t],f,x,hhor,y);
lg:matrix([exp(4*p)*'diff(f,x)^2*exp(q),0,0],
  [0,exp(4*p)*f^2*exp(q),0],
  [0,0,exp(4*p)*f^2*sin(y)^2]);
/*depends([a,c],[x,y,t]);
lg:matrix([exp(4*p)*a,0,0],[0,exp(4*p)*b,0],[0,0,exp(4*p)*d*sin(y)^2]);
lk:matrix([exp(4*p)*ha,exp(4*p)*hc,0],
  [exp(4*p)*hc,exp(4*p)*hb,0],
  [0,0,exp(4*p)*hd*sin(y)^2]);*/
/*lg:matrix([exp(4*p)*a,exp(4*p)*c,0],
  [exp(4*p)*c,exp(4*p)*b,0],
  [0,0,exp(4*p)*d*sin(y)^2]);*/
lk:matrix([exp(4*p)*'diff(f,x)^2*ha*exp(q),exp(4*p)*'diff(f,x)^2*exp(q)*hc,0],
  [exp(4*p)*'diff(f,x)^2*exp(q)*hc,exp(4*p)*f^2*hb*exp(q),0],
  [0,0,exp(4*p)*f^2*hd*sin(y)^2]);
uv:[v1,v2,0];
ametric:lg[1,1]/exp(4*p);
bmetric:lg[2,2]/exp(4*p);
dmetric:lg[3,3]/(exp(4*p)*sin(y)^2);
gcont();
kmix(true);
kcont(true);
lowerv(true);
christof(mcs);
lricci(true);
mricci(true);
scurvature();
tracek();
dgdt(true);
dmkdt(true);
dkdt(true);
hamiltonian();
lmomenta(true);
momenta(true);
maxslice();

sahoreqns(true);
lapsederivs();
expand(mkdot[2,1]*f^2/diff(f,x)^2-mkdot[1,2]);
\end{verbatim}

\fancyhead[RO,LE]{\thepage}
\fancyfoot{} 
\chapter{Appendix: Alternate and Additional Equations for Numerical Evolution}\label{appendix:eqns}
\bigskip

\section{Covariant Extrinsic Curvature Version of Equations}\label{sec:coveqns}
If instead of choosing the covariant extrinsic curvature to be of the form given in equation (\ref{eqn:origcovkijtensor}) we use the alternate form:
\begin{eqnarray}\label{eqn:covkijtensor}
K_{ij} & = & \left[\begin{array}{ccc}
\tilde{H}_a & \tilde{H}_c & 0 \\
\tilde{H}_c & \tilde{H}_b  & 0 \\
 0 & 0 & \tilde{H}_d \end{array} \right]
\end{eqnarray}

The metric evolution equations (\ref{eqn:3p1:gammadot}) become
\begin{equation}
\frac{\dot{\gamma}_{11}}{\left(f_{\eta}^2 e^{q+4\phi}\right)} = \dot{q}+4\dot{\phi} = v_2q_{\theta} + 4\phi_{\theta}v_2 +2{v_1}_{\eta} + q_{\eta}v_1 + 4\phi_{\eta}v_1 + 2\frac{f_{\eta\eta}}{f_{\eta}}v_1 - \frac{2\alpha \tilde{H}_a}{e^{q+4\phi}f_\eta^2}
\label{eqn:covgam11dot}\end{equation}

\begin{equation}
\frac{\dot{\gamma}_{22}}{\left(f^2 e^{q+4\phi} \right)} = \dot{q}+4\dot{\phi} = 2{v_2}_{\theta} + v_2q_{\theta} + 4\phi_{\theta}v_2 + q_{\eta}v_1 + 4\phi_{\eta}v_1 + 2\frac{f_{\eta}}{f}v_1 - \frac{2\alpha \tilde{H}_b}{e^{q+4\phi}f^2}
\label{eqn:covgam22dot}\end{equation}

\begin{equation}
\frac{\dot{\gamma}_{12}}{\left(f_{\eta}^2 e^{q+4\phi}\right)} = \frac{\dot{\gamma}_{21}}{f_{\eta}^2} = 0 = \left(\frac{f}{f_\eta}\right)^2{v_2}_{\eta} + {v_1}_{\theta} - \frac{2 \alpha \tilde{H}_c}{e^{q+4\phi}f_\eta^2}
\label{eqn:covgam12dot}\end{equation}

\begin{equation}
\frac{\dot{\gamma}_{33}}{4f^2e^{4\phi}\sin^2\theta} = \dot{\phi} = \frac{v_2}{2} \cot\theta + \phi_{\theta}v_2 + \phi_{\eta}v_1 + \frac{v_1}{2}\frac{f_{\eta}}{f} - \frac{\alpha \tilde{H}_d}{2\,e^{4\phi}\,f^2\,\sin^2\theta}
\label{eqn:covgam33dot}\end{equation}

this leads to
\begin{equation}\label{eqn:qdotcov}
\dot{q} = 2{v_2}_{\theta} + v_2 q_{\theta} + q_{\eta}v_1 + \frac{2\alpha}{e^{4\phi}\,f^2}\left(\frac{\tilde{H}_d}{\sin^2\theta}-\frac{\tilde{H}_b}{e^q}\right) - 2v_2\cot\theta
\end{equation}

These lead to the new shift vector potential equations:
\begin{eqnarray}\label{eqn:deccovshiftvecs}
\left(\frac{f}{f_{\eta}}\right)^2 \chi_{\eta\eta} + \chi_{\theta\theta} + \left(\frac{f}{f_{\eta}}\right)\partial_{\eta}\left(\frac{f}{f_{\eta}}\right)\chi_{\eta} & = & \frac{\alpha}{e^{q+4\phi}}\left(\frac{\tilde{H}_a}{f_\eta^2} - \frac{\tilde{H}_b}{f^2}\right) \nonumber \\ \mbox{}
\left(\frac{f}{f_{\eta}}\right)^3\Phi_{\eta\eta} + \left(\frac{f}{f_{\eta}}\right) \Phi_{\theta\theta} + \left(\frac{f}{f_{\eta}}\right)^2\partial_{\eta}\left(\frac{f}{f_{\eta}}\right)\Phi_{\eta} & = & \frac{2\, \alpha\,\tilde{H}_c}{e^{q+4\phi}f_\eta^2}
\end{eqnarray}

Expanding (\ref{eqn:gencovkevol}) by using the covariant derivatives defined in section \ref{sec:covdiff} we end up with
\begin{eqnarray}\label{eqn:covkevolvacuum} \partial_t K_{ij}&=&-\partial_i\partial_j\alpha+\Gamma^l_{ji}\partial_l\alpha
+\alpha\left[R_{ij}-2K_{il}K^l_j+K_{ij} K\right]
+\beta^l\left[\partial_l K_{ij} - \Gamma^k_{li} K_{kj} 
- \Gamma^k_{lj} K_{ik}\right]
\nonumber \\ & & \mbox{}
+K_{il}\left[\partial_j \beta^l + \Gamma^l_{jk} \beta^k\right]
+K_{lj}\left[\partial_i \beta^l + \Gamma^l_{ik} \beta^k\right]
\end{eqnarray}
using the conventions of this thesis.  Solving these yields four equations for the four extrinsic curvature variables:
\begin{eqnarray}\label{eqn:hacovevol}
\frac{\partial \tilde{H}_a}{\partial t} & = & \left(\frac{f_{\eta}}{f}\right)^2\left[
-\frac{\alpha\, q_{\theta} \,\cot\theta }{2}
-2\,\alpha\, \phi_{\theta} \,\cot\theta
-\frac{\alpha\, q_{\theta\theta} }{2}
-\alpha\, \phi_{\theta} \, q_{\theta} \right. \nonumber \\ \mbox{} & & \left.
-\frac{ {\alpha}_{\theta} \, q_{\theta} }{2}
-2 \alpha\, \phi_{\theta\theta}
-4 \alpha\, \phi_{\theta}^2
-2 \alpha_\theta\, \phi_{\theta}\right]
-\frac{\alpha\, q_{\eta\eta}}{2}
+ \alpha\, \phi_\eta\, q_\eta
+\frac{\alpha_\eta\, q_{\eta}}{2}
\nonumber \\ \mbox{} & &
+\frac{\alpha\, q_{\eta}}{2}\frac{f_{\eta\eta}}{ f_\eta}
-4\alpha\, \phi_{\eta\eta}
+2 \alpha_\eta\, \phi_{\eta}
-4\alpha\, \phi_\eta\left(\frac{f}{f_\eta}\right)\partial_\eta\left(\frac{f}{f_\eta}\right)
-\alpha_{\eta\eta}
+ \alpha_{\eta} \frac{f_{\eta\eta}}{f_\eta}\nonumber \\ \mbox{} & &
+\frac{\alpha}{e^{q+4\phi}}\left[\frac{-2 \tilde{H}_c^2}{f^2} + \frac{\tilde{H}_a\, \tilde{H}_b}{f^2} - \frac{\tilde{H}_a^2}{f_\eta^2} \right]
+ \frac{\tilde{H}_a \tilde{H}_d \alpha}{e^{4\phi} f^2 \sin^2\theta}\nonumber \\ \mbox{} & &
+2 \tilde{H}_c\, {v_2}_{\eta}
+\tilde{H}_{a,\theta}\, v_2
+2 \tilde{H}_a\, {v_1}_{\eta}
+\tilde{H}_{a,\eta}\, v_1
\end{eqnarray}

\begin{eqnarray}\label{eqn:hccovevol}
\frac{\partial \tilde{H}_c}{\partial t} & = &
\frac{\alpha\, q_{\eta} \,\cot\theta }{2 }
+ \frac{\tilde{H}_c \tilde{H}_d \alpha}{e^{4\phi} f^2 \sin^2\theta}
+\alpha\, \phi_{\eta} \, q_{\theta}
+\frac{ {\alpha}_{\eta} \, q_{\theta} }{2}
+\alpha\, \phi_{\theta} \, q_{\eta} \nonumber \\ \mbox{} & &
+\frac{ {\alpha}_{\theta} \, q_{\eta} }{2}
+4\,\alpha\, \phi_{\eta} \, \phi_{\theta}
+2\, {\alpha}_{\eta} \, \phi_{\theta}
-2\,\alpha\, \phi_{\eta\theta} \nonumber \\ \mbox{} & &
+2\, {\alpha}_{\theta} \, \phi_{\eta}
-{\alpha}_{\eta\theta}
+\left(\frac{f_\eta}{f}\right)\left[{\alpha}_{\theta}+2\,\alpha\, \phi_{\theta}
+\frac{\alpha\, q_{\theta}}{2} \right]
\nonumber \\ \mbox{} & &
+\tilde{H}_c\, {v_2}_{\theta} + \tilde{H}_b \,{v_2}_{\eta} + \tilde{H}_{c,\theta}\, v_2 + \tilde{H}_a\, {v_1}_{\theta} + \tilde{H}_c\, {v_1}_{\eta} + \tilde{H}_{c,\eta} \,v_1
\nonumber \\ \mbox{} & & + \frac{\alpha}{e^{q+4\phi}}\left[-\frac{\tilde{H}_b\,\tilde{H}_c}{f^2}-\frac{\tilde{H}_a\,\tilde{H}_c}{f_\eta^2}\right]
\end{eqnarray}

\begin{eqnarray}\label{eqn:hbcovevol}
\frac{\partial \tilde{H}_b}{\partial t} & = & 
\frac{\alpha\, q_{\theta} \cot\theta }{2}
-2\,\alpha\, \phi_{\theta} \cot\theta
-\frac{\alpha\, q_{\theta\theta} }{2}
+\alpha\, \phi_{\theta} \, q_{\theta}
+\frac{ {\alpha}_{\theta} \, q_{\theta} }{2}
-4\,\alpha\, \phi_{\theta\theta}
+2\, {\alpha}_{\theta} \, \phi_{\theta}
-{\alpha}_{\theta\theta} \nonumber \\ \mbox{} & &
+ \frac{f}{f_\eta}\left\{-\frac{\alpha\, q_\eta}{2}\left[1+\partial_\eta\left(\frac{f}{f_\eta}\right)\right] - 2\alpha\, \phi_\eta\left[2+\partial_\eta\left(\frac{f}{f_\eta}\right)\right] - \alpha_\eta\right\} \nonumber \\ \mbox{} & &
+ \left(\frac{f}{f_\eta}\right)^2\left[
-\frac{\alpha\, q_{\eta\eta}}{2}
-\alpha\, \phi_{\eta} \, q_{\eta}
-\frac{ {\alpha}_{\eta} \, q_{\eta}}{2}
-2\,\alpha\, \phi_{\eta\eta}
-4\,\alpha\, \phi_{\eta}^2
-2\, {\alpha}_{\eta} \, \phi_{\eta}
\right] 
\nonumber \\ \mbox{} & &
+ \frac{\tilde{H}_b \tilde{H}_d \alpha}{e^{4\phi} f^2 \sin^2\theta}
+\frac{\alpha}{e^{q+4\phi}}\left[\frac{-2 \tilde{H}_c^2}{f_\eta^2} - \frac{\tilde{H}_b^2}{f^2} + \frac{\tilde{H}_a \,\tilde{H}_b }{f_\eta^2} \right] \nonumber \\ \mbox{} & &
+2 \tilde{H}_b\, {v_2}_{\theta}
+\tilde{H}_{b,\theta}\, v_2
+2 \tilde{H}_c\, {v_1}_{\theta}
+\tilde{H}_{b,\eta}\, v_1
\end{eqnarray}

\begin{eqnarray}\label{eqn:hdcovevol} \frac{\partial \tilde{H}_d}{\partial t} & = &
\frac{\sin^2\theta}{e^{q}}\left\{\frac{}{}
 -4 \alpha\, \phi_{\theta}\, \cot\theta
-\alpha_{\theta}\, \cot\theta
-2 \alpha\, \phi_{\theta\theta}
-4 \alpha\, \phi_{\theta}^2
-2 \alpha_{\theta}\, \phi_{\theta} \right. \nonumber \\ \mbox{} & & \left.
+\left(\frac{f}{f_\eta}\right)^2\left[
-2 \alpha\, \phi_{\eta\eta}
-4 \alpha\, \phi_{\eta}^2
-2 \alpha_{\eta}\, \phi_{\eta}
\right] \right. \nonumber \\ \mbox{} & & \left.
-2 \alpha\, \phi_\eta\, \left(\frac{f}{f_\eta}\right)\left[2+\partial_\eta\left(\frac{f}{f_\eta}\right)\right]
-\alpha_{\eta}\frac{f}{f_\eta}\right\}  \nonumber \\ \mbox{} & &
+\tilde{H}_{d,\theta} v_2 + \tilde{H}_{d,\eta} v_1 
\nonumber \\ \mbox{} & &
- \frac{\tilde{H}_d^2 \alpha}{e^{4\phi} f^2 \sin^2\theta}
+\frac{\alpha}{e^{q+4\phi}}\left[\frac{\tilde{H}_b\, \tilde{H}_d}{f^2} + \frac{\tilde{H}_a\, \tilde{H}_d}{f_\eta^2}\right]
\end{eqnarray}

The Hamiltonian constraint becomes:
\begin{eqnarray}\label{eqn:hamconcovphiearly}
 & & \left(\frac{f}{f_{\eta}}\right)^2\phi_{\eta\eta} + \phi_{\theta\theta} +  \frac{f}{f_{\eta}}\left[1+\partial_{\eta}\left(\frac{f}{f_{\eta}}\right)\right] \phi_{\eta} + \cot(\theta) \phi_{\theta} + \left(\frac{f}{f_{\eta}}\right)^2 \phi_{\eta}^2 + \phi_{\theta}^2 \nonumber \\ \mbox{} & &
 = \frac{1}{4}\left[\frac{\tilde{H}_b\, \tilde{H}_d}{f^2\sin^2\theta} + \frac{\tilde{H}_a\, \tilde{H}_d}{f_\eta^2\sin^2\theta} - \frac{\tilde{H}_c^2}{e^q \,f_\eta^2} +\frac{\tilde{H}_a\, \tilde{H}_b}{e^q \,f_\eta^2}\right] e^{-4\phi} \nonumber \\ \mbox{} & &
- \frac{1}{8}\left[q_{\eta\eta}\left(\frac{f}{f_{\eta}}\right)^2+q_{\eta}\left(\frac{f}{f_{\eta}}\right)\partial_{\eta}\left(\frac{f}{f_{\eta}}\right)+q_{\theta\theta}\right]
\end{eqnarray}

The momentum constraints become\footnote{after multiplying through by $f^2e^{q+4\phi}$ - so a rescaling}
\begin{eqnarray}\label{eqn:covmomcons}
0 & =& \tilde{H}_c\,\cot\theta
+\frac{e^q}{\sin^2\theta}\left[2\, \tilde{H}_d\, \phi_\eta
-\tilde{H}_{d,\eta}+ \tilde{H}_d \frac{f_\eta}{f}\right]
+\frac{\tilde{H}_b\, q_{\eta} }{2}
+\frac{\tilde{H}_a\, q_{\eta} }{2}\left(\frac{f}{f_\eta}\right)^2
+2\, \tilde{H}_c\, \phi_\theta \nonumber \\ \mbox{} & &
+2\, \tilde{H}_b\, \phi_\eta
+4\, \tilde{H}_a\, \phi_\eta\left(\frac{f}{f_\eta}\right)^2
+\tilde{H}_{c,\theta}
-\tilde{H}_{b,\eta}
+\tilde{H}_b\frac{f_\eta}{f}
+2\, \tilde{H}_a\frac{f}{f_\eta}
\nonumber \\ \mbox{}
0 & = & \tilde{H}_b\,\cot\theta
+\frac{e^q}{\sin^2\theta}\left[\frac{}{}2\, \tilde{H}_d\, \phi_\theta
-\tilde{H}_{d,\theta} + \tilde{H}_d\,\cot\theta\right]
+\frac{\tilde{H}_b\, q_\theta}{2}
+\frac{\tilde{H}_a\, q_\theta}{2}\left(\frac{f}{f_\eta}\right)^2 \nonumber \\ \mbox{} & &
+4\, \tilde{H}_b\, \phi_\theta
+2\, \tilde{H}_a\, \phi_\theta\left(\frac{f}{f_\eta}\right)^2
+2\, \tilde{H}_c\, \phi_\eta\left(\frac{f}{f_\eta}\right)^2
+\tilde{H}_{c,\eta} \left(\frac{f}{f_\eta}\right)^2 \nonumber \\ \mbox{} & &
+\tilde{H}_c\left(\frac{f}{f_\eta}\right)\left[1+\partial_\eta\left(\frac{f}{f_\eta}\right)\right]
-\tilde{H}_{a,\theta}\left(\frac{f}{f_\eta}\right)^2
\end{eqnarray}

The maximal slicing equation is unchanged as it does not rely on the extrinsic curvature terms.

The trapped surface equation only changes slightly as none of the derivative terms involve extrinsic curvature terms, so setting 
\begin{eqnarray}\label{eqn:Hcovtransform}
H_a & \rightarrow & \frac{\tilde{H}_a}{e^{q+4\phi}\,f_\eta^2} \nonumber \\ \mbox{}
H_b & \rightarrow & \frac{\tilde{H}_b}{e^{q+4\phi}\,f^2}  \nonumber \\ \mbox{}
H_c & \rightarrow & \frac{\tilde{H}_c}{e^{q+4\phi}\,f_\eta^2} \nonumber \\ \mbox{}
H_d & \rightarrow & \frac{\tilde{H}_d}{e^{4\phi}\,f^2\,\sin^2\theta}
\end{eqnarray}
We arrive at
\begin{eqnarray}
\mathbf{h_{\theta\theta}} + \left[\cot\theta+\frac{q_\theta}{2}+4\phi_\theta\right]\mathbf{h_\theta} & & \nonumber \\ 
- \left[\frac{q_\eta}{2} + 4\phi_\eta + \left(\frac{f_\eta}{f}\right)\left(2+\partial_\eta\left(\frac{f}{f_\eta}\right)\right)\right]\mathbf{h_\theta^2} & &  \nonumber \\
+ \left(\frac{f_\eta}{f}\right)^2\left[\cot\theta + \frac{q_\theta}{2}+4\phi_\theta\right]\mathbf{h_\theta^3} & & \nonumber \\
+ \frac{f^2e^{2\phi-q}}{f_\eta^4}\left[\left(\frac{\tilde{H}_d}{e^{4\phi}\,f^2\,\sin^2\theta}\right) + \left(\frac{\tilde{H}_b}{e^{q+4\phi}\,f^2}\right) + \left(\frac{\tilde{H}_a}{e^{q+4\phi}\,f_\eta^2}\right)\right]\mathbf{\delta^3} & & \nonumber \\
+ e^{2\phi} \left[-\left(\frac{\tilde{H}_b}{e^{q+4\phi}\,f^2}\right) h_\theta^2 + 2\left(\,\frac{\tilde{H}_c}{e^{q+4\phi}\,f_\eta^2}\right)\,h_\theta-\left(\frac{f}{f_\eta}\right)^2\left(\frac{\tilde{H}_a}{e^{q+4\phi}\,f_\eta^2}\right)\right]\mathbf{\delta} & & \nonumber \\
-\left(\frac{f}{f_\eta}\right)^2\left[\frac{q_\eta}{2}+4\phi_\eta\right]-2\left(\frac{f}{f_\eta}\right) & = & 0
\end{eqnarray}
or
\begin{eqnarray}\label{eqn:sahorcov}
\mathbf{h_{\theta\theta}} + \left[\cot\theta+\frac{q_\theta}{2}+4\phi_\theta\right]\mathbf{h_\theta} & & \nonumber \\ 
- \left[\frac{q_\eta}{2} + 4\phi_\eta + \left(\frac{f_\eta}{f}\right)\left(2+\partial_\eta\left(\frac{f}{f_\eta}\right)\right)\right]\mathbf{h_\theta^2} & &  \nonumber \\
+ \left(\frac{f_\eta}{f}\right)^2\left[\cot\theta + \frac{q_\theta}{2}+4\phi_\theta\right]\mathbf{h_\theta^3} & & \nonumber \\
+ \frac{e^{-2\phi-q}}{f_\eta^4}\left[\left(\frac{\tilde{H}_d}{\sin^2\theta}\right) + \left(\frac{\tilde{H}_b}{e^{q}}\right) + \frac{\tilde{H}_a}{e^{q}}\left(\frac{f}{f_\eta}\right)^2\right]\mathbf{\delta^3} & & \nonumber \\
+ e^{-2\phi} \left[-\left(\frac{\tilde{H}_b}{e^{q}\,f^2}\right) h_\theta^2 + 2\left(\,\frac{\tilde{H}_c}{e^{q}\,f_\eta^2}\right)\,h_\theta-\left(\frac{f}{f_\eta}\right)^2\left(\frac{\tilde{H}_a}{e^{q}\,f_\eta^2}\right)\right]\mathbf{\delta} & & \nonumber \\
-\left(\frac{f}{f_\eta}\right)^2\left[\frac{q_\eta}{2}+4\phi_\eta\right]-2\left(\frac{f}{f_\eta}\right) & = & 0
\end{eqnarray}

\section{Weyl Curvature Components}\label{sec:PSINeqns}
As Bernstein \cite{bernstein} has slightly different definitions for his metric\footnote{compare to equation \ref{eqn:3dmetric}} 
\begin{eqnarray}
\gamma_{ab} & = & \psi^4 \left( \begin{array}{ccc}
A & 0 & 0 \\
0 & B & 0 \\
0 & 0 & D \sin^2\theta \\
\end{array} \right)
\end{eqnarray}
and extrinsic curvature tensors\footnote{compare to equation \ref{eqn:origcovkijtensor}}
\begin{eqnarray}
K_{ab} & = & \psi^4 \left( \begin{array}{ccc}
\tilde{H}_a & \tilde{H}_c & 0 \\
\tilde{H}_c & \tilde{H}_b & 0 \\
0 & 0 & \tilde{H}_d \sin^2\theta \\
\end{array} \right)
\end{eqnarray}
We must make the transformations
\begin{equation}
\psi=e^\phi \;;\; 
\frac{\psi_\theta}{\psi}=\phi_\theta \;;\; 
\frac{\psi_{\theta\theta}}{\psi}=\phi_\theta^2+\phi_{\theta\theta} \;;\; 
\frac{\psi_\eta}{\psi}=\phi_\eta \;;\; 
\frac{\psi_{\eta\eta}}{\psi}=\phi_\eta^2+\phi_{\eta\eta} \;;\; 
\frac{\psi_{\eta\theta}}{\psi}=\phi_{\eta\theta}+\phi_\eta\phi_\theta
\nonumber \end{equation}
\begin{equation}
A=f_\eta^2 e^q \;;\; 
\frac{A_\theta}{A}=q_\theta \;;\; 
\frac{A_\eta}{A}=q_\eta+2\frac{f_{\eta\eta}}{f_\eta} \;;\; 
\frac{A_{\theta\theta}}{A}=q_\theta^2+q_{\theta\theta}
\nonumber \end{equation}
\begin{equation}
\frac{A_{\eta\eta}}{A}=q_\eta^2+q_{\eta\eta}+4 q_\eta \frac{f_{\eta\eta}}{f_\eta} + 2\left(\frac{f_{\eta\eta}}{f_\eta}\right)^2 + \frac{f_{\eta\eta\eta}}{f_\eta}
\nonumber \end{equation}
\begin{equation}
B=f^2 e^q \;;\; 
\frac{B_\theta}{B}=q_\theta \;;\; 
\frac{B_\eta}{B}=q_\eta+2\frac{f_{\eta}}{f} \;;\; 
\frac{B_{\theta\theta}}{B}=q_\theta^2+q_{\theta\theta}
\nonumber \end{equation}
\begin{equation}
\frac{B_{\eta\eta}}{B}=q_\eta^2+q_{\eta\eta}+4 q_\eta \frac{f_{\eta}}{f} + 2\left(\frac{f_{\eta}}{f}\right)^2 + \frac{f_{\eta\eta}}{f}
\nonumber \end{equation}
\begin{equation}
D=f^2\;;\; 
\frac{D_\theta}{D}=0 \;;\; 
\frac{D_\eta}{D}=2\frac{f_{\eta}}{f} \;;\; 
\frac{D_{\theta\theta}}{D}=0
\nonumber \end{equation}
\begin{equation}
\frac{D_{\eta\eta}}{D}=2\left(\frac{f_{\eta}}{f}\right)^2 + \frac{f_{\eta\eta}}{f}
\nonumber \end{equation}
\begin{equation}
\tilde{H}_a=f_\eta^2 e^q H_a \;;\;
\tilde{H}_{a,\eta}=f_\eta^2 e^q \left(2 H_a \frac{f_{\eta\eta}}{f_\eta} + H_a q_\eta + H_{a,\eta}\right) \;;\;
\tilde{H}_{a,\theta}=f_\eta^2 e^q \left(q_\theta H_a + H_{a,\theta} \right)
\nonumber \end{equation}
\begin{equation}
\tilde{H}_b=f^2 e^q H_b \;;\;
\tilde{H}_{b,\eta}=f^2 e^q \left(2 H_b \frac{f_{\eta}}{f} + H_b q_\eta + H_{b,\eta}\right) \;;\;
\tilde{H}_{b,\theta}=f^2 e^q \left(q_\theta H_b + H_{b,\theta} \right)
\nonumber \end{equation}
\begin{equation}
\tilde{H}_c=f_\eta^2 e^q H_c \;;\;
\tilde{H}_{c,\eta}=f_\eta^2 e^q \left(2 H_c \frac{f_{\eta\eta}}{f_\eta} + H_c q_\eta + H_{c,\eta}\right) \;;\;
\tilde{H}_{c,\theta}=f_\eta^2 e^q \left(q_\theta H_c + H_{c,\theta} \right)
\nonumber \end{equation}
\begin{equation}
\tilde{H}_d=f^2 H_d \;;\;
\tilde{H}_{d,\eta}=f^2 \left(2 H_d \frac{f_{\eta}}{f} + H_{d,\eta}\right) \;;\;
\tilde{H}_{d,\theta}=f^2 H_{d,\theta}
\nonumber \end{equation}
Then with the aid of the equations derived in \cite{bernstein} (Appendix G), we can find the five independent components of the $4$-D Weyl curvature.  We reproduce, without prejudice, those equations here where the $H_i$ in the equations above are the extrinsic curvature variables of this thesis.  These $H_i$ map to the $\tilde{H}_i$ in the equations below for the $\Psi_i$ via the transformations above.
\begin{eqnarray}\label{eqn:psi0bern}
\Psi_0 & = & \frac{\psi_\theta \cot\theta}{B \psi^5}
+\frac{{v_2}_\eta \cot\theta}{4 \sqrt{A}\alpha\psi^2}
+\frac{\sqrt{A}{v_1}_\theta \cot\theta}{4 \alpha B \psi^2}
+\frac{A_\theta \cot\theta}{4 A B \psi^4}
-\frac{\psi_{\theta\theta}}{B \psi^5}
+\frac{3 \psi_\theta^2}{B \psi^6}
\nonumber \\ \mbox{} & &
-\frac{{v_2}_\eta \psi_\theta}{2 \sqrt{A}\alpha\psi^3}
-\frac{\sqrt{A}{v_1}_\theta \psi_\theta}{2 \alpha B \psi^3}
+\frac{D_\theta \psi_\theta}{2 B D \psi^5}
+\frac{B_\theta \psi_\theta}{2 B^2 \psi^5}
-\frac{\tilde{H}_d \psi_\eta}{\sqrt{A} D \psi^3}
+\frac{\tilde{H}_b \psi_\eta}{\sqrt{A} B \psi^3}
\nonumber \\ \mbox{} & &
+\frac{D_\eta \psi_\eta}{2 A D \psi^5}
-\frac{B_\eta \psi_\eta}{2 A B \psi^5}
-\frac{\tilde{H}_{d,\eta}}{2 \sqrt{A} D \psi^2}
+\frac{D_\eta \tilde{H}_d}{4 \sqrt{A} D^2 \psi^2}
+\frac{\tilde{H}_{b,\eta}}{2 \sqrt{A} B \psi^2}
-\frac{B_\eta \tilde{H}_b}{4 \sqrt{A} B^2 \psi^2}
\nonumber \\ \mbox{} & &
+\frac{D_\eta \tilde{H}_a}{4 A^{3/2} D \psi^2}
-\frac{B_\eta \tilde{H}_a}{4 A^{3/2} B \psi^2}
+\frac{{v_2}_\eta D_\theta}{8 \sqrt{A}\alpha D \psi^2}
+\frac{\sqrt{A} {v_1}_\theta D_\theta}{8 \alpha B D \psi^2}
-\frac{{v_2}_{\eta\theta}}{4 \sqrt{A} \alpha \psi^2}
-\frac{{v_2}_\eta B_\theta}{8 \sqrt{A}\alpha B \psi^2}
\nonumber \\ \mbox{} & &
+\frac{\alpha_\theta {v_2}_\eta}{4 \sqrt{A} \alpha^2 \psi^2}
-\frac{A_\theta {v_2}_\eta}{8 A^{3/2} \alpha \psi^2}
-\frac{\sqrt{A} {v_1}_{\theta\theta}}{4 \alpha B \psi^2}
+\frac{\sqrt{A} B_\theta {v_1}_{\theta}}{8 \alpha B^2 \psi^2}
+\frac{\sqrt{A} \alpha_\theta {v_1}_{\theta}}{4 \alpha^2 B \psi^2}
-\frac{3 A_\theta {v_1}_\theta}{8 \sqrt{A} \alpha B \psi^2}
\nonumber \\ \mbox{} & &
+\frac{A_\theta D_\theta}{8 A B D \psi^4}
+\frac{D_{\eta\eta}}{4 A D \psi^4}
-\frac{D_\eta^2}{8 A D^2 \psi^4}
-\frac{A_\eta D_\eta}{8 A^2 D \psi^4}
+\frac{A_\theta B_\theta}{8 A B^2 \psi^4}
-\frac{B_{\eta\eta}}{4 A B \psi^4}
\nonumber \\ \mbox{} & &
+\frac{B_\eta^2}{8 A B^2 \psi^4}
+\frac{A_\eta B_\eta}{8 A^2 B \psi^4}
-\frac{A_{\theta\theta}}{4 A B \psi^4}
+\frac{A_\theta^2}{8 A^2 B \psi^4}
-\frac{\tilde{H}_a \tilde{H}_d}{2 A D}
-\frac{\tilde{H}_c^2}{2 A B}
\nonumber \\ \mbox{} & &
+\frac{{v_1}_\theta \tilde{H}_c}{2 \alpha B}
+\frac{\tilde{H}_a \tilde{H}_b}{2 A B}
-\frac{{v_1}_\theta {v_2}_\eta}{4 \alpha^2}
-\frac{A \; {v_1}_\theta^2}{4 \alpha^2 B}
\end{eqnarray}

\begin{eqnarray}\label{eqn:psi1bern}
\Psi_1 & = & -\frac{\tilde{H}_d \cot\theta}{2 \sqrt{B} D \psi^2}
+\frac{\tilde{H}_b \cot\theta}{2 B^{3/2} \psi^2}
+\frac{D_\eta \cot\theta}{4 \sqrt{AB} D \psi^4}
-\frac{B_\eta \cot\theta}{4 \sqrt{A} B^{3/2} \psi^4}
-\frac{3 \psi_\eta \psi_\theta}{\sqrt{AB} \psi^6}
-\frac{\tilde{H}_d \psi_\theta}{\sqrt{B} D \psi^3}
\nonumber \\ \mbox{} & &
+\frac{\tilde{H}_b \psi_\theta}{B^{3/2} \psi^3}
-\frac{B_\eta \psi_\theta}{2 \sqrt{A} B^{3/2} \psi^5}
+\frac{\psi_{\eta\theta}}{\sqrt{AB} \psi^5}
+\frac{\sqrt{B} {v_2}_\eta \psi_\eta}{2 A \alpha \psi^3}
+\frac{{v_1}_\theta \psi_\eta}{2 \alpha \sqrt{B} \psi^3}
-\frac{A_\theta \psi_\eta}{2 A^{3/2} \sqrt{B} \psi^5}
\nonumber \\ \mbox{} & &
-\frac{\tilde{H}_{d,\theta}}{2 \sqrt{B} D \psi^2}
+\frac{D_\theta \tilde{H}_d}{4 \sqrt{B} D^2 \psi^2}
+\frac{D_\theta \tilde{H}_b}{4 B^{3/2} D \psi^2}
+\frac{\sqrt{B} {v_2}_\eta D_\eta}{8 A \alpha D \psi^2}
+\frac{{v_1}_\theta D_\eta}{8 \alpha \sqrt{B} D \psi^2}
-\frac{D_\eta D_\theta}{8 \sqrt{AB} D^2 \psi^4}
\nonumber \\ \mbox{} & &
-\frac{B_\eta D_\theta}{8 \sqrt{A} B^{3/2} D \psi^4}
+\frac{D_{\eta\theta}}{4 \sqrt{AB} D \psi^4}
-\frac{A_\theta D_\eta}{8 A^{3/2} \sqrt{B} D \psi^4}
-\frac{\sqrt{B} {v_2}_\eta \tilde{H}_d}{4 \sqrt{A} \alpha D }
-\frac{\sqrt{A} {v_1}_\theta \tilde{H}_d}{4 \alpha \sqrt{B} D }
\end{eqnarray}

\begin{eqnarray}\label{eqn:psi2bern}
\Psi_2 & = & \frac{\psi_\theta \cot\theta}{B \psi^5}
+\frac{D_\theta \cot\theta}{2 B D \psi^4}
-\frac{B_\theta \cot\theta}{4 B^2 \psi^4}
+\frac{\psi_{\theta\theta}}{B \psi^5}
-\frac{\psi_{\theta}^2}{B \psi^6}
+\frac{D_\theta \psi_{\theta}}{2 B D \psi^5}
\nonumber \\ \mbox{} & &
-\frac{B_\theta \psi_{\theta}}{2 B^2 \psi^5}
+\frac{2 \psi_{\eta}^2}{A \psi^6}
+\frac{D_\eta \psi_{\eta}}{2 A D \psi^5}
+\frac{B_\eta \psi_{\eta}}{2 A B \psi^5}
+\frac{D_{\theta\theta}}{4 B D \psi^4}
+\frac{D_{\theta}^2}{8 B D^2 \psi^4}
\nonumber \\ \mbox{} & &
-\frac{B_\theta D_{\theta}}{8 B^2 D \psi^4}
+\frac{B_\eta D_\eta}{8 A B D \psi^4}
-\frac{1}{2 B \psi^4}
-\frac{\tilde{H}_b \tilde{H}_d}{2 B D}
\end{eqnarray}

\begin{eqnarray}\label{eqn:psi3bern}
\Psi_3 & = & -\frac{\tilde{H}_d \cot\theta}{2 \sqrt{B} D \psi^2}
+\frac{\tilde{H}_b \cot\theta}{2 B^{3/2} \psi^2}
-\frac{D_\eta \cot\theta}{4 \sqrt{AB} D \psi^4}
+\frac{B_\eta \cot\theta}{4 \sqrt{A} B^{3/2} \psi^4}
+\frac{3 \psi_\eta \psi_\theta}{\sqrt{AB} \psi^6}
-\frac{\tilde{H}_d \psi_\theta}{\sqrt{B} D \psi^3}
\nonumber \\ \mbox{} & &
+\frac{\tilde{H}_b \psi_\theta}{B^{3/2} \psi^3}
+\frac{B_\eta \psi_\theta}{2 \sqrt{A} B^{3/2} \psi^5}
-\frac{\psi_{\eta\theta}}{\sqrt{AB} \psi^5}
+\frac{\sqrt{B} {v_2}_\eta \psi_\eta}{2 A \alpha \psi^3}
+\frac{{v_1}_\theta \psi_\eta}{2 \alpha \sqrt{B} \psi^3}
+\frac{A_\theta \psi_\eta}{2 A^{3/2} \sqrt{B} \psi^5}
\nonumber \\ \mbox{} & &
-\frac{\tilde{H}_{d,\theta}}{2 \sqrt{B} D \psi^2}
+\frac{D_\theta \tilde{H}_d}{4 \sqrt{B} D^2 \psi^2}
+\frac{D_\theta \tilde{H}_b}{4 B^{3/2} D \psi^2}
+\frac{\sqrt{B} {v_2}_\eta D_\eta}{8 A \alpha D \psi^2}
+\frac{{v_1}_\theta D_\eta}{8 \alpha \sqrt{B} D \psi^2}
+\frac{D_\eta D_\theta}{8 \sqrt{AB} D^2 \psi^4}
\nonumber \\ \mbox{} & &
+\frac{B_\eta D_\theta}{8 \sqrt{A} B^{3/2} D \psi^4}
-\frac{D_{\eta\theta}}{4 \sqrt{AB} D \psi^4}
+\frac{A_\theta D_\eta}{8 A^{3/2} \sqrt{B} D \psi^4}
+\frac{\sqrt{B} {v_2}_\eta \tilde{H}_d}{4 \sqrt{A} \alpha D }
+\frac{\sqrt{A} {v_1}_\theta \tilde{H}_d}{4 \alpha \sqrt{B} D }
\end{eqnarray}

\begin{eqnarray}\label{eqn:psi4bern}
\Psi_4 & = & \frac{\psi_\theta \cot\theta}{B \psi^5}
-\frac{{v_2}_\eta \cot\theta}{4 \sqrt{A}\alpha\psi^2}
-\frac{\sqrt{A}{v_1}_\theta \cot\theta}{4 \alpha B \psi^2}
+\frac{A_\theta \cot\theta}{4 A B \psi^4}
-\frac{\psi_{\theta\theta}}{B \psi^5}
+\frac{3 \psi_\theta^2}{B \psi^6}
\nonumber \\ \mbox{} & &
+\frac{{v_2}_\eta \psi_\theta}{2 \sqrt{A}\alpha\psi^3}
+\frac{\sqrt{A}{v_1}_\theta \psi_\theta}{2 \alpha B \psi^3}
+\frac{D_\theta \psi_\theta}{2 B D \psi^5}
+\frac{B_\theta \psi_\theta}{2 B^2 \psi^5}
+\frac{\tilde{H}_d \psi_\eta}{\sqrt{A} D \psi^3}
-\frac{\tilde{H}_b \psi_\eta}{\sqrt{A} B \psi^3}
\nonumber \\ \mbox{} & &
+\frac{D_\eta \psi_\eta}{2 A D \psi^5}
-\frac{B_\eta \psi_\eta}{2 A B \psi^5}
+\frac{\tilde{H}_{d,\eta}}{2 \sqrt{A} D \psi^2}
-\frac{D_\eta \tilde{H}_d}{4 \sqrt{A} D^2 \psi^2}
-\frac{\tilde{H}_{b,\eta}}{2 \sqrt{A} B \psi^2}
+\frac{B_\eta \tilde{H}_b}{4 \sqrt{A} B^2 \psi^2}
\nonumber \\ \mbox{} & &
-\frac{D_\eta \tilde{H}_a}{4 A^{3/2} D \psi^2}
+\frac{B_\eta \tilde{H}_a}{4 A^{3/2} B \psi^2}
-\frac{{v_2}_\eta D_\theta}{8 \sqrt{A}\alpha D \psi^2}
-\frac{\sqrt{A} {v_1}_\theta D_\theta}{8 \alpha B D \psi^2}
+\frac{{v_2}_{\eta\theta}}{4 \sqrt{A} \alpha \psi^2}
+\frac{{v_2}_\eta B_\theta}{8 \sqrt{A}\alpha B \psi^2}
\nonumber \\ \mbox{} & &
-\frac{\alpha_\theta {v_2}_\eta}{4 \sqrt{A} \alpha^2 \psi^2}
+\frac{A_\theta {v_2}_\eta}{8 A^{3/2} \alpha \psi^2}
+\frac{\sqrt{A} {v_1}_{\theta\theta}}{4 \alpha B \psi^2}
-\frac{\sqrt{A} B_\theta {v_1}_{\theta}}{8 \alpha B^2 \psi^2}
-\frac{\sqrt{A} \alpha_\theta {v_1}_{\theta}}{4 \alpha^2 B \psi^2}
+\frac{3 A_\theta {v_1}_\theta}{8 \sqrt{A} \alpha B \psi^2}
\nonumber \\ \mbox{} & &
+\frac{A_\theta D_\theta}{8 A B D \psi^4}
+\frac{D_{\eta\eta}}{4 A D \psi^4}
-\frac{D_\eta^2}{8 A D^2 \psi^4}
-\frac{A_\eta D_\eta}{8 A^2 D \psi^4}
+\frac{A_\theta B_\theta}{8 A B^2 \psi^4}
-\frac{B_{\eta\eta}}{4 A B \psi^4}
\nonumber \\ \mbox{} & &
+\frac{B_\eta^2}{8 A B^2 \psi^4}
+\frac{A_\eta B_\eta}{8 A^2 B \psi^4}
-\frac{A_{\theta\theta}}{4 A B \psi^4}
+\frac{A_\theta^2}{8 A^2 B \psi^4}
-\frac{\tilde{H}_a \tilde{H}_d}{2 A D}
-\frac{\tilde{H}_c^2}{2 A B}
\nonumber \\ \mbox{} & &
+\frac{{v_1}_\theta \tilde{H}_c}{2 \alpha B}
+\frac{\tilde{H}_a \tilde{H}_b}{2 A B}
-\frac{{v_1}_\theta {v_2}_\eta}{4 \alpha^2}
-\frac{A \; {v_1}_\theta^2}{4 \alpha^2 B}
\end{eqnarray}

\fancyhead[RO,LE]{\thepage}
\fancyfoot{} 
\chapter{Appendix: Miscellaneous Algorithms}\label{appendix:misc_algorithms}
\bigskip

\section[Algorithm Design]{Algorithm Design and Speed}\label{sec:algorithmdesign}
The topic of numerical algorithm design and optimisation is a rich and well-investigated field, for example \cite{burden,golub,press}.  As an example of how this relates to the speed of this particular code, let us investigate algorithms for performing a summation of all the elements of a matrix, that minimize the possible errors when the terms in the summation can vary by many orders of magnitude.
$$\sum_{i,j}A_{ij}$$
As discussed in section \ref{subsec:addterms}, the order in which we perform summation of the terms is important in a numerical algorithm.  Let us now consider two such algorithms for performing this summation.  For the first algorithm, we calculate the logs of the individual terms to get bounds on how far apart the terms are and what order they should be added in, then iterate through the known powers to calculate the final result.

\begin{verbatim}
c find logs and min/max powers
      minpower=500; maxpower=-500
      do i=limin,limax; do j=ljmin,ljmax
        if(f(i,j).ne.0.d0) then
          logf(i,j)=int(dlog(dabs(f(i,j))))
          minpower=min(minpower,logf(i,j))
          maxpower=max(maxpower,logf(i,j))
        else
          logf(i,j)=-500
        endif
      enddo; enddo
c iterate through powers and find answer
      ans=0.d0
      do p=minpower,maxpower
        temp=0.d0
        do i=limin,limax; do j=ljmin,ljmax
c add up all values of a similar power first
          if(logf(i,j).eq.p) temp=temp+f(i,j)
        enddo; enddo
c then add to other terms
        ans=ans+temp
      enddo
\end{verbatim}

This algorithm produces a correct result, however if we assume that there are $M$ radial grid points and $N$ angular grid points, and we have a maximum difference in logarithms between the values inside our matrix $A_{ij}$ of $D$, then we require 
$$(D+1)\times M\times N$$
looping iterations to calculate the result\footnote{I am leaving out a consideration of the computational cost of addition, logarithm calculations, matrix value lookups, etc. as there are approximately the same number of each in both algorithms}.

In this second algorithm we assume that all values are between the maximum and minimum expressible $64$ bit double precision values $(\sim 10^{-308} \rightarrow 10^{308})$, assign a ``bin'' to each integer logarithm value and add on the fly, then calculate the total at the end:

\begin{verbatim}
c assume logs live between -999 -> +999
      do p=1,2000
         sumf(p)=0.d0
      enddo

c find logs and min/max powers
      do i=limin,limax; do j=ljmin,ljmax
        if(f(i,j).ne.0.d0) then
          logf=int(dlog(dabs(f(i,j))))
          if(abs(logf).le.999) then
            sumf(1000+logf)=sumf(1000+logf)+f(i,j)
          else
            print*,'logf too large=,',logf,i,j; stop
          endif
        endif
      enddo; enddo
c iterate through powers and find answer
      ans=0.d0
      do p=1,2000
        ans=ans+sumf(p)
      enddo
\end{verbatim}

This algorithm requires
$$2\times 2000+M\times N$$
iterations\footnote{and has no calls to MIN and MAX}, which can be significantly less than the previous algorithm given a $200 \times 60$ grid size.  In fact, if $D=2$ (we only have two distinct ``bins''), which is the worst case scenario for the second algorithm and best case for the first, the second algorithm is still computationally cheaper.

After changing the above addition algorithm, the time to perform one sweep through the full BiCGStab algorithm changed from $\sim 17$s to $\sim 12$s for a particular IVP, a significant speedup when over $90\%$ of the code's run time is spend in the BiCGStab subroutine.

\section{Miscellaneous numerical methods}

\subsection{Finding roots of a cubic equation}\label{subsec:cubicroots}
As described by \cite{mcallister}, we can find the roots of a cubic equation of the form
$$z^3+A_2 z^2+A_1 z +A_0=0$$
by defining $b$ and $c$ as
$$b=\frac{A_1}{3}-\frac{A_2^2}{9}$$
$$c=\frac{A_1A_2-3A_0}{6}-\frac{A_2^3}{27}$$
and the (potentially) complex variables $s_1$ and $s_2$ as
$$s_1=\left(c+\sqrt{b^3+c^2}\right)^{\frac{1}{3}}$$
$$s_2=\left(c-\sqrt{b^3+c^2}\right)^{\frac{1}{3}}$$
Our roots are then given by
$$z_1=(s_1+s_2)-\frac{A_2}{3}$$
$$z_2=-\frac{(s_1+s_2)}{2}-\frac{A_2}{3} + \frac{i\sqrt{3}}{2}(s_1-s_2)$$
$$z_3=-\frac{(s_1+s_2)}{2}-\frac{A_2}{3} - \frac{i\sqrt{3}}{2}(s_1-s_2)$$
We can classify the root structure by defining $\delta=b^3+c^2$ and noting that
\begin{itemize}
\item $\delta>0$ corresponds to one real root and two complex (conjugate) roots
\item $\delta=0$ means (at least) two roots are equal.  All three are equal in the case of a degenerate cubic.
\item $\delta<0$ means all roots are real
\end{itemize}
and McAllister \cite{mcallister} notes the following relations:
$$z_1+z_2+z_3=-A_2$$
$$z_1z_2+z_1z_3+z_2z_3=A_1$$
$$z_1z_2z_3=-A_0$$

\subsection{Finding the inverse of a ''small'' matrix}
While in general our matrix equations are too large to solve analytically, there are some cases where we need to invert a $2\times 2$ or $3\times 3$ matrix\footnote{commonly inside fitting algorithms where we are using two or three stencil points to fit a polynomial or inverse function and need to solve for the fitting coefficients}.  For completeness we include the analytic expressions for these cases.  Given the $2\times 2$ matrix
\begin{equation}A_{ij}=\left[\begin{array}{cc}
a_{11} & a_{12} \\
a_{21} & a_{22}
\end{array}\right]\end{equation}
then
\begin{equation}A^{-1}_{ij}=\frac{1}{\Delta}\left[\begin{array}{cc}
a_{22} & -a_{12} \\
-a_{21} & a_{11}
\end{array}\right]\end{equation}
where
$$\Delta=a_{11}a_{22}-a_{12}a_{21}$$
For a $3\times 3$ system we find that given
\begin{equation}B_{ij}=\left[\begin{array}{ccc}
b_{11} & b_{12} & b_{13} \\
b_{21} & b_{22} & b_{23} \\
b_{31} & b_{32} & b_{33}
\end{array}\right]\end{equation}
the inverse is given by
\begin{equation}B^{-1}_{ij}=\frac{1}{det(B)}\left[\begin{array}{ccc}
b_{33}b_{22}-b_{32}b_{23} & -(b_{33}b_{12}-b_{32}b_{13}) & b_{23}b_{12}-b_{22}b_{13} \\
-(b_{33}b_{21}-b_{31}b_{23}) & b_{33}b_{11}-b_{31}b_{13} & -(b_{23}b_{11}-b_{21}b_{13} \\
b_{32}b_{21}-b_{31}b_{22} & -(b_{32}b_{11}-b_{31}b_{12}) & b_{22}b_{11}-b_{21}b_{12}
\end{array}\right]\end{equation}
where
$$ det(B) =  b_{11}(b_{33}b_{22}-b_{32}b_{23})-b_{21}(b_{33}b_{12}-b_{32}b_{13})+b_{31}(b_{23}b_{12}-b_{22}b_{13})$$

\fancyhead[RO,LE]{\thepage}
\fancyfoot{} 
\chapter{Appendix: Testing Numerical Methods and Formulations in 1+1}\label{chap:testgridcoord}
\bigskip

Much credit must be given to a number of early researchers in the field of numerical relativity and GR simulations (i.e. \cite{bernstein2,evans,frontiers,hobill:slicing}) for one of the lessons learned in the early days of higher-dimensional simulations was that if you can't get a numerical method, gridding scheme, physical situation, etc. to work on a simple system, it will never work on a complicated one.  This is also computer programming axiom: start simple and build the complexity gradually so that you can debug and troubleshoot at a reasonable scale.

It is easier to program \emph{and test} a new coordinate system or a new algorithm in 1+1 dimensions than it is in 2+1 or 3+1 for that matter\footnote{``N+1'' refers to N spatial dimensions and one temporal dimension}.  Many problems become tractable when the number of dimensions is reduced to view if the phenomenon is the result of increased dimensionality or is really a problem with the algorithm/coordinates.

We will use this approach to test two different choices available to us:
\begin{itemize}
\item investigate some different numerical gridding algorithms (staggered leap-frog vs. strict CTCS\footnote{Centered in Time and Centered in Space})
\item an alternate formulation (BSSN\footnote{Baumgarte-Shapiro-Shibata-Nakamura \emph{et al.} \cite{BSSN,BSSN2}}) to the Cauchy evolution problem compared to ADM
\end{itemize}

\section{1+1 formulations}
We will now present the mathematical foundations of the 1+1 formalisms we wish to test (ADM, BSSN).  1+1 in this context refers to ``one temporal and one spatial dimension'', i.e. time evolution of a spherically symmetric spacetime.  All variables are assumed to be functions of $(t,r)$.

\subsection{BSSN - an alternate Cauchy formulation}
One alternative we investigate in 1+1 before heading to 2+1 is an alternate method of looking at the Cauchy problem discussed in section (\ref{sec:cauchyivp}).

The BSSN set of evolution equations is based upon the work of Baumgarte-Shapiro-Shibata-Nakamura \emph{et al.} \cite{BSSN,BSSN2}, which gives an alternative to the popular ADM manner of posing Einstein's equations in a ``3-space plus time-evolution'' manner.  The BSSN formulation still splits the spacetime into spatial 3-hypersurfaces and evolves along time like ADM, however, it uses an alternate set of dynamical variables.

The major ``benefits'' of the BSSN formulation are discussed in \cite{BSSN,BSSN2}, as the formulation moves some of the non-linearity to a different part of the evolution process.  There has been some success in using the BSSN formulation to disperse numerical error \cite{BSSN,Alcubierre,Thornburg:cartesian}.  Its major failing is the necessity of employing a conformally flat 3-metric\footnote{A conformally flat metric is one which varies from flat space by only a conformal factor, $\psi$, meaning we can write our metric as $\tilde{\gamma}_{ij}=\psi^n \gamma_{ij}$, where $\gamma_{ij}$ is the flat-space metric.} which, if one examines the Brill conditions in section (\ref{subsec:brillformalism}), is incompatible.  Some thought was given to attempting to modify the original assumptions of BSSN to include a Brill-compatible metric, so we will investigate the performance of BSSN relative to ADM in 1+1 first.

\subsection{ADM and BSSN equations in a 1+1 spacetime}\label{sec:admbssneqns}
We have already covered the basic formulation of the ADM framework in section (\ref{subsec:admoverview}), and here we present it in spherical symmetry (see \cite{bernstein2} or various texts for a formal derivation).

The formulation of the BSSN framework is covered in \cite{BSSN2} and we present the spherically symmetric equations here, however it is important to note that they use $\psi=e^\phi$ in their formulation.

In the metric variables presented below a number of logarithmic derivative terms of the type $\frac{V_{\eta}}{V}$ appear throughout the equations.  This is partially the motivation for the introduction of exponential variables in section (\ref{sec:aximetric}) and the presentation of the mixed Christoffel symbols in section (\ref{sec:christoffel}).  As both sets of equations (ADM and BSSN) were initially formed with non-exponential metric variables\footnote{Except for $\phi$ in BSSN.}, we leave them as such for our comparative analysis.

The spatial part of the Schwarzschild metric using a conformal decomposition is:
\begin{eqnarray}\label{eqn:1dschwmetric}
dl^2 & = & \psi^4[a \;dr^2 + b (d\theta^2 + \sin^2\theta \; d\phi^2)] 
\end{eqnarray}
where our metric variables are $a$, $b$ and $\psi$.  We set the shift quantity $\beta=0$.
The Ricci tensor quantities for 1-D ADM coordinates are given by:
\begin{eqnarray}R_{11} & = & \frac{4 \frac{{\partial}^2 \psi}{\partial{r}^2}}{\psi}
+ \frac{4 \left(\frac{\partial \psi}{\partial{r}}\right)^2}{\psi^2}
- \frac{2 \frac{\partial b}{\partial{r}}\frac{\partial\psi}{\partial{r}}}{b\psi}
+ \frac{2 \frac{\partial a}{\partial{r}}\frac{\partial\psi}{\partial{r}}}{a\psi}
- \frac{\frac{\partial^2 b}{\partial{r}^2}}{b}
- \frac{\left(\frac{\partial b}{\partial{r}}\right)^2}{2b^2}
+ \frac{\frac{\partial a}{\partial{r}}\frac{\partial b}{\partial{r}}}{2ab}
\end{eqnarray}

\begin{eqnarray}R_{22} & = & -\frac{2b\frac{\partial^2\psi}{\partial{r}^2}}{a\psi}
-\frac{2b\left(\frac{\partial\psi}{\partial{r}}\right)^2}{a\psi^2}
- \frac{3 \frac{\partial b}{\partial{r}}\frac{\partial\psi}{\partial{r}}}{a\psi}
- \frac{3 \frac{\partial a}{\partial{r}}b\frac{\partial\psi}{\partial{r}}}{a^2\psi}
-\frac{\frac{\partial^2b}{\partial{r}^2}}{2a}
+ \frac{\frac{\partial a}{\partial{r}}\frac{\partial b}{\partial{r}}}{4a^2} + 1
\end{eqnarray}

The Hamiltonian Constraint for the 1-D ADM coordinates is given by:
\begin{eqnarray} 0 & = & \frac{2 R_{22}}{b\psi^4} + \frac{R_{11}}{a\psi^4} + \frac{2H_b^2 }{b^2} + \frac{4 H_a H_b}{ab}
\end{eqnarray}

The single Momentum Constraint for the 1-D ADM coordinates is given by:
\begin{eqnarray} 0 & = & \frac{4 H_b \frac{\partial\psi}{\partial{r}}}{b\psi}
- \frac{4 H_a \frac{\partial\psi}{\partial{r}}}{a\psi}
+ \frac{2 \frac{\partial H_b}{\partial{r}}}{b}
- \frac{\frac{\partial b}{\partial{r}}H_b}{b^2}
- \frac{\frac{\partial b}{\partial{r}}H_a}{ab}
\end{eqnarray}

The evolution equations for the 1-D ADM coordinates are:
\begin{eqnarray}\label{eqn:1DADMevol} \dot{a} & = & -\frac{4 \beta \frac{\partial\psi}{\partial{r}}}{\psi^5}
+\frac{2 \frac{\partial\beta}{\partial{r}}}{\psi^4}
-\frac{\frac{\partial a}{\partial{r}} \beta}{a\psi^4}
-2\alpha H_a \nonumber \\
 & = & -2\alpha H_a \nonumber \\
\dot{b} & = & \frac{4 b \beta\frac{\partial\psi}{\partial{r}}}{a\psi^5}
+ \frac{\frac{\partial b}{\partial{r}} \beta}{a\psi^4} - 2\alpha H_b \nonumber \\
& = & - 2\alpha H_b \nonumber \\
\dot{H_a} & = & \frac{\alpha R_{11}}{\psi^4}
- \frac{4 \beta H_a \frac{\partial\psi}{\partial{r}}}{a\psi^5}
+ \frac{2 \frac{\partial\alpha}{\partial{r}}\frac{\partial\psi}{\partial{r}}}{\psi^5}
+ \frac{\beta \frac{\partial H_a}{\partial{r}}}{a\psi^4}
+ \frac{2 \frac{\partial\beta}{\partial{r}} H_a}{a\psi^4}
- \frac{2 \frac{\partial a}{\partial{r}}\beta H_a}{a^2\psi^4}
- \frac{\frac{\partial^2\alpha}{\partial{r}^2}}{\psi^4} \nonumber \\ \mbox{}
& & + \frac{\frac{\partial a}{\partial{r}} \frac{\partial\alpha}{\partial{r}}}{2a\psi^4}
+ \frac{2\alpha H_a H_b}{b}
- \frac{\alpha H_a^2}{a} \nonumber \\
 & = & \frac{\alpha R_{11}}{\psi^4}
+ \frac{2 \frac{\partial\alpha}{\partial{r}}\frac{\partial\psi}{\partial{r}}}{\psi^5}
- \frac{\frac{\partial^2\alpha}{\partial{r}^2}}{\psi^4}
+ \frac{\frac{\partial a}{\partial{r}} \frac{\partial\alpha}{\partial{r}}}{2a\psi^4}
+ \frac{2\alpha H_a H_b}{b}
- \frac{\alpha H_a^2}{a} \nonumber \\
\dot{H_b} & = & \frac{\alpha R_{22}}{\psi^4}
+ \frac{4 \beta H_b \frac{\partial\psi}{\partial{r}}}{a\psi^5}
- \frac{2 \frac{\partial a}{\partial{r}} b \frac{\partial\psi}{\partial{r}}}{a\psi^5}
+ \frac{\beta \frac{\partial H_b}{\partial{r}}}{a\psi^4}
- \frac{\frac{\partial\alpha}{\partial{r}}\frac{\partial b}{\partial{r}}}{2a\psi^4}
+ \frac{\alpha H_a H_b}{a} \nonumber \\
 & = & \frac{\alpha R_{22}}{\psi^4}
- \frac{2 \frac{\partial a}{\partial{r}} b \frac{\partial\psi}{\partial{r}}}{a\psi^5}
- \frac{\frac{\partial\alpha}{\partial{r}}\frac{\partial b}{\partial{r}}}{2a\psi^4}
+ \frac{\alpha H_a H_b}{a}
\end{eqnarray}

The maximal slicing equation for the 1-D ADM coordinates is:
\begin{eqnarray} \alpha_{{r}{r}} + \alpha_{{r}}\left[2\frac{\psi_{r}}{\psi} + \frac{b_{r}}{b}-\frac{a_{r}}{2a} \right] + & & \nonumber \\ \mbox{}
 \alpha\left[8\frac{\psi_{r}}{\psi} \left(\frac{b_{r}}{b}-\frac{a_{r}}{2a}\right) + 8\frac{\psi_{{r}{r}}}{\psi} + 2\frac{b_{{r}{r}}}{b}-\frac{b_{r}^2}{2b^2}-\frac{a_{r} b_{r}}{ab} - \frac{2a}{b} \right] & = & 0
\end{eqnarray}

The Ricci tensor quantities for the 1-D BSSN coordinates are given by:
\begin{eqnarray}
R_{11}^{BSSN} & = & -4 \phi_{rr} - 2\frac{b_r}{b}\phi_r + 2 \frac {a_r}{a}\phi_r
-\frac{1}{2}\left(\frac{b_r}{b}\right)^2
-\frac{1}{2}\left(\frac{a_{rr}}{a}\right)
+\frac{3}{4}\left(\frac{a_r}{a}\right)^2
+\frac{\tilde{\Gamma}^1 a_r}{2}
+\tilde{\Gamma}^1_r a \nonumber \\
R_{22}^{BSSN} & = & -2\frac{b}{a}\phi_{rr} - 4\frac{b}{a}\phi_r^2 
-3\frac{b_r}{a}\phi_r + \frac{a_r b}{a^2}\phi_r
-\frac{1}{2}\left(\frac{b_{rr}}{a}\right)
+\frac{1}{2}\left(\frac{b_{r}^2}{ab}\right)
+ \frac{1}{2}\tilde{\Gamma}^1 b_r + 1
\end{eqnarray}
where $\tilde{\Gamma}^1$ is an auxiliary variable introduced in the BSSN formulation that is calculated from the Christoffel coefficients via
$$\tilde{\Gamma}^1=\tilde{\gamma}^{jk}\tilde{\Gamma}^1_{\;jk}$$
  The evolution equations for metric and extrinsic curvature variables in 1-D BSSN are given by:
\begin{eqnarray}
\partial_ta & = & -2 \alpha H_a \nonumber \\
\partial_tb & = & -2 \alpha H_b \nonumber \\
\partial_tH_a & = & \left(\frac{e^{-4\phi}}{3}\right) \left[-4\alpha\phi_{rr}
+ 8\alpha\phi_r^2 + 8\alpha_r\phi_r + 2\alpha\phi_r\left(\frac{b_r}{b}+\frac{a_r}{a}\right)
- 2\alpha_{rr} + \left(\frac{b_r}{b}+\frac{a_r}{a}\right)\alpha_r\right.
\nonumber \\ & & \mbox{} 
\left.+\left(\frac{b_{rr}}{b} -2\left(\frac{b_r}{b}\right)^2 -\tilde{\Gamma}^1\frac{b_r}{b}a -2\frac{a}{b} - \frac{a_{rr}}{a} + \frac{3}{2}\left(\frac{a_r}{a}\right)^2 + \tilde{\Gamma}^1a_r + 2\tilde{\Gamma}^1_ra \right)\alpha\right] -2\frac{H_a^2}{a}\alpha \nonumber \\
\partial_tH_b & = & \left(\frac{e^{-4\phi}}{3}\right) \left[\frac{b\alpha}{a}\left(2\phi_{rr}-4\phi_r^2\right) \left(-4\frac{b\alpha_r}{a}-\frac{b_r\alpha}{a}-\frac{a_r}{a}\frac{b\alpha}{a}\right)\phi_r + \frac{b\alpha_{rr}}{a} - \frac{b_r\alpha_r}{2a} - \frac{a_r}{a}\frac{b\alpha_r}{2a}\right.
\nonumber \\ & & \mbox{}
\left.+\left(\frac{b_{rr}}{2a}+\frac{b_r}{b}\frac{b_r}{a} + \tilde{\Gamma}^1\frac{b_r}{2} + \frac{a_{rr}b}{2a^2} -\frac{3}{4}\left(\frac{a_r}{a}\right)^2\frac{b}{a} - \tilde{\Gamma}^1\frac{a_r}{a}\frac{b}{2} -\tilde{\Gamma}^1_rb + 1\right)\alpha \right] + 2\frac{H_b^2}{b}\alpha \nonumber \\
\partial_t\tilde{\Gamma}^1 & = & \left( 12\alpha\phi_r -2\alpha_r \right)\frac{H_a}{a^2} -2\frac{b_r}{b}\frac{H_b\alpha}{ab} + \frac{a_r}{a}\frac{H_a\alpha}{a^2} \end{eqnarray}
The Maximal Slicing equation for the 1-D BSSN coordinates is:
\begin{eqnarray}
0 & = & -\left(\frac{e^{-4\phi}}{a}\right)\alpha_{rr}
+ \left(\frac{e^{-4\phi}}{a}\right)\left[-2\phi_r - \frac{b_r}{b}+\frac{a_r}{2a}\right]\alpha_r
+ \left[\frac{6H_b^2}{b^2} + \frac{3H_a^2}{a^2}\right]\alpha
\end{eqnarray}
The Hamiltonian Constraint for the 1-D BSSN coordinates is:
\begin{eqnarray}
0 & = & R^{BSSN} + K^2 - K^{ij}K_{ij} \nonumber \\
 & = & \frac{R_{11}^{BSSN}}{e^{4\phi}a} + 2\frac{R_{22}^{BSSN}}{e^{4\phi}b} - \left(\frac{H_a}{a}\right)^2 - 2\left(\frac{H_b}{b}\right)^2 \nonumber \\
0 & = & -\frac{8\phi_{rr}}{a} -\frac{8 \phi_r^2}{a} -\frac{8b_r\phi_r}{ab} + \frac{4a_r\phi_r}{a^2} + \frac{b_{rr}}{ab} + \frac{b_r^2}{2ab^2} + \tilde{\Gamma}^1\frac{b_r}{b} + \frac{2}{b} \nonumber \\
& & - \frac{a_{rr}}{2a^2} + \frac{3 a_r^2}{4a^3} + \frac{\tilde{\Gamma}^1a_r}{2a} + \tilde{\Gamma}^1_r - \left(\frac{H_a}{a}\right)^2 - 2\left(\frac{H_b}{b}\right)^2
\end{eqnarray}
and the Momentum Constraint for 1-D BSSN coordinates is:
\begin{eqnarray}
\frac{-4H_b\phi_r}{ab} + \frac{4H_a\phi_r}{a^2} - \frac{b_rH_b}{ab^2}  + \frac{{H_a}_r}{a^2} + \frac{b_rH_a}{a^2b} - \frac{a_rH_a}{a^3}  & = & 0
\end{eqnarray}

\section[Numerical Tests in 1+1]{Numerical Testing of BSSN, Staggered Leap-Frog (SLF) and Strict CTCS in a 1+1 spacetime}

\subsection{Gridding methods}
When discretising a system of differential equations for time evolution via numerical methods there are a variety of choices available.  One major decision is how one is going to calculate derivative terms, for the Taylor expansion can be taken around a number of different grid points to potentially give an equivalent convergence for the order of the error in calculating the derivative at a point (see also chapter \ref{chap:nummethod}).

\begin{figure}
\centering
%\psfrag{D}{$\Delta$}
\includegraphics{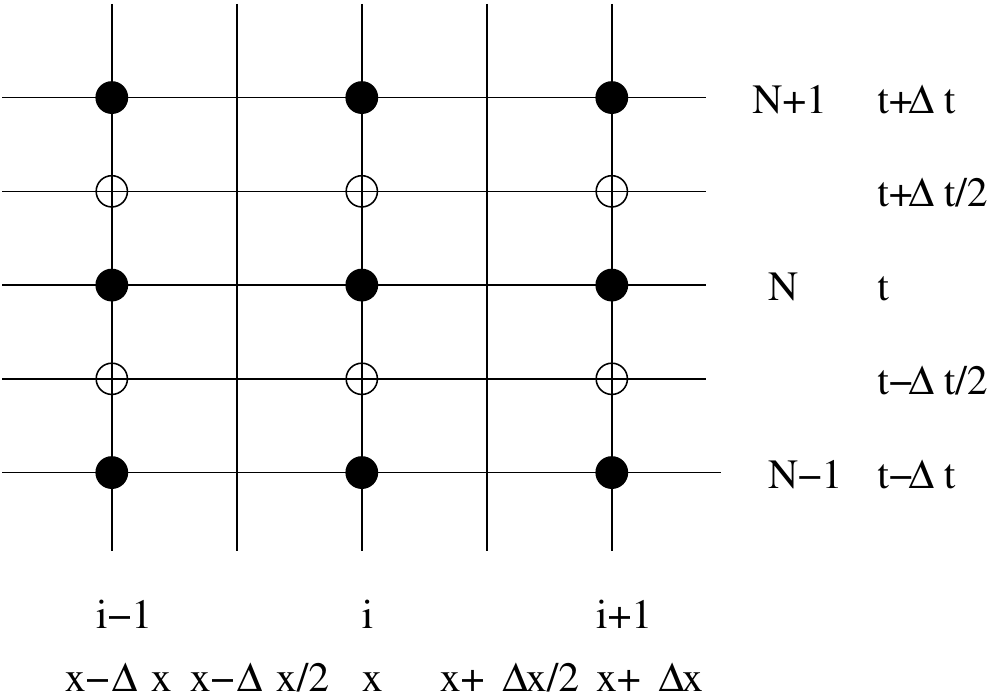}
\caption[A discretised 2-D grid]{A diagram showing where grid points are located on a 2-D grid with a discrete time counter $N$ and grid spacing $\Delta t$, and a discrete spatial counter $i$ with grid spacing $\Delta x$}
\label{fig:CTCS_SLF_coords}
\end{figure}

For our \emph{spatial} coordinates, to help avoid any directional dissipation and due to the highly non-linear nature of the equations we will be solving it is preferable to choose the central difference operator.
While there are some physical cases where the other difference operators are tailored better to the problem at hand\footnote{i.e. heat equation \cite{recktenwald} and fluid dynamics \cite{CFD} or when matter is present and dominates the physics}, there is insufficient motivation at this time to use a different scheme for our purely \emph{vacuum} GR problems.  Sometimes a preferred direction of wave motion can motivate a directional scheme, or perturbation analysis will uncover that conditioning with one type of operator is better than another\footnote{Frequently called Von Neumann stability analysis.}.

The decision to use a central difference operator for spatial dimensions is also based on years of using various methods to reduce the space and memory requirements that simple, non-directional coupling demands as staggering your grid or using lower order operators can potentially decrease the processing time or memory requirements by a factor of one half or more, at the cost of precision and accuracy.

To motivate the choice of our time coordinate difference operator, there have been several studies on the efficacity of various finite difference techniques in different spacetimes.  We reference especially \cite{bernstein2}, which does a comparative analysis of FTCS (forward in time, centered in space), Lax, Lax-Wendroff, MacCormack, Brailovskaya and various Leapfrog predictor-corrector techniques in a Schwarzschild spacetime.

What we term a ``strict'' CTCS (centered in time and centered in space) schema is one that simply uses the central difference operators to define first and higher order derivatives in both spatial and temporal directions, with no interpolations, extrapolations or predictor/corrector techniques.  It is also very straightforward to program.  The spacetime variables are evaluated on the black dots in figure \ref{fig:CTCS_SLF_coords}.  For example the first spatial derivative at $(t_N,x_i)$ of a function $f(t,x)$ would be calculated to second order by
$$\frac{f(t_N,x_{i+1})-f(t_N,x_{i-1})}{2 \Delta x}$$
and similarly the discretised first time derivative is given by
$$\frac{f(t_{N+1},x_{i})-f(t_{N-1},x_{i})}{2 \Delta t}$$

The second numerical scheme we compare to is staggered leap-frog with half-step extrapolation.  This method is useful in various predictor-corrector techniques and was in the original code implementation \cite{orig_code,paul_thesis}.  After troubleshooting some issues that arose in the 2+1 code it was decided to revisit this method in 1+1 in order to track down numerical issues on a simpler system.

The staggered leap-frog (SLF) scheme we consider here is still centered in space, but the metric variables are evaluated on the integer time steps (solid dots in \ref{fig:CTCS_SLF_coords}) and the extrinsic curvature variables are evaluated on the half-steps (open dots in \ref{fig:CTCS_SLF_coords}).  We use a second order correct half-step extrapolation technique to compute the variables needed on alternating steps:
$$f_{n+\frac{1}{2}}=1.5 f_n - 0.5 f_{n-1}$$

Our 1+1 code will investigate the properties of these two numerical methods in the context of time evolution to help motivate our choice of scheme for the 2+1 code.

\subsection{The Code}
To perform a comparison of these alternatives two 1+1 numerical evolution codes were written to implement (1) ADM leap-frog and ADM CTCS and (2) ADM CTCS and BSSN CTCS evolutions in a Schwarzschild spacetime.

Just as a coordinate transformation can change the regularity 
of the simulation via the numerical coupling between variables and conditioning of the characteristic matrices\footnote{For one example of many in electrical engineering see \cite{ieeecoordtrans}}, a numerical algorithm for stepping through any variable (in this case time) can produce exactly the same effect.

\section{Testing strict CTCS and SLF}\label{sec:CTCSSLFtests}
Firstly a program was created to compute four types of evolution in the SLF and CTCS frameworks (with ADM slicing) in a 1+1 vacuum Schwarzschild spacetime\footnote{We set the mass of the Schwarzschild black hole $M=1$}.  While the Schwarzschild solution is a static solution, we can choose coordinates that freely fall into the black hole, remain static, or undergo alternative motions.  The physical spacetime does not evolve, it is the coordinate locations on the slices that we have chosen - so it is understood that when we discuss an ``evolution'' of the Schwarzschild solution that we are really looking at the progression of the coordinate points onto subsequent slices.

The four conditions we test\footnote{Details of previous work and motivation can be found in \cite{bernstein2}} with $M=1$ are:
\begin{enumerate}
\item Static spacetime slices (Schwarzschild Slicing): $\alpha=\tanh(\frac{{\eta}}{2})$, $\beta=0$.  The coordinates are static because the RHS of the time evolution equations (\ref{eqn:1DADMevol}) all vanish exactly on the initial slice.  There is no ``evolution''.
\item Geodesic slicing (free-falling coordinates):
$$\alpha = 1 \rightarrow dt=d\tau$$
As coordinate time is equal to proper time the coordinates are falling along radial geodesics.  The initial slice covers the region exterior to the black hole, so our innermost radial grid point starts at rest at $r=2M$.
The inner radial grid point falls into the black hole when we start the simulation and then hits the origin in $\tau=\pi$ units (this is an analytic result that falls out of the Schwarzschild solution \cite{MTW} pg. 824).  This is because we can parametrise the radius $\eta$ and proper time $\tau$ of the innermost coordinate grid point as follows:
$$r=(1+\cos\lambda)$$
$$\tau=(\lambda + \sin\lambda)$$
so when our parametrisation $\lambda=\pi$, $r=0$ and $\tau=\pi$
\item Maximal slicing ($\alpha$ ``evolving'' via $Tr K = 0$) - our spatial slices should ``evolve'' to a state where the coordinates are mostly static once the lapse collapses ($\alpha \rightarrow 0$) at the origin.
\item $AB^2$ slicing ($\alpha = \sqrt{ab^2}$).  As $\sqrt{ab^2}$ represents the volume element this slows down the evolution in areas of small proper volume.  This is useful for avoiding the origin as volume decreases when the grid points fall into $r=0$.  It also slows down the evolution in areas where grid points are focusing (which is unlikely to happen here).
\end{enumerate}
In all these cases the conformal factor has an analytic solution \cite{bernstein2}
$$\psi = \sqrt{2M} \cosh\left(\frac{\eta}{2}\right)$$
These simulations were run on a Linux machine running gfortran, which was built with off-the shelf components.\footnote{Intel Q9550 processor, 8GB ram, 64-bit Fedora 11} 

\subsection{Test \#1) Static spacetime slices}\label{sec:CDCSSLFstatictest}
This is a simple test to check for errors in the code, like typographical errors (human) and differencing operator errors (numerical).  As all variables are known exactly on the first time step and all the terms on the right hand side of the evolution equations should cancel to zero this simulation should, in theory, run indefinitely with no evolution.  See table \ref{tbl:schwtest1crash} for results\footnote{in the FORTRAN programming language NANQ represents ``Not A Number in $\mathbb{Q}$'', the set of rational numbers.}, which are used to give a \emph{relative} measure of how well the algorithms are performing.

\begin{table}\begin{centering}\begin{tabular}{|c|c|c|} \hline
Algorithm & \# bits for real number & \# time steps before NANQ \\ \hline
SLF & 32 & 139782 \\ \hline
SLF & 64 & 58740 \\ \hline
CTCS & 32 & 139838 \\ \hline
CTCS & 64 & 58405 \\ \hline
\end{tabular}\caption[Schwarzschild numerical test \#1]{Number of time steps before test \#1 causes NANQ (not a number) FORTRAN numerical errors}\label{tbl:schwtest1crash}\end{centering}\end{table}

These results indicate that the time stepping algorithm is not the limiting factor, rather the crashing of the code is due to accumulated errors in the spatial differences.  The errors start manifesting in the $H_a$ and $H_b$ terms first due to calculating numerical derivatives of the lapse $\alpha$, then eventually become large enough to show up in $a$ and $b$ and cause them to crash.

It is interesting to note how increasing the precision of a floating point variable from single to double precision ($32$ to $64$ bit) actually \emph{decreased} the time to crash.  So increased precision is not always better.  This was due to the fact that the accumulated error in the extrinsic curvature variables was so small that it was less than numerical precision when fed back into the metric variables for quite a while during the 32 bit evolution, whereas it gets picked up faster in the 64 bit evolution.

This also provides insight into why it is that things being ``equal to $0$'' numerically is problematic, as any amount of error will get picked up and causes infinite relative deviation from theory.

\subsection{Test \#2) Free-falling coordinates}\label{sec:CTCSSLFfreetest}
Free-falling coordinates have a theoretical time limit \cite{MTW} after which the interior most coordinate point should ``run into'' the origin ($t=\pi$)\cite{orig_code,bernstein2,MTW}.  There are various ways of checking this, but we consider $a<0$, $b<0$, or NANQ values to be ``running into'' the physical singularity at origin as they are all non-physical.  In both simulations the innermost grid point was capable of hitting the origin, but CTCS hit it within numerical accuracy, whereas SLF was incapable of hitting it within numerical accuracy.  See table \ref{tbl:schwtest2crash} for results.

\begin{table}\begin{centering}\begin{tabular}{|c|c|c|c|} \hline
Algorithm & Time to hit origin  & $\Delta t$ & Matches theory \\ \hline
SLF & 3.15888 & .0004 & no \\ \hline
CTCS & 3.14128 & .0004 & yes \\ \hline
\end{tabular}\caption[Schwarzschild numerical test \#2]{Total proper time before test \#2 causes NANQ (not a number) FORTRAN numerical errors.  Theoretical value is time=$\pi$.}\label{tbl:schwtest2crash}\end{centering}\end{table}

\subsection{Test \#3) Maximal Slicing}\label{sec:CTCSSLFmaximaltest}
Maximal Slicing is a method of ensuring that the \emph{numerical} evolution of the spacetime slows down/stops in areas of large curvature.

Furthermore, we know that our Schwarzschild radius is given by comparing (\ref{eqn:1dschwmetric}) with the standard Schwarzschild metric, i.e.:
$$r_{schw}=\sqrt{\psi^4b} = \psi^2 \sqrt{b}$$
and using the result that $r_{schw} \rightarrow \frac{3}{2}M = \frac{3}{2}$ as $t \rightarrow \infty$ \cite{estabrook} in a Schwarzschild spacetime with $M=1$, we can determine if the limiting value of our calculated $r_{schw}$ is what we expect.

The CTCS results are as seen in table \ref{tbl:schwtest3maxslice}, however a stable SLF implementation was not achieved.  The CTCS simulation ran at $\sim$900 loops per second, and ran for 66273 iterations before giving a non-physical value $\alpha<0$.  For interest the simulation was left to run and after 3 hours and 10,000,000 iterations, was still running without crashing\footnote{On the old ACS system CTCS was capable of reaching $\sim$2,651,000 time steps before finally crashing.}.  The evolution of $a$ and $b$ were frozen by $N=66273$ time steps, and $\alpha$ had collapsed to $<10^{-50}$ at the origin. ($\Delta t = .005$)

\begin{table}\begin{centering}\begin{tabular}{|c|c|c|}\hline
$\psi$ & $b$ & $r_{schw}$ \\ \hline
$1.4142135623730951$ & $0.56269516386420659$ & $1.5002601959183040$ \\ \hline
\multicolumn{3}{c}{} \\ \hline
$t$ & \# steps & $\alpha$ \\ \hline
$331.365$ &  $66273$ & $1.17497372086159688 \times  10^{-52}$ \\ \hline
$50000$ & $10^7$ & $-3.94959747091917039 \times  10^{-71}$ \\ \hline
\end{tabular}\caption[Schwarzschild numerical test \#3]{Values at the inner grid point for CTCS maximal slicing evolution before $\alpha$ goes negative, and after 10 million time steps.  A stable SLF implementation was not achieved.  Theoretical value for $r_{schw}$ is $\frac{3}{2}$.}\label{tbl:schwtest3maxslice}\end{centering}\end{table}

\subsection{Test \#4) $\alpha = \sqrt{ab^2}$}
Both CTCS and SLF ran for 3924 time steps ($t=19.62$ with $dt=0.005$) before crashing, indicating that once again the accumulated error from finite differencing to second order is the limiting factor in this test.

\subsection{SLF vs CTCS}
From the results of these four tests in 1+1 ADM, there is evidence that one should use a CTCS framework instead of SLF for this problem\footnote{SLF also encountered other strange coupling problems that were difficult to replicate.}.

\section{Tests with various coordinates systems - BSSN vs. ADM}\label{sec:bssnvsadm}
Let us now examine the BSSN versus ADM results (i.e. different formulations of the 3+1 splitting of the Einstein equations as opposed to the different finite difference schemes studied in the previous section).  We use a CTCS framework for both after the results from the previous section.

To this end a code was created to evolve the ADM and BSSN frameworks side by side (i.e. at the same time and in the same program to eliminate any discrepancies from coding) for easy comparison and we discuss some results in the following sections.

\subsection{Static space-time test}
As in section \ref{sec:CDCSSLFstatictest}, we tested to see how many iterations we could achieve in a simulation with a known, analytic, static solution before it crashed\footnote{We use 64 bit double-precision floating point variables unless otherwise stated}.  The BSSN code went 22026 time steps before the accumulated error in the extrinsic curvature variables caused the code to crash.  The ADM code ran for 100,000 iterations with no issues.

This test indicates that the ADM code is much more stable than BSSN in this configuration.

\subsection{Maximal Slicing test}
After evolving the two coordinate systems for $80,000$ time steps, we look at the Schwarzschild radius of the inner grid point as a measure of the accuracy solution (in theory it should be $\frac{3}{2}$ when $M=1$- see section \ref{sec:CTCSSLFmaximaltest}).  A comparison of the metric quantities and lapse function can be seen in table \ref{tbl:onedadmvsbssn}.

It is obvious that the solutions have diverged significantly and that the BSSN solution has experienced significant drift away from the theoretical values.  Leaving the evolution to run for more time steps does not change the results significantly as the lapse $\rightarrow 0$ at the innermost grid point, so the evolution is ``frozen'' at that point.

From this analysis it has become apparent that the BSSN code presented here would need significant work to approach the theoretical and ADM results.

\begin{table}\begin{centering}\begin{tabular}{|c|c|c|}\hline
& ADM & BSSN \\ \hline
$a$ & 3.16049383519459237 & 1.83469103161665914 \\
$b$ & 0.562500005939911008 & 0.738275626466401991 \\
$\alpha$ & 0.292535732885255978E-09 & 0.483064506323913317E-10 \\
$r_{schw}$ & 1.50000000791988186 & 1.71845934076591056 \\ \hline
\end{tabular}\caption[ADM vs. BSSN Maximal Slicing test]{A comparison of the metric and lapse variables in a maximal slicing 1-D Schwarzschild spacetime with ADM and BSSN variables, at the innermost grid point ($i=2$) after $80,000$ time steps.  $r_{schw}$ should be $1.5$ for $M=1$.}\label{tbl:onedadmvsbssn}\end{centering}\end{table}

\subsection{Free-falling coordinates}
As discussed in section \ref{sec:CTCSSLFfreetest}, we know that the inner most coordinate point should ``hit'' the origin at $t=\pi$ (for $M=1$) and cause the code to crash. We examine the actual performance of ADM and BSSN in table \ref{tbl:freefalladmvsbssn}.

\begin{table}\begin{centering}\begin{tabular}{|c|c|c|c|}\hline
Algorithm & \# iterations & time & matches theory \\ \hline
BSSN & 7859 & 3.1436 & no \\ \hline
ADM & 7854 & 3.1416 & yes \\ \hline
\end{tabular}\caption[ADM vs. BSSN Free-fall test]{A comparison of the time to have the innermost grid point hit the origin with free-fall conditions for ADM and BSSN. In theory this should be $\pi \pm \Delta t=0.0004$. }\label{tbl:freefalladmvsbssn}\end{centering}\end{table}

Recalling that this is using the exact same gridding, numerical solvers/precision, parameters, etc. BSSN is not capable of delivering the expected results within the numerical precision we require.

\section{Conclusion on Gridding and Alternate Coordinates}
Firstly, these results provided a good case for switching to a CTCS finite differencing scheme instead of using the SLF \cite{orig_code} scheme.
Furthermore, from these results, there does not seem to be a compelling reason to switch to BSSN coordinates over ADM.

While it might be possible to overcome the conformally flat metric requirement in BSSN with some reformulation, these results suggest that the effort is not warranted.

%\bibliography
\bibliographystyle{plainurl}

\begin{thebibliography}{888}

\bibitem{Abrahams} A.M. Abrahams and J.W. York Jr. \emph{3+1 General Relativity in Hyperbolic Form}. pp 179-190 in ``Relativistic Gravitation and Gravitational Radiation'' (editors: J-A Marck and J-P Lasota).  Cambridge University Press, Cambridge, UK, 1997.  Preprint: arXiv:gr-qc/9601031v1

\bibitem{Abrahams2} Andrew M. Abrahams, Karen R. Heiderich, Stuart L. Shapiro, and Saul A. Teukolsky.  \emph{Vacuum initial data, singularities, and cosmic censorship}.  Phys. Rev. D 46, 2452, 1992.  

\bibitem{ADM} R. Arnowitt, S. Deser, C. W. Misner. \emph{The Dynamics of General Relativity}. pp. 227-265 in ``Gravitation: An Introduction to Current Research'' (editor: L. Witten), John Wiley \& Sons, New York, 1962.  Reprint: arXiv:gr-qc/0405109v1

\bibitem{Alcubierre} Miguel Alcubierre \emph{et al}. \emph{Gravitational collapse of gravitational waves in 3D numerical relativity}.  Phys.Rev.D61:041501,2000, Preprint: arXiv:gr-qc/9904013v1, 1999.

%\bibitem{alcubierre:cartesian} M. Alcubierre et al., (1999), gr-qc/9904013.

\bibitem{alcubierre:3dbrill} \emph{Evolution of Brill waves in 3D progress report}, presented by Miguel Alcubierre, Max-Planck-Institut, Poster from GR15 meeting in Puna/India, available at
\verb+http://svn.cactuscode.org/arrangements/CactusArchive/cvs_cactus/+
\verb+trunk/CactusWebSite/Papers/BRILL_GR15.ps.gz+

\bibitem{alcubierre:3p1num} Miguel Alcubierre, \emph{Introduction to 3+1 Numerical Relativity}, Oxford University Press, Oxford, UK, 2008.

\bibitem{alcubierre:testbed} Miguel Alcubierre \emph{et al}. \emph{Towards standard testbeds for numerical relativity}, Classical and Quantum Gravity, vol 21 no 2, pg 589, 2004, Preprint: arXiv:gr-qc/0305023v1.

\bibitem{allen} Gabrielle D. Allen and Bernard F. Schutz. \emph{An ADI scheme for a black hole problem}. pp. 292-296 in ``Approaches to Numerical Relativity''  (Ray d'Inverno editor), Cambridge University Press, Cambridge, UK, 1992.

\bibitem{ashtekar} Abhay Ashtekar and Badri Krishnan, \emph{Isolated and Dynamical Horizons and Their Applications}, Living Rev. Relativity 7,  (2004),  10. URL (cited on Nov 2013): \verb+http://www.livingreviews.org/lrr-2004-10+

\bibitem{Baker} John Baker. \emph{Lazarus Approach to Binary Black Hole Modeling}, Building Bridges: CGWA Inauguration, 15 December 2003, via the internet, \verb+http://www.ebookpp.com/la/lazev-ppt.html+

\bibitem{Blau} Matthias Blau.  \emph{Lecture Notes on General Relativity}, Albert Einstein Center for Fundamental Physics, July 31, 2014, via the internet, \verb+http://www.blau.itp.unibe.ch/newlecturesGR.pdf+

\bibitem{Brown}David Brown and Lisa Lowe. \emph{AMRMG (Adaptive Mesh Refinement MultiGrid Code)}, NC State University, Penn State Numerical Relativity Lunch Talk, February 21, 2003, via the internet
\verb+http://www.astro.psu.edu/nr/nrlunch/2003/2003_02_21_Brown/+
\verb+Transparencies.pdf+

\bibitem{bardeen} James M. Bardeen and Tsvi Piran.  \emph{General Relativistic Axisymmetric Rotating Systems: Coordinates and Equations}.  Physics Reports 96, No. 4, pp 205-250, 1983.

\bibitem{bernstein} D.H. Bernstein.  \emph{A Numerical Study of the Black Hole Plus Brill Wave Spacetime}, Doctoral Thesis, University of Illinois at Urbana-Champaign, 1993.

\bibitem{bernstein2}  Berstein, Hobill and Smarr.  \emph{Black Hole Spacetimes: Testing Numerical Relativity}. pp 57-73 in ``Frontiers in Numerical Relativity'', Cambridge University Press, Cambridge, UK, 1989.

\bibitem{BHSS} David Berstein, David Hobill, Edward Seidel, Larry Smarr. \emph{Initial data for the black hole plus Brill wave spacetime}.  Phys Rev. D, Vol 50, No. 6, 1994.

\bibitem{ABHSS} Abrahams, Bernstein, Hobill, Seidel and Smarr.  \emph{Numerically generated axisymmetric black-hole spacetimes: Interaction with gravitational waves}.  Phys. Rev. D 45, 3544–3558, 1992.

\bibitem{BHSST} Bernstein, Hobill, Seidel, Smarr and Towns.  \emph{Numerically generated axisymmetric black hole spacetimes: Numerical methods and code test}. Phys. Rev. D 50, 5000–5024, 1994.

\bibitem{bernuzzi} Sebastiano Bernuzzi and David Hilditch.  \emph{ Constraint violation in free evolution schemes: comparing BSSNOK with a conformal decomposition of Z4}, Phys.Rev.D81:084003,2010, Preprint: arXiv:0912.2920v2 [gr-qc], 2010.

\bibitem{Brill} D.R. Brill. \emph{On the positive definite mass of the Bondi-Weber-Wheeler time-symmetric gravitational waves.} pp 466-483 in ``Annals of Physics'' Volume 7, Issue 4 (August 1959)

\bibitem{BSSN} Thomas Baumgarte and Stuart Shapiro. \emph{On the Numerical Integration of Einstein's Field Equations}, Phys.Rev. D59 (1999) 024007. Preprint: arXiv:gr-qc/9810065v1, 1998.

\bibitem{BSSN2} Yoneda, Gen and Shinkai, Hisa-aki. \emph{Advantages of modified ADM formulation: constraint and propogation analysis of Baumgarte-Shapiro-Shibata-Nakamura system}, Phys.Rev. D66 (2002) 124003, Preprint: arXiv:gr-qc/0204002v3.

\bibitem{choptuik1} Matthew Choptuik, Eric Hirschmann , Steven Liebling and Frans Pretorius. \emph{An axisymmetric gravitational collapse code}, Classical Quantum Gravity 20 (2003) 1857-1878, Preprint: arXiv:qr-qc/0301006v1.

\bibitem{choptuik2} Matthew Choptuik, Eric Hirschmann, Steven Liebling and Frans Pretorius.  \emph{Critical Collapse of the Massless Scalar Field in Axisymmetry}. Phys.Rev. D68 (2003) 044007 Preprint: arXiv:gr-qc/0305003v1, 2003.

\bibitem{CK-Minkowski-stability} Demetrios Christodoulou and Sergiu Klainerman.
\emph{The Global Nonlinear Stability of the Minkowski Space}, Princeton University Press, Princeton, NJ, 1994.

\bibitem{cook} Gregory B. Cook.  \emph{Initial Data for Numerical Relativity} Living Rev. Relativity 3,  (2000), 5.  \verb+http://www.livingreviews.org/lrr-2000-5 +

\bibitem{cookphd} Gregory B. Cook.  \emph{Initial Data for the Two-Body Problem of General Relativity}, PhD Thesis, Chapel Hill, 1990. Available via the web at \verb+http://users.wfu.edu/cookgb/Thesis1side.pdf+

\bibitem{critical} Gundlach, Carsten.  \emph{Critical Phenomena in Gravitational Collapse}, Living Reviews in Relativity, 1999.
\verb+http://www.livingreviews.com/Articles/Volume2/1999-4gundlach+

\bibitem{d'inverno} Ray D'Inverno. \emph{Introducing Einstein's Relativity},  Oxford University Press, New York, 1996.

\bibitem{deadman}  Edvin Deadman. \emph{Outer Boundary Conditions in Numerical Relativity}. 2010.92, PhD thesis, Cambridge University. \verb+http://eprints.ma.man.ac.uk/1534/+

\bibitem{eppley}Kenneth Eppley. \emph{Pure Gravitational Waves in Sources of Gravitational Radiation}. pp 275-291 (Editor: Larry Smarr), Cambridge University Press, Cambridge, UK, 1979.

\bibitem{etoolkit} Frank L\"{o}ffler, et al.  \emph{The Einstein Toolkit: A Community Computational Infrastructure for Relativistic Astrophysics}, arXiv:1111.3344v1 [gr-qc], 2011

\bibitem{estabrook} Frank Estabrook and Hugo Wahlquist. \emph{Maximally Slicing a Black Hole}, Phys. Rev. D 7, 2814–2817 (1973).

\bibitem{evans} C. Evans. \emph{A Method For Numerical Relativity: Simulation of Axisymmetric Gravitational Collapse and Gravitational Radiation Generation}, Doctoral Thesis, University of Texas at Austin, 1984.

\bibitem{ARH}C. Evans, L Smarr and J. Wilson.  \emph{Numerical Relativistic Gravitational Collapse With Spatial Time Slices} pp. 491-525 in ``Astrophysical Radiation Hydronamics'', D. Reidel Publishng Company, 1986.

\bibitem{friedrich} Friedrich, Helmut. \emph{On the hyperbolicity of Einstein's and other gauge field equations}, Comm. Math. Phys. Volume 100, Number 4 (1985), 525-543.

\bibitem{frontiers}C. Evans, L. Finn and D. Hobill. \emph{Frontiers in Numerical Relativity}, Cambridge University Press, Cambridge, UK, 1989.

\bibitem{garfinkle}Garfinkle, David and Duncan, G. Comer.  \emph{Numerical Evolution of Brill Waves}, Phys.Rev. D63 (2001) 044011, Preprint: arXiv:gr-qc/0006073, 2000.

\bibitem{gentle}Gentle, Adrian P.  \emph{Simplicial Brill wave initial data}, Classical and Quantum Gravity vol. 16, no. 6, pg 1987, 1999.  Preprint: gr-qc/9901071v1, 1999.

\bibitem{gentle2}A. Gentle, D. Holz, W. Miller and J. Wheeler. \emph{Apparent horizons in simplicial Brill wave initial data}, Classical and Quantum Gravity vol. 16, no. 6, pg 1979, 1999.  Preprint: gr-qc/9812057v1, 1998.

\bibitem{hawking}S. Hawking and G. Ellis. \emph{The Large Scale Structure of Space-Time}, Cambridge University Press, Cambridge, UK, 1991.

\bibitem{hawkingmass} S. Hawking. \emph{Gravitational Radiation in an Expanding Universe}  J. Math. Phys. 9, 598 (1968)

\bibitem{harmonic:var} Béla Szilágyi, Denis Pollney, Luciano Rezzolla, Jonathan Thornburg, and Jeffrey Winicour.  \emph{An explicit harmonic code for black-hole evolution using excision}.
Classical and Quantum Gravity 24(12), S275-S293 [2007 June 21]  Preprint: arXiv:gr-qc/0612150

\bibitem{hobill:slicing} D. Hobill. \emph{Computational Methods for Vacuum Spacetimes}. pp 98-112 in ``Gravitation: A Banff Summer Institute'', World Scientific, Singapore, 1991.

\bibitem{HRC}Ryan, Michael P. Jr. and Shepley, Lawrence C.  \emph{Homogeneous Relativistic Cosmologies}, Princeton University Press, Princeton, NJ, 1975.

\bibitem{jackson_em} Jackson, John D. \emph{Classical Electrodynamics (3rd ed.)}, Wiley, New York, 1998.

\bibitem{kar_raych} Sayan Kar, and Soumitra SenGupta. \emph{The Raychaudhuri equations: a brief review}, Indian Academy of Sciences, vol. 69, no. 1, pp 49-76, 2007.  Preprint: arXiv:gr-qc/0611123v1, 2006.

\bibitem{Lehner} Luis Lehner.  \emph{Numerical Relativity: A review}, Classical Quantum Gravity 18:R25-R86,2001. Preprint: arXiv:gr-qc/0106072v3, 2001.

\bibitem{Lichnerowitz} Lichnerowitz, A. \emph{L'integration des $\acute{e}$quations de la gravitation relativiste et le probl$\grave{e}$me des n corps}, J. Math. Pures Appl.,23, 37-63, (1944)

\bibitem{lindblad-minkowski-stability} H. Lindblad, I. Rodnianski. \emph{The global stability of the Minkowski space-time in harmonic gauge}, Annals of Mathematics, vol. 171, no. 3, pp1401-1477, 2010.  Preprint: arXiv:math/0411109v2, 2010.

\bibitem{Lousto} Carlos Lousto, et al.  \emph{Full Numerical Simulation of Black Holes}, Presentation at Madeira, Sep 2, 2011, via the internet: \verb+http://blackholes.ist.utl.pt/nrhep/Lousto.pdf+

\bibitem{lindblom:harmonic} Lee Lindblom, Mark A Scheel, Lawrence E Kidder, Robert Owen and Oliver Rinne.  \emph{A New Generalized Harmonic Evolution System}.  Classical Quantum Gravity vol. 23, no. 16, S447, 2006.

\bibitem{marion_em} Mark Heald and Jerry Marion. \emph{Classical Electromagnetic Radiation, 3 ed.}, Saunders College Publishing, Pacific Grove, CA, 1995.

\bibitem{mastersonmsc}Masterson, Andrew.  \emph{The Brill Gravitational Wave Initial Value Problem}.  Master's Thesis, University of Calgary, 2002.

\bibitem{matzner} R.A. Matzner and L.C. Shepley. \emph{Spacetime and Geometry}, University of Texas Press, Austin, 1982.

\bibitem{mcallister} McAllister, Howard C. \emph{Roots of the Cubic Equation}, 1997 (MacOS), via the internet: \verb+http://www.hawaii.edu/suremath/jrootsCubic.html+

\bibitem{millman} R.S. Millman. and G.D. Parker. \emph{Elements of Differential Geometry}, Prentice-Hall Inc., Upper Saddle River, New Jersey, 1977.

\bibitem{miyama} Miyama, Shoken M.  \emph{Time Evolution of Pure Gravitational Waves}, Progress of Theoretical Physics, Vol. 65, No. 3, March 1981, pp 894-909.

\bibitem{MTW} C. Misner, K. Thorne, J. Wheeler. \emph{Gravitation}, W.H. Freeman and Co., San Francisco, CA, 1974.

\bibitem{Murchadha} Niall \'O Murchadha, \emph{Brill Waves}, in ``Directions in General Relativity'', 1993.  Preprint: arXiv:gr-qc/9302023v1

\bibitem{Murchadha_trap} R. Beig and N. \'O Murchadha, \emph{Trapped Surfaces in Vacuum Spacetimes}, Classical Quantum Gravity 11:419-430, 1994. Preprint: arXiv:gr-qc/9304034v1

\bibitem{Murchadha_trap2} R. Beig and N. \'O Murchadha, \emph{Vacuum Spacetimes with Future Trapped Surfaces}, Classsical Quantum Gravity 13 (1996) 739-752. Preprint: arXiv:gr-qc/9511070v1

\bibitem{newmanpenrose} Newman, E.T. and R. Penrose. \emph{An Approach to Gravitational Radiation by the Method of Spin Coefficients}, Journal of Mathematical Physics 3, pg 566-578, 1962.

\bibitem{newmantod} E.T. Newman and K.P. Tod.  \emph{Asymptotically Flat Space-Times}, from ``General Relativity and Gravitation'', Vol. 2, ed. Held, Plenum Press, NY, 1980.

\bibitem{NINJA} \emph{Testing gravitational-wave searches with numerical relativity waveforms: Results from the first Numerical INJection Analysis (NINJA) project}, Classical and Quantum Gravity, vol. 26, no. 16, id. 165008 (2009).  Preprint: arXiv:0901.4399v2 [gr-qc]

\bibitem{oliveira}H. P. de Oliveira and E. L. Rodrigues. \emph{Brill wave initial data: Using the Galerkin-collocation method}, Phys. Rev. D 86, 064007, 2012.

\bibitem{orig_code} P. Anninos, D. Bernstein, D. Hobill, E. Seidel, L. Smarr and J. Towns, \emph{NCSA}, 1993, 1994, original evolution code.

\bibitem{paul_thesis} Webster, Paul S. \emph{Black Holes and Radiative Fields in General Relativity}, Doctoral Thesis, University of Calgary, 1999.

\bibitem{penrosemass} R. Penrose.  \emph{Quasi-Local Mass and Angular Momentum in General Relativity},  Proc. R. Soc. Lond. A 8 May 1982 vol. 381 no. 1780 53-63.

\bibitem{pfeifferIVP} Harald P. Pfeiffer \emph{et al}.  \emph{Initial data for Einstein's equations with superposed gravitational waves.}  Phys Rev D, 71, 024020 (2005)

\bibitem{physrevd4510} Andrew Abrahams, David Bernstein, David Hobill, Edward Seidel and Larry Smarr.  \emph{Numerically generated black-hole spacetimes: Interaction with gravitational waves}, Phys Rev. D 45 No. 10, 3544-3558 (May 1992).

\bibitem{physrevd5006a} David Bernstein, David Hobill, Edward Seidel and Larry Smarr.  \emph{Initial data for the black hole plus Brill wave spacetime}, Phys Rev. D 50 No. 06, 3760-3782 (Sep 1994).

\bibitem{physrevd5006b} Peter Anninos, David Bernstein, Steven R. Brandt, David Hobill, Edward Seidel and Larry Smarr.  \emph{Dynamics of black hole apparent horizons}, Phys Rev. D 50 No. 06, 3801-3815 (Sep 1994).

\bibitem{physrevd5008} David Bernstein, David Hobill, Edward Seidel, Larry Smarr and John Towns.  \emph{Numerically generated axisymmetric black hole spacetimes: Numerical methods and code tests}, Phys Rev. D 50 No. 8, 5000-5024 (Oct 1994).

\bibitem{Pollney} Pollney D et al. \emph{Recoil velocities from equal-mass binary black-hole mergers: a systematic investigation of spin-orbit aligned configurations}, 2007 Phys. Rev. D 76 124002. Preprint: arXiv:0707.2559v1 [gr-qc]

\bibitem{raychaudhuri2003} A.K. Raychaudhuri, S. Banerji and A. Banerjee. \emph{General Relativity, Astrophysics, and Cosmology}, Springer, NY, 2003.

\bibitem{richter} Ronny Richter and Christian Lubich. \emph{Free and constrained symplectic integrators for numerical general relativity}, Classical Quantum Gravity 25:225018, 2008. Preprint: arXiv:0807.0734v2 [gr-qc]

\bibitem{RinnePHD} Rinne, Oliver. \emph{Axisymmetric Numerical Relativity}, Doctoral Thesis, Trinity College, Cambridge, 13 September 2005.

\bibitem{Rinne} Rinne, Oliver.  \emph{Gravitational collapse of prolate Brill waves}, Spanish Relativity Meeting, University of Salamanca, 16 September 2008, via the internet.
\verb+http://www.usal.es/ere2008/website/modules/tinyd0/content/pdf/+
\verb+Rinne-1.pdf+

\bibitem{RinneCQG} Rinne, Oliver.  \emph{Constrained evolution in axisymmetry and the gravitational collapse of prolate Brill waves}, Classical and Quantum Gravity, 25 (13). Art. No. 135009, Preprint: arXiv:0802.3791 [gr-qc]

\bibitem{shahar} Shahar Hod and Tsvi Piran.  \emph{Critical Behaviour and Universality in Gravitational Collapse of a Charged Scalar Field}. Phys. Rev. D 55, 3485–3496 (1997) 

\bibitem{Smarr} L. Smarr and J.W. York. \emph{Kinematical conditions in the construction of spacetime}, Phys. Rev. D Vol.17, Issue 10, pp 2529-2551, 1978.

\bibitem{sorkin}  Evgeny Sorkin.  \emph{On critical collapse of gravitational waves}, Classical Quantum Gravity 28:025011, 2011. Preprint: arXiv:1008.3319v2 [gr-qc]

\bibitem{sorkin:code} Evgeny Sorkin.  \emph{An axisymmetric generalized harmonic evolution code}, Phys. Rev. D81:084062, 2010. Preprint: arXiv:0911.2011v2 [gr-qc]

\bibitem{Thornburg:AH} Jonathan Thornburg.  \emph{Event and Apparent Horizon Finders for 3+1 Numerical Relativity}, Living Rev. Relativity, 10, 2007, 3. \verb+http://relativity.livingreviews.org/Articles/lrr-2007-3+

\bibitem{Thornburg:patches} Jonathan Thornburg. \emph{Coordinates and Boundary Conditions for the General Relativistic Initial Data Problem}.  Classical and Quantum Gravity 4(5), 1119-1131 [Sep.1987]

\bibitem{Thornburg:cartesian} Miguel Alcubierre, Steve Brandt, Bernd Bruegmann, Daniel Holz, Ed Seidel, Ryoji Takahashi, and Jonathan Thornburg.  \emph{Symmetry without Symmetry: Numerical Simulation of Axisymmetric Systems using Cartesian Grids}, International Journal of Modern Physics D 10(3) [June 2001], 273-289.  Preprint: arXiv:gr-qc/9908012

\bibitem{Thornburg:sixangles} Christian Reisswig, Nigel T. Bishop, Chi Wai Lai, Jonathan Thornburg, and Béla Szilágyi.  \emph{Numerical Relativity with Characteristic Evolution, Using Six Angular Patches}, Classical and Quantum Gravity 24(12), S327-S339 [2007 June 21].  Preprint: arXiv:gr-qc/0610019

\bibitem{VandenBergh:2003yd} Van den Bergh, Norbert. \emph{Tidal effects cannot be absent in a vacuum: Letter to the editor}, Classical Quantum Gravity, 20, L165-L168, 2003. Preprint: arXiv:gr-qc/0303056.

\bibitem{Wald} Robert M. Wald.  \emph{General Relativity}. University of Chicago Press, Chicago, 1984.

\bibitem{walsh} D. M. Walsh.  \emph{Non-uniqueness in conformal formulations of the Einstein constraints}, Class.Quant.Grav.24:1911-1926,2007, Preprint: arXiv:gr-qc/0610129v2

\bibitem{York} J.W. York. \emph{Kinematics and Dynamics of General Relativity}, Phys. Rev. Letters 26, p 1656-1658 (1971)

\bibitem{York2} J.W. York. \emph{Role of Conformal Three-Geometry in the Dynamics of Gravitation}, Phys. Rev. Letters 28, p 1082-1085 (1972)

\bibitem{Zakhary_elecweyl} Zakhary, E. and Carminati, J. \emph{On purely gravito-magnetic vacuum space-times}, General Relativity and Gravitation, vol. 37 no. 3, pp. 605-613, 2005.

\bibitem{Zhang} Xiao Zhang. \emph{On the relation between ADM and Bondi energy-momenta}, Adv.Theor.Math.Phys.10:261-282,2006. Preprint: arXiv:gr-qc/0511036v4

\bibitem{jdnorton} John D. Norton.  \emph{Spaces of Variable Curvature}, lecture notes from the internet
\verb+http://www.pitt.edu/~jdnorton/teaching/HPS_0410/chapters/+
\verb+non_Euclid_variable/index.html+

\bibitem{anton} Howard Anton.  \emph{Calculus, Brief Edition, 6th ed.} John Wiley \& Sons, New York, 1990.

\bibitem{mathphys} Arfken, George and Weber, Hans.  \emph{Mathematical Methods for Physicists 4th ed.}, Academic Press, 1995.

\bibitem{bergamaschi} Luca Bergamaschi.  \emph{Iterative Methods for Sparse Linear Systems}, Lecture Notes, University of Pandova, available at
\verb+http://www.dmsa.unipd.it/~berga/2ndweek.pdf+

\bibitem{burden} Richard Burden and J. Douglas Faires.  \emph{Numerical Analysis, 6th Edition}.  Brooks/Cole Publishing, Pacific Grove, CA, 1997.

\bibitem{dijkstra} W. Dijkstra, R.M.M. Mattheij.  \emph{The condition number of the BEM-matrix arising from Laplace's equation}, Electronic Journal of Boundary Elements, Vol 4, No 2 (2006).  Available at \href{http://www.win.tue.nl/analysis/reports/rana06-13.pdf}{http://www.win.tue.nl/analysis/reports/rana06-13.pdf}.

\bibitem{F77} Larry Nyhoff and Sanford Leestma.  \emph{Fortran 77 for Engineers and Scientists 3rd ed.}, Maxwell Macmillan Canada, Toronto, 1988.

\bibitem{golub} Gene Golub and Charles Loan. \emph{Matrix Computations, 3rd Edition}. John Hopkins University Press, Baltimore, 1996.

\bibitem{goldberg} David Goldberg.  \emph{What Every Computer Scientist Should Know About Floating-Point Arithmetic}, ACM Computing Surveys, Volume 23 Issue 1, Pages 5-48, 1991.

\bibitem{press} Press, Flannery, Teukolsky and Vetterling.  \emph{Numerical Recipes in Fortran: The Art of Scientific Computing}, Cambridge University Press, Cambridge, UK, 1992.

\bibitem{recktenwald} Gerald W. Recktenwald.  \emph{Finite-Difference Approximations to the Heat Equation}.  From the internet
\verb+http://www.f.kth.se/~jjalap/numme/FDheat.pdf+

\bibitem{reula} Oscar A. Reula.  \emph{Hyperbolic Methods for Einstein's Equations}, Living Reviews in Relativity 1 (1998), 3, \verb+http://www.livingreviews.org/lrr-1998-3+

\bibitem{ross_ode} Shepley Ross. \emph{Introduction to Ordinary Differential Equations 4th ed.}, John Wiley and Sons, New York, 1989.

\bibitem{saad_sparse} Yousef Saad.  \emph{Iterative Methods for Sparse Linear Systems, 2nd ed.}, SIAM, 2003.  Available at
\verb+http://www-users.cs.umn.edu/~saad/IterMethBook_2ndEd.pdf+

\bibitem{bicgstab} H. A. van der Vorst. \emph{Bi-CGSTAB: A Fast and Smoothly Converging Variant of Bi-CG for the Solution of Nonsymmetric Linear Systems}, SIAM J. Sci. and Stat. Comput. 13-2 , pp. 631-644, 1992.

\bibitem{CFD} J.H Ferziger and M Peric.  \emph{Computational Methods for Fluid Dynamics}, Springer, 2002.

\bibitem{ieeecoordtrans} L. Miguel Silveira, et al.  \emph{A coordinate-transformed Arnoldi algorithm for generating guaranteed stable reduced-order models of RLC circuits}.  IEEE/ACM International Conference on Computer-Aided Design, ICCAD-96. Digest of Technical Papers, pp 288-294, 1996.

\end{thebibliography}

\end{document}